\documentclass[BCOR=5mm,DIV=13,chapterprefix = true,appendixprefix = true]{scrbook}

\usepackage[export]{adjustbox} 
\usepackage[UKenglish,ngerman]{babel} 
\usepackage[bf]{caption}
\usepackage[T1]{fontenc} 
\usepackage[utf8]{inputenc} 
\usepackage[nottoc,numbib]{tocbibind}
\usepackage[table]{xcolor}
\usepackage{longtable} 
\usepackage{amsfonts, 
            amsmath, 
            amssymb, 
            amsthm, 
            arydshln, 
            bbding, 
            booktabs, 
            calligra, 
            calc, 
            cite, 
            dsfont, 
            empheq, 
            etex, 
            fancyvrb, 
            float, 
            lastpage, 
            listings, 
            mathrsfs, 
            microtype, 
            multirow, 
            pbox, 
            scrhack, 
            scrlayer-scrpage, 
            setspace, 
            siunitx, 
            subfig, 
            textcomp, 
}

\usepackage{tikz} 
\usetikzlibrary{arrows,matrix,calc,plotmarks,fit,decorations.markings, decorations.pathmorphing, decorations.pathreplacing} 

\tikzset{
particle/.style={draw=black, postaction={decorate},decoration={markings,mark=at position .5 with {\arrow[black]{triangle 45}}}},
photon/.style={draw=black, decorate, decoration={snake}}
}

\usepackage{pgfplots}
\pgfplotsset{compat=1.9}

\tikzstyle{vecArrow} = [thick, decoration={markings,mark=at position 
   1 with {\arrow[semithick]{open triangle 60}}},
   double distance=1.4pt, shorten >= 5.5pt,
   preaction = {decorate},
   postaction = {draw,line width=1.4pt, white,shorten >= 4.5pt}]

\usepackage[pagebackref=true]{hyperref} 
\usepackage{cleveref} 

\hypersetup{
    pdftitle={Cohomologies of coherent sheaves and massless spectra in F-theory},
    pdfauthor={Martin Bies},
    pdfsubject={F-Theory, Massless Spectra, Coherent Sheaves, Sheaf Cohomology}
}

\definecolor{promptColor}{rgb}{0.0,0.0,0.589}
\definecolor{brkpromptColor}{rgb}{0.589,0.0,0.0}
\definecolor{gapinputColor}{rgb}{0.589,0.0,0.0}
\definecolor{gapoutputColor}{rgb}{0.0,0.0,0.0}
\definecolor{darkgreen}{rgb}{0.05,0.6,0.1}

\renewcommand{\tilde}{\widetilde}
\renewcommand{\hat}{\widehat}

\newcommand{\vect}[1]{\boldsymbol{#1}}
\newcommand{\Mint}{\int \limits}
\newcommand{\ie}{i.e.\ }
\newcommand{\eg}{e.g.\ }
\newcommand{\fp}{f.p.\ }
\newcommand{\cf}{c.f.\ }
\newcommand{\cE}{\mathcal{E}}
\newcommand*\widefbox[1]{\fbox{\hspace{2em}#1\hspace{2em}}}

\DeclareMathOperator{\Hom}{\mathscr{H}\text{\kern -3pt {\calligra\large om}}\,}

\newcommand{\xtwoheadrightarrow}[2][]{%
  \xrightarrow[#1]{#2}\mathrel{\mkern-14mu}\rightarrow
}

\renewcommand{\[}{\begin{equation}}
\renewcommand{\]}{\end{equation}}
\newcommand{\ebox}[1]{
\begin{center}
\fbox{\begin{minipage}[c]{0.9 \textwidth} #1 \end{minipage}}
\end{center}
}

\makeatletter
\def\enumfix{%
\if@inlabel
 \noindent \par\nobreak\vskip-\topsep\hrule\@height\z@
\fi}
\let\olditemize\itemize
\def\itemize{\enumfix\olditemize}
\makeatother

\makeatletter
\def\enumfix{%
\if@inlabel
 \noindent \par\nobreak\vskip-\topsep\hrule\@height\z@
\fi}
\let\oldenumerate\enumerate
\def\enumerate{\enumfix\oldenumerate}
\makeatother

\newtheoremstyle{break}  
  {\topsep}   
  {\topsep}   
  {}  
  {0pt}       
  {\bfseries} 
  {:}         
  {\newline}  
  {}          

\theoremstyle{break}

\newtheorem*{remark}{Remark}

\newtheorem*{note}{Note}

\newtheorem{defi}{Definition}[section]
\newtheorem{examp}{Example}[section]

\newtheorem{theorem}{Theorem}[section]
\newtheorem{lemma}{Lemma}[section]
\newtheorem{cor}{Corollary}[section]

\newtheorem{conj}{Conjecture}[section]
\newtheorem*{construc}{Construction}
\newtheorem*{conseq}{Consequence}

\newtheorem{proposition}[theorem]{Proposition}

\makeatletter
\let\@addpunct\@gobble
\makeatother
\newenvironment{myproof}[1][\bfseries \proofname]{%
  \begin{proof}[#1]$ $\par\nobreak\ignorespaces
}{%
  \end{proof}
}

\lstnewenvironment{mat}
{\lstset{language=mathematica,mathescape,columns=flexible}}
{}

\sbox0{\small\ttfamily A}
\edef\mybasewidth{\the\wd0 }
\lstnewenvironment{gap}
{\lstset{language=GAP,
	frame=single,
	basicstyle=\ttfamily\small,
	numbers=left,
	numberstyle=\small,
	keywordstyle=\color{red},
	stringstyle=\color{blue},
	commentstyle=\color{green!70!black},
	columns=fullflexible,
	basewidth=\mybasewidth,
	keepspaces=true,
	stepnumber=1,
	showstringspaces=false,
	breaklines=true}}
{}

\newenvironment{gapConsole}{\VerbatimEnvironment \begin{Verbatim}[commandchars=!@&,fontsize=\small,frame=single,xleftmargin = -.3em, xrightmargin=-0.3em,label=Example]}{\end{Verbatim}}

\clearpairofpagestyles
\rehead[]{\rightmark}        
\lohead[]{\leftmark}        
\lehead[]{}         
\rohead[]{}         

\KOMAoptions{headsepline = true}

\cefoot[\pagemark / \pageref*{LastPage}]{\pagemark / \pageref*{LastPage}}     
\cofoot[\pagemark / \pageref*{LastPage}]{\pagemark / \pageref*{LastPage}}     

\renewcommand*{\backref}[1]{}
\renewcommand*{\backrefalt}[4]{({%
    \ifcase #1 Not cited.%
          \or Page~#2.%
          \else Pages #2.%
    \fi%
    })}

\begin{document}


\pagenumbering{roman}
\thispagestyle{empty}

\begin{center}

{\normalsize  \setstretch{1.5} \textbf{
Dissertation \\
submitted to the \\
Combined Faculties of the Natural Sciences and Mathematics \\
of the Ruperto-Carola-University of Heidelberg, Germany \\
for the degree of \\
Doctor of Natural Sciences \\
}}

\vfill

{\normalsize \setstretch{1.5}
\textbf{Put forward by} \\
\textit{Martin Bies \\
born in: Merzig, Germany \\
Oral examination: 1 February 2018} \\
}

\end{center}

\newpage
\thispagestyle{empty}
\null

\newpage
\thispagestyle{empty}

\begin{center}
 
{\Huge \setstretch{1.5} \textbf{
Cohomologies of coherent sheaves and \\ 
massless spectra in F-theory \\
}}

\vfill

{\setstretch{2}
\textit{ Referees: \\ }
\textsc{
Prof. Dr. Timo Weigand \\
Prof. Dr. Johannes Walcher} \\
}

\end{center}

\newpage
\thispagestyle{empty}
\begin{center}
  
  \begin{minipage}[c][0.48\textheight][t]{\textwidth}

    \selectlanguage{UKenglish}
    \textbf{Cohomologies of coherent sheaves and massless spectra in F-Theory}
    
    In this PhD thesis we investigate the significance of Chow groups for zero mode counting and anomaly cancellation in F-theory vacua.
    
    The major part of this thesis focuses on zero mode counting. We explain that elements of Chow group describe a subset of gauge backgrounds and give rise to a line bundle on each matter curve. The sheaf cohomologies of these line bundles are found to encode the chiral and anti-chiral localised zero modes in this compactification. Therefore, it is of prime interest to compute these sheaf cohomologies. Unfortunately, the line bundles in question are in general non-pullback line bundles. In particular, this is the case for the hypercharge flux employed in F-theory models of \emph{grand unified theories} (GUTs). Consequently, existing methods, such as the \emph{cohomCalg}-algorithm, cannot be applied. In collaboration with the mathematician Mohamed Barakat, we have therefore implemented algorithms which determine the sheaf cohomologies of all coherent sheaves on toric varieties. These algorithms are provided by the \texttt{gap}-package \texttt{SheafCohomologiesOnToricVarieties} which extends the \emph{homalg\_project} of Mohamed Barakat. We exemplify these algorithms in explicit (toy-)models of F-theory GUTs.

    As a spin-off of this analysis, we proved that in an entire class of F-theory vacua, the \emph{matter surface fluxes} satisfy a number of relations in the Chow ring, which we related to anomaly cancellation. Based on this evidence we conjecture that the well-known anomaly cancellation conditions in F-theory -- typically phrased as intersections in the cohomology ring -- can be extended even to relations in the Chow ring.
    
  \end{minipage}\par

  \vfill

  \begin{minipage}[c][0.48\textheight][c]{\textwidth}

    \selectlanguage{ngerman}
    \textbf{Kohomologien kohärenter Garben und masselose Spektren in F-Theorie}
    
    In dieser Doktorarbeit untersuchen wir die Signifikanz von Chow-Gruppen für das Zählen von Nullmoden sowie für Anomaliekürzungen in F-Theorie-Vakua.

    Im Großteil dieser Arbeit konzentieren wird uns auf das Zählen von Nullmoden. Wir erklären, dass Elemente der Chow-Gruppe eine Teilmenge der Eichhintergründe beschreiben und auf jeder Materiekurve ein Geradenbündel induzieren. Die Garbenkohomologien dieser Geradenbündel enkodieren die chiralen und anti-chiralen lokalisierten Nullmoden dieser Kompaktifizierung. Folglich ist es von zentralem Interesse diese Garbenkohomologien zu berechnen. Unglücklicherweise sind die zu betrachtenden Geradenbündel im Allgemeinen keine Rückzugsgeradenbündel. Dies trifft insbesondere auf den Hyperladungsfluss zu, welcher in F-Theorie-Modelle von \emph{großen vereinheitlichten Theorien} (GUT) verwendet wird. Daher konnten existierende Methoden, wie der \emph{cohomCalg}-Algorithmus, nicht angewendet werden. In Zusammenarbeit mit dem Mathematiker Mohamed Barakat haben wir daher Algorithmen implementiert, welche die Garbenkohomologien aller kohärenten Garben auf torischen Varietäten berechnen. Diese Algorithmen werden durch das \texttt{gap}-Paket \texttt{SheafCohomologiesOnToricVarieties} bereitgestellt, welches das \emph{homalg\_project} von Mohamed Barakat ergänzt. Beispielhaft wenden wir diese Algorithmen in expliziten F-Theorie-GUT-Modellen an.

    Als Nebenprodukt dieser Analyse, konnten wir zeigen, dass in einer gesamten Klasse von F-Theorie-Vakua die Materieoberflächenflüsse einige Relationen im Chow-Ring erfüllen, welche wir mit Anomaliekürzungen in Beziehung setzen konnten. Basierend auf dieser Beobachtung vermuten wir, dass die wohlbekannten Anomaliekürzungsbedingungen in F-Theorie -- normalerweise durch Schnitte im Kohomologiering ausgedrückt -- sogar zu Bedingungen im Chow-Ring erweitert werden können.
    
  \end{minipage}

\end{center}

\newpage
\pagestyle{empty}

\phantom{something}
\vspace{17em}

\begin{flushright}
\LARGE{\emph{If you are receptive and humble, \\
mathematics will lead you by the hand.}} \\
\vspace{1em}
\normalsize{Paul Dirac}
\end{flushright}

\newpage
\pagestyle{empty}
\selectlanguage{UKenglish}
\setcounter{tocdepth}{2} 
\addtocontents{toc}{\protect\thispagestyle{empty}}
\tableofcontents
\cleardoublepage

\pagenumbering{arabic}
\setcounter{page}{1}
\allowdisplaybreaks
\pagestyle{scrheadings}


\chapter{Motivation}

\paragraph{From Isaac Newton to the \emph{Standard Model} of Particle Physics}
In 1687 Isaac Newton published his book \emph{Philosophiae Naturalis Principia Mathematica Mathematica} \cite{newton1687philosophiae}. In this work Newton laid the foundations of classical mechanics in formulating his \emph{laws of motion}. In addition, he proposed his \emph{law of universal gravitation} by which he derived \emph{Kepler's laws of planetary motion} \cite{kepler2015astronomia}. He could even explain, to good approximation, the trajectories of celestial objects in general, the tides of the oceans, the precession of the equinoxes and many other phenomena from these foundational works on classical mechanics and gravity. Hence, Newton found a unified mathematical description for all of these phenomena. Likewise, James Clark Maxwell observed in \cite{1865RSPT..155..459C} that electric and magnetic phenomena allow a unified description in the theory of electromagnetism.

Ever since, physicists have followed the example of Newton and Maxwell, to find a unified description of many different natural phenomena. To date, it is commonly believed that our physical reality can, to very good approximation, be explained by combined effects of four \emph{fundamental forces}. Newton's \emph{law of universal gravitation} has been replaced by Albert Einstein's theory of \emph{general relativity} \cite{einstein1916grundlage}. 

Around 1900, Max Planck's work on black-body radiation triggered the quantum revolution in physics \cite{2058-7058-13-12-34}. The outreach of this work was awarded with the Nobel Prize in physics for Max Planck in 1918. Eventually, it lead Richard Feynman, Julian Schwinger and Sin-Itiro Tomonaga to formulate a quantum analogue of Maxwell's theory of electromagnetism -- \emph{quantum electrodynamics} (QED). In this description of nature, the effect of the electromagnetic force between electrically charged particles is mediated by photons. In this sense, QED can be summarised as the interactions between light and matter \cite{feynman1990q}. Richard Feynman, Julian Schwinger and Sin-Itiro Tomonaga were honoured with the Nobel Prize in physics in 1965. 

Only a few years later, Sheldon Glashow, Abdus Salam and Steven Weinberg, extended QED to the \emph{electro-weak theory}. It extends early work of Enrico Fermi \cite{1934ZPhy...88..161F} and describes the combined effect of QED and the \emph{weak interaction}. Among other things, it predicts the existence of massive vector bosons, namely the $W^\pm$ and $Z$-bosons. In analogy to photons mediating the electromagnetic force in QED, it is these bosons which mediate the \emph{weak interaction}. For the detection of the $W^\pm$ and $Z$-boson, Carlo Rubbia and Simon van der Meer were awarded the Nobel Prize in physics in 1984. Even before that, in 1979, Sheldon Glashow, Abdus Salam and Steven Weinberg were honoured with the Nobel Prize in physics for the formulation of the \emph{electro-weak theory}.

In 1973, Harald Fritzsch, Heinrich Leutwyler and Murray Gell-Mann formulated a quantum field theory, which today is known as \emph{Quantum Chromodynamics} (QCD) \cite{1973PhLB...47..365F}. It employed the field-theory machinery developed by Chen Ning Yang and Robert Mills \cite{PhysRev.96.191}. QCD is a (special) \emph{Yang-Mills theory} and describes the interaction of the quarks by the \emph{strong force}. In the spirit of having the photon as a force particle of QED and $W^{\pm}$, $Z$ as force particles of the weak interaction, also the strong force is mediated by force particles. The gauge bosons in question are called \emph{gluons}.

To date, the combined effects of electromagnetism, weak and strong interaction are described by a quantum field theory with $SU(3) \times SU(2) \times U(1)_{\mathrm{Y}}$ gauge symmetry. It is known as the \emph{standard model of particle physics} \cite{oerter2006theory}. To the extent of current experiments, \ie to energies of up to 
\SI[parse-numbers=false]{14}{\tera \electronvolt}, this model proves to be a very accurate description of particle physics. Only recently, in 2013, the successful history of the \emph{standard model} was crowned with the Nobel prize awarded to Peter Higgs and Fran\c{c}ois Englert for their works on the Higgs boson \cite{1964PhRvL..13..508H, 1964PhRvL..13..321E}.

\paragraph{Open Questions and Grand Unification}
Despite its successes, the \emph{standard model of particle physics} leaves open quite a number of questions. For example, it requires at least 19 input parameters, including the masses of the elementary particles. These parameters however are not predicted by the theory! Likewise, the gauge group $SU(3) \times SU(2) \times U(1)_{\mathrm{Y}}$ is not explained by the \emph{standard model}. Even more, the \emph{standard model} treats electromagnetism, weak and strong interaction as three distinct forces -- each with its own coupling constant -- which happen to share a common mathematical formulation. In the spirit of Newton and Maxwell, a unified description of these forces would rather consist of a single unified force with \emph{one} coupling constant only. Can we find a quantum field theory of such a unified force, which explains the combined phenomena of the electromagnetism, the weak force and the strong interaction?

Often, the behaviour of quantum objects radically differs from our everyday experience. Among others this includes the concept of so-called \emph{running coupling constants}. In contrast to what the term \emph{constant} indicates, these are by no means constants. Rather, they depend on an energy scale $\Lambda$: Naively, the formalism of quantum field theory leads to physical observables of infinite values. These divergences stem from particles of very high energies. Suppose that in an experiment the maximally available energy is $\Lambda$, then it seems meaningful to take into account only contributions of particles with energies below $\Lambda$. This is the so-called \emph{cut-off regularisation}. It is followed by renormalization -- the now finite expressions are used to relate the \emph{bare parameters}, which enter the Lagrangian of the quantum field theory, to the physical observables. Of course, the latter include the coupling constants, so that these will in general depend on the energy scale $\Lambda$ used for the regularisation. This energy dependence of the coupling constants is captured in the so-called $\beta$-functions \cite{peskin1995introduction}.

The dependence of the coupling strength of the electromagnetic, weak and strong interactions on the energy scale $\Lambda$ has been computed to very high precision. They tend to yield similar values at an incredibly high energy scale $\Lambda_{\mathrm{GUT}} \sim \SI[parse-numbers=false]{10^{16}}{\giga \electronvolt}$ \cite{ross1984grand}. This is known as \emph{gauge coupling unification}. Could it be that at energies greater than $\Lambda_{\mathrm{GUT}}$ those three forces are actually described by one unified force, which is broken into electromagnetic, weak and strong phenomena at energies below $\Lambda_{\mathrm{GUT}}$? Among others, it was this question that triggered people to work on \emph{grand unified theories} (GUTs). Such a GUT-model comes with a gauge group $G_{\mathrm{GUT}}$, which contains the \emph{standard model} gauge group as a subgroup, and a mechanism to break this bigger gauge group $G_{\mathrm{GUT}}$ at low energies to $SU(3) \times SU(2) \times U(1)_Y$ or slight extensions thereof. Examples include the Pati-Salam model with $G_{\mathrm{GUT}} = SU(4) \times SU(2) \times SU(2)$ \cite{1994spas.book..343P} and the Georgi-Glashow model with $G_{\mathrm{GUT}} = SU(5)$ \cite{1974PhRvL..32..438G}. In this thesis, we will be particularly interested in the latter.

\paragraph{A Theory of Quantum Gravity -- String Theory}
GUT-models bring us a lot closer to a unified description of all physical phenomena. However, they are obviously incomplete as they do not include gravity. Whilst many experiments are nicely approximated by gravity or the \emph{standard model}, our understanding of black holes or the big bang hinges on a consistent theory of quantum gravity. Naively, we could try to formulate Einstein's theory of \emph{general relativity} \cite{einstein1916grundlage} in the language of quantum field theory. In doing so, gravitational interactions are mediated by a gauge boson which is called the \emph{graviton}. It is a massless spin-two particle \cite{weinberg1995quantum}. Unfortunately, in the resulting quantum field theory, the physical observables suffer from divergences in an uncontrollable manner \cite{GOROFF198581}. Therefore, a different approach is required.

From classical mechanics to \emph{general relativity} the point-particle-concept is omnipresent. Even the divergences encountered in quantum field theory are believed to be tied to this concept. String theory breaks with the point-particle philosophy and models the fundamental entities of nature as 1-dimensional strings. As such strings have not been observed experimentally, they must be of incredibly short lengths $l_s$. It is believed that $l_s$ is of the order of the Planck length $l_{\mathrm{Planck}} \sim \SI[parse-numbers=false]{10^{-35}}{\meter}$. Equivalently, the string scale $M_s = l_s^{-1}$ is of the order of $M_{\mathrm{Planck}} \sim \SI[parse-numbers=false]{10^{19}}{\giga \electronvolt}$.

To date, \emph{string theory} is arguably one of the best candidates of a theory of quantum gravity \cite{polchinski1998string, polchinski2001string, green1988superstring1, green1988superstring2, Ibanez2014string, blumenhagen2012basic}: It was originally proposed as a description of the strong force. With rising popularity of QCD in the 1970s it was almost forgotten. Luckily, people found that \emph{string theory} gives rise to a massless spin-two particle -- the defining property of the graviton \cite{weinberg1995quantum}. Additionally the extended nature of the fundamental strings `smears out interaction vertices'. In working out the string perturbation theory to first orders, UV-finite scattering amplitudes were found \cite{polchinski1998string}. To date, it is commonly believed that \emph{string theory} is a UV-finite quantum description including a graviton -- this is what makes this framework such a natural and interesting candidate for a description of quantum gravity.

If we include bosonic and fermionic excitations in \emph{string theory}, then the consistency of this theory requires a 10-dimensional spacetime and supersymmetry. In this context one speaks of \emph{superstring theory}. As it turns out, the formulation of this 10-dimensional theory is not unique, rather there exist five different, consistent superstring theories in flat 10-dimensional Minkowski spacetime $\mathcal{M}_{1,9}$. These are termed type I, type IIA, type IIB, heterotic $E_8 \times E_8$ and heterotic $SO ( 32 )$ \emph{string theory}. Connections between these different formulations exist and go by the name of \emph{T-duality} and \emph{S-duality} \cite{polchinski2001string}.

When taking \emph{string theory} seriously, one of the main phenomenological challenges arises from connecting the 10-dimensional spacetime of \emph{string theory} with the 4-dimensional spacetime used in formulating the \emph{standard model} of particle physics or \emph{general relativity}. The bridge between these geometries is called \emph{compactification}. This two-step procedure starts by taking the 10-dimensional spacetime $M_{10}$, on which we formulate superstring theory, as a product $M_{10} = \mathcal{E}_4 \times \mathcal{I}_6$. Here $\mathcal{E}_4$ is the 4-dimensional spacetime of the \emph{standard model} or \emph{general relativity}, and $\mathcal{I}_6$ is the so-called \emph{internal} space. $\mathcal{I}_6$ is taken so very small that it has not yet been observed experimentally. In the second step we follow the examples of Theodor Kaluza \cite{Kaluza1921} and Oskar Klein \cite{1926ZPhy...37..895K, 1926Natur.118..516K} to obtain an effective 4-dimensional theory. The latter involves various integrals over the internal space $\mathcal{I}_6$ and leads to additional field excitations on $\mathcal{E}_4$. In honour of Theodor Kaluza and Oskar Klein, these excitations are termed \emph{Kaluza-Klein-modes} -- or short \emph{KK-modes}.

In 1985 Philip Candelas, Gary Horowitz, Andrew Strominger and Edward Witten discussed superstring compactifications on Calabi--Yau manifolds \cite{CANDELAS198546}. They considered a compactification of the (perturbative) heterotic $E_8 \times E_8$ superstring theory. This theory has an $E_8 \times E_8$ gauge symmetry. In \cite{CANDELAS198546} a Yang-Mills connection $V$ on $M_{10}$ was employed to spontaneously break this gauge symmetry to $E_6 \times E_8$. The $E_8$ group constitutes the so-called \emph{hidden sector}, whilst the observable $E_6$ was used to furnish a GUT-model. As one of the first string compactifications, this work underlines both the importance of Calabi--Yau manifolds and the significance of GUTs in superstring model building. Offsprings of this work were \cite{GREENE1986667, doi:10.1142/9789812798411_0026}, where Calabi--Yau manifolds were classified to fit physical needs in such compactifications. More recent works in the field of heterotic superstring compactifications include \cite{Nilles:2008gq, Anderson:2008ex, Anderson:2011ns, Anderson:2012yf}.

Of course, compactifications of other superstring theories were investigated also, see \eg \cite{Douglas:2006es} for an overview of the different approaches.  
In particular objects called \emph{Dp-branes} feature prominently in string compactifications. Let us explain what these are. A string has two endpoints. If these are distinct, we term the string \emph{open} and otherwise \emph{closed}. The subspace of $M_{10}$ at which open strings end forms the world-volume of a Dp-brane. The established terminology is that a Dp-brane has a $p+1$-dimensional world volume, consisting of $p$-spatial dimensions and the time-dimension of the spacetime $M_{10}$.\footnote{Let us mention that also instantonic branes are being considered in the \emph{string theory} literature. As the name indicates, these branes occupy just a single point along the time direction, and are in this sense instantaneous.} It was a crucial observation that these Dp-branes form non-perturbative and dynamical objects by themselves \cite{doi:10.1142/S0217732389002331}. Even more important, stacks of coincident Dp-branes realise gauge theories \cite{polchinski1998string, polchinski2001string, green1988superstring1, green1988superstring2, Ibanez2014string, blumenhagen2012basic}.

For example, in (perturbative) type IIA superstring theory one can consider a stack of three coincident D6-branes and a second stack of two coincident D6-branes. If placed correctly, they support a $U(3)$ and $U(2)$ gauge theory, respectively. On the intersection of the stacks, massless matter localises in the bifundamental representation $(\overline{\mathbf{3}}, \mathbf{2})$. Based on this, one can construct so-called \emph{intersecting D6-brane models} which realise the \emph{standard model} in \emph{string theory} without the need for a GUT-construction. Many more details and examples can be found in \cite{Ibanez:2001nd, MarchesanoBuznego:2003hp, Blumenhagen:2006ci, Blumenhagen:2005mu, Cvetic:2001tj, Cvetic:2002wh, Cvetic:2002qa, Berasaluce-Gonzalez:2016kqb, Honecker:2016gyz}.

Perturbative type IIB superstring theory is closely related to its type IIA cousin. In particular, in type IIB compactifications the \emph{standard model} gauge group can easily be realised in compactifications with D3- and D7-branes \cite{Blumenhagen:2006ci, Blumenhagen:2005mu, Lust:2004ks}. Unfortunately, backreactions of the D7-branes cause a break down of the perturbative theory as reviewed in \cite{Weigand:2010wm}. Consequently, non-perturbative techniques are required to handle such compactifications. This can be seen as the birth place of \emph{F-theory}.

\paragraph{F-Theory and the goal of this Thesis}

The backreaction of the D7-branes is described by the profile of a particular field on $M_{10}$ -- the axio-dilaton $\tau$. It was realised around 1996 that this field exhibits the same properties as the complex-structure modulus of a torus \cite{Vafa:1996xn, Morrison:1996na, Morrison:1996pp}. Since this complex structure modulus uniquely fixes the shape of a 2-dimensional torus, a way to encode the axio-dilaton geometrically is to attach to each point $p$ of spacetime the torus with complex structure modulus $\tau ( p )$. Such a geometric construction is known as a torus-fibration. Such fibrations are an integral part of \emph{F-theory}, as their geometry encodes many physical properties \cite{Donagi:2008ca, Heckman:2010bq}. For example, the singularities of this torus fibration encode the gauge groups along the D7-branes, the matter fields -- which are localised at the intersections of such D7-branes -- and their Yukawa interactions \cite{Bershadsky:1996nh, Katz:1996xe, Donagi:2008ca, Beasley:2008dc, Beasley:2008kw}.

As we have just motivated, \emph{F-theory} can be understood as a non-perturbative limit of type IIB \emph{string theory}. However, it also bears close connections to heterotic \emph{string theory}, see \cite{Hayashi:2008ba, Vafa:1996xn, Morrison:1996na, Morrison:1996pp, Friedman:1997yq, Heckman:2013sfa, Anderson:2014gla}. Much of the understanding of \emph{F-theory} is obtained from this duality. Yet another approach to \emph{F-theory} comes from M-theory. The latter is an 11-dimensional physical theory, of which the above mentioned five superstring theories are believed to be different 10-dimensional incarnations. The most precise understanding of \emph{F-theory} follows from this last approach, as we will explain in the next chapter.

Of course, it is possible to formulate GUT-models within \emph{F-theory} \cite{Beasley:2008kw, Beasley:2008dc, Donagi:2008kj, Marsano:2009wr, Dolan:2011aq, Palti:2012dd}. In any such model, the mechanism to break the GUT gauge group $G_{\mathrm{GUT}}$ to $SU(3) \times SU(2) \times U(1)_{\mathrm{Y}}$ is crucial. One way to achieve this is by exploiting the Higgs effect. This however requires control over the Higgs potential, and in \emph{string theory} this potential derives from the geometry of the compactification. Since the compactification spaces are usually quite involved, both knowledge and control over this potential are in general lacking. An alternative way of breaking the GUT-group is established by giving a vacuum expectation value (VEV) to the so-called \emph{hypercharge flux}. In \emph{F-theory}, and more generally type IIB superstring compactifications, we can achieve this reduction of gauge symmetry without giving a mass to the hypercharge gauge field. The precise conditions for such a mass to be absent were originally worked out in type IIB compactifications \cite{Buican:2006sn} (see also \cite{Tatar:2008zj, Blumenhagen:2008zz}) and were later put to use in F-theory \cite{Beasley:2008dc, Beasley:2008kw, Donagi:2008kj, Donagi:2008ca, Marsano:2009ym, Dudas:2010zb, Marsano:2010sq, Palti:2012dd, Mayrhofer:2013ara, Braun:2014pva}.

In phenomenological applications of \emph{F-theory} compactifications -- and also superstring compactifications more generally -- we wish to tell if a compactification can serve as a model of our universe. To this end, we need to extract physical observables from these compactification. Usually, these observables are encoded in the geometry of the internal space $\mathcal{I}_6$, and hence we need to investigate its geometry in detail. As the geometry of $\mathcal{I}_6$ is usually fairly involved, it is beneficial to start off simple. Let us therefore first recall that at the intersection of D7-branes massless chiral fermions localise. By massless, we mean that they have no masses prior to gauge group breaking, but can obtain masses \eg from the Higgs mechanism. A very basic parameter to judge the phenomenological relevant of a specific compactification is the difference of the number of such chiral fermions and the number of their anti-chiral cousins. This number is called the chiral index. It happens to be a topological invariant and is therefore often computable with minimal effort. 

To date, experimental evidence indicates that the \emph{standard model} does not contain pairs of chiral and anti-chiral fermions in the same representation. To check if the massless spectrum of a \emph{F-theory} compactifications satisfies this property also, we have to go beyond the chiral index and actually compute the number of chiral and anti-chiral fermions. Even more, the running couplings and the Yukawa interactions strongly depend on these numbers. Finally, supersymmetric extension of the \emph{standard model} must contain (at least) one vector-like pair of superfields, namely the Higgs doublets $H_u$ and $H_d$. Based on this motivation, the goal of this thesis is to present a solution to the following task:
\begin{center}
\fbox{\begin{minipage}[c]{0.9 \textwidth} \begin{center} Count the localised, massless (anti-)chiral fermions in \emph{F-theory} GUT-models. \end{center} \end{minipage}}
\end{center}

\paragraph{Outline}

In \cref{chapter:Warmup} we start our journey with a revision of foundational material. As we aim at building \emph{F-theory} GUT-models, it seems appropriate to start with a revision of the \emph{standard model of particle physics} and \emph{grand unified theories} in \cref{sec:GUTs}. Subsequently, we turn to \emph{string theory} in \cref{sec:StringTheory} and finally \emph{F-theory} in  \cref{sec:F-theory}. In particular, we will explain how Deligne cohomology can be used to model the gauge backgrounds in F-theory compactifications. We complete this chapter by reviewing zero mode counting in type IIB orientifold compactifications. As we explain in 
\cref{sec:RevisionOnSheavesAndSheafCohomology}, these zero modes are counted by sheaf cohomologies of line bundles located at D-brane intersections. Therefore, this section contains a revision of basic material on sheaves and their sheaf cohomologies. To illustrate these formal terms, we exemplify these notions for line bundles on compact, connected Riemann surfaces.

We pointed out in \cite{Bies:2014sra} that Chow groups can describe a subset of Deligne cohomology and that they lend themselves nicely to explicit computations. The
Chow group is formed from equivalence classes of algebraic cycles -- two algebraic cycles are equivalent if their difference can be understood as the divisor of a rational function. In this sense Chow groups classify algebraic cycles up to rational equivalence. This definition manifestly involves rational functions, which are defined on varieties only. So in order to apply Chow groups as parametrisation of a subset of Deligne cohomology, we have to break with analytic geometry. Therefore, we explain varieties and their Chow group in detail in \cref{sec:TheChowRing}. Subsequently, we argue in \cref{sec:SystGaugeBack} that the zero modes in F-theory vacua are counted by sheaf cohomologies of line bundles $L_{\mathbf{R}}$ on the so-called \emph{matter curves} $C_{\mathbf{R}}$. 

In phenomenological applications one wishes to compute these quantities for large numbers of explicit examples. Consequently, we set out to design computer algorithms which compute the required sheaf cohomologies. To this end, we can only allow for special geometries and will, in this thesis, focus on toric varieties. The required tools are introduced in \cref{sec:ToolsForCompuationalF-theoryVacua}. Subsequently, we study a class of toric \emph{F-theory} compactifications, in \cref{sec:ToricFTheoryGUTModels}. These geometries are derived from a so-called $SU(5) \times U(1)_X$-top, which was originally introduced in \cite{Krause:2011xj} and has been analysed with more refined techniques in \cref{sec:SU5xU1Top}. \footnote{Indeed \cref{chapter:ToricTops} summarises details on related geometries, which are employed in this thesis.}

In \cref{sec:ToricFTheoryGUTModels} we derive the line bundles $L_{\mathbf{R}}$ for this entire class of F-theory compactifications. To actually compute the sheaf cohomologies of these line bundles, we need more refined techniques. These we introduce in \cref{chapter:DetailsOnFPGradedSModules}. As we explain in \cref{subsec:CoherentSheavesOnVarieties}, on affine varieties coherent sheaves can be modelled by \emph{modules} over the coordinate ring of the variety.\footnote{A module $M$ over a ring $R$ is essentially the same as vector space $V$ over a field $k$, except that we do not restrict to `ground fields' $k$ but allow for more general `ground rings' $R$.} For toric varieties there exists a similar construction -- here \emph{finitely presented (f.p.) and graded} modules over the coordinate ring give rise to coherent sheaves. We detail these modules in \cref{sec:ProjSmodule} and \cref{subsec:FPGradedSModules}. Subsequently, we explain in \cref{sec:Sheafification} how they encode a coherent sheaf. Finally, we apply this knowledge in \cref{sec:ComputingTheSpectra} to study an F-theory toy model. The technical steps which allow us to relate the defining data of an \fp graded module to the sheaf cohomologies of the associated sheaf are based on \cref{mytheorem}, which is derived in \cref{chapter:MathDetailsSheafCohomologies}. The subsequent \cref{chapter:GUTModels} studies far more realistic F-theory GUT models than the toy model discussed in \cref{sec:ComputingTheSpectra}.

While putting the finishing touches to \cite{Bies:2017fam}, we realised that Chow groups can also help to deepen our understanding of local anomalies in F-theory. We complete this thesis by explaining this interplay in \cref{chapter:LocalAnomaliesInF-Theory}.

\paragraph{Published material}

In \cref{subsec:SheafAndSheafCohomology} and \cref{subsec:LineBundlesOnRiemannSurfaces} we follow \cite{BiesMaster} closely. The material presented in \cref{subsec:GaugeBackgroundsFromDeligneCohomology} is inspired by \cite{Bies:2014sra}. The contents of \cref{chapter:MasslessSpectraAndSheafCohomology}, \cref{chapter:DetailsOnFPGradedSModules} and parts of \cref{sec:SU5xU1Top} were published in \cite{Bies:2017fam}. \Cref{chapter:LocalAnomaliesInF-Theory}, \cref{sec:DetailsOfSU4Top} and parts of \cref{sec:SU5xU1Top} are taken from \cite{Bies:2017abs}. The computation of zero modes in explicit examples was performed with the help of the implementations \cite{CAPCategoryOfProjectiveGradedModules, CAPPresentationCategory, PresentationsByProjectiveGradedModules, TruncationsOfPresentationsByProjectiveGradedModules, SheafCohomologyOnToricVarieties} on the computer \texttt{plesken} at the \emph{University of Siegen}.

\chapter{Towards Zero Mode counting in F-Theory} \label{chapter:Warmup}

Much of the material presented in this chapter is textbook material. Background on the \emph{standard model} and \emph{grand unified theories} is for example provided in \cite{langacker2009standard, ross1984grand}. More details on \emph{string theory} can be found in \cite{polchinski1998string, polchinski2001string, green1988superstring1, green1988superstring2, Ibanez2014string, blumenhagen2012basic}. Neat expositions of \emph{F-theory} are given in \cite{Weigand:2010wm,Lin:2016zha}. Additional information on the mathematics discussed in \cref{sec:RevisionOnSheavesAndSheafCohomology} and later in \cref{sec:TheChowRing}, \cref{sec:ToolsForCompuationalF-theoryVacua} can be found in \cite{griffiths2011principles, hartshorne1977algebraic, FreitagRiemann, FreitagComplexSpaces,lane1998categories, cox2011toric, fulton1993introduction, FultonInt}.

\section{Grand Unified Theories} \label{sec:GUTs}

\subsection{Overview of the Standard Model} \label{subsec:OverviewOfStandardModel}

\paragraph{Gauge Group}
To date, experimental evidence in particle physics is accurately predicted by a physical framework known as the \emph{standard model of particle physics}. It describes the dynamics of a number of matter particles, gauge bosons and of the famous Higgs boson under the influence of the strong, the weak and the electromagnetic force. To this end, it employs the language of quantum field theory and formulates a gauge theory with gauge group $G_{\mathrm{SM}} = SU( 3 )_C \times SU(2)_W \times U(1)_Y$.

\paragraph{Matter Particles}
The matter particles are fermions. They feel the electromagnetic, strong and weak force as listed in \cref{my_old_table_1}. All matter particles appear in replicas which carry identical charge under all forces but differ quite drastically in their rest masses. Such replicas are termed \emph{generations}. Current experiments indicate that every matter particle appears in precisely three such generations.

Being fermions, every matter particle comes with a spin. The spin component along the direction of motion is termed the \emph{helicity} or \emph{handedness} of this fermion. A fermion is said to be \emph{right-handed} precisely if its helicity and direction of motion are parallel. Conversely, if helicity and direction of motion are anti-parallel, then the fermion is termed \emph{left-handed}.

For massive particles, helicity is not a well-defined concept. This is because there always exists a Lorentz boost which inverts the direction of motion of the particle but does not affect its spin. Consequently, any such boost will interchange left-handed and right-handed matter particles. For massless particles, however, helicity is a meaningful concept since such particles move at the speed of light. In particular, for such particles \emph{helicity} and \emph{chirality} coincide.

Prior to the electro-weak symmetry breaking, which we will discuss below, all matter particles in the \emph{standard model of particle physics} are chiral and massless fermions. This is crucial since the weak force acts chirally -- it affects left-handed particles only. The electro-weak symmetry breaking eventually generates masses for all matter particles. This process involves the famous Higgs boson.

The \emph{standard model of particle physics} is formulated as quantum field theory. It is well known that in such a formulation, all particles furnish irreducible representations of the underlying gauge group \cite{weinberg1995quantum}. Therefore, the fermionic particles listed in \cref{my_old_table_1} and the bosonic particles, to which we come momentarily, transform in certain representations of $G_{\mathrm{SM}}$. We list the representations for the fermionic particles in \cref{my_old_table_2}. Note that the group $U(1)_Y$ which appears in $G_{\mathrm{SM}}$ is not the electromagnetic $U(1)_{\mathrm{em}}$. The latter is a subgroup of the electro-weak gauge symmetry group $SU(2)_W \times U(1)_Y$, as we will discuss momentarily. \Cref{my_old_table_2} lists both the charge $Q_Y$ under $U(1)_Y$ and the charge $Q_{\mathrm{em}}$ under $U(1)_\mathrm{em}$.

\begin{table}[tbp]
\centering
\begin{tabular}{l@{\hskip 25pt}ccc@{\hskip 25pt}c@{\hskip 25pt}ccc}
\toprule
& \multicolumn{3}{c@{\hskip 25pt}}{Generation} & Electric charge & \multicolumn{3}{c@{\hskip 25pt}}{Feels the force of} \\
& $1^{\text{st}}$ & $2^{\text{nd}}$ & $3^{\text{rd}}$ & in units of $e$ & Strong & Weak & EM \\
\midrule
U-Type Quarks & u & c & t & $+ \frac{2}{3}$ & \checkmark & \checkmark & \checkmark \\
D-Type Quarks & d & s & b & $- \frac{1}{3}$ & \checkmark & \checkmark & \checkmark \\
Charged Leptons & $e$ & $\mu$ & $\tau$ & $-1$ & & \checkmark & \checkmark \\
Neutral Leptons & $\nu_e$ & $\nu_\mu$ & $\nu_\tau$ & $0$ & & \checkmark & \\
\bottomrule
\end{tabular}
\caption[The matter particles of the \emph{standard model}.]{The matter particles in the \emph{standard model}. This table is inspired by \cite{BiesBachelor}.}
\label{my_old_table_1}
\end{table}

\begin{table}[tb]
\centering
\begin{tabular}{ccccccr}
\toprule
Symbol & Particle & Type & Chirality & Representation of $G_{\mathrm{SM}}$ & $Q_{em}$ & $Q_Y$ \\
\midrule
$Q$ & quarks & fermion & L & $\left( \mathbf{3}, \mathbf{2} \right)_{1/6}$ & $+ \frac{2}{3}$, $- \frac{1}{3}$ & $+\frac{1}{6}$ \\
$U$ & up-quarks & fermion & R & $\left( \mathbf{3}, \mathbf{1} \right)_{2/3}$ & $+\frac{2}{3}$ & $+\frac{2}{3}$ \\
$D$ & down-quarks & fermion & R & $\left( \mathbf{3}, \mathbf{1} \right)_{-1/3}$ & $-\frac{1}{3}$ & $-\frac{1}{3}$ \\
\vspace{-0.5em} & \\
$Q^c$ & quarks & fermion & R & $\left( \mathbf{\overline{3}}, \mathbf{\overline{2}} \right)_{-1/6}$ & $-\frac{2}{3}$, $+\frac{1}{3}$ & $-\frac{1}{6}$ \\
$U^c$ & up-quarks & fermion & L & $\left( \mathbf{\overline{3}}, \mathbf{1} \right)_{-2/3}$ & $-\frac{2}{3}$ & $-\frac{2}{3}$ \\
$D^c$ & down-quarks & fermion & L & $\left( \mathbf{\overline{3}}, \mathbf{1} \right)_{1/3}$ & $+\frac{1}{3}$ & $+\frac{1}{3}$ \\
\midrule
$L$ & leptons & fermion & L & $\left( \mathbf{1}, \mathbf{2} \right)_{-1/2}$ & $-1,0$ & $-\frac{1}{2}$ \\
$E$ & charged leptons & fermion & R & $\left( \mathbf{1}, \mathbf{1} \right)_{-1}$ & $-1$ & $-1$ \\
$N$ & neutral leptons & fermion & R & $\left( \mathbf{1}, \mathbf{1} \right)_0$ & $0$ & $0$ \\
\vspace{-0.5em} & \\
$L^c$ & leptons & fermion & R & $\left( \mathbf{1}, \mathbf{\overline{2}} \right)_{1/2}$ & $+1,0$ & $+\frac{1}{2}$ \\
$E^c$ & charged leptons & fermion & L & $\left( \mathbf{1}, \mathbf{1} \right)_{1}$ & $+1$ & $+1$ \\
$N^c$ & neutral leptons & fermion & L & $\left( \mathbf{1}, \mathbf{1} \right)_0$ & $0$ & $0$ \\
\midrule
$H$ & Higgs doublet & boson & \XSolidBrush & $\left( \mathbf{1}, \mathbf{2} \right)_{1/2}$ & 0 & $+\frac{1}{2}$ \\
$H^c$ & Higgs doublet & boson & \XSolidBrush & $\left( \mathbf{1}, \mathbf{\overline{2}} \right)_{-1/2}$ & 0 & $-\frac{1}{2}$ \\
\bottomrule
\end{tabular}
\caption[The irreducible representations of the matter particles in the \emph{standard model}.]{Irreducible representations for the matter particles and the Higgs doublet in the \emph{standard model}. The representations are written as $SU \left( 3 \right)_C \times SU \left( 2 \right)_W \times U \left( 1 \right)_Y$. This table is inspired by \cite{BiesBachelor}.}
\label{my_old_table_2}
\end{table}

\paragraph{Gauge Bosons}

Besides the matter particles, the \emph{standard model} also describes the dynamics of gauge bosons. These bosons mediate the forces between matter particles. They transform in the adjoint representation of the gauge group in question. For example, there are eight force particles for the strong force with gauge group $SU(3)$, since its adjoint is of dimension eight. These force carriers are called gluons. Similarly, there are three force particles for the weak force with $SU(2)$ gauge group -- the $W^{\pm}$ and $Z$ bosons. Finally, $U(1)_Y$ has a single force particle.

\subsection{Electro-Weak Symmetry Breaking} \label{subsec:ElectroWeakBreaking}

The electro-weak symmetry breaking `reduces' the electro-weak gauge group according to
\[ SU(2)_W \times U(1)_Y \to U(1)_{\mathrm{em}} \, . \]
Along this breaking, one gauge boson of $SU(2)_W \times U(1)_Y$ is required to remain massless. This remaining massless force carrier is the photon, which mediates the electromagnetic force. We will momentarily discuss \emph{grand unified theories} (GUTs) and how we break their large gauge groups to the one of the \emph{standard model}. To understand this process better, it is instructive to take a closer look at the electro-weak symmetry breaking. Suffice it here to phrase our analysis in the language of \emph{classical} field theory.

\paragraph{Electro-Weak Gauge Theory}

The group $SU(2)$ is a matrix Lie group in the sense of \cite{hall2003lie}. The generators of its Lie algebra are $T^a = \frac{1}{2} \sigma^a$ where $\sigma^a$ denotes the Pauli matrices
\[ \sigma^1 = \left( \begin{array}{cc} 0 & i \\ -i & 0 \end{array} \right) \, , \qquad \sigma^2 = \left( \begin{array}{cc} 0 & 1 \\ 1 & 0 \end{array} \right) \, , \qquad \sigma^3 = \left( \begin{array}{cc} 1 & 0 \\ 0 & -1 \end{array} \right) \, . \label{equ:PauliMatrices} \]
We denote the generator of the group $U(1)_Y$ by $Y$. Since $SU(2) \times U(1)_Y$ is a compact and connected matrix Lie group, every element of this group can be expressed as \cite{hall2003lie}
\[ g ( \boldsymbol{\alpha}, \beta ) = \mathrm{exp} \left( i \boldsymbol{\alpha} \cdot \mathbf{T} \right) \cdot \mathrm{exp} \left( i \beta \cdot Y \right) \, . \label{equ:GroupParametrisation} \]
The $SU(2)$ gauge field is given by $\mathbf{A}_\mu \cdot \mathbf{T}$ and that of $U(1)_Y$ by $B_\mu$. They transform in the adjoint representation. So for 
$U = \mathrm{exp} ( i \cdot \boldsymbol{\alpha}( x ) \cdot \mathbf{T} )$ and $g$, $g^\prime$ the coupling constants of $SU(2)$ and $U(1)_Y$, respectively, the transformations are
\[ \mathbf{A} \cdot \mathbf{T} \mapsto U \cdot \left( \mathbf{A} \cdot \mathbf{T} \right) U^{-1} + \frac{i}{g} \cdot \left( \partial_\mu U \right) U^{-1} \, , \qquad B_{\mu} \mapsto B_{\mu} - \frac{\partial_\mu \beta}{g^\prime} \, . \]
The field strengths satisfy
\[ \mathbf{F}_{\mu \nu} = \partial_\mu A^a_\nu T^a - \partial_\nu A^a_{\mu} T^a + i g A^b_\mu A^c_\nu \left[ T^b, T^c \right] \, , \qquad G_{\mu \nu} = \partial_\mu B_{\nu} - \partial_{\nu} B_{\mu} \, , \]
where summation over repeated indices is understood implicitly. This notation in place, the gauge theory underlying the electro-weak theory is described by the Lagrangian
\[ \mathcal{L}_{\text{gauge}} = - \frac{1}{2} \mathrm{tr} \left( \mathbf{F}_{\mu \nu} \mathbf{F}^{\mu \nu} \right) - \frac{1}{4} G_{\mu \nu} G^{\mu \nu} \, . \]

\paragraph{Preparation for Gauge Group Breaking}

Our aim is to break the $SU(2)_W \times U(1)_Y$ gauge symmetry to $U(1)_{\mathrm{em}}$. To this end, we fix an embedding of this group into $SU(2)_W \times U(1)_Y$. At the level of their Lie algebras we express this embedding as
\[ \mathfrak{u} \left( 1 \right)_{\mathrm{em}} \hookrightarrow \mathfrak{su} \left( 2 \right) \oplus \mathfrak{u} \left( 1 \right)_Y \; , \; Q \mapsto T_3 \oplus Y \, , \qquad Y = \left( \begin{array}{cc} 1/2 & 0 \\ 0 & 1/2 \end{array} \right) \, . \]
The desired breaking is now a special instance of \emph{spontaneous symmetry breaking} followed by the \emph{Goldstone theorem} and the \emph{Higgs effect}. As a first step we extend $\mathcal{L}_{\mathrm{gauge}}$ by a Lagrangian which is used to trigger the gauge symmetry breaking. We take
\[ \mathcal{L}_{\mathrm{total}} = \mathcal{L}_{\mathrm{gauge}} + \mathcal{L}_{\phi} \, , \qquad \mathcal{L}_{\phi} = \frac{1}{2} \left( D_{\mu} \phi \right)^{\dagger} \left( D^\mu \phi \right) - \frac{\lambda}{2} \left( \phi^\dagger \phi - \frac{v^2}{2} \right)^2 \label{LagrangianForHiggs} \]
where $\lambda$ and $v$ are real constants. This Lagrangian captures the entire dynamics of the electro-weak theory. The field $\phi$ is chosen to transform in the representation $\mathbf{2}_{1/2}$ of $SU(2) \times U(1)_Y$, \ie it transforms in the fundamental representation of $SU(2)$ and has $U(1)_Y$ charge $\frac{1}{2}$. Its \emph{covariant derivative} is consequently given by
\[ D_\mu \phi = \left[ \partial_\mu \cdot \mathds{1} + i g \mathbf{A}_\mu \cdot \frac{\boldsymbol{\sigma}}{2} + i g^\prime B_\mu \cdot \frac{\mathds{1}}{2} \right] \phi \, . \]
Note that $\phi$ is taken to transform trivially under the $SU(3)_C$ group in $G_{\mathrm{SM}}$. So in terms of $G_{\mathrm{SM}}$ the field $\phi$ transforms in the representation $( \mathbf{1}, \mathbf{2} )_{1/2}$. This is one of the so-called \emph{Higgs doublet}, which was listed in \cref{my_old_table_2}.

\begin{figure}[tbp]
\centering
\includegraphics[width = 0.6\textwidth]{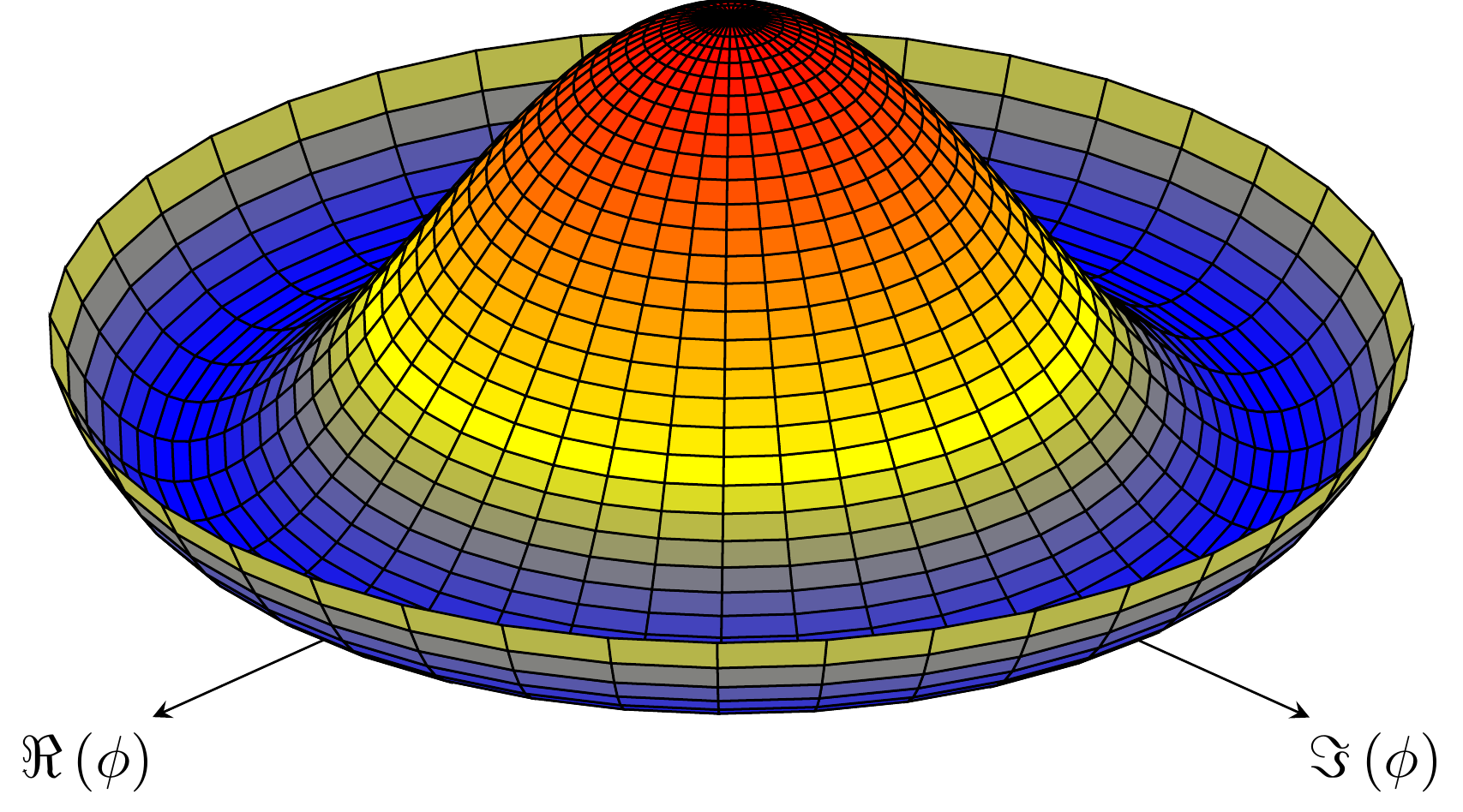}
\caption[A schematic picture of the Higgs potential.]{The potential $V( \phi ) = \frac{\lambda}{2} ( \phi^\ast \phi - \frac{v^2}{2} )^2$ for a complex valued scalar field $\phi$.}
\label{figure:HiggsPotential}
\end{figure}

\paragraph{Spontaneous Symmetry Breaking}

A ground state for this Higgs doublet is a field profile $\phi_0$ which has minimal potential energy $V( \phi_0 )$. In the current setup this minimal potential energy is zero, and from \cref{figure:HiggsPotential} we can see that there are many possible field profiles with $V( \phi_0 ) = 0$. They form the ground state manifold
\[ \mathcal{M}_0 = \left\{ \phi_0 \left| \phi_0^\dagger \phi_0 = \frac{v^2}{2} \right. \right\} \, . \]
Given this redundancy, the physical system will \emph{spontaneously} pick a ground state from $\mathcal{M}_0$. As we can envision from \cref{figure:HiggsPotential}, any such choice will break the gauge symmetry of $\mathcal{L}_{\mathrm{total}}$. This is the rationale behind spontaneous symmetry breaking -- a spontaneous choice of a ground state will in general break the gauge symmetry of the system.

One can easily generalise these findings to spontaneously break a system with gauge group $G$ down to a gauge group $H \subseteq G$. Under moderate assumptions, the Goldstone theorem states that the broken theory exhibits $\text{dim} ( G ) - \text{dim} ( H )$ massless bosons. These bosons are termed \emph{Goldstone bosons} \cite{PhysRev.127.965, Goldstone1961, 1960PhRv..117..648N}. Each of these bosons corresponds to a Lie algebra generator $T_{\hat{a}}$, whose associated group element $\mathrm{exp} ( i T_{\hat{a}} )$ does \emph{not} leave the spontaneously chosen ground state invariant.

\paragraph{The Higgs Effect}

We can now employ a gauge transformation which involves the broken Lie group generators $T_{\hat{a}}$ only. This is the so-called \emph{unitary gauge}, in which the physical particle content of the field theory is apparent.\footnote{In the quantum field theory analogue of our classical discussion, renormalizability of the theory must be discussed. This can in general not be judged in the unitary gauge. For this reason one then employs a different gauge. This one is sometimes termed the \emph{renormalization gauge} or \emph{renormalizable gauge}.} It turns out that in the unitary gauge the massless Goldstone bosons are absent. They are replaced by \emph{massive} vector bosons. This is the \emph{Higgs effect} -- the massless Goldstone bosons are `eaten' up by vector bosons, which thereby become massive \cite{1964PhRvL..13..508H, 1964PhRvL..13..321E}.

\paragraph{Explicit Analysis of the Electro-Weak Symmetry Breaking}

Let us demonstrate this mechanism explicitly for the electro-weak theory. Recall that it is our goal to leave the $U(1)_{\mathrm{em}}$ subgroup of $SU(2)_W \times U(1)_Y$, generated by $Q = T_3 \oplus Y$ invariant. Therefore, we should pick our ground state correspondingly. The choice
\[ \phi_0 = \left( \begin{array}{c} 0 \\ \frac{v}{\sqrt{2}} \end{array} \right) \,  \]
indeed satisfies this need. To justify this statement, let us work out which gauge transformations leave the ground state $\phi_0$ invariant. To this end, we parametrise the group $SU(2)_W \times U(1)_Y$ according to \cref{equ:GroupParametrisation}. For $\alpha \in [ 0, 2 \pi )$, $\beta \in \mathbb{R}$ and $\mathbf{n} \in \mathbb{R}^3$ with $\left| \mathbf{n} \right| = 1$ it is readily verified that
\[ \mathrm{exp} \left( i \alpha \cdot \frac{\mathbf{n} \cdot \boldsymbol{\sigma}}{2} \right) \cdot \mathrm{exp} \left( i \beta Y \right) = \left[ \mathrm{cos} \left( \frac{\alpha}{2} \right) \mathds{1} + \mathrm{sin} \left( \frac{\alpha}{2} \right) \cdot \left( \begin{array}{cc} i n_3 & i n_2 - n_1 \\ i n_2 + n_1 & - i n_3\end{array} \right) \right] e^{i \beta/2} \, . \]
From this it follows that only the matrix exponential of (real multiples of) $T_3 \oplus Y \in \mathfrak{su} ( 2 ) \oplus \mathfrak{u} ( 1 )$ leaves $\phi_0$ invariant. 

As a next step we install the unitary gauge. As the group elements generated from $T_1$, $T_2$ and $T_3 \oplus ( - Y )$ do not leave $\phi_0$ invariant, we perform a gauge transformation which involves these generators. It is not too hard to see that in the resulting unitary gauge the field $\phi$ is given by
\[ \phi = \left( \begin{array}{cc} 0 \\ \frac{v}{\sqrt{2}} \end{array} \right) + \left( \begin{array}{c} 0 \\ \frac{H}{\sqrt{2}}  \end{array} \right) \, \]
where $H$ is a real scalar field. It then follows
\begin{align}
\begin{split}
D_\mu \phi &= \left[ \left( \begin{array}{cc} \partial_\mu & 0 \\ 0 & \partial_{\mu} \end{array} \right)
                     + \frac{i g}{2} \left( \begin{array}{cc} A_{3,\mu} & A_{2,\mu} + i A_{1,\mu} \\ A_{2,\mu} - i A_{1,\mu} & - A_{3,\mu} \end{array} \right)
                     + \frac{i g^\prime B_\mu}{2} \left( \begin{array}{cc} 1 & 0 \\ 0 & 1 \end{array} \right) \right] \phi \\
          &= \frac{1}{\sqrt{2}} \left( \begin{array}{c} 0 \\ \partial_\mu H \end{array} \right) + \frac{i g}{2 \sqrt{2}} \cdot \left( v + H \right)  \cdot \left( \begin{array}{c} i \left[ A_{1,\mu} - i A_{2,\mu} \right] \\ - \left[ A_{3,\mu} - \frac{g^\prime}{g} B_\mu \right] \end{array} \right) \, .
\end{split}
\end{align}
In terms of
\[ W_{\mu} = \frac{1}{\sqrt{2}} \left( A_{\mu,1} - i A_{\mu,2} \right) \, , \qquad \tan \theta_W = \frac{g^\prime}{g} \, , \qquad Z_{\mu} = \cos \theta_W \left( A_{3, \mu} - \tan \theta_W B_\mu \right) \, . \]
the Lagrangian $\mathcal{L}_{\phi}$ takes the form
\begin{align}
\begin{split}
\mathcal{L}_{\phi} &= \frac{1}{2} \, \partial_\mu H \, \partial^\mu H + \frac{g^2 \left( v + H \right)^2}{4} \cdot W_\mu^\ast W^\mu + \frac{g^2 \left( v + H \right)^2}{8 
\cos^2 \theta_W} \cdot Z_\mu Z^\mu - \frac{\lambda}{8} \left( 2 v H + H^2 \right)^2 \, .
\end{split}
\end{align}
This Lagrangian describes a number of massive fields: First the real scalar field $H$ with mass $m_H^2 = \lambda v^2$. This is the famous \emph{Higgs boson}. Secondly there are 3 vector bosons -- the vector bosons $W_\mu$ and $W_\mu^\ast$ both have mass $m_W^2 = \frac{g^2 v^2}{4}$ and the vector boson $Z_\mu$ satisfies $m_Z^2 = \frac{g^2 v^2}{4 \cos^2 \theta_W}$. Indeed as promised, the gauge field $A_\mu = \sin \Theta_W A_{3,\mu} + \cos \Theta_W B_\mu$ -- corresponding to $Q = T_3 \oplus Y$ -- remained massless. This is the gauge field of the electromagnetic $U(1)_{\mathrm{em}}$ gauge theory. We summarise these findings in \cref{table:GaugeFieldPriorAndAfterHiggsEffect}.

\begin{table}[tbp]
\centering
\begin{tabular}{l@{\hskip 20pt}lc}
\toprule
Before Higgs effect & After Higgs effect & $\mathrm{Mass}^2$ from Higgs effect \\
\midrule
$A_{1,\mu}$ & $W_\mu = \frac{1}{\sqrt{2}} \left( A_{1,\mu} - i A_{2,\mu} \right)$ & $\frac{g^2 v^2}{4}$ \\
$A_{2,\mu}$ & $W_\mu^\ast = \frac{1}{\sqrt{2}} \left( A_{1,\mu} + i A_{2,\mu} \right)$ & $\frac{g^2 v^2}{4}$ \\
\vspace{-0.5em} \\
$A_{3,\mu}$ & $Z_\mu = \cos \theta_W A_{3,\mu} - \sin \theta_W B_{\mu}$ & $\frac{g^2 v^2}{4 \cos^2 \theta_W}$ \\
$B_\mu$ & $A_\mu = \sin \theta_W A_{3,\mu} + \cos \theta_W B_{\mu}$ & -- \\
\bottomrule
\end{tabular}
\caption{The gauge fields prior and after the Higgs effect.}
\label{table:GaugeFieldPriorAndAfterHiggsEffect}
\end{table}

\subsection{Running Couplings} \label{subsec:RunningCouplings}

\paragraph{Successes and Open Questions of the Standard Model}
The \emph{standard model of particle physics} is an extraordinary successful framework. For example the electro-weak theory accurately describes experimental findings. In particular, the $W^\pm$ and $Z$ bosons were detected by Carlo Rubbia and Simon van der Meer, for which they were awarded the Nobel Prize in physics in 1984. The measured masses of these bosons are \cite{PhysRevD.86.010001}
\[ m_W = \SI[parse-numbers=false]{80.385 \pm 0.015}{\giga \electronvolt}, \qquad m_Z = \SI[parse-numbers=false]{91.1876 \pm 0.0021}{\giga \electronvolt} \, . \]
The Higgs boson was only detected a few years ago with a mass \cite{2015PhRvL.114s1803A}
\[ m_H = \SI[parse-numbers=false]{125.09 \pm 0.21 (stat) \pm 0.11 (syst)}{\giga \electronvolt} \, . \]
This lead to the Nobel Prize in physics for Peter Higgs \cite{1964PhRvL..13..508H} and Fran\c{c}ois Englert \cite{1964PhRvL..13..321E} in 2013.

Despite its successes, it also leaves open lots of questions. For example, the masses $m_W$, $m_Z$ or $m_H$ cannot be predicted from the theory, but rather must be inferred from measurements. For the entire \emph{standard model of particle physics}, including the strong interaction, at least 19 such input parameters are required. Why these parameters have the observed values cannot be answered by this framework.

Compared to Newton's law of universal gravity, which merely requires the \emph{gravitational constant} $g_N \simeq \SI[parse-numbers=false]{6.67 \cdot 10^{-11}}{\meter^3 \kilo \gram^{-1} \second^{-2}}$ as input parameter, 19 input parameters seem quite excessive. We can therefore wonder, if it is possible to formulate a physical theory which, at low energies, is equivalent to the \emph{standard model of particle physics}, but obtains corrections at high energies. An indication into that direction is obtained from taking a quantum field theory look at the coupling constants $g_S$ of the strong interaction, $g$ of the weak interaction and $e$ of electromagnetism -- all of them turn out to be energy dependent! This effect is not paralleled in classical physics and is termed \emph{running coupling}.

\paragraph{Basics of QED}

\begin{figure}[tb]
\begin{center}

\subfloat[No scattering.]{\label{figure-Bhabha1}
\includegraphics{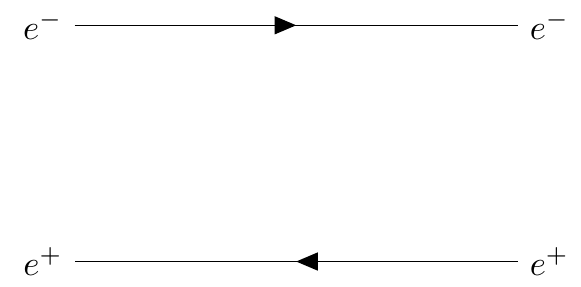}
}
\hspace{3em}
\subfloat[Vaccum transition.]{\label{figure-Bhabha2}
\includegraphics{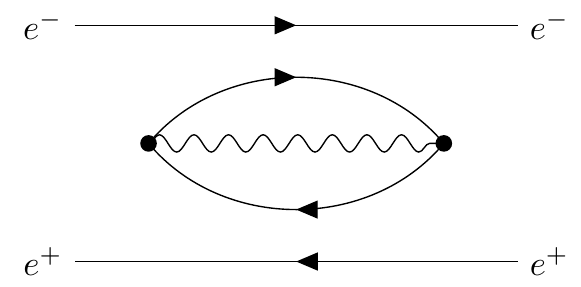}
}

\subfloat[Electron self-energy.]{\label{figure-Bhabha3}
\includegraphics{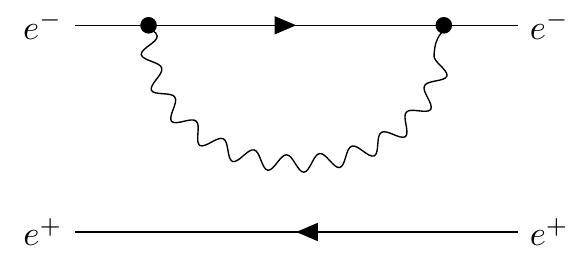}
}
\hspace{3em}
\subfloat[Proton self-energy.]{\label{figure-Bhabha4}
\includegraphics{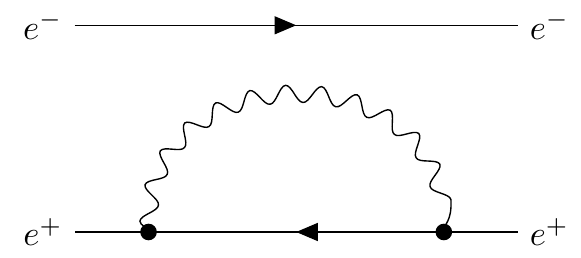}
}

\subfloat[Leading contribution to `real' scattering.]{\label{figure-Bhabha5}
\includegraphics{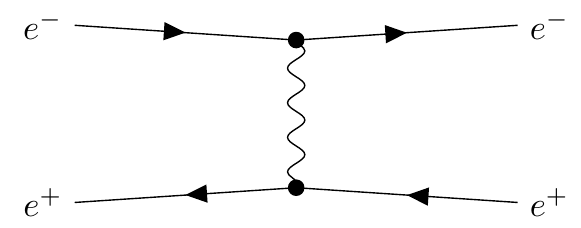}
}

\caption[Bhabha scattering to 2nd order in perturbation theory.]{Feynman diagrams to 2nd order in perturbation theory for Bhabha scattering.}
\label{figure:FeynmanTo2ndOrderOfBhabhaScattering}
\end{center}
\end{figure}

\begin{figure}[tb]
\begin{center}

\subfloat[No scattering.]{\label{figure-Photon1}
\includegraphics{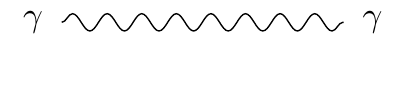} 
}
\hspace{2em}
\subfloat[Photon self-energy.]{\label{figure-Photon2}
\includegraphics{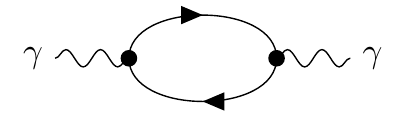} 
}
\hspace{2em}
\subfloat[Vacuum transition.]{\label{figure-Photon3}
\includegraphics{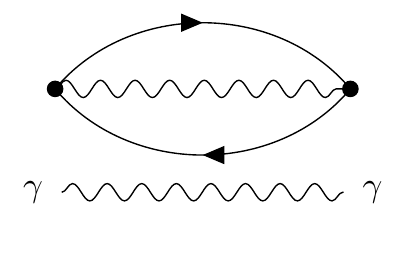} 
}

\caption[Propagation of a photon to 2nd order in perturbation theory.]{Feynman diagrams for photon progagation to 2nd order in perturbation theory.}
\label{figure:FeynmanTo2ndOrderOfPhotonPropagation}
\end{center}
\end{figure}

To obtain a basic idea on these running couplings, let us look at the quantum field description of the electromagnetic interaction -- quantum electrodynamics. In this framework, the photon mediates the electromagnetic force between electrically charged particles. The corresponding processes are instructively summarised by Feynman diagrams. We denote photons by wiggled lines and electrons by straight lines. The arrows along straight lines indicate the follow of negative electric charge. The annihilation of an electron $e^-$ and a positron $e^+$ to a photon $\gamma$ is then summarised by the following diagram:
\[
\begin{tikzpicture}[scale = 1.5]

\draw[particle] (-1.732,0.8) -- (0,0);
\draw[particle] (0,0) -- (-1.732,-0.8);
\draw[photon] (1.5,0) -- (0,0);

\fill[black] (0,0) circle (1.6pt);

\node[right] at (1.5,0) {$\gamma$};
\node[left] at (-1.732,0.8) {$e^-$};
\node[left] at (-1.732,-0.8) {$e^+$};

\end{tikzpicture}
\]
The perturbation theory of QED classifies Feynman diagrams according to the number of such interactions. For example for \emph{Bhabha scattering} of an electron $e^-$ and a positron $e^+$, all Feynman diagrams that contribute up to second order in perturbation theory are summarised in \cref{figure:FeynmanTo2ndOrderOfBhabhaScattering}. Naively we would consider \cref{figure-Bhabha1} and \cref{figure-Bhabha5} only. However there are three more such diagrams -- \cref{figure-Bhabha2} contributes to the vacuum energy and \cref{figure-Bhabha3}, \cref{figure-Bhabha4} give contributions to the \emph{self-energy} of the electron and positron respectively. Similarly, the propagation of a photon is altered by such diagrams. Up to second order in perturbation theory, the corresponding Feynman diagrams are given in
\cref{figure:FeynmanTo2ndOrderOfPhotonPropagation}.

\paragraph{An Overview of the Feynman Rules of QED}

The Feynman rules turn Feynman diagrams into algebraic expressions. Let us motivate this process by looking at electrons: 
\begin{itemize}
 \item In classical field theory on Minkowski spacetime $\mathcal{M}_{1,3}$ with metric $\eta_{\mu \nu}$, electrons are described as Dirac spinor fields $\psi$. These 
      are solutions to the Dirac equation
      \[ 0 = \left( \partial_\mu \gamma^\mu - m \cdot \mathds{1}_{4 \times 4}\right) \psi \,   \label{equ:DiracEquation} \]
      where $m$ is understood as mass of the field. The matrices $\gamma^0$, $\gamma^1$, $\gamma^2$, $\gamma^3$ furnish a representation of the Clifford algebra. Explicitly they are given by
      \[ \gamma^0 = \left( \begin{array}{cccc} \mathds{1}_{2 \times 2} & \\ & \mathds{1}_{2 \times 2} \end{array} \right) \, , \qquad \gamma^i =  \left( \begin{array}{cccc} & \sigma_i \\ - \sigma_i & \end{array} \right) \, \]
      where $\sigma_i$ are the Pauli matrices of \cref{equ:PauliMatrices}.
 \item In quantum field theory, a classical solution $\psi$ is turned into an operator valued function $\hat{\psi}$ on $\mathcal{M}_{1,3}$. For every $x^\mu \in 
      \mathcal{M}_{1,3}$, the operator $\hat{\psi} ( x^\mu )$ acts on a Hilbert space $\mathcal{H}$ which contains a vector $|0 \rangle$. In the absence of interactions with other particles, this vector $|0 \rangle$ is interpreted as the vacuum of the theory. The propagation of a \emph{non-interacting} electron from the spacetime point $x_1^\mu$ to $x_2^\mu$ is then measured by the \emph{Feynman propagator}
      \[ G \left( x_1^\mu - x_2^\mu \right) = \left. \left\langle \left. 0 \left. \left| T \left( \hat{\psi} \left( x_1^\mu \right) \hat{\overline{\psi}} \left( x_2^\mu \right) \right) \right. \right| 0 \right. \right. \right\rangle \, , \]
      where $\overline{\psi} = \psi^{\dagger} \gamma^0$ and $T$ is the so-called \emph{time-ordering operator}. One can show that \footnote{The parameter $\epsilon$ comes from a particular type of integration in the complex plane. Depending on the chosen path, this expression can then describe retarded, advanced or causal Feynman propagators. Once the dust has settled, the limit $\epsilon \to 0$ is considered. Thereby, this parameter disappears from physical observables. More details can be found in \cite{mandl1993quantum, weinberg1995quantum, peskin1995introduction}.}
      \[ G \left( x_1^\mu - x_2^\mu \right) = \Mint_{\mathcal{M}_{1,3}}{\frac{d^4p}{\left( 2 \pi \right)^4} \cdot \frac{i \left( p_\mu \gamma^\mu + m \right)}{p^2 - m^2 + i \epsilon} \cdot e^{-i p_\mu \left( x_1^\mu - x_2^\mu \right) }} \, . \]
 \item The momentum space Feynman rule, finally states to identify the propagation of an (internal) electron of momentum $p^\mu$ with the factor $\frac{i ( p_\mu 
      \gamma^\mu + m )}{p^2 - m^2 + i \epsilon}$. The exponentials $e^{-i p_\mu ( x_1^\mu - x_2^\mu )}$ lead to delta distributions, which eventually ensure the classical momentum conservation laws at interaction vertices.
\end{itemize}
Along these lines one derives the Feynman rules for QED. Among others these state \cite{mandl1993quantum}:
\begin{subequations}
\label{equ:FeymnanRules}
\begin{align}
\text{internal fermionic line of momentum $p^\mu$} \qquad & \leftrightarrow \qquad \frac{i \left( p_\mu \gamma^\mu + m \right)}{p^2 - m^2 + i \epsilon} \, , \\
\text{internal photonic line of momentum $k^\mu$} \qquad & \leftrightarrow \qquad - \frac{i \eta_{\mu \nu}}{k^2 + i \epsilon} \, , \\
\text{vertex} \qquad & \leftrightarrow \qquad - i e \gamma^{\mu} \, , \\
\text{closed fermionic loop with momentum $p^\mu$} \qquad & \leftrightarrow \qquad (-1) \Mint_{\mathcal{M}_{1,3}}{\frac{d^4p}{\left( 2 \pi \right)^4}} \, .
\end{align}
\end{subequations}

\paragraph{A First Encounter with Renormalization}

Crucially, the parameter $e$ and $m$ above are charge and mass of electrons as they enter the Lagrangian. These are called the \emph{bare parameters}. Whilst classically we would directly infer to measure these values in experiments, the conversion between bare parameters and experimentally observed quantities is far from 1-to-1. We exemplify this by looking at the propagation of a photon.\footnote{We follow the analysis in \cite{peskin1995introduction} closely.} We ignore the vacuum transition in \cref{figure-Photon3}. According to the Feynman rules, these diagrams then correspond to the algebraic expression
\[ \mathcal{A} = - \frac{i \eta_{\mu \nu}}{k^2 + i \epsilon} + \left( - \frac{i \eta_{\mu \alpha}}{k^2 + i \epsilon} \right) \cdot \left( i \Pi_2^{\alpha \beta} \left( k^\mu \right) \right) \cdot \left( - \frac{i \eta_{\beta \nu}}{k^2 + i \epsilon} \right) + \mathcal{O} \left( e^4 \right) \label{equ:AToFirstOrder} \]
where $i \Pi_2^{\alpha \beta} ( k^\mu )$ corresponds to the closed fermion loop in \cref{figure-Photon2}. We will find momentarily that this expression is logarithmically divergent -- it is an example of the famous infinities encountered in quantum field theories. This divergence originates from electrons which run with very high momenta $q^\mu$ in this loop. A first approach to resolve such divergences is to focus on electrons with energies below a certain energy $\Lambda$. This is called \emph{cut-off regularisation}. In the case at hand it unfortunately leads to an infinite photon mass \cite{peskin1995introduction}. A procedure which fits physical desires better is \emph{dimensional regularisation}. In this case one computes the integral formally in $4 + \epsilon$ dimensions. Thereby, one obtains an expression $\Pi_2^{\alpha \beta} ( k^\mu, \epsilon )$ which is divergent for $\epsilon \to 0$. As this is a rather technical task, suffice it for now to quote from \cite{peskin1995introduction, mandl1993quantum, weinberg1995quantum} that one finds $\Pi_2^{\alpha \beta} ( k^\mu ) = ( k^2 \eta^{\alpha \beta} - k^\alpha k^\beta ) \Pi_2 ( k^2 )$, which allows us to write
\[ \mathcal{A} = \frac{-i}{k^2 + i \epsilon} \cdot \left( \frac{\eta_{\mu \nu} - \frac{k_\mu k_\nu}{k^2} }{1 - \Pi_2 \left( k^2 \right)} \right) + \frac{-i}{k^2 + i \epsilon} \cdot \left( \frac{k_\mu k_\nu}{k^2} \right) + \mathcal{O} \left( e^4 \right) \, . \]
As a consequence of the Ward–Takahashi identities we may ignore the terms that contain $k_\mu$ -- see \eg \cite[chapter 5]{mandl1993quantum} for more details. Hence, we have
\[ \mathcal{A} = \frac{-i \eta_{\mu \nu}}{k^2 + i \epsilon} \cdot \left( \frac{1}{1 - \Pi_2 \left( k^2 \right)} + \mathcal{O} \left( e^4 \right) \right) \, . \]
This expression describes the propagator of a physical photon. As just argued, to 2nd order in perturbation theory, it contains the interaction with two electrons, which run in the loop displayed in \cref{figure-Photon2}. Therefore, we can interpret
\[ e_{\text{exp}}^2 \left( k^2 \right) = e^2 \cdot \left( \frac{1}{1 - \Pi_2 \left( k^2 \right)} + \mathcal{O} \left( e^4 \right) \right) \label{equ:RenormalisedCharge} \]
as the electric charge measured in experiments. It is the so-called \emph{renormalized charge}.

\paragraph{Renormalised Perturbation Theory}

\begin{figure}[tbp]
\begin{center}

\subfloat[Electron self-energy.]{\label{figure-electron-self-energy}
\includegraphics{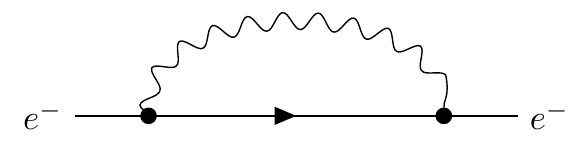} 
}
\hspace{3em}
\subfloat[Vertex modification.]{\label{figure-vertex-modification}
\includegraphics{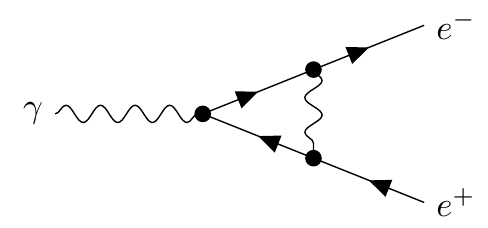} 
}

\caption[Charge renormalization from electron self-energy and vertex modification.]{Charge renormalization from electron self-energy and vertex modification.}
\label{figure:ChargeRenormalisationOthers}
\end{center}
\end{figure}

A systematic study of such renormalization processes leads to the so-called \emph{renormalized perturbation theory}. In particular, it is found that the electron self-energy diagram in \cref{figure-electron-self-energy} and a vertex modification in \cref{figure-vertex-modification} give contributions to the renormalized charge. Fortunately, in QED one can show that these contributions cancel exactly \cite{peskin1995introduction, weinberg1995quantum, mandl1993quantum}, and so \cref{equ:RenormalisedCharge} is the final result for the renormalized electric charge.

\paragraph{From Regularisation to the Beta Function}

Now we come back to compute $\Pi^{\alpha \beta}_2 (  k^\mu, \Lambda )$. To this end, we employ the Feynman rules \cref{equ:FeymnanRules}. They imply
\begin{align}
\begin{split}
i \Pi^{\alpha \beta}_2 \left( k^\mu \right) &= - e^2 \Mint_{\mathcal{M}_{1,3}}{\frac{d^4q}{\left( 2 \pi \right)^4} \cdot \frac{\mathrm{Tr} \left[ \left( q_\mu \gamma^\mu + m \right) \gamma^\alpha \left( q_\nu \gamma^\nu + k_\nu \gamma^\nu + m \right) \gamma^\beta \right]}{\left( q^2 - m^2 + i \epsilon \right) \cdot \left( \left( q + k \right)^2 - m^2 + i \epsilon \right)} } \, .
\end{split}
\end{align}
This expression is logarithmically divergent and can be regularised by cut-off regularisation. Another such means is \emph{dimensional regularisation}, which amounts to solving this integral \emph{formally} in $(4 + \eta)$-dimensions where $\eta$ is a \emph{real} parameter. Along these lines one finds \cite{peskin1995introduction}
\begin{align}
\begin{split}
i \Pi^{\alpha \beta}_2 \left( k^\mu, \epsilon \right) &= \left( k^2 \eta^{\alpha \beta} - k^\alpha k^\beta \right) \cdot i \Pi_2 \left( k^2 , \epsilon \right) \, , \\
\Pi_2 \left( k^2 , \epsilon \right) &= - \frac{2 \alpha}{6 \pi} \left( \frac{2}{\epsilon} - \gamma + \mathrm{ln} \left( 4 \pi \right) \right) + \frac{2 \alpha}{\pi} \cdot \Mint_{0}^{1}{dx \, x \cdot \left( 1 - x \right) \cdot \mathrm{ln} \left( m^2 - k^2 x \cdot \left( 1 - x \right) \right)} \\
                                          &= - \frac{2 \alpha}{3 \pi \epsilon} + \Delta \left( k^2 \right) \, .
\end{split}
\end{align}
In this expression $\alpha = e^2 / 4 \pi$ is the fine structure constant. The constant $\gamma \sim 0.577$ is known as the \emph{Euler-Mascheroni constant}. We recover the original divergence of this expression in the limit $\epsilon \to 0$. Thereby, we can rewrite \cref{equ:RenormalisedCharge} to order $\alpha$ as
\[ \alpha_{\text{exp}} \left( k^2 \right) \simeq \frac{\alpha}{1 - \left( \Pi_2 \left( k^2 \right) - \Pi_2 \left( 0 \right) \right)} = \frac{\alpha}{1 + \frac{2 \alpha}{\pi} \Mint_{x = 0}^{1}{dx \, x \cdot \left( 1 - x \right) \cdot \mathrm{log} \left( \frac{m^2}{m^2 - x \cdot \left( 1 - x \right) \cdot k^2} \right) } } \, , \label{equ:RenormalisedChargeII} \]
which does not not depend on $\epsilon$ any more. 

For a physical photon, $k^2 = 0$. However, in the expressions this is not the case. One expresses this observation by saying that this photon is \emph{not on shell}, or equivalently terms the photon in question a \emph{virtual photon}. That said, let us consider the relativistic limit of $k^2$. This corresponds to high velocities or equivalently to $- k^2 \gg m^2$. Under this assumption we can evaluate \cref{equ:RenormalisedChargeII} further. As pointed out in \cite{peskin1995introduction}, this yields
\[ \alpha_{\text{exp}} \left( k^2 \right) \simeq \frac{\alpha}{1 - \frac{\alpha}{3\pi} \cdot \left[ \mathrm{log} \left( \frac{-k^2}{m^2} \right) - \frac{5}{3} \right]} = \frac{\alpha}{1 - \frac{\alpha}{3\pi} \cdot \mathrm{log} \left( \frac{-k^2}{A m^2} \right)} \]
where $A = \mathrm{exp} ( 5/3 )$. So to leading order the coupling constant increases in this limit with
\[ \beta \left( \Lambda \right) = \left( \frac{\partial \alpha_{\text{exp}}}{\partial \text{log} \left( -k^2 \right)} \right) \left( \Lambda \right) \simeq \frac{2 \alpha^2}{3 \pi} \, . \label{equ:BetaForQED} \]
This is the well-known result for the $\beta$-function of QED.

\paragraph{Running Couplings in the Standard Model}

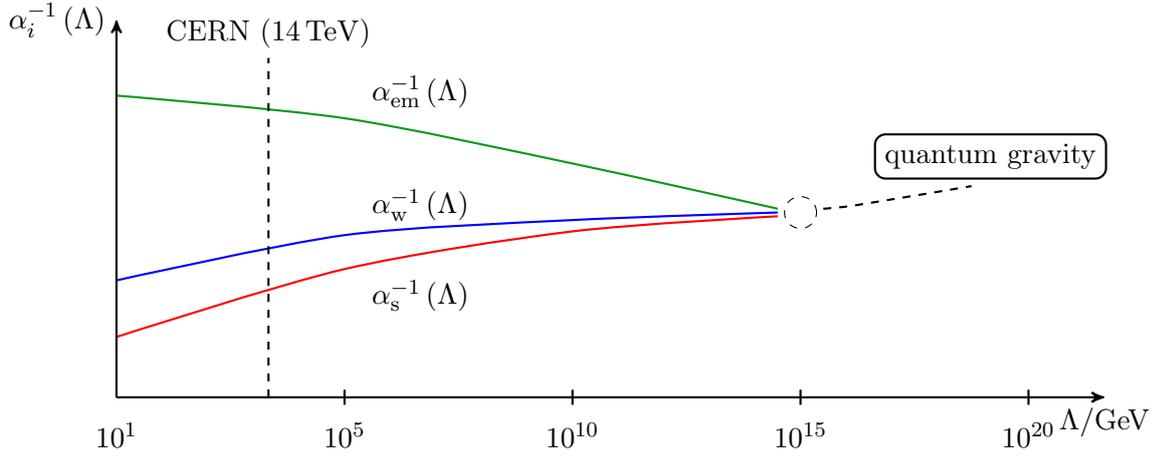
\begin{figure}[tbp]
\centering
\begin{tikzpicture}[thick,>=stealth',dot/.style = {draw,fill = white,circle,inner sep = 0pt,minimum size = 4pt}]
  
  \draw (3,0.1)--(3,-0.1);
  \draw (6,0.1)--(6,-0.1);
  \draw (9,0.1)--(9,-0.1);
  \draw (12,0.1)--(12,-0.1);
  \node[below] at (0,-0.2) {$10^1$};
  \node[below] at (3,-0.2) {$10^5$};
  \node[below] at (6,-0.2) {$10^{10}$};
  \node[below] at (9,-0.2) {$10^{15}$};
  \node[below] at (12,-0.2) {$10^{20}$};

  \node[below,draw,rounded corners] at (11.5,3.5) {quantum gravity};
  
  \draw[thin, dashed] (9,2.45) circle (6pt);

  \draw[darkgreen] plot[smooth] coordinates {(0,4) (3,3.7) (6,3.1) (8.7,2.5)};
  \draw[blue] plot[smooth] coordinates {(0,1.55) (3,2.15) (6,2.35) (8.7,2.45)};
  \draw[red] plot[smooth] coordinates {(0,0.8) (3,1.7) (6,2.2) (8.7,2.4)};

  \draw[dashed] plot[smooth] coordinates {(9.3,2.5) (9.75,2.55) (11.25,2.8)};
  
  \node[above] at (4,3.7) {$\alpha_{\mathrm{em}}^{-1} \left( \Lambda \right)$};
  \node[above] at (4,2.2) {$\alpha_{\mathrm{w}}^{-1} \left( \Lambda \right)$};
  \node[below] at (4,1.7) {$\alpha_{\mathrm{s}}^{-1} \left( \Lambda \right)$};
  
  \draw[dashed] (2,0) -- (2,4.5);
  \node[above] at (2,4.5) {CERN ($\SI[parse-numbers=false]{14}{\tera \electronvolt}$)};
  
  \coordinate (O) at (0,0);
  \draw[->] (0,0) -- (13,0) coordinate[label = {below:$\Lambda / \text{GeV}$}] (xmax);
  \draw[->] (0,0) -- (0,5) coordinate[label = {left:$\alpha_i^{-1} \left( \Lambda \right)$}] (ymax);
  
\end{tikzpicture}
\caption[Schematic picture of gauge coupling unification in the MSSM.]{Schematic picture of gauge coupling unification in the (minimally supersymmetric) \emph{standard model}. Note that at the Planck scale $\Lambda \sim \SI[parse-numbers=false]{10^{19}}{\giga \electronvolt}$ aspects of quantum gravity become relevant. We will come to discuss this in more detail in \cref{sec:StringTheory}.}
\label{figure-GaugeCouplingUnification}
\end{figure}

One can repeat the above analysis for the coupling strength of the weak and the strong interaction in the \emph{standard model of particle physics}. Thereby, it is found that the couplings tend to yield similar values at an energy scale $\Lambda_{\mathrm{GUT}} \sim \SI[parse-numbers=false]{10^{16}}{\giga \electronvolt}$. Schematically, this is pictured in \cref{figure-GaugeCouplingUnification}. \footnote{A more accurate such picture is given in \cite{peskin1995introduction}.} For a minimal supersymmetric extension of the \emph{standard model}, \ie models which include a supersymmetry partner for all matter particles, this matching becomes even better \cite{1981PhLB..105..439I, 1982PhRvD..25.3092M, 1981PhRvD..24.1681D}.

We use this observation as motivation to study \emph{grand unified theories} (GUTs). The energy scale $\Lambda_{\mathrm{GUT}} \sim \SI[parse-numbers=false]{10^{16}}{\giga \electronvolt}$ is known as the \emph{GUT-scale}. Such a model is typically governed by a Yang Mills theory with coupling constant $g$. The gauge group of such a GUT model is larger than $G_{\mathrm{SM}}$ and contains the latter as subgroup. Examples include the Pati-Salam model with $G_{\mathrm{GUT}} = SU(4) \times SU(2) \times SU(2)$ \cite{1994spas.book..343P} and the Georgi-Glashow model with $G_{\mathrm{GUT}} = SU(5)$ \cite{1974PhRvL..32..438G}. As latter will be important later in this thesis, let us discuss this type of GUT in more detail.

\subsection{The Georgi-Glashow Model} \label{subsec:Georgi-Glashow}

This model was originally proposed by Howard Georgi and Sheldon Glashow in \cite{1974PhRvL..32..438G}. It is based on a Yang-Mills theory which admits an $SU(5)$ gauge  symmetry.

\paragraph{Embedding of $\mathbf{G_{\text{SM}}}$ into $\mathbf{SU(5)}$}

Ultimately, our goal is to break the $SU(5)$ gauge theory down to the \emph{standard model} gauge group $G_{\mathrm{SM}}$. This process is very similar to the electro-weak symmetry breaking, which we discussed in \cref{subsec:ElectroWeakBreaking}. To this we first have to specify an embedding of $G_{\mathrm{SM}}$ into $SU(5)$. Since the exponential map for $SU(5)$ is surjective, we may as well look at the Lie algebra $\mathfrak{su} ( 5 )$ and embed $\mathfrak{su} ( 3 ) \oplus \mathfrak{su} ( 2 ) \oplus \mathfrak{u} ( 1 )$ into it. Let us recall that $\mathfrak{su}(5)$ has 24 generators $T^a$, each of which is a traceless and Hermitian $5 \times 5$-matrix. We normalise the generators such that $\mathrm{tr} ( T^a T^b ) = \frac{1}{2} \delta^{ab}$. Then we can write $T^a = \frac{1}{2} \lambda^a$ and the matrices $\lambda^a$ are as follows:
\begin{itemize}
 \item $T^1$ generates a $U(1)$ subgroup of $SU(5)$ and $\lambda^1 = \frac{1}{\sqrt{15}} \cdot \mathrm{diag} ( -2, -2, -2, 3, 3 )$.
 \item The generators $T^2$, $T^3$, $T^4$ exponentiate to furnish an $SU(2)$ subgroup of $SU(5)$, where
      $$
      \begin{array}{ccc}
      \lambda^2 = \begin{psmallmatrix} \cdot & \cdot & \cdot & \cdot & \cdot \\ \cdot & \cdot & \cdot & \cdot & \cdot \\ \cdot & \cdot & \cdot & \cdot & \cdot \\ \cdot & \cdot & \cdot & \cdot & 1 \\ \cdot & \cdot & \cdot & 1 & \cdot \end{psmallmatrix} \, ,
      &
      \lambda^3 = \begin{psmallmatrix} \cdot & \cdot & \cdot & \cdot & \cdot \\ \cdot & \cdot & \cdot & \cdot & \cdot \\ \cdot & \cdot & \cdot & \cdot & \cdot \\ \cdot & \cdot & \cdot & \cdot & -i \\ \cdot & \cdot & \cdot & i & \cdot \end{psmallmatrix} \, ,
      &
      \lambda^4 = \begin{psmallmatrix} \cdot & \cdot & \cdot & \cdot & \cdot \\ \cdot & \cdot & \cdot & \cdot & \cdot \\ \cdot & \cdot & \cdot & \cdot & \cdot \\ \cdot & \cdot & \cdot & 1 & \cdot \\ \cdot & \cdot & \cdot & \cdot & -1 \end{psmallmatrix} \, .
      \end{array}
      $$
 \item The generators $T^5$, \dots, $T^{12}$ generate an $SU(3)$ subgroup of $SU(5)$, where
      $$
      \begin{array}{cccl}
      
      \lambda^5 = \begin{psmallmatrix} \cdot & 1 & \cdot & \cdot & \cdot \\ 1 & \cdot & \cdot & \cdot & \cdot \\ \cdot & \cdot & \cdot & \cdot & \cdot \\ \cdot & \cdot & \cdot & \cdot & \cdot \\ \cdot & \cdot & \cdot & \cdot & \cdot \end{psmallmatrix} \, ,

      &

      \lambda^6 = \begin{psmallmatrix} \cdot & -i & \cdot & \cdot & \cdot \\ i & \cdot & \cdot & \cdot & \cdot \\ \cdot & \cdot & \cdot & \cdot & \cdot \\ \cdot & \cdot & \cdot & \cdot & \cdot \\ \cdot & \cdot & \cdot & \cdot & \cdot \end{psmallmatrix} \, ,

      &

      \lambda^7 = \begin{psmallmatrix} 1 & \cdot & \cdot & \cdot & \cdot \\ \cdot & -1 & \cdot & \cdot & \cdot \\ \cdot & \cdot & \cdot & \cdot & \cdot \\ \cdot & \cdot & \cdot & \cdot & \cdot \\ \cdot & \cdot & \cdot & \cdot & \cdot \end{psmallmatrix} \, ,

      & 
      
      \lambda^8 = \begin{psmallmatrix} \cdot & \cdot & 1 & \cdot & \cdot \\ \cdot & \cdot & \cdot & \cdot & \cdot \\ 1 & \cdot & \cdot & \cdot & \cdot \\ \cdot & \cdot & \cdot & \cdot & \cdot \\ \cdot & \cdot & \cdot & \cdot & \cdot \end{psmallmatrix} \, ,

      \\ \vspace{-0.5em} & \\

      \lambda^9 = \begin{psmallmatrix} \cdot & \cdot & -i & \cdot & \cdot \\ \cdot & \cdot & \cdot & \cdot & \cdot \\ i & \cdot & \cdot & \cdot & \cdot \\ \cdot & \cdot & \cdot & \cdot & \cdot \\ \cdot & \cdot & \cdot & \cdot & \cdot \end{psmallmatrix} \, ,

      &

      \lambda^{10} = \begin{psmallmatrix} \cdot & \cdot & \cdot & \cdot & \cdot \\ \cdot & \cdot & 1 & \cdot & \cdot \\ \cdot & 1 & \cdot & \cdot & \cdot \\ \cdot & \cdot & \cdot & \cdot & \cdot \\ \cdot & \cdot & \cdot & \cdot & \cdot \end{psmallmatrix} \, ,

      &
      
      \lambda^{11} = \begin{psmallmatrix} \cdot & \cdot & \cdot & \cdot & \cdot \\ \cdot & \cdot & -i & \cdot & \cdot \\ \cdot & i & \cdot & \cdot & \cdot \\ \cdot & \cdot & \cdot & \cdot & \cdot \\ \cdot & \cdot & \cdot & \cdot & \cdot \end{psmallmatrix} \, ,

      &

      \lambda^{12} = \frac{\begin{psmallmatrix} 1 & \cdot & \cdot & \cdot & \cdot \\ \cdot & 1 & \cdot & \cdot & \cdot \\ \cdot & \cdot & -2 & \cdot & \cdot \\ \cdot & \cdot & \cdot & \cdot & \cdot \\ \cdot & \cdot & \cdot & \cdot & \cdot \end{psmallmatrix}}{\sqrt{3}} \, .

      \end{array}
      $$
 \item The remaining twelve generators are given in terms of the following matrices:
      $$
      \begin{array}{cccc}
      
      \lambda^{13} = \begin{psmallmatrix} \cdot & \cdot & \cdot & 1 & \cdot \\ \cdot & \cdot & \cdot & \cdot & \cdot \\ \cdot & \cdot & \cdot & \cdot & \cdot \\ 1 & \cdot & \cdot & \cdot & \cdot \\ \cdot & \cdot & \cdot & \cdot & \cdot \end{psmallmatrix} \, ,

      &

      \lambda^{14} = \begin{psmallmatrix} \cdot & \cdot & \cdot & -i & \cdot \\ \cdot & \cdot & \cdot & \cdot & \cdot \\ \cdot & \cdot & \cdot & \cdot & \cdot \\ i & \cdot & \cdot & \cdot & \cdot \\ \cdot & \cdot & \cdot & \cdot & \cdot \end{psmallmatrix} \, ,

      &

      \lambda^{15} = \begin{psmallmatrix} \cdot & \cdot & \cdot & \cdot & 1 \\ \cdot & \cdot & \cdot & \cdot & \cdot \\ \cdot & \cdot & \cdot & \cdot & \cdot \\ \cdot & \cdot & \cdot & \cdot & \cdot \\ 1 & \cdot & \cdot & \cdot & \cdot \end{psmallmatrix} \, ,
      
      &
      
      \lambda^{16} = \begin{psmallmatrix} \cdot & \cdot & \cdot & \cdot & -i \\ \cdot & \cdot & \cdot & \cdot & \cdot \\ \cdot & \cdot & \cdot & \cdot & \cdot \\ \cdot & \cdot & \cdot & \cdot & \cdot \\ i & \cdot & \cdot & \cdot & \cdot \end{psmallmatrix} \, ,

      \\ \vspace{-0.5em} & \\
     
      \lambda^{17} = \begin{psmallmatrix} \cdot & \cdot & \cdot & \cdot & \cdot \\ \cdot & \cdot & \cdot & 1 & \cdot \\ \cdot & \cdot & \cdot & \cdot & \cdot \\ \cdot & 1 & \cdot & \cdot & \cdot \\ \cdot & \cdot & \cdot & \cdot & \cdot \end{psmallmatrix} \, ,

      &

      \lambda^{18} = \begin{psmallmatrix} \cdot & \cdot & \cdot & \cdot & \cdot \\ \cdot & \cdot & \cdot & -i & \cdot \\ \cdot & \cdot & \cdot & \cdot & \cdot \\ \cdot & i & \cdot & \cdot & \cdot \\ \cdot & \cdot & \cdot & \cdot & \cdot \end{psmallmatrix} \, ,
      
      &
      
      \lambda^{19} = \begin{psmallmatrix} \cdot & \cdot & \cdot & \cdot & \cdot \\ \cdot & \cdot & \cdot & \cdot & 1 \\ \cdot & \cdot & \cdot & \cdot & \cdot \\ \cdot & \cdot & \cdot & \cdot & \cdot \\ \cdot & 1 & \cdot & \cdot & \cdot \end{psmallmatrix} \, ,

      &

      \lambda^{20} = \begin{psmallmatrix} \cdot & \cdot & \cdot & \cdot & \cdot \\ \cdot & \cdot & \cdot & \cdot & -i \\ \cdot & \cdot & \cdot & \cdot & \cdot \\ \cdot & \cdot & \cdot & \cdot & \cdot \\ \cdot & i & \cdot & \cdot & \cdot \end{psmallmatrix} \, ,

      \\ \vspace{-0.5em} & \\

      \lambda^{21} = \begin{psmallmatrix} \cdot & \cdot & \cdot & \cdot & \cdot \\ \cdot & \cdot & \cdot & \cdot & \cdot \\ \cdot & \cdot & \cdot & 1 & \cdot \\ \cdot & \cdot & 1 & \cdot & \cdot \\ \cdot & \cdot & \cdot & \cdot & \cdot \end{psmallmatrix} \, ,
      
      &
      
      \lambda^{22} = \begin{psmallmatrix} \cdot & \cdot & \cdot & \cdot & \cdot \\ \cdot & \cdot & \cdot & \cdot & \cdot \\ \cdot & \cdot & \cdot & -i & \cdot \\ \cdot & \cdot & i & \cdot & \cdot \\ \cdot & \cdot & \cdot & \cdot & \cdot \end{psmallmatrix} \, ,

      &

      \lambda^{23} = \begin{psmallmatrix} \cdot & \cdot & \cdot & \cdot & \cdot \\ \cdot & \cdot & \cdot & \cdot & \cdot \\ \cdot & \cdot & \cdot & \cdot & 1 \\ \cdot & \cdot & \cdot & \cdot & \cdot \\ \cdot & \cdot & 1 & \cdot & \cdot \end{psmallmatrix} \, ,

      &

      \lambda^{24} = \begin{psmallmatrix} \cdot & \cdot & \cdot & \cdot & \cdot \\ \cdot & \cdot & \cdot & \cdot & \cdot \\ \cdot & \cdot & \cdot & \cdot & -i \\ \cdot & \cdot & \cdot & \cdot & \cdot \\ \cdot & \cdot & i & \cdot & \cdot \end{psmallmatrix} \, .
      
      \end{array}
      $$
\end{itemize}
Thus, we employ $T^a$, $1 \leq a \leq 12$, to generate the $SU(3) \times SU(2) \times U(1)$ subgroup of $SU(5)$.

\paragraph{The Lagrangian}

The $SU(5)$ gauge field is given by $\mathbf{A}_\mu \mathbf{T} = A^a_\mu T^a$ and the field strength satisfies
\[ \mathbf{F}_{\mu \nu} = \partial_\mu \mathbf{A}_\nu - \partial_\nu \mathbf{A}_\mu + i g \sum_{b,c = 1}^{24}{A^b_\mu A^c_\nu \left[ T^b, T^c \right]} \, . \]
Consequently, the dynamics of the underlying classical field theory is governed by the Lagrangian
\[ \mathcal{L}_{\mathrm{gauge}} = - \frac{1}{2} \cdot \mathrm{tr} \left( \mathbf{F}_{\mu \nu} \mathbf{F}^{\mu \nu} \right) \label{equ:LagrangianGUT} \, . \]

\paragraph{Gauge Group Breaking}

To break the gauge invariance of this classical Yang-Mills theory along the embedding
\[ \iota \colon SU \left( 3 \right) \times SU \left( 2 \right) \times U \left( 1 \right) \hookrightarrow SU \left( 5 \right) \]
we follow closely to the discussion of the Higgs effect in \cref{subsec:ElectroWeakBreaking}. In the case at hand, our `Higgs doublet' $\phi$ is a field in the adjoint representation of $SU(5)$. It can be written as
\[ \phi \left( x \right) = \sum_{a = 1}^{24}{\phi^a \left( x \right) T^a} \, . \]
where the fields $\phi^a$ are real scalar fields. The entire system is now governed by the Lagrangian $\mathcal{L}_{\mathrm{total}} = \mathcal{L}_{\mathrm{gauge}} + \mathcal{L}_\phi$ where
\[ \mathcal{L}_\phi = \frac{1}{2} \cdot \mathrm{tr} \left( \left( D_\mu \phi \right) \left( D^\mu \phi \right) \right) - V \left( \phi \right) \, , \qquad D_\mu \phi = \partial_\mu \phi -i g A^a_\mu \left[ T^a, \phi \right] \, .  \]
We look at the potential $V( \phi )$ given by
\[ V \left( \phi \right) = - \frac{\mu^2}{2} \cdot \text{tr} \left( \phi^2 \right) + \frac{\lambda_1}{4} \cdot \mathrm{tr} \left( \phi^4 \right) + \frac{\lambda_2}{4} \cdot \mathrm{Tr} \left( \phi^2 \right)^2 \, . \]
Not only does it enjoy an $SU(5)$ symmetry, but it even respects the additional $\mathbb{Z}_2$-symmetry $\phi \leftrightarrow - \phi$. Indeed, this is the only potential $V( \phi )$ which has these symmetries and in addition leads to a renormalizable quantum field theory. The latter means that we can follow the example of QED and `absorb' all divergences into conversion factors between the bare parameters of the Lagrangian and the measured physical properties.

An explicit inspection shows that $[ T^1, T^a ] = 0$ only for $1 \leq a \leq 12$. Therefore, the ground state $\phi_0 = v \cdot T^1 $ is left invariant only by operations with the subgroup $SU(3) \times SU(2) \times U(1)$ of $SU(5)$. Consequently, the Higgs effect generates masses for the gauge fields $A_\mu^a$ with $13 \leq a \leq 24$. The corresponding mass terms again stem from the kinetic Lagrangian for $\phi$, namely
\begin{align}
\begin{split}
\mathcal{L}_\phi &\supset \frac{1}{2} \left( - ig \right)^2 \mathrm{tr} \left( \sum_{a = 1}^{24}{A^a_\mu \left[ T^a, \phi_0 \right]} \cdot \sum_{b = 1}^{24}{A^b_\mu \left[ T^b, \phi_0 \right]} \right) \\
                 &\supset - \sum_{a = 1}^{24}{ \frac{g^2}{2} \cdot \mathrm{tr} \left( \left[ T^a, \phi_0 \right]^2 \right) \cdot A^a_\mu A^{a \mu} } \, .
\end{split}
\end{align}
This shows that $A^a_\mu$ ($13 \leq a \leq 24$) acquires mass $m_a^2 = g^2 \mathrm{tr} ( [ T^a, \phi_0 ]^2 )$, but the gauge fields of the $SU(3) \times SU(2) \times U(1)$ subgroup of $SU(5)$ remain massless.

\paragraph{Reduction of representations}

The particles of the \emph{standard model}, as discussed in \cref{subsec:OverviewOfStandardModel}, can be organised in the representations $\mathbf{\overline{5}}$, $\mathbf{10}$ and $\mathbf{1}$ of $SU(5)$. Let us first look at the anti-fundamental representation $\mathbf{\overline{5}}$ and write a vector $\psi$ in this representation as
\[ \psi = \left( \begin{array}{c} d^c \\ s^c \\ b^c \\ \hline e \\ \nu_e \end{array} \right) \, . \label{equ:DecompositionOfOverline5} \]
Recall that $^c$ denotes charge conjugation. Therefore, the first three entries denote the first generations of the three down quarks, which are right-handed and charge conjugated in this description. Indeed we have the decomposition $\mathbf{\overline{5}} \to \left( \mathbf{\overline{3}}, \mathbf{1} \right)_{1/3} \oplus \left( \mathbf{1}, \mathbf{2} \right)_{-1/2}$ into representations of $G_{\mathbf{SM}}$ \cite{1974PhRvL..32..438G}. By comparing with \cref{my_old_table_2} we identify these as $D^c$ and $L$. Likewise the anti-symmetric rank-2 tensor representation $\mathbf{10}$ decomposes. Note that a general such tensor $T_{ab}$ can be written as
\[ T_{ab} = \left( \begin{array}{ccc|cc} 
0 & t^c & - c^c & - u & - d \\
- t^c & 0 & u^c & - c & - s \\
c^c & - u^c & 0 & - t & - b \\
\hline
u & c & t & 0 & - e^c \\
d & s & b & e^c & 0
\end{array} \right) \, . \]
Indeed it can be shown \cite{1974PhRvL..32..438G}:
\[ \mathbf{10} \to \left( \mathbf{3}, \mathbf{2} \right)_{1/6} \oplus \left( \mathbf{\overline{3}}, \mathbf{1} \right)_{-2/3} \oplus \left( \mathbf{1}, \mathbf{1} \right)_1 \]
and from \cref{my_old_table_2} these representations are found to represent $Q$, $U^c$ and $E^c$. Finally,the singlet representation $\mathbf{1}$ of $SU(5)$ turns into $\left( \mathbf{1}, \mathbf{1} \right)_0$ representing right-handed neutrinos $N^c$. In this sense the representation $\overline{\mathbf{5}} \oplus \mathbf{10} \oplus \mathbf{1}$ of $SU(5)$ contains precisely one generation of \emph{standard model} fermions. We summarise these findings in \cref{table-BrokenSpectrum}.

\begin{table}
\centering
\begin{tabular}{ccc}
\toprule
Rep. of $SU(5)$ & Rep. of $G_{\mathbf{SM}}$ & \emph{standard model} particles \\
\midrule
$\overline{\mathbf{5}}$ & $\left( \overline{\mathbf{3}}, \mathbf{1} \right)_{1/3} \oplus \left( \mathbf{1}, \mathbf{2} \right)_{-1/2}$ & $D^c \oplus L$ \\
$\mathbf{5}$ & $\left( \mathbf{3}, \mathbf{1} \right)_{-1/3} \oplus \left( \mathbf{1}, \mathbf{2} \right)_{1/2}$ & $D \oplus L^c$ \\
\vspace{-0.5em} & \\
$\mathbf{10}$ & $\left( \mathbf{3}, \mathbf{2} \right)_{1/6} \oplus \left( \mathbf{\overline{3}}, \mathbf{1} \right)_{-2/3} \oplus \left( \mathbf{1}, \mathbf{1} \right)_{1}$ & $Q \oplus U^c \oplus E^c$ \\
\vspace{-0.5em} & \\
$\mathbf{1}$ & $\left( \mathbf{1}, \mathbf{1} \right)_0$ & $N$ \\
\midrule
$\mathbf{24}$ & \parbox{0.3\textwidth}{\centering$\left( \mathbf{8}, \mathbf{1} \right)_0 \oplus \left( \mathbf{1}, \mathbf{3} \right)_0 \oplus \left( \mathbf{1}, \mathbf{1} \right)_0$ \\ $ \oplus \left( \mathbf{3}, \mathbf{2} \right)_{5/6} \oplus \left( \mathbf{\overline{3}}, \mathbf{2} \right)_{-5/6}$} & \parbox{0.3\textwidth}{\centering gluons, $W^\pm$, $Z$, $B$ \\ exotic $X$ and $Y$ bosons } \\
\bottomrule
\end{tabular}
\caption{Representations of $SU(5)$ lead to the \emph{standard model} particles (\cf \cref{subsec:OverviewOfStandardModel} ).}
\label{table-BrokenSpectrum}
\end{table}

Of course also the $SU(5)$ gauge bosons can be decomposed along these lines. The adjoint representation $\mathbf{24}$ of $SU(5)$ decomposes according to
\[ \mathbf{24} \to \left( \mathbf{8}, \mathbf{1} \right)_0 \oplus \left( \mathbf{1}, \mathbf{3} \right)_0 \oplus \left( \mathbf{1}, \mathbf{1} \right)_0 \oplus \left( \mathbf{3}, \mathbf{2} \right)_{5/6} \oplus \left( \mathbf{\overline{3}}, \mathbf{2} \right)_{-5/6} \, . \]
The first three irreducible representations correspond to the gluons of $SU(3)$, the $W^\pm$ and $Z$ boson of $SU(2)$ and the gauge field $B$ of the hypercharge $U(1)_Y$, respectively. The last two terms denote the gauge fields, which acquire masses by breaking $SU(5)$ with the Higgs mechanism. As discussed above, these are the twelve gauge fields $A^a_\mu$ with $13 \leq a \leq 24$. More conventionally they are grouped into two groups of six gauge bosons each, which are referred to as $X$ and $Y$ bosons respectively. Collectively they are also termed \emph{leptoquarks}.

\paragraph{Challenges of the Georgi-Glashow Model}

Above we have explained that one generation of matter particles in the standard model nicely fits into the representation $\overline{5} \oplus \mathbf{10} \oplus \mathbf{1}$ of $SU(5)$. We have also seen that the adjoint representation $\mathbf{24}$ decomposes upon reduction of $SU(5)$ to $SU(3) \times SU(2) \times U(1)$ to give us all gauge bosons of the \emph{standard model}, including however also the exotic gauge bosons $X$ and $Y$ in the bifundamental representation $\left( \mathbf{3}, \mathbf{2} \right)_{5/6}$ and $\left( \mathbf{\overline{3}}, \mathbf{2} \right)_{-5/6}$ respectively. This picture is apparently incomplete without fitting the Higgs doublet into an representation of $SU(5)$. As we have already seen that $\overline{5}$ decomposes into a doublet and triplet (\cf \cref{equ:DecompositionOfOverline5}) it is natural to take a closer look at the fundamental representation $\overline{5}$. And indeed
\[ \mathbf{\overline{5}} \to \left( \mathbf{1}, \mathbf{2} \right)_{1/2} \oplus \left( \mathbf{3}, \mathbf{1} \right)_{-1/3} \,  \]
gives us the Higgs doublet in representation $( \mathbf{1}, \mathbf{2} )_{1/2}$ together with a triplet $( \mathbf{3}, \mathbf{1} )_{-1/3}$. The latter is termed a \emph{Higgs triplet} $T$.

\begin{figure}[tb]
\centering
\includegraphics[valign = c]{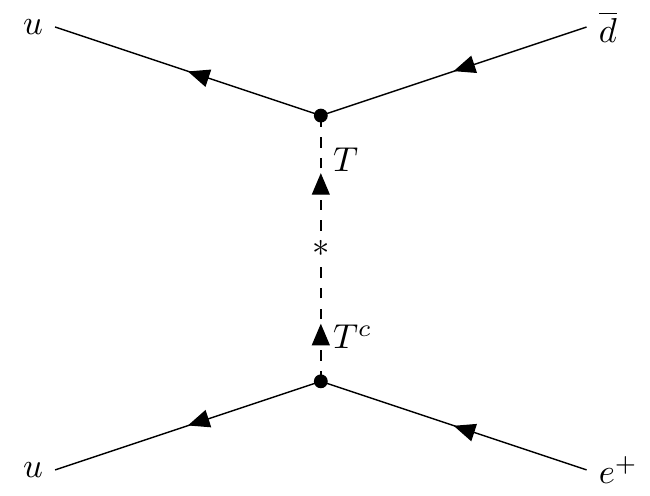}
\caption[Proton decay induced by the Higgs triplet in Georgi-Glashow GUT.]{The Higgs triplet $T$ and its charge conjugate $T^c$ allow for a process which turns a pair of up-quarks $u$ into an anti-down quark $\overline{d}$ and a positron $e^+$.}
\label{figure:ProtonDecay}
\end{figure}

In the \emph{standard model} we generate masses for the leptons and quarks from Yukawa interactions. These correspond to interaction terms in the Lagrangian consisting of two leptons/quarks and the Higgs doublet. After electroweak symmetry breaking, \ie imposing a VEV for the Higgs doublet, this interaction term turns into a sum consisting of a mass term for these leptons/quarks and an interaction term with the Higgs field $H$. Any such triplet interaction terms must be gauge invariant, which restricts the allowed choices. The presence of the Higgs triplet $T$ allows for additional interactions. For example the process in \cref{figure:ProtonDecay} is now allowed and the proton $p^+$ can decay into a neutral pion $\Pi^0 = d \overline{d}$ and a positron $e^+$ via
\[ p^+ = u \, u \, d \to \overline{d} \, e^+ \, d = \Pi^0 + e^+ \, . \]

Therefore, the Georgi-Glashow model predicts proton decay. Even more, not only the Higgs triplet $T$ triggers this decay, but the $X$ and $Y$ bosons induce additional such decay processes. Unfortunately however, to date proton decay has not been witnessed experimentally \cite{Abe:2014mwa}. This puts severe tension on this model and  leads for example to the doublet-triplet splitting problem. To suppress the proton decay channels triggered by the Higgs triplet, this field $T$ must be very massive. Typically its mass is assumed of the order of the GUT-scale, \ie $m_T^2 \sim \SI[parse-numbers=false]{10^{16}}{\giga \electronvolt}$. On the other hand the Higgs doublet $\phi$ should have a mass of the weak scale, \ie roughly $m_\phi^2 \sim \SI[parse-numbers=false]{10^{2}}{\giga \electronvolt}$. The doublet-triplet splitting problem asks to explain what keeps the Higgs doublets $\phi$ so light whilst producing such heavy Higgs triplets. For more background on the challenge of proton decay and doublet-triplet splitting to GUTs -- including supersymmetric extensions of the Georgi-Glashow model -- see for example \cite{PhysRevD.26.287, 1982PhLB..112..133D, Randall:1995sh, Senjanovic:2009kr}.

\paragraph{Towards String Theory}
Whilst both the proposed proton decay and the doublet-triplet splitting problem require attention, conceptionally there is an even bigger major drawback. Namely, motivated by the Newtonian example we have set out to find a unified description of all phenomena in nature. The Georgi-Glashow GUT however is apparently incomplete, as it does not account for gravity. 

To combine quantum field theory with gravity, we need a theory of quantum gravity. Naively, we could simply try to formulate Einstein's theory of \emph{general relativity} \cite{einstein1916grundlage} in the language of quantum field theory. In doing so, gravitational interactions are mediated by a gauge boson which is called the \emph{graviton}, which is a massless spin-two particle \cite{weinberg1995quantum}. Unfortunately, in the resulting quantum theory the physical observables suffer from divergences in an uncontrollable manner \cite{GOROFF198581}. Therefore, a different approach is required. To date, \emph{string theory} is a very promising candidate for a theory of quantum gravity. Consequently, a natural next step is to discuss GUT-models in \emph{string theory}. 

In such phenomenological applications of \emph{string theory} we wish to tell if a compactification can make close contact to the \emph{standard model}. A first indication is obtained from looking at the fields, which are massless prior to GUT-group breaking. A very basic parameter to judge the phenomenological relevant of such a massless spectrum is the difference of the number of chiral and anti-chiral fermions in the same representation. This number is called the chiral index. It happens to be a topological invariant and can therefore often be computed with minimal effort.

To date experimental evidence indicates that the \emph{standard model} does not contain pairs of chiral and anti-chiral fermions in the same representation. To check if the massless spectrum of a \emph{string theory} compactifications satisfies this property also, we have to go beyond the chiral index and actually compute the number of chiral and anti-chiral fermions. Even more, the running couplings and the Yukawa interactions strongly depend on these numbers. Finally, supersymmetric extension of the \emph{standard model} must contain (at least) one vector-like pair of superfields, namely the Higgs doublets $H_u$ and $H_d$. 

Hence, before we come to discuss GUT-models in \emph{string theory}, let us develop the necessary machinery to compute these zero modes in special \emph{string theory} compactification which are known as \emph{F-theory vacua}. It turns out that this is quite a delicate question, and we will not come back to constructing GUT-models before \cref{chapter:GUTModels}.

\section{String Theory} \label{sec:StringTheory}

\subsection{From point particles to quantum strings} \label{subsec:PointsToStrings}

\paragraph{Nambu-Goto action}

From classical mechanics to general relativity the point-particle-concept is omnipresent. Even the divergences in quantum field theory, which we encountered in \cref{subsec:RunningCouplings}, are believed to be tied to this concept. String theory breaks with the point-particle philosophy and models the fundamental entities of nature as one-dimensional strings. 

A propagating string sweeps out a two-dimensional surface $\Sigma$. This surface $\Sigma$ is termed the \emph{string-worldsheet} and its (local) coordinates are commonly  denoted as $\sigma^a = ( \sigma^0, \sigma^1 ) = ( \tau, \sigma )$. Given a $d$-dimensional spacetime $M_d$ with metric tensor $G_{\mu \nu}$, we consider the embedding
\[ \iota \colon \Sigma \hookrightarrow M_d \; , \; \sigma^a \mapsto X^\mu ( \sigma^a ) \, . \label{equ:EmbeddingOfStringworldsheet} \]
The pullback of $G_{\mu \nu}$ along $\iota$ gives a metric tensor $\gamma_{ab}$ on $\Sigma$. At $p \in \Sigma$ it is given by
\[ \gamma_{ab} \left( p \right) = \left( \frac{\partial X^\mu}{\partial \sigma^a} \right) \left( p \right) \cdot \left( \frac{\partial X^\nu}{\partial \sigma^b} \right) \left( p \right) \cdot G_{\mu \nu} \left( \iota \left( p \right) \right) \, . \]
The dynamics of the string is encoded in the Nambu-Goto action
\[ S_{\mathrm{NG}} = - T \cdot \Mint_{\Sigma}{d \tau d \sigma \sqrt{- \mathrm{det} \gamma_{ab}}} \, . \label{equ:NGAction} \]
$S_{\mathrm{NG}}$ measures the area of the worldsheet $\Sigma$ as embedded into $M_d$ via \cref{equ:EmbeddingOfStringworldsheet}. The prefactor $T$ is known as the string tension. It satisfies $T = 2 \pi / \alpha^\prime$ where $l_s = 2 \pi \sqrt{\alpha^\prime}$ is the string length, \ie the typical length of strings. Until now, there exists no direct experimental evidence for \emph{string theory}. Consequently, the string scale $M_s = l_s^{-1}$ must be far beyond the currently probed energy scales. As aspects of quantum gravity become relevant at the Planck scale $M_{\mathrm{Planck}} \sim \SI[parse-numbers=false]{10^{19}}{\giga \electronvolt}$, it is believed that $M_s$ is of the order of $M_{\mathrm{Planck}}$.

\paragraph{String theory as field theory on the world sheet}

\begin{figure}[tbp]
\centering
\includegraphics[width = 0.75\textwidth]{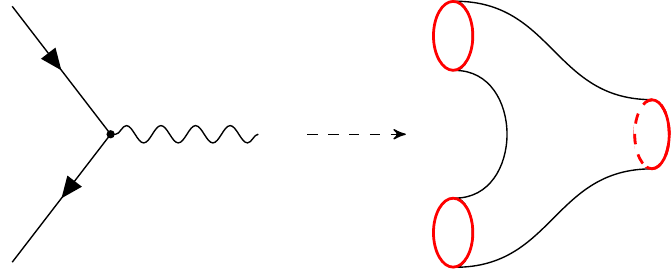}
\caption[Feynman diagram in QFT and string joining in \emph{string theory}.]{In quantum field theory the scattering occurs in the vertices of Feynman diagrams. In \emph{string theory} these vertices are absent, leading to UV-finiteness of \emph{string theory}.}
\label{TransitionFromFeynmanDiagramsToPairOfPants}
\end{figure}

The action \cref{equ:NGAction} depends on $d$ real scalar fields $X^\mu ( \sigma )$. As they describe the motion of the string, they constitute the dynamics. In particular, we can interpret \cref{equ:NGAction} as a field theory on the worldsheet $\Sigma$ with bosonic fields $X^\mu$ as dynamical objects -- such a \emph{string theory} is termed \emph{bosonic \emph{string theory}}. To include fermionic degrees of freedom, it is necessary to introduces supersymmetric partners $\psi^\mu$ to the fields $X^\mu$. Appealing features of the so-obtained \emph{superstring theory} include the following:
\begin{itemize}
 \item The classical theory is invariant under conformal transformations. This symmetry is fairly important -- it is the sole reason why we can 
      fully solve the quantum string. Remarkably, demanding this symmetry to persist at the quantum level predicts the spacetime dimension: $d = 26$ for bosonic \emph{string theory} and $d = 10$ for superstring theory. Throughout this thesis we are interested in superstring theory, so $d = 10$ from now on.
 \item As a consequence of supersymmetry, the quantum theory on $M_{10}$ is free of unstable tachyonic ground states. As of now there is no direct experimental evidence 
      for supersymmetry. Hence, it is assumed that supersymmetry exists at high energies and is broken at the currently probed ones. Although mechanisms to break supersymmetry will be of no importance to this thesis, let us point the interested reader to \cite{Blumenhagen:2006ci, Ibanez2014string} for more information.
 \item Strings come in two types -- the string is closed if the endpoints are identical and open otherwise. Interactions between strings can be pictured as joining of 
      strings. In \cref{TransitionFromFeynmanDiagramsToPairOfPants} we picture the joining of two closed strings to a new closed string. Due to the extended 
      nature of strings, scattering does not occur at a single point, as opposed to quantum field theory. This ultimately leads to UV-finiteness of \emph{string theory} \cite{polchinski1998string}.
\end{itemize}

\paragraph{Why zero mode counting?}

Quantisation of \cref{equ:NGAction} \footnote{The so-called \emph{Polyakov action} is easier to quantise than the Nambu-Goto action. Since both actions agree on shell, \ie upon use of the equations of motions, the quantum string is usually based on the Polyakov action.} turns every excitation state of the classical string into a particle. Hence, one obtains an infinite tower of particles. As an example consider a closed bosonic string in 10-dimensional Minkowski spacetime $\mathcal{M}_{1,9}$. Its excitations are labelled by $N \in \mathbb{Z}$ and the associated particles have mass
\[ m_N^2 = \frac{4 \left( N - 1 \right)}{\alpha} = \left( \frac{4 \pi}{l_s} \right)^2 \cdot \left( N - 1 \right) \, \label{equ:MassBosonicStringInFlatSpace} .\]
In general, the masses of string excitations exhibit $l_s^{-1}$-dependencies. Therefore, all but the massless particles are far too heavy to be detected at current experiments. With an eye towards low energy (as compared to the string scale $M_s = 1 / l_s$) physics, we may thus discard all but the massless excitations. In consequence, phenomenological applications of \emph{string theory} as presented in this thesis, seek to count these zero modes.

\paragraph{D-branes}

Closed strings can move freely in $M_{10}$. Hence, their excitations are referred to as \emph{bulk states}. Conversely, the equations of motion force open strings to end on the world volume of a Dp-brane. This world volume of  a Dp-brane consists of $p$ spatial and one time dimension, \ie is a $p+1$-dimensional subspace of the spacetime $M_{10}$.

It is possible to generalise \cref{equ:NGAction} to an action for Dp-branes. This is the DBI-action. Classically one can derive from this action the existence of a Yang-Mills gauge theory on the world volume of a Dp-brane. In this sense Dp-branes realise gauge theories in \emph{string theory}, for which reasons they enjoy ample attention in string phenomenology.

Dp-branes must not be confused with fundamental strings, \ie strings subject to \cref{equ:NGAction}. Although the world volume of a $D1$-brane looks like the world sheet $\Sigma$ of a fundamental string, they are fundamentally different objects obeying different actions. To make this distinction explicit, fundamental strings are referred to as F1-strings. For example D1-strings and F1-strings are charged differently as we will discuss in \cref{subsec:TypeIIBStringTheory} in more detail.

Whilst we can formulate the DBI-action to describe the motion of a classical Dp-brane, its quantisation is hard, owing to the lack of an analogue of the conformal symmetry of \cref{equ:NGAction}. This obstacle can be understood as result of the non-perturbative nature of Dp-branes.

\paragraph{Compactification}

Consistency of superstring theory requires ten spacetime dimensions, whilst our everyday experience indicates four spacetime dimensions. The bridge between these seemingly contradictory statements is achieved by a procedure called \emph{compactification}. It rests upon two assumptions on the 10-dimensional spacetime $M_{10}$ of \emph{string theory}:
\begin{enumerate}
 \item $M_{10}$ contains a 4-dimensional, non-compact subspace $\mathcal{E}_4$ which resembles the spacetime of our everyday-experience.
 \item The `complement' of $\mathcal{S}_4$ in $M_{10}$ -- the \emph{internal space} $\mathcal{I}_6$ -- is so small that current experiments cannot (yet) observe it.
\end{enumerate}
One can envision very complicated geometries which fit these criteria. To simplify matters it is common practise to consider direct products $M_{10} = \mathcal{E}_4 \times \mathcal{I}_6$. Often one uses $\mathcal{E}_4 = \mathcal{M}_{1,3}$, \ie models the spacetime of our everyday-experience by the 4-dimensional Minkowski space $\mathcal{M}_{1,3}$.

In the absence of fluxes, a necessary condition for conformal invariance of \emph{string theory} is \cite{blumenhagen2012basic}
\[ 0 = \alpha^\prime R_{\mu \nu} + \frac{1}{2} \left( \alpha^\prime \right)^2 R_{\mu \lambda \rho \sigma} R^{\lambda \rho \sigma}_{\nu} + \dots \label{equ:CYCondition}\]
where $R_{\mu \nu \lambda \sigma}$ is the Riemann-tensor of the metric $G_{\mu \nu}$ on $M_{10}$. Let us assume that $M_{10} = \mathcal{E}_4 \times \mathcal{I}_6$ and that $\alpha^\prime$ is small (\ie we are considering the so-called \emph{supergravity limit}). Then \cref{equ:CYCondition} demands that the metric of the internal space $\mathcal{I}_6$ be Ricci-flat. This leads to compactifications on Calabi--Yau manifolds. Let us be very clear about these manifolds -- in this thesis we reserve the term \emph{Calabi--Yau} manifold for compact, connected complex-n-dimensional Kähler manifolds $X$ which satisfy the following two conditions:
\[ 
c_1 ( T_X ) = 0 \, , \hspace{5em} X \text{ has \emph{full} } SU(n) \text{-holonomy.} \label{equ:StrictCalabiYauCondition}
\]
In this expression $c_1 ( T_X )$ denotes the so-called first Chern class of the tangent bundle of $X$. It can be shown that any such manifold satisfies $h^{i,0} ( X ) = 0$ for $1 \leq i \leq n-1$. By the theorem of Yau \cite{Yau:1977ms}, on such manifolds there indeed exists a Ricci flat metric.

A field theory defined on $M_{10} = \mathcal{E}_4 \times \mathcal{I}_6$ leads to an effective field theory on $\mathcal{E}_4$. As a simple example consider a real  scalar field $\phi$ on a 5-dimensional spacetime $M_5$ with action
\[ S = \Mint_{M_5}{d^5x \left( - \frac{1}{2} \partial_A \phi \, \partial^A \phi \right)} \, . \label{equ:BosonicFieldAction}\]
Now take $M_5 = \mathcal{M}_{1,3} \times S^1_R$, \ie compactify $M_5$ on a circle $S^1_R$ of radius $R$. Let the coordinates of Minkowski spacetime $\mathcal{M}_{1,3}$ be $x^\mu$ and the ones of $S^1_R$ be denoted by $y$. Our strategy is to express $\phi ( x^\mu, y )$ in terms of a complete set of functions on $S^1_R$. Once we achieve this, we can plug the result back into \cref{equ:BosonicFieldAction} and work out the $y$-integration. Consequently, we are then left with an action which depends solely on the coordinates $x^\mu$ of $\mathcal{M}_{1,3}$, \ie we have obtained an effective 4-dimensional field theory.

But what set of functions should we pick? To come up with a neat choice, let us take a look at the equation of motion induced from \cref{equ:BosonicFieldAction}. This is the Klein-Gordon equation
\[ 0 = \Box \phi = \left( - \partial_{x^0}^2 + \partial_{x^1}^2 + \partial_{x^2}^2 + \partial_{x^3}^2 \right) \phi + \partial_{y}^2 \phi \, . \]
Hence, it could be useful to expand $\phi ( x^\mu, y )$ in terms of the eigenfunctions of the Laplace operator on the circle $S^1_R$. This amounts to a Fourier-transformation and we find
\[ \phi \left( x^\mu, y \right) = \sum_{k \in \mathbb{Z}}{\phi_k \left( x^\mu \right) e^{\frac{i k y}{R}} } \, . \]
We can now perform the $y$-integration in \cref{equ:BosonicFieldAction}. By setting $\phi_k^\ast = \phi_{-k}$ we then obtain
\[ S = \Mint_{\mathcal{M}_{1,3}}{d^4x \left[ - \frac{1}{2} \partial_\mu \phi_0 \, \partial^\mu \phi_0 - \sum_{k = 1}^{\infty}{\left( \partial_\mu \phi_k \, \partial^\mu \phi_k^\ast + \frac{k^2}{R^2} \phi_k \phi_k^\ast \right)} \right]} \, . \label{equ:4DEffectiveActionKKExample} \]
This 4-dimensional effective action contains a real, massless scalar field $\phi_0$ and for every $k \in \mathbb{N}_{> 0}$ a complex scalar field of mass $\frac{k}{R}$.

Examples along these lines were first investigated by \emph{Theodor Kaluza} \cite{Kaluza1921} and \emph{Oskar Klein} \cite{1926ZPhy...37..895K, 1926Natur.118..516K}. In their honour, the above scalar fields carry the name \emph{KK-modes}. Below the so-called KK-scale $1/R$, the effective field theory in \cref{equ:4DEffectiveActionKKExample} detects only the massless scalar field $\phi_0$. The fifth dimension of $S^1_R$ then remains completely unnoticed.

We can slightly extend this example to discuss \emph{string theory} on $\mathcal{M}_{1,8} \times S^1_R$. In this geometry we must allow for strings which wrap the circle $S^1_R$ multiple times. As a consequence the mass formula in \cref{equ:MassBosonicStringInFlatSpace} is altered -- \eg the mass of the N-th excitation state of a closed bosonic string is given by
\[ m_{(N,w,n)}^2 = \left( \frac{n}{R} \right)^2 - \frac{2 n w}{\alpha^\prime} + \left( \frac{w R}{\alpha^\prime} \right)^2 + \frac{4}{\alpha^\prime} \left( N - 1 \right) \, . \label{equ:StringMassIncludingWindingAndMomenta}\]
The additional contributions stem from winding number $w$ and momentum number $n$:
\begin{itemize}
 \item A string has tension $T$, so energy is required to wind it $w$-times around $S^1_R$.
 \item Momenta along $S^1_R$ are quantised as $n/R$, due to compactness of $S^1_R$.
\end{itemize}
Crucially, \cref{equ:StringMassIncludingWindingAndMomenta} is invariant under $n \leftrightarrow w$ and $R \leftrightarrow R^\prime = \alpha^\prime R^{-1}$. One can show that this symmetry extends to the entire spectrum of the \emph{string theory}. In this sense \emph{string theory} on $\mathcal{M}_{1,8} \times S^1_R$ and $\mathcal{M}_{1,8} \times S^1_{R^\prime}$ are equivalent. This duality is a special instance of a more general phenomenon called \emph{T-duality}. As we will discuss momentarily, T-dualities relate different types of string theories. In particular, the definition of \emph{F-theory} which we give in \cref{subsec:DefinitionOfFTheory}, is essentially just a sequence of T-dualities.

Let us now come back to \emph{string theory} compactifications in general. Of course the effective theory on $\mathcal{E}_4$ strongly depends on the shape of $\mathcal{I}_6$. Collectively all mathematically consistent compactifications are termed the \emph{string landscape}. It consists of many compactification spaces $\mathcal{I}_6$ -- predictions ranging from $10^{500}$ to $10^{2700}$.

We already mentioned that D-branes support gauge theories in \emph{string theory}. As gauge theories are of special importance to physics, the existence of such D-branes and their location are of ample importance in string compactifications. A common choice is to consider so-called \emph{spacetime-filling D-branes}. A Dp-brane (with $p \geq 3$) is spacetime-filling precisely if its world volume covers the external space $\mathcal{E}_4$ completely and wraps a cycle in the internal space $\mathcal{I}_6$. For example a D7-brane with world volume $\mathcal{D}_7$ is spacetime filling precisely if $\mathcal{D}_7 = \mathcal{E}_4 \times \Sigma_4$, where $\Sigma_4 \subseteq \mathcal{I}_6$ is a cycle of (real) dimension $4$.

\paragraph{M-theory}

\begin{figure}[tb]
\centering
\includegraphics{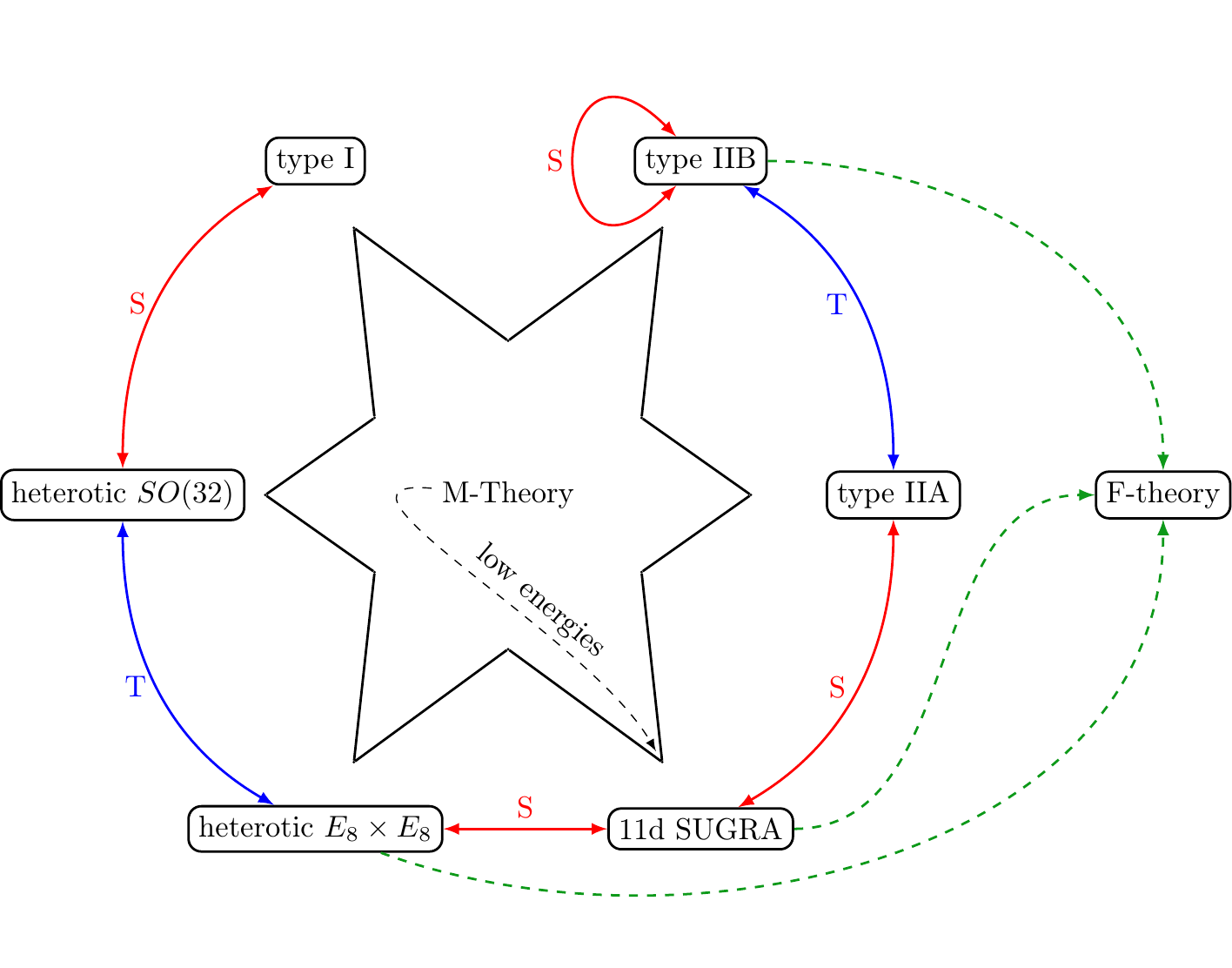}
\caption[String theory, \emph{F-theory} and \emph{M-theory}.]{There are five consistent formulations of \emph{string theory}, which are connected by S- and T-dualities. It is believed that they are different incarnations of the 11-dimensional \emph{M-theory}. To this date, only the low-energy effective description of \emph{M-theory} is known -- 11-dimensional supergravity. \emph{F-theory} can be approached from heterotic $E_8 \times E_8$ \emph{string theory}, type IIB \emph{string theory} (\cref{subsec:FTheoryFromIIB}) or 11-dimensional supergravity (\cref{subsec:DefinitionOfFTheory}).}
\label{figure-1098130985}
\end{figure}

The action \cref{equ:NGAction} allows to understand \emph{string theory} as field theory on the worldsheet $\Sigma$. As it turns out this formulation is not unique and there are five inequivalent and consistent superstring theories in flat 10-dimensional Minkowski spacetime $\mathcal{M}_{1,9}$. These are referred to as type I, type IIA, type IIB, heterotic $E_8 \times E_8$ and heterotic $SO ( 32 )$ \emph{string theory}. Some of these string theories are related by the above mentioned T-duality, other by another interrelation called S-duality. In \cref{figure-1098130985} we summarise the overall structure.

It is believed that these five string theories are different incarnations of an 11-dimensional theory -- \emph{M-theory}. The low energy limit of \emph{M-theory} is the 11-dimensional supergravity. It satisfies $\mathcal{N} = 1$ supersymmetry. Let us denote the 11-dimensional space as $M_{11}$, its metric tensor as $G_{\mu \nu}$ and the associated Ricci scalar as $R$. Then the bosonic fields of the 11-dimensional supergravity are the metric tensor $G$ and a 3-form field $C_3$. Set $G_4 = d C_3$, then the dynamics is governed by the action
\[ S_{11D} = \frac{M_{11D}^9}{2} \Mint_{M_{11}}{d^{11}x \left( \sqrt{- \mathrm{det} G} R - \frac{1}{2} G_4 \wedge \ast G_4 - \frac{1}{6} C_3 \wedge G_4 \wedge G_4 \right)} \, . \]
In analogy to D-branes in \emph{string theory}, there exist extended objects in \emph{M-theory} which are charged under $C_3$ -- M2-branes are electrically and M5-branes magnetically charged under $C_3$.

\paragraph{Special \emph{String Theory} GUT-Models}

As motivated in \cref{sec:GUTs}, we wish to study GUT models in string theory. In principle we could do this in any of the five 10-dimensional string theories. Still, we choose to focus on a special form of type IIB string theory, which is known as \emph{F-theory}. It is a non-perturbative framework in the following sense: By definition, the five string theories are understood as the supergravity limit of \emph{M-theory}. This supergravity limit is the limit of small $\alpha^\prime$. In addition, the joining and splitting of strings is accounted for by a coupling constant $g_s$, very much along the example set by quantum field theories. The value of this coupling constant $g_s$ is set by the so-called dilaton field $\phi$. More explicitly, the value $\phi_0$ of this dilaton field $\phi$ far away from Dp-branes sets the string coupling constant to the value $g_s = e^{\phi_0}$ \cite{green1988superstring1, green1988superstring2, polchinski1998string, polchinski2001string}. Consequently, a perturbative expansion in string theory involves both $\alpha^\prime$ and $g_s$. \emph{F-theory} is different -- as it is defined by the supergravity limit of M-theory,  it assumes small $\alpha^\prime$, but takes into account strong couplings in $g_s$. In this sense \emph{F-theory} allows us to study non-perturbative corners of \emph{string theory}, which is the reason why we set out to study GUT-models in \emph{F-theory}.

\emph{F-theory} can be approached from type IIB string theory, heterotic $E_8 \times E_8$ \emph{string theory} and \emph{M-theory}. Indeed much of our understanding of \emph{F-theory} stems from investigations of the heterotic/F-theory duality \cite{Vafa:1996xn, Morrison:1996na, Morrison:1996pp, Friedman:1997yq, Heckman:2013sfa, Anderson:2014gla}. Also structures from type IIB \emph{string theory} and of \emph{M-theory} feature prominently in \emph{F-theory}. For example, M-branes are prominent features of \emph{F-theory} discussions as well as $G_4 = dC_3$. One very important step in this thesis, which we will discuss in \cref{sec:TheChowRing}, is to parametrise a subset of $G_4$-fluxes by certain elements of the so-called \emph{Chow ring} \cite{Bies:2014sra}.

For illustrative reasons, we will approach \emph{F-theory} from type IIB \emph{string theory} in \cref{subsec:FTheoryFromIIB}. Subsequently, we define \emph{F-theory} via 11-dimensional supergravity in \cref{subsec:DefinitionOfFTheory}. In anticipation of this, we will now complete this section with a review on type IIB \emph{string}.

\subsection{Type IIB \emph{string theory} and backreaction of D7-branes} \label{subsec:TypeIIBStringTheory}

\paragraph{\texorpdfstring{$\mathbf{SL(2, \mathbb{Z})}$}{SL(2,Z)}-symmetry}

At the massless level, type IIB \emph{string theory} admits an effective supergravity description. In flat 10-dimensional Minkowski space $\mathcal{M}_{1,9}$, this effective field theory has $\mathcal{N} = ( 2,0 )$ supersymmetry and its bosonic fields are listed in \cref{table-N1}. Of these fields the dilaton $\phi$ plays a particularly important role, as its value $\phi_0$ far away from Dp-branes settles the string coupling constant as $g_s = e^{\phi_0}$ \cite{green1988superstring1, green1988superstring2, polchinski1998string, polchinski2001string}.

\begin{table}[tbp]
\centering
\begin{tabular}{ccccc}
\toprule
field & symbol & type & electric BPS state & magnetic BPS state \\
\midrule
dilaton & $\phi$ & scalar & -- & -- \\
metric & $G_{\mu \nu}$ & symmetric 2-tensor & -- & -- \\
B-field & $B_2$ & 2-form & F1-string & NS5-brane \\
RR $0$-form & $C_0$ & 0-form & D(-1) instanton & D7-brane \\
RR $2$-form & $C_2$ & 2-form & D1-string & D5-brane \\
RR $4$-form & $C_4$ & 4-form & D3-brane & D3-brane \\
\bottomrule
\end{tabular}
\caption{Bosonic field content of 10-dimensional type IIB supergravity -- based on \cite{Lin:2016zha}.}
\label{table-N1}
\end{table}

It is useful to define the axio-dilaton $\tau := C_0 + i e^{- \phi}$, the field strength $F_{n+1} := dC_n$ for the p-form fields $C_n$ and $H_3 := d B_2$ for the Kalb-Ramond field $B_2$. Finally, let us set $G_3 := F_3 - \tau H_3$ and $\tilde{F_5} := F_5 - \frac{1}{2} C_2 \wedge H_3 + \frac{1}{2} B_2 \wedge F_3$. In terms of these fields, the bosonic part of the type IIB supergravity action takes the form \cite{polchinski1998string}
\[ S_{IIB} = \frac{2 \pi}{l_s^8} \Mint_{M_{10}}{d^{10}x \left[ \sqrt{-G} R - \frac{d\tau \wedge \star d \overline{\tau}}{2 \left( \Im \tau \right)^2} + \frac{dG_3 \wedge \star d \overline{G_3} }{\Im \tau} + \frac{1}{2} \tilde{F_5} \wedge \tilde{F_5} + C_4 \wedge H_3 \wedge F_3 \right]} \, ,\]
where $G$ is the determinant of the metric $G_{\mu \nu}$ on $\mathcal{M}_{1,9}$ and $R$ its Ricci scalar. This action enjoys an $SL ( 2, \mathbb{R} )$-symmetry mediated by
\[ 
\resizebox{0.9\textwidth}{!}{$
\begin{psmallmatrix} C_4 \\ G \end{psmallmatrix} \mapsto \begin{psmallmatrix} C_4 \\ G \end{psmallmatrix} \, , \quad \tau \mapsto \frac{a \tau + b}{c \tau + d} \, , \quad \begin{psmallmatrix} C_2 \\ B_2 \end{psmallmatrix} \mapsto \begin{psmallmatrix} a & b \\ c & d \end{psmallmatrix} \begin{psmallmatrix} C_2 \\ B_2 \end{psmallmatrix} \, , \quad \begin{psmallmatrix} a & b \\ c & d \end{psmallmatrix} \in SL( 2, \mathbb{R} ) \, . \label{equ:SL2Z-transformation}$}
\]

Upon quantisation only an $SL( 2, \mathbb{Z} )$-symmetry remains. This breaking occurs since $D(-1)$ instantons contribute a factor $\exp ( 2 \pi i \tau )$ to the partition function, which is only invariant under $SL( 2, \mathbb{Z} )$ transformations of $\tau$. It is believed that this $SL( 2, \mathbb{Z} )$-symmetry persists as a symmetry of the quantised type IIB \emph{string theory}.

\paragraph{Electric and magnetic charges of D-branes}

A Dp-brane with world volume $\mathcal{D}_{p+1}$ couples electrically to a $(p+1)$-form field $C_{p+1}$ via
\[ S_{\mathrm{electric}} = \mu_p \Mint_{\mathcal{D}_{p+1}}{C_{p+1}} \, , \qquad \mu_p = \frac{\left( 2 \pi \sqrt{\alpha^\prime} \right)^{3-p}}{\left( 2 \pi \right)^3 \left( \alpha^\prime \right)^2} \, . \]
We associate to $C_{p+1}$ the field strength $F_{p+2} = dC_{p+1}$ and apply the Hodge star operator. This yields a $(10-p-2)$-form field $\ast F_{p+2}$ and we can find $\tilde{C}_{10-p-3}$ with $\ast F_{p+2} = d \tilde{C}_{10-p-3}$.\footnote{Every class in $H^{p+1}_{\mathrm{dR}} ( \mathcal{I}_6 )$ is represented by exactly one harmonic form. In particular, the form fields $C_{p+1} \in \Omega^{p+1} ( M_{10} )$ are chosen to be harmonic, because the Hodge star operator does not map closed forms to closed forms,  but harmonic forms to harmonic forms. For an example of a closed form which is not mapped to a closed form by the Hodge star operator, consider the real line $\mathbb{R}$ and $\omega = f ( x ) dx \in \Omega^1 ( \mathbb{R} )$. Then $d \omega = 0$ for dimensional reasons. However $\ast \omega = f ( x )$ which is not closed unless the function $f$ is constant.} We say that a D(10-p-4)-brane with world volume $\mathcal{D}_{10-p-3}$ couples magnetically to $C_{p+1}$ via the action
\[ S_{\mathrm{magnetic}} = \mu_{10-p-4} \Mint_{\mathcal{D}_{10-p-3}}{\tilde{C}_{10-p-3}} \, . \]
From this we obtain the electric and magnetic couplings listed in \cref{table-N1}.

\paragraph{The Need for Orientifolds}

Thus, a D7-brane is charged magnetically under $C_0$ via
\[ S_{\mathrm{magnetic}} = \mu_7 \Mint_{\mathcal{D}_8}{\tilde{C}_8} = \mu_7 \Mint_{M_{10}}{\tilde{C}_{8} \wedge \mathrm{p} \left( \mathcal{D}_8 \right)} \, , \qquad \mu_7 = \frac{1}{\left( 2 \pi \right)^7 \left( \alpha^\prime \right)^4} \, , \]
where $p ( \mathcal{D}_8 )$ is the Poincaré-dual of the world volume $\mathcal{D}_8$ of the D7-brane in question. The equations of motions for the form field $\tilde{C}_8$ follow from variation of the Chern-Simons action
\begin{align}
S_{CS} = - \frac{1}{4 \kappa_{10}} \cdot \Mint_{M_{10}}{d \tilde{C}_8 \wedge \ast d \tilde{C}_8 } + \mu_7 \Mint_{M_{10}}{\tilde{C}_{8} \wedge \mathrm{p} \left( \mathcal{D}_8 \right)} \, ,
\end{align}
where $\kappa_{10}^2 = \left( 2 \pi \right)^7 \left( \alpha^\prime \right)^4 / 2$ is the 10-dimensional gravitational constant. The vanishing of the variation of $S_{CS}$ with respect to $\tilde{C}_8$ is equivalent to
\[ 0 \stackrel{!}{=} - \frac{1}{4 \kappa_{10}} \frac{\delta \left( d \tilde{C}_8 \wedge \ast d \tilde{C}_8\right)}{\delta \tilde{C}_8} + \mu_7 \cdot p \left( \mathcal{D}_8 \right) \, . \label{equ:TadpoleCancellationIIB} \]
Since $\delta ( d \tilde{C}_8 \wedge \ast d \tilde{C}_8 )$ is proportional to $d^2 \tilde{C}_8 = 0$, we thus find $p ( \mathcal{D}_8 ) = 0$. By repeating this logic for numbers of (stacks of) D7-branes in $M_{10}$, we find that such configurations are inconsistent with the equation of motion for $\tilde{C}_8$, unless such a configuration has trivial Poincaré-dual. Fortunately we can circumvent this strong demand with so-called \emph{orientifold-planes} O7, which are charged negatively under $\tilde{C}_8$ \cite{polchinski1998string, polchinski2001string}. 

Such O7-planes exist in orbifolds only. Let us therefore consider an involution $\sigma$ on the internal space $\mathcal{I}_6$ and replace $\mathcal{I}_6$ by $\mathcal{B}_6 := \mathcal{I}_6 / \sigma$. The \emph{string theory} spectrum on $\mathcal{B}_6$ is formed from the states of superstring theory on $\mathcal{I}_6$, which are invariant under application of both the parity operator $\Omega$ and $(-1)^{F_L}$. The so-obtained physical theory is termed an \emph{orientifold theory}. Details on such constructions and explicit examples can be found \eg in \cite{Cremades:2002cs, MarchesanoBuznego:2003hp, Blumenhagen:2006ci}.

\paragraph{Backreaction of D7-branes}

Consider an electron in an electric field $\mathbf{E}$. Then the electric charge of the electron will alter the electric field $\mathbf{E}$ -- we say it backreacts with the field $\vect{E}$. Similarly, a (charged) D7-brane backreacts with $C_0$. Let us analyse this phenomenon in the complex plane orthogonal to the D7-brane. In this complex plane with coordinate $z$, let $r$ denote the distance from the D7-brane located at $z_0$. 

Heuristically the backreaction is described by a Poisson equation. In two dimensions its solution is proportional to $\log r$. Consequently, this backreaction does not trail off as we move away from the D7-brane, and so it cannot be neglected.\footnote{Note that for a Dp-brane with $p < 7$ similar backreaction trails off with $r^{7-p}$ and can consequently be ignored sufficiently far away from the Dp-brane.} Via $\tau ( z ) = C_0 ( z ) + i e^{-\phi ( z )}$, this backreaction on $C_0$ also influences the profile of the axio-dilaton. In the vicinity of the D7-brane one can write 
\[ \tau \left( z \right) = \frac{1}{2 \pi i} \log \left( z - z_0 \right) + \dots \, . \label{equ:AxioDilatonProfileNearD7brane} \]

In the complex plane, the logarithm suffers a branch cut and with every passing thereof, the value of $\tau$ increases by $1$. This is a so-called monodromy of the axio-dilaton profile. Luckily the transformation $\tau \mapsto \tau + 1$ corresponds to an $SL( 2, \mathbb{Z} )$-transformation in \cref{equ:SL2Z-transformation} with the matrix $\begin{psmallmatrix} 1 & 1 \\ 0 & 1 \end{psmallmatrix}$. A careful analysis reveals that one can indeed determine the location of a stack of D7-branes from the monodromies of the axio-dilaton $\tau$ \cite{Sen:1996vd}.

We are now forced to look for a framework, which describes type IIB \emph{string theory} with these backreactions on the axio-dilaton $\tau$. One way is to study so-called (p,q)-strings and (p,q)-branes by string junctions \cite{Gaberdiel:1997ud, Gaberdiel:1998mv, DeWolfe:1998zf}. In this thesis we will follow an alternative approach, which leads to a geometric description of such non-perturbative type IIB \emph{string theory} vacua -- \emph{F-theory} \cite{Greene:1989ya,Vafa:1996xn}.

\section{F-theory} \label{sec:F-theory}

\subsection{From \texorpdfstring{$\mathbf{\mathrm{SL} ( 2, \mathbb{Z} )}$}{SL(2,Z)} invariance to a geometric book-keeping device} \label{subsec:FTheoryFromIIB}

\paragraph{Torus Surfaces}

\begin{figure}[tb]
\centering
\includegraphics{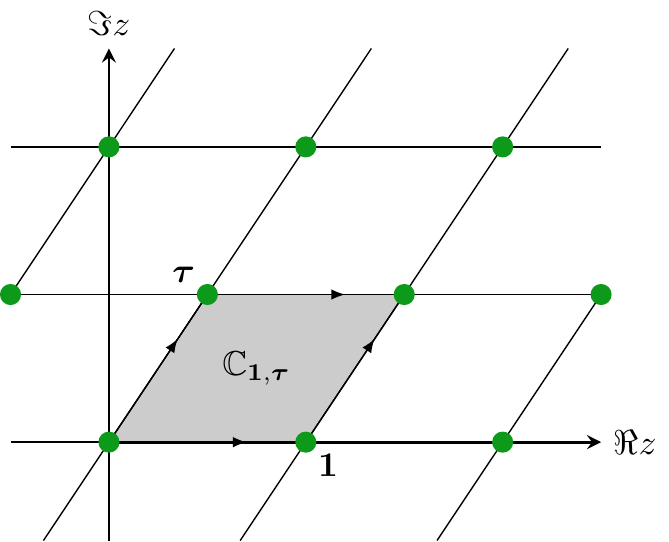}
\caption[Torus surfaces as $\mathbb{C} / \Lambda_{\vect{a},\vect{b}}$ where $\Lambda_{\vect{a},\vect{b}} \cong \mathbb{Z}^2$ is a lattice.]{The lattice $\Lambda = \mathrm{Span}_{\mathbb{Z}} \left( \vect{1}, \vect{\tau} \right)$ is formed from the green dots. The torus surface $\mathbb{C}_{\vect{1},\vect{\tau}}$ is obtained from $\mathbb{C}$ by identifying $z \sim z + 1 \sim z + \tau$ for all $z \in \mathbb{C}$. We can therefore picture $\mathbb{C}_{\vect{1},\vect{\tau}}$ by the grey parallelogram.}
\label{fig-5610351095328}
\end{figure}

A torus surface can be understood as the quotient $\mathbb{C}_{\vect{a},\vect{b}} = \mathbb{C} / \Lambda_{\vect{a},\vect{b}}$ of the complex plane $\mathbb{C}$ by a lattice $\Lambda_{\vect{a},\vect{b}} = \mathrm{Span}_{\mathbb{Z}} \left( \vect{a}, \vect{b} \right)$ of rank 2, where $\vect{a}, \vect{b} \in \mathbb{C}$. Without loss of generality we choose $\vect{a} = 1$, set $\vect{b} = \vect{\tau}$ and picture $\mathbb{C}_{\vect{1},\vect{\tau}}$ in \cref{fig-5610351095328}: The size of $\mathbb{C}_{\vect{1},\vect{\tau}}$ is related to the area of the grey parallelogram. It is termed the Kähler modulus of $\mathbb{C}_{\vect{1},\vect{\tau}}$. The shape of $\mathbb{C}_{\vect{1},\vect{\tau}}$ is determined by $\vect{\tau}$, which is its complex structure modulus.

The transformations $\tau \mapsto \tau +1 $ and $\tau \mapsto - \frac{1}{\tau}$ generate the group $SL( 2, \mathbb{Z} )$ and leave the lattice $\Lambda_{\vect{1},\vect{\tau}} = \mathrm{Span}_{\mathbb{Z}} (\vect{1},\vect{\tau})$ unaltered. Consequently, $SL( 2, \mathbb{Z} )$ is the symmetry group of $\mathbb{C}_{\vect{1},\vect{\tau}}$. By use of such symmetry transformations, any $\tau \in \mathbb{C}$ can be mapped to the fundamental domain
\[ \mathcal{F} = \left\{ \tau \in \mathbb{C} \; \left| \; \left| \tau \right| \geq 1 \, , \, \left| \Re \tau \right| < \frac{1}{2} \right. \right\} \, , \]
which is the set of inequivalent complex structure moduli of $\mathbb{C}_{\vect{1},\vect{\tau}}$.

\begin{figure}[tb]
\centering
\subfloat[A torus surface $\mathbb{C}_{1, \tau} \cong \mathbb{C} / \mathrm{Span}_{\mathbb{Z}} ( 1, \tau )$. The generators $1$, $\tau$ of the lattice in \cref{fig-5610351095328} correspond to the 1-cycles indicated in red.]{\label{figure-1592356} \includegraphics{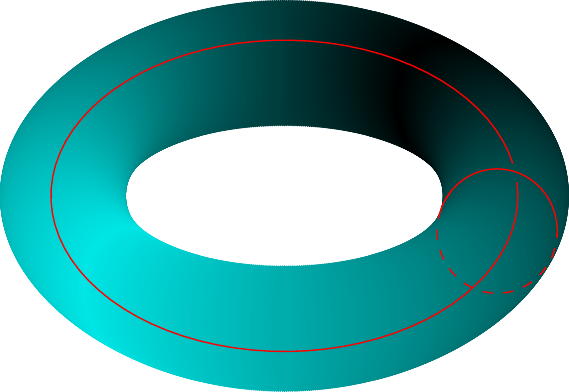} }
\hspace{2em}
\subfloat[Schematic picture of a singular torus surface -- a 1-cycle shrinks to zero and causes a singularity, indicated by the red ball.]{\label{figure-159235235} \includegraphics{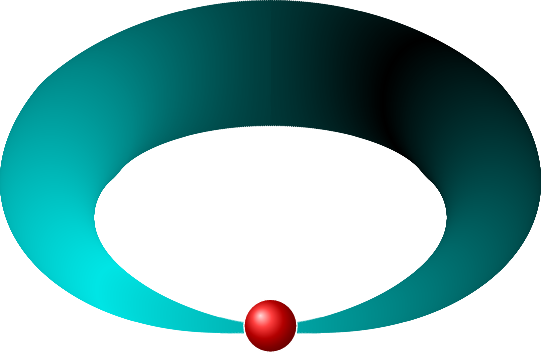} }
\caption{A smooth and a singular torus surface.}
\label{figure-1235123515325}
\end{figure}

The lattice vectors $\vect{1}$ and $\vect{\tau}$ correspond to two non-trivial $1$-cycles in $\mathbb{C}_{\mathbf{1},\boldsymbol{\tau}}$ as pictured in \cref{figure-1592356}. An infinite value of $\tau$ tells us that one of these two 1-cycles has shrunk to size zero -- the torus surface is singular. We give a schematic picture of this phenomenon in \cref{figure-159235235}.

\paragraph{Elliptic fibrations as book-keeping devices}

\begin{figure}[tbp]
\centering
\includegraphics{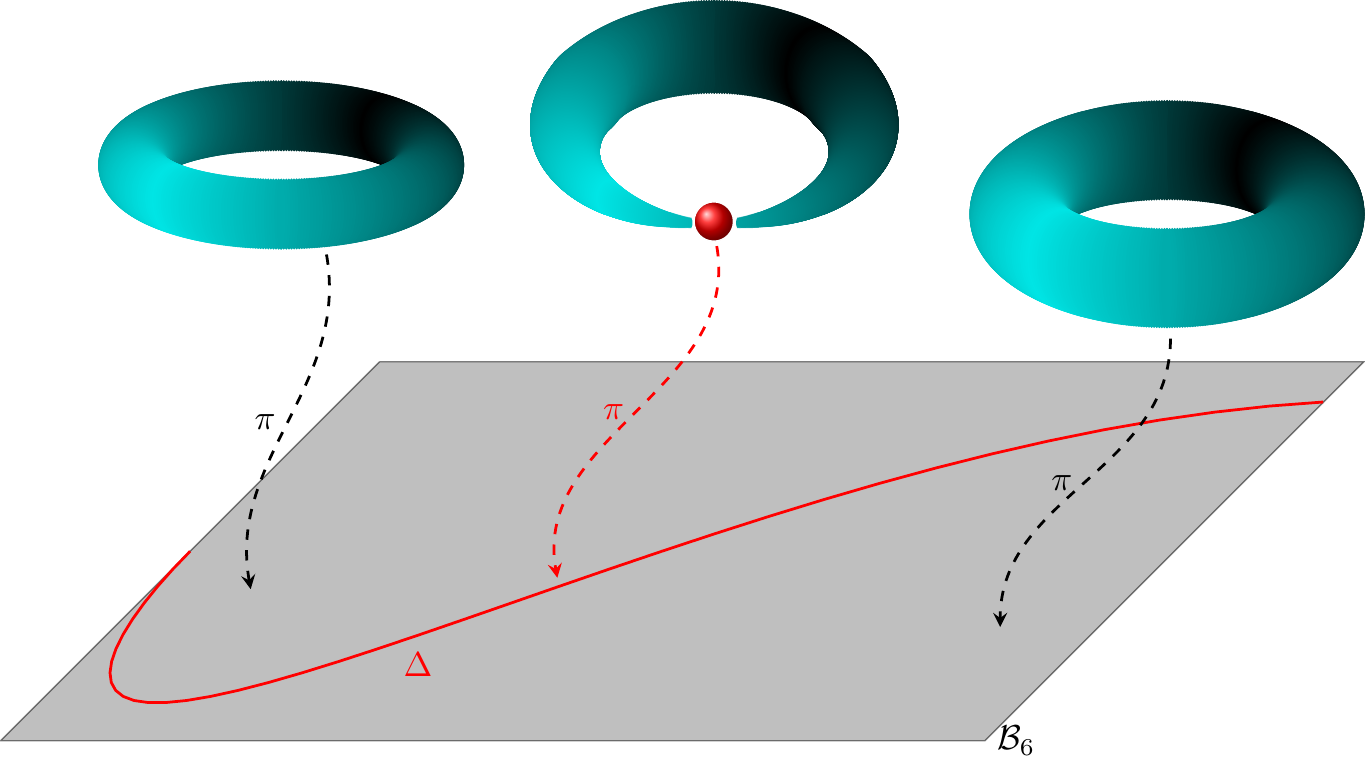}
\caption[Schematic picture of a singular torus fibration $Y_4 \twoheadrightarrow \mathcal{B}_6$.]{Schematic picture of a torus fibration $Y_4 \twoheadrightarrow \mathcal{B}_6$: singular tori are located over the so-called singular locus $\Delta \subseteq \mathcal{B}_6$.}
\label{figure-12362153}
\end{figure}

Let us now come back to compactifications of type IIB \emph{string theory}. Recall that the axio-dilaton is given by
\[ \tau \colon \mathcal{E}_4 \times \mathcal{B}_6 \to \mathbb{C} \; , \; x^\mu \mapsto C_0 \left( x^\mu \right) + i e^{- \phi \left( x^\mu \right)} \, . \]
On $\mathcal{E}_4$ the axio-dilaton is constant. So we focus on its behaviour over $\mathcal{B}_6$. We interpret the value of $\tau$ at $x^\mu \in \mathcal{B}_6$ as the complex structure modulus of a torus surface $\mathbb{C}_{\mathbf{1}, \boldsymbol{\tau ( x^\mu )}}$. A geometry which captures all this information is obtained from attaching  $\mathbb{C}_{\mathbf{1}, \boldsymbol{\tau ( x^\mu )}}$ to $x^\mu \in \mathcal{B}_6$. Such a construction forms a space $Y_4$ `over' $\mathcal{B}_6$ -- `over' means that there exists a projection map $\pi \colon Y_4 \twoheadrightarrow \mathcal{B}_6$ such that $\pi^{-1} ( x^\mu ) = \mathbb{C}_{\mathbf{1}, \boldsymbol{\tau ( x^\mu )}}$ for every $x^\mu \in \mathcal{B}_6$. In this way the geometry of $Y_4$ encodes the axio-dilaton profile. We give a schematic picture of $Y_4$ in \cref{figure-12362153}.

Recall that the axio-dilaton profile near a D7-brane diverges (\cf \cref{equ:AxioDilatonProfileNearD7brane}). Geometrically, the divergence of the complex-structure modulus $\tau$ of $\mathbb{C}_{\mathbf{1}, \boldsymbol{\tau}}$ signals a singularity of this torus surface. Consequently, the loci $\Delta \subseteq \mathcal{B}_6$ over which the torus fibration $\pi \colon Y_4 \twoheadrightarrow \mathcal{B}_6$ becomes singular encode the location of D7-branes. Even more, we will see momentarily that the type of singularity over $\Delta$ encodes the gauge dynamics on the D7-branes in question. Hence, the locus $\Delta$ is of ample importance for the physics encoded by $Y_4$.

An elliptic curve is a torus surface $\mathbb{C}_{\mathbf{1},\boldsymbol{\tau}}$ with a distinguished point \cite{Connell:1999}. Hence, if we fibre elliptic curves over $\mathcal{B}_6$, then for every $x^\mu \in \mathcal{B}_6$ there exists a distinguished point $P ( x^\mu ) \in \pi^{-1} ( x^\mu ) = \mathbb{C}_{\mathbf{1}, \boldsymbol{\tau ( x^\mu )}}$. We can thus consider the mapping $s \colon \mathcal{B}_6 \to Y_4 \, , \; x^\mu \mapsto P ( x^\mu )$, which is known as section of the fibration $\pi \colon Y_4 \twoheadrightarrow \mathcal{B}_6$. Such a fibration is termed an \emph{elliptic fibration}.

\paragraph{The Weierstrass Model}
Up to birational equivalence, the generic fibre of an elliptic fibre is given by a \emph{Weierstrass model} \cite{Delign1975, milne2006}. To this end, we look at global sections $f \in H^0( \mathcal{B}_6, \overline{K}_{\mathcal{B}_6}^{\otimes 4})$, $g \in H^0( \mathcal{B}_6, \overline{K}_{\mathcal{B}_6}^{\otimes 6} )$ of powers of the anti-canonical bundle $\overline{K}_{\mathcal{B}_6}$ of the space $\mathcal{B}_6$. The fibre over $p \in \mathcal{B}_6$ is then given as the hypersurface
\[ \mathcal{C} \left( p \right) := \left\{ \left[ x,y,z \right] \in \mathbb{P}^{2,3,1}_{\mathbb{C}} \; \left| \; y^2 - x^3 - f ( p ) xz^4 - g ( p ) z^6 = 0 \right. \right\} \, . \]
This shows that $f(p)$ and $g(p)$ determine the shape of this curve $\mathcal{C} ( p )$. Indeed they are related to its complex structure modulus $\tau ( p )$ by the famous \emph{$j$-invariant} \cite{milne2006, freitag2009complex}
\[ j \left( \tau \left( p \right) \right) = 55296 \cdot f^3 \left( p \right) \cdot \Delta^{-1} \left( p \right) \, , \]
where $\Delta ( p ) = 27 g^2 ( p ) + 4 f^3( p )$ cuts out the singular locus $\Delta \subseteq \mathcal{B}_6$ of this elliptic fibration, \ie 
\[ \Delta = \left\{ p \in \mathcal{B}_6 \; \left| \; \Delta \left( p \right) = 0 \right. \right\} \, . \]

\paragraph{Global Tate Models}

We have already pointed out the significance of the singularity locus $\Delta$ for \emph{F-theory} compactifications -- among others it encodes the location of D7-branes in this compactifications. Even more, we will explain in \cref{subsec:GaugeTheoriesInFTheory} how the singularity structure encodes non-Abelian gauge theories. Hence, a study of these singularities is of ample importance for our understanding of \emph{F-theory} compactifications.

The singularities of $\pi \colon Y_4 \twoheadrightarrow \mathcal{B}_6$ are local properties. Therefore, let us consider a point $p \in \mathcal{B}_6$. It was pointed out in \cite{Bershadsky:1996nh} that every such point admits an open neighbourhood $p \in U \subseteq \mathcal{B}_6$ such that for every $q \in U$ it holds
\[ \pi^{-1} \left( \tilde{ q } \right) = \left\{ \left[ x, y, z \right] \in \mathbb{P}^{2,3,1}_{\mathbb{C} } \, \left| \, P_T ( q, x, y, z ) = 0 \right. \right\} \, \]
where the Tate polynomial $P_T$ is given by
\[ P_T \left( q, x, y, z \right) = x^3 - y^2 + xyz a_1 \left( q \right) + x^2 z^2 a_2 \left( q \right) + y z^3 a_3 \left( q \right) + x z^4 a_4 \left( q \right) + z^6 a_6 \left( q \right) \, . \]
In this expression the $a_i$ are local sections of $\overline{K}_{\mathcal{B}_6}^{\otimes i}$ over $U$. The advantage of this description over a Weierstrass model is that one can easily classify the singularities over $\Delta \cap U$. A systematic analysis leads to the results summarised in \cite[Table 2]{Bershadsky:1996nh}. To extend this powerful approach to the entire base space $\mathcal{B}_6$, one simply cuts out the elliptic fibre over every point $p \in \mathcal{B}_6$ from the Tate polynomial. This construction is known as a \emph{global Tate model}.

\paragraph{An Example of a Global Tate Model}

Let us use this opportunity to introduce a global Tate model, which we will discuss frequently in this thesis. While we are at it, we also exemplify a number of structure, which we will discuss in far more generality in \cref{subsec:GaugeTheoriesInFTheory}. The global Tate model that we have in mind has $SU(5)$ gauge symmetry  over a hypersurface
\[ W = \left\{ p \in \mathcal{B}_6 \; \left| \; w \left( p \right) = 0 \right. \right\} \subset \mathcal{B}_6 \, . \]
As the appearance of $SU(5)$ signals, these geometries will have application in F-theory GUT models. In anticipation of this, we denote the hypersurface $W$ over which this $SU(5)$-gauge symmetry is present as the \emph{GUT-surface}. 

As we will explain in more detail in \cref{subsec:GaugeTheoriesInFTheory}, the singularity structure of the elliptic fibration over $W$ encodes this gauge symmetry. By the analysis of \cite{Bershadsky:1996nh}, the singularity type in turn is encoded in the vanishing orders of the sections $a_i$ on $W$. For example, for an $SU(5)$ gauge symmetry these sections must factor according to
\[ a_2 = a_{2,1} \cdot w, \quad a_3 = a_{3,2} \cdot w^2, \quad a_4 = a_{4,3} \cdot w^3, \quad a_6 \equiv a_{6,5} \cdot w^5 \,  \]
where $a_{i,j} \in H^0( \mathcal{B}_6, \overline{K}_{\mathcal{B}_6}^{\otimes i} \otimes \mathcal{O}_{\mathcal{B}_6} ( - W )^{\otimes j} )$ does not vanish on $W$.

Phenomenologically it is quite appealing, to consider so-called flipped GUT-models. In these models the gauge group contains additional Abelian gauge group factors, and the resulting selection rules can help to avoid proton decay. In F-theory, such Abelian gauge symmetries originate from additional sections of the fibration. We will explain this in more detail in \cref{subsec:GaugeTheoriesInFTheory}, but suffice it for now to state that arguable, one of the simplest ways to achieve a single  additional section, is by considering the so-called \emph{$U(1)$-restricted model}.\footnote{See for example \cite{Cvetic:2013uta, Borchmann:2013hta} for F-theory with multiple $U(1)$-factors.} The latter was originally introduced in \cite{Grimm:2010ez} and requires $a_6 \equiv 0$. Altogether, this shows that we can install an $SU(5) \times U(1)_X$-gauge symmetry in such a global Tate model by factoring the sections $a_i$ according to
\[ a_2 = a_{2,1} \cdot w \, , \quad a_3 = a_{3,2} \cdot w^2 \, , \quad a_4 = a_{4,3} \cdot w^3 \, , \quad a_6 \equiv 0 \, . \]

Over special subloci of $W$, the sections $a_i$ vanish to even higher orders, which correspond by the works of \cite{Bershadsky:1996nh} to more severe singularities. This in turn signals gauge enhancement in F-theory. In the case at hand this was analysed in \cite{Krause:2011xj, oai:arXiv.org:1202.3138}. It was found that such more severe singularities are located over the so-called \emph{matter curves}:
\begin{align}
\begin{split}
 C_{\mathbf{10}_1} &= \left\{ p \in \mathcal{B}_6 \; \left| \; w \left( p \right) = a_1 \left( p \right) = 0 \right. \right\} \, , \\
 C_{\mathbf{5}_3} &= \left\{ p \in \mathcal{B}_6 \; \left| \; w \left( p \right) = a_{3,2} \left( p \right) = 0 \right. \right\} \, , \\
 C_{\mathbf{5}_{-2}} &= \left\{ p \in \mathcal{B}_6 \; \left| \; w \left( p \right) = a_{4,3} \left( p \right) \cdot a_1 \left( p \right) - a_{3,2} \left( p \right) a_{2,1} \left( p \right) = 0 \right. \right\} \, . \\
\end{split}
\end{align}
Intuitively, in type IIB language these curves correspond to the intersections of D7-branes, at which charged, massless matter is known to localise. Indeed, we will explain in \cref{subsec:GaugeTheoriesInFTheory} that these curves are intricately related to massless matter in F-theory. In this picture, the Yukawa interactions of these matter states happen at the so-called \emph{Yukawa-loci}, which are the co-dimensional 2-loci of $W$, over which even more severe singularity enhancements occur.
Therefore the Yukawa loci are always collections of points. In the case at hand, there are precisely three such loci, namely
\begin{align}
\begin{split}
Y_1 &= \left\{ p \in \mathcal{B}_6 \; \left| \; w \left( p \right) = a_1 \left( p \right) = a_{2,1} \left( p \right) = 0 \right. \right\} \, , \\
Y_2 &= \left\{ p \in \mathcal{B}_6 \; \left| \; w \left( p \right) = a_1 \left( p \right) = a_{3,2} \left( p \right) = 0 \right. \right\} \, , \\
Y_3 &= \left\{ p \in \mathcal{B}_6 \; \left| \; w \left( p \right) = a_{3,2} \left( p \right) = a_{4,3} \left( p \right) = 0 \right. \right\} \, .
\end{split}
\end{align}
For example $Y_1$ encodes the Yukawa interaction $\mathbf{10}_1$ $\mathbf{10}_1$ $\mathbf{5}_{-2}$. More details can be found in \cite{Denef:2008wq, Donagi:2008ca, Weigand:2010wm, Krause:2011xj, oai:arXiv.org:1202.3138}.

\subsection{A definition of F-Theory} \label{subsec:DefinitionOfFTheory}

\begin{figure}[tbp]
\centering
\includegraphics{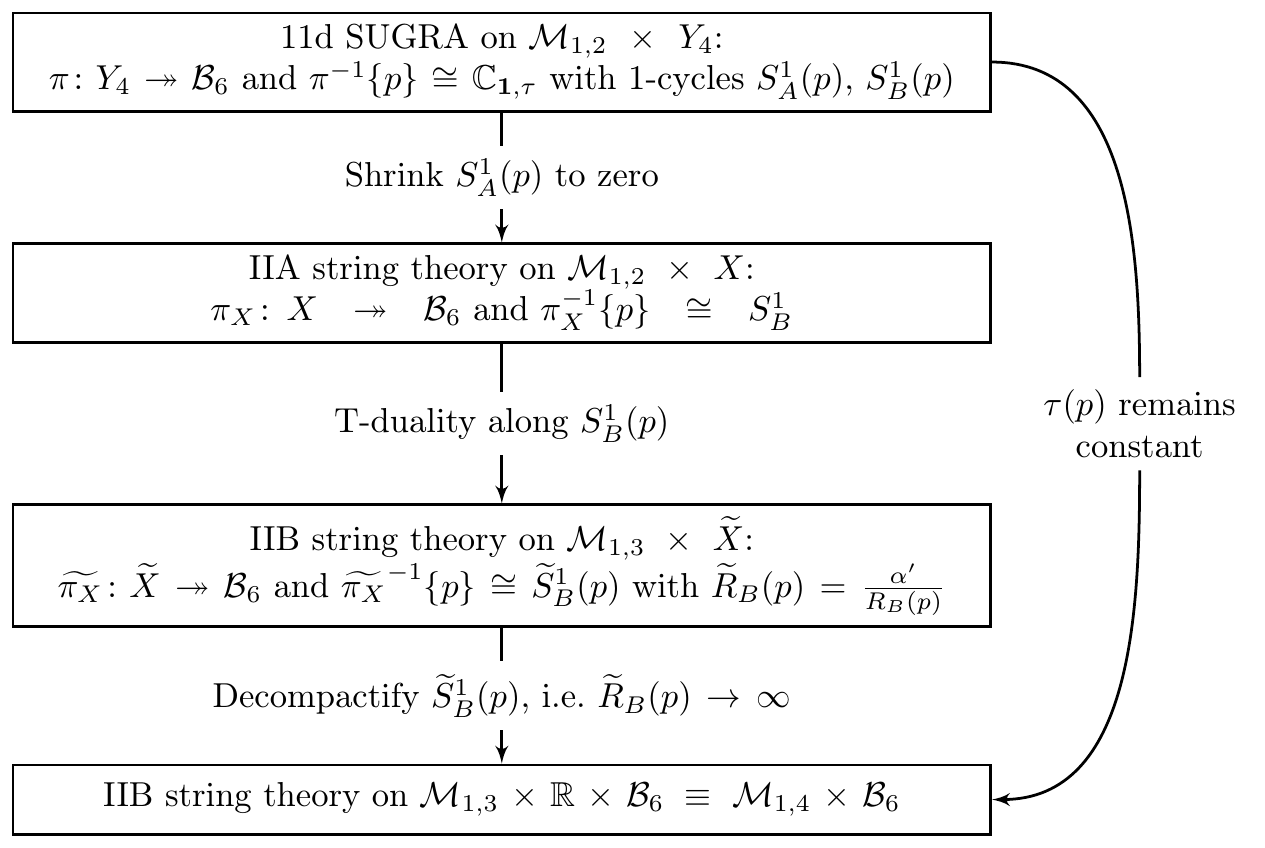}
\caption[Definition of \emph{F-theory} as special \emph{M-theory} limit.]{F-theory on $\mathcal{M}_{1,3} \times Y_4$ is understood as the type IIB \emph{string theory} dual, in the above sense, to \emph{M-theory} on $\mathcal{M}_{1,2} \times Y_4$.}
\label{figure-9876545234}
\end{figure}

Thus far, we have given an intuitive geometric description which encodes the axio-dilaton profile of a type IIB compactification. Unfortunately this picture lacks rigorous structures to deduce physical quantities from. This is overcome by approaching \emph{F-theory} from \emph{M-theory} as follows (see \cite{Denef:2008wq} for more details):
\begin{enumerate}
 \item Compactify the low-energy limit of \emph{M-theory}, \ie 11-dimensional supergravity, on $\mathcal{M}_{1,8} \times T^2$. Let the radii of the two 1-cycles of this 
      torus $T^2$ be $R_A$ and $R_B$ and its complex structure modulus be $\tau$.
 \item In the limit $R_A \to 0$, this 11-dimensional field theory turns into weakly coupled ($g_s \simeq \frac{R_A}{l_s}$) type IIA \emph{string theory} on $\mathcal{M}_{1,8} 
      \times S^1_B$.
 \item Next perform a T-duality along $S^1_B$, \ie (\cf \cref{equ:StringMassIncludingWindingAndMomenta})
      \[ \mathcal{M}_{1,8} \times S^1_B \to \mathcal{M}_{1,8} \times \tilde{S^1_B} \, , \qquad \tilde{R_B} = \frac{\alpha^\prime}{R_B} \, . \]
      In consequence, we recover type IIB \emph{string theory} on $\mathcal{M}_{1,8} \times \tilde{S^1_B}$.
 \item Finally, the limit $R_B \to 0$ decompactifies $\tilde{S^1_B}$ and gives type IIB \emph{string theory} on $\mathcal{M}_{1,9}$.
\end{enumerate}
Crucially, the complex structure modulus $\tau$ of the torus $T^2$ remains constant while passing through all these limiting processes -- only its area is subject to change. It is a simple matter to generalise this procedure to the compactification of \emph{M-theory} on $\mathcal{M}_{1,3} \times Y_4$ with $Y_4 \twoheadrightarrow \mathcal{B}_6$ an elliptic fibration. This leads to the following definition of \emph{F-theory}:
\ebox{F-theory on $\mathcal{M}_{1,3} \times Y_4$ ($\pi \colon Y_4 \twoheadrightarrow \mathcal{B}_6$ an elliptic fibration) is the type IIB \emph{string theory} which is dual, in the sense summarised in \cref{figure-9876545234}, to \emph{M-theory} on $\mathcal{M}_{1,2} \times Y_4$.}
Based on this definition one can now match \emph{M-theory} properties with geometric quantities of $\pi \colon Y_4 \twoheadrightarrow \mathcal{B}_6$. Many results along these lines have been obtained, see \eg \cite{Morrison:1996na, Morrison:1996pp, Denef:2008wq, Weigand:2010wm}. Shortly we will revise a selection of results, which are  of relevance to this thesis.

\subsection{F-theory Vacua in this Thesis} \label{subsec:ConditionsOnBaseAndFibration}

\paragraph{Conditions on The Torus Fibration}

Before we do this, let us point out that it is possible to analyse \emph{F-theory} from mere torus-fibrations along the lines of \cite{Witten:1996bn, Braun:2014oya, Anderson:2014yva, Klevers:2014bqa, Garcia-Etxebarria:2014qua, Mayrhofer:2014haa, Mayrhofer:2014laa, Morrison:2014era, Cvetic:2015moa, Lin:2015qsa, Kimura:2016crs}. As anticipated in the previous section already, we will not follow this route in this thesis. Rather we impose additional conditions, which we motivate and explain now.

As the singular locus $\Delta$ of $Y_4$ resembles the location of D7-branes, the space $Y_4$ is singular for non-trivial \emph{F-theory} compactifications. Therefore, this geometric approach forces us to embrace singularities as `good quantities'. And indeed there are ideas on how to perform \emph{F-theory} compactifications directly on such singular spaces -- see \cite{Collinucci:2014qfa, Collinucci:2014taa} and references therein. Still, it is far more common a practise in \emph{F-theory} to resolve the singular space $Y_4$ and then to work with a smooth resolution $\hat{Y}_4$. In particular, this includes the analysis in \cite{Krause:2011xj, oai:arXiv.org:1202.3138}. As we will eventually wish to apply these results, we follow this philosophy and demand the existence a smooth resolution $\hat{\pi} \colon \hat{Y}_4 \twoheadrightarrow \mathcal{B}_6$.

Whilst the generic fibre of the singular fibration $\pi \colon Y_4 \twoheadrightarrow \mathcal{B}_6$ is a smooth curve, the fibres $\pi^{-1} ( d )$ for $d \in \Delta$ are singular curves. There exist situations, in which the fibre $\pi^{-1} ( b )$ over certain points $b \in \mathcal{B}_6$ is no longer a complex curve, but an object of strictly higher dimension. Such a fibration is termed \emph{non-flat}. The examples considered in \emph{F-theory} almost exclusively consider \emph{flat} fibrations, as many mathematical results apply to such fibrations only. In this thesis we will follow this philosophy and demand that $\pi \colon Y_4 \twoheadrightarrow \mathcal{B}_6$ and its smooth resolution $\hat{\pi} \colon \hat{Y}_4 \twoheadrightarrow \mathcal{B}_6$ are flat fibrations.

We already pointed out the importance of Calabi--Yau manifolds for string compactifications. Whilst required for consistency of a compactification without fluxes, they are phenomenologically appealing as they preserve a fraction of the original supersymmetry. Also \emph{F-theory} is no exception to this rule. If the singular torus fibration $Y_4$ is Calabi--Yau, then the \emph{F-theory} compactification on $\mathcal{M}_{1,3} \times Y_4$ leads to an $\mathcal{N} = 1$ effective 4-dimensional theory -- see \cite{Weigand:2010wm} and references therein. We will hence demand that $Y_4$ is Calabi--Yau. The process of resolution of $Y_4$ can lead to a smooth space $\hat{Y}_4$ with different first Chern class. Resolutions that however leave $c_1 ( T_{\hat{Y}_4} )$ unaltered, are termed \emph{crepant resolutions}. We will consequently focus on singular spaces $Y_4$ which are Calabi--Yau and can be resolved crepantly. The latter then leads to a smooth resolution $\hat{\pi} \colon \hat{Y}_4 \twoheadrightarrow \mathcal{B}_6$ which is Calabi--Yau also.\footnote{This cannot be satisfied in the presence of $\mathbb Q$-factorial terminal singularities as studied systematically in \cite{Arras:2016evy} and references therein. In such instances, we are forced to work on a singular space if the Calabi--Yau property is to be maintained.}

The studies of \cite{Krause:2011xj, oai:arXiv.org:1202.3138} focused on elliptic fibrations, as this allows for a study of global Tate models (\cf \cref{subsec:FTheoryFromIIB}). The section of this elliptic fibration is typically referred to as the \emph{zero section} $S_0$. Its associated (co)homology class will be denoted by $[ S_0 ] \in H^{1,1} ( \hat{Y}_4 )$.\footnote{The existence of a section as such is not a restriction because one can always pass to the associated Jacobian fibration even if a given fibration has no section. However, typically the Weierstrass model describing the Jacobian has $\mathbb Q$-factorial terminal singularities in codimension-2 which cannot be resolved crepantly \cite{Braun:2014oya}. In this sense, the prime assumption is really the absence of terminal singularities such that a smooth Calabi--Yau fibration exists.} As we wish to apply these results, we focus on elliptic fibrations. In summary we impose the following conditions on the singular torus fibration $Y_4$:
\begin{itemize}
 \item The singular torus fibration $Y_4$ is an elliptic fibration which satisfies the Calabi--Yau condition in \cref{equ:StrictCalabiYauCondition}.
 \item $Y_4$ can be resolved smooth, flat and crepantly.
\end{itemize}
Resolutions satisfying these criteria have been constructed for four-dimensional \emph{F-theory} compactifications for instance in \cite{Blumenhagen:2009yv, Grimm:2009yu, Chen:2010ts, oai:arXiv.org:1011.6388, Knapp:2011wk, Esole:2011sm, Marsano:2011hv, Krause:2011xj, oai:arXiv.org:1111.1232, oai:arXiv.org:1202.3138, Lawrie:2012gg, Hayashi:2013lra, Hayashi:2014kca, Esole:2014bka} and references therein.

\paragraph{Conditions on the Base Space}

Not every base space $\mathcal{B}_6$ admits an elliptic fibration $\pi \colon Y_4 \twoheadrightarrow \mathcal{B}_6$ with vanishing first Chern class. Let us first note that the singular locus $\Delta$ of such a fibration is in general the union of irreducible components, \ie
\[ \Delta = \bigcup_{I \in \mathcal{I}}{\Delta_I} \, . \]
It was shown in \cite{KodairaCompactAnalyticSufraces} that the first Chern class of the tangent bundle $T_{Y_4}$ of $Y_4$ satisfies \cite{KodairaCompactAnalyticSufraces}
\[ c_1 \left( T_{Y_4} \right) \cong \pi^\ast \left[ c_1 \left( T_{\mathcal{B}_6} \right) - \sum_{I \in \mathcal{I}}{\frac{\delta_I}{12} \, p \left( \Delta_I \right)} \right] \]
where $p \left( \Delta_I \right)$ denotes the Poincaré dual in $\mathcal{B}_6$ of the irreducible components $\Delta_I$ of the singular locus $\Delta$. The vanishing order of $\Delta$ along $\Delta_I$ is denoted by $\delta_I$. This shows that $Y_4$ has vanishing first Chern class precisely if
\[ \sum_{I \in \mathcal{I}}{\delta_I \, p \left( \Delta_I \right)} = 12 c_1 \left( T_{\mathcal{B}_6} \right) \]
which strongly resembles the tadpole cancellation condition in type IIB \emph{string theory}, which we discussed around \cref{equ:TadpoleCancellationIIB}.
      
If $c_1 \left( T_{\mathcal{B}_6} \right) = 0$, then the demand $c_1 ( T_{Y_4} ) = 0$ implies that $\Delta$ must not vanish to any order. Hence, the fibration $\pi \colon Y_4 \twoheadrightarrow \mathcal{B}_6$ has no singularity. However, we will point out momentarily that these singularities encode the gauge group of the \emph{F-theory} compactification, and no singularities correspond to trivial group symmetry. Therefore, this scenario is not desirable, and we conclude that for non-trivial \emph{F-theory} compactifications $\mathcal{B}_6$ must not be Calabi--Yau.
 
In this thesis we focus on connected base spaces $\mathcal{B}_6$. Then necessary conditions for elliptic fibrations to exist over $\mathcal{B}_6$ include $h^{1,0} ( \mathcal{B}_6 ) = h^{2,0} ( \mathcal{B}_6 ) = 0$. This parallels the situations studied in \cite{Couzens:2017way, Bonetti:2011mw}. More results on the intricate relationship between the elliptic fibration $\hat{Y}_4$ and $\mathcal{B}_6$ follow for example from \cite{2016arXiv160802997D}.

In \cref{sec:ToricFTheoryGUTModels} we will work out (self-)intersection numbers of some curves. This in turn is achieved by expressing some $\mathbb{Z}$-
Cartier divisors $D$ as sums of proper $\mathbb{Q}$-Cartier divisors, \eg as
\[ D = \frac{5}{7} \cdot D + \frac{2}{7} \cdot D = \frac{1}{7} \left( 5 D + 2 D \right) \, . \label{equ:RewriteDivisor} \] 
This of course is meaningful only if $5 D$ and $2 D$ are not linearly equivalent to the trivial divisor. More generally one says that a divisor class $D$ is torsion-free if for any $n \in \mathbb{N}_{>0}$, $n D$ is not linearly equivalent to zero. We should hence demand that the entire Picard group of $\mathcal{B}_6$ be torsion-free. 

Let us now formulate a criterion under which this torsion is absent. To this end, we recall that, as explained in \cite{Bies:2014sra} and references therein, the Picard group $\mathrm{Pic} ( \mathcal{B}_6 )$ fits into an exact sequence
\[ 0 \to H^{1,0} \left( \mathcal{B}_6 \right) / H^1 \left( \mathcal{B}_6, \mathbb{Z} \right) \hookrightarrow \mathrm{Pic} \left( \mathcal{B}_6 \right) \xtwoheadrightarrow {c_1} H^{1,1}_{\mathbb{Z}} \left( \mathcal{B}_6 \right) \to 0 \, . \label{equ:DeligneCohomologyForLineBundles} \]
The map $c_1$ assigns to a line bundle its first Chern class. Physically, this first Chern class is understood as the field strength of a $U(1)$-gauge theory encoded by the gauge connection of the corresponding line bundle. The quotient $H^{1,0} ( \mathcal{B}_6 ) / H^1 ( \mathcal{B}_6, \mathbb{Z} )$ is known as the so-called \emph{intermediate Jacobian}. In physics lingo it describes Wilson-line degrees of freedom. We can now see from \cref{equ:DeligneCohomologyForLineBundles} that the Picard group is torsion-free iff $H_1( \mathcal{B}_6, \mathbb{Z} )$ vanishes. 

In \cref{sec:ComputingTheSpectra} and \cref{chapter:GUTModels}, we analyse geometries derived from an $SU(5) \times U(1)_X$-top which was first introduced in \cite{Krause:2011xj} and which we analyse with more refined techniques in \cref{sec:SU5xU1Top}. Also these studies involve manipulations of the type \cref{equ:RewriteDivisor} and we should consequently demand that the Picard group of $\mathcal{B}_6$ be torsion-free. Also in \cref{chapter:LocalAnomaliesInF-Theory} we investigate such geometries. Despite our entirely different focus in that chapter, the applied manipulations again require the absence of torsional divisors. All that said, we will demand throughout this thesis that $h_1( \mathcal{B}_6, \mathbb{Z} ) = 0$. So we focus in this thesis on F-theory base spaces which satisfy the following:
\ebox{$\mathcal{B}_6$ is a connected 3-fold with $h^{1,0} ( \mathcal{B}_6 ) = h^{2,0} ( \mathcal{B}_6 ) = 0$ and $h_1 ( \mathcal{B}_6, \mathbb{Z} ) = 0$.}

\subsection{Gauge Theories from Geometry} \label{subsec:GaugeTheoriesInFTheory}

\paragraph{A Picture of Blow-Up Resolutions}

\begin{figure}[tb]
\centering
\includegraphics{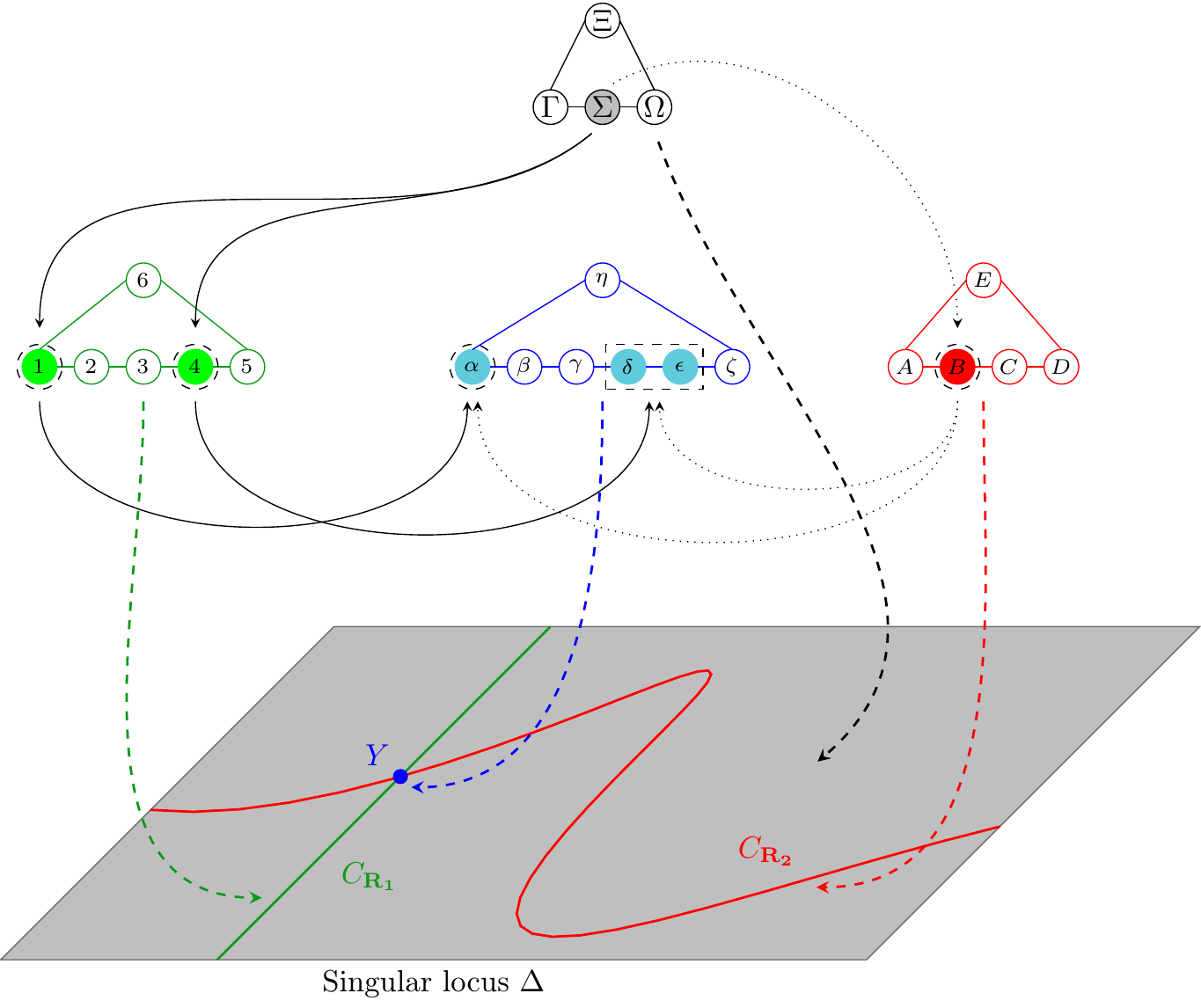}
\caption{Schematic picture of the resolved fibre structure over the singular locus $\Delta$.}
\label{figure-asdf}
\end{figure}

Let us now look at how we resolve a singular elliptic fibration $\pi \colon Y_4 \twoheadrightarrow \mathcal{B}_6$, of which the singular locus $\Delta \subseteq \mathcal{B}_6$ encodes the location of D7-branes. We achieve this smooth resolution $\hat{\pi} \colon \hat{Y}_4 \twoheadrightarrow \mathcal{B}_6$ by replacing the singular fibres over $\Delta$ by a finite number of complex projective spaces. In this sense we `blow-up' every singularity into a finite number of $\mathbb{P}^1_{\mathbb{C}}$s, for which reason this procedure is known as \emph{blow-up resolution}. To gain some intuition for this procedure let us assume that $\Delta$ is irreducible. Then \cref{figure-asdf} captures a generic fibre structure found after a blow-up resolution.

Over a generic point of the singular locus $\Delta$, we resolved the singular elliptic fibre by use of three $\mathbb{P}^1_{\mathbb{C}}$s $\Gamma$, $\Sigma$, $\Omega$. The $\mathbb{P}^1_{\mathbb{C}}$ $\Xi$ represents the original singular elliptic curve. It is a general feature that the type of singularity becomes more severe on special subloci of $\Delta$. Such loci of codimension-1 are termed \emph{matter curves}. In \cref{figure-asdf} we have sketched two matter curves $C_{\mathbf{R}_1}$ and $C_{\mathbf{R}_2}$. Owing to the enhanced singularity type over $C_{\mathbf{R}_1}$ and $C_{\mathbf{R}_2}$, additional $\mathbb{P}^1_{\mathbb{C}}$s were required to resolve the singular fibre -- five over $C_{\mathbf{R}_1}$ and four over $C_{\mathbf{R}_2}$. The original singular elliptic fibres over $C_{\mathbf{R}_1}$ and $C_{\mathbf{R}_2}$ are given by the fibre-$\mathbb{P}^1_{\mathbb{C}}$s $6$ and $E$, respectively. Likewise, over special subloci of codimension-$1$ of the matter curves, even more severe singularities occur. These codimension-2 subloci of $\Delta$ are known as the \emph{Yukawa loci}. In \cref{figure-asdf} we have sketched one such locus $Y$, whose singularity required the use of six $\mathbb{P}^1_{\mathbb{C}}$s for blow-up resolution.

We can envision a continuous process of moving a point $p \in \Delta \backslash \{ C_{\mathbf{R}_1, \mathbf{R}_2} \}$ to one of the matter curves and subsequently onto the Yukawa locus $Y$. For example, we can move $p$ to $C_{\mathbf{R}_1}$ first and then to $Y$. In doing so, the fibre-$\mathbb{P}_{\mathbb{C}}^1$ $\Sigma$ changes. First, once $p$ arrives at $C_{\mathbf{R}_1}$ it splits into the sum $1+4$ of fibre-$\mathbb{P}_{\mathbb{C}}^1$s. Likewise, once $p$ is moved to the Yukawa locus $Y$, these $\mathbb{P}^1_{\mathbb{C}}$s split into $\alpha$ and $\delta + \epsilon$, respectively, which are fibre-$\mathbb{P}^1_{\mathbb{C}}$s over $Y$. Likewise we have the splitting process
\[ \Sigma \xrightarrow{p \to C_{\mathbf{R}_2}} B \xrightarrow{p \to Y} \alpha + \delta + \epsilon \, . \]

Of ample importance to F-theory are the intersection numbers of the $\mathbb{P}^1_{\mathbb{C}}$s in the resolved fibre $\hat{\pi}^{-1} \{ p \}$, $p \in \Delta$. In \cref{figure-asdf}, we indicate a single intersection by a connecting line between two fibre-$\mathbb{P}^1_{\mathbb{C}}$s. Hence, given a point $p \in \Delta \backslash \{ C_{\mathbf{R}_1}, C_{\mathbf{R}_2} \}$, $\Gamma$ and $\Sigma$  intersect with multiplicity $+1$ in the fibre $\pi^{-1} \{ p \}$. If in addition we assume that all fibre-$\mathbb{P}^1_{\mathbb{C}}$s have self-intersection number $-2$, then we pattern formed from $\Gamma$, $\Sigma$, $\Omega$, $\Xi$ is well-known in representation theory of Lie algebras. It is termed the affine Dynkin diagrams 
$\tilde{A}_4$ of $\mathfrak{su(4)}$. This is how non-Abelian gauge symmetries arise in F-theory! 

Before we continue with the physics discussion, it is therefore worth to take a brief detour into the theory of Lie algebras. Our revision follows \cite{hall2003lie}, where the interested reader can find additional information.

\paragraph{Roots and Dynkin Diagrams}

Recall that a Lie algebra is a vector space $\mathfrak{g}$ together with the Lie bracket $\mathfrak{g} \times \mathfrak{g} \to \mathfrak{g} \, , \; ( x,y ) \mapsto [ x, y ]$. A subspace $\mathfrak{i} \subseteq \mathfrak{g}$ which satisfies $[ \mathfrak{g}, \mathfrak{i} ] \subseteq \mathfrak{i}$ is termed an ideal of $\mathfrak{g}$. A Lie algebra is by definition simple if its only ideals are $\{ 0 \}$ and $\mathfrak{g}$ itself. Finally, a Lie algebra is semi-simple precisely if it (isomorphic to) the direct sum of simple Lie algebras. For the physics applications in this thesis, it suffices to focus on finite-dimensional, semi-simple complex Lie algebras. Unless stated differently, this is implicitly assumed from now on.

A major result concerning Lie algebras is that they are classified by Dynkin diagrams \cite{dynkin2000selected}. The novel story behind this classification is as follows: Given a Lie algebra $\mathfrak{g}$, there exists a special, maximally commutative subalgebra -- the Cartan subalgebra $\mathfrak{h}$ of $\mathfrak{g}$. A \emph{root} of $\mathfrak{g}$ is a linear functional $\alpha$ on $\mathfrak{h}$, such that there exists a non-zero $X \in \mathfrak{g}$ with
\[ \left[ H, X \right] = \alpha \left( H \right) \cdot X \, \quad \forall H \in \mathfrak{h} \, . \]
It is possible and convenient, to identify a root $\alpha$ with an element $H^\alpha \in \mathfrak{h}$, which is subject to the condition that for all $H \in \mathfrak{h}$ it holds $\alpha ( H ) = \langle H^\alpha, H \rangle$, where $\langle \cdot, \cdot \rangle$ denotes an inner-product on $\mathfrak{h}$. For example one can consider the Hilbert-Schmidt inner product given by $\langle X, Y \rangle = \mathrm{tr} ( X^\ast Y )$. Along these lines, we will from now on understand a root $\alpha$ as an element of the Cartan subalgebra $\mathfrak{h}$. The collection $R = \{ \alpha_i \}$ of all these roots forms a \emph{root system}, which means that they satisfy the following rules:
\begin{itemize}
 \item The roots span $\mathfrak{h}$.
 \item For every $\alpha_i \in R$, the only scalar multiplies contained in $R$ are $\pm \alpha_i$.
 \item For all $\alpha_i, \alpha_j \in R$, also $\alpha_j - 2 \cdot \frac{\langle \alpha_i, \alpha_j \rangle}{\langle \alpha_i, \alpha_i \rangle} \cdot \alpha_i$ is 
      contained in $R$.
 \item The Cartan matrix $C_{ij} = 2 \frac{\langle \alpha_i, \alpha_j \rangle}{\langle \alpha_i, \alpha_i \rangle}$ is integer valued.
\end{itemize}
Any root system even admits a set of positive roots $R^+$, subject to the following:
\begin{itemize}
 \item For any $\alpha_i \in R$, exactly one of the roots $\pm \alpha_i$ is in $R^+$,
 \item For any two distinct $\alpha_i, \alpha_j \in R^+$ with $\alpha_i + \alpha_j \in R$, also $\alpha_i + \alpha_j \in R^+$.
\end{itemize}
A positive root $\alpha \in R^+$ is \emph{simple}, if it cannot be written as sum of other positive roots. In general, a root system $R$ admits many choices of simple roots. We denote such a choice by $\Delta$. Then we can visualise the associated root system by a Dynkin diagram. In such a diagram each element $d_i \in \Delta$ is represented by a vertex $v_i$. Two vertices $v_i$ and $v_j$ are connected by edges, which represent the angle formed by the roots $\alpha_i$ and $\alpha_j$ (as well as their relative lengths). It is a non-trivial result that this visualisation actually classifies the Lie algebras in question \cite{dynkin2000selected}. For example the finite-dimensional, simple, complex Lie algebras fall in four families $A_n$, $B_n$, $C_n$, $D_n$ with the five exceptions $F_4$, $G_2$, $E_6$, $E_7$ and $E_8$. Of these the Lie algebras of type $A$, $D$, $E_6$, $E_7$ and $E_8$ are known as simply-laced, which among others means that all their roots have the same length.

\begin{figure}[tbp]
\centering
\begin{tikzpicture}[scale = 3, >=stealth]

    \def\l{2.5}
    \def\h{1}
    
    \draw[->] (-0.9,0) -- (1.2,0);
    \draw[->] (0,-1.1) -- (0,1.2);
    
    \draw (-0.5,0.05) -- (-0.5,-0.05);
    \draw (0.5,0.05) -- (0.5,-0.05);
    \draw (-0.05,1) -- (0.05,1);
    \draw (-0.05,0.5) -- (0.05,0.5);
    \draw (-0.05,-0.5) -- (0.05,-0.5);
    
    \node at (0.5,-0.1) [below] {$0.5$};    
    \node at (-0.05,0.5) [left] {$0.5$};
    
    \draw[thick, darkgreen,->] (0,0) -- (-1,0);
    \draw[thick, darkgreen,->] (0,0) -- (0.5,-0.866);
    \draw[thick, darkgreen,->] (0,0) -- (-0.5,-0.866);
    \draw[thick, darkgreen,->] (0,0) -- (0.5,0.866);
    \draw[thick, red,->] (0,0) -- (1,0);
    \draw[thick, red,->] (0,0) -- (-0.5,0.866);
    
    \node at (1,0) [below] {$\alpha_1$};
    \node at (-0.5,0.866) [left] {$\alpha_2$};

    \def\shift{2.5}
    \draw[black] (-0.5+\shift, 0) -- (0.5+\shift,0);
    \draw[red, fill] (-0.5+\shift,0) circle (0.7pt);
    \node at (-0.5+\shift,0) [below] {$\alpha_1$};
    \draw[red, fill] (0.5+\shift,0) circle (0.7pt);
    \node at (0.5+\shift,0) [below] {$\alpha_2$};
    
\end{tikzpicture}
\caption{The root system $A_2$, a choice of simple roots $\alpha_1$, $\alpha_2$ and its Dynkin diagram.}
\label{figure-RootAndDynkinOfA2}
\end{figure}
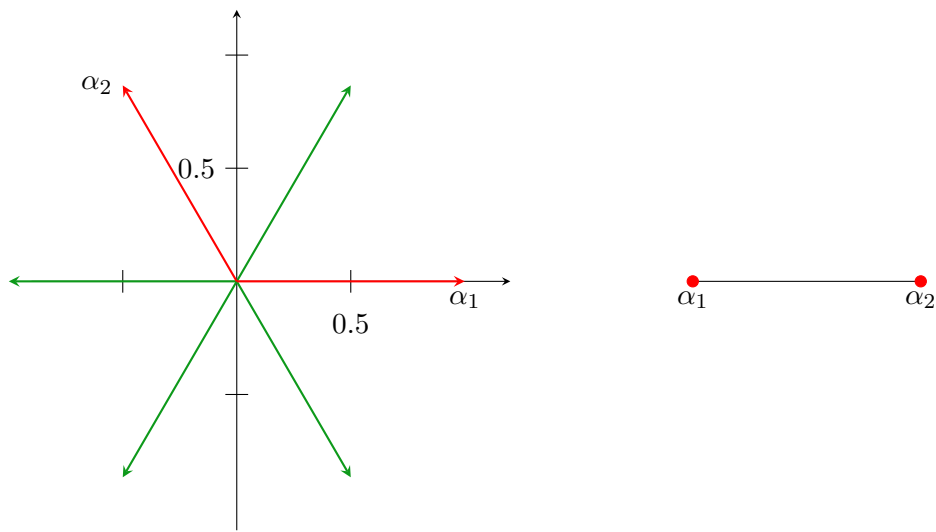

An example for a Lie algebra of type $A$ is $\mathfrak{su} ( 3 )$. Its root system is known as $A_2$ and can be represented by the following collection of roots
\[ R = \left\{ \pm \left( \begin{array}{c} 1 \\ 0 \end{array} \right), \pm \left( \begin{array}{c} -1/2 \\ \sqrt{3} / 2 \end{array} \right), \pm \left( \begin{array}{c} 1/2 \\ \sqrt{3} / 2 \end{array} \right) \right\} \, . \label{equ:RootSystemA2} \]
We picture these roots together with a choice of simple roots $\Delta = \{ \alpha_1, \alpha_2 \}$ in \cref{figure-RootAndDynkinOfA2}. The associated Dynkin diagram is presented in this figure also. In this particular case $\alpha$ and $\beta$ form an angle of 120°, which by convention corresponds to an undirected single edge. This diagram closely resembles the fibre structure in \cref{figure-asdf}, expect for the absence of the affine node.

\paragraph{Non-Abelian Gauge Theories from Singularities}

With this background, let us now come back to the discussion of \cref{figure-asdf}. We follow \cite{Bies:2017abs} closely and first note that in general the singular locus $\Delta$ of $\pi \colon Y_4 \twoheadrightarrow \mathcal{B}_6$ is the union of irreducible components $\Delta_I$,
\[ \Delta = \bigcup_{I \in {\cal I}}{\Delta_I} \, . \]
After resolution, we have over each irreducible component a number of fibre $\mathbb{P}^1_{\mathbb{C}}$, which intersect in the affine Dynkin diagram of a Lie algebra $\mathfrak{g}_I$.\footnote{If the fibre is of Kodaira type I$_1$ or II, no resolution is required and the gauge algebra $\mathfrak{g}_I$ is trivial.} So again, the type of singularity over $\Delta_I$ encodes the gauge theory on the D7-brane stack located at $\Delta_I$. For the following discussion we assume that the resulting gauge group is a direct product of non-Abelian gauge groups $G_I$, \ie
\[ G_{\mathrm{tot}} = \prod_{I}{G_I} \,. \]
Hence, we consider an elliptic fibration $\pi \colon Y_4 \twoheadrightarrow \mathcal{B}_6$ with codimension-1 singularities $\Delta_I$, each of which is understood as locus of a D7-brane carrying the non-Abelian gauge group $G_I$. For ease of notation, let us label the fibre-$\mathbb{P}^1_{\mathbb{C}}$s over $\Delta_I$ by $\mathbb{P}^1_{i_I}$, $i = 1, \ldots, \mathrm{rk}(G_I)$. Each of these gives rise to a divisor class in $\hat{Y}_4$ -- the associated resolution divisor $E_{i_I}$, $i = 1, \ldots, \mathrm{rk}(G_I)$ is understood as fibration of $\mathbb{P}^1_{i_I}$ over $\Delta_I$. Group theoretically, we identify these resolution divisors $E_{i_I}$ with the the generators of the Cartan subgroups of $G_I$, \ie the simple roots of the Lie algebra $\mathfrak{g}_I$. The gauge potential $\mathbf{A}_{i_I}$ associated with $E_{i_I}$ is obtained by expanding the \emph{M-theory} 3-form as
\[ \label{C3expansion}  C_3 = \mathbf{A}_{i_I} \wedge \left[ E_{i_I} \right] + \ldots \, . \]

Let us explain in more detail how we arrive at identifying the $E_{i,I}$ with the simple roots of $\mathfrak{g}_I$. To this end, let us point out that the projection $\hat{\pi}$ of the fibration $\hat{Y}_4$ induces the pushforward $\hat{\pi}_*: H_k(\hat{Y}_4) \rightarrow H_{k}(\mathcal{B}_6)$. Let us denoted by $[ D^{\mathrm{b}}_\alpha ]$ the Poincaré dual cohomology class associated with the divisor $D^{\mathrm{b}}_\alpha \in \mathrm{Cl} ( \mathcal{B}_6 )$. Then this pushforward satisfies
\[ [ E_{i,I} ] \cdot_{\hat{Y}_4} [ E_{j,J} ] \cdot_{\hat{Y}_4} [ D_\alpha^\mathrm{b}] \cdot_{\hat{Y}_4} [ D_\beta^\mathrm{b} ] = \hat{\pi}_\ast \left( \left[ E_{i,I} \right] \cdot_{\hat{Y}_4} \left[ E_{j,J} \right] \right) \cdot_{\mathcal{B}_6} [ D_\alpha^\mathrm{b} ] \cdot_{\mathcal{B}_6} [ D_\beta^\mathrm{b} ] \, . \]
for all $[ D^{\mathrm{b}}_\alpha ] \in H^{1,1}(\mathcal{B}_6)$. In this expression $\cdot_{\hat{Y}_4}$ denotes the intersection product of cohomology classes on $\hat{Y}_4$ and $\cdot_{\mathcal{B}_6}$ the corresponding intersection product on $\mathcal{B}_6$. Whenever the context allows for it, we will drop the subscripts. Let $C_{i_I k_I}$ be the Cartan matrix of $\mathfrak{g}_I$. Then the group theoretic interpretation of the $E_{i_I}$ follows from the important relation
\[ \hat{\pi}_* \left( \left[ E_{i,I} \right] \cdot \left[ E_{k,K} \right] \right) = - \delta_{IK} \, C_{i_I k_I} \, \left[ \Delta_I \right] \,, \label{intersectionEiEj} \]

Up to this point in the discussion, we have silently assumed that over each irreducible component $\Delta_I$, the complex projective spaces $\mathbb{P}^1_{i_I}$ intersect in affine Dynkin diagram of simply-laced Lie algebras. For non-simply laced Lie algebras, the resolution process introduces exceptional divisors $E_i$, such that the fibre of the $E_i$'s split into several $\mathbb{P}_{\mathbb{C}}^1$s, which are exchanged by monodromies along $\Delta_I$ \cite{Bershadsky:1996nh}. This is nicely explained in \cite{Park:2011ji} and references therein. In this case $\mathbb{P}^1_{i_I}$ labels a basis of linearly independent rational curves in the fibre over $\Delta_I$. To account for these features, \cref{intersectionEiEj} must be modified to take the form
\[ \hat{\pi}_* \left( \left[ E_{i,I} \right] \cdot \left[ E_{k,K} \right] \right) = - \delta_{IK} \, \mathfrak{C}_{i_I k_I} \, \left[ \Delta_I \right] \,, \label{intersectionEiEjNonSimplyLaced} \]
where
\[ \mathfrak{C}_{i_I k_I} = \frac{\langle \alpha, \alpha\rangle_\mathrm{max}}{\langle\alpha_{k_I},\alpha_{k_I} \rangle} \frac{2 \langle \alpha_{i_I}, \alpha_{k_I} \rangle}{\langle \alpha_{k_I}, \alpha_{k_I} \rangle} = \frac{2}{\lambda} \frac{1}{\langle\alpha_{k_I},\alpha_{k_I} \rangle} C_{i_I k_I} \, . \label{DefcalC} \]
The object $\langle \alpha, \alpha\rangle_\mathrm{max}$ denotes the length of the longest simple root. It is related to the so-called \emph{Dynkin index} $\lambda$ of the fundamental representation via 
\[ \lambda = \frac{2}{\left\langle \alpha, \alpha \right\rangle_\mathrm{max}} \, . \label{lambdadef}  \]
The matrix $\mathfrak{C}_{ij}$ also governs the normalization of the coroot basis via $\label{trace-coroot} \mathrm{tr}_{\mathbf{R}}  \mathcal{T}_i \mathcal{T}_j = \lambda \, \mathfrak{C}_{ij} \, c^{(2)}_{\mathbf{R}}$, where the group theoretic constants $c_{\mathbf{R}}^{(n)}$ interpolate between the trace in representation $\mathbf{R}$ and the trace in the fundamental representation via
\[ \label{cRn-def} \mathrm{tr}_{\mathbf{R}} F^n = c^{ \left( n \right)}_{\mathbf{R}} {\mathrm{tr}}_{\mathrm{fund}} F^n \,,\qquad n=2,3 \, . \]
For more details we refer to the group theoretic discussion in \cite{Park:2011ji}.

\paragraph{Abelian Gauge Theories from Sections}

Let us now extend the gauge group by Abelian factors, \ie let us look at a 4-dimensional \emph{F-theory} compactification with gauge group
\[ \label{Gtot}
G_{\mathrm{tot}} = \prod_{I}{G_I} \times \prod_{A}{U(1)_A} \,.
\]
which not only contains non-Abelian gauge groups $G_I$ but also Abelian gauge group factors $U(1)_A$. These Abelian gauge group factors $U(1)_A$ are related to the existence of extra rational sections $S_A$ in addition to the zero-section $S_0$ of the fibration. To each rational section $S_A$ we assign a divisor \footnote{Eventually we will understand such an object as (the class of) an algebraic cycle in the Chow group ${\mathrm{CH}}^1 ( \hat{Y}_4 )$.} 
\[ \label{UAdefintion} U_A = S_A - S_0 - D^{\mathrm{b}} + \sum_{i_I} k_{i_I} E_{i_I} \,. \]
Here the divisor class $D^{\mathrm{b}}$ on $\mathcal{B}_6$ and the coefficients $k_{i_I}$ are to be chosen such that $U_A$ satisfies the following transversality conditions for all $\alpha$, $\beta$, $\gamma$:
\begin{align}
0 = [ U_A ] \cdot [D_\alpha^{\mathrm{b}}] \cdot [D_\beta^{\mathrm{b}}] \cdot [D_\gamma^{\mathrm{b}}] = [ U_A ] \cdot [D_\alpha^{\mathrm{b}}] \cdot [D_\beta^{\mathrm{b}}] \cdot  \left[S_0\right] = [ U_A ] \cdot [D_\alpha^{\mathrm{b}}] \cdot [D_\beta^{\mathrm{b}}] \cdot \left[E_{i_I}\right] \, .
\end{align}
In these equations $[D^{\mathrm{b}}_\alpha]$ represents basis elements of $H^{1,1}(\mathcal{B}_6)$. Such a $U_A$ serves as the generator of the Abelian gauge group $U(1)_A$. The $U(1)_A$ gauge potential $\mathbf{A}_A$ arises by replacing $E_{i_I}$ in the expansion \cref{C3expansion} with $U_A$ \cite{Morrison:1996na, Morrison:1996pp, Grimm:2010ez}. Such constructions have been studied extensively in \emph{F-theory} model building, for example in \cite{Marsano:2009gv, Marsano:2009wr, Dolan:2011aq, Palti:2012dd}.

\paragraph{Weights and Irreducible Representations}

We complete this exposition on gauge theories in F-theory by studying massless matter. It is well-known that in quantum field theory, the matter states furnish irreducible representations of the gauge group. As F-theory is not different, it is worth recalling a few fact of representation theory. Under the assumption that the gauge groups are Lie groups, we suffice it to look at representations of their Lie algebras. More details can be found in \cite{hall2003lie}. 

\begin{figure}[tb]
\centering
\includegraphics{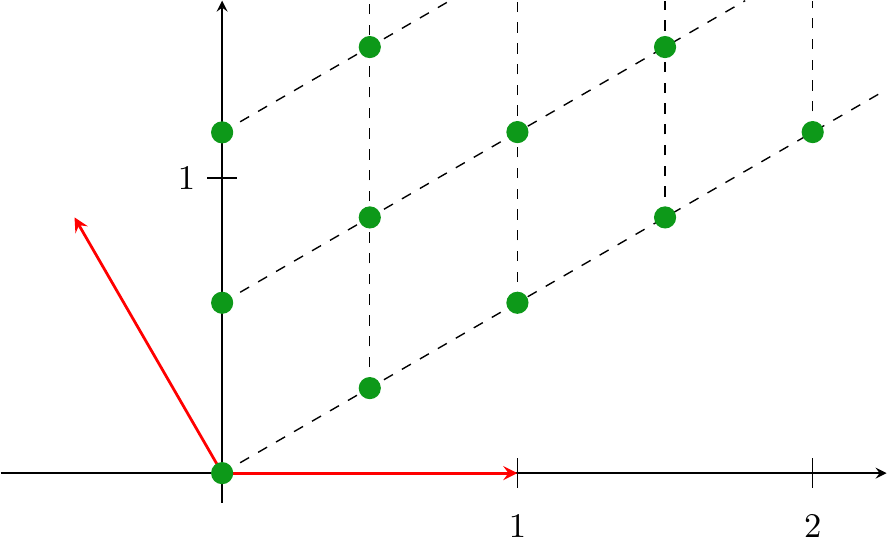}
\caption{The integral dominant elements $\hat{\beta}$ for the root system $A_2$ of $\mathfrak{su} ( 3 )$ are denoted by green bullets.}
\label{figure-IntegralDominantElements}
\end{figure}

The term \emph{weight} is central to representation theory of Lie algebras. It is defined as follows: Take a finite dimensional representation $\mathbf{R} \colon \mathfrak{g} \to \mathrm{End} ( V )$. Then $\beta \in \mathfrak{h}$ is a \emph{weight} precisely  if there exists a non-zero $v \in V$ with the property
\[ \mathbf{R} \left( H \right) \cdot v = \left\langle \beta, H \right\rangle \cdot v \, \quad \forall H \in \mathfrak{h} \, . \label{equ:Weight} \]
By the \emph{theorem of highest weight}, the irreducible representations of $\mathfrak{g}$ are one-to-one to special such weights, namely the \emph{integral dominant elements}. Given a choice $\Delta$ of simple roots \footnote{It can be shown that any two such choices can be mapped to each other by the operation of the so-called Weyl group. In this sense, the explicit choice does not matter.}, an element $\hat{\beta} \in \mathfrak{h}$ is an integral dominant element precisely if for all simple roots $\alpha$, it holds
\[ 2 \cdot \frac{\left\langle \hat{\beta}, \alpha \right\rangle}{\left\langle \alpha, \alpha \right\rangle} \in \mathbb{Z}_{\geq 0} \, . \label{equ:IntegralDominant} \]
This expression tells us that the orthogonal projection of $\hat{\beta}$ onto any simple root $\alpha_i$ is a half-integer $w_i / 2$. In this sense, we can understand integral dominant elements as collection $\{ w_1, \dots, w_{| \Delta |} \}$ of non-negative integers. These will resurface as wrapping numbers of M2-branes, momentarily.
We picture the corresponding situation for the root system $A_2$ of $\mathfrak{su} ( 3 )$ in \cref{figure-IntegralDominantElements}. 

Given an irreducible representation $\mathbf{R} \colon \mathfrak{g} \to \mathrm{End} ( V )$, we see from \cref{equ:Weight} that for a given weight $\beta$ there exists a non-zero vector $v$ with eigenvalue $\langle \beta, H \rangle$ under $\mathbf{R} ( H )$. Of course, the endomorphism $\mathbf{R} ( H )$ will in general allow for many distinct such eigenvalues. Therefore, a given representation $\mathbf{R} \colon \mathfrak{g} \to \mathrm{End} ( V )$ with highest weight $\hat{\beta}$ will admit other weights too. As it turns out, the admissible weights are obtained from $\hat{\beta}$ by the operation of the Weyl group, and gives rise to weights $\beta^a ( \mathbf{R} )$ with $a = 1, \ldots, \mathrm{dim} ( \mathbf{R} )$.\footnote{This parallels a familiar situation from quantum mechanism. Here $\mathfrak{g} = \mathfrak{su} ( 2 )$ and all irreducible representations are classified by non-negative integers $m$ -- the angular momentum in z-direction. The irreducible representation associated to $m \in \mathbb{Z}_{\geq 0}$ is of dimension $m+1$, and the admissible weights are $\{-m, -m+2, -m, \dots, m-2, m \}$.}

\paragraph{Massless Matter}

There are two types of massless matter in such \emph{F-theory} compactifications: The so-called bulk matter arises from states which are free to propagate over the divisors $\Delta_I$ and which transform, in absence of gauge flux, in the adjoint representation of $G_I$. In addition, there exists matter, which is localised in codimension two in the base $\mathcal{B}_6$ and transforms in some representation $\mathbf{R}$ of the full gauge group $G$. It is this matter, which we will primarily focus on during this thesis. We will denote the matter curve on $\mathcal{B}_6$ on which these massless states localise by $C_{\mathbf{R}}$. 

The origin of such matter states are M2-branes wrapping a suitable combination of rational curves in the fibre of $\hat{Y}_4$ over $C_{\mathbf{R}}$. In the physics lingo, the entries of the weight $\beta^a_{i_I}(\mathbf{R})$, $a=1, \ldots, \mathrm{dim}(\mathbf{R})$, are termed the \emph{charges of the $a$-th state in the representation under the Cartan $U(1)$ with generator $E_{i_I}$}. To each state in representation $\mathbf{R}$ we have an associated matter surface $S^a_{\mathbf{R}}$, given by a linear combination of fibral curves over $C_{\mathbf{R}}$. M2-branes wrapping this linear combination of fibral curves $S^a_{\mathbf{R}}$ give rise to the $a$-th state. Thereby, the weight is matched with the intersection number\footnote{In order to ensure for \cref{weightdefinition} we allow $S^a_{\mathbf{R}}$ to be either a linear combinations of holomorphic fibral curves with non-negative coefficients or a linear combination of anti-holomorphic fibral curves with non-negative coefficients. Alternatively, we could stick to linear combinations of holomorphic fibral curves with non-negative coefficients only, but then we would have to include suitable minus signs in \cref{weightdefinition}, depending on the phase of the resolution \cite{Intriligator:1997pq,oai:arXiv.org:1111.1232,Hayashi:2014kca,Esole:2014bka}.}
\[ \label{weightdefinition} \beta^a_{i_I} \left( \mathbf{R} \right) \, \left[ C_{\mathbf{R}} \right] = \hat{\pi}_{\ast} \left( \left[ E_{i_I} \right] \cdot \left[ S^a_{\mathbf{R}} \right] \right) \,. \]
Note that in some cases it may be necessary to allow for fractional coefficients in the definition of $S^a(\mathbf{R})$ in order to achieve \cref{weightdefinition}. An example of this phenomenon is presented in \cref{sec:DetailsOfSU4Top}. Finally, the state encoded by $S^a_{\mathbf{R}}$ has $U(1)_A$ charge
\[ \label{qAdefinition} q_A \left( \mathbf{R} \right) \, \left[ C_{\mathbf{R}} \right] = \hat{\pi}_{\ast} \left( \left[ U_A \right] \cdot \left[ S^a_{\mathbf{R}} \right] \right) \, . \]

\subsection{Gauge Backgrounds from Deligne Cohomology} \label{subsec:GaugeBackgroundsFromDeligneCohomology}

\paragraph{\texorpdfstring{$\mathbf{G_4}$}{G4}-Fluxes}

Whilst the geometry of the smooth resolution $\hat{\pi} \colon \hat{Y}_4 \twoheadrightarrow \mathcal{B}_6$ encodes much information about a four-dimensional \emph{F-theory} compactification, it does not contain it all. Rather such a compactification relies also on a specified gauge background. By duality with \emph{M-theory}, this gauge background is encoded by the \emph{M-theory} 3-form $C_3$ and its field strength $G_4$. 

Recall from Maxwell's theory of electrodynamics that the field strength $F$ encodes only part of the gauge information: the full information is provided by the gauge field $A$ which identifies the field strength via $F = dA$. Similarly, keeping track of $G_4$ only, merely accounts for \emph{part} of the information of the gauge background. Momentarily we will explain that the `gauge field' of a $G_4$-flux can be understood as an element of the Deligne cohomology. Although incomplete in this sense, we can still extract a number of physical quantities from $G_4$ alone. This we revise now. First of all, a $G_4$-flux is specified by an element $H^{2,2} ( \hat{Y}_4 )$. This space enjoys a decomposition
\[ H^{2,2} ( \hat{Y}_4 ) = H^{2,2}_{\mathrm{vert}} ( \hat{Y}_4 ) \oplus H^{2,2}_{\mathrm{hor}} ( \hat{Y}_4 ) \oplus H^{2,2}_{\mathrm{rem}} ( \hat{Y}_4 ) \label{H22decompo} \, \]
where elements of the three subspaces are mutually orthogonal with respect to the homological intersection pairing. This leads to a classification of $G_4$-fluxes, which we will explain in much detail in \cref{sec:SystGaugeBack}.

A $G_4$-flux is supposed to have `one leg along the fibre' \cite{oai:arXiv.org:hep-th/9908088}. Phrased explicitly, this means that the two transversality constraints 
\[ G_4 \cdot \pi^*\omega_4 = 0 \,, \qquad  G_4 \cdot [S_0] \cdot \pi^*\omega_2 =0 \label{transversality-gen1} \]
must be enforced for every element $\omega_4 \in H^4(\mathcal{B}_6)$, $\omega_2 \in H^2(\mathcal{B}_6)$.\footnote{Recall that $[S_0] \in H^{1,1}(\hat{Y}_4)$ is the class of the zero section of the fibration $\pi \colon \hat{Y}_4 \twoheadrightarrow \mathcal{B}_6$, whose existence we assumed as stated above. The generalisation of this condition in absence of a section has been described in \cite{Lin:2015qsa}.} Furthermore, in combination with cohomology classes, the operation $\cdot$ denotes the intersection product in the cohomology ring, here on $\hat{Y}_4$. A $G_4$-flux does not to break the non-Abelian gauge group $G_I$ if
\[ G_4 \cdot [ E_{i_I} ] \cdot \iota^\ast ( \omega_2 ) = 0 \label{gaugeinvariantflux} \, \]
for all $\omega_2 \in H^2(\mathcal{B}_6)$ and $i = 1, \ldots, \mathrm{rk} ( G_I )$.

As a matter of fact, a $G_4$-flux is actually required to be an element of $H^{2,2}_{{\mathbb Z}/2}(\hat{Y}_4)$ as a consequence of the quantisation condition \cite{oai:arXiv.org:hep-th/9609122}
\[ G_4 + \frac{1}{2} c_2(\hat{Y}_4) \in H^{2,2}_{{\mathbb Z}}(\hat{Y}_4) \, . \label{FW2} \]
Once such a $G_4 \in H^{2,2}_{ \mathbb{Z}/2} ( \hat{Y}_4 )$ is specified, the chiral index of massless matter state with weight $\beta^a(\mathbf{R})$ of representation $\mathbf{R}$ localised on the matter curve $C_\mathbf{R}$ can be obtained by evaluating \cite{Donagi:2008ca, oai:arXiv.org:0904.1218, Braun:2011zm, Marsano:2011hv, Krause:2011xj, oai:arXiv.org:1111.1232, oai:arXiv.org:1202.3138, oai:arXiv.org:1203.6662},
\[ \chi_{\mathbf{R}} = \Mint_{S^a_\mathbf{R}}{G_4} \, . \label{chiRsecgen}\]
If the gauge flux does not break the non-Abelian gauge algebra, this result is independent of the matter surface $S^a_\mathbf{R}$, $a=1, \ldots, \mathbf{R}$. Note that $\chi_\mathbf{R} \in \mathbb{Z}$ since $\frac{1}{2} \int_{S^a_\mathbf{R}}{c_2(\hat{Y}_4)} \in {\mathbb Z}$ \cite{oai:arXiv.org:1011.6388, oai:arXiv.org:1202.3138, oai:arXiv.org:1203.4542}.

\paragraph{\texorpdfstring{$\mathbf{G_4}$}{G4}-Fluxes from Deligne Cohomology}

Let us now come to discuss the `gauge field' associated to a $G_4$-flux. As it turns out, there exists a natural mathematical choice -- the so-called \emph{Deligne cohomology} $H^4_D(\hat{Y}_4, \mathbb Z(2))$ \cite{BeilinsonConjectures}.\footnote{Note that there is also an equivalent formulation in terms of the theory of Cheeger-Simons differential characters \cite{Cheeger-Simons}, as employed \eg in \cite{oai:arXiv.org:1203.6662} in F/M-theory.} It has been explored to describe the necessary gauge data both in \emph{M-theory} and \emph{F-theory} in \cite{Diaconescu:2003bm,Freed:2004yc,oai:arXiv.org:hep-th/0409158,oai:arXiv.org:1203.6662} and  \cite{Curio:1998bva,Donagi:1998vw,Donagi:2011jy, Clingher:2012rg,Anderson:2013rka,Bies:2014sra}, respectively. To motivate the use of Deligne cohomology, let us start by looking at the 3-form gauge background $C_3$. To this 3-form field $C_3$ we associate the following:
\begin{itemize}
 \item A quantised field strength $G_4 = dC_3 \in H^{2,2}_{\mathbb{Z}} ( \hat{Y}_4 )$.
 \item The holonomies of $C_3$ around non-trivial 3-cycles. These correspond to flat gauge backgrounds or Wilson `lines'. This is known as the intermediate Jacobian $J^2 
      ( \hat{Y}_4 )$.
\end{itemize}
Hence, the full gauge data should fit into the short exact sequence
\[ 0 \to J^2 ( \hat{Y}_4 ) \hookrightarrow \text{full gauge data} \xtwoheadrightarrow{\hat{c}_2} H_{\mathbb{Z}}^{2,2} ( \hat{Y}_4 ) \to 0 \,. \]
It can be inferred from the Deligne-Beilinson complex \cite{BeilinsonConjectures} that indeed such an object exists -- it is Deligne cohomology which satisfies
\[ 0 \to J^2 ( \hat{Y}_4 ) \hookrightarrow H_D^4 ( \hat{Y}_4, \mathbb{Z} ( 2 ) ) \xtwoheadrightarrow{\hat{c}_2} H_{\mathbb{Z}}^{2,2} ( \hat{Y}_4 ) \to 0 \,. \label{SESDeligne} \]
Note that this exact sequence generalises the situation found for line bundles in \cref{equ:DeligneCohomologyForLineBundles}. In particular, we can interpret the objects in this sequence analogously: Given an element $H_D^4 ( \hat{Y}_4, \mathbb{Z} ( 2 ) )$ we may thus employ the surjective map $\hat{c}_2$ to map it to an integral $( 2,2 )$-form. This cohomology class is identified with the 4-form flux $G_4 \in H^{2,2}_{\mathbb{Z}} ( \hat{Y}_4 )$ \cite{oai:arXiv.org:hep-th/9605053,oai:arXiv.org:hep-th/9606122,oai:arXiv.org:hep-th/9908088}. Note however that in general $\hat{c}_2$ is not injective -- its kernel is the intermediate Jacobian $J^2 ( \hat{Y}_4 )$. This is the space of flat 3-form connections which, by definition, have vanishing gauge flux even though such configurations are non-trivial gauge backgrounds. In particular,  $H_D^4 ( \hat{Y}_4, \mathbb{Z} ( 2 ) )$ contains strictly more information than $H_{\mathbb{Z}}^{2,2} ( \hat{Y}_4 )$. More details can be found in \cite{Bies:2014sra}.

\section{Zero Mode Counting in Type IIB String Theory} \label{sec:RevisionOnSheavesAndSheafCohomology}

\subsection{A Physics Motivation for Sheaves and Sheaf Cohomologies}

\paragraph{The Ehresmann connection -- a global Gauge Field}
Before we dive into zero modes in type IIB string theory, let us start simpler and think about gauge theories on a manifold $M_d$. Let us assume that the symmetry operations be given by the elements of a compact, connected Lie group $G$. In physics, often only the \emph{local} situation in an affine patch $U$ with  coordinates $\left\{ x^\mu \right\}_{1 \leq \mu \leq d}$ is studied. Among others this makes it conceivable to employ the gauge fields $\mathbf{A}_U = A_\mu^a T^a dx^\mu$ where $T^a$ are the generators of the Lie algebra $\mathfrak{g}$ of the group $G$. What however is the global picture?

As reviewed in \cite{Nakahara:2003nw}, the global analogue of these gauge fields is an \emph{Ehresmann connection} on a principal $G$ bundle over $M_d$. First of all, the construction of a principal $G$-bundle $\mathcal{P} ( M_d, G )$ is very similar to the torus fibrations, expect that we replace the torus by a Lie group $G$. Hence, we attach to every point $p \in M_d$ this group $G$. As the latter admits a topology, the result of this operation is a new topological space -- the so-called \emph{total space} $\mathcal{P}$ of this principal bundle. The projection $\pi \colon \mathcal{P} \twoheadrightarrow \mathcal{B}_6$ identifies the fibre over $p \in M_d$ as $\pi^{-1} ( p )$. For every $u \in \mathcal{P}$ we can now separate the tangent space $T_u \mathcal{P}$ into a direct sum of a \emph{vertical subspace} $V_u \mathcal{P}$ -- which is parallel to the fibre $\pi^{-1} ( u )$ -- and its complement in $T_u \mathcal{P}$, which is known as the \emph{horizontal subspace} $H_u \mathcal{P}$:
\[ T_u \mathcal{P} = V_u \mathcal{P} \oplus H_u \mathcal{P} \, . \]
A connection on $\mathcal{P} ( M_d, G )$ is such a separation of $T_u \mathcal{P}$ for every $u \in \mathcal{P}$. It can be specified by a \emph{connection one-form} $\omega \in \mathfrak{g} \otimes T^\vee \mathcal{P}$, which projects $T_u \mathcal{P}$ onto $V_u \mathcal{P} \simeq \mathfrak{g}$ via
\[ H_u \mathcal{P} := \left\{ \left. X \in T_u \mathcal{P} \, \right| \, \omega ( X ) = 0 \right\} \, . \]
Such a connection one-form $\omega$ is known as an \emph{Ehresmann connection}. Over a local patch $U$, $\omega$ can be expressed as $\mathbf{A}_U = A_\mu^a T^a dx^\mu$. By changing to another open patch $V$, the transition functions $t_{UV}$ of $\mathcal{P} ( M_d, G )$ must be taken into account. Then for every $q \in U \cap V$, the local pictures satisfy
\[ \mathbf{A}_V \left( q \right) = t_{UV}^{-1} \left( q \right) A_U \left( q \right) t_{UV} \left( q \right) + t_{UV}^{-1} \left( q \right) dt_{UV} \left( q \right) \]
which is just a gauge transformation in the adjoint representation of $G$. In this sense, an Ehresmann connection is the global analogue of the local gauge fields $\mathbf{A}_\mu$. Many more details can be found \eg in \cite{Nakahara:2003nw}.

\paragraph{Classical Fields as Global Sections Of Vector Bundles}
From this approach we can now also understand fields in classical field theory better. Namely giving a representation $\mathbf{R}$ means to specify a group homomorphism
\[ \rho_{\mathbf{R}} \colon G \to \mathrm{Aut} \left( V \right) \]
where $\text{Aut} ( V )$ denotes the automorphisms on a vector space $V$, \ie the linear maps $V \to V$. Given a principal $G$-bundle $\mathcal{P} ( M_d, G )$, let us now replace the fibre $\pi^{-1} ( p ) \simeq \mathfrak{g}$ by the vector space $V$. In addition, we employ $\rho_{\mathbf{R}}$ to turn the $G$-valued transition functions $t_UV$ of $\mathcal{P} ( M_d, G )$ into $\text{Aut} ( V )$. Thereby, we obtain a vector bundle -- the so-called \emph{associated vector bundle} $V( M_d, \mathcal{P}, \rho_{\mathbf{R}} )$ of $\mathcal{P}$ under $\rho_{\mathbf{R}}$. Similar to the local gauge fields $\mathbf{A}_U$, all sections of this vector bundle now transform in the representation $\mathbf{R}$ of $G$. In particular, the global sections provide us with such globally defined fields. We are therefore led to conclude that in classical field theory, the fields are actually sections of associated vector bundles. But the global sections of a vector bundle $V$ form its sheaf cohomology group $H^0 ( M_d, V )$.

\paragraph{From Vertex Operators To Hodge Numbers}
So far, we have looked at classical field theories only. To arrive at quantum field theory, they must undergo a quantisation process. In string theory for example, we consider a conformal field theory on the world sheet $\Sigma$, where so-called \emph{vertex operators} can be used to create or destroy particles \cite{green1988superstring1, green1988superstring2, polchinski1998string, polchinski2001string}. This is a special instance of the creation and annihilation operators employed throughout quantum field theory.

In 1985 Philip Candelas, Gary Horowitz, Andrew Strominger and Edward Witten formulated a compactification of heterotic $E_8 \times E_8$ string theory on a Calabi--Yau 3-fold $M_6$ \cite{CANDELAS198546}. In this geometry they broke $E_8$ to $E_6 \times SU(3)$, with the intention to construct a GUT-model from the group $E_6$. Under this breaking, the adjoint representation of $E_8$ decomposes as
\[ \mathbf{248} \to \left( \mathbf{27}, \mathbf{3} \right) \oplus \left( \overline{\mathbf{27}}, \mathbf{\overline{3}} \right) \oplus \left( \mathbf{78}, \mathbf{1} \right) \oplus \left( \mathbf{1}, \mathbf{8} \right) \, . \]
The representations $( \mathbf{27}, \mathbf{3} )$ and $( \overline{\mathbf{27}}, \mathbf{\overline{3}} )$ give rise to 4-dimensional chiral fermions. Consequently, their number is of prime interest. As explained in \cite{green1988superstring2} these massless chiral multiplets are determined by the Dolbeault cohomology groups as
\[ n_{(\mathbf{27},\mathbf{3})} = h^{2,1} \left( M_6 \right) \, , \qquad n_{(\mathbf{\overline{27}},\mathbf{\overline{3}})} = h^{1,1} \left( M_6 \right) \, . \]
This triggered interest in constructing Calabi--Yau 3-folds and lead in its early days for example to the works in \cite{GREENE1986667, doi:10.1142/9789812798411_0026}. For our applications it is important to note that the Hodge numbers are given by sheaf cohomologies also! Namely let $\Omega^p$ denote the sheaf of holomorphic $p$-forms on $M_6$, then it holds \cite{huybrechts2005complex}
\[ H^{p,q} \left( M_6 \right) \cong H^q \left( M_6, \Omega^p \right) \, . \]

This early result already signals the importance of sheaf cohomologies in string compactifications. As reviewed in \cite{Aspinwall:2004jr} for example, in topological string theory so-called B-branes on a Calabi--Yau manifold $M_6$ are found to correspond to elements in the derive, bounded category of coherent sheaves $D^b ( \mathfrak{Coh} ( B_6 ) )$. Morphisms in this category now correspond to open strings between the associated B-branes. These can in turn be understood as so-called extension group of coherent sheaves. See \cite{Katz:2002gh} for additional background. Let us also mention that the category $D^b ( \mathfrak{Coh} ( B_6 ) )$ is one of the major ingredients in \emph{homological mirror symmetry}, which proposes that this category is equivalent to the so-called \emph{derived Fukaya category}, see for example \cite{Aspinwall:2004jr} for a review.

\paragraph{Outlook}
These points emphasise that coherent sheaves and their cohomology groups are central ingredients in various areas of physics. Not only in classical field theory, but 
also in string theory. In fact, the central result of \cref{chapter:MasslessSpectraAndSheafCohomology} will be to relate zero modes in F-theory to the cohomology classes of certain coherent sheaves. Therefore, let us use this opportunity to gain some familiarity with the topic. To this end, we focus on zero mode counting in type IIB to exemplify results from \cite{Katz:2002gh, Donagi:2011jy, Donagi:2011dv, Donagi:2010pd}.

Before we discuss this in detail, let us however start simpler. Whilst coherent sheaves become particularly important in the remainder of this thesis and are explained in \cref{subsec:CoherentSheavesOnVarieties}, we will not introduce them just yet. Rather for the purposes of this section it suffices to review sheaves as well as their sheaf cohomologies. Our discussion follows \cite{FreitagRiemann}. In this generality, sheaves are fairly abstract objects, and so the employed language can be hard to swallow on a first encounter. This the reason why we provide explicit examples and present a survey of line bundles on compact, connected Riemann surfaces in \cref{subsec:LineBundlesOnRiemannSurfaces}. This revision follows a similar discussion in \cite{BiesMaster} and prepares us for the discussions in \cref{chapter:MasslessSpectraAndSheafCohomology}, where we will find that the sheaf cohomologies of line bundles on matter curves $C_{\mathbf{R}}$ count the number of chiral and anti-chiral zero modes in F-theory compactifications.

\subsection{Sheaves and Sheaf Cohomology} \label{subsec:SheafAndSheafCohomology}

\paragraph{Presheaves}

Let $( X, \tau )$ be a topological space. A presheaf $\mathcal{F}$ of Abelian groups on $( X, \tau )$ consists of Abelian groups $\mathcal{F} ( U )$ for all open $U \subseteq X$ and group homomorphisms
\[ \left( \mathrm{res}_{\mathcal{F}} \right)^U_V \colon \mathcal{F} \left( U \right) \to \mathcal{F} \left( V \right) \]
for all open $V \subseteq U$, such that the following conditions hold true:
\begin{enumerate}
 \item[(P1)] $\mathcal{F} \left( \emptyset \right) = 0$ -- the trivial group.
 \item[(P2)] $\left( \mathrm{res}_{\mathcal{F}} \right)^U_U = \mathrm{id}_{\mathcal{F} \left( U \right)}$.
 \item[(P3)] For $W \subseteq V \subseteq U$ we have $\left( \mathrm{res}_{\mathcal{F}} \right)^U_W = \left( \mathrm{res}_{\mathcal{F}} \right)^V_W \circ \left( \mathrm{res}_{\mathcal{F}} \right)^U_V$.
\end{enumerate}
The elements of $\mathcal{F} ( U )$ are termed (local) sections of $\mathcal{F}$ over $U$, the maps $\left( \mathrm{res}_{\mathcal{F}} \right)^U_V$ its \emph{restriction maps}. If no confusion is possible, we denote the restrictions as $( \mathrm{res}_{\mathcal{F}} )^U_V ( s ) = s |_V$ for $s \in \mathcal{F} ( U )$.

A simple example of a presheaf is the following: For every open $U \subseteq X$ consider the set $\mathcal{F} ( U ) := \{ f \colon U \to \mathbb{R} \; , \; f \text{ is constant} \}$. $\mathcal{F} ( U )$ is an Abelian group upon the composition
\begin{align}
\begin{split}
+ \colon \mathcal{F} \left( U \right) \times \mathcal{F} \left( U \right) \to \mathcal{F} \left( U \right) \; , \; \left( f, g \right) \mapsto & f + g \, , \\
& f + g \colon U \to \mathbb{R} \; , \; x \mapsto f ( x ) + g( x ) \, . \label{equ:PointwiseAdditionOfFunctions}
\end{split}
\end{align}
Consequently, $\mathcal{F}( \emptyset )$ is the trivial group and so (P1) is satisfied. As restriction maps we pick the ordinary restriction of functions, so that (P2) and (P3) are fulfilled.

Given presheaves $\mathcal{F}$ and $\mathcal{G}$ of Abelian groups on a topological space $(X, \tau)$, a presheaf homomorphism $f \colon \mathcal{F} \to \mathcal{G}$ is a collection $\{ f_U \colon \mathcal{F} ( U ) \to \mathcal{G} ( U ) \; , \; U \subseteq X \text{ open } \}$ of group homomorphisms, such that for any two $V \subseteq U \subseteq X$ open, the following diagram commutes:
\[ \includegraphics[valign = c]{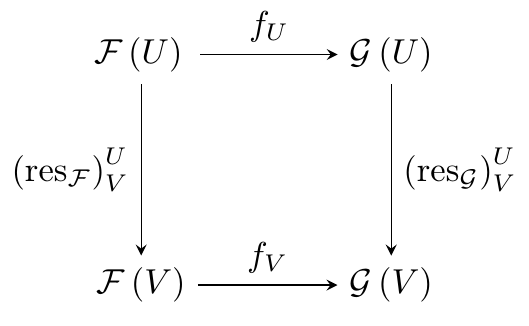} \]

For a presheaf homomorphism $f \colon \mathcal{F} \to \mathcal{G}$, the kernel $\mathrm{ker} ( f )$ is a presheaf given by
\begin{itemize}
 \item $\mathrm{ker} ( f ) ( U ) = \mathrm{ker} ( f_U ) \subseteq \mathcal{F} ( U )$ for all $U \subseteq X$ open,
 \item $( \mathrm{res}_{\mathrm{ker} ( f )} )^U_V \colon \mathrm{ker} ( f_U ) \to \mathrm{ker} ( f_V ) \; , \; x \mapsto ( \mathrm{res}_{\mathcal{F}} )^{U}_{V} ( x )$.
\end{itemize}
Similarly, one can define image and cokernel of a presheaf homomorphism. Both happen to be presheaves.

The local behaviour of presheaves is described in terms of germs and stalks. To define these pick a point $p \in X$ and consider pairs $( U, s )_p$ where $U$ is an open neighbourhood of $p$ and $s \in \mathcal{F} ( U )$. For such pairs we define the equivalence relation
\[ \left( U, s \right)_p \sim \left( V, t \right)_p \; \Leftrightarrow \; \exists \text{ an open set $W$ such that } p \in W \subseteq U \cap V \text{ and } \left. s \right|_W = \left. t \right|_W \, . \]
We term the equivalence class $[ U, s ]_p := \{ ( V, t )_p \; , \; ( V, t )_p \sim ( U, s )_p \}$ \emph{the germ} of the section $s$ at the point $p$. The union of all such germs
\[ \mathcal{F}_a := \left\{ \left. \left[ U, s \right]_p \; \right| \; p \in U \subseteq X \text{ open and } s \in \mathcal{F} \left( U \right) \right\} \]
is the \emph{stalk} of the presheaf $\mathcal{F}$ at the point $p$. $\mathcal{F}_p$ is an Abelian group.

\paragraph{Sheaves}

A sheaf $\mathcal{F}$ of Abelian groups on a topological space $(X, \tau)$ is a presheaf of Abelian groups $\mathcal{F}$ which satisfies the following two conditions:
\begin{itemize}
 \item[(S1)] Let $\mathcal{U} = \{ U_i \}_{i \in I}$ be an open cover of $U \subseteq X$ open and $s,t \in \mathcal{F} ( U )$ such that for all $i \in I$ it holds
      $s |_{U_i} = t |_{U_i}$. Then it holds $s = t$.
 \item[(S2)] Be $\mathcal{U} = \left\{ U_i \right\}_{i \in I}$ an open cover of $U \subseteq X$ open and $\left\{s_i \in \mathcal{F} ( U_i ) \right\}_{i \in I}$ a 
      family with the property $s_i |_{U_i \cap U_j} = s_j |_{U_i \cap U_j}$ for all $i,j \in I$. Then there exists $s \in \mathcal{F} ( U )$ with $s |_{U_i} = s_i$ for all $i \in I$.
\end{itemize}

A standard example of a sheaf is as follows: Take $(X, \tau)$ to be a topological manifold. For $U \subseteq X$ open consider the set
$\mathcal{O}_X ( U ) = \{ \left. f \colon U \to \mathbb{R} \; \right| \; \text{$f$ is continuous } \}$. This set $\mathcal{O}_X ( U )$ is an Abelian group (compare the discussion around \cref{equ:PointwiseAdditionOfFunctions}). The restriction maps $\mathrm{res}^U_V$ are taken to be the ordinary restriction of functions. Then $\mathcal{O}_X$ is readily found to be a presheaf. In addition, (S1) and (S2) are satisfied, since continuity is a local property. Consequently, $\mathcal{O}_X$ is a sheaf, the sheaf of continuous functions $\mathcal{O}_X$ on the topological manifold $(X, \tau )$.

In the above writing, we can replace the word \emph{Abelian group} by \emph{ring}, \emph{module}, \emph{algebra} and alike. In doing so one defines the concept of (pre)sheaves of \emph{rings}, \emph{modules}, \emph{algebras}. For example, the sheaf of continuous functions $\mathcal{O}_X$ on a topological manifold $(X, \tau)$ is a sheaf of rings -- the ring structure on $\mathcal{O}_X ( U )$ is mediated by the multiplication
\begin{align}
\begin{split}
\cdot \colon \mathcal{O}_X \left( U \right) \times \mathcal{O}_X \left( U \right) \to \mathcal{O}_X \left( U \right) \; , \; \left( f, g \right) \mapsto & f \cdot g \, , \\
& f \cdot g \colon U \to \mathbb{R} \; , \; x \mapsto f \left( x \right) \cdot g \left( x \right) \, .
\end{split}
\end{align}
and the ordinary restriction of functions respects this multiplication.

More examples of sheaves include the sheaf of smooth functions on smooth manifolds. On a complex manifold the sheaf of holomorphic functions can be considered. As continuous/ smooth/ holomorphic functions are characteristic to the structure of topological/ smooth/ complex manifolds these sheaves are referred to as the \emph{structure sheaf of the manifold} in question.

All of the above example sheaves are sheaves of rings. This observation leads to the concept of a \emph{ringed space}. A ringed space is a pair $( X, \mathcal{O}_X )$ consisting of a topological space $X$ and a sheaf of rings $\mathcal{O}_X$ on $X$. On a ringed space it is possible to consider sheaves of $\mathcal{O}_X$-modules. Such a sheaf $\mathcal{F}$ on $X$ assigns to every open $U \subseteq X$ an $\mathcal{O}_X ( U )$-module $\mathcal{F} ( U )$. In addition, the restriction maps of $\mathcal{F}$ respect this module structure. The coherent sheaves, whose sheaf cohomologies will turn out to be one of the most important objects of study in this thesis, are special instances of such sheaves of $\mathcal{O}_X$-modules. We will study this point further in \cref{subsec:CoherentSheavesOnVarieties}.

It is crucial to note that not every presheaf is a sheaf. For a simple example consider a topological space $(X, \tau )$ such that $X = U \cup V$ but $U \cap V = \emptyset$, \ie $X$ is not connected. As discussed previously, the real-valued \emph{constant} functions on $X$ form a presheaf $\mathcal{F}$. However they do not form a sheaf, since the sheaf property (S2) is violated. To see this consider
\[ \mathcal{F} \left( U \right) \ni s_0 \colon U \to \mathbb{R} \; , \; x \mapsto 0 \, , \qquad \mathcal{F} \left( V \right) \ni s_1 \colon V \to \mathbb{R} \; , \; x \mapsto 1 \, .\]
Since $U \cap V = \emptyset$ we have $\left. s_0 \right|_{U \cap V} = \left. s_1 \right|_{U_V}$. If property (S2) were satisfied, it would imply the existence of
a constant function $s \colon X \to \mathbb{R}$ such that $\left. s \right|_{U_0} = s_0$ and $\left. s \right|_{U_1} = s_1$. Since $0 \neq 1$ this is impossible.

However, there is a cure to turn the presheaf of constant functions on $X = U \cup V$ with $U \cap V = \emptyset$ into a sheaf: merely consider the \emph{locally} constant functions on $X$. This is an example of the so-called \emph{generated sheaf} which we will discuss momentarily.

Before we do so, let us give yet another example of a presheaf which fails to be a sheaf. To this end, we first note that for sheaves $\mathcal{F}$ and $\mathcal{G}$ on a topological space $(X, \tau)$, a sheaf homomorphism $f \colon \mathcal{F} \to \mathcal{G}$ is simply a homomorphism of the underlying presheaves. Given a sheaf homomorphism $f \colon \mathcal{F} \to \mathcal{G}$, we can thus study its kernel, image and cokernel presheaves. In particular, we can wonder if they happen to be sheaves. Whilst the answer is affirmative for the kernel presheaf, both image and cokernel presheaf fail to be sheaves in general. Here is an example: Consider the complex manifold $X = \mathbb{C}$ (with standard topology) and the sheaf homomorphism
\[ \varphi \colon \mathcal{O}_X \to \mathcal{O}_X^\ast \; , \; f \mapsto \exp \left( 2 \pi i f \right) \label{equ:SheafHomoWithImagePresheafNotASheaf} \]
of the sheaf of holomorphic functions $\mathcal{O}_X$ to the sheaf of non-vanishing, holomorphic functions $\mathcal{O}_X^\ast$. We will argue that the image presheaf $\mathrm{im} ( \varphi )$ violates the sheaf property (S2). To this end, pick the open subset $U = \{ z \in \mathbb{C} \; , \; |z | > 1 \}$ and cover it by $U_1 = \{ z \in U \; , \; \mathrm{Re} ( z ) < \frac{1}{2} \}$ and $U_2 = \{ z \in U \; , \; \mathrm{Re} ( z ) > - \frac{1}{2} \}$. Next consider $f_1 \in \mathcal{O}^\ast ( U_1 )$ and $f_2 \in \mathcal{O}^\ast (U_2)$ given by
\[ f_1 \colon U_1 \to \mathbb{C}^\ast \; , \; z \mapsto \frac{1}{z}, \qquad f_2 \colon U_2 \to \mathbb{C}^\ast \; , \; z \mapsto \frac{1}{z}. \]
Since $U_1$ and $U_2$ are both simply-connected, $f_1$ and $f_2$ admit a holomorphic logarithm. Hence, they are in the image of \cref{equ:SheafHomoWithImagePresheafNotASheaf}, \ie $f_1 \in \mathrm{im} ( \varphi ) ( U_1 )$ and $f_2 \in \mathrm{im} ( \varphi ) ( U_2 )$.

$f_1$ and $f_2$ apparently agree on $U_1 \cap U_2$. If $\mathrm{im} ( \varphi )$ were to satisfy (S2), then this would imply the existence of
$f \in \mathrm{im} ( \varphi )( U )$ with $\left. f \right|_{U_1} = f_1$ and $\left. f \right|_{U_1} = f_2$ and, by definition of \cref{equ:SheafHomoWithImagePresheafNotASheaf}, this function $f$ would admit a holomorphic logarithm. This however is impossible since $U$ is \emph{not} simply connected.

This observation shows that, in order to define image and cokernel of a sheaf homomorphism $\mathcal{F} \to \mathcal{G}$, we need a means to turn a given presheaf into a `minimal' sheaf. We will elaborate on the meaning of `minimal' momentarily. For now suffice to point out that this demand leads to the concept of the so-called \emph{generated sheaf}: Given a presheaf $\mathcal{F}$ on a topological space $( X, \tau )$, we define the local sections of the generated sheaf $\hat{\mathcal{F}}$ over $U \subseteq X$ as
\[ \hat{\mathcal{F}} \left( U \right) = \left\{ s \colon U \mapsto \mathop{\dot{\bigcup}}_{p \in U}{\mathcal{F}_p} \; , \; \text{ functions with special property} \right\} \, . \label{equ:GeneratedSheaf} \]
The `special property' means to satisfy the following two conditions:
\begin{enumerate}
 \item $p \in U$ is mapped to $s ( p ) \in \mathcal{F}_p$, \ie is mapped to an element in the stalk $\mathcal{F}_p$.
 \item Be $p \in U$, then there exists an open neighbourhood $p \in V \subseteq U$ and $t \in \mathcal{F} ( V )$ such that for all $q \in V$ it holds that $s ( q ) = t_q := ( V, t )_q$.
\end{enumerate}
Let us now elaborate on the meaning of `minimal'. First, there exists a homomorphism of presheaves $\Theta \colon \mathcal{F} \to \hat{\mathcal{F}}$. But this is not just any kind of presheaf homomorphism. It resembles the close relationship between $\mathcal{F}$ and $\hat{\mathcal{F}}$. With a view towards category theory, this relationship is most easily stated in a so-called \emph{universal property}: Let $\mathcal{G}$ be a sheaf on $X$ and $\varphi \colon \mathcal{F} \to \mathcal{G}$ a homomorphism of presheaves, then there exists a \emph{unique} sheaf homomorphism $\psi \colon \hat{\mathcal{F}} \to \mathcal{G}$ such that the following diagram commutes:
\[ \includegraphics[valign = c]{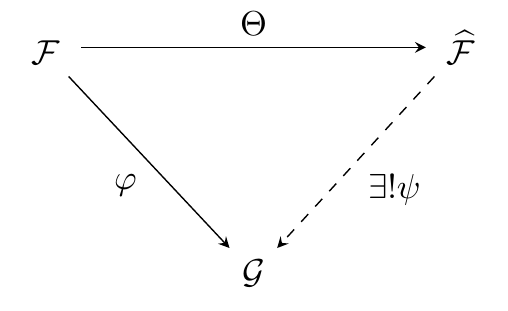} \]
In fact it follows from this universal property alone that the generated sheaf $\hat{\mathcal{F}}$ is unique up to sheaf isomorphism. In this sense  \cref{equ:GeneratedSheaf} defines a special instance of the generated sheaf. Irrespective of the chosen representant, we have the following two results:
\begin{itemize}
 \item For any $p \in X$ it holds that $\hat{\mathcal{F}}_p \cong \mathcal{F}_p$.
 \item If $\mathcal{F}$ is a sheaf, then $\mathcal{F} \cong \hat{\mathcal{F}}$.
\end{itemize}

Finally, we state that for a sheaf homomorphism $f \colon \mathcal{F} \to \mathcal{G}$ of sheaves on a topological space $(X, \tau )$, the generated sheaves $\hat{\mathrm{im} ( f )}$ and $\hat{\mathrm{coker} ( f )}$ form the image and cokernel sheaves of $f$. That point finally settled, we can define (co)homologies of (co)complexes of sheaves: Consider a family of sheaves $\{ \mathcal{F}_i \}_{i \in \mathbb{Z}}$ and sheaf homomorphism $\{ f_{i+1} \colon \mathcal{F}_{i+1} \to \mathcal{F}_i \}_{i \in \mathbb{Z}}$:
\[ \dots \to \mathcal{F}_{i+1} \xrightarrow{f_{i+1}} \mathcal{F}_i \xrightarrow{f_i} \mathcal{F}_{i-1} \to \dots \]
Provided that $\mathrm{im} ( f_{i-1} ) \subseteq \mathrm{ker} ( f_i )$ for all $i \in \mathbb{Z}$, we term this family a complex $\mathcal{F}_\bullet$ of sheaves. A complex of sheaves is exact at $i \in \mathbb{Z}$ precisely if $\mathrm{im} ( f_{i+1} ) = \mathrm{ker} ( f_i )$. We can measure by how much a complex differs from being exact by considering the quotient sheaves
\[ H_i \left( \mathcal{F}_\bullet \right) := \mathrm{ker} \left( f_i \right) / \mathrm{im} \left( f_{i+1} \right). \]

\paragraph{Sheaf Cohomology}

Given a sheaf $\mathcal{F}$ of Abelian groups on a topological space $(X, \tau)$, we construct the so-called \emph{Godement sheaf} $\mathcal{F}^{(0)}$. Its local sections over $U \subseteq X$ open are given by 
\[ \mathcal{F}^{(0)} ( U ) = \prod_{p \in U}{\mathcal{F}_p} \, . \]
For open $V \subseteq U$, the restriction map
\[ \left( \mathrm{res}_{\mathcal{F}^{(0)}} \right)^U_V \colon \prod_{p \in U}{\mathcal{F}_p} \to \prod_{p \in V}{\mathcal{F}_p} \; , \; \left( s_p \right)_{p \in U} \mapsto \left( s_p \right)_{p \in V} \]
simply `forgets' all $p \in U \backslash V$. $\mathcal{F}^{(0)}$ indeed forms a sheaf and the family
\[ \left\{ \iota_U \colon \mathcal{F} \left( U \right) \to \mathcal{F}^{(0)} \left( U \right) \; , \; s \mapsto \left( s_p \right)_{p \in U} \right\}_{U \subseteq X \mathrm{ open}} \]
forms a sheaf homomorphism $\iota \colon \mathcal{F} \to \mathcal{F}^{(0)}$. The sheaf property (S1) of $\mathcal{F}$ implies $\mathrm{ker} ( \iota ) = 0$. Hence, the sequence $0 \to \mathcal{F} \xhookrightarrow{\iota} \mathcal{F}^{(0)}$ is exact at $\mathcal{F}$. As a next step we consider the cokernel sheaf $\hat{\mathrm{coker} ( \iota )}$ and embed this sheaf into its associated Godement sheaf $\mathcal{F}^{(1)}$. In following this logic we extend $0 \to \mathcal{F} \xhookrightarrow{\iota} \mathcal{F}^{(0)}$ to obtain the so-called \emph{Godement resolution} of $\mathcal{F}$:
\[ 0 \to \mathcal{F} \xhookrightarrow{\iota} \mathcal{F}^{(0)} \to \mathcal{F}^{(1)} \to \mathcal{F}^{(2)} \to \dots \label{equ:GodementResolutionOfSheafF} \, . \]
To define the sheaf cohomologies of $\mathcal{F}$, we `subtract' $\mathcal{F}$ from this complex and consider the induced sequence of global sections, \ie look at
\[ 0 \to \mathcal{F}^{(0)} \left( X \right) \to \mathcal{F}^{(1)} \left( X \right) \to \mathcal{F}^{(2)} \left( X \right) \to \dots \label{equ:ResolutionForSheafCohomologyOfSheafF} \, . \]
The sheaf cohomologies of $\mathcal{F}$ are the cohomologies of this sequence -- the cohomology at position $i$ is denoted by $H^i ( X, \mathcal{F} )$. As an immediate consequence we have $H^0 ( X, \mathcal{F} ) = \mathcal{F} ( X )$.

\subsection{Line Bundles on compact, connected Riemann Surfaces} \label{subsec:LineBundlesOnRiemannSurfaces}

Given these abstract words, some explicit examples are in order. As a very common task in the remainder of this thesis will concern line bundles on compact and connected, Riemann surfaces (\ie a compact, connected, smooth, complex manifolds of complex dimension one), let us exemplify the sheaf terminology on these line bundles. Consequently, this section summarises classical results about holomorphic line bundles on compact, connected Riemann surfaces. This is a fairly well-known topic and many good references exist. Historical references include \cite{atiyah1957vector, narasimhan1976vector}. Of the many available textbooks we would like to highlight \cite{forster1981lectures, gunning1967lectures, FreitagRiemann}. Background on Riemann surfaces is given in \cite{freitag2009funktionentheorie}.

\paragraph{Classification by Genus}

\begin{figure}[tbp]
\centering
\subfloat[$g = 0$]{\label{figure-g0} \includegraphics[width = 0.15\textwidth]{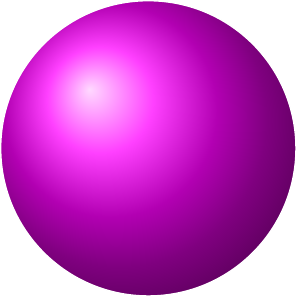} }
\hspace{2em}
\subfloat[$g = 1$]{\label{figure-g1} \includegraphics[width = 0.23\textwidth]{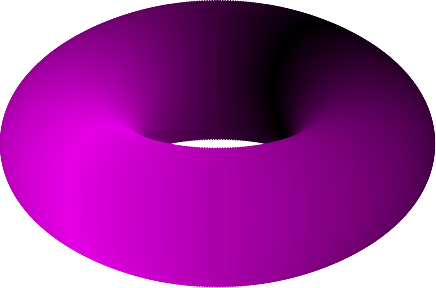} }
\hspace{2em}
\subfloat[$g = 2$]{\label{figure-g2} \includegraphics[width = 0.4\textwidth]{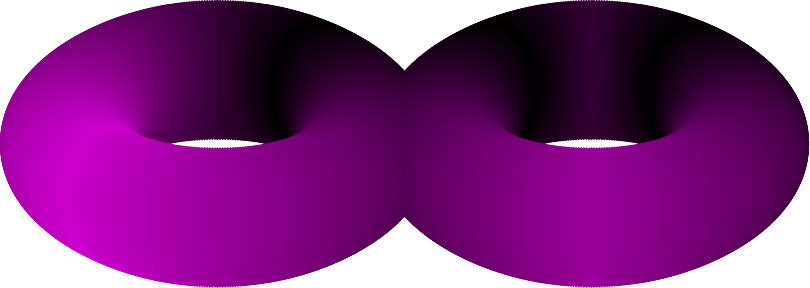} }
\caption{The Riemann surface $M_g$ for $g = 0, 1, 2$.}
\label{figure-g012}
\end{figure}

A topological classification of compact, connected Riemann surfaces exists. Every compact, connected Riemann surface can be transformed continuously into a `doughnut with $g$ holes' (\cf \cref{figure-g012}). A proof of this topological classification is given in \cite{freitag2009funktionentheorie}. The simplest examples of compact connected Riemann surfaces include the sphere $S^2$ and a torus surface $T^2$. They correspond to $g = 0$ and $g = 1$, respectively. The integer $g$ is termed \emph{the genus} of the surface in question. A compact, connected Riemann surface of genus $g$ is denoted by $M_g$ throughout this section.

\paragraph{Divisors}

\begin{figure}[tbp]
\centering
\subfloat[The trivial divisor $D_0$.]{\label{figure-spin0OnM1} \includegraphics{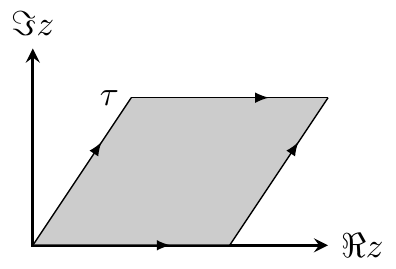} }
\hspace{1.8em}
\subfloat[${D_1 = (-1) [ 0 ] + (+1) [ 1/2 ] }$.]{\label{figure-spin1OnM1} \includegraphics{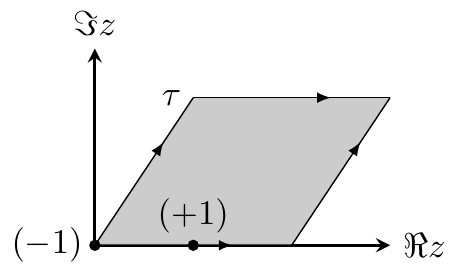} }

\subfloat[${D_2 = (-1) [ 0 ] + (+1) [ \tau/2 ] }$.]{\label{figure-spin2OnM1} \includegraphics{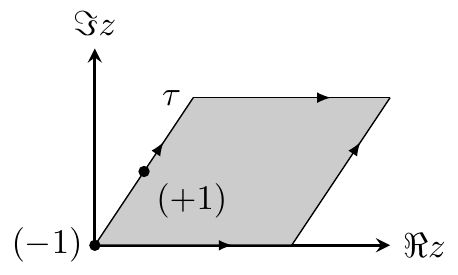} }
\hspace{1.8em}
\subfloat[${D_3 = (-1) [ 0 ] + (+1) [ ( 1 + \tau ) /2 ] }$.]{\label{figure-spin3OnM1} \includegraphics{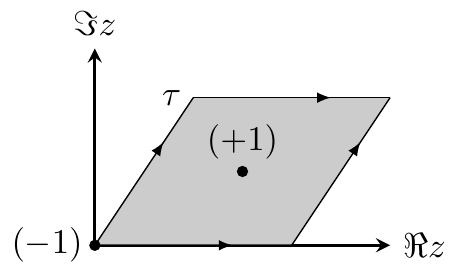} }
\caption[The four spin divisors on the torus surface $\mathbb{C}_{\mathbf{1}, \boldsymbol{\tau}}$.]{As discussed in \cref{sec:F-theory} we can visualise the torus surface $\mathbb{C}_{\mathbf{1}, \boldsymbol{\tau}}$ by a parallelogram in $\mathbb{C}$. The above shows the four spin divisors on $\mathbb{C}_{\mathbf{1}, \boldsymbol{\tau}}$.}
\label{figure-SpinDivisorsOnM1}
\end{figure}

A divisor $D \in \mathrm{Div} ( M_g )$ is a formal linear sum
\[ D := \sum_{p \in M_g}{n ( p ) \cdot p} \; , \qquad n ( p ) \in \mathbb{Z} \, . \label{equ:divisor} \]
Divisors can be added pointwise. Thereby, the set $\mathrm{Div} ( M_g )$ of all divisors on $M_g$ becomes an Abelian group. We give examples of divisors on $M_1$ in 
\cref{figure-SpinDivisorsOnM1}.

We can associate to a meromorphic function $f \colon M_g \to \mathbb{C}$ a divisor. In this case the points $p$ are the zeros and poles of $f$ and the integers $n ( p )$ represent the order of vanishing and the order of pole respectively. This principal divisor of the meromorphic function $f$ is denoted by $\mathrm{div} ( f )$. Of course not every divisor is the divisor of a meromorphic function. For example, we can define for any divisor of the form \cref{equ:divisor} the degree
\[ \mathrm{deg} ( D ) = \sum_{p \in M_g}^{k}{n( p )} \, . \]
Every principal divisor has vanishing degree. Put differently -- a meromorphic function on $M_g$ has as many zeros as poles (counted with multiplicities). However, the converse is not true -- not every divisor of vanishing degree is a principal divisor! See \eg \cite{BiesMaster} and references therein for more details.

\paragraph{Divisor Classes}

We define an equivalence relation among $\mathrm{Div} ( M_g )$ by
\[ D_1 \sim D_2 \; \Leftrightarrow \; D_1 - D_2 \text{ is a principal divisor} \, . \]
This equivalence relation is known as \emph{linear equivalence} and defines the divisor class group $\mathrm{Cl} ( M_g ) = \mathrm{Div} ( M_g ) / \sim$. Since the degree of principal divisors vanishes, we can use the assignment $\mathrm{deg} ( [ D ] ) = \mathrm{deg} ( D )$ to establish a well-defined notion of degree of divisor classes.

\paragraph{From Divisor Classes to Line Bundles}

We can associate to a divisor class $C \in \mathrm{Cl} ( M_g )$ a line bundle, \ie a locally free $\mathcal{O}_{M_g}$-module of rank one. To make this connection let us first define another relation among divisors. Given
\[ D := \sum_{p \in M_g}{n ( p ) \cdot p} \; , \qquad E := \sum_{q \in M_g}{m ( q ) \cdot q} \]
we compare $D$ and $E$ pointwise. Hence, $D \geq E$ precisely if for all $q \in M_g$ we have $n( q ) \geq m ( q )$. That said we define for $C \in \mathrm{Cl} ( M_g )$ the sheaf $\mathcal{O}_{M_g}( C )$ as follows:
\begin{enumerate}
 \item Pick a divisor $D \in \mathrm{Div} ( M_g )$ which represents the divisor class $C$.
 \item Over open $U \subseteq M_g$ the sheaf $\mathcal{O}_{M_g}( C )$ has local sections
      \[ \mathcal{O}_{M_g} ( C ) ( U ) := \left\{ \left. f \colon U \to \mathbb{C} \text{ meromorphic } \right| \mathrm{div} \left( f \right) \geq - D \right\} \cup \left\{ 0 \right\} \ . \]
\end{enumerate}
This sheaf $\mathcal{O}_{M_g} ( C )$ indeed happens to be a holomorphic line bundle. Also the above construction does not depend on the chosen representant of the divisor class $C$ since there is an isomorphism between $\mathrm{Cl} ( M_g )$ and the set $\mathrm{Pic} ( M_g )$ of all (isomorphisms classes of) holomorphic line bundles on $M_g$. In particular, for every holomorphic line bundle $L$ there exists $C \in \mathrm{Cl} ( M_g )$ such that $L \cong \mathcal{O}_{M_g} ( C )$. Therefore, we can define $\mathrm{deg} ( L ) = \mathrm{deg} ( C )$. $\mathrm{Pic} ( M_g )$ is an Abelian group upon
\[ \mathcal{O}_{M_g} \left( D_1 \right) \otimes_{\mathcal{O}_{M_g}} \mathcal{O}_{M_g} \left( D_2 \right) = \mathcal{O}_{M_g} \left( D_1 + D_2 \right) \, . \]
We use $L^\vee \cong \mathcal{O}_{M_g} \left( -D \right)$ to denote the inverse of the holomorphic line bundle $L \cong \mathcal{O}_{M_g} \left( D \right)$.

\paragraph{The Theorem of Riemann-Roch}

A simple consequence of the construction of $\mathcal{O}_{M_g} ( D )$ is
\[ H^0 \left( M_g, \mathcal{O}_{M_g} \left( D \right) \right) = \left\{ \left. f \colon M_g \to \mathbb{C} \text{ meromorphic } \right| \mathrm{div} ( f ) \geq - D \right\} \cup \left\{ 0 \right\}\, . \]
It is an important fact that for every holomorphic line bundle $L$ on $M_g$, the sheaf cohomologies $H^0 \left( M_g, L \right)$ and $H^1 \left( M_g, L \right)$ happen to be finite-dimensional complex vector spaces. In addition, all higher sheaf cohomologies vanish identically. Let us hence introduce the abbreviation
\[ h^i \left( M_g, L \right) := \mathrm{dim}_\mathbb{C} \left( H^i \left( M_g, L \right) \right) \, . \]
One of the most important results regarding $h^i( M_g, L )$ is the Riemann-Roch theorem.\footnote{This result holds as long as $M_g$ is smooth. If $M_g$ is singular, this result must be altered. We will give examples along these lines later in this thesis.} It says that the cohomology dimensions of every holomorphic line bundle $L$ on $M_g$ satisfy
\[ h^0 \left( M_g, L \right) - h^1 \left( M_g, L \right) = \mathrm{deg} \left( L \right) + 1 - g \, . \]

\paragraph{The Canonical Bundle and Consequences}

All compact, connected Riemann surfaces $M_g$ admits a non-trivial meromorphic 1-form $\omega$. To this 1-form one can associate a divisor -- the so-called \emph{canonical divisor $K_g$}. The holomorphic line bundle $\mathcal{K}_g := \mathcal{O}_{M_g}( K_g )$ is termed the \emph{canonical line bundle}. One can prove 
$h^0 \left( M_g, \mathcal{K}_g \right) = g$ and $\mathrm{deg} \left( \mathcal{K}_g \right) = 2g- 2$. By means of $\mathcal{K}_g$ one can formulate a duality theorem for every holomorphic line bundle $L$ on $M_g$:
\[ h^1 \left( M_g, L \right) = h^0 \left( M_g , \mathcal{K}_g \otimes L^\vee \right) \, . \]
This is a special form of the so-called \emph{Serre-duality}.

\paragraph{Examples}

Let us consider a holomorphic line bundle $L$ on $M_g$ and differ a number of cases regarding its degree. By combining the above results one readily verifies:
\begin{enumerate}
 \item $\mathrm{deg} ( L ) < 0$, then $ h^0 ( M_g, L ) = 0$.
 \item $\mathrm{deg} ( L ) = 0$, then $h^0 \left( M_g, L \right) = 1$ iff $L \cong \mathcal{O}_{M_g}$ and otherwise $h^0 \left( M_g, L \right) = 0$.
 \item $\mathrm{deg} ( L ) \geq 2g-1$, then $h^1 ( M_g, L ) = 0$.
\end{enumerate}
The last result is a special instance of the \emph{Kodaira vanishing theorem}.

For $g = 0$ or $g = 1$ these results are strong enough to deduce all sheaf cohomology dimensions of holomorphic line bundles. Let us set $d = \mathrm{deg} ( L )$, then we have for the Riemann sphere $M_0$:
\[ h^0 \left( M_0, L \right) = \begin{cases} d+1 & d \geq 0 \\ 0 & d < 0 \end{cases}, \qquad h^1 \left( M_0, L \right) = \begin{cases} 0 & d \geq 0 \\ -(d+1) & d < 0 \end{cases} \]
Similarly, we find for a torus surface $M_1$:
\[
h^0 \left( M_1, L \right) = \begin{cases} d & d > 0 \\ 1 & \text{$d = 0$ and $L \cong \mathcal{O}_{M_1}$} \\ 0 & \text{$d = 0$ and $L \not \cong \mathcal{O}_{M_1}$} \\ 0 & d < 0 \end{cases}, \quad
h^1 \left( M_1, L \right) = \begin{cases} 0 & d > 0 \\ 1 & \text{$d = 0$ and $L \cong \mathcal{O}_{M_1}$} \\ 0 & \text{$d=0$ and $L \not \cong \mathcal{O}_{M_1}$} \\ -d & d < 0 \end{cases} \]

Note that for any divisor $D$ on $M_1$ of degree $0$ it holds $h^0 \left( M_1, \mathcal{O}_{M_1} ( D ) \right) = h^1 \left( M_1, \mathcal{O}_{M_1} ( D ) \right)$. If $D$ is the trivial divisor both cohomology dimensions match $1$, otherwise they are zero. This shows that the cohomology dimensions can jump as we alter the divisor $D$.
Examples of such jumps will be presented later in this thesis.

\paragraph{Spin Bundles}

Let us complete this review on holomorphic line bundles on compact, connected Riemann surface by turning our attention to spin bundles on $M_g$. For physical applications these bundles are important as the existence of spinors hinges on the existence of spin bundles which admit global sections. 

Spin bundles exist on spaces for which the first two Stiefel-Whitney classes vanish \cite{Nakahara:2003nw}. Indeed this is the case for $M_g$. A spin divisor $D \in \mathrm{Div} \left( M_g \right)$ satisfies $2D = K_g$ and a holomorphic line bundle $L \cong \mathcal{O}_{M_g} ( D )$ associated to a spin divisor is termed a spin bundle on $M_g$. It was proven in \cite{atiyah1971riemann, mumford1971theta} that on $M_g$ there exist $2^{2g}$ linearly independent spin divisors. They have the following property:
\begin{itemize}
 \item $2^{g-1} \cdot \left( 2^g - 1 \right)$ spin bundles have at least one non-trivial global section.
 \item $2^{g-1} \cdot \left( 2^g + 1 \right)$ spin bundles either have no or at least two non-trivial global sections.
\end{itemize}
Consequently, the choice of a spin bundle on $M_g$ is far from unique. Luckily, in many physical situations there are additional constraints which can help to fix the bundle. An example of such constraints can be found in \cite[pp. 58]{Beasley:2008dc}.

Let us recall from \cref{subsec:FTheoryFromIIB} that a torus surface $M_1$ can be understood as the quotient $\mathbb{C}_{\vect{a},\vect{b}} = \mathbb{C} / \Lambda_{\vect{a},\vect{b}}$ of the complex plane $\mathbb{C}$ by a lattice $\Lambda_{\vect{a},\vect{b}} = \mathrm{Span}_{\mathbb{Z}} \left( \vect{a}, \vect{b} \right)$ of rank 2, where $\vect{a}, \vect{b} \in \mathbb{C}$. Without loss of generality we chose $\vect{a} = 1$, set $\vect{b} = \vect{\tau}$ and pictured $\mathbb{C}_{\vect{1},\vect{\tau}}$ in \cref{fig-5610351095328}. For $\mathbb{C}_{\mathbf{1}, \boldsymbol{\tau}}$ it holds $\mathcal{K}_g = \mathcal{O}_{\mathbb{C}_{\mathbf{1}, \boldsymbol{ \tau}}}$ and there exist four spin divisors, which are illustrated in \cref{figure-SpinDivisorsOnM1} \cite{BiesMaster}. The spin divisor $D_0$ satisfies
\[ h^0 ( \mathbb{C}_{\mathbf{1}, \boldsymbol{ \tau}}, \mathcal{O}_{\mathbb{C}_{\mathbf{1}, \boldsymbol{ \tau}}} \left( D_0 ) \right) = h^1 ( \mathbb{C}_{\mathbf{1}, \boldsymbol{ \tau}}, \mathcal{O}_{\mathbb{C}_{\mathbf{1}, \boldsymbol{ \tau}}} \left( D_0 ) \right) = 1. \]
For the other spin divisors $D_1$, $D_2$ and $D_3$ these sheaf cohomology dimensions vanish identically. This nicely reproduces the results from \cite{atiyah1971riemann, mumford1971theta}.

\subsection{Zero Mode Counting in type IIB \emph{String Theory}}

Finally, let us come back to zero mode counting in type IIB string theory. As explained in \cref{subsec:TypeIIBStringTheory}, we have to focus on an orientifold compactification to allow for non-trivial D7-brane configurations. Let us assume that each D7-brane is \emph{spacetime filling}, \ie covers the external space $\mathcal{E}_4$ completely and wrap a holomorphic 4-cycle $D_i \subseteq \mathcal{B}_6$. In this setup, a large class of gauge background is specified by a family $\left\{ V_i \right\}$ of vector bundles on the holomorphic 4-cycles $D_i$ \cite{Aspinwall:2004jr}. More generally, one can consider a family $\left\{ F_i \right\}$ of coherent sheaves on the 4-cycles $D_i$. For now we suffice it to state our analysis in terms of vector bundles, but we come back to the topic of coherent sheaves in \cref{sec:ToolsForCompuationalF-theoryVacua}.

Let us focus on $U(1)$ gauge theories on two D7-brane stacks $D_a$, $D_b$. Such gauge theories are encoded in the gauge connections of holomorphic line bundles $L_i$ on the 4-cycles $D_a$, $D_b$. We assume that $D_a$ and $D_b$ intersect in $C_{ab} = D_a \cap D_b$. On this locus we can consider the line bundle 
\[ L_{ab} :=\left. L_a \right|_{C_{ab}} \otimes \left. L_b^\vee \right|_{C_{ab}} \, . \label{equ:NaiveGaugeEnhancement} \]
The sections of this line bundle are then seen to transform in the representation $( \mathbf{1}_a, \overline{\mathbf{1}}_b )$ of the gauge group $U(1)_a \times U(1)_b$. This is a special instance of the more general phenomenon of gauge enhancement at D-brane intersections.

Whilst intuitive, this picture is incomplete for essentially two reasons:
\begin{itemize}
 \item \textbf{No gluing morphisms:} Charged open strings between D7-branes $D_i$ are given by sheaf homomorphisms $\alpha_{ij} \colon 
      V_i \to V_j$. If these strings are given non-zero VEVs, then the \emph{gluing morphisms} $\alpha_{ij}$ must be taken into account \cite{Donagi:2010pd}. Ultimately, these morphisms tell us how to join $\left. V_a \right|_{C_{ab}}$ and $\left. V_b \right|_{C_{ab}}$ on the intersection $C_{ab}$. A mere tensor product, as used in \cref{equ:NaiveGaugeEnhancement}, correctly accounts for situations without gluing morphisms only. See \cite{Donagi:2011jy, Donagi:2011dv} for setups with non-trivial gluing morphisms.
 \item \textbf{The spin bundle:} Ultimately, we wish to count the localised, charged, chiral \emph{fermions} at D7-brane intersections. The above picture however 
      accounts for bosonic states only, as we are missing the spin bundle.
      
      Let us assume that $C_{ab}$ is a compact and connected Riemann surface. These we discussed in the previous section, where we learned that the spin structures are given by those holomorphic line bundles $S$ with $S^{\otimes 2} \cong K_{C_{ab}}$ where $K_{C_{ab}}$ is the canonical bundle of $C_{ab}$ \cite{atiyah1971riemann, mumford1971theta}. Schematically we can thus describe the admissible spin structures as the `roots' $\sqrt{K_{C_{ab}}}$ of the canonical bundle. Unfortunately these `roots' are not unique -- on a compact and connected Riemann surface of genus $g$ there are $2^{2g}$ inequivalent spin structures \cite{atiyah1971riemann, mumford1971theta}. We have seen an example of this in \cref{figure-SpinDivisorsOnM1}, which lists the four spin divisors on a torus $\mathbb{C}_{1, \tau}$.
      
      Which spin bundle should be picked? A canonical choice comes from using the embedding $C_{ab} \hookrightarrow D_a$ \cite[pp. 58]{Beasley:2008dc}. If we assume $C_{ab}$ to be a complete intersection of $D_a$ and $D_b$, then the adjunction formula \cite{griffiths2011principles} implies
      \[ 
      \resizebox{0.84\textwidth}{!}{$
      K_{C_{ab}} = \left. K_{D_a} \right|_{C_{ab}} \otimes N_{C_{ab} / D_a} = \left. \left( K_{\mathcal{B}_6} \otimes \mathcal{O}_{\mathcal{B}_6} \left( D_a \right) \otimes \mathcal{O}_{\mathcal{B}_6} \left( D_b \right) \right) \right|_{C_{ab}} := \left. \mathcal{O}_{\mathcal{B}_6} \left( M \right) \right|_{C_{ab}} \, . 
      $}
      \]
      The Freed-Witten-quantisation condition ensures that $\left. D_a \right|_{C_{ab}} + \left. D_b \right|_{C_{ab}} + 1/2 \cdot \left. M \right|_{C_{ab}}$ is linearly equivalent to a $\mathbb{Z}$-Cartier divisor $D_{ab}$ on $C_{ab}$. This finally leads us to consider the line bundle \cite{Bies:2014sra}
      \[ L_{ab} := \mathcal{O}_{C_{ab}} \left( D_{ab} \right) \cong \mathcal{O}_{C_{ab}} \left( \left. D_a \right|_{C_{ab}} + \left. D_b \right|_{C_{ab}} + 1/2 \cdot \left. M \right|_{C_{ab}} \right) \, . \label{equ:EducatedGaugeEnhancement} \]
\end{itemize}

If the gauge data on the cycles $D_i$ is given by vector bundles $V_i$ (or even coherent sheaves $F_i$) then global extension groups of (combinations of) vector bundles/coherent sheaves count the localised zero modes \cite{Katz:2002gh}. However, for a simplified setup without gluing morphisms and Abelian gauge groups only, it suffices to study sheaf cohomologies of line bundles. Therefore, the localised, charged, (anti-)chiral fermions at the intersection $C_{ab}$ are counted as follows:
\[ \text{chiral} \leftrightarrow H^0 \left( C_{ab} , L_{ab} \right) \, , \qquad  \text{anti-chiral} \leftrightarrow H^1 \left( C_{ab} , L_{ab} \right) \, . \label{equ:ZeroModesInTypeIIB}\]

\section{Summary} \label{sec:SummaryChapter2}

In this chapter we have given an introduction to \emph{GUTs}, \emph{string theory} and \emph{F-theory}. In particular, we found that a singular elliptic fibration $\pi \colon Y_4 \twoheadrightarrow \mathcal{B}_6$ encodes a lot of information about the associated \emph{F-theory} compactification. For example, the singular locus $\Delta \subseteq \mathcal{B}_6$ of $Y_4$ encodes the location of D7-branes, and the singularities over $\Delta$ the gauge symmetry. In this thesis we restrict to singular elliptic fibrations which satisfy the following conditions (\cf \cref{sec:F-theory}):
\begin{enumerate}
 \item There exists a flat, smooth, crepant resolution $\hat{\pi} \colon \hat{Y}_4 \twoheadrightarrow \mathcal{B}_6$ which is a Calabi--Yau.
 \item $\hat{Y}_4$ admits a section, which is referred to as the \emph{zero section} $S_0$.
 \item $\mathcal{B}_6$ is a connected 3-fold with $h^{1,0} ( \mathcal{B}_6 ) = h^{2,0} ( \mathcal{B}_6 ) = 0$ and $h^1( \mathcal{B}_6, \mathbb{Z} ) = 0$.
\end{enumerate}
Of particular importance to our discussion was the Deligne cohomology $H^4_D ( \hat{Y}_4, \mathbb{Z} ( 2 ) )$ which encodes gauge backgrounds in the \emph{F-theory} compactifications at hand. 

The revision on zero mode counting in type IIB \emph{string theory} showed that sheaf cohomologies can be used to count zero modes in string compactifications. In the next chapter we will analyse the corresponding situation in \emph{F-theory}. It is known that a blow-up resolution corresponds to the Coulomb branch of the dual 3-dimensional \emph{M-theory} vacuum. The non-Abelian gauge symmetry is restored in the dual 4-dimensional \emph{F-theory} vacuum after taking the fibre volume to zero. Nonetheless, this means that by working with $\hat{Y}_4$ we are only able to detect Abelian gauge backgrounds. Non-Abelian gauge bundles on the D7-branes, by contrast, cannot be encoded in the Deligne cohomology. As a consequence, we will derive in the next chapter a result very similar to \cref{equ:ZeroModesInTypeIIB}.

\chapter{Zero Mode Counting via Sheaf Cohomology} \label{chapter:MasslessSpectraAndSheafCohomology}
\section{The Chow Ring Of Varieties} \label{sec:TheChowRing}

Thus far, we have considered analytic geometries to describe the geometry of an \emph{F-theory} compactification. In particular, we have explained that that elements of the Deligne cohomology $H_D^4 ( \hat{Y}_4, \mathbb{Z} ( 2 ) )$ describe gauge backgrounds in \emph{F-theory} vacua. Unfortunately Deligne cohomology is hard to handle in practise. Therefore, we look for a more convenient parametrisation of (subsets of) gauge backgrounds. 

On varieties one can define the Chow ring $\mathrm{CH}^{\bullet} ( \hat{Y}_4 )$. As it turns out, this ring can be used to parametrise at least a subset of $H_D^4 ( \hat{Y}_4, \mathbb{Z} ( 2 ) )$. In addition, $\mathrm{CH}^{\bullet} ( \hat{Y}_4 )$ lends itself much better for explicit computations. Unfortunately however, the definition of the Chow ring explicitly rests on rational functions. This makes it necessary to break with analytic geometry and dive into algebraic geometry.

\subsection{Complex Varieties} \label{subsec:AffineVarietiesBasics}

\paragraph{Affine Varieties}

We consider the ring $R = \mathbb{C} [ x_1, \dots, x_n ]$. An affine algebraic set $S \subseteq \mathbb{C}^n$ is the vanishing locus of finitely many polynomials $f_1, \dots, f_N \in R$, \ie
\[ S = V ( f_1, \dots, f_N ) = \left\{ \left. p \in \mathbb{C}^n \right| f_1 \left( p \right) = \ldots = f_N ( p ) = 0 \right\} \, . \]
Examples include $\mathbb{C}^n = V( 0 )$, $\emptyset = V( 1 )$ or $\left\{ 1 \right\} = V ( x - 1 ) \subseteq \mathbb{C}$. 

An affine algebraic set $S \subseteq \mathbb{C}^n$ admits two different topologies. For once there is the standard analytic topology. On the other hand we can define the  Zariski topology: the Zariski closed subsets of $S$ are those affine algebraic sets of $\mathbb{C}^n$ which are contained in $S$. The Zariski open subsets of $S$ are the complements of these Zariski closed subsets of $S$.

In any topological space $( X, \tau )$ a non-empty subset $Y$ is termed irreducible if it cannot be written as $Y = Y_1 \cup Y_2$ where $Y_1$, $Y_2$ are two proper and closed (with respect to $\tau)$) subsets of $Y$. That said we can wonder under what condition an affine algebraic set $S \subseteq \mathbb{C}^n$ is irreducible with respect to the Zariski topology on $\mathbb{C}^n$. This property is nicely encoded in the ideal
\[ I ( S ) := \left\{ \left. f \in R \right| f ( p ) = 0 \text{ for all } p \in S \right\} \, . \]
Namely $S$ is irreducible precisely if $I ( S ) \subseteq R$ is a prime ideal. So for example the affine space $\mathbb{C}^n$ is irreducible (with respect to the Zariski topology). 

An affine variety $V \subseteq \mathbb{C}^n$ is an irreducible (affine) algebraic set. Examples of affine algebraic varieties include $\mathbb{C}$ or $\left\{ 1 \right\} = V ( x - 1 ) \subseteq \mathbb{C}$. Note that $\emptyset = V( 1 )$ is not an affine variety, since $I ( \emptyset ) = R$ and this is not a prime ideal -- prime ideals are (special) proper ideals. By use of $I( V )$ we can define the coordinate ring of $V$ as $\mathbb{C} [ V ] = R / I( V )$. The elements of $\mathbb{C} [ V ]$ can be understood as complex-valued polynomial functions on $V$. An affine algebraic variety $V$ is irreducible precisely if its coordinate ring $\mathbb{C} [ V ]$ is an integral domain. Even more, it can  be seen that a point $p \in V$ defines a maximal ideal
\[ \mathfrak{m}_p := \left\{ \left. f \in \mathbb{C} \left[ V \right] \right| f \left( p \right) = 0 \right\} \subseteq \mathbb{C} \left[ V \right] \, . \]
We can therefore understand $V$ as the set of maximal ideals in $\mathbb{C} [ V ]$. This finding is expressed as $V = \mathrm{Specm} ( \mathbb{C} [ V ] )$ and resembles approaches to affine schemes closely. We will have much more to say about this in \cref{subsec:CoherentSheavesOnVarieties}.

Let us complete our discussion on affine algebraic varieties by discussing rational functions. Be $Y \subseteq V$ a Zariski open subset of an affine algebraic variety $V$. Then a function $f \colon Y \to \mathbb{C}$ is regular at $p \in Y$ if there exists an open neighbourhood $p \in U \subseteq Y$ and polynomials $g,h \in R$ such that $h$ is nowhere zero on $U$ and $f = \frac{g}{h}$ on $U$. $f \colon Y \to \mathbb{C}$ is regular if it is regular at every point of $Y$.

\paragraph{Abstract Varieties}

An abstract variety can be understood as a topological space $( X, \tau )$, such that every point $p \in X$ admits an (in $\tau$) open neighbourhood $p \in U \subseteq X$ such that $( U, \left. \tau \right|_U )$ is isomorphic to an affine variety. For example, the projective line $\mathbb{P}^1_{\mathbb{C}}$ with homogeneous coordinates $[ u_1 : u_2 ]$ is an abstract variety. We will explain in detail in \cref{subsec:TowardsToricVarieties} that an affine open cover is given by
\[ U_1 = \mathrm{Specm} \left( \mathbb{C} \left[ \frac{x_1}{x_2} \right] \right), \qquad U_2 = \mathrm{Specm} \left( \mathbb{C} \left[ \frac{x_2}{x_1} \right] \right) \, . \]

On an abstract variety $Y$ one can consider the so-called \emph{rational functions}. Collectively these functions form the function field $K( Y )$ of the variety $Y$ which is formed from equivalence classes of pairs $\left( U, f \right)$. Explicitly we have
\[ K( Y ) = \left\{ \; \left. \left[ \left( U ,f \right) \right] \; \right| \emptyset \neq U \subseteq Y \text{ open subset and } f \text{ a regular function on } U \right\} \, , \]
where $( U, f ) \sim (V, g)$ precisely if $f = g$ on $U \cap V$.

\subsection{The Chow Group}

Let us now come back to \emph{F-theory} compactifications. We use the above knowledge to describe $\hat{Y}_4$ as abstract variety over $\mathbb{C}$. This enables us to define the group of algebraic cycles $Z^p ( \hat{Y}_4 )$ of complex codimension $p$ in $\hat{Y}_4$.\footnote{In this thesis we use the symbol $Z^p ( \hat{Y}_4 )$ to denote the group of algebraic cycle of complex \textbf{co}dimension $p$ in $\hat{Y}_4$. In contrast, $Z_p ( \hat{Y}_4 )$ is to denote the group of algebraic cycles of complex dimension $p$ in $\hat{Y}_4$. We adopt this very notation also for the Chow rings below, \ie $\mathrm{CH}^2 ( \hat{Y}_4 )$ will represent classes of algebraic cycles of codimension $2$ in $\hat{Y}_4$, whilst \eg $\mathrm{CH}_1 ( \hat{Y}_4 )$ is for classes of algebraic cycle of dimension $1$ in $\hat{Y}_4$.} This group is formed from elements of the form
\[ C = \sum_{i = 1}^{N}{n_i C_i} \]
where $N \in \mathbb{N}$, $n_i \in \mathbb{Z}$ and $C_i$ are \emph{not necessarily smooth} but irreducible subvarieties of $\hat{Y}_4$. We say that an algebraic cycle $C \in Z_p ( \hat{Y}_4)$ is rationally equivalent to zero, $C \sim 0$, if and only if
\[ C = \sum_{i = 1}^{N}{\left[ \mathrm{div} \left( r_i \right) \right]} \]
for suitable $N \in \mathbb{N}$ and invertible rational functions $r_i \in \mathbb{C} ( W )^\ast$ on some $( p + 1 )$-dimensional subvarieties $W_i$ of $\hat{Y}_4$. Let $\mathrm{Rat}_p ( \hat{Y}_4 )$ be the subgroup of $Z_p ( \hat{Y}_4 )$ formed from all algebraic cycles which are rationally equivalent to $0$. Then the Chow group $\mathrm{CH}_p ( \hat{Y}_4 )$ is defined as the quotient
\[ \mathrm{CH}_p ( \hat{Y}_4 ) := Z_p ( \hat{Y}_4 ) / \mathrm{Rat}_p ( \hat{Y}_4 ) \, . \]

\begin{figure}[tbp]
{\centering
\includegraphics{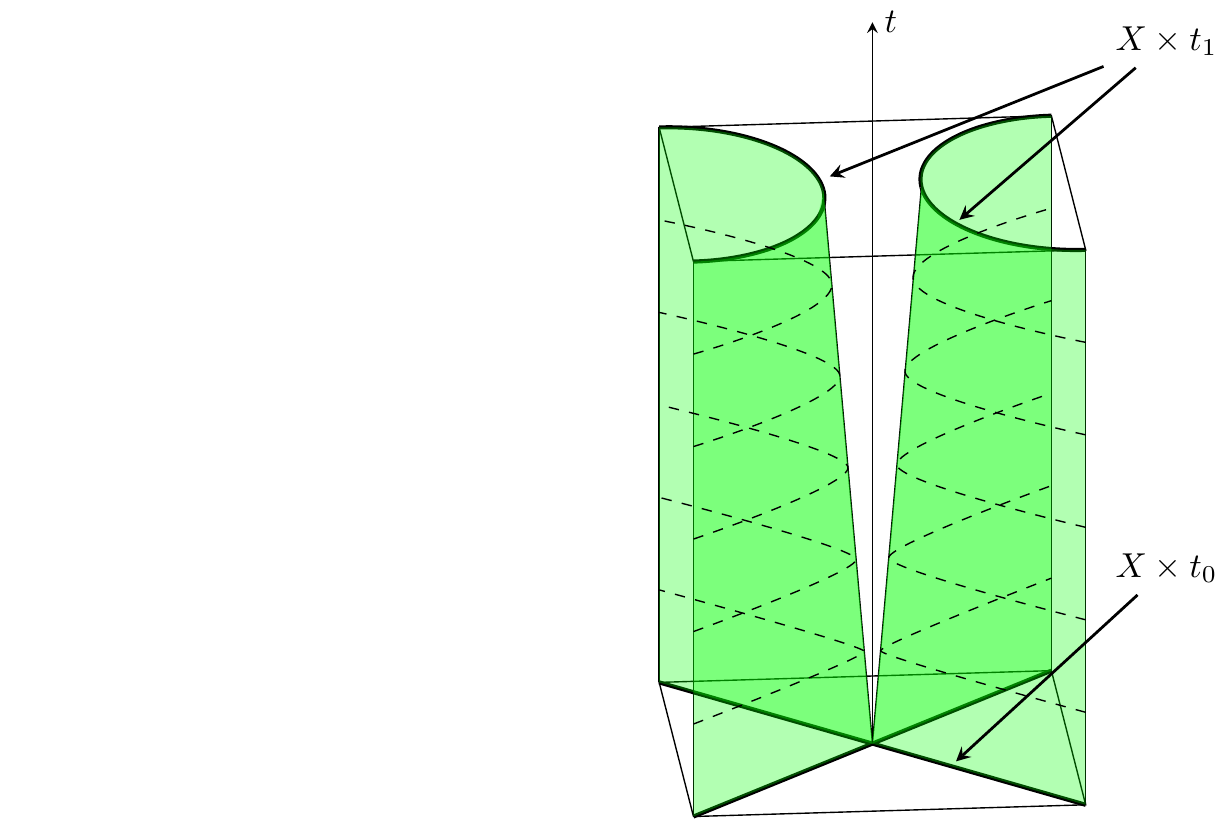}}
\caption[Intuitive picture of rational equivalence of algebraic cycles.]{Rational equivalence between the union of two lines in $\mathbb{CP}^2$ and a hyperbola. This picture is inspired by \cite{ATIT} and taken from \cite{Bies:2014sra}.}
\label{figure-takenFrom2014Paper}
\end{figure}

Let us mention that an intuitive understanding of rational equivalence can be obtained as follows: Two cycles $C_1, C_2$ are rationally equivalent if and only if one can find a rationally parametrized family of cycles which interpolates between $C_1$ and $C_2$. This means that we can find a cycle $\Gamma(t)$ on $\mathbb{P}^1 \times X$ such that 
\[ \Gamma \left( t=t_1 \right) = C_1, \qquad \quad \Gamma \left( t=t_2 \right) = C_2 \]
with $t \in \mathbb{P}^1$ parametrising the interpolation between $C_1$ and $C_2$. In other words, $C_1$ and $C_2$ are related by a `rational homotopy'. We sketch this finding in \cref{figure-takenFrom2014Paper}.

\subsection{From Chow Groups to Deligne Cohomology}

To any algebraic cycle in $Z^p ( \hat{Y}_4 )$ one can associate a cocycle in $H_{\mathbb{Z}}^{p,p} ( \hat{Y}_4 )$ via a group homomorphism $\gamma_{\hat{Y}_4, p} \colon Z^p ( \hat{Y}_4 ) \to H^{p,p}_{\mathbb{Z}} ( \hat{Y}_4 )$ which is termed the \emph{cycle map}. The description of 3-form data in $H^4_D ( \hat{Y}_4, \mathbb{Z} ( 2 ) )$ by a class of algebraic cycles in $\mathrm{CH}^2 ( \hat{Y}_4 )$ hinges on a refined version of this statement: There exists a so-called \emph{refined cycle map} (see \eg p.~123 in \cite{GreenMurreVoisin})
\[ \hat{\gamma}_{\hat{Y}_4, p} \colon Z^p ( \hat{Y}_4 ) \to H^4_D ( \hat{Y}_4, \mathbb{Z} ( 2 ) ) \, . \]
This homomorphism respects rational equivalence -- for any two $C_1, C_2 \in Z^p ( \hat{Y}_4 )$ it holds
\[ C_1 \sim C_2 \; \Rightarrow \; \hat{\gamma}_{\hat{Y}_4, p} \left( C_1 \right) = \hat{\gamma}_{\hat{Y}_4, p} \left( C_2 \right) \, . \]
In consequence, the refined cycle map extends to a map $\hat{\gamma}_{\hat{Y}_4,p} \colon \mathrm{CH}^p ( \hat{Y}_4 ) \to H^{2p}_D ( \hat{Y}_4, \mathbb{Z} ( 2 ) ) \label{equ:DeligneForP}$ which gives rise to a group homomorphism
\[ \hat{\gamma} \colon \mathrm{CH}^{\bullet} ( \hat{Y}_4 ) \to H^{\bullet}_D ( \hat{Y}_4, \mathbb{Z} ( 2 ) ) \; , \; x \mapsto \bigoplus_{p \in \mathbb{N}_{\geq 0}}{\hat{\gamma_p} \left( x \right)} \, . \]
Gauge backgrounds in $H^4_D(\hat{Y}_4, \mathbb Z(2))$ can thus be described by $A \in \mathrm{CH}^2 ( \hat{Y}_4 )$ along the map
\[ \hat{\gamma}_2 \colon \mathrm{CH}^2 ( \hat{Y}_4 ) \to H^{4}_D ( \hat{Y}_4, \mathbb{Z} ( 2 ) ) \label{refcyclemap} \, . \]

Fortunately, the concrete form of $\hat{\gamma}$ is not required for our computations. This map is surjective over $\mathbb{Q}$ if and only if the Hodge conjecture holds true. In general it is not injective. This means that two different Chow classes may in principle map to the same element in $H^4_D ( \hat{Y}_4, \mathbb{Z} ( 2 ) )$. 
This is not a big drawback for our purposes. What will be crucial is rather the fact that two algebraic cycles which are rationally equivalent are guaranteed to map to the same element in $H^4_D ( \hat{Y}_4, \mathbb{Z} ( 2 ) )$. This means that manipulations modulo rational equivalence do not change the 3-form background parametrised by an algebraic 2-cycle. This fact comes in particularly handy whenever the 2-cycles representing the gauge background are in turn obtained by pullback from an ambient space on which the Chow group is explicitly known. 

A particularly common situation is one, in which $\hat{Y}_4$ is embedded into a smooth toric variety $X_\Sigma$. On such varieties, rational equivalence and homological equivalence coincide, as we will outline in \cref{subsec:TowardsToricVarieties}. Hence, we can manipulate the underlying 2-cycle as long as we keep its homology class on $X_\Sigma$ fixed, and are guaranteed that the gauge data remains unchanged. As this simplifies computations enormously, we will encounter this situation frequently. To avoid too heavy notation, let us not distinguish algebraic cycles $A \in Z^2 ( \hat{Y}_4 )$ from their classes in $\mathrm{CH}^2 ( \hat{Y}_4 )$. Rather we use capital and upright letters to denote both the relevant class and (if necessary) an explicit representant of this class. Similarly, $\mathcal{A}$ will denote elements of $\mathrm{CH}^2 ( \hat{Y}_\Sigma )$ and $Z^2 ( \hat{Y}_\Sigma)$ whenever a toric ambient space $Y_\Sigma$ of $\hat{Y}_4$ is considered. Likewise the symbol $\hat{Y}_5$ is reserved for a (not necessarily toric) 5 complex-dimensional ambient space of $\hat{Y}_4$.

\subsection{The Intersection Product}

The group $\mathrm{CH}^{\bullet} ( \hat{Y}_4 )$ has even more structure. Namely there exists an intersection product, which turns this group into the \emph{Chow ring}. In particular, the discussion in \cref{chapter:MasslessSpectraAndSheafCohomology} heavily relies on this intersection product. Let us therefore briefly revise this intersection product in the Chow ring $\mathrm{CH}^\bullet ( X )$ of a (not necessarily smooth) variety $X$: First look at a morphism $f \colon X \to Y$ from the variety $X$ to a smooth variety $Y$ of dimension $n$. Now the graph morphism $\gamma_f \colon X \to X \times Y \; , \; x \mapsto ( x, f ( x ) )$ induces a Gysin morphism $\gamma_f^\ast$ such that
\[ \mathrm{CH}_k \left( X \right) \otimes \mathrm{CH}_l \left( Y \right) \stackrel{\times}{\rightarrow} \mathrm{CH}_{k+l} \left( X \times Y \right) \stackrel{\gamma^\ast_f}{\rightarrow} \mathrm{CH}_{k+l-n} \left( X \right) \]
is a well-defined mapping. Given $\alpha \in \mathrm{CH}_k ( X )$, $\beta \in \mathrm{CH}_l ( Y )$ we denote the image of $\alpha \otimes \beta$ under the above map as $\alpha \cdot_f \beta$. We term this the intersection product of $\alpha$, $\beta$ under the morphism $f$. Often we will omit $f$.

\section{Zero Modes in \emph{F-Theory} via Super-Yang-Mills Theories} \label{sec:SheafCohomologyAndMasslessMatter}

The gauge theory on D7-branes locally enjoys a description as partially topologically twisted 8-dimensional Super-Yang-Mills theory \cite{Donagi:2008ca, Beasley:2008dc}. In this language, the definition of the gauge background and the computation of the exact massless matter content is well established. To prepare ourselves to extract the local gauge data from Deligne cohomology, we therefore find it illustrative to briefly revise this (local) approach via a topologically twisted gauge theory and the description of its massless matter.

\subsection{Bulk Matter} \label{subsec:BulkMatter}

Let us begin by recalling the nature of the so-called bulk 7-brane matter on a non-Abelian brane stack. A stack of 7-branes wrapping the component $\Delta_I$ of the discriminant locus carries, in addition to a gauge multiplet, massless 4-dimensional $\mathcal{N}=1$ chiral multiplets in the adjoint representation of the non-Abelian gauge algebra $\mathfrak{g}_I$ underlying the gauge group $G_I$. As stressed at the end of the previous section, we restrict ourselves to Abelian gauge backgrounds. Locally, these are described by a line bundle $L_I$ in the Picard group $\mathrm{Pic}(\Delta_I)$ on the 7-brane stack on $\Delta_I$.
Embedding the structure group $U(1)_I$ of $L_I$ into $G_I$ breaks $G_I \rightarrow H_I \times U(1)_I$, which induces the decomposition 
\[ {\textbf{adj}}(G_I) \rightarrow \bigoplus_{m_I} {\mathbf{r}}_{  m_I} \label{adjdecom-gen} \]
into irreducible representations of $H_I$. Then the massless bulk matter in representation ${\mathbf{r}}_{m_I}$ on $\Delta_I$ is counted by the cohomology groups \cite{Katz:2002gh,Donagi:2008ca,Beasley:2008dc,Blumenhagen:2008zz}
\begin{align}
\begin{split} \label{cohoms-local-bulk}
H^0(\Delta_I, L_{m_I}) \quad \phantom{\oplus} \qquad &  \\
H^1(\Delta_I, L_{m_I}) \quad \oplus \qquad &H^0(\Delta_I, L_{m_I} \otimes K_{\Delta_I}) \\
H^2(\Delta_I, L_{m_I}) \quad \oplus \qquad &H^1(\Delta_I, L_{m_I} \otimes K_{\Delta_I})  \\
                              &         H^2(\Delta_I, L_{m_I} \otimes K_{\Delta_I}) \, .
\end{split}
\end{align}
Here the relevant line bundle $L_{m_I}$ is given by $L_{m_I} = L_I^{q_I({\mathbf{r}}_{m_I})}$ with $q_I({\mathbf{r}}_{m_I})$ the $U(1)_I$ charge of representation ${\mathbf{r}}_{m_I}$. The second and third line count chiral and, respectively, anti-chiral $\mathcal{N}=1$ multiplets in representation $\mathbf{r}_{m_I}$. The first and fourth line vanish for supersymmetric fluxes \cite{Blumenhagen:2008zz} and can hence be discarded.
The chiral index resulting from this spectrum is $\chi_{{\mathbf{r}}_{m_I}} = - c_1 ( \Delta_I ) \cdot_{\Delta_I} c_1 ( L_{m_I} )$.

\subsection{Localised Matter} \label{subsec:LocalisedMatterAndTheSpinBundle}

Consider next the massless matter localised on a curve $C_{\mathbf{R}}$, and denote by $L_{\mathbf{R}} \in \mathrm{Pic} ( C_{\mathbf{R}} )$ the local gauge background to which such matter in the twisted theory on $\mathbb R^{1,3} \times C_{\mathbf{R}}$ is coupled. The number of massless matter multiplets in representation ${\mathbf{R}}$ is given by the dimensions of the cohomology groups \cite{Donagi:2008ca, Beasley:2008dc}
\[ \label{mattercohoma}
H^{i}(C_{\mathbf{R}}, L_{\mathbf{R}} \otimes \sqrt{K_{C_{\mathbf{R}}}}) 
\]
with $i=0$ and $i=1$ counting chiral and anti-chiral $\mathcal{N}=1$ multiplets, respectively. In this expression $\sqrt{K_{C_{\mathbf{R}}}}$ is \textbf{a} spin bundle on $C_{\mathbf{R}}$. Recall from our discussion in \cref{subsec:LineBundlesOnRiemannSurfaces} that the choice of this bundle is in general far from unique. In this current context, global consistency lead us to assert in \cite{Bies:2014sra} that we should take $\sqrt{K_{C_{\mathbf{R}}}}$ as the spin bundle induced by the embedding of $C_{\mathbf{R}}$ into the base $\mathcal{B}_6$. This comes about as follows: Let $D_a, D_b \in \mathrm{Cl} ( \mathcal{B}_6 )$ and $C = D_a \cap D_b \subseteq \mathcal{B}_6$ a curve, then the adjunction formula for $C$ and $D_a$ gives
\[ K_C = \left. K_{D_a} \right|_C \otimes N_{C/D_a} = \left. \left( K_{\mathcal{B}_6} \otimes \mathcal{O}_{\mathcal{B}_6}(D_a) \otimes \mathcal{O}_{\mathcal{B}_6}(D_b) \right) \right|_C \equiv \left. M \right|_C \,. \]
The line bundle $\mathcal{O}_{\mathcal{B}_6} ( M ) \in \mathrm{Pic} ( \mathcal{B}_6 )$ is uniquely determined by its first Chern class since $b^1 ( \mathcal{B}_6 ) = 0$.\footnote{Recall that this is a necessary condition for an elliptically fibred Calabi--Yau 4-fold $\hat{Y}_4 \twoheadrightarrow \mathcal{B}_6$ to exist, for which $b^1( \hat{Y}_4) =0$.} If both $D_a$ and $D_b$ are spin\footnote{This means there exists a $\mathbb{Z}$-Cartier divisor $k_a$ such that $2 k_a$ is the canonical divisor on $D_a$.} then $c_1( \mathcal{O}_{\mathcal{B}_6} ( M ) )$ is an even class in $H^2(\mathcal{B}_6,\mathbb Z)$ and $\sqrt{M}$ is the unique line bundle on $\mathcal{B}_6$ with first Chern class $\frac{1}{2} c_1( \mathcal{O}_{\mathcal{B}_6} ( M ) ) \in H^2(\mathcal{B}_6,\mathbb Z)$. The spin structure to take in \cref{mattercohoma} is then 
$\sqrt{ K_{C_\mathbf{R}}} = \sqrt{M} |_{C_{\mathbf{R}}}$.

If $D_a$ or $D_b$ are not spin, then the Freed-Witten quantisation condition ensures that $L_{\mathbf{R}} \otimes \sqrt{K_{C_{\mathbf{R}}}}$ can be split up as a product of two integral bundles $L_1 \otimes L_2$ such that $L_2$ is a product of line bundles obtained as the pullback of well-defined bundles on $\mathcal{B}_6$.  The pullback bundles underlying $L_1$ and $L_2$ then include the contribution from the spin structure.

\section{From Gauge Backgrounds in \emph{F-Theory} to Line Bundles} \label{sec:FromC3toL}

Our next task is to extract the line bundles appearing in the cohomologies \cref{cohoms-local-bulk} and \cref{mattercohoma} from the gauge background on the globally defined 4-fold $\hat{Y}_4$. The general idea is that the matter states are associated with the quantised moduli space of M2-branes \cite{Witten:1996qb} wrapping suitable components of the fibre, as reviewed in \cref{subsec:GaugeTheoriesInFTheory}. An M2-brane couples in a standard way to the 3-form background via the Chern-Simons action $S_{\mathrm{CS}} = 2 \pi \int_{\mathrm{M2}}{C_3}$. We therefore need to integrate the 3-form gauge background over the fibral curve wrapped by the M2-brane associated with a given state. The resulting object then describes the Abelian gauge background to which the M2-brane excitations along the base couple. 

The formalism of Chow groups allows us to perform this operation of integration over the fibre in a manner which is guaranteed to keep the full amount of information about the gauge background \cite{Bies:2014sra}. Indeed, by using intersection theory within the Chow ring we are able to extract a line bundle either on the 7-brane $\Delta_I$ or on the matter curve $C_{\mathbf{R}}$ on the base.

\begin{figure}[tbp]
\centering
\includegraphics{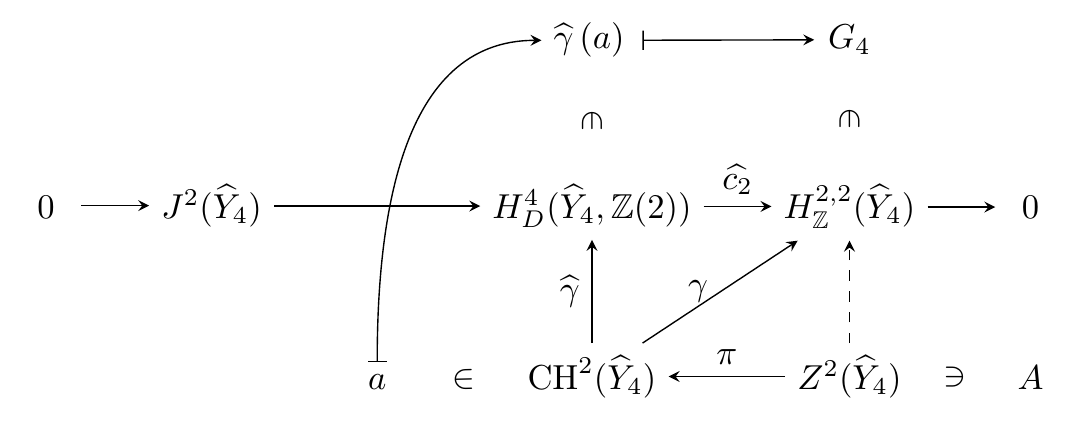}
\caption[Algebraic cycles $A \in \mathrm{CH}^2 ( \hat{Y}_4 )$ encode gauge backgrounds in $H^4_D ( \hat{Y}_4, \mathbb{Z} ( 2 ) )$.]{Summary on how (classes of) algebraic cycles $A \in \mathrm{Z}^2 ( \hat{Y}_4 )$ encode gauge backgrounds in $H^4_D ( \hat{Y}_4, \mathbb{Z} ( 2 ) )$.}
\label{fig:C3BackgroundFromChow2}
\end{figure}

Our approach to gauge backgrounds in \emph{F-theory} via Chow groups is summarised in the commutative diagram in \cref{fig:C3BackgroundFromChow2}. The compatibility of the refined cycle map $\hat{\gamma}$ with intersections in the Chow ring, as explained in \cref{sec:F-theory} and with the pushforward under the fibration $\pi \colon \hat{Y}_4 \twoheadrightarrow \mathcal{B}_6$ ensure that our procedure correctly recovers the information both about the first Chern class and about the flat holonomies of the line bundle on the base.

\subsection{Localised Matter}

For localised matter this formalism has already been carried out in \cite{Bies:2014sra}. Consider a matter surface $S^a_\mathbf{R}$, given by a fibration of rational curves over the matter curve $C_\mathbf{R}$ on the base $\mathcal{B}_6$. Furthermore fix a 2-cycle class $A \in \mathrm{CH}^2(\hat{Y}_4)$ to represent the 3-form background as explained in \cref{sec:F-theory} and summarised in \cref{fig:C3BackgroundFromChow2}. We can then form the intersection product $S^a_\mathbf{R} \cdot_{\iota_{\mathbf{R},a}} A$, loosely speaking by pulling back $A$ to $S^a_\mathbf{R}$ (see \cref{sec:F-theory} for more details). Since the pullback is compatible with rational equivalence and preserves codimension, we can view this as an object in $\mathrm{CH}^2(S^a_\mathbf{R})$, \ie an element of the class of points on $S^a_\mathbf{R}$. Integration along the fibre, as motivated above, thus corresponds to projection onto the base. This leads us to consider the object
\[ D \left( S^a_\mathbf{R}, A\right) := \pi_{\mathbf{R} \ast} (S^a_\mathbf{R} \cdot_{\iota_{\mathbf{R},a}} A) \in \mathrm{CH}_0(C_\mathbf{R}) \,. \label{projectionformula1} \]
Here $\pi_{\mathbf{R},a}$ denotes the projection from the surface $S^a_\mathbf{R}$ to $C_\mathbf{R}$. We have therefore obtained a Chow class of points on $C_{\mathbf{R}}$. This defines a line bundle $\mathcal{O}_{C_\mathbf{R}} ( D ( S^a_\mathbf{R}, A ) )$ on $C_\mathbf{R}$ because
\[ \mathrm{CH}_0(C_\mathbf{R}) \simeq \mathrm{CH}^1(C_\mathbf{R}) \simeq \mathrm{Pic}(C_\mathbf{R}) \, . \]
By construction, the quantum excitations of the M2-brane wrapping the fibre of $S^a_\mathbf{R}$ couple to this very line bundle in the presence of the gauge background $A \in \mathrm{CH}^2(\hat{Y}_4)$. If the gauge background preserves the non-Abelian gauge symmetry in the \emph{F-theory} limit, then for fixed $\mathbf{R}$ this construction gives the same line bundle for each choice of $S^a_\mathbf{R}$, $a= 1, \ldots, \mathrm{dim}(\mathbf{R})$. In this case we can omit the index $a$. By comparison with \cref{mattercohoma}, we conclude that the massless chiral matter encoded by $S^a_{\mathbf{R}}$, \ie in the $a$-th state of the representation $\mathbf{R}$, is counted in the presence of the gauge background $A \in \mathrm{CH}^2(\hat{Y}_4)$ by 
\[ H^i \left(C_\mathbf{R}, L \left( S^a_\mathbf{R}, A \right) \right) \,, \qquad \quad L \left( S^a_\mathbf{R}, A \right) = \mathcal{O}_{C_\mathbf{R}} \left( D \left( S^a_\mathbf{R}, A \right) \right) \otimes \sqrt{K_{C_\mathbf{R}}} \, . \label{cohomoswithflux}\]
A formalisation of these steps based on an accurate application of the intersection product within the Chow ring has been given in \cite{Bies:2014sra}.\footnote{
\Cref{cohomoswithflux} counts the exact massless matter spectrum modulo potential spacetime instanton corrections in \emph{F-theory}: For instance, if the matter is charged only under an Abelian gauge group which acquires a St\"uckelberg mass (\cf \cref{subsec:StueckelbergMasses}), then D3/M5-instantons can generate non-perturbative mass terms which are exponentially suppressed by the K\"ahler moduli in the sense of \cite{Blumenhagen:2006xt, Ibanez:2006da, Florea:2006si, Haack:2006cy, Blumenhagen:2009qh}. These corrections are not accounted for by the topological field theory derivation of \cref{cohomoswithflux}.}

\subsection{Bulk Matter}\label{subsec:bulk-global}

In a similar manner, we can now define the line bundles on a stack of 7-branes wrapping $\Delta_I$ to which the bulk matter couples. The fibre of $\hat{Y}_4$ over a generic point on $\Delta_I$ takes the form of a connected sum of rational curves $\mathbb P^1_{i_I}$ intersecting like the nodes of the affine Dynkin diagram of $\mathfrak{g}_I$. The fibration of $\mathbb P^1_{i_I}$ over $\Delta_I$ defines the so-called resolution divisor $E_{i_I}$. 

Let us now consider the Chow class $E_{i_I} \in \mathrm{CH}^1(\hat{Y}_4)$ associated with $E_{i_I}$ and fix a gauge background $A \in \mathrm{CH}^2(\hat{Y}_4)$. By considering the intersection product we can pull $A$ back to $E_{i_I}$. This produces an element $E_{i_I} \cdot_{\iota_{i_I}} A \in \mathrm{CH}^2(E_{i_I}) \simeq \mathrm{CH}_1(E_{i_I})$. Integration over the fibre amounts to projection onto the base of $E_{i_I}$ via the pushforward of $\pi_{i_I} \colon E_{i_I} \, {\twoheadrightarrow} \, \Delta_I$. This results in
\[ \pi_{i_I \ast}(E_{i_I} \cdot_{\iota_{i_I}} A) \in \mathrm{CH}_1(\Delta_I) \,. \]
On the complex 2-cycle $\Delta_I$ we have that $\mathrm{CH}_1(\Delta_I) \simeq \mathrm{CH}^1(\Delta_I) \simeq \mathrm{Pic}(\Delta_I)$, which identifies
\[ L_{i_I} = \mathcal{O}\left( \pi_{i_I \ast}(E_{i_I} \cdot_{\iota_{i_I}} A)\right) \]
as the line bundle on $\Delta_I$ induced by the gauge background with structure group $U(1)_{i_I}$. This process is to be repeated for all $E_{i_I}$, $i_I=1, \ldots, \mathrm{rk}(\mathfrak{g}_I)$.
If the structure group $U(1)_I$ of the line bundle $L_I$ presented in \cref{subsec:BulkMatter} is a linear combination $U ( 1 )_I = \sum_{i_I} a_{i_I} \,  U(1)_{i_I}$, then
\[ L_I = \bigotimes_{i_I} L_{i_I}^{a_{i_I}} \in \mathrm{Pic}(\Delta_I) \,. \]

\section{A Physical `Classification' of Gauge Backgrounds} \label{sec:SystGaugeBack}

So far we have established that elements $A \in \mathrm{CH}^2(\hat{Y}_4)$ describe gauge backgrounds in \emph{F-theory}. Of course, different such algebraic cycles come with different properties. In particular, this effects the physical significance of such cycles. Therefore, we wish to characterise these algebraic cycles in this section.

A first systematics of such gauge background comes from looking at the cohomology class $G_4 := [A] \in H^{2,2}(\hat{Y}_4)$ which defines a 4-form flux on $\hat{Y}_4$. The cohomology group $H^{2,2}(\hat{Y}_4)$ enjoys a decomposition into three orthogonal subspaces
\[ H^{2,2}(\hat{Y}_4,\mathbb R) =  H^{2,2}_\mathrm{vert}(\hat{Y}_4,\mathbb R)  \oplus H^{2,2}_\mathrm{hor}(\hat{Y}_4,\mathbb R)  \oplus H^{2,2}_\mathrm{rem}(\hat{Y}_4,\mathbb R) \,. \]
This leads to the following systematics of gauge backgrounds:
\begin{itemize}
 \item Vertical fluxes: \\
      If $G_4$ is contained in the primary vertical subspace $H^{2,2}_\mathrm{vert}(\hat{Y}_4,\mathbb R)$, then $A$ is termed a \emph{vertical flux}. 
      $H^{2,2}_\mathrm{vert}(\hat{Y}_4,\mathbb R)$ is generated by elements of the form $H^{1,1}(\hat{Y}_4) \wedge H^{1,1}(\hat{Y}_4)$. An example of such a vertical flux is the so-called $U(1)_X$-flux introduced in \cite{Krause:2011xj, oai:arXiv.org:1202.3138}. The computation of its zero modes was discussed in \cite{Bies:2014sra} already. We will explain this very result in \cref{sec:ToricFTheoryGUTModels} in more detail.
 \item Horizontal fluxes: \\
      If $G_4$ is an element of the primary horizontal subspace $H^{2,2}_\mathrm{hor}(\hat{Y}_4,\mathbb R)$, then $A$ is termed a \emph{horizontal flux}. 
      $H^{2,2}_\mathrm{hor}(\hat{Y}_4,\mathbb R)$ is the subspace of $H^{2,2}(\hat{Y}_4)$ obtained by variation of Hodge structure starting from the unique $(4,0)$ form \cite{Greene:1993vm}. Such fluxes are much harder to exemplify than the vertical fluxes above.
 \item Remainder fluxes: \\
      Finally, $H^{2,2}_\mathrm{rem}(\hat{Y}_4,\mathbb R)$ denotes the remainder piece which is neither vertical nor horizontal \cite{Braun:2014xka}. We name such 
      fluxes \emph{remainder fluxes}. A particularly prominent example of such a remainder flux is the hypercharge flux in \emph{F-theory} GUT-models \cite{Donagi:2008kj,Mayrhofer:2013ara,Braun:2014pva,Beasley:2008kw}. We will study this very flux in much detail in \cref{chapter:GUTModels}.
\end{itemize}

Of course we can also enforce physical conditions such as transversality and gauge invariance to classify gauge backgrounds. Usually these conditions are phrased in terms of the 4-form $G_4$ and not the gauge background $A \in \mathrm{CH}^2 ( \hat{Y}_4 )$. Consequently, we revisit these conditions in \cref{subsec:GaugeInvTrans} and replace them by constraints directly at the level of algebraic cycles. Vertical gauge backgrounds allow fir a finer classification, on which we elaborate in \cref{subsec:systematicsvertical}. In particular, we introduce a special family of vertical fluxes, which we term \emph{matter surface fluxes}. Except for the hypercharge flux and the $U(1)_X$-flux mentioned above, these are the fluxes which we investigate in detail in this thesis. For vertical fluxes one can specialise the strategy of \cref{sec:FromC3toL}, which leads to quite an intuitive idea about the relevant intersection theoretic operations. We explain this approach in \cref{subsec:NewStrategyForMasslessSpectra}.

\subsection{Gauge Invariance and Transversality} \label{subsec:GaugeInvTrans}

\paragraph{Gauge Invariance}

The condition for a flux $G_4 \in H^{2,2}_{\mathrm{vert}}(\hat{Y}_4)$ not to break non-Abelian gauge symmetries in \emph{F-theory} is typically formulated in the literature as
\cite{Borchmann:2013hta}
\[ G_4 \cdot [E_{i_I}] \cdot [D^\mathbf{b}_\alpha] = 0 \qquad \forall \, [D^\mathbf{b}_\alpha] \in H^{1,1}(\mathcal{B}_6) \, . \label{gauge-inv-verta} \]
However, this condition is not sufficient to ensure gauge invariance of a gauge flux in $H^{2,2}_{\mathrm{rem}}(\hat{Y}_4)$: Due to the orthogonality of $H^{2,2}_{\mathrm{rem}}(\hat{Y}_4)$ and $H^{2,2}_{\mathrm{vert}}(\hat{Y}_4)$ any flux in $H^{2,2}_{\mathrm{rem}}(\hat{Y}_4)$ satisfies \cref{gauge-inv-verta} even though it might break the non-Abelian gauge group on a 7-brane. A prime example of such a flux in $H^{2,2}_{\mathrm{rem}}(\hat{Y}_4)$ is the above-mentioned hypercharge flux.

Based on this observation alone, we are lead to alter \cref{gauge-inv-verta}. We wish to replace this demand by a condition which involves the gauge background $A \in \mathrm{CH}^2 ( \hat{Y}_4 )$. From the perspective of the topologically twisted theory on the 7-brane $\Delta_I$, the condition for gauge invariance is that the gauge bundle embedded into the structure group of the non-Abelian group should be the trivial bundle on $\Delta_I$. Hence, we impose
\[ \pi_{i_I\ast} \left(  E_{i_I} \cdot_{i_I} A  \right) = 0 \in \mathrm{CH}_1(\Delta_I) \simeq \mathrm{CH}^1(\Delta_I) \qquad \forall \, E_{i_I} \, . \label{condition-gauge-weaker} \]

\paragraph{Transversality}

The condition for $G_4 \in H^{2,2}(\hat{Y}_4)$ to descend to a well-defined gauge flux in \emph{F-theory} is typically formulated as the transversality conditions \cite{Borchmann:2013hta, oai:arXiv.org:1202.3138}
\begin{align}
G_4 \cdot [D^\mathbf{b}_\alpha] \cdot [D^\mathbf{b}_\beta] &= 0 \qquad \forall \, [D^\mathbf{b}_\alpha], [D^\mathbf{b}_\beta] \in H^{1,1}(\mathcal{B}_6) \, ,  \label{verticality-hom-1} \\
G_4 \cdot [D^\mathbf{b}_\alpha] \cdot [S_0] &= 0 \qquad \forall \, [D^\mathbf{b}_\alpha] \in H^{1,1}(\mathcal{B}_6) \, . \label{verticality-hom-2}
\end{align}
One possible derivation of these constraints is via the observation that they are equivalent to the absence of certain Chern-Simons-terms in the dual 3-dimensional \emph{M-theory} vacuum of the form $\int_{\mathcal{M}_{1,2}}{A_\alpha \wedge F_\beta}$ and $\int_{\mathcal{M}_{1,2}}{A_\alpha \wedge F_0}$ respectively. Here $A_\alpha$ and its field strength refer to the $h^{1,1}(\mathcal{B}_6)$ vector multiplets associated with the K\"ahler moduli of the base $\mathcal{B}_6$ in the 3-dimensional $\mathcal{N}=2$ theory, and $A_0$, $F_0$ refer to the vector multiplets associated with the Kaluza-Klein $U(1)$ associated with circle reduction of the 4-dimensional \emph{F-theory} to the 3-dimensional \emph{M-theory} \cite{oai:arXiv.org:1111.1232}. Now, since these Chern-Simons terms are not generated at one loop in the transition from the 4-dimensional \emph{F-theory} compactification to the 3-dimensional \emph{M-theory} vacuum, and they do not descend from classical terms in four dimensions upon circle reduction either, they must be absent in the \emph{M-theory} effective action in order for the 3-dimensional vacuum to lift to a Poincar\'e invariant \emph{F-theory} vacuum \cite{oai:arXiv.org:1111.1232}.

While conceptually very clear, this derivation is only sensitive to the intersection product in cohomology. In particular, these conditions are again trivially satisfied by a gauge flux $G_4$ not in $H^{2,2}_{\mathrm{vert}}(\hat{Y}_4)$. An alternative derivation of condition \cref{verticality-hom-1} is to require that $G_4$ do not affect the chirality of states wrapping the full fibre as these correspond to the higher KK modes in \emph{M-theory}. Their chirality must therefore equal that of the zero-mode in order for the 4-dimensional \emph{F-theory} vacuum to be Lorentz invariant. In particular, consider a matter surface $S^a(\mathbf{R})$ over a curve $C_\mathbf{R}$ in the base. The surface $S^a_n(\mathbf{R})$ is defined by adding $n$ multiples of the full fibre $F$ to the fibral curves over $C_\mathbf{R}$. M2-branes wrapping the fibre of $S^a_n(\mathbf{R})$ correspond to the $n$-th KK state associated with the 4-dimensional \emph{F-theory} multiplet with weight $\beta^a(\mathbf{R})$. The condition \cref{verticality-hom-1} guarantees that
\[ G_4 \cdot [S^a_n(\mathbf{R})] = G_4 \cdot [S^a(\mathbf{R})]   \qquad \forall \, n \, . \]
Hence, the chirality of the 3-dimensional spectrum of KK modes with weight $\beta^a(\mathbf{R})$ equals that of the KK zero-mode in the 3-dimensional $\mathcal{N}=2$ \emph{M-theory} vacuum.\footnote{Clearly there is no intrinsic notion of chirality in three dimensions in the sense of Weyl spinors. However, the 3-dimensional $\mathcal{N}=2$ theories under consideration here lift to 4-dimensional $\mathcal{N}=1$ theories in the \emph{F-theory} limit. Such a 3-dimensional $\mathcal{N}=2$ theory can be defined to be vector-like if in the lifted theory the number of 4-dimensional $\mathcal{N}=1$ chiral multiplets in representation $\mathbf{R}$ and $\mathbf{\overline{R}}$ coincide, and chiral otherwise.}

However, once we specify the gauge background beyond its flux, we require not only that the net chirality of the 3-dimensional spectrum of KK modes should agree, but rather that the exact number of 3-dimensional KK states in representation $\mathbf{R}$ and $ \mathbf{\overline{R}}$ must match that of the zero modes. This is guaranteed if $\pi_*( S^a_n(\mathbf{R}) \cdot A)$ is independent of the index $n$, \ie \footnote{Here and in the sequel, whenever the precise meaning is clear by the context,  we abbreviate the intersection product within the Chow ring simply by $\cdot$ and the projection just by $\pi$ to avoid too heavy notation.}
\[ \pi_*( S^a_n(\mathbf{R}) \cdot A) = \pi_*( S^a(\mathbf{R}) \cdot A ) \qquad \forall n \, . \]
Given the construction of $S^a_n(\mathbf{R})$ described above, a sufficient condition generalizing \cref{verticality-hom-1} is to require that
\[ \pi_* \left( \pi^{-1} \left( C \right) \cdot A \right) = 0 \qquad \forall \, C \in \mathrm{CH}_1 \left( \mathcal{B}_6 \right) \, . \label{TransverseChow1} \]
From a conceptual point of view we therefore propose this condition to replace \cref{verticality-hom-1} at the level of Chow groups. It would be interesting to investigate if there exist gauge backgrounds whose associated flux satisfies \cref{verticality-hom-1} even though its underlying Chow class violates \cref{TransverseChow1}.

Less clear is the correct interpretation of \cref{verticality-hom-2} at the level of Chow groups. In view of \cref{condition-gauge-weaker} and \cref{TransverseChow1}, a natural generalisation would be to require 
\[ \pi_*( S_0 \cdot A  ) = 0 \, . \label{TransverseChow2} \]
We leave it for future investigations to determine if there are any non-trivial examples of gauge backgrounds which distinguish between \cref{TransverseChow2} and \cref{verticality-hom-2}. For the gauge backgrounds considered in this thesis, both conditions lead to equivalent constraints.

\subsection{Systematics of Vertical Gauge Backgrounds} \label{subsec:systematicsvertical}

Except for the hypercharge flux, we will focus in this thesis on gauge backgrounds whose associated $G_4$ flux lies in $H^{2,2}_{\mathrm{vert}}(\hat{Y}_4,\mathbb R)$ and is subject to the transversality conditions \cref{transversality-gen1}.\footnote{For an incomplete list of more recent works on aspects of fluxes in $H^{2,2}_{\mathrm{hor}}(\hat{Y}_4,\mathbb R)$ and the induced superpotential see \eg \cite{Grimm:2009sy, Grimm:2009ef, Alim:2010za, Braun:2011zm, oai:arXiv.org:1203.6662, Bizet:2014uua} and references therein.} These can be classified, on a fibration over a generic base space $\mathcal{B}_6$, as follows:

\paragraph{Matter Surface Flux}

Consider a matter surface and its associated element $S^a_\mathbf{R} \in \mathrm{CH}^2(\hat{Y}_4)$. By construction $G_4=[S^a_\mathbf{R}]$ satisfies \cref{TransverseChow1} and \cref{TransverseChow2}. If we are interested in describing a gauge background which does not break the non-Abelian gauge algebra in the \emph{F-theory} limit, we must modify the Chow class $S^a_\mathbf{R}$ by adding suitable terms of the form
\begin{align}
\begin{split}\label{def-mattersurfaceflux}
A({\mathbf{R}}) &= S^a_\mathbf{R} + \Delta^a(\mathbf{R}) \in \mathrm{CH}^2(\hat{Y}_4) \\
\Delta^a(\mathbf{R}) &= \left(\beta^a({\mathbf{R}})^T_{i_I} \mathfrak{C}^{-1}_{i_I k_K} \right) \, \left. E_{k_K} \right|_{  C_{\mathbf{R}} } \,.
\end{split}
\end{align}
Here $E_{i_I}$ denotes the resolution divisors for the gauge algebra $\mathfrak{g}_I$ and $E_{k_K} |_{  C_{\mathbf{R}} }$ is their restriction to the curve $C_{\mathbf{R}}$. Furthermore, $\mathfrak{C}^{-1}_{i_I k_K} = \delta_{IK} \, \mathfrak{C}^{-1}_{i_I k_I}$ where the matrix $\mathfrak{C}^{-1}_{i_I j_I}$ is related to the inverse of the Cartan matrix $C_{i_I j_I}$ of $\mathfrak{g}_I$. Recall from \cref{DefcalC} that these matrices differ in general, but match for simply-laced Lie algebras.

\Cref{def-mattersurfaceflux} is indeed gauge invariant because the intersection of any $E_{i_I}$ with $S^a(\mathbf{R})$ in the fibre reproduces the entry $\beta^a({\mathbf{R}})_{i_I}$ in the weight vector, and likewise intersection of $E_{i_I}$ with the component $E_{k_K}$ in the fibre reproduces the negative of the corresponding Cartan matrix entry. 

One can convince oneself that the final result after adding the correction terms $\Delta^a(\mathbf{R})$ to $S^a_\mathbf{R}$ is independent of the index $a = 1, \ldots, \mathrm{dim}(\mathbf{R})$, which is why  $A_{\mathbf{R}} $ carries no such index. We term gauge backgrounds of this form \emph{matter surface fluxes}. As long as $[S^a_\mathbf{R}] \in H^{2,2}_{\mathrm{vert}}(\hat{Y}_4)$\footnote{This property is satisfied for most models discussed so far in the literature, an exception being \cite{Braun:2014pva}.}, also $A(\mathbf{R}) \in H^{2,2}_{\mathrm{vert}}(\hat{Y}_4)$. The first example of such a flux has been given, at the level of cohomology, in \cite{Marsano:2011hv}, and this approach to fluxes has been developed systematically in \cite{Borchmann:2013hta}. 

\paragraph{$\mathbf{U(1)}$ Flux}
A second type of vertical gauge background arises in the presence of extra independent rational sections $S_X$. Via the Shioda map, each independent such section is associated with a divisor class $U_X \in \mathrm{CH}^1(\hat{Y}_4)$ such that $C_3 = A_X \wedge [U_X] + \ldots$ defines a $U(1)_A$ gauge potential $A_X$. Given any divisor class ${ F} \in \mathrm{ CH}^1(\mathcal{B}_6)$ the object
\[ A_X ( F) = { F} \cdot U_X \in \mathrm{CH}^2(\hat{Y}_4) \label{AX-general} \]
defines a gauge background which is automatically vertical and respects the non-Abelian gauge algebras. At the level of fluxes these backgrounds have been introduced in \cite{Grimm:2010ez,Krause:2011xj} (see also \cite{Braun:2011zm}).

\paragraph{Cartan Flux}
Third, we can consider the gauge background for the Cartan $U(1)_{i_I}$ by restricting the resolution divisor $E_{i_I}$ to any curve $C \subseteq \Delta_I$. This defines the element
\[ A_{i_I}(C) = \left. E_{i_I}\right|_{C} \in \mathrm{CH}^2(\hat{Y}_4) \, . \label{AiC-def} \]
While automatically vertical, this gauge background clearly breaks the gauge algebra $\mathfrak{g}_I$. If the curve class $[C]$ is in the image of $H^{1,1}(\mathcal{B}_6)$ restricted to $\Delta_I$, the class $[A_{i_I}(C)]$ lies in $H^{2,2}_\mathrm{vert} (\hat{Y}_4)$. Indeed, if we denote by $[D_C] \in H^{1,1}(\mathcal{B}_6)$ the divisor class such that $[C] = [\Delta_I] \cdot [D_C]$, then  $[A_{i_I}(C)] = [D_C] \cdot [E_{i_I}]$. More generally, the curve class $[C] \in H_{2}(\Delta_I)$ might contain components trivial in $H_2(\mathcal{B}_6)$, but this still defines a valid flux \cite{Mayrhofer:2013ara,Braun:2014pva}. In this case, $[A_{i_I}(C)]$ contains a component in $H^{2,2}_{\mathrm{rem}}(\hat{Y}_4)$ \cite{Braun:2014xka}.

Let us stress that these three types of vertical fluxes and their underlying elements in $\mathrm{CH}^2(\hat{Y}_4)$ are the ones which exist for a generic choice of base $\mathcal{B}_6$. However, they are not all independent due to a number of non-trivial relations within the Chow group, which descend to corresponding relations in $H^{2,2}(\hat{Y}_4)$ via the cycle map. In particular, in \cref{chapter:LocalAnomaliesInF-Theory} we prove that anomaly cancellation implies that the matter surface fluxes satisfy a set of linear relations in $H^{2,2}(\hat{Y}_4)$, thereby generalising previous observations in \cite{Lin:2016vus}. We furthermore exemplify that this relation holds at the level of the Chow group, and conjecture this to hold true more generally.

An equivalent approach to classifying gauge backgrounds with $G_4 \in H^{2,2}_{\mathrm{vert}}(\hat{Y}_4)$ is by systematically forming all intersections of two elements in $\mathrm{CH}^1(\hat{Y}_4)$ and determining the linearly independent combinations whose cohomology classes satisfy verticality and gauge invariance. This approach requires finding a basis of $H^{2,2}_{\mathrm{vert}}(\hat{Y}_4)$ by analysing the relations in the intersection ring. The first systematics of such a classification for \emph{F-theory} fibrations over general base has been carried out in \cite{oai:arXiv.org:1202.3138}. This approach is widely used in the literature, including \cite{Krause:2011xj, oai:arXiv.org:1111.1232, Cvetic:2013uta, Bizet:2014uua, Cvetic:2015txa, Lin:2015qsa, Lin:2016vus}.

\subsection{Intersection Theory for Vertical Gauge Backgrounds} \label{subsec:NewStrategyForMasslessSpectra}

We now describe an approach to perform the programme outlined in \cref{sec:FromC3toL} for vertical gauge fluxes. For such fluxes transverse intersections factorise into a piece in the fibre and a piece in the base, and the projection onto the base is easily evaluated. Non-transverse intersections, on the other hand, can be expressed as sums or differences of transverse intersections by making use of the relations \emph{within the Chow group} between the Chow classes representing the relevant gauge background. It is here where the formalism of \cref{sec:TheChowRing} becomes most crucial: The fact that the refined cycle map \cref{refcyclemap} is defined at the level of the Chow groups guarantees that using such relations within $\mathrm{CH}^2(\hat{Y}_4)$ does not alter the gauge background and is hence permissible in evaluating the intersections.

\begin{figure}[tbp]
\centering
\includegraphics{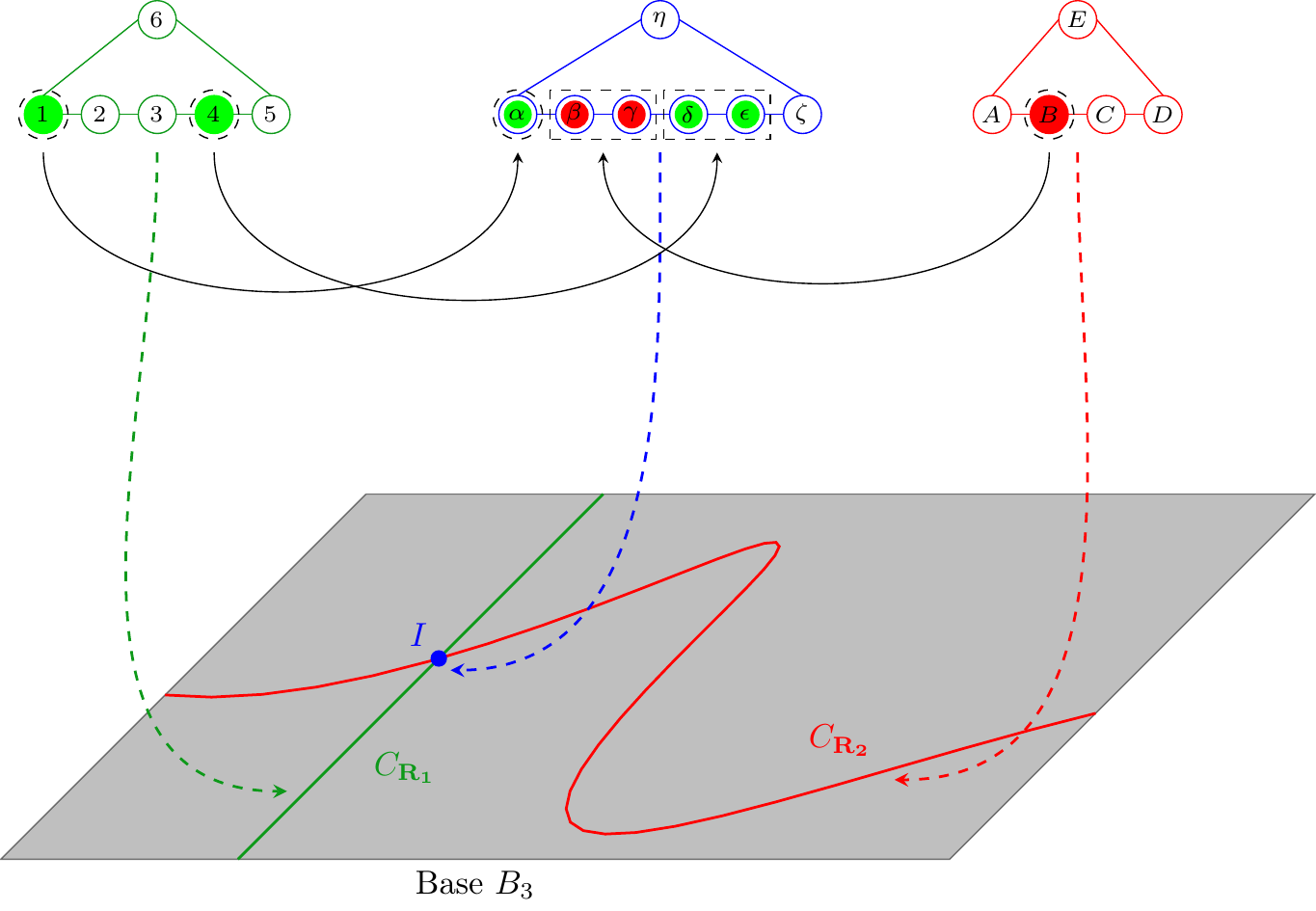}
\caption{Schematics of the identification of massless spectra of matter surface fluxes.}
\label{figure-0}
\end{figure}

\paragraph{Strategy for matter surface fluxes} For a matter surface flux $A$ we proceed as follows:
\begin{enumerate}
 \item Matter surface flux: \\
      Consider a matter curve $C_{\mathbf{R_1}} \subseteq \mathcal{B}_6$, a matter surface $S^a(\mathbf{R_1})$ and the associated matter surface flux $A(\mathbf{R_1}) \in \mathrm{CH}^2(\hat{Y}_4)$ which is given by a formal linear sum of $\mathbb{P}^1$-fibrations over $C_{\mathbf{R_1}}$. The latter is depicted in green colour in \cref{figure-0}.
 \item State: \\
      Next consider a state over the matter curve $C_{\mathbf{R_2}}$. Such a state is encoded by a matter surface $S^{a}_{\mathbf{R_2}}$, whose fibre structure 
      is indicated in red colour in \cref{figure-0}.
 \item 'Intersection' of $S^{a}_{\mathbf{R_2}}$ and $A(\mathbf{R_1})$: \\
      We assume that the curves $C_{\mathbf{R_1}}$ and $C_{\mathbf{R_2}}$ intersect \emph{transversely} in $\mathcal{B}_6$, and discuss non-transverse intersections below. For simplicity we assume in \cref{figure-0} that the two curves intersect in one point $I$ only, with multiplicity $m ( C_{\mathbf{R_1}} \cap C_{\mathbf{R_2}}, I )$.\footnote{Note that in general multiple such intersection points will exist. Then our strategy has to be repeated for every intersection point.} From knowledge of the splittings in the fibre we have
      \[ \left. A (\mathbf{R_1}) \right|_{I} = \alpha + \delta + \epsilon, \qquad \left. S^a_{{\mathbf{R_2}}} \right|_{I} = \beta + \gamma \, . \]
      Consequently, we can compute the intersection number $n_f$ in the fibre over $I$ as
      \[ n_f \left( I \right) = \left( \left. A (\mathbf{R_1}) \right|_{I} \right) \cdot \left( \left. S^{a}_{\mathbf{R_2}} \right|_{I} \right) = \left( \alpha + \delta + \epsilon \right) \left( \beta + \gamma \right) \, . \label{MSFprojform-transverse} \]
      Based on this we can form the divisor $D\left(S^a_{\mathbf{R_2}}, A(\mathbf{R_1})\right) \in \mathrm{CH}^1 \left( C_{\mathbf{R_2}} \right)$ given by
      \[ D\left(S^a_{\mathbf{R_2}}, A(\mathbf{R_1})\right) = \pi_{\mathbf{R_2} \ast} \left(S^a_{\mathbf{R_2}} \cdot_{\iota_{\mathbf{R_2},a}} A(\mathbf{R_1}) \right) 
      = \underbrace{n_f \left( I \right) \cdot m \left( C_{\mathbf{R_1}} \cap C_{\mathbf{R_2}}, I \right)}_{:= m_I } \cdot I  \label{intersectionMSF} 
      \]
      and its associated line bundle $L\left(S^a_{\mathbf{R_2}}, A(\mathbf{R_1}) \right) = \mathcal{O}_{C_{\mathbf{R_2}}} \left( m_I \cdot I \right)$. The sheaf cohomologies 
      \[ H^i\left(C_{\mathbf{R_2}},  L\left(S^a_\mathbf{R_2}, A(\mathbf{R_1})\right)  \otimes \sqrt{K_{C_{\mathbf{R_2}}}}\right) \]
      now count the massless excitations of the state $S^a_{\mathbf{R_2}}$ in the presence of $A(\mathbf{R_1})$.
 \item 'Non-Transverse-Intersections': \\
      If the curves $C_{\mathbf{R_1}}$ and $C_\mathbf{R_2}$ do not intersect transversely, then we use a different representant of the Chow class $A( \mathbf{R}_1 )$, for which the relevant intersections are indeed transverse. We will demonstrate this explicitly in \cref{subsec:ExampleMasslessSpectrumOfUniversalFlux}, where this replacement is achieved by a particular set of linear relations.
\end{enumerate}

\paragraph{Strategy for $\mathbf{U(1)_X}$-fluxes} For a $U(1)_X$ gauge background described by $A_X(F) = F \cdot U_X$, a very similar logic has already been applied in \cite{Bies:2014sra} to deduce the relevant line bundles on the base. Consider a state in representation $\mathbf{R}$ with $U(1)_X$ charge $q_X(\mathbf{R})$. This state is associated with a matter surface $S_\mathbf{R} $ over a curve $C_\mathbf{R}$ on the base $\mathcal{B}_6$.\footnote{Note that we are suppressing the superscript $a$ in $S_\mathbf{R}$ since the $U(1)_X$ background is gauge invariant.} We are also introducing the notation
\[ \iota_{C_\mathbf{R}}: C_{\mathbf{R}} \hookrightarrow \mathcal{B}_6 \]
for the embedding of the curve $C_\mathbf{R}$ into $\mathcal{B}_6$. Then the line bundle $L ( S_\mathbf{R},A_X(F) )$ on $C_\mathbf{R}$ to which the zero modes couple is associated to the divisor
\[ D \left( S_\mathbf{R},A_X \left( F \right) \right) = \pi_{\mathbf{R} \ast} \left( S_\mathbf{R} \cdot_{\iota_{\mathbf{R}}} A_X \left( F \right) \right) = q_X \left( \mathbf{R} \right) \left( C_\mathbf{R} \cdot_{\iota_{C_\mathbf{R}}} F \right) \in \mathrm{CH}^1 \left( C_\mathbf{R} \right) \, . \label{U1Xint-gen} \]
The number of zero modes is counted by $h^i ( C_\mathbf{R}, L ( S_\mathbf{R},A_X(F) ) \otimes \sqrt{K_{C_\mathbf{R}}})$. This result follows from the fact that, as in the construction above, the intersection between the gauge background $A_X(F)$ and the matter surface $S_\mathbf{R}$ factors into an intersection in the fibre and in the base. Projecting onto $C_\mathbf{R}$ gives a multiplicity, which by construction of $U_X$ is exactly the $U(1)_X$ charge $q_X ( \mathbf{R} )$. The term in brackets in \cref{U1Xint-gen} describes the intersection on the base.

\paragraph{Strategy for Cartan fluxes} As a somewhat special case this reasoning can be applied to the Cartan gauge backgrounds \cref{AiC-def}, where we now invoke the embedding of the matter curve $\iota_{C_\mathbf{R}}: C_\mathbf{R} \hookrightarrow \Delta_I$ directly into the divisor $\Delta_I$ wrapped by the 7-brane stack to obtain
\[ \pi_{\mathbf{R} \ast} \left(S^a_\mathbf{R} \cdot_{\iota_{\mathbf{R}}} E_{i_I}|_C \right) = \beta^a(\mathbf{R})_{i_I} \left( C_\mathbf{R} \cdot_{\iota_{C_\mathbf{R}}} C  \right) \in \mathrm{CH}^1(C_\mathbf{R}) \, . \label{Cartanfluxintersection} \]
The weight $\beta^a(\mathbf{R})_{i_I}$ is of course by definition the charge of the state under the Cartan $U(1)_{i_I}$.

\paragraph{Alternative strategy} All of the above approaches focus on the geometric interpretation of the Chow group elements, which are used to model the gauge background in question. Alternatively one can try to perform the intersections in the Chow ring directly and project the result down onto $\mathcal{B}_6$. In general it is very hard to compute intersections in the Chow ring. This task simplifies significantly if the gauge backgrounds $A \in \mathrm{CH}^2 ( \hat{Y}_4 )$ in question can be expressed as pullbacks of algebraic cycles $\mathcal{A} \in Z^2 ( X_\Sigma )$ on a simplicial toric ambient space $X_\Sigma$. Namely, as explained in \cref{subsec:TowardsToricVarieties}, we can then alter $\mathcal{A}$ by homological relations without changing the Chow class of $\mathcal{A}$. These modifications can be used to simplify the relevant computations enormously. We will demonstrate this in the example geometry studied in the next section. More details of these computations are listed in \cref{subsec:MasslessSpectraMSFSU5xU1}.

\section{Tools for Computational \emph{F-theory} Vacua} \label{sec:ToolsForCompuationalF-theoryVacua}

Thus far, we have connected the zero modes of \emph{F-theory} vacua to sheaf cohomologies of line bundles on the matter curves $C_{\mathbf{R}}$. As our motivation of study derives from phenomenology, we intend to compute these sheaf cohomologies in explicit examples of \emph{F-theory} compactifications. Of course we do not wish to perform these computations for general varieties -- this would be far too much a challenge. Rather we look for a family of varieties, which admits enough structure to allow for corresponding computer implementations.

Toric varieties feature prominently in physics. For example, Edward Witten introduced in the context of heterotic \emph{string theory} a gauged linear sigma model (GLSM), whose vacuum configurations are given toric varieties \cite{Witten:1993yc}. See also \cite{RahnPhd} for a neat exposition of this material. Another application of toric varieties comes from studying moduli stabilisation in type II \emph{string theory} as outlined in \cite{Larfors:2010wb}. So maybe, toric varieties can help in \emph{F-theory} compactifications also? Indeed, the Chow ring of toric varieties is typically accessible \cite{cox2011toric} and the intersection theory is well enough understood to allow for computer implementations \eg in \emph{Sage} \cite{sage}.

Yet another attractive feature of toric varieties derives from Chow's theorem. To see this, let us first recall that an \emph{analytic variety} $V \subset U$ of an open subset $U$ of $\mathbb{C}^n$ is such that any $p \in U$ admits an open neighbourhood $p \in U^\prime$ such that $V \cap U^\prime$ is the common zero locus of a finite collection of holomorphic functions $f_1, \dots, f_k$ on $U^\prime$ \cite{griffiths2011principles}. The number of holomorphic functions, which cut out this analytic variety need not be constant. If it is, one terms $V$ a \emph{pure-dimensional variety}. This shows that analytic varieties form a very big class of complex spaces. For the complex projective space $\mathbb{P}^n_{\mathbb{C}}$ the situation is different. For homogeneous polynomials $Q_1, \dots, Q_n$ we can consider the \emph{algebraic variety}
\[ V = \left\{ \left. p \in \mathbb{P}^n_{\mathbb{C}} \; \right| \; Q_1 \left( p \right) = \ldots = Q_n \left( p \right) = 0 \right\} \, \]
and Chow's theorem tells us that every analytic subvariety of $\mathbb{P}^n_{\mathbb{C}}$ -- which is closed in the analytic topology of $\mathbb{P}^n_{\mathbb{C}}$ -- is of this form \cite{ChowsTheorem}. A major simplification!

This all motivates the use of toric varieties for phenomenological purposes in \emph{F-theory}. It is therefore quite common to engineer a toric ambient space $\hat{Y}_\Sigma$ of $\hat{Y}_4$ and to realise $\hat{Y}_4$ as complete intersection in $\hat{Y}_\Sigma$. Often $\hat{Y}_\Sigma$ will be 5-dimensional and therefore $\hat{Y}_4$ be realised as hypersurface therein.\footnote{We reserve the symbol $\hat{Y}_5$ for 5-dimensional ambient spaces of $\hat{Y}_4$, which need not be toric.} Likewise a toric ambient space $X_\Sigma$ of $\mathcal{B}_6$ can be used to model $\mathcal{B}_6$ and even the matter curves $C_{\mathbf{R}}$ as complete intersections therein.

In such a setup, let us now come back to the zero mode counting. We mentioned already that this is equivalent to computing the sheaf cohomologies of certain line bundles $L_{\mathbf{R}}$ on the matter curves $C_{\mathbf{R}}$. We can thus wonder if we can relate $L_{\mathbf{R}} \in \mathrm{Pic} ( C_{\mathbf{R}} )$ to some data on the surrounding base toric ambient space $X_\Sigma$. In the next section we will find that in general the line bundles in question are not simply pullbacks of line bundles from $X_\Sigma$. Hence, we cannot simply apply the \emph{cohomCalg}-algorithm \cite{Blumenhagen:2010pv, Blumenhagen:2010ed, Blumenhagen:2011xn, KoszulExtensionManual, cohomCalg:Implementation, 2011JMP....52c3506J, Rahn:2010fm} to count zero modes in \emph{F-theory} vacua. Our approach is therefore to simply extend the lines bundles $L_{\mathbf{R}}$ by zero outside of $C_{\mathbf{R}}$. This gives rise to a so-called \emph{coherent sheaf} on $X_\Sigma$. We are thus lead to wonder if we can model all coherent sheaves on toric varieties, and if we can compute their sheaf cohomologies. 

Indeed this is possible, and we will explain the details in \cref{chapter:DetailsOnFPGradedSModules}. To prepare us for this, and to perform a case study on \emph{F-theory} compactifications derived from toric geometry in \cref{sec:ToricFTheoryGUTModels}, we now revise the necessary material. In \cref{subsec:CoherentSheavesOnVarieties} we revise affine varieties -- in contrast to \cref{subsec:AffineVarietiesBasics} however in the lingo of affine schemes. In particular, we will explain how modules give rise to coherent sheaves on such varieties. For toric varieties the story will be quite similar -- so-called \emph{finitely-presented, graded} modules will then give rise to coherent sheaves. To this we turn in \cref{chapter:DetailsOnFPGradedSModules}, and suffice it to revise basics of toric varieties in \cref{subsec:TowardsToricVarieties}.

\subsection{Coherent Sheaves on Varieties} \label{subsec:CoherentSheavesOnVarieties}

An affine variety can, with a view towards algebraic geometry, be seen as a topological space associated to a (commutative and unitial) ring $R$. As it turns out, there is natural way to turn an $R$-module $M$ into a sheaf on this space. This process is termed \emph{sheafification}. In \cref{chapter:DetailsOnFPGradedSModules} it will turn out that the story is not so very different for toric varieties. Therefore, let us revise the sheafification of $R$-modules on affine varieties in this section. A central ingredient in this construction is the so-called \emph{localisation of rings}, which we begin with.

\paragraph{Localisation of Rings}

We consider a commutative unitial ring $R$ and a multiplicatively closed subset $1 \in S \subseteq R$. We now construct a new ring $R_S$ from this data. To this end, define the following equivalence relation on $R \times S$:
\[ \left( r_1, s_1 \right) \sim \left( r_2, s_2 \right) \; \Leftrightarrow \; \exists t \in S \colon t \cdot \left( r_1 s_2 - r_2 s_1 \right) = 0 \, .\]
Then consider the equivalence classes
\[ \frac{r_1}{s_1} := \left[ \left( r_1, s_1 \right) \right] := \left\{ \left. \left( r_2, s_2 \right) \in R \times S \; \right| \; \left( r_1, s_1 \right) \sim \left( r_2, s_2 \right) \right\} \]
and denote the collection of all these equivalence classes by $S^{-1} R \equiv R_S$. It is readily verified that the binary compositions
\begin{align}
\begin{split}
+ \colon R_S \times R_S \to R_S \; , \; & \left( \frac{r_1}{s_1}, \frac{r_2}{s_2} \right) \mapsto \frac{r_1 s_2 + r_2 s_1}{s_1 s_2}\, , \\
\cdot \colon R_S \times R_S \to R_S \; , \; & \left( \frac{r_1}{s_1}, \frac{r_2}{s_2} \right) \mapsto \frac{r_1 r_2}{s_1 s_2}
\end{split}
\end{align}
turn this set $R_S$ into a ring. This ring $R_S$ is termed the \emph{localisation of $R$ at $S$}.

Often one localises rings at prime ideals $\mathfrak{p} \subseteq R$. Recall that a prime ideal $\mathfrak{p} \subseteq R$ in a commutative ring is a proper ideal (\ie $\mathfrak{p} \neq R$) such that for all $a,b \in R$ with $a b \in \mathfrak{p}$ it holds $a \in \mathfrak{p}$ or $b \in \mathfrak{p}$. In particular, $1 \notin \mathfrak{p}$ for otherwise $\mathfrak{p} = R$ in contradiction to $\mathfrak{p}$ being proper. However, for the localisation $R_S$ we assumed $1 \in S$! This is the reason why, by convention, localisation at a prime ideal $\mathfrak{p} \subseteq R$ means to localise at the multiplicatively closed set $R - \mathfrak{p}$, \ie
\[ R_{\mathfrak{p}} := \left( R - \mathfrak{p} \right)^{-1} R = R_{R - \mathfrak{p}} \, . \]

Another common type of localisation is at an element $0 \neq f \in R$. By this we mean to form the set $\{ 1, f, f^2, f^3, \dots \}$ and then to localise at this set $S$. The localisation at such a set is denoted by $R_f$. Yet another common situation is to start with a graded ring $R$. Given a multiplicatively closed set $1 \in S$ of homogeneous elements, one can perform the so-called \emph{homogeneous localisation}. To this end, we first define the degree of the equivalence class $\frac{r_1}{r_2}$ as
\[ \mathrm{deg} \left( \frac{r_1}{r_2}\right) = \mathrm{deg} \left( r_1 \right) - \mathrm{deg} \left( r_2 \right) \, . \]
Thereby, one realises that the localisation $R_S$ is a graded ring. The homogeneous localisation $R_{(S)}$ is now defined by $R_{(S)} := \left( R_S \right)_{0}$, \ie $R_{(S)}$ consists of all elements of $R_S$ that have vanishing degree. We will make use of this homogeneous localisation in \cref{chapter:DetailsOnFPGradedSModules}.

\paragraph{Affine Varieties}

Let us recall the notion of an affine variety with a view towards algebraic geometry. To this end, let $R$ be a commutative unitial ring. An ideal $\mathfrak{m} \subseteq R$ is a \emph{maximal ideal} precisely if for every proper ideal $\mathfrak{a} \subseteq R$ with $\mathfrak{m} \subseteq \mathfrak{a}$ it holds $\mathfrak{m} = \mathfrak{a}$.

Whilst every maximal ideal is a prime ideal, the converse is not quite true. To see this consider the ring $k [ x ]$ where $k$ is an algebraically closed field. For every $a \in k$ the ideal
\[ \mathfrak{m}_{a} := \left\langle x - a \right\rangle \subseteq k \left[ x \right] \]
is a maximal ideal, and -- since $k$ is algebraically closed -- all maximal ideals of $k [ x ]$ are of this form. In addition, the field $k$ is free of zero divisors \footnote{This means that for every $a,b \in R - 0$ it holds $a b \neq 0$.} and therefore the trivial ideal $\langle 0 \rangle$ is a prime ideal as well. Clearly this ideal is not maximal since $\langle 0 \rangle \subsetneq \langle x \rangle$!

Let us construct a topological space from $R$. A scheme-theoretic approach would consider
\[ \mathrm{Spec} \left( R \right) = \left\{ \left. \mathfrak{p} \subseteq R \; \right| \; \mathfrak{p} \text{ a prime ideal } \right\} \]
and equip it with the Zariski topology. For an algebraically closed field $k$ this means 
\[ \mathrm{Spec} \left( k \left[ x \right] \right) = \left\{ \left. \left\langle x - a \right\rangle \; \right| \; a \in k \right\} \cup \left\{ \left\langle 0 \right\rangle \right\} \cong k \cup \left\{ \left\langle 0 \right\rangle \right\} \, . \]
So besides containing all `points' of $k$, this affine scheme contains also the so-called \emph{generic point} $\langle 0 \rangle$. Such points distinguish the scheme-theoretic approach from a variety-theoretic approach. Whilst it is indeed possible to formulate all findings in the language of (toric) schemes as presented in \cite{RohrerDissertation, 2011arXiv1107.2483R, 2012arXiv1212.3956R, 2011arXiv1107.2713R}, we follow \cite{cox2011toric} to describe our results in the language of (toric) varieties over $\mathbb{Q}$. Consequently, we consider the set of maximal ideals
\[ \mathrm{Specm} \left( R \right) = \left\{ \left. \mathfrak{p} \subseteq R \; \right| \; \mathfrak{p} \text{ a maximal ideal} \right\} \, . \]
The Zariski topology is defined by saying that for every ideal $\mathfrak{a} \subseteq R$ the following set is closed
\[ V \left( \mathfrak{a} \right) = \left\{ \left. \mathfrak{p} \in \mathrm{Specm} \left( R \right) \; \right| \; \mathfrak{a} \subseteq \mathfrak{p} \right\} \subseteq \mathrm{Specm} \left( R \right) \, . \]
For every $f \in R \backslash \{ 0 \}$ we can consider the open set
\[ D \left( f \right) := \mathrm{Specm} \left( R \right) - V \left( \left\langle f \right\rangle \right) = \left\{ \left. \mathfrak{p} \in \mathrm{Specm} \left( A \right) \; \right| \; f \notin \mathfrak{p} \right\} \, . \]
$\mathcal{Z} := \{ D ( f ) \}_{f \in R \backslash \{ 0 \}}$ is a basis of the Zariski topology. Moreover $D ( f ) \cong \mathrm{Specm} ( R_f )$ where $R_f$ denotes the localisation of the ring $R$ at $f \in R \backslash \{ 0 \}$.

As an example let us consider the ring $\mathbb{C} [ x ]$. Since $\mathbb{C}$ is algebraically closed we have an isomorphism of sets $\mathrm{Specm} ( \mathbb{C} [ x ] ) \cong \mathbb{C}$. Next recall that every $f \in \mathbb{C} [ x ]$ has finitely many zeros. In addition, by the Hilbert Nullstellensatz, every ideal $\mathfrak{a} \subseteq \mathbb{C} [ x ]$ is finitely generated. Consequently, every (Zariski) closed subset of $\mathrm{Specm} ( \mathbb{C} [ x ] )$ is a finite subset.\footnote{Algebraic minds will prefer to reach this conclusion by recalling that $\mathbb{C} [ x ]$ is a principal ideal domain (PID).} Or conversely, an open subset of $\mathrm{Specm} ( \mathbb{C} [ x ] )$ consists of the entire space $\mathrm{Specm} ( \mathbb{C} [ x ] )$ up to finitely many exceptions. Therefore, for any two $\mathfrak{p}, \mathfrak{q} \in \mathrm{Specm} ( \mathbb{C} [ x ] )$, any open neighbourhood $U$ of $\mathfrak{p}$ and any open neighbourhood $V$ of $\mathfrak{q}$ intersect non-trivially, \ie $U \cap V \neq \emptyset$. This shows that the Zariski topology is not Hausdorff in general. For this reason it defies intuition from Euclidean or complex geometry.

\paragraph{$\mathbf{R}$-modules}

A left $R$-module is an Abelian group $(M,+)$ together with a map
\[ R \times M \to M \; , \; ( r, m ) \mapsto r \cdot m \]
such that for all $r_1, r_2 \in R$ and all $m_1, m_2 \in M$ it holds
\[ \begin{array}{rclcrcl}
s_1 \cdot ( s_2 \cdot m_1 ) &=& ( s_1 \cdot s_2 ) \cdot m_1 \, , & & ( s_1 + s_2 ) \cdot m_1 &=& s_1 \cdot m_1 + s_2 \cdot m_1 \\
s_1 \cdot ( m_1 + m_2 ) &=& s_1 \cdot m_1 + s_1 \cdot m_2 \, , & & 1 \cdot m_1 &=& m_1 \, .
\end{array} \]
Consequently, a left $R$-module $M$ looks very much like a `vector space over $R$', except for the fact that $R$ need not be a field. For ease of notation we will drop the term `left' for now (more details in \cref{chapter:DetailsOnFPGradedSModules}). Examples of $R$-module include the ring $R$ itself, ideals $I \subseteq R$ and quotient rings $R / I$.

Let $J$ be an indexing set. Then an $R$-module of the form $M = \bigoplus_{j \in J}{R}$ is called \emph{free}. The indexing set $J$ need not be finite. However, if $J$ is finite, then $| J |$ is termed the \emph{rank of $M$}. The ring $R$ is therefore a free $R$-module of rank $1$.

One of the main differences between $R$-modules and $k$-vector spaces is that $R$-modules need not have a basis. A standard example is the $\mathbb{Z}$-module $\mathbb{Q}$ -- we can argue that any $\mathbb{Z}$-generating set $\mathcal{G}$ of $\mathbb{Q}$ is linearly dependent over $\mathbb{Z}$. First suppose that $\mathcal{G}$ consisted of a single element. Then $\mathrm{Span}_{\mathbb{Z}} ( \mathcal{G} ) \neq \mathbb{Q}$. Hence, $\mathcal{G}$ must contain at least two different elements $a,b \in \mathbb{Q}$. We can write $a = n_1 / d_1$ and $b = n_2 / d_2$ for $n_1, n_2 \in \mathbb{Z}$ and $d_1, d_2 \in \mathbb{Z} \backslash \{ 0 \}$. Consequently, $n_2 d_1 a - n_1 d_2 b = 0$ which shows that indeed $a,b \in \mathbb{Q}$ are linearly dependent over $\mathbb{Z}$.

As an explicit example consider $\mathrm{Span}_{\mathbb{Z}} ( 1, \frac{3}{2} )$. Then note:
\[ \mathrm{Span}_{\mathbb{Z}} ( 1 ) = \{ \textcolor{red}{0}, \pm 1, \pm 2, \pm \textcolor{red}{3}, \pm 4, \dots \} \, , \quad 
\mathrm{Span}_{\mathbb{Z}} \left( \frac{3}{2} \right) = \{ \textcolor{red}{0}, \pm \frac{3}{2}, \pm \textcolor{red}{3}, \pm \frac{9}{2}, \dots \} \, . \]
Hence, $\mathrm{Span}_{\mathbb{Z}} ( 1, \frac{3}{2} ) \not \cong \mathbb{Z}^2$ since the red numbers are contained both in $\mathrm{Span}_{\mathbb{Z}} ( 1 )$ and $\mathrm{Span}_{\mathbb{Z}} ( \frac{3}{2} )$. We can remove this redundancy by removing the relation $0 = 3 \cdot 1 + (-2) \cdot 3/2$ of the generators $1$ and $\frac{3}{2}$. Heuristically we then obtain
\[ \mathrm{Span}_{\mathbb{Z}} \left( 1, \frac{3}{2} \right) / \{ 3 \cdot 1 + (-2) \cdot \frac{3}{2} = 0 \} = \left\{ 0, \pm 1, \pm \frac{3}{2}, \pm 2, \pm 4, \pm \frac{9}{2}, \pm 5, \dots \right\} \cong \mathbb{Z} \, . \]
A clearer way is to consider the cokernel of the following map of $\mathbb{Z}$-modules:
\[ \varphi \colon \mathbb{Z} \xrightarrow{( 3, -2 )} \mathbb{Z}^2 \, . \]
This is a so-called presentation of the `naive' $\mathrm{Span}_{\mathbb{Z}} ( 1, \frac{3}{2} ) / \{ 3 \cdot 1 + (-2) \cdot \frac{3}{2} = 0 \}$.

Cokernels of maps of $R$-modules provide yet more examples of $R$-modules. If source and range of the $R$-module homomorphism in question are both free and of finite rank, then this presentation is termed finite. Modules that can be presented in this way are termed finitely presented (f.p.) $R$-modules. In \cref{chapter:DetailsOnFPGradedSModules} we will investigate the category of \fp \emph{graded} modules in much detail. In particular, we will make use of commutative diagrams to write down module presentations and morphisms between them. In particular, it will become clear that the embedding of the `naive' $\mathrm{Span}_{\mathbb{Z}} ( 1, \frac{3}{2} ) / \{ 3 \cdot 1 + (-2) \cdot \frac{3}{2} = 0 \}$ into the $\mathbb{Z}$-module $\mathbb{Q}$ can be summarised in the following commutative diagram:
\[ \includegraphics[valign = c]{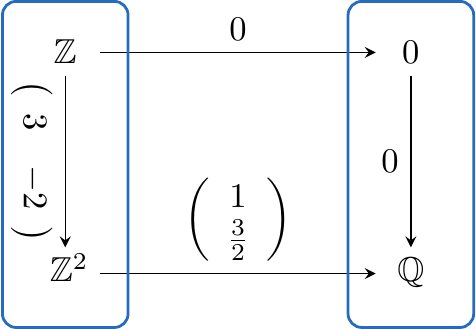} \]
On (affine) varieties finitely presented $R$-modules encode coherent sheaves upon a process called \emph{sheafification}. This we will discuss next.

\paragraph{Sheafification of Modules on Affine Varieties}

The general idea of sheafification on affine varieties is very simple: Given an affine variety $X$ with coordinate ring $R$ and an $R$-module $M$, we define a sheaf $\tilde{M}$ associated to $M$. This sheaf $\tilde{M}$ is defined by stating its (local) sections over the open subsets which form the basis of topology $\mathcal{Z}$. 
These open sets are of the form $D(f) = X - V(\langle f \rangle)$ with $f \in R - 0$, \ie are formed from the total space $X$ minus the vanishing locus of $f$. On these open sets, the sheaf $\tilde{M}$ has the (local) sections given by
\[ \tilde{M} \left( D \left( f \right) \right) = M \otimes_R R_f \equiv M_f \, , \]
where $R_f$ denotes the localisation of $R$ at $f$. Intuitively, $R_f$ consists of all (rational) functions on $D(f)$. Of course one has to verify that these assignments indeed form a sheaf as defined in \cref{sec:RevisionOnSheavesAndSheafCohomology}. Details on this can be found \eg in \cite{hartshorne1977algebraic}. Useful properties of $\tilde{M}$ include:
\begin{itemize}
 \item $H^0 ( X, \tilde{M} ) = M$,
 \item Be $\mathfrak{p} \in \mathrm{Specm} ( R ) = X$. Then the stalk $\tilde{M}_{\mathfrak{p}}$ of $\tilde{M}$ in $\mathfrak{p}$ is isomorphic to the 
      localisation $M_{\mathfrak{p}}$ of the module $M$ at $\mathfrak{p}$.
\end{itemize}
A particular neat example is the ring $R$ itself. It gives rise to the sheaf $\tilde{R}$ with $\tilde{R} \cong \mathcal{O}_{\mathrm{Specm} ( R )}$. So in particular $H^0 ( \mathrm{Specm} ( R ), \mathcal{O}_{\mathrm{Specm} ( R )} ) \cong R$. In addition, since $M \otimes_R R_f$ is an $R_f$-module, it follows that the sheaf $\tilde{M}$ is a sheaf of $\mathcal{O}_{\mathrm{Specm} ( R )}$-modules.

\paragraph{Coherent Sheaves on Varieties}

Let $X$ be a topological space and $\mathcal{O}_X$ a sheaf of rings on $X$. The pair $( X, \mathcal{O}_X )$ is termed a \emph{locally ringed space} if for every $p \in X$ the stalk $\mathcal{O}_{X,p}$ is a local ring.\footnote{This means that the ring $\mathcal{O}_{X,p}$ has a unique maximal ideal.} An abstract variety is a locally ringed space $( X, \mathcal{O}_X )$ such that for every $p \in X$ there exists an open neighbourhood $p \in U \subseteq X$ such that $( U, \left. \mathcal{O}_X \right|_U )$ is isomorphic (as locally ringed space) to $( \mathrm{Specm} ( R ), \tilde{R})$ for a suitable commutative unitial ring $R$.

For a sheaf $\mathcal{F}$ of $\mathcal{O}_X$-modules on a variety $( X, \mathcal{O}_X )$ we define the following notions:
\begin{itemize}
 \item For an open subset $U \subseteq X$, the restriction $\mathcal{F} |_U$ of $\mathcal{F}$ to $U$ is the sheaf of $\mathcal{O}_U$-modules 
      obtained from $ \mathcal{F} |_U ( V ) = \mathcal{F} ( V )$ for every $V \subseteq U$ open.
 \item $\mathcal{F}$ is \emph{quasi-coherent} precisely if $X$ admits an open affine cover \footnote{A variety $( X, \mathcal{O}_X )$ may admit various open affine 
      covers!} $\mathcal{U} = \{ U_\alpha \}_{\alpha \in I}$, \ie
      \[ \left( U_\alpha, \left. \mathcal{O}_X \right|_{U_\alpha} \right) \cong \left( \mathrm{Specm} \left( R_{\alpha} \right), \tilde{R_\alpha} \right) \, , \]
      such that for every $\alpha \in I$ there exists an $R_\alpha$-module $M_\alpha$ with the property $\tilde{M_\alpha} \cong \mathcal{F} |_{U_\alpha}$.
 \item $\mathcal{F}$ is \emph{coherent} if in addition for every $\alpha \in I$ the module $M_\alpha$ is finitely presented.
\end{itemize}

\subsection{Toric Varieties} \label{subsec:TowardsToricVarieties}

\paragraph{Generalities of toric varieties}

Toric varieties are special instances of abstract varieties. The geometry of such a variety is encoded in a fan $\Sigma$. We follow the exposition in \cite{cox2011toric} and first recall the notion of a fan: First, a rational polyhedral cone $\sigma \subseteq \mathbb{R}^n$ is a cone generated by finitely many $p_i \in \mathbb{R}^n$ as
\[ \sigma = \left\{ \alpha_1 p_1 + \dots + \alpha_l p_l \in \mathbb{R}^n \; , \; \alpha_i \geq 0 \right\} = \mathrm{Span}_{\mathbb{R}_{\geq 0}} \left( \left\{ p_1, \dots, p_l \right\} \right) \, . \]
Such a rational polyhedral cone $\sigma$ is strongly convex precisely if $\sigma \cap ( - \sigma ) = \{ 0 \}$. For ease of notation will abbreviate a strongly convex rational polyhedral cone as \emph{scrapc}. The dual cone of $\sigma$ is a cone $\sigma^\vee \subseteq \mathbb{R}^n$ defined by $\sigma^\vee = \{ m \in \mathbb{R}^n \; , \; \langle m, u \rangle \geq 0 \text{ for all } u \in \sigma \}$. In this expression $\langle m,u \rangle$ is the ordinary dot product on $\mathbb{R}^n$.

A particularly neat generating set of a \emph{scrapc} $\sigma$ is constructed as follows: Since $\sigma$ is strongly convex, any edge $\rho$ is a ray. Rationality ensures that the semigroup $\rho \cap \mathbb{Z}^n$ has a unique generator $u_\rho \in \rho \cap \mathbb{Z}^n$. This generator $u_\rho$ is termed \emph{the ray generator} of $\rho$. Collectively, these ray generators $u_\rho$ form a neat generating set of $\sigma$, which we will come to use momentarily.

A fan $\Sigma$ in $\mathbb{R}^n$ is a finite collection of scrapcs in $\mathbb{R}^n$ such that the following is satisfied:
\begin{itemize}
 \item $\sigma \in \Sigma$ and $\tau$ a face of $\sigma$, then $\tau \in \Sigma$.
 \item $\sigma, \tau \in \Sigma$, then $\sigma \cap \tau \in \Sigma$ is a face of $\sigma$ and $\tau$.
\end{itemize}
The support of a fan $| \Sigma |$ is the (set-theoretic) union of all $\sigma \in \Sigma$. 

That all said, let us construct the variety $X_\Sigma$ associated to the fan $\Sigma$. For this construction we will only allow fans $\Sigma$ such that 
$\text{Span}_{\mathbb{Z}} \{ u_\rho \, | \, \rho \in \Sigma ( 1 ) \} = \mathbb{Z}^n$. This ensures that the variety $X_\Sigma$ has no torus factor. The affine patches of $X_\Sigma$ are obtained from associating to $\sigma \in \Sigma$ the affine variety \footnote{In this expression $M$ is the character lattice of the toric variety $X_\Sigma$. It should not be confused with any kind of $S$-module $M$ which we might intend to sheafify.}
\[ U_\sigma \cong \mathrm{Specm} ( \mathbb{Q} [ \sigma^\vee \cap M ] ) \, . \]
These affine varieties indeed glue to form an abstract variety -- the normal toric variety $X_\Sigma$ with class group $\mathrm{Cl} ( X_\Sigma )$.

Next note that each $\sigma \in \Sigma$ has a generating set consisting of its ray generators. This generating set is finite. Since $\Sigma$ is a finite union of scrapcs, the union of all the rays generators of scrapcs $\sigma \in \Sigma$ is finite. Let us label this collection of ray generators as $\{ \rho_1, \dots \rho_m \}$. To each ray generators we associate an indeterminate $x_i$ in the polynomial ring $S = \mathbb{Q} [ x_1, \dots, x_m ]$. This ring $S$ is known as the \emph{Cox ring}. We may assume $x_1 \leftrightarrow \rho_1$, $x_2 \leftrightarrow \rho_2$ and so on. 

To each ray $\rho_i \in \Sigma$, there is an associated divisor $D_{\rho_i} \in \mathrm{Div} ( X_\Sigma )$.\footnote{Upon homogenisation of $X_\Sigma$ one may think of this divisor as the locus $V ( x_i ) \subseteq X_\Sigma$.} Since by assumption $X_\Sigma$ has no torus factor, there exists an exact sequence $0 \to M \cong \mathbb{Z}^n \to \bigoplus_{\rho \in \Sigma ( 1 )}{\mathbb{Z} D_\rho} \xrightarrow{\mathrm{deg}} \mathrm{Cl} ( X_\Sigma ) \to 0$ where
\begin{align}
m \in M \mapsto \mathrm{div} \left( \chi^m \right) = \sum_{\rho \in \Sigma \left( 1 \right)}{\left\langle m, u_\rho \right\rangle \cdot D_\rho} \, .
\end{align}
The divisor $D_{\rho_i}$ is an element of $\bigoplus_{\rho \in \Sigma ( 1 )}{\mathbb{Z} D_\rho}$. Hence, we may apply the map $\mathrm{deg}$ to obtain an element in $\mathrm{Cl} ( X_\Sigma )$. Since $D_{\rho_i}$ is associated to the indeterminate $x_i \in S$, we may thus grade $x_i \in S$ by $\mathrm{deg} ( D_{\rho_i} ) \in \mathrm{Cl} ( X_\Sigma )$. From this it follows that the Cox ring $S = \mathbb{Q} [ x_1, \dots, x_m ]$ is graded under $\mathrm{Cl} ( X_\Sigma )$. For any cone $\sigma \in \Sigma$, we now form the monomial
\[ x^{\hat{\sigma}} := \prod_{\rho \notin \sigma \left( 1 \right)}{x_\rho} \in S \, . \]
In consequence, there exists an isomorphism of rings $\pi_\sigma^\ast \colon \mathbb{Q} [ \sigma^\vee \cap M ] \xrightarrow{\sim} ( S_{\hat{\sigma}} )_{0}$. Crucially, for a face $\tau = \sigma \cap m^\bot$ of $\sigma$ it holds $( S_{x^{\hat{\tau}}} )_0 = ( ( S_{x^{\hat{\sigma}}} )_0 )_{\pi_\sigma^\ast ( \chi^m )}$ and the following diagram commutes:
\[ \centering
\begin{tikzpicture}[baseline=(current  bounding  box.center),>=stealth]

  \matrix (m) [matrix of math nodes,row sep=4em,column sep=4em,minimum width=2em]
  {  \left( S_{x^{\hat{\sigma}}} \right)_0 & \left( \left( S_{x^{\hat{\sigma}}} \right)_0 \right)_{\pi_\sigma^\ast \left( \chi^m \right)} \\
     \mathbb{Q} \left[ \sigma^\vee \cap M \right] & \mathbb{Q} \left[ \tau^\vee \cap M \right]_{\chi^m} \\
  };

  \path[->] (m-1-1.east |- m-1-2) edge node {} (m-1-2);
  \path[->] (m-1-1) edge node {} (m-2-1);
  \path[->] (m-1-2) edge node {} (m-2-2);
  \path[->] (m-2-1.east |- m-2-2) edge node {} (m-2-2);

\end{tikzpicture}
\label{equ:Gluing} \]
Hence, we can understand the affine variety associated to the cone $\sigma$ as $U_\sigma = \mathrm{Specm} ( ( S_{\hat{\sigma}} )_{0} )$. In addition, \cref{equ:Gluing} ensures that these affine varieties glue in a meaningful fashion. The discussion given in \cref{subsec:CoherentSheavesOnVarieties} now lends itself nicely to provide an understanding of the affine patches of $X_\Sigma$.

Many geometric properties of $X_\Sigma$ can be read off from the geometry and combinatorics of the fan $\Sigma$. Before we state a number of such relations, recall the following: 
\begin{itemize}
 \item A \emph{scrapc} $\sigma$ is \emph{smooth} precisely if it is generated by a subset of a basis of $\mathbb{Z}^n$.
 \item A \emph{scrapc} $\sigma$ is \emph{simplicial} precisely if it is generated by a subset of a basis of $\mathbb{R}^n$.
 \item A fan $\Sigma$ is smooth/simplicial precisely if every $\sigma \in \Sigma$ is smooth/simplicial. 
\end{itemize}
One can now prove the results stated in \cref{table-N2}. In this thesis we will focus on normal toric varieties $X_\Sigma$ which are smooth, complete or simplicial, projective. So in particular all toric varieties of interest are compact. If then follows from \cref{table-N2} that they are simply connected and $\mathrm{Pic} ( X_\Sigma )$ is a free Abelian group.

\begin{table}[tbp]
\centering
\begin{tabular}{ccc}
\toprule
Property of $\Sigma \subseteq \mathbb{R}^n$ & \phantom{Implication} & Property of $X_\Sigma$ \\
\midrule
$\left| \Sigma \right| = \mathbb{R}^n$ & $\Leftrightarrow$ & $X_\Sigma$ is compact \\
\\[-0.5em]
$\Sigma$ is smooth & $\Leftrightarrow$ & $X_\Sigma$ is smooth \\
\\[-0.5em]
$\Sigma$ is simplicial & $\Leftrightarrow$ & $X_\Sigma$ has at worst finite quotient singularities \\
\\[-0.5em]
$\exists \sigma \in \Sigma$ with $\mathrm{dim}_{\mathbb{R}} \left( \sigma \right) = n$ & $\Rightarrow$ & \pbox{\textwidth}{$X_\Sigma$ is simply connected and \\ $\mathrm{Pic} ( X_\Sigma )$ is a free Abelian group} \\
\bottomrule
\end{tabular}
\caption[The geometry of a toric variety $X_\Sigma$ is encoded in its fan $\Sigma$.]{Geometric properties of a normal toric variety $X_\Sigma$ are encoded in the combinatorics of its fan $\Sigma$. The above table lists a number of such relations.}
\label{table-N2}
\end{table}

\paragraph{Chow Ring of Toric Varieties}

It is useful to define the irrelevant ideal $B_\Sigma$ and the Stanley-Reisner ideal $I_{\mathrm{SR}}$:
\begin{align}
\begin{split}
B_\Sigma &= \left\langle \left. x^{\hat{\sigma}} \right| \sigma \in \Sigma \right\rangle \, , \\
I_{\mathrm{SR}} &= \left\langle \left. x_{i_1} \cdots x_{i_s} \right| i_j \text{ are distinct and } \rho_{i_1} + \dots + \rho_{i_s} \text{ is not a cone of } \Sigma \right\rangle \, .
\end{split}
\end{align}
In addition, we define the \emph{ideal of linear relations} $I_{\mathrm{LR}}$, which is generated by the linear forms
\[ \sum_{\rho_i \in \Sigma ( 1 )}{\left\langle m, u_{\rho_i} \right\rangle x_{i}} \]
where $m$ ranges over the character lattice $M$. We define $R_{\mathbb{Q}} \left( \Sigma \right) = \mathbb{Q} \left[ x_1, \dots, x_m \right] / \left( I_{\mathrm{SR}} + I_{\mathrm{LR}} \right)$ and have the following results:
\begin{itemize}
 \item For smooth $X_\Sigma$: $\mathrm{CH}^{\bullet} \left( X_\Sigma \right) \cong H^{\bullet} \left( X_\Sigma, \mathbb{Z} \right) \cong R_{\mathbb{Q}} \left( \Sigma \right)$.
 \item For simplicial $X_\Sigma$: $\mathrm{CH}^{\bullet} \left( X_\Sigma \right)_{\mathbb{Q}} \cong H^\bullet \left( X_\Sigma, \mathbb{Q} \right) \cong R_{\mathbb{Q}} \left( \Sigma \right)_{\mathbb{Q}}$.
\end{itemize}
Hence, for complete, simplicial toric varieties the Chow ring is easily accessible. We will make use of this fact in explicit computations.

\paragraph{Chow Ring of \texorpdfstring{$\mathbf{\hat{Y}_4}$}{Y4} from a Toric Ambient Space}

Recall that an \emph{F-theory} compactification relies on the geometry of (the smooth resolutions of) an elliptic fibration $\hat{\pi} \colon \hat{Y}_4 \twoheadrightarrow \mathcal{B}_6$. Explicit examples in turn rely on modelling $\hat{Y}_4$ as subset of a toric variety $X_\Sigma$ -- in the simplest case $\hat{Y}_4$ is a hypersurface in $X_\Sigma$. Recall that the Chow ring of a simplicial, complete toric variety is easy to handle whilst that of $\hat{Y}_4$ can be hard to come by. We can thus hope to model at least a subset of $\mathrm{CH} ( \hat{Y}_4 )$ from pullbacks from $X_\Sigma$.

\begin{figure}[tb]
\centering
\includegraphics{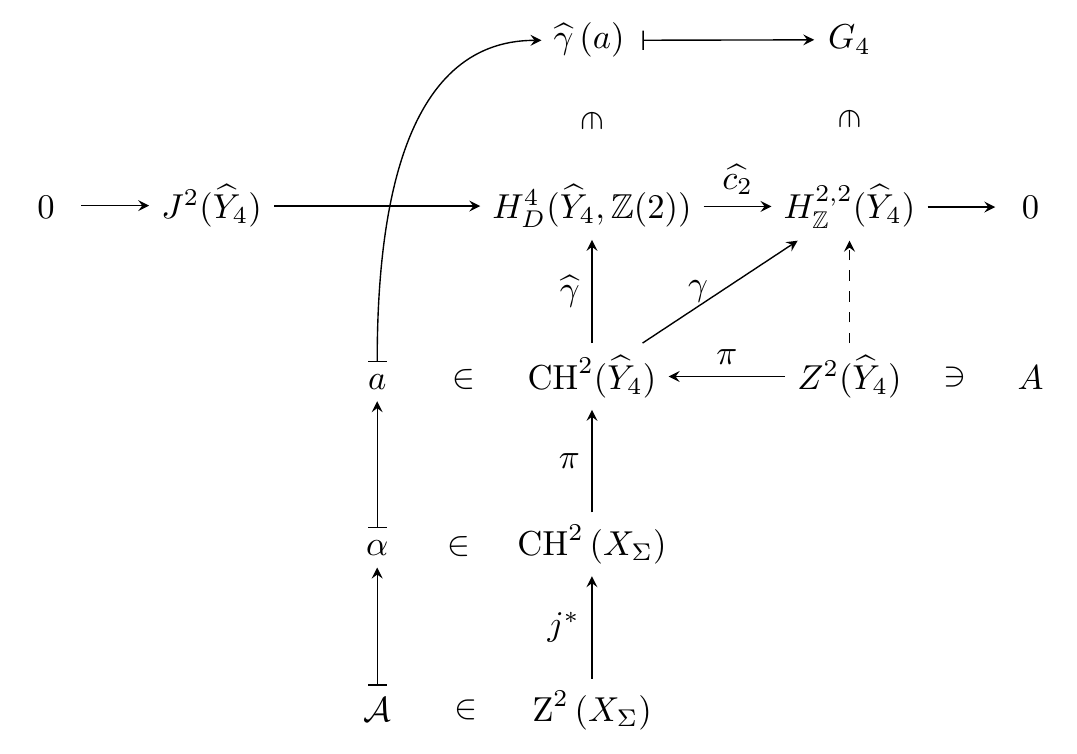}
\caption[Algebraic cycles $A \in \mathrm{CH}^2 ( \hat{Y}_4 )$ encode gauge backgrounds in $H^4_D ( \hat{Y}_4, \mathbb{Z} ( 2 ) )$.]{Classes of algebraic cycles $A \in \mathrm{Z}^2 ( \hat{Y}_4 )$ encode gauge backgrounds $H^4_D ( \hat{Y}_4, \mathbb{Z} ( 2 ) )$.}
\label{fig:C3BackgroundFromChow}
\end{figure}

Indeed, it is possible to express an algebraic cycle $A \in Z^2 ( \hat{Y}_4) $ in terms of $X_\Sigma$ whenever $X_\Sigma$ is smooth. Then the embedding $j \colon \hat{Y}_4 \hookrightarrow X_\Sigma$ induces pullback maps
\[ j^\ast \colon \mathrm{CH}^2 ( X_\Sigma ) \cong H^2 \left( X_\Sigma, \mathbb{Z} \right) \to \mathrm{CH}^2 ( \hat{Y}_4 ) \, . \] 
Consequently, modifications of the pre-image of $A$ on $X_\Sigma$ which leave its class in $H^2 ( X_\Sigma, \mathbb{Z} )$ unchanged, do not alter the gauge background described by $A$ via the refined cycle map. We thus approach $\mathrm{CH}^2 ( \hat{Y}_4 )$ as follows (\cf \cref{fig:C3BackgroundFromChow}):
\begin{enumerate}
 \item Specify $\mathcal{A} \in Z^2 ( X_\Sigma )$. Use manipulations respecting the homology class associated with $\mathcal{A}$ to simplify this algebraic cycle or 
      represent it differently whenever necessary.
 \item $\mathcal{A}$ induces $\mathcal{\alpha} = \pi ( \mathcal{A} ) \in \mathrm{CH}^2 ( X_\Sigma )$ and $a \in \mathrm{CH}^2 ( \hat{Y}_4 )$. 
 \item $\hat{\gamma} ( a ) \in H^4_D ( \hat{Y}_4, \mathbb{Z} ( 2 ) )$ is the 3-form data that we are after in the first place.
 \item $\gamma ( a ) = \hat{c_2} \circ \hat{\gamma} ( a ) = G_4$ is the (class of the) differential form usually referred to as $G_4$-flux.
\end{enumerate}

In many practical applications, including the model presented in \cref{subsec:Simplifying}, it can happen that the toric ambient space $X_\Sigma$ of $\hat{Y}_4$ is not smooth, but rather a complete toric orbifold. Since such a toric variety is simplicial we recall $\mathrm{CH}^{\bullet} \left( X_\Sigma \right)_{\mathbb{Q}} \cong H^\bullet \left( X_\Sigma, \mathbb{Q} \right)$ and are lead to wonder if the pullback $j^\ast \colon \mathrm{CH} ( X_\Sigma ) \to \mathrm{CH} ( \hat{Y}_4 )$ is well-defined in such a geometry. This very pullback amounts to computing intersection products of elements of $\mathrm{CH} ( X_\Sigma )$ and $\hat{Y}_4$. This intersection is well-defined as long as the embedding $\iota \colon \hat{Y}_4 \hookrightarrow X_\Sigma$ is a closed embedding and $\hat{Y}_4$ a \emph{local} complete intersection \cite{Bies:2014sra}, as these conditions guarantee a well-defined Gysin-homomorphism. Hence, under these assumptions even for a complete toric orbifold $X_\Sigma$ we can start with elements $\mathcal{A} \in Z^2 ( X_\Sigma )$, modify them by manipulations which respect the homology class of $\mathcal{A}$ in $X_\Sigma$ to simplify computations, and use the pullback of the so-obtained cycle to $\hat{Y}_4$ as model for $G_4$-fluxes on $\hat{Y}_4$.

\paragraph{Defining a Toric Variety the `Physics Way'}

Our presentation makes it tempting to specify a normal toric variety $X_\Sigma$ by a fan $\Sigma$. However, in the literature on string phenomenology people follow an alternative approach -- they state the following two pieces of information:
\begin{enumerate}
 \item The grading of the Cox ring $S$ under $\mathrm{Cl} ( X_\Sigma ) \cong \mathbb{Z}^n$.
 \item The generators of the ideal $I_{\mathrm{SR}}$.
\end{enumerate}
This in motivated by the introduction of gauged linear sigma models by Witten in \cite{Witten:1993yc}, where the grading of the Cox ring is interpreted as the so-called \emph{GLSM-charges}. In particular, the \emph{toric tops} which were investigated \eg in \cite{Krause:2011xj, Borchmann:2013hta} are specified in this way. The geometries relevant to this thesis are summarised in \cref{chapter:ToricTops}.

We can reconstruct the fan in question as follows: First note that for normal toric varieties without torus factor, there is an exact sequence
\[ 0 \to \mathbb{Z}^k \to \bigoplus_{\rho \in \Sigma \left( 1 \right)}{\mathbb{Z} D_\rho} \xrightarrow{\mathrm{deg}} \mathrm{Cl} \left( X_\Sigma \right) \cong \mathbb{Z}^n \to 0 \, . \label{equ:ClassGroupForToricVarietiesWithTorusFactor} \]
The mapping $\mathrm{deg} \colon \bigoplus_{\rho \in \Sigma \left( 1 \right)}{\mathbb{Z} D_\rho} \to \mathrm{Cl} \left( X_\Sigma \right)$ is provided as defining information of the variety. In addition, the mapping $\mathbb{Z}^n \to \bigoplus_{\rho \in \Sigma \left( 1 \right)}{\mathbb{Z} D_\rho}$ is mediated by
\[ m \in M \cong \mathbb{Z}^n \mapsto \mathrm{div} \left( \chi^m \right) = \sum_{\rho \in \Sigma \left( 1 \right)}{\left\langle m, u_\rho \right\rangle \cdot D_\rho} \, .\]
By exactness of \cref{equ:ClassGroupForToricVarietiesWithTorusFactor} we can thus use the mapping $\mathrm{deg}$ to identify a set of ray generators $\left\{ u_\rho \right\}_{\rho \in \Sigma ( 1 )}$ of the fan $\Sigma$. It remains to tell which subsets of $\left\{ u_\rho \right\}_{\rho \in \Sigma ( 1 )}$ generate cones in the fan $\Sigma$. This information is provided by the Stanley-Reisner ideal $I_{\mathrm{SR}}$.

\section{Toric \emph{F-Theory} Vacua with \texorpdfstring{$\mathbf{SU ( 5 ) \times U ( 1 )}$}{SU(5) x U(1)}-Symmetry} \label{sec:ToricFTheoryGUTModels}

Our next goal is to demonstrate our formalism of computing the line bundles induced by gauge backgrounds $A \in \mathrm{CH}^2 ( \hat{Y}_4 )$ in an explicit example. We shall design the example to be as simple as possible while at the same time exhibiting all ingredients of our general discussion so far. In order to exemplify the role of Abelian fluxes of type \cref{AX-general}, we need at least one extra rational section. The arguably simplest type of such fibrations is obtained by a $U(1)$ restricted Tate model \cite{Grimm:2010ez}. To study the behaviour of massless matter charged also under a non-Abelian gauge algebra, we model an extra non-Abelian gauge factor, the simplest class being represented by Tate models with an $I_n$ singularity. The existence of additional non-trivial vertical gauge fluxes in such models requires the non-Abelian gauge group to be at least $SU(5)$ \cite{oai:arXiv.org:1202.3138}. In this sense a $U(1)$ restricted Tate model with gauge group $G = SU ( 5 ) \times U ( 1 )_X$ is indeed minimal. The top describing the resolved fibre of this model was originally introduced in \cite{Krause:2011xj}. We will now provide a brief review on the topic. The interested reader is referred to the above references and \cref{sec:SU5xU1Top} for further details.

\subsection{\texorpdfstring{$\mathbf{SU ( 5 ) \times U ( 1 )_X}$}{SU(5) x U(1)} Fibration} \label{subsec:SpecialFTheoryGUTModel}

Let us start by considering the total space $X_5$ of a $\mathbb{P}_{2,3,1}$-fibration over a smooth space $\mathcal{B}_6$ and pick sections 
$a_i \in H^0 ( \mathcal{B}_6, \overline{K}_{\mathcal{B}_6}^{\otimes i} )$. We denote the homogeneous coordinates of the $\mathbb{P}_{2,3,1}$-fibre by $[ x \colon y \colon z ]$ and define $Y_4 = V ( P_T ) \subseteq X_5$ where
\[ P_T = y^2 + a_1 \left( p \right) x y z + a_3 \left( p \right) y z^3 - x^3 - a_2 \left( p \right) x^2 z^2 - a_4 \left( p \right) x z^4 - a_6 \left( p \right) z^6 \]
is the so-called \emph{Tate-polynomial}. Such a construction is referred to as a \emph{global Tate model} in the \emph{F-theory} literature.

Recall that by the \emph{F-theory} philosophy, the locus $\Delta \subseteq \mathcal{B}_6$ over which the elliptic fibre becomes singular encodes the location of D7-branes in the associated \emph{F-theory} compactification. In addition, the singularity structure over $\Delta$ encodes the gauge degrees of freedom. Therefore, it is both by desire and construction that $Y_4$ is singular of a particular type. Recall from \cref{sec:SummaryChapter2} that we are interested in $\mathbb{P}_{2,3,1}$-fibrations over $\mathcal{B}_6$, such that there exists at least one smooth resolution $\hat{Y}_4$ which satisfies the Calabi--Yau condition. Consequently, $H^{1,0} (\mathcal{B}_6) = H^{2,0} (\mathcal{B}_6) = 0$. In addition, we focus on base spaces $\mathcal{B}_6$ without torsional 1-cycles, \ie $H^1(\mathcal{B}_6, \mathbb Z) = 0$.

To have an $SU ( 5 ) \times U ( 1 )_X$ gauge theory in this \emph{F-theory} compactification we specify a locus
\[ W = V(w) := \{ w=0 \} \subseteq \mathcal{B}_6 \,  \]
and subsequently choose the sections $a_i$ to match the pattern
\[ a_1 = a_1, \qquad a_2 = a_{2,1} \, w, \qquad a_3 = a_{3,2} \, w^2, \qquad a_4 = a_{4,3} \, w^3 \, . \]
This choice establishes an $SU(5)$-singularity over $W$. In addition, let us demand $a_6 \equiv 0$. This requirement implements an extra section of the $\mathbb{P}_{2,3,1}$-fibration over $\mathcal{B}_6$, which in turn establishes the Abelian gauge group factor $U(1)_X$. 

This singular space $Y_4$ encodes the geometry of the \emph{F-theory} compactification that we are interested in. To simplify computations we wish to work on a smooth resolution of $Y_4$. Our strategy is to \emph{blow up} the singularities in the fibre one after another. For the geometry in question this has been worked out in detail in \cite{Krause:2011xj}: First one introduces blow-up coordinates $e_i$, $i=1, \ldots, 4$ and $s$. The coordinate $s$ is associated to the additional section of this fibration installed by the demand $a_6 = 0$ \cite{Krause:2011xj}. The resolved fibration $\hat{\pi} \colon \hat{Y}_4 \twoheadrightarrow \mathcal{B}_6$ can be described as hypersurface $V( P_T^\prime ) \subseteq \hat{Y}_5$, where $\hat{Y}_5$ is a new ambient space and $P_T^\prime$ the so-called \emph{proper transform} of the Tate polynomial. Explicitly it holds
\begin{align}
\begin{split}
P_T^\prime &= y^2 s e_3 e_4 + a_1 \left( p \right) x y z s + a_{3,2} \left( p \right) y z^3 e_0^2 e_1 e_4 - x^3 s^2 e_1 e_2^2 e_3 \\
           & \hspace{13em} - a_{2,1} \left( p \right) x^2 z^2 s e_0 e_1 e_2 - a_{4,3} \left( p \right) x z^4 e_0^3 e_1^2 e_2 e_4 \, , \label{equ:properTransformOfPT}
\end{split}
\end{align}
where due to the blow-ups $\hat\pi^* ( w ) = e_0 e_1 e_2 e_3 e_4$. 

Let $p \in \mathcal{B}_6$. Then the resolved fibre $\hat{\pi}^{-1} ( p )$ can be understood as the vanishing locus of $P_T^\prime ( p )$ in a particular toric ambient space. This very toric space is referred to as a \emph{top} in the \emph{F-theory} lingo. Such \emph{tops} exist for many types of gauge groups -- see for example \cite{Borchmann:2013hta}. For the gauge group $G = SU ( 5 ) \times U( 1 )_X$, the top in question was introduced in \cite{Krause:2011xj}. Its homogeneous coordinates and their degrees are given in \cref{table-N3}.

\begin{table}[tbp]
\centering
\begin{tabular}{c@{\hskip 20pt}ccccc@{\hskip 20pt}cccc}
\toprule
           & $e_0$   & $e_1$   & $e_2$   & $e_3$   & $e_4$   & x & y & z & s \\
\midrule
$\overline{K}_{\mathcal{B}_6}$ & $\cdot$ & $\cdot$ & $\cdot$ & $\cdot$ & $\cdot$ & 2 & 3 & $\cdot$ & $\cdot$ \\
W                              & 1       & $\cdot$ & $\cdot$ & $\cdot$ & $\cdot$ & $\cdot$ & $\cdot$ & $\cdot$ & $\cdot$ \\
\midrule
$E_1$ & -1        & 1 & $\cdot$ & $\cdot$ & $\cdot$ & -1 & -1 & $\cdot$ & $\cdot$ \\
$E_1$ & -1        & $\cdot$ & 1 & $\cdot$ & $\cdot$ & -2 & -2 & $\cdot$ & $\cdot$ \\
$E_1$ & -1        & $\cdot$ & $\cdot$ & 1 & $\cdot$ & -2 & -3 & $\cdot$ & $\cdot$ \\
$E_1$ & -1        & $\cdot$ & $\cdot$ & $\cdot$ & 1 & -1 & -2 & $\cdot$ & $\cdot$ \\
\midrule
Z & $\cdot$   & $\cdot$ & $\cdot$ & $\cdot$ & $\cdot$ & 2 &  3 & 1 & $\cdot$ \\
S & $\cdot$   & $\cdot$ & $\cdot$ & $\cdot$ & $\cdot$ & -1 & -1 & $\cdot$ & 1 \\
\bottomrule
\end{tabular}
\caption{Homogeneous coordinates and grading of the Cox ring of the $SU(5) \times U(1)_X$-top.}
\label{table-N3}
\end{table}
As pointed out in \cite{Krause:2011xj}, there exist a number of possible ways to resolve the $SU(5) \times U(1)_X$-singularity and still remain within the realm of \emph{tops}. These correspond to different triangulations of the above toric space. Of the many possible choices we will stick to triangulation $T_{11}$ of \cite{Krause:2011xj} throughout this entire thesis. This choice is equivalent to stating that the Stanley-Reisner ideal of the resolved fibre ambient space takes the form
\begin{align}
\begin{split} \label{SRideal}
I_{\mathrm{SR}} \left( \mathrm{top} \right) = & \left\{ xy, x e_0 e_3, x e_1 e_3, x e_4, y e_0 e_3, y e_1, y e_2, z s, z e_1 e_4, z e_2 e_4, \right. \\
                & \hspace{9em} \left. z e_3, s e_0, s e_1, s e_4, e_0 e_2, z e_4, z e_1, z e_2, s e_2, e_0 e_3, e_1 e_3 \right\} \, .
\end{split}
\end{align}

This data finally specifies a toric variety `the physics way' as explained in \cref{subsec:TowardsToricVarieties}.

By $E_i \in \mathrm{CH}^1(\hat{Y}_4)$ we denote the class of algebraic cycles rationally equivalent to the vanishing locus $V ( e_i ) \subseteq \hat{Y}_5$.\footnote{We apply similar notations for the vanishing loci associated to the other homogeneous coordinates of the \emph{top}.} These classes correspond to the generators of the Cartan $U (1 )$ symmetries of $SU(5)$. We use $S$ and $Z$ to denote the classes of the extra rational section $V ( s )$ and the zero section $V(z)$ in $\mathrm{CH}^1 ( \hat{Y}_4 )$, respectively. These allow us to express the generator of the $U ( 1 )_X$ gauge symmetry as
\[ U_X := - 5 \left( S- Z- \overline{K} \right) - 2 E_1 - 4 E_2 - 6 E_3 - 3 E_4 \in \mathrm{CH}^1(\hat{Y}_4) \, . \label{equ:Ux} \]

Matter in the representations $\mathbf{10}_{1}$, ${\mathbf{5}_{3}}$, ${\mathbf{5}_{-2}}$ and ${\mathbf{1}_{5}}$ localises on the following curves of $\mathcal{B}_6$:
\begin{equation} \label{mattercurvesSU5equ}
  \begin{aligned}
     C_{\mathbf{10}_{1}} &= V \left(w, a_{1,0} \right), \qquad & C_{\mathbf{5}_{3}} &= V \left(w, a_{3,2} \right), \\
     C_{\mathbf{5}_{-2}} &= V \left(w, a_1 a_{4,3} - a_{2,1} a_{3,2} \right), \qquad & C_{\mathbf{1}_{5}} &= V \left(a_{4,3}, a_{3,2} \right) \, .
  \end{aligned}
\end{equation}
The subscript denote the $U(1)_X$-charge of the respective $SU(5)$ representation. The matter surfaces $S^a(\mathbf{R}) \in \mathrm{CH}^2 ( \hat{Y}_4 )$ over these matter curves can be obtained by analysing the fibre structure of $\hat{\pi} \colon \hat{Y}_4 \twoheadrightarrow \mathcal{B}_6$ \cite{Krause:2011xj, oai:arXiv.org:1202.3138}. We review this subject in \cref{subsec:FibreStructureSU5xU1}. In particular, note that the $U ( 1 )_X$-charge of $S^a ( \mathbf{R} )$ is encoded in its intersection with $U_X$ in the (generic) elliptic fibre over $C_{\mathbf{R}}$.

For explicit computation it will be crucial to make use of the so-called \emph{linear relations} of the $SU ( 5 ) \times U ( 1 )_X$-top. These are the generators of the ideal of linear relations as introduced in \cref{subsec:TowardsToricVarieties}. In this very geometry there are three generators, namely
\begin{align}
\begin{split} \label{linearrelationsambient}
\mathcal{X} - \mathcal{Y} + \mathcal{Z} + \mathcal{E}_0 + \mathcal{E}_1 + \mathcal{E}_2 - \mathcal{W} + \overline{\mathcal{K}}_{\mathcal{B}_6} &= 0 \in \mathrm{CH}^1 ( 
      \hat{Y}_5 ) \, ,\\
-3 \mathcal{X} + 2 \mathcal{Y} - \mathcal{S} - \mathcal{E}_1 -2 \mathcal{E}_2 + \mathcal{E}_4 &= 0 \in \mathrm{CH}^1 ( \hat{Y}_5 ) \, , \\ 
2 \mathcal{X} - \mathcal{Y} - \mathcal{Z} + \mathcal{S} + \mathcal{E}_1 + 2 \mathcal{E}_2 + \mathcal{E}_3 - \overline{\mathcal{K}}_{\mathcal{B}_6} &= 0 \in \mathrm{CH}^1 
      ( \hat{Y}_5 ) \, ,
\end{split}
\end{align}
where $\mathcal{X} \in \mathrm{CH}^1 ( \hat{Y}_5)$ is the class of algebraic cycles rationally equivalent to $V ( x ) \subseteq \hat{Y}_5$, and similarly for the other divisors. In particular, we have $ \mathcal{X} |_{\hat{Y}_4} \equiv X \in \mathrm{CH}^1 ( \hat{Y}_4 )$, the latter being the class of algebraic cycles in $\hat{Y}_4$ which are rationally equivalent to $ V ( x ) |_{\hat{Y}_4} = V ( x, P_T^\prime ) \subseteq \hat{Y}_4$.

\subsection{(Self-)Intersections of Matter Curves} \label{subsec:SelfIntersectionsOfMatterCurves}

To evaluate expressions such as \cref{intersectionMSF} we need the (self-)intersections of the matter curves $C_{\mathbf{10}_{1}}$, $C_{\mathbf{5}_{3}}$ and $C_{\mathbf{5}_{-2}}$ in the divisor $W \subseteq \mathcal{B}_6$. In anticipation of this application, let us derive the relevant results in this section.

Two curves $C_i, C_j \subseteq W$ can be regarded as divisors on $W$, \ie elements of $\mathrm{CH}^1(W)$. Hence, their intersection product is a special case of the general intersection product within the Chow ring as introduced in \cref{sec:F-theory}. Throughout this section we will use the canonical embedding $\iota_{C_i}: C_i \hookrightarrow W$. Let us therefore abbreviate $C_i \cdot_{\iota_{C_i}} C_j $ as $C_i \cdot C_j$. This intersection is given by an element in $\mathrm{CH}_0(C_i) \simeq \mathrm{CH}^1(C_i)$. In picking a particular representant of this Chow class we can thus write
\[ C_i \cdot C_j = \sum_{k = 1}^{N}{m_k \cdot p_k} \in \mathrm{CH}_0(C_i) \, \]
where $N \in \mathbb{N}_{\geq 0}$, $m_k \in \mathbb{Z}$ and $p_k \in C_j$. The line bundle associated to this very divisor is $\mathcal{O}_W (C_j) |_{C_i}$. That said, the intersection number $|C_i \cdot C_j|$ simply takes the form \cite{ATIT, cox2011toric}
\[ |C_i \cdot C_j | = \mathrm{deg} \left( \left. \mathcal{O}_{W} \left( C_j \right) \right|_{C_i} \right) = \Mint_{C_i}{c_1 \left( \left. \mathcal{O}_{W} \left( C_j \right) \right|_{C_i} \right)} = \sum_{k = 1}^{N}{m_k} \, . \]
In this sense, the intersection $D = C_i \cdot C_j$ occurs over the points $p_k$ with multiplicities $m_k$. Note however that this notion of `intersection points' is only well-defined up to linear equivalence of divisors - pick a divisor $D^\prime \neq D$ but with $D^\prime \sim D$, then $D^\prime$ will denote different intersection points with different multiplicities. However, the intersection number computed from $D^\prime$ is identical to the one computed from $D$.

\begin{figure}[tbp]
\centering
\includegraphics{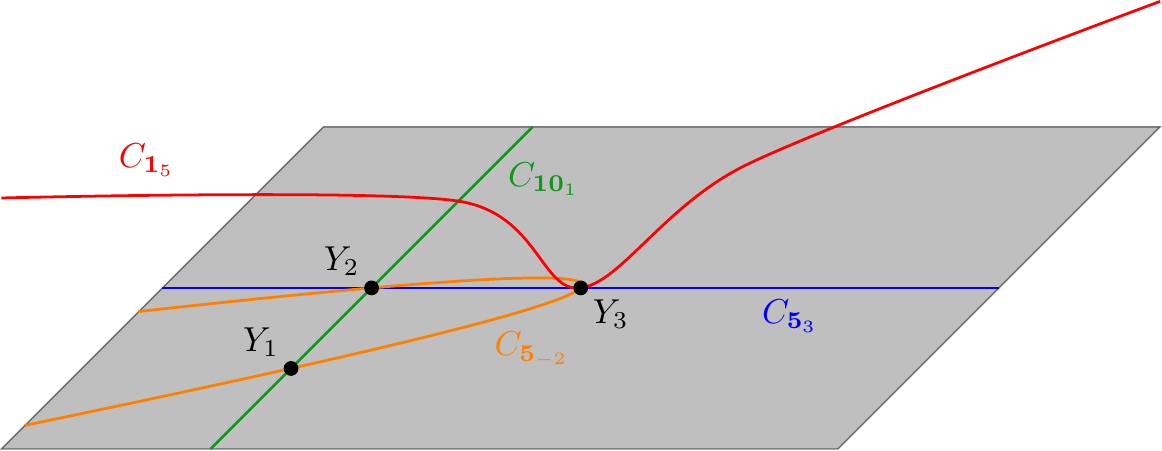}
\caption[Intersections of matter curves in $SU(5) \times U(1)_X$-top.]{Schematic picture of the intersections of the matter curves $C_{\mathbf{10}_{1}}$, $C_{\mathbf{5}_{3}}$, $C_{\mathbf{5}_{-2}}$ and $C_{\mathbf{1}_{5}}$ in the GUT surface $W$, which is displayed as they grey rectangle.}
\label{figure-1b}
\end{figure}

Bearing this in mind, let us work out intersection points and their associated multiplicities for the intersections of the matter curves $C_{\mathbf{10}_{1}}$, $C_{\mathbf{5}_{3}}$ and $C_{\mathbf{5}_{-2}}$ in $W$. First recall that the Yukawa loci are given by (\cf \cref{subsec:FibreStructureSU5xU1})
\[ Y_1 = V \left( w, a_{1,0}, a_{2,1} \right), \quad Y_2 = V \left( w, a_{1,0}, a_{3,2} \right), \quad Y_3 = V \left( w, a_{4,3}, a_{3,2} \right) \, . \label{Yidefinition} \]
From the explicit realisation of these curves in \cref{mattercurvesSU5equ} the transverse intersections follows as
\[\label{transverseY}
C_{\mathbf{5}_{3}} \cdot C_{\mathbf{10}_{1}}= Y_2 \, , \qquad C_{\mathbf{5}_{-2}} \cdot C_{\mathbf{10}_{1}} = Y_1 + Y_2 \, , \qquad C_{\mathbf{5}_{-2}} \cdot C_{\mathbf{5}_{3}} = Y_2 + Y_3 \, . \]
This structure is depicted schematically in \cref{figure-1b}. Note that the loci $Y_i$ are in general reducible and consist of the following number of points:
\begin{align}
\begin{split}
n_{Y_1} &=\left[ \overline{K}_{\mathcal{B}_6} \right] \cdot_W \left[ 2 \overline{K}_{\mathcal{B}_6} - {W} \right] \, , \\
n_{Y_2} &=\left[ \overline{K}_{\mathcal{B}_6} \right] \cdot_W \left[ 3 \overline{K}_{\mathcal{B}_6} - 2 {W} \right] \, , \\
n_{Y_3} &= \left[ 4 \overline{K}_{\mathcal{B}_6} - 3 {W} \right] \cdot_W \left[ 3 \overline{K}_{\mathcal{B}_6} - 2{W} \right] \, .
\end{split}
\end{align}
For the self-intersections we proceed very similarly. Let us start with $C_{\mathbf{10}_{1}}$ and note that
\begin{align}
\begin{split}
N_{C_{\mathbf{10}_{1}} \subseteq W} &= \left. \mathcal{O}_{W} \left( \overline{K}_{\mathcal{B}_6} \right) \right|_{C_{\mathbf{10}_{1}}} \\
                                    &\cong \left. \mathcal{O}_{W} \left( 2 \cdot \left( 2 \overline{K}_{\mathcal{B}_6} - {W} \right) \right) \right|_{C_{\mathbf{10}_{1}}} \otimes \left. \mathcal{O}_{W} \left( -1 \cdot \left( 3 \overline{K}_{\mathcal{B}_6} - 2 {W} \right) \right) \right|_{C_{\mathbf{10}_{1}}} \\
                                    &\cong \mathcal{O}_{C_{\mathbf{10}_{1}}} \left( 2 Y_1 - Y_2 \right) \, .
\end{split}
\end{align}
Consequently, up to linear equivalence, $C_{\mathbf{10}_{1}}$ self-intersects at $Y_1$, $Y_2$ with multiplicities $+2$ and $-1$ respectively. Similarly, we have for $C_{\mathbf{5}_{3}}$
\begin{align}
\begin{split}
 N_{C_{\mathbf{5}_{3}} \subseteq W} &= \left. \mathcal{O}_{W} \left( 3 \overline{K}_{\mathcal{B}_6} - 2 {W} \right) \right|_{C_{\mathbf{5}_{3}}} \\
 &\cong \left. \mathcal{O}_{W} \left( \frac{1}{3} \cdot \overline{K}_{\mathcal{B}_6} \right) \right|_{C_{\mathbf{5}_{3}}} \otimes \left. \mathcal{O}_{W} \left( \frac{2}{3} \cdot \left( 4 \overline{K}_{\mathcal{B}_6} - 3 {W}  \right) \right) \right|_{C_{\mathbf{5}_{3}}} \\
 &\cong \mathcal{O}_{C_{\mathbf{5}_{3}}} \left( \frac{1}{3} Y_2 + \frac{2}{3} Y_2 \right)
\end{split}
\end{align}
Finally we find for $C_{\mathbf{5}_{-2}}$
\begin{align}
\begin{split}
N_{C_{\mathbf{5}_{-2}} \subseteq W} &= \left. \mathcal{O}_{W} \left( 5 \overline{K}_{\mathcal{B}_6} - 3 {W} \right) \right|_{C_{\mathbf{5}_{-2}}} \\
&\cong \left. \mathcal{O}_{W} \left( \frac{3}{2} \left( 3 \overline{K}_{\mathcal{B}_6} - 2 {W} \right) \right) \right|_{C_{\mathbf{5}_{-2}}} \otimes \left. \mathcal{O}_{W} \left( \frac{1}{2} \cdot \overline{K}_{\mathcal{B}_6} \right) \right|_{C_{\mathbf{5}_{-2}}} \\
&\cong \mathcal{O}_{C_{\mathbf{5}_{-2}}} \left( \frac{1}{2} Y_1 + 2 Y_2 + \frac{3}{2} Y_3 \right) \, .
\end{split}
\end{align}
The above manipulations manifestly involve rational coefficients. This is correct by our assumption that $\mathcal{B}_6$ does not contain torsional divisors (\cf \cref{sec:SummaryChapter2}). Irrespective of the appearance of rational coefficients, note that by construction the above normal bundles  are integer quantised. We summarise our findings in \cref{table-N4}.

\begin{table}[tbp]
\centering
\begin{tabular}{c@{\hskip 20pt}ccc@{\hskip 20pt}c}
\toprule
& $Y_1$ & $Y_2$ & $Y_3$ & net intersection number \\
\midrule
$\mathbf{10}_{1} \cdot \mathbf{10}_{1}$ & $2 \cdot n_{Y_1}$ & $\left(-1 \right) \cdot n_{Y_2}$ & $\cdot$ & $\left[ \overline{K}_{\mathcal{B}_6} \right] \cdot \left[ \overline{K}_{\mathcal{B}_6} \right]$ \\
$\mathbf{10}_{1} \cdot \mathbf{5}_{3}$ & $\cdot$ & $1 \cdot n_{Y_2}$ & $\cdot$ & $\left[ \overline{K}_{\mathcal{B}_6} \right] \cdot \left[ 3 \overline{K}_{\mathcal{B}_6} - 2 {W} \right]$ \\
$\mathbf{10}_{1} \cdot \mathbf{5}_{-2}$ & $1 \cdot n_{Y_1}$ & $1 \cdot n_{Y_2}$ & $\cdot$ & $\left[ \overline{K}_{\mathcal{B}_6} \right] \cdot \left[  5 \overline{K}_{\mathcal{B}_6} - 3 {W} \right]$ \\ 
\midrule
$\mathbf{5}_{3} \cdot \mathbf{5}_{3}$ & $\cdot$ & $\frac{1}{3} \cdot n_{Y_2}$ & $\frac{2}{3} \cdot n_{Y_3}$ & $\left[ 3 \overline{K}_{\mathcal{B}_6} - 2 {W} \right] \cdot \left[ 3 \overline{K}_{\mathcal{B}_6} - 2 {W} \right]$ \\
\vspace{-1em} & \\
$\mathbf{5}_{3} \cdot \mathbf{5}_{-2}$ & $\cdot$ & $1 \cdot n_{Y_2}$ & $ 1 \cdot n_{Y_3}$ & $\left[ 5 \overline{K}_{\mathcal{B}_6} - 3 {W} \right] \cdot \left[ 3 \overline{K}_{\mathcal{B}_6} - 2 {W} \right]$ \\
\midrule
$\mathbf{5}_{-2} \cdot \mathbf{5}_{-2}$ & $\frac{1}{2} \cdot n_{Y_1}$ & $2 \cdot n_{Y_2}$ & $\frac{3}{2} \cdot n_{Y_3}$ & $\left[ 5 \overline{K}_{\mathcal{B}_6} - 3 {W} \right] \cdot \left[ 5 \overline{K}_{\mathcal{B}_6} - 3 {W} \right]$ \\
\bottomrule
\end{tabular}
\caption[Intersection points and multiplicities of matter curves in $SU(5) \times U(1)_X$-top.]{Intersection points and multiplicities of the matter curves $C_{\mathbf{10}_{1}}$, $C_{\mathbf{5}_{3}}$, $C_{\mathbf{5}_{-2}}$ in $W$.}
\label{table-N4}
\end{table}

\subsection{Vertical and Gauge Invariant Matter Surface Fluxes over Matter Curves} \label{subsec:VerticalAndGaugeInvariantMatterSurfaceFluxes}

We now construct the matter surface fluxes introduced in \cref{def-mattersurfaceflux} over the matter curves $C_{\mathbf{10}_{1}}$, $C_{\mathbf{5}_{3}}$, $C_{\mathbf{5}_{-2}}$ and $C_{\mathbf{1}_{5}}$. We begin with the flux associated to a matter surface $S^a(\mathbf{10}_{1})$ over $C_{\mathbf{10}_1}$. The 2-cycle $A(\mathbf{10}_{1})$ obtained by subtracting suitable correction terms will be a linear combination of rational curves fibred over  $C_{\mathbf{10}_{1}}$. These rational curves arise from the splitting of the fibres of the resolution divisors $E_i \subseteq \hat{Y}_4$, $i=1,\ldots 4,$ over $C_{\mathbf{10}_{1}}$. For example, the fibre of $E_2$ splits into two rational curves $\mathbb{P}_{24}^1$ and $\mathbb{P}_{2B}^1$. The 2-cycles obtained by fibring these over $C_{\mathbf{10}_{1}}$ will be denoted by  $\mathbb{P}_{24}^1 ( \mathbf{10}_{1} )$  and $\mathbb{P}_{2B}^1 ( \mathbf{10}_{1} )$, respectively. An overview of such fibral curves and their associated 2-cycles is given in \cref{subsec:FibreStructureSU5xU1}. More details can be found in \cite{Krause:2011xj, oai:arXiv.org:1202.3138}.

By evaluating the general expression \cref{def-mattersurfaceflux}, we find that the complex 2-cycle $A ( \mathbf{10}_{1} ) \in \mathrm{CH}^2 ( \hat{Y}_4 )$ is given by\footnote{Note that this expression for gauge-invariant matter surface fluxes is valid only for simply-laced Lie algebras. More generally one can replace the Cartan matrix $C$ by the matrix $\mathfrak{C}$ defined in \cref{DefcalC}. They agree for simply-laced Lie algebras, but will differ for non-simply-laced ones.}
\[ \label{10flux1}
A \left( \mathbf{10}_{1} \right) = S^{a} \left( \mathbf{10}_{1} \right) + {\beta}^a \left( \mathbf{10}_1 \right)^T \cdot C^{-1} \cdot \left( \begin{array}{c} \mathbb{P}_{14}^1 \left( \mathbf{10}_{1} \right) \\ \mathbb{P}_{24}^1 \left( \mathbf{10}_{1} \right) + \mathbb{P}_{2B}^1 \left( \mathbf{10}_{1} \right) \\ \mathbb{P}_{3C}^1 \left( \mathbf{10}_{1} \right) \\ \mathbb{P}_{14}^1 \left( \mathbf{10}_{1} \right) + \mathbb{P}_{24}^1 \left( \mathbf{10}_{1} \right) + \mathbb{P}_{4D}^1 \left( \mathbf{10}_{1} \right) \end{array} \right) \, . \]
This expression uses the inverse of the $SU(5)$ Cartan matrix 
\[ C = \left( \begin{array}{cccc} 2 & -1 & 0 & 0 \\ -1 & 2 & -1 & 0 \\ 0 & -1 & 2 & -1 \\ 0 & 0 & -1 & 2 \end{array} \right) \, . \]
Note that the column vector appearing in \cref{10flux1} contains precisely those $\mathbb{P}^1$-fibrations over $C_{\mathbf{10}_{1}}$ which arise from restriction of $E_1$, $E_2$, $E_3$ and $E_4$ to $C_{\mathbf{10}_{1}}$. Since this procedure gives the same cycle class for all matter surfaces $S^{a} ( \mathbf{10}_{1} )$ we have dropped the superscript `$a$' in $A ( \mathbf{10}_{1})$. The result can be expressed as \footnote{The fractions $\frac{1}{5}$ originate from the \emph{inverse} of the Cartan matrix $C$.}
\[ A \left( \mathbf{10}_{1} \right) \left( \lambda \right) := A \left( \mathbf{10}_{1} \right) \left( 0, 0, - \frac{2\lambda}{5}, \frac{\lambda}{5}, -\frac{\lambda}{5}, \frac{2 \lambda}{5} \right) \in \mathrm{CH}^2(\hat{Y}_4) \, . \label{equ:flux10Final} \]
In this expression we are using a rational coefficient $\lambda \in \mathbb{Q}$ and apply the short-hand notation
\begin{align}
\begin{split}
A \left( \mathbf{10}_{1} \right) \left( a_0, a_1, a_2, a_3, a_4, a_5 \right) &= a_0 \cdot \mathbb{P}^1_{0A} \left( \mathbf{10}_{1} \right) + a_1 \cdot \mathbb{P}^1_{14} \left( \mathbf{10}_{1} \right) + a_2 \cdot \mathbb{P}^1_{24} \left( \mathbf{10}_{1} \right) \\
&+ a_3 \cdot \mathbb{P}^1_{2B} \left( \mathbf{10}_{1} \right) + a_4 \cdot \mathbb{P}^1_{3C} \left( \mathbf{10}_{1} \right) + a_5 \cdot \mathbb{P}^1_{4D} \left( \mathbf{10}_{1} \right) \, .
\end{split}
\end{align}
The coefficient $\lambda \in \mathbb{Q}$ has to be chosen subject to the flux quantisation condition \cref{FW2}. It is not too hard to repeat this analysis for matter surface fluxes over $C_{\mathbf{5}_{3}}$ and $C_{\mathbf{5}_{-2}}$. We list the relevant $\mathbb{P}^1$-fibrations  in \cref{subsec:FibreStructureSU5xU1}. This enables us to state the result as
\begin{align}
\begin{split} \label{5flux32}
A \left( \mathbf{5}_{3} \right) \left( \lambda \right) &= A \left( \mathbf{5}_{3} \right) \left( 0, - \frac{\lambda}{5}, - \frac{2 \lambda}{5}, - \frac{3 \lambda}{5}, \frac{2 \lambda}{5}, \frac{\lambda}{5} \right)\in \mathrm{CH}^2(\hat{Y}_4), \\
A \left( \mathbf{5}_{-2} \right) \left( \lambda \right) &= A \left( \mathbf{5}_{-2} \right) \left( 0, - \frac{\lambda}{5}, - \frac{2 \lambda}{5}, - \frac{3 \lambda}{5}, \frac{2 \lambda}{5}, \frac{\lambda}{5} \right)\in \mathrm{CH}^2(\hat{Y}_4) \, ,
\end{split}
\end{align}
where we use analogous notation for the respective fibral curves in the order listed in \cref{53curvesplitting} and \cref{52curvesplitting}. By similar arguments it can be shown that
\[ A \left( \mathbf{1}_{5} \right) \left( \lambda \right) = \lambda \,  \mathbb{P}^1_A \left( \mathbf{1}_{5} \right) = \lambda \,  V \left( P_T^\prime, s, a_{3,2}, a_{4,3} \right)\in \mathrm{CH}^2(\hat{Y}_4) \]
is both vertical and gauge invariant.\footnote{The fluxes described above are the ones which exist generically for every choice of $\mathcal{B}_6$. In addition, there can be extra gauge invariant matter surface fluxes if $\mathcal{B}_6$ has special properties, for example if the matter curves are forced to split.} As remarked in \cref{subsec:NewStrategyForMasslessSpectra} it is sometimes convenient to express these cycles as elements $\mathcal{A} \in \mathrm{CH}^2 ( \hat{Y}_5 )$ such that $\mathcal{A} |_{\hat{Y}_4} = A$. Here these take the form
\begin{align}
\begin{split}
\label{equ:MatterSurfaceFluxes-X5}
\mathcal{A} \left( \mathbf{10}_{1} \right) \left( \lambda \right) &= - \frac{\lambda}{5} \cdot \left( 2 \mathcal{E}_1 - \mathcal{E}_2 + \mathcal{E}_3 - 2 \mathcal{E}_4 \right) \cdot \overline{\mathcal{K}}_{\mathcal{B}_6} - \lambda \cdot \mathcal{E}_2 \cdot \mathcal{E}_4 \, , \\
\mathcal{A} \left( \mathbf{5}_{3} \right) \left( \lambda \right) &= - \frac{\lambda}{5} \cdot \left( \mathcal{E}_1 + 2 \mathcal{E}_2 - 2 \mathcal{E}_3 - \mathcal{E}_4 \right) \cdot \left( 3 \overline{\mathcal{K}}_{\mathcal{B}_6} - 2 \mathcal{W} \right) - \lambda \cdot \mathcal{E}_3 \cdot \mathcal{X} \, , \\
\mathcal{A} \left( \mathbf{5}_{-2} \right) \left( \lambda \right) &= - \frac{\lambda}{5} \cdot \left( \mathcal{E}_1 + 2 \mathcal{E}_2 + 3 \mathcal{E}_3 - \mathcal{E}_4 \right) \cdot \left( 5 \overline{\mathcal{K}}_{\mathcal{B}_6} - 3 \mathcal{W} \right) \\
& \hspace{15em} + \lambda \cdot \left( \mathcal{E}_3 \cdot \overline{\mathcal{K}}_{\mathcal{B}_6} + \mathcal{E}_3 \cdot \mathcal{Y} - \mathcal{E}_3 \cdot \mathcal{E}_4 \right) \, , \\
\mathcal{A} \left( \mathbf{1}_{5} \right) \left( \lambda \right) &= \lambda \cdot \mathcal{S} \cdot \left( 3 \overline{\mathcal{K}}_{\mathcal{B}_6} - 2 \mathcal{W} \right) - \lambda \cdot \mathcal{S} \cdot \mathcal{X} \, .
\end{split}
\end{align}
In writing down the equality for the flux $\mathcal{A} \left( \mathbf{5}_{-2} \right) \left( \lambda \right)$ we have used that 
on $\hat{Y}_5$ 
\[ \mathbb{P}^1_{3H} \left( \mathbf{5}_{-2} \right) = V \left( P_T^\prime, e_3, a_{2,1} e_0 x z e_1 e_2 - a_{1,0} y \right) - V \left( P_T^\prime, e_3, e_4 \right) \, . \]
 Also note that the $U \left( 1 \right)_X$-flux -- as introduced in \cite{Krause:2011xj} -- can be expressed in terms of $\hat{Y}_5$ as\footnote{Strictly speaking, we should call this $\mathcal{A}_X \left( \frac{1}{5} \mathcal{F} \right)$. The normalization factor of $\frac{1}{5}$ has been introduced for historical reasons.}  
\[ \mathcal{A}_X \left( \mathcal{F} \right) = - \frac{1}{5} \cdot \mathcal{F} \cdot  \left( 5 \mathcal{S} - 5 \mathcal{Z} - 5 \overline{\mathcal{K}}_{\mathcal{B}_6} + 2 \mathcal{E}_1 + 4 \mathcal{E}_2 + 6 \mathcal{E}_3 + 3 \mathcal{E}_4 \right) \, . \label{equ:U1X-Flux-X5} \]
In this expression $F \in \mathrm{CH}^1 ( \mathcal{B}_6 )$ and $\mathcal{F} \in \mathrm{CH}^1 ( \hat{Y}_5 )$ satisfy $\mathcal{F} |_{\hat{Y}_4} = \hat\pi^*F$. 

Now that we have identified the gauge background, as a final step we wish to identify the lines bundles which encode the massless matter content of these fluxes. Following our general formalism we therefore have to compute the intersection product of the relevant 2-cycle class, which defines the gauge background, with the relevant matter surfaces, and then project the result onto the matter curves in question. There are two ways to perform these computations in practice: In \cref{subsec:ExampleMasslessSpectrumOfUniversalFlux} we follow the strategy outlined in \cref{subsec:NewStrategyForMasslessSpectra}. Equivalently, though computationally more involved, we can perform the intersection computations with the help of the presentations \cref{equ:MatterSurfaceFluxes-X5} of the gauge data as pullbacks from the ambient space. In \cref{subsec:ExampleOnMasslessSpectrumComputation} we will demonstrate this approach for the $U(1)_X$ gauge background. In addition, we present this computation for all matter surface fluxes in \cref{subsec:MasslessSpectraMSFSU5xU1}. The results of these two approaches match precisely.

\subsection{Zero Modes of \texorpdfstring{$\mathbf{A ( \mathbf{10}_{1} ) ( \lambda )}$}{A(10)(lambda)} via Transverse Intersections} \label{subsec:ExampleMasslessSpectrumOfUniversalFlux}

To derive the massless spectrum for the gauge background $\mathcal{A} ( \mathbf{10}_{1} ) ( \lambda )$, we first notice that the matter curve $C_{\mathbf{10}_{1}}$ intersects the matter curves $C_{\mathbf{5}_{3}}$, $C_{\mathbf{5}_{-2}}$ and $C_{\mathbf{1}_{5}}$ transversely. Hence, to compute the massless spectra induced by ${A} ( \mathbf{10}_{1} ) ( \lambda )$ of states localised on the latter three matter curves, we can directly apply \cref{intersectionMSF}. The transverse intersection numbers in the base have already been computed in \cref{transverseY}, and it merely remains to determine the intersection numbers in the fibre over the intersection points in question. These intersection numbers follow from the results listed in \cref{subsec:FibreStructureSU5xU1}. As explained in more detail therein, due to a seeming $\mathbb{Z}_2$-orbifold singularity in the top over the Yukawa locus $Y_1$ (despite $\hat{Y}_4$ being smooth) some of the intersection numbers of the rational curves in the top over the locus $Y_1$ are in fact fractional, see \cref{tab:IntersectionsOfFibralCurvesOverY1SU(5)xU(1)}. By use of the information displayed in
\cref{subsec:FibreStructureSU5xU1} the intersection products in \cref{equ:MatterSurfaceFluxes-X5} eventually take the form \footnote{Since their precise meaning is now clear, we omit the subscripts in the symbols for the projection and the intersection product.}
\[ \begin{array}{rclcrcl}
 \pi_\ast \left(S_{\mathbf{5}_{3}} \cdot A \left(\mathbf{10}_{1} \right)(\lambda) \right) &=& - \frac{2 \lambda}{5} \cdot Y_2 \, ,
 & & \pi_\ast \left(S_{\mathbf{1}_{5}} \cdot  A\left(\mathbf{10}_{1} \right)(\lambda) \right) &=& 0 \, , \\
 \pi_\ast \left(S_{\mathbf{5}_{-2}} \cdot  A\left(\mathbf{10}_{1} \right)(\lambda) \right) &=& - \frac{2 \lambda}{5} \cdot Y_2 + \frac{3 \lambda}{5} Y_1 \, .
\end{array} \label{transverseintersections10-example} \]
As an example consider the intersection in the second line. The 2-cycle ${A} ( \mathbf{10}_{1} ) ( \lambda)$ is given explicitly in \cref{equ:flux10Final} in terms of $\mathbb P^1$-fibrations over the curve $C_{{\mathbf{10}}_{1}}$. As for $S_{\mathbf{5}_{-2}}$, since the gauge background respects the $SU(5)$ gauge symmetry, we can pick any of the matter surfaces corresponding to the five states in $\mathbf{5}_{-2}$ as listed in \cref{app:MatterSurfaces52} of \cref{subsec:FibreStructureSU5xU1}. For instance, take $S^{(4)}_{\mathbf{5}_{-2}} = \mathbb P^1_{3H}(C_{\mathbf{5}_{-2}})$. Now, from \cref{transverseY} we read off that $C_{\mathbf{10}_{1}}$ and $C_{\mathbf{5}_{-2}}$ intersect over the two point sets $Y_1$ and $Y_2$. We hence need to study the splitting of the fibres of ${A} ( \mathbf{10}_{1} ) ( \lambda )$ and of $S^{(4)}_{\mathbf{5}_{-2}}$ over these two loci and compute the intersection numbers in the fibre. The relevant fibre splittings over $Y_1$ are listed in \cref{tab:Splitting10ToY1SU(5)xU(1)}, \cref{tab:Splitting5M2ToY1SU(5)xU(1)}, the ones over $Y_2$ in \cref{tab:Splitting10ToY2SU(5)xU(1)} and \cref{tab:Splitting5M2ToY2SU(5)xU(1)}. For instance, over $Y_1$ we have $S^{(4)}_{\mathbf{5}_{-2}} |_{Y_1} = \mathbb{P}^1_{3J}(Y_1)$ and
\begin{align}
\begin{split}
\left.{A} \left( \mathbf{10}_{1} \left( \lambda \right) \right)\right|_{Y_1} =& - \frac{2 \lambda}{5} \mathbb{P}^1_{24}(Y_1) + \frac{\lambda}{5} \mathbb{P}^1_{23}(Y_1) -  \frac{\lambda}{5} (\mathbb{P}^1_{34}(Y_1) + \mathbb{P}^1_{3J}(Y_1)) \\
& \quad + \frac{2 \lambda}{5} \left( \mathbb{P}^1_{24}(Y_1) + \mathbb{P}^1_{34}(Y_1) \right) \, ,  \\
\end{split}
\end{align}
The only non-zero intersection numbers between the fibral curves involved, as tabulated in \cref{tab:IntersectionsOfFibralCurvesOverY1SU(5)xU(1)}, are
\[ \left|\mathbb P^1_{3J}(Y_1) \cdot \mathbb P^1_{3J}(Y_1)\right| = -2, \qquad  \left|\mathbb P^1_{3J}(Y_1) \cdot \mathbb P^1_{34}(Y_1) \right|= 1 \, . \]
These are to be viewed as intersection numbers between two rational curves in the complex two dimensional top over $Y_1$. Summing everything up explains the term proportional to $Y_1$ in the second line of \cref{transverseintersections10-example}. A similar analysis is to be performed over $Y_2$. Finally, $- \frac{2 \lambda}{5} \cdot Y_2 + \frac{3 \lambda}{5} Y_1 \in \mathrm{CH}_0(C_{\mathbf{5}_{-2}}) \simeq \mathrm{Pic}^1(C_{\mathbf{5}_{-2}})$ defines a line bundle on $C_{\mathbf{5}_{-2}}$ given by \footnote{This is the case provided that $\lambda$ has been quantised properly. For general $\lambda \in \mathbb{Q}$, this will of course not happen to be the case.}
\[ L_{\mathbf{5}_{-2}} = \mathcal{O}_{C_{\mathbf{5}_{-2}}} \left( - \frac{2 \lambda}{5} \, Y_2 + \frac{3 \lambda}{5}\, Y_1\right) \, . \]
This line bundle has the property that (in general) it cannot be obtained by restriction of another line bundle on $W$ to the curve $C_{\mathbf{5}_{-2}}$. This is equivalent to the statement that the divisor class of $- \frac{2 \lambda}{5} Y_2 + \frac{3 \lambda}{5} Y_1$ does not arise as a complete intersection of the divisor class of $C_{\mathbf{5}_{-2}}$ with another divisor class on $W$. This will be important when it comes to evaluating the sheaf cohomologies of this line bundle, which count the massless matter states on $C_{\mathbf{5}_{-2}}$.

The computation of $ \pi_\ast (S_{\mathbf{10}_{1}} \cdot  A(\mathbf{10}_{1})(\lambda) )$ is more involved due to the self-intersection of $C_{\mathbf{10}_{1}}$. 
However, as pointed out before, there exist non-trivial linear relations between the 2-cycle representing the gauge backgrounds which allow us to treat non-transverse intersections of this type for a linear combination of transverse ones. In \cref{chapter:LocalAnomaliesInF-Theory} we will prove that in the model at hand these relations take the form 
\begin{align}
\begin{split} \label{relationsforAsinChow}
A \left( \mathbf{10}_{1} \right) \left(  \lambda \right) &= A \left( \mathbf{5}_{3} \right) \left( -\lambda \right) + A \left( \mathbf{5}_{-2} \right) \left( - \lambda \right) \, , \\
A \left( \mathbf{5}_{-2} \right) \left( \lambda \right) &= A_X \left( \lambda W \right) \, , \\
A \left( \mathbf{1}_{5} \right) \left( \lambda \right) &= A_X \left( - \lambda \left[ 6 \overline{K}_{\mathcal{B}_6} - 5 W \right] \right) + A \left( \mathbf{10}_{1} \right) \left( \lambda \right) \, .
\end{split}
\end{align}
These are the manifestation of a more general set of relations between 2-cycle classes which in fact follow, at a general level, from absence of gauge and gravitational anomalies in \emph{F-theory}. With the help of the first relation, it is readily verified that\footnote{The first relation in \cref{relationsforAsinChow} is equivalent to $\mathbb{P}^1_{3x} ( \mathbf{5}_{3} ) + \mathbb{P}^1_{3G} ( \mathbf{5}_{-2} ) = \mathbb{P}^1_{24} ( \mathbf{10}_{1} ) \in \mathrm{CH}^2( \hat{Y}_4)$. Alternatively, this enables us to rewrite the relevant matter surface such that the intersection in question is given as sum of two transverse intersections, leading to the same result.}
\begin{align}
\begin{split}
 \pi_\ast \left(S_{\mathbf{10}_{1}} \cdot  A\left(\mathbf{10}_{1}\right)(\lambda) \right) &= \pi_\ast \left(S_{\mathbf{10}_{1}} \cdot A\left(\mathbf{5}_{3} \right)(-\lambda) \right) + \pi_\ast \left(S_{\mathbf{10}_{1}} \cdot  A\left(\mathbf{5}_{-2} \right)(-\lambda) \right) \\
  & = - \frac{3 \lambda}{5} Y_1 + \frac{4 \lambda}{5} Y_2 \, ,
\end{split}
\end{align}
where the two transverse intersections appearing in the first line are computed analogously and the result is tabulated in \cref{table-N6}. This table also contains all other intersections between 2-cycles for matter surface backgrounds and matter surfaces. 

Another straightforward application of this formalism is a derivation of the line bundle induced by a Cartan flux on the matter curves. Such a gauge background is expressed as
\[ A \left( C \right) := \sum_{i = 1}^{4}{a_i E_i|_C} \in \mathrm{CH}^2 \left( \hat{Y}_4 \right) \, , \label{equ:GeneralCartanFlux} \] 
which automatically satisfies the transversality conditions \cref{TransverseChow1} and \cref{TransverseChow2}  but of course violates, by construction, \cref{condition-gauge-weaker}. Here $C \in \mathrm{CH}_1(W)$ denotes any curve on the 7-brane divisor. For instance, the hypercharge flux takes the form \cite{Donagi:2008kj,Mayrhofer:2013ara,Braun:2014pva}
\[ A_Y \left( C \right) =  \left. \left(2 E_1 + 4 E_2 + 6 E_3 + 3 E_4\right) \right|_{C} \, . \label{equ:HyperchargeFlux} \]
The intersections in the fibre of $A \left( C \right) |_{C_{\mathbf{R}}}$ with the matter surfaces $S^{a}_{\mathbf{R}}$ are readily worked out from the results of \cref{subsec:FibreStructureSU5xU1} for the representations present in this model and explicitly confirm the result \cref{Cartanfluxintersection}.

      \begin{table}[tbp]
      \centering
      \begin{tabular}{c@{\hskip 15pt}ccccc}
      \toprule
      & $A_X \left( F \right)$ & $A \left( \mathbf{10}_{1} \right) \left( \lambda \right)$ & $A \left( \mathbf{5}_{3} \right) \left( \lambda \right)$ & $A \left( \mathbf{5}_{-2} \right) \left( \lambda \right)$ & $A \left( \mathbf{1}_{5} \right) \left( \lambda \right)$ \\
      \midrule
      $C_{\mathbf{10}_{1}}$ & $\frac{1}{5} \left. F \right|_{C_{\mathbf{10}_{1}}}$ & $- \frac{3\lambda}{5} Y_1 + \frac{4 \lambda}{5} Y_2$ & $- \frac{2 \lambda}{5} Y_2$ & $ \frac{3 \lambda}{5} Y_1 - \frac{2 \lambda}{5} Y_2$ & 0 \\
      $C_{\mathbf{5}_{3}}$ & $\frac{3}{5} \left. F \right|_{C_{\mathbf{5}_{3}}}$ & $- \frac{2\lambda}{5} Y_2$ & $- \frac{2 \lambda}{5} Y_2 + \frac{\lambda}{5} Y_3$ & $ - \frac{\lambda}{5} Y_3 + \frac{4 \lambda}{5} Y_2$ & $- \lambda Y_3$ \\
      $C_{\mathbf{5}_{-2}}$ & $- \frac{2}{5} \left. F \right|_{C_{\mathbf{5}_{-2}}}$ & $\frac{3 \lambda}{5} Y_1 - \frac{2 \lambda}{5} Y_2$ & $\frac{4 \lambda}{5} Y_2 - \frac{\lambda}{5} Y_3$ & $- \frac{3 \lambda}{5} Y_1 - \frac{2 \lambda}{5} Y_2 + \frac{\lambda}{5} Y_3$ & $+ \lambda Y_3$ \\
      $C_{\mathbf{1}_{5}}$ & $\left. F \right|_{C_{\mathbf{1}_{5}}}$ & $0$ & $- \lambda Y_3$ & $\lambda Y_3$ & $- \left. \lambda a_{65} \right|_{C_{\mathbf{1}_{5}}}$ \\
      \bottomrule
      \end{tabular}
      \caption[Line bundles on matter curves whose cohomologies encode the massless spectrum.]{Divisors on the matter curves induced by the fluxes $A_X \left( F \right)$, $A \left( \mathbf{10}_{1} \right) \left( \lambda \right)$, $A \left( \mathbf{5}_{3} \right) \left( \lambda \right)$, $A \left( \mathbf{5}_{-2} \right) \left( \lambda \right)$ and $A \left( \mathbf{1}_{5} \right) \left( \lambda \right)$. The tensor product of the associated line bundles  $\mathcal{O}_{C_{\mathbf{R}}}$ with the spin bundle on the matter curve $C_{\mathbf{R}}$ is a line bundle, whose sheaf cohomologies count the massless matter localised on $C_{\mathbf{R}}$. We are using $a_{65} := 6 \overline{K}_{\mathcal{B}_6} - 5 W$.}
      \label{table-N5}
      \end{table}

      \begin{table}[tbp]
      \centering
      \resizebox{\textwidth}{!}{
      \begin{tabular}{c@{\hskip 15pt}ccccc}
      \toprule
      & $A_X \left( F \right)$ & $A \left( \mathbf{10}_{1} \right) \left( \lambda \right)$ & $A \left( \mathbf{5}_{3} \right) \left( \lambda \right)$ & $A \left( \mathbf{5}_{-2} \right) \left( \lambda \right)$ & $A \left( \mathbf{1}_{5} \right) \left( \lambda \right)$ \\
      \midrule
      $C_{\mathbf{10}_{1}}$ & $\frac{1}{5} \left[ F \right] \left[ \overline{K}_{\mathcal{B}_6} \right]$ & $\frac{\lambda}{5} \left[ \overline{K}_{\mathcal{B}_6} \right]  \left[ a_{65} \right]$ & $- \frac{2 \lambda}{5} \left[ \overline{K}_{\mathcal{B}_6} \right] \left[ a_{32} \right]$ & $\frac{\lambda}{5} \left[ \overline{K}_{\mathcal{B}_6} \right] \left[ W \right]$ & 0 \\
      $C_{\mathbf{5}_{3}}$ & $\frac{3}{5} \left[ F \right] \left[ a_{32} \right]$ & $- \frac{2 \lambda}{5} \left[ \overline{K}_{\mathcal{B}_6} \right] \left[ a_{32} \right]$ & $\frac{\lambda}{5} \left[ a_{32} \right] \left[ a_{23} \right]$ & $\frac{3 \lambda}{5} \left[ a_{32} \right] \left[ W \right]$ & $-\lambda \left[ a_{32} \right] \left[ a_{43} \right]$ \\
      $C_{\mathbf{5}_{-2}}$ & $- \frac{2}{5} \left[ F \right] \left[ a_{53} \right]$ & $ \frac{\lambda}{5} \left[ \overline{K}_{\mathcal{B}_6} \right] \left[ W \right]$ & $\frac{3 \lambda}{5} \left[ a_{32} \right] \left[ W \right]$ & $- \frac{2 \lambda}{5} \left[ W \right] \left[ a_{53} \right]$ & $\lambda \left[ a_{32} \right] \left[ a_{43} \right]$ \\
      $C_{\mathbf{1}_{5}}$ & $\left[ F \right] \left[ a_{32} \right] \left[ a_{43} \right]$ & 0 & $- \lambda \left[ a_{32} \right] \left[ a_{43} \right]$ & $\lambda \left[ a_{32} \right] \left[ a_{43} \right]$ & $- \lambda \left[ a_{65} \right] \left[ a_{43} \right] \left[ a_{32} \right]$ \\
      \bottomrule
      \end{tabular}
      }
      \caption[Chiralities of zero modes localised on matter curves.]{Chiralities of the massless spectra of the fluxes $A_X \left( F \right)$, $A \left( \mathbf{10}_{1} \right) \left( \lambda \right)$, $A \left( \mathbf{5}_{3} \right) \left( \lambda \right)$, $A \left( \mathbf{5}_{-2} \right) \left( \lambda \right)$ and $A \left( \mathbf{1}_{5} \right) \left( \lambda \right)$. These chiralities \emph{include} the contributions from the spin bundle. We have set $[a_{ij}]= i \left[ K_{\mathcal{B}_6} \right] - j \left[ W \right]$.}
      \label{table-N6}
      \end{table}

\subsection{Zero Modes of \texorpdfstring{$\mathbf{A_X( F )}$}{AX(F)} via Intersections in Ambient Space}\label{subsec:ExampleOnMasslessSpectrumComputation}

In this section we exemplify the explicit evaluation of the projection formula \cref{projectionformula1} using a different method. It is based on the representation of the 2-cycles describing the gauge background as the pullback of elements of $\mathrm{CH}^2(\hat{Y}_5)$ with $\hat{Y}_5$ the (non-toric) ambient space of $\hat{Y}_4$. Let us make this concrete for the $U ( 1 )_X$-background $A_X( F)$, postponing the analogous computation for the other types of gauge backgrounds to \cref{subsec:MasslessSpectraMSFSU5xU1}. In \cref{subsec:VerticalAndGaugeInvariantMatterSurfaceFluxes} we pointed out that this background can be described by restriction to $\hat{Y}_4$ of
\[ \mathcal{A}_X \left( F \right) = - \frac{1}{5} \mathcal{F} \cdot \left( 2 \mathcal{E}_1 + 4 \mathcal{E}_2 + 6 \mathcal{E}_3 + 3 \mathcal{E}_4 + 5 \mathcal{S} - 5 \mathcal{Z} - 5 \overline{\mathcal{K}}_{\mathcal{B}_6} \right) \in \mathrm{CH}^2 ( \hat{Y}_5 ) \, . \]
Our task is to compute the intersections $S^{a}_{\mathbf{R}} \cdot_{\iota_\mathbf{R}} {A}_X ( F ) $ for the matter curves $C_{\mathbf{10}_{1}}$, $C_{\mathbf{5}_{3}}$, $C_{\mathbf{5}_{-2}}$ and $C_{\mathbf{1}_{5}}$.  We do so by interpreting also $S^{a}_{\mathbf{R}}$ as an element $\mathcal{S}^{a}_{\mathbf{R}} \in \mathrm{CH}^3(\hat{Y}_5)$ and working entirely on $\hat{Y}_5$. By construction $\mathcal{A}_X ( F )$ is gauge invariant. Consequently, the result does not depend on which of the matter surfaces listed in \cref{app_ S10}, \cref{app_S53}, \cref{app:MatterSurfaces52} of \cref{subsec:FibreStructureSU5xU1} over a given matter curve we pick for each representation. For example, focus on intersections with the following matter surfaces \footnote{In \cref{app_S53} $S^{(4)}_{\mathbf{5}_{3}}$ is given as $\mathbb{P}^1_{3F}( \mathbf{5}_{3} )$. Since $\mathbb P^1_3$ splits into $\mathbb P^1_{3x}$ and $\mathbb P^1_{3F}$ over $C_{\mathbf{5}_{3}}$ and since we only consider intersections with 2-cycles representing gauge invariant backgrounds here, we can represent the matter state in this way.}
\begin{align*}
S^{(6)}_{\mathbf{10}_{1}} &= \mathbb{P}^1_{4D} \left( \mathbf{10}_{1} \right) = V \left( a_{10}, e_4, xs e_2 e_3 + a_{21} z^2 e_0 \right), \\
S^{(4)}_{\mathbf{5}_{3}} &= - \mathbb{P}^1_{3x} \left( \mathbf{5}_{3} \right) = - V \left( a_{32}, e_3, x \right), \\
S^{(4)}_{\mathbf{5}_{-2}} &=\mathbb{P}_{3H}^1 \left( \mathbf{5}_{-2} \right) = V \left( a_{3,2} a_{2,1} - a_{4,3} a_{1,0}, e_3, a_{4,3} e_0 x z e_1 e_2 - a_{3,2} y, a_{2,1} e_0 x z e_1 e_2 - a_{1,0} y  \right) \\
&= V \left( e_3, a_{43} e_0 e_1 e_2 x z - a_{32} y, a_{21} e_0 x z e_1 e_2 - a_{10} y \right) \\
S_{\mathbf{1}_{5}} &= \mathbb{P}^1_{A} \left( \mathbf{1}_{5} \right) =  V \left( a_{32}, a_{43}, s \right).
\end{align*}

Let us compute $\mathbb{P}^1_{4D} ( \mathbf{10}_{1} ) \cdot_{\iota_{\mathbf{10}_{1}}} \mathcal{A}_X ( F )$ term by term by first determining the vanishing ideal representing the intersection points in $\hat{Y}_5$. In determining this ideal, we are free to use the relations in $\mathrm{CH}(\hat{Y}_5)$ without changing the result up to rational equivalence. The key point is now that, as far as the toric fibre ambient space is concerned, rational and homological equivalence agree. We are therefore free to use the linear relations \cref{linearrelationsambient} and the relations encoded in the SR-ideal \cref{SRideal} of the ambient space. Furthermore, in the following $f$ will denote a polynomial in the coordinate ring of $\hat{Y}_5$ in the same Chow class as $\mathcal{F}$. With this in mind, we find
\begin{align}
\begin{split}
\mathcal{F} \cdot \mathcal{E}_1 \cdot \mathbb{P}^1_{4D} \left( \mathbf{10}_{1} \right) &= V \left( a_{1,0}, f, e_1, e_4, e_2 + a_{2,1} e_0 \right) \, , \\
\mathcal{F} \cdot \mathcal{E}_2 \cdot \mathbb{P}^1_{4D} \left( \mathbf{10}_{1} \right) &= V \left( a_{1,0}, f, e_2, e_4, a_{2,1} \right) \, , \\
\mathcal{F} \cdot \mathcal{E}_3 \cdot \mathbb{P}^1_{4D} \left( \mathbf{10}_{1} \right) &= V \left( a_{1,0}, f, e_3, e_4, a_{2,1} \right) \, , \\
\mathcal{F} \cdot \mathcal{E}_4 \cdot \mathbb{P}^1_{4D} \left( \mathbf{10}_{1} \right) &= V \left( a_{1,0}, f, e_1, e_4, e_2 + a_{2,1} e_0 \right) - 2 V \left( a_{1,0}, f, y, e_4, e_3 + a_{2,1} e_0 \right), \\
\mathcal{F} \cdot \overline{\mathcal{K}}_{\mathcal{B}_6} \cdot \mathbb{P}^1_{4D} \left( \mathbf{10}_{1} \right) &= V \left( a_{1,0}, f, e_1, e_4, e_2 + a_{2,1}    
      e_0 \right) - V \left( a_{1,0}, f, y, e_4, e_3 + a_{2,1} e_0 \right) \, .
\end{split}
\end{align}
In the fourth line we used the linear relation $\mathcal{E}_4 = \mathcal{E}_1 + 2 \mathcal{E}_2 + \mathcal{S} + 3 \mathcal{X} - 2 \mathcal{Y}$, and for the last line $\overline{\mathcal{K}}_{\mathcal{B}_6} = \mathcal{E}_1 + 2 \mathcal{E}_2 + \mathcal{E}_3 + \mathcal{S} + 2 \mathcal{X} - \mathcal{Y} - \mathcal{Z}$. Concerning the vanishing locus $V ( a_{1,0}, f, e_2, e_4, a_{2,1} )$, note that $e_2 = 0$ implies that this is a sublocus of the fibration over the GUT-surface. Inside $\mathcal{B}_6$ this locus is described by the intersection of the divisors $F$ (associated to $f$), $\overline{K}_{\mathcal{B}_6}$ and $2 \overline{K}_{\mathcal{B}_6} - W$ inside this surface. This intersection is the empty set. Therefore, we can discard this vanishing locus. Along the same lines we can discard $V ( a_{1,0}, f, e_3, e_4, a_{2,1} )$. Finally, note
\[ \mathcal{F} \cdot \mathcal{S} \cdot \mathbb{P}^1_{4D} \left( \mathbf{10}_{1} \right) = \emptyset \, , \qquad \mathcal{F} \cdot \mathcal{Z} \cdot \mathbb{P}^1_{4D} \left( \mathbf{10}_{1} \right) = \emptyset \, .  \]
Summing up all contributions we obtain
\[ \mathbb{P}^1_{4D} \left( \mathbf{10}_{1} \right) \cdot_{\iota_{\mathbf{10}_{1}}} \mathcal{A}_X \left( F \right) = \frac{1}{5} \cdot V \left( a_{10}, f, y, e_4, e_3 + a_{21} e_0 \right) \in \mathrm{CH}_0 \left(  \hat{Y}_4 |_{C_{\mathbf{10}_{1}}} \right) \, . \]
Upon use of the projection $\pi_{\mathbf{10}_{1}} \colon \hat{Y}_4 |_{C_{\mathbf{10}_{1}}} \twoheadrightarrow C_{\mathbf{10}_{1}}$ this yields
\[ \pi_{\mathbf{10}_{1} \ast} \left( \mathbb{P}^1_{4D} \left( \mathbf{10}_{1} \right) \cdot_{\iota_{\mathbf{10}_{1}}} \mathcal{A}_X \left( F \right) \right) = \frac{1}{5} \cdot \left. \mathcal{F} \right|_{C_{\mathbf{10}_{1}}} \in \mathrm{CH}_0 \left( C_{\mathbf{10}_{1}} \right) \, . \]
Along these lines one arrives at the results listed in \cref{table-N5}.

\subsection{Zero Modes of the Hypercharge Flux} \label{subsec:ZeroModesOfHyperchargeFlux}

In addition, to the above fluxes, we will consider the so-called \emph{hypercharge flux} in \cref{chapter:GUTModels}. The algebraic cycle $A^Y( \mathcal{H} )$ defining this gauge background depends on a curve $\mathcal{H} \subseteq W$. This curve is subject to a number of subtle constraints which we will discuss in detail in \cref{sec:ChoiceOfHyperchargeFluxAndExotics}. For the time being it suffices to demand that $\mathcal{H}$ is distinct from the matter curves $C_{\mathbf{10}_1}$, $C_{\mathbf{5}_3}$ and $C_{\mathbf{5}_{-2}}$ and that the self-intersections of $\mathcal{H}$ do not occur over the Yukawa loci $Y_1$, $Y_2$, $Y_3$ introduced in \cref{subsec:SelfIntersectionsOfMatterCurves}. As mentioned already it holds \cite{Donagi:2008kj,Mayrhofer:2013ara,Braun:2014pva}
\[ A_Y \left( \mathcal{H} \right) =  - \left. \left(2 E_1 + 4 E_2 + 6 E_3 + 3 E_4\right) \right|_{\mathcal{H}} \, . \]
The intersections in the fibre of $A \left( \mathcal{H} \right) |_{C_{\mathbf{R}}}$ with the matter surfaces $S^{a}_{\mathbf{R}}$ follow from the results presented in \cref{subsec:FibreStructureSU5xU1} and lead to the line bundles listed in \cref{table-N10}.

Based on this results one can anticipate that this flux breaks the $SU(5) \times U(1)_X$-symmetry. Indeed this is precisely the purpose of these fluxes in \emph{F-theory} GUT-models. We will discuss examples of such constructions in \cref{sec:ChoiceOfHyperchargeFluxAndExotics}.

\begin{table}[tbp]
\centering
\begin{tabular}{|c|c|c|c|c|}
\hline
 & $C_{\mathbf{10}_1}$ & $C_{\mathbf{5}_3}$ & $C_{\mathbf{5}_{-2}}$ & $C_{\mathbf{1}_{5}}$ \\
\hline \hline
$S_{*}^{(1)}$ & \multirow{3}{*}[-0pt]{$- 4 \cdot \left. \mathcal{H} \right|_{C_{\mathbf{10}_1}} + \frac{1}{2} K_{\mathbf{10}_1}$} 
              & \multirow{3}{*}[-0pt]{$- 2 \cdot \left. \mathcal{H} \right|_{C_{\mathbf{5}_3}} + \frac{1}{2} K_{\mathbf{5}_3}$}
              & \multirow{3}{*}[-0pt]{$- 2 \cdot \left. \mathcal{H} \right|_{C_{\mathbf{5}_{-2}}} + \frac{1}{2} K_{\mathbf{5}_{-2}}$}
              & $\frac{1}{2} K_{\mathbf{1}_5}$ \\ \cline{5-5}
$S_{*}^{(2)}$ & & & \\
$S_{*}^{(3)}$ & & & \\ \cline{2-4}
$S_{*}^{(4)}$ & \multirow{6}{*}[-0pt]{$1 \cdot \left. \mathcal{H} \right|_{C_{\mathbf{10}_1}} + \frac{1}{2} K_{\mathbf{10}_1}$}
              & \multirow{2}{*}[-0pt]{$3 \cdot \left. \mathcal{H} \right|_{C_{\mathbf{5}_3}} + \frac{1}{2} K_{\mathbf{5}_3}$}
              & \multirow{2}{*}[-0pt]{$3 \cdot \left. \mathcal{H} \right|_{C_{\mathbf{5}_{-2}}} + \frac{1}{2} K_{\mathbf{5}_{-2}}$} \\
$S_{*}^{(5)}$ & & & \\\cline{3-4}
$S_{*}^{(6)}$ & \\
$S_{*}^{(7)}$ & \\
$S_{*}^{(8)}$ & \\
$S_{*}^{(9)}$ & \\ \cline{2-2} 
$S_{*}^{(10)}$ & $6 \cdot \left. \mathcal{H} \right|_{C_{\mathbf{10}_1}} + \frac{1}{2} K_{\mathbf{10}_1}$ \\
\cline{1-2}
\end{tabular}
\caption[Zero modes of the hypercharge flux $A_Y ( \mathcal{H} )$.]{The zero modes of the hypercharge flux $A_Y ( \mathcal{H} )$ \cite{Donagi:2008kj,Mayrhofer:2013ara,Braun:2014pva} are counted by the sheaf cohomologies of the line bundles associated to the above divisors.}
\label{table-N10}
\end{table}

\section{Summary} \label{sec:ComputingMasslessSpectraWithGAP}

\paragraph{Zero Mode Counting and the Need for Coherent Sheaves}

In this chapter we have explained how an element $A \in \mathrm{CH}^2 ( \hat{Y}_4 )$ encodes the massless spectrum of a $G_4$-flux. Given such a Chow class $A$ and a matter surface $S^a_{\mathbf{R}}$ over a matter curve $C_{\mathbf{R}}$, we explained the computation of a divisor $D ( S^a_{\mathbf{R}}, A ) \in \mathrm{CH}^1 ( C_{\mathbf{R}} )$ and argued that the sheaf cohomologies of the following line bundle encode the massless spectrum
\[ L \left( S^a_\mathbf{R}, A \right) =  \mathcal{O}_{C_{\mathbf{R}}} \left( D ( S^a_\mathbf{R}, A) \right) \otimes \sqrt{K_{C_\mathbf{R}}} \in \mathrm{Pic} \left( C_\mathbf{R} \right) \, . \label{calLbundletocompute} \]
In consequence, the task at hand is to compute the sheaf cohomologies of this line bundle $L \left( S^a_\mathbf{R}, A \right)$.

For line bundles on $C_\mathbf{R}$ obtained by pullback, the computation of the associated sheaf cohomologies can often be achieved by use of \emph{cohomCalg}
\cite{Blumenhagen:2010pv, Blumenhagen:2010ed, Blumenhagen:2011xn, KoszulExtensionManual, cohomCalg:Implementation, 2011JMP....52c3506J, Rahn:2010fm} or techniques employed in heterotic string compactifications for complete intersection Calabi--Yau manifolds (CICYs) \cite{Anderson:2008ex}. Unfortunately though, in general the line bundles $L \left( S^a_\mathbf{R}, A \right)$  cannot be obtained as pullbacks. For instance, this is the case for the line bundle induced by the gauge background $A ( \mathbf{10}_{1} ) ( \lambda )$ on $C_{\mathbf{5}_{-2}}$, \ie \footnote{The flux $A ( \mathbf{10}_1 ) ( \lambda )$ is gauge invariant. Consequently, $D( S^a_{\mathbf{5}_{-2}}, A( \mathbf{10}_1 ) ( \lambda ) )$ is the same for all matter surfaces $S^a_{\mathbf{5}_{-2}}$. In this spirit we suppress the superscript `a' in this section.}
\[ \forall L \in \mathrm{Pic} \left( W \right): \qquad  \left. L \right|_{C_{\mathbf{5}_{-2}}} \not \cong L \left( S_{\mathbf{5}_{-2}}, A \left( \mathbf{10}_{1} \right) \left( \lambda \right) \right) \, . \]
A similar obstruction exists for the hypercharge gauge background in \emph{F-theory} GUT-Models on the divisor $W$, which cannot be obtained as a pullback bundle from the base $\mathcal{B}_6$ \cite{Donagi:2008kj,Mayrhofer:2013ara,Braun:2014pva,Beasley:2008kw}. Given these obstructions, \emph{cohomCalg} 
\cite{Blumenhagen:2010pv, Blumenhagen:2010ed, Blumenhagen:2011xn, KoszulExtensionManual, cohomCalg:Implementation, 2011JMP....52c3506J, Rahn:2010fm} and the techniques in \cite{Anderson:2008ex} are not applicable in our case, and we need to develop a new framework.

Even though $L ( S_\mathbf{R}, A )$ does in general not descend from a line bundle on an ambient space, we can extend this line bundle by zero outside of $C_{\mathbf{R}}$. The so-obtained object is a coherent sheaf $\mathcal{F}(S_\mathbf{R}, A)$ on the space into which $C_\mathbf{R}$ is embedded. In case $C_{\mathbf{R}}$ is embedded into a toric ambient space $X_\Sigma$, we thus obtain elements in $\mathfrak{Coh} ( X_\Sigma )$, the category of coherent sheaves on the toric ambient space $X_\Sigma$. Note that $\mathfrak{Coh} ( X_\Sigma )$ not only includes these non-pullback line bundles, but is far bigger than that. Also vector bundles which are not direct sums of line bundles, quotients thereof, T-branes in the language of \cite{Collinucci:2014qfa} or skyscraper sheaves are coherent sheaves. In this sense, the task at hand is to compute the sheaf cohomology groups for coherent sheaves. Our methods are based on \cite{2010arXiv1003.1943B, 2012arXiv1202.3337B, 2012arXiv1210.1425B, 2012arXiv1212.4068B, 2014arXiv1409.6100B, BL_GabrielMorphisms}. These methods, which we are extending further, apply as long as $X_\Sigma$ is a normal toric variety which is smooth and complete. Note however that smoothness of the matter curves is not required.

\begin{figure}[tb]
\centering
\begin{tikzpicture}

 \tikzstyle{every node}=[font=\small]

 \matrix [column sep=0em,row sep=1em] {
      \node [rectangle, draw=black, thick, fill=white, text width=0.8 \textwidth, text centered, minimum height=2em] (Chow) {Specify gauge background $A \in \mathrm{CH}^2 ( \hat{Y}_4 )$  (\cref{sec:F-theory}) }; \\
      \node [rectangle, draw=none, thick, fill=none, text width=0.8 \textwidth, text centered, minimum height=1em] (Intersect) {Intersection theory in the Chow ring}; \\
      \node [rectangle, draw=black, thick, fill=white, text width=0.8 \textwidth, text centered, minimum height=2em] (Divisor) {Divisor $D$ on matter curve $C_{\mathbf{R}}$ (\cref{sec:FromC3toL}): zero modes \, $\leftrightarrow$  \, $H^i(C_\mathbf{R}, \mathcal{O}_{C_{\mathbf{R}}}(D))$ }; \\
      \node [rectangle, draw=none, thick, fill=none, text width=0.4 \textwidth, text centered, minimum height=1em] (DivisorAsZeroLocus) { $D = V ( f_1, \dots, f_n ) \subseteq C_{\mathbf{R}}$ }; \\
      \node [rectangle, draw=black, thick, fill=white, text width=0.8 \textwidth, text centered, minimum height=2em] (RelationToIdealSheaf) {Line bundle $\mathcal{O}_{C_{\mathbf{R}}} ( D ) \cong \Hom_{\mathcal{O}_{C_{\mathbf{R}}}} ( \mathcal{I}_{C_{\mathbf{R}}} ( f_1, \dots, f_n ), \mathcal{O}_{C_{\mathbf{R}}} )$}; \\
      \node [rectangle, draw=black, dashed, thick, fill=none, text width=0.8 \textwidth, text ragged, minimum height=1em] (SwitchToToric) {
      \vspace{-0.7em}
      \begin{itemize}
       \item Toric variety $X_\Sigma$ over $\mathbb{Q}$ (\cref{subsec:TowardsToricVarieties})
       \item $C_{\mathbf{R}} = V ( g_1, \dots, g_k )$, \, $D = V ( f_1, \dots, f_n )$ in $X_\Sigma$, \, $S ( C_{\mathbf{R}} ) = S / \langle g_1, \dots, g_k 
            \rangle$.
      \end{itemize}
      }; \\    
      \node [rectangle, draw=black, thick, fill=white, text width=0.8 \textwidth, text centered, minimum height=2em] (SheafifyIdeal) {
       F.p. graded $S ( C_{\mathbf{R}} )$-module $M_{C_{\mathbf{R}}}$ with $\tilde{M_{C_{\mathbf{R}}}} \cong \mathcal{O}_{C_{\mathbf{R}}} ( D )$ (\cref{sec:Sheafification}, \cref{subsec:LineBundlesFromModules})}; \\
      \node [rectangle, draw=none, thick, fill=none, text width=0.8 \textwidth, text centered, minimum height=1em] (ExtensionByZero) {Extend $\tilde{M_{C_{\mathbf{R}}}}$ by zero outside of $C_{\mathbf{R}}$.}; \\    
      \node [rectangle, draw=black, thick, fill=white, text width=0.8 \textwidth, text centered, minimum height=2em] (Module) { $M \in S \mathrm{\textnormal{-}fpgrmod}$ s.t. $\tilde{M} \in \mathfrak{Coh} ( X_\Sigma )$ satisfies $\left. \tilde{M} \right|_{C_{\mathbf{R}}} \cong \mathcal{O}_{C_{\mathbf{R}}} ( D )$  (\cref{subsec:LineBundlesFromModules})}; \\
      \node [rectangle, draw=none, thick, fill=none, text width=0.8 \textwidth, text centered, minimum height=1em] (IdentifyIdealForSheafCohomologyComputation) {Analyse $M$ with the \texttt{gap}-package \cite{SheafCohomologyOnToricVarieties}.}; \\
      \node [rectangle, draw=black, thick, fill=white, text width=0.8 \textwidth, text centered, minimum height=2em] (Compute) {
      $I \subseteq S$ such that $H^i ( X_\Sigma, \tilde{M} ) \cong \mathrm{Ext}^i_S ( I, M )_0$   via \cite{SheafCohomologyOnToricVarieties}  \,(\cref{subsec:SheafCohomologyFromFPGradedSModules}) }; \\
      };

    \path[draw, thick] (Chow) -- (Intersect);
    \path[draw, thick, -latex'] (Intersect) -- (Divisor);
    \path[draw, thick] (Divisor) -- (DivisorAsZeroLocus);
    \path[draw, thick, -latex'] (DivisorAsZeroLocus) -- (RelationToIdealSheaf);
    \path[draw, thick] (RelationToIdealSheaf) -- (SwitchToToric);
    \path[draw, thick, -latex'] (SwitchToToric) -- (SheafifyIdeal);
    \path[draw, thick] (SheafifyIdeal) -- (ExtensionByZero);
    \path[draw, thick, -latex'] (ExtensionByZero) -- (Module);
    \path[draw, thick] (Module) -- (IdentifyIdealForSheafCohomologyComputation);
    \path[draw, thick, -latex'] (IdentifyIdealForSheafCohomologyComputation) -- (Compute);
    
\end{tikzpicture}
\caption[Towards computing zero modes via the \texttt{gap}-package \cite{SheafCohomologyOnToricVarieties}.]{The sheaf cohomologies $H^i(C_\mathbf{R}, \mathcal{O}_{C_{\mathbf{R}}}(D)) \simeq H^i ( X_\Sigma, \tilde{M} )$ encode the zero modes. We compute them with the above algorithm by use of the \texttt{gap}-package \cite{SheafCohomologyOnToricVarieties}.}
\label{figure-987654}
\end{figure}
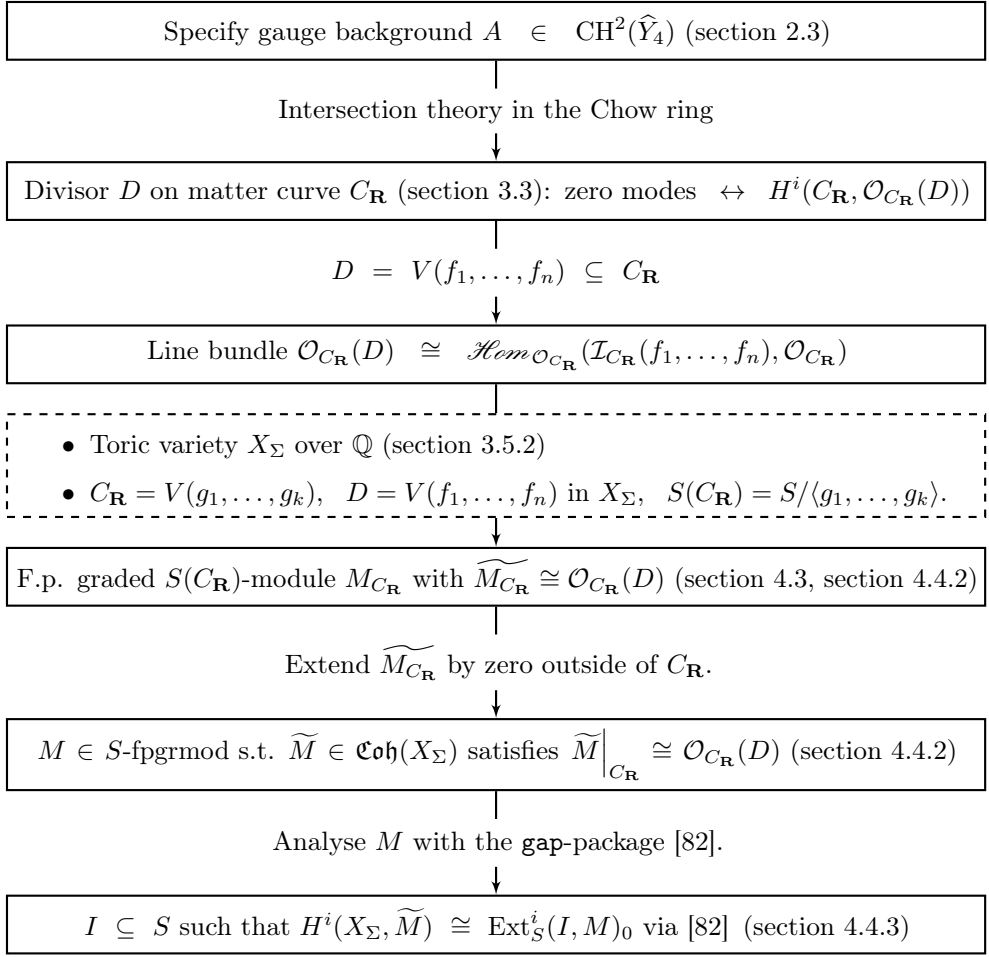

In \cref{sec:ComputingTheSpectra} we will employ these methods. As preparation to the technical discussions in \cref{chapter:DetailsOnFPGradedSModules}, we found it instructive to summarise our overall strategy in \cref{figure-987654}. Let us therefore use this opportunity, and discuss the individual steps in this flow chart briefly.

\paragraph{Line Bundles as Computer Input}

Let us return to our task of computing the sheaf cohomology groups of the line bundle $L ( S_{\mathbf{R}}, A )$ defined, as in \cref{calLbundletocompute}, via a divisor 
\[D  = D ( S_\mathbf{R}, A ) + \sqrt{K_{C_\mathbf{R}}} \in \mathrm{CH}^1(C_\mathbf{R}) \, . \]
Here $\sqrt{K_{C_{\mathbf{R}}}}$ denotes, by slight abuse of notation, the divisor on $C_{\mathbf{R}}$ associated with the spin bundle induced by the embedding of the curve $C_\mathbf{R}$ into $\mathcal{B}_6$. From \cref{subsec:LineBundlesOnRiemannSurfaces} we know that such a Cartier divisor on a complex variety gives rise to a holomorphic line bundle $\mathcal{O}_X ( D )$. This is also true in algebraic geometry on a variety $X$ over $\mathbb{Q}$. Then over Zariski open $U \subseteq X$, the local sections of $\mathcal{O}_X ( D )$ are given by
\[ \mathcal{O}_X \left( D \right) \left( U \right) = \left\{ \left. f \colon U \to \mathbb{Q} \text{ rational } \; \right| \; \left( \mathrm{div} \left( f \right) + D \right)_{U} \geq 0 \right\} \cup \{ 0 \} \, , \label{equ:DefinitionOfOX(D)} \]
where $\mathrm{div}(f)$ denotes the (principal) divisor associated with the not identically vanishing rational function $f$. If $D |_U = V \left( d_U \right)$ for a suitable function $d_U$ -- which strictly speaking is only possible if $U$ is `small enough' -- then the requirement $(\mathrm{div}(f) + D)|_{U} \geq 0$ means that the product $f \, d_U$ has no poles on $U$.

This definition is rather abstract. In addition, it is not at all obvious at this stage how we can actually encode this data in a form understandable for computers. To bridge this gap, let us also recall the notion of a so-called \emph{ideal sheaf}. To this end, we first look at the sheaf of regular functions $\mathcal{O}_X$ on an algebraic variety $X$, which assigns to every open subset $U \subseteq X$ the set $\mathcal{O}_X ( U ) = \{ f \colon U \to \mathbb{Q} \; , \; f \text{ regular } \}$. Note that $\mathcal{O}_X ( U )$ forms a (commutative and unitial) ring. Now let us consider global sections $f_i \in H^0 \left( X, \mathcal{O}_X \left( D_i \right) \right)$ for suitable divisors $D_i \in \mathrm{Cl} ( X )$ and $1 \leq i \leq n$. In addition, let $\mathcal{U} = \{ U_j \}_{j \in J}$ be an affine open cover of $X$. Consequently, $\left. f_i \right|_{U_j} \in \mathcal{O}_X ( U_j )$. Therefore, these global sections $f_i$ cut out an algebraic subvariety $Y \subseteq X$ defined by
\[ Y \cap U_j = V \left( \left. f_1 \right|_{U_j}, \dots, \left. f_n \right|_{U_j} \right) \, . \]
For every open $W \subseteq U_j$ we have the ideal $\left\langle \left. f_1 \right|_W, \left. f_2 \right|_W, \dots, \left. f_n \right|_W \right\rangle \subseteq \mathcal{O}_X \left( W \right)$. This assignment of ideals forms a sheaf on $U_j$. Finally, these sheaves on the affine patches $U_j$ glue to form a sheaf on $X$ -- \emph{the ideal sheaf $\mathcal{I}_X ( f_1, \dots, f_n )$ of $Y$}.\footnote{More generally, one can use any closed embedding $\iota \colon Y \hookrightarrow X$ to define the morphism of sheaves $\iota^\sharp \colon \mathcal{O}_X \to \iota_\ast \mathcal{O}_Y$. The kernel of $\iota^\sharp$ is then termed the ideal sheaf of $Y$ in $X$.}

Now let us look at a Cartier divisor $D \subseteq C_{\mathbf{R}}$, which we assume to be given by $D = V ( f_1, \dots, f_n )$ for global sections $f_1, \dots, f_n$. \footnote{The divisor $D$ need not be a complete intersection. Consequently, there is no contradiction between $D$ being of codimension $1$ and it being cut out by more than one global section.} We can then wonder if there is a relation between the ideal sheaf $\mathcal{I}_{C_{\mathbf{R}}} ( f_1, \dots, f_n )$ and the line bundle $\mathcal{O}_{C_{\mathbf{R}}} ( D )$. And indeed, by proposition 6.18 of \cite{hartshorne1977algebraic},
\[ \mathcal{O}_{C_{\mathbf{R}}} ( -D ) \cong \mathcal{I}_{C_{\mathbf{R}}} \left( f_1, \dots, f_n \right) \, . \label{Idealshaefdef} \]
Up to an important $-1$, this brings us close to handling the line bundles $\mathcal{O}_{C_{\mathbf{R}}} ( D )$ in a form understandable to computers.

To overcome the additional $-1$, let us recall that line bundles $\mathcal{O}_{C_{\mathbf{R}}} ( D )$ are invertible sheaves. This means that there exists another sheaf $\mathcal{F}$ on $C_{\mathbf{R}}$ with the property $\mathcal{O}_{C_{\mathbf{R}}} ( D ) \otimes_{\mathcal{O}_{C_{\mathbf{R}}}} \mathcal{F} \cong \mathcal{O}_{C_{\mathbf{R}}}$. We can describe this sheaf quite explicitly. Namely we consider all sheaf homomorphisms from the sheaf $\mathcal{O}_{C_{\mathbf{R}}}$ to the line bundle $\mathcal{O}_{C_{\mathbf{R}}} ( D )$. It is a well-known fact that these homomorphisms form a sheaf, the so-called \emph{sheaf-Hom} $\Hom_{\mathcal{O}_{C_{\mathbf{R}}}} ( \mathcal{O}_{C_{\mathbf{R}}}, \mathcal{O}_{C_{\mathbf{R}}} ( D ) )$. It can be shown that $\Hom_{\mathcal{O}_{C_{\mathbf{R}}}} ( \mathcal{O}_{C_{\mathbf{R}}} ( D ), \mathcal{O}_{C_{\mathbf{R}}} )$ is isomorphic to $\mathcal{O}_{C_{\mathbf{R}}} ( -D )$. Hence, we reach the conclusion
\[ \mathcal{O}_{C_{\mathbf{R}}} \left( D \right) \cong \Hom_{\mathcal{O}_{C_{\mathbf{R}}}} \left( \mathcal{I}_{C_{\mathbf{R}}} \left( f_1, \dots, f_n \right), \mathcal{O}_{C_{\mathbf{R}}} \right) \, . \label{equ:OXDAndIdealSheaf} \]
This formula connects the defining data of the divisor $D  = D(S_\mathbf{R}, A) + \sqrt{K_{C_\mathbf{R}}} \in \mathrm{CH}^1(C_\mathbf{R})$ far more explicitly to the line bundle $L ( S_{\mathbf{R}}, A )$ than \eg \cref{equ:DefinitionOfOX(D)}. But we can do even better.

\paragraph{Ideal Sheaves on Toric Varieties}

To achieve an even more explicit description, we now turn our attention to toric varieties $X_\Sigma$. Every toric variety $X_\Sigma$ comes equipped with a coordinate ring $S$ -- typically referred to as the Cox ring -- which is graded by $\mathrm{Cl} ( X_\Sigma )$. This ring is very similar to the coordinate ring $R$ of an affine variety $\mathcal{A}$. Recall from \cref{subsec:CoherentSheavesOnVarieties} that every $R$-module $M$ encodes a coherent sheaf $\tilde{R}$ on $\mathcal{A}$. It turns out that there exists a similar sheafification process for toric varieties: As we will explain in detail in \cref{chapter:DetailsOnFPGradedSModules} there exists the so-called \emph{sheafification functor}
\[ \widetilde{\phantom{m}} \colon S \mathrm{\textnormal{-}fpgrmod} \to \mathfrak{Coh} X_\Sigma \, . \label{equ:SheafificationFunctorMainText} \]
This functor turns a so-called \emph{finitely presented (f.p.) graded $S$-module} into a coherent sheaf on $X_\Sigma$. Homogeneous ideals $I \subseteq S$ are special examples of \fp graded $S$-modules. Hence, for homogeneous polynomials $f_1, \dots, f_n \in S$, we can turn the ideal $I = \langle f_1, \dots, f_n \rangle \subseteq S$ into a coherent sheaf $\tilde{I}$ on $X_\Sigma$. And indeed $\tilde{I} \cong \mathcal{I}_{X_\Sigma} ( f_1, \dots, f_n )$, \ie the sheafification of the ideal $I$ provides nothing but the ideal sheaf on $X_\Sigma$ generated by $f_1, \dots, f_n$. In this sense, our next best model for the sheaf $\mathcal{I}_{X_\Sigma} ( f_1, \dots, f_n )$ is the ideal $\langle f_1, \dots, f_n \rangle$ itself, which provides a very explicit description for this sheaf.

\paragraph{Sheafification of F.P. Graded Modules on Toric Varieties}

For the presentation of the ideal it turns out more practical to specify the relations satisfied by its generators than specifying the generators themselves. Such relations are conveniently expressed as a linear map $M$ acting on (finite) direct sums of modules over $S$ respecting the grading. This is explained further in \cref{sec:ProjSmodule}). Such a homomorphism $M$ is a \emph{finitely presented (f.p.) graded $S$-module}, in the sense defined in \cref{subsec:FPGradedSModules}.

The rough idea behind the functor \cref{equ:SheafificationFunctorMainText} is as follows: The toric variety $X_\Sigma$ is defined by the combinatorics of a fan $\Sigma$. For every cone $\sigma \in \Sigma$ there is an affine patch $U_\sigma$ and a monomial $x^{\hat{\sigma}} \in S$. Given an \fp graded $S$-module $M$, we can perform a so-called homogeneous localisation of $M$ with respect to $x^{\hat{\sigma}}$, which we revised in \cref{subsec:CoherentSheavesOnVarieties}. The result is an \fp graded $S_{(x^{\hat{\sigma}})}$-module $M_{(x^{\hat{\sigma}})}$, which defines a unique coherent sheaf $\tilde{M_{(x^{\hat{\sigma}})}}$ on $U_\sigma$ with the property \cite{hartshorne1977algebraic}
\[ \tilde{M_{(x^{\hat{\sigma}})}} \left( U_\sigma \right) = M_{(x^{\hat{\sigma}})} \, . \]
This last step employs the sheafification process on affine varieties as explained in \cref{subsec:CoherentSheavesOnVarieties}. We have thus obtained a coherent sheaf on every affine patch $U_\sigma$ of $X_\Sigma$. As a consequence of \cref{equ:Gluing} these sheaves glue together to form a sheaf on the entire variety $X_\Sigma$ \cite{cox2011toric}.

\paragraph{Line Bundles on Matter Curves in Toric Varieties as Computer Input}

Given a divisor $D = V ( f_1, \dots, f_n ) \subseteq X_\Sigma$, we see from \cref{equ:OXDAndIdealSheaf} that we need to invert the sheaf $\tilde{I}$ associated with $I = \langle f_1, \dots, f_n \rangle$ to describe $\mathcal{O}_X ( D )$. We are thus looking for an analogue of this equation in terms of \fp graded $S$-modules. Motivated by the fact $\tilde{S} \cong \mathcal{O}_X$, which can of course be proven rigorously, the analogue in question is $M = \mathrm{Hom}_S ( I, S )$ (c.f. \cref{sec:ExtOfFPModules}). The so-defined \fp graded $S$-module $M$ now satisfies $\tilde{M} \cong \mathcal{O}_{X_\Sigma} ( D )$. We provide more details in \cref{subsec:LineBundlesFromModules}.

Finally note that in general the matter curve $C_{\mathbf{R}}$ is not a divisor in a toric ambient space $X_\Sigma$ but of higher codimension. Suppose that $C_{\mathbf{R}} = V ( g_1, \dots, g_k )$ and $D = V ( f_1, \dots, f_n ) \subseteq C_{\mathbf{R}}$ for homogeneous polynomials $g_i, f_i \in S$. Then we can consider the  ring $S ( C_{\mathbf{R}} ) = S / \langle g_1, \dots, g_k \rangle$ and construct from $f_1, \dots, f_n$ an \fp graded $S ( C_\mathbf{R} )$-module $M_{C_\mathbf{R}}$ such that $\tilde{M_{C_\mathbf{R}}} \cong \mathcal{O}_{C_{\mathbf{R}}} ( D )$. To make use of the structure of the toric ambient space $X_\Sigma$, where we can for example apply the \emph{cohomCalg}-algorithm \cite{Blumenhagen:2010pv, Blumenhagen:2010ed, Blumenhagen:2011xn, KoszulExtensionManual, cohomCalg:Implementation, 2011JMP....52c3506J, Rahn:2010fm}, we now turn this module $M_{C_\mathbf{R}}$ into an \fp graded $S$-module $M$ such that $\tilde{M} \in \mathfrak{Coh} ( X_\Sigma )$ is the coherent sheaf on $X_\Sigma$ which is zero outside of $C_{\mathbf{R}}$ and matches $\mathcal{O}_{C_{\mathbf{R}}} ( D )$ on the matter curve $C_{\mathbf{R}}$. We explain the transition from $M_{C_\mathbf{R}}$ to $M$ in \cref{subsec:LineBundlesFromModules}.

\paragraph{Sheaf Cohomology of Coherent Sheaves from F.P. Graded Modules}

This now brings us to the final question: Given an \fp graded $S$-module $M$, how do we compute the sheaf cohomology dimension of $\tilde{M}$ from the data defining $M$?
Before we answer this question, let us mention that for any two \fp graded $S$-modules $M$, $N$ one can compute extension groups -- denoted by $\mathrm{Ext}^i_S ( M, N )$ -- which are \fp graded $S$-modules themselves. Details are given in \cref{sec:ExtOfFPModules}. Note that $\mathrm{Ext}^0_S ( M, N ) \cong \mathrm{Hom}_S ( M, N )$, which we already encountered in the dualisation process above. In particular, we can truncate $\mathrm{Ext}^i_S ( M, N )$ to any $d \in \mathrm{Cl} ( X_\Sigma )$. Now, to compute the sheaf cohomologies of $\tilde{M}$, we have designed algorithms which compute an ideal $I \subseteq S$ such that 
\[ H^i ( X_\Sigma, \tilde{M} ) \cong \mathrm{Ext}^i_S ( I, M )_0 \, . \]
Our implementations \cite{CAPCategoryOfProjectiveGradedModules, CAPPresentationCategory, PresentationsByProjectiveGradedModules, TruncationsOfPresentationsByProjectiveGradedModules, SheafCohomologyOnToricVarieties}, whose mathematical and algorithmic foundations are explained in \cref{chapter:MathDetailsSheafCohomologies}, are optimised in the following ways:
\begin{itemize}
 \item We employ \emph{cohomCalg} \cite{Blumenhagen:2010pv, Blumenhagen:2010ed, Blumenhagen:2011xn, KoszulExtensionManual, cohomCalg:Implementation, 2011JMP....52c3506J, 
      Rahn:2010fm} to compute vanishing sets $V^i$ on smooth, complete and simplicial, projective toric varieties. These vanishing sets identify 
      all line bundles $L$ on $X_\Sigma$ for which $h^i ( X_\Sigma, L ) = 0$. These vanishing sets serve as properly refined versions of the semigroup $\mathbb{\mathcal{K}}^{\mathrm{sat}}$ introduced in \cite{Maclagan03multigradedcastelnuovo-mumford} which was put to use \eg in \cite{Oberwolfach} to propose a means to compute sheaf cohomology of coherent sheaves on smooth, projective toric varieties. Our algorithm employs these refined vanishing sets and can thereby compute the sheaf cohomologies of all coherent sheaves on smooth and complete toric varieties.
 \item The computation of generators of $H^i ( X_\Sigma, \tilde{M} )$ requires the use of Gr\"obner basis computations. A first approximation to this data is the mere 
      number of these generators, or equivalently the $\mathbb{Q}$-dimension of $H^i ( X_\Sigma, \tilde{M} )$. We have found that for the computation of these cohomology dimensions, it is possible to replace some Gr\"obner basis computations by Gau{\ss} eliminations. The use of these Gau{\ss} eliminations allowed us to compute sheaf cohomology dimensions significantly faster than a set of generators.
 \item Finally, the implementations of our algorithms use of parallel computing.
\end{itemize}

\paragraph{Outlook}
In the next chapter, we will explain the category $S \mathrm{\textnormal{-}fpgrmod}$ of \fp graded $S$-modules and the computation of sheaf cohomologies in more detail.
Eventually, this leads us to study an \emph{F-theory} toy models in \cref{sec:ComputingTheSpectra}, in which we exemplify the powers of these algorithms. Subsequently, in \cref{chapter:GUTModels}, we discuss more refined examples of \emph{F-theory} GUT-models. In these setups, a hypercharge flux is used for the breaking of the GUT group. This mechanism is very similar to the Higgs effect, which we employed for the gauge group breaking in \cref{subsec:Georgi-Glashow}. Crucially though, hypercharge fluxes must not be formed from pullback line bundles \cite{Braun:2014pva, Beasley:2008dc, Beasley:2008kw, Donagi:2011jy, Donagi:2008ca, Donagi:2011dv}. Consequently, the computation of zero modes in such \emph{F-theory} GUT-models is made possible only by use of our implementations \cite{CAPCategoryOfProjectiveGradedModules, CAPPresentationCategory, PresentationsByProjectiveGradedModules, TruncationsOfPresentationsByProjectiveGradedModules, SheafCohomologyOnToricVarieties}.

\chapter{From F.P. Graded \texorpdfstring{$\mathbf{S}$}{S}-Modules to Zero Mode Counting} \label{chapter:DetailsOnFPGradedSModules}
The following chapter is taken in large parts from \cite{Bies:2017fam}. It provides an introduction to the mathematics underlying the sheafification functor
\[ \widetilde{\phantom{m}} \colon S \mathrm{\textnormal{-}fpgrmod} \to \mathfrak{Coh} X_\Sigma \, , \label{equ:SheafificationFunctorIntroduction} \]
which turns an \fp graded $S$-module $M$ into a coherent sheaf $\tilde{M}$ on $X_\Sigma$, summarises how we compute the sheaf cohomologies of $\tilde{M}$ from $M$ and how we can apply this to zero mode counting in \emph{F-theory}.

We begin with a general review of the category of \fp graded modules in \cref{sec:ProjSmodule} and \cref{subsec:FPGradedSModules}. Based on the material introduced in 
\cref{sec:RevisionOnSheavesAndSheafCohomology}, \cref{subsec:CoherentSheavesOnVarieties} and \cref{subsec:TowardsToricVarieties} we then describe the sheafification functor on toric varieties in \cref{sec:Sheafification}. In particular, we point out that the functor in \cref{equ:SheafificationFunctorIntroduction} is no equivalence of categories. This redundancy is actually important to the algorithm underlying our sheaf cohomology computations. The theoretical foundation for this algorithm comes from a mathematical theorem, which we prove in \cref{chapter:MathDetailsSheafCohomologies}. It shows that one can compute the sheaf cohomologies of interest from extension modules of \fp graded $S$-modules. We explain the computation of these extension modules in \cref{sec:ExtOfFPModules}. The details underlying our algorithm are all summarised in \cref{chapter:MathDetailsSheafCohomologies}. For convenience to the reader we summarise the general outline in \cref{subsec:SheafCohomologyFromFPGradedSModules}.

We conclude this chapter with an example application to an \emph{F-theory} toy model in \cref{sec:ComputingTheSpectra}. The geometry of this \emph{F-theory} vacuum is derived from the $SU(5) \times U(1)_X$-top discussed in \cref{subsec:SpecialFTheoryGUTModel}. In particular, all results which we obtained on this geometry can be put to use in this sample geometry. 

With a view towards massless spectra of \emph{F-theory}, recall from \cref{chapter:MasslessSpectraAndSheafCohomology} that we are essentially facing the following task:
\ebox{Given a toric variety $X_\Sigma$ with Cox ring $S$, a matter curve $C \subseteq X_\Sigma$ and a divisor $D \in \mathrm{Div} ( C )$, construct \fp graded $S$-modules $M_{\pm}$ such that $\tilde{M_{\pm}} \in \mathfrak{Coh} \left( X_\Sigma \right)$ are supported on $C$ only and satisfy $\left. \tilde{M_{\pm}} \right|_C \cong \mathcal{O}_{C} ( \pm D )$. Finally, compute $H^i ( X_\Sigma, \tilde{M_{\pm}} )$.}
In \cref{subsec:LineBundlesFromModules} we will explain how we identify such \fp graded $S$-modules $M_{\pm}$. Subsequently, we compute their sheaf cohomologies in \cref{subsec:Spectrum1} for different values of the complex structure moduli of the matter curves. As anticipated in \cref{subsec:LineBundlesOnRiemannSurfaces} already, we expect jumps in the cohomology dimensions as we move along this moduli space of complex structure moduli. Indeed our explicit computations confirm this expectation in this very example geometry.

Some words of caution before we start:
In a supersymmetric context, it is common practice to work with \emph{analytic} geometry in physics. Unfortunately since the complex numbers $\mathbb{C}$ are not suitable to be modelled in a computer, it is necessary for us to switch to a finite field extension of the rational numbers $\mathbb{Q}$. The same limitations hold for holomorphic functions, \ie absolutely convergent power series. Hence, computer applications require us to switch to \emph{algebraic} geometry over (finite field extensions of) the rational numbers $\mathbb{Q}$. Therefore, unless stated explicitly, all explicit computations in this thesis are performed in (toric) varieties over $\mathbb{Q}$ with Cox ring $S$. We will provide a basic introduction to these spaces from the point of view of algebraic geometry. 

Whilst we formulate our approach in the language of toric varieties, a scheme-theoretic approach is indeed possible. The interested reader may wish to consult \cite{RohrerDissertation, 2011arXiv1107.2483R, 2012arXiv1212.3956R, 2011arXiv1107.2713R} and references therein.

\section{The Category of Projective Graded \texorpdfstring{$\mathbf{S}$}{S}-Modules} \label{sec:ProjSmodule}

We assume that $X_\Sigma$ is a normal toric variety over $\mathbb{Q}$ which is smooth and complete. For such varieties $X_\Sigma$ the coordinate ring $S$ -- termed the Cox ring -- is a polynomial ring $S = \mathbb{Q} [ x_1, \dots, x_m ]$ which is graded by $\mathrm{Cl} ( X_\Sigma ) \cong \mathbb{Z}^n$. This means that there is a homomorphism of monoids
\[ \mathrm{deg} \colon \mathrm{Mon} ( S ) \to \mathbb{Z}^n \]
such that the images of the monomials $x_1, \dots, x_m$ generate $\mathbb{Z}^n$ as a group. Here $\mathrm{Mon} ( S )$  denotes the set of monomials in $S$. For a monomial $f \in \mathrm{Mon} ( S )$ we term $\mathrm{deg} ( f )$ the \emph{degree of $f$}. A polynomial $P \in S$ for which all its monomials have identical degree $d \in \mathbb{Z}^n$ is termed a homogeneous polynomial (of degree $d$). The homogeneous elements of degree $d$ form a subgroup $S_d$ of $S$. As a group, the ring $S$ therefore admits a direct sum decomposition
\[ S = \bigoplus_{d \in \mathbb{Z}^n}{S_d} \]
such that the multiplication in $S$ satisfies $S_d \cdot S_e \subseteq S_{d+e}$ for all $d,e \in \mathbb{Z}^n$. We term the group $S_d$ the \emph{degree $d$ layer of $S$}.

Given a $\mathbb{Z}^n$-graded Cox ring $S = \mathbb{Q} [ x_1, \dots, x_m ]$, we can define for every $d \in \mathbb{Z}^n$ a degree-shift of this ring. Namely $S ( d )$ is the $\mathbb{Z}^n$-graded ring with $S ( d )_e = S ( 0 )_{e+d} \equiv S_{e+d}$.

As an example consider $\mathbb{P}^2_{\mathbb{Q}}$. This toric variety has $\mathrm{Cl} ( \mathbb{P}^2_{\mathbb{Q}} ) = \mathbb{Z}$. Its Cox ring $\mathbb{Q} [ x_1, x_2, x_3 ]$ is $\mathbb{Z}$-graded upon $\mathrm{deg} ( x_1 ) = \mathrm{deg} ( x_2 ) = \mathrm{deg} ( x_3 ) = 1$. In particular, $1 \in S ( 0 )$ satisfies $\mathrm{deg} ( 1 ) = 0$. Now consider the ring $S ( -1 )$. By definition it satisfies
\[ S \left( -1 \right)_1 = S \left( 0 \right)_{1 + \left( -1 \right)} = S \left( 0 \right)_0 \, . \]
Consequently, those $x \in S ( 0 )$ which have degree $0$ are considered elements of degree $1$ in $S ( -1 )$. For example, $1 \in S ( - 1 )$ therefore satisfies $\mathrm{deg} ( 1 ) = +1$.

For a $\mathbb{Z}^n$-graded ring $S = \mathbb{Q} [ x_1, \dots, x_m ]$, we now wish to give a brief introduction to the \emph{category of projective graded $S$-modules}. Recall from \cref{subsec:CoherentSheavesOnVarieties} that a (left) $S$-module $M$ is an Abelian group $( M,+ )$ together with a scalar multiplication $S \times M \to M \; , \; \left( s, m \right) \mapsto s \cdot m$ such that for all $s_1, s_2 \in S$ and all $m_1, m_2 \in M$ it holds
\[ \begin{array}{rclcrcl}
s_1 \cdot \left( s_2 \cdot m_1 \right) &=& \left( s_1 \cdot s_2 \right) \cdot m_1 \, , & & \left( s_1 + s_2 \right) \cdot m_1 &=& s_1 \cdot m_1 + s_2 \cdot m_1 \, , \\
s_1 \cdot \left( m_1 + m_2 \right) &=& s_1 \cdot m_1 + s_1 \cdot m_2 \, , & & 1 \cdot m_1 &=& m_1 \ .
\end{array} \]
Therefore (left) $S$-modules look very much like vector spaces over $S$, except that $S$ need not be a field. A free $S$-module
\[ M = \bigoplus_{d \in I}{S \left( d \right)}, \qquad I \subseteq \mathbb{Z}^n \]
is called \emph{graded} precisely if $S_i M_j \subseteq M_{i+j}$. Note that the indexing set $I$ need not be finite. However, if $I$ is finite, then $| I |$ is termed the \emph{rank of $M$}. For reasons that will become clear eventually, we will refer to such modules as \emph{projective graded (left) $S$-modules}. For ease of notation we will often drop the term `left'.

The morphisms in the \emph{category of projective graded (left) $S$-modules} are module homomorphisms which respect the grading. For example, $\varphi \colon S ( -1 ) \xrightarrow{( x_1 )} S ( 0 )$ is such a morphism since 
\[ \underbrace{S \left( -1 \right) \ni 1}_{\mathrm{degree 1}} \mapsto \varphi \left( 1 \right) = \underbrace{x_1 \in S \left( 0 \right)}_{\mathrm{degree 1}} \, . \]

Given projective graded $S$-modules $M$ and $N$ of finite rank $m$ and $n$, a morphism $M \to N$ is given by a matrix $A$ with entries from $S$. It is now a pure matter of convention to express elements of $M$, $N$ either as columns or rows of polynomials from $S$. Suppose that we express $e \in M$, $f \in N$ as rows of polynomials, then $A$ has to be a matrix with $m$ rows and $n$ columns. In particular, we multiply $e \in M$ from the left to the matrix $A$ to obtain its image $e \cdot A \in N$. As this multiplication is performed from the left, this convention applies to \emph{left}-modules.

Of course, one can also choose to represent elements of $M$ and $N$ as columns. In this case $A$ must be a matrix with $n$ rows and $m$ columns and we multiply $e \in M$ from the right to obtain its image $A \cdot e \in N$. In this case one deals with \emph{projective graded right $S$-modules} and \emph{projective graded right $S$-module homomorphisms}.

For historical reasons, it is tradition in algebra to use left-modules in papers, and we follow this tradition here. Hence, elements of projective graded $S$-modules are always expressed as rows of polynomials in $S$.

Given a $\mathbb{Z}^n$-graded ring $S$, the \emph{category of projective graded $S$-modules} happens to be an additive monoidal category, which is both strict and rigid symmetric closed \cite{CAP, PosurDoktor, GutscheDoktor}. We provide an implementation of this category in the language of \emph{CAP} \cite{CAP, PosurDoktor, GutscheDoktor} in the software package \cite{CAPCategoryOfProjectiveGradedModules}.

\section{The Category \texorpdfstring{$\mathbf{S\text{\textnormal{-}fpgrmod}}$}{fpgrmod}} \label{subsec:FPGradedSModules}

Based on a $\mathbb{Z}^n$-graded ring $S = \mathbb{Q} [ x_1, \dots, x_m ]$ and its associated \emph{category of projective graded $S$-modules}, we now wish to build a new category -- the category of \fp graded $S$-modules ($S \mathrm{\textnormal{-}fpgrmod}$). The importance of this construction for us is that it allows us to describe ideals (or vanishing loci) via the relations enjoyed by the generators of the ideal. In order to understand this, we need a bit of preparation. The basic idea however is very simple:
\ebox{Objects in $S \mathrm{\textnormal{-}fpgrmod}$ are presented by morphisms of projective graded $S$-modules of finite rank.}
For an example let us pick $\mathbb{P}^2_{\mathbb{Q}}$ again and look at the following two morphisms of projective graded $S$-modules (of finite rank):
\[ \varphi \colon 0 \to S \left( 0 \right) \, , \qquad \psi \colon S \left( -2 \right)^{\oplus 3} \xrightarrow{R} S \left( -1 \right)^{\oplus 3} \, , \qquad R = \begin{psmallmatrix} 0 & - x_3 & x_2 \\ x_3 & 0 & - x_1 \\ - x_2 & x_1 & 0 \end{psmallmatrix} \, . \]
Abstractly we intend to describe the modules
\[ M_\varphi \equiv \mathrm{coker} \left( \varphi \right) = \mathrm{codomain} \left( \varphi \right) / \mathrm{im} \left( \varphi \right), \qquad M_{\psi} \equiv \mathrm{coker} \left( \psi \right) = \mathrm{codomain} \left( \psi \right) / \mathrm{im} \left( \psi \right) \, . \]
A means to present these modules $M_\varphi$, $M_\psi$ is indeed provided by the morphisms $\varphi$, $\psi$. Hence, we term the codomain of $\varphi, \psi$ the \emph{generators} of $M_{\varphi}$ and $M_\psi$ respectively. Similarly, the domain of $\varphi, \psi$ is given the name \emph{relations} of $M_{\varphi}$ and $M_\psi$ respectively. To make explicit these terms in the commutative diagrams to follow, we box morphisms of projective graded $S$-modules (of finite rank) in blue colour if they are to present an \fp graded $S$-module. Consequently, we picture for example $M_{\psi}$ and $M_{\varphi}$ as follows:
\[ \includegraphics[valign = c]{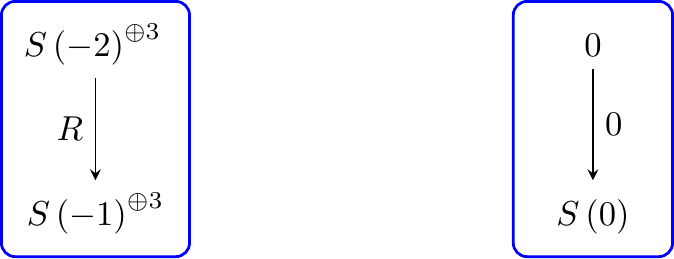} \]

Obviously the \emph{relations} of $M_\varphi$ are $0$. Consequently, $M_\varphi$ is canonically isomorphic to the projective graded $S$-module $S ( 0 )$. $M_\psi$ however is not quite so simple -- its generators have to satisfy 3 relations. Let us work out these relations in detail. To this end, we first identify generating sets:
\begin{itemize}
 \item $S \left( -2 \right)^{\oplus 3}$ is (freely) generated over $S$ by $\mathcal{R} = \left\{ \left( 1,0,0 \right), \left( 0,1,0 \right), \left( 0,0,1 \right) 
      \right\} \equiv \left\{ r_1, r_2, r_3 \right\}$.
 \item $S \left( -1 \right)^{\oplus 3}$ is (freely) generated over $S$ by $\mathcal{G} = \left\{ \left( 1,0,0 \right), \left( 0,1,0 \right), \left( 0,0,1 \right) 
      \right\} \equiv \left\{ g_1, g_2, g_3 \right\}$. 
\end{itemize}
Consequently, we have
\[\label{eq:relation_1} \mathcal{R} \ni r_1 = \left( 1,0,0 \right) \mapsto \left( 1,0,0 \right) \cdot R = \left( 0, -x_3, x_2 \right) = - x_3 g_2 + x_2 g_3 \, . \]
Now let us think of the cokernel of $\psi$ in terms of classes. Then the representants of these classes are not unique, but can be chosen up to addition of elements of the form $-x_3 g_2 + x_2 g_3$. In fact, there are two more such redundancies, which follow from the images of $r_2$ and $r_3$. Namely
\[\label{eq:relation_2_and_3} r_2 \mapsto  x_3 g_1 - x_1 g_3, \qquad r_3 \mapsto -x_2 g_1 + x_1 g_2 \, . \]
In this sense there are the 3 relations \eqref{eq:relation_1} and \eqref{eq:relation_2_and_3} for the generators $g_1, g_2, g_3$ of $M_\psi$.

Let us now turn to the morphisms in $S\mathrm{\textnormal{-}fpgrmod}$. A morphism $M_\psi \to M_\varphi$ of \fp graded $S$-modules is a commutative diagram of the following form:
\[ \includegraphics[valign = c]{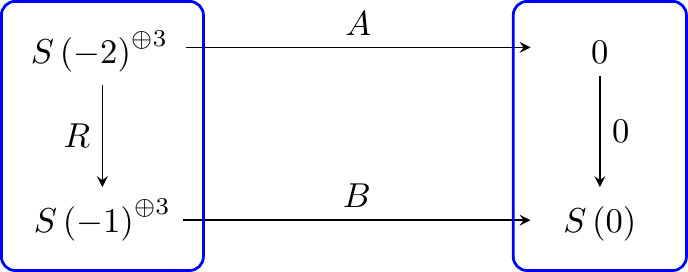} \]
Recall that we are working with projective graded \emph{left} $S$-module homomorphisms. Therefore, the commutativity is to say that $A \cdot 0 = R \cdot B$. We say that the above morphism is congruent to the zero morphism \footnote{Note that there is a slight difference between the notion of a \emph{classical category} and a \emph{CAP-category}. The latter comes equipped with the additional datum of congruence of morphisms. Upon factorisation of this congruence, a \emph{CAP-category} turns into the corresponding \emph{classical category}. For ease of computer implementations, the congruences are added as additional datum. See \cite{CAP, PosurDoktor, GutscheDoktor} for more information.} precisely if there exists a morphism of projective graded $S$-modules $\gamma \colon S ( -1 )^{\oplus 3} \xrightarrow{D} 0$ such that the following diagram commutes:
\[ \includegraphics[valign = c]{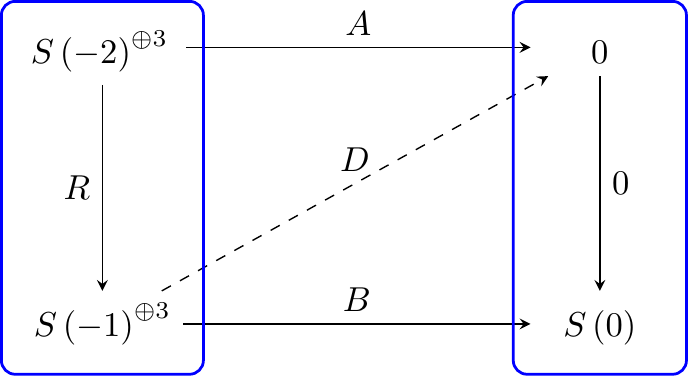} \]
Intuitively, the existence of such a morphism implies that all generators of $M_\psi$ can be thought of as relations of $M_\varphi$. A particular example of such a morphism of \fp graded $S$-modules is as follows:
\[ \includegraphics[valign = c]{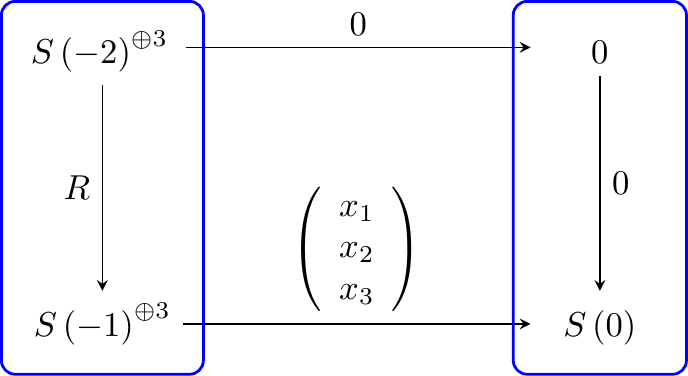} \label{equ:IdealEmbedding} \]
It is readily seen that this morphism is not congruent to the zero morphism. To gain some intuition, let us investigate this morphism in more detail. To this end, recall the generating sets $\mathcal{R}$ and $\mathcal{G}$ of domain and codomain of $\psi$ as introduced above. In particular, we can apply the displayed mapping of projective graded modules to the elements $g_i \in S ( -1 )^{\oplus 3}$, \eg
\[ \mathcal{G} \ni g_1 = \left( 1,0,0 \right) \mapsto \left( 1,0,0 \right) \cdot \left( \begin{array}{c} x_1 \\ x_2 \\ x_3 \end{array} \right) = x_1 \, . \]
Similarly, $g_2$ maps to $x_2$ and $g_3$ to $x_3$. We denote these images by $\mathcal{H} = \{ h_1, h_2, h_3 \}$. Note that the images of the relations map to zero, \eg $- x_3 g_2 + x_2 g_3$ turns into
\[ - x_3 h_2 + x_2 h_3 = - x_3 x_2 + x_2 x_3 = 0 \, . \]
In fact it turns out that the map $S ( -2 )^{\oplus 3} \stackrel{R}{\rightarrow} S ( -1 )^{\oplus 3}$ is the kernel embedding of the following morphism of projective graded $S$-modules
\[ S \left( -1 \right)^{\oplus 3} \xrightarrow{\begin{psmallmatrix} x_1 \\ x_2 \\ x_3 \end{psmallmatrix}} S \left( 0 \right) \, . \]
For this very reason, the morphism in \cref{equ:IdealEmbedding} is a monomorphism of \fp graded $S$-modules. Consequently, it describes the embedding of an ideal into $S ( 0 )$, and this very ideal is the one generated by $x_1, x_2, x_3$. Thus, $M_\psi$ is nothing but a presentation of the irrelevant ideal $B_\Sigma = \langle x_1, x_2, x_3 \rangle \subseteq S$ of $\mathbb{P}_{\mathbb{Q}}^2$ and the morphism in \cref{equ:IdealEmbedding} is its standard embedding $B_\Sigma \hookrightarrow S$. In the language of \fp graded $S$-modules, an ideal is thus encoded by the relations satisfied by its generators and their degrees. To emphasise this important finding we extend the above diagram to take the following shape:
\[ \includegraphics[valign = c]{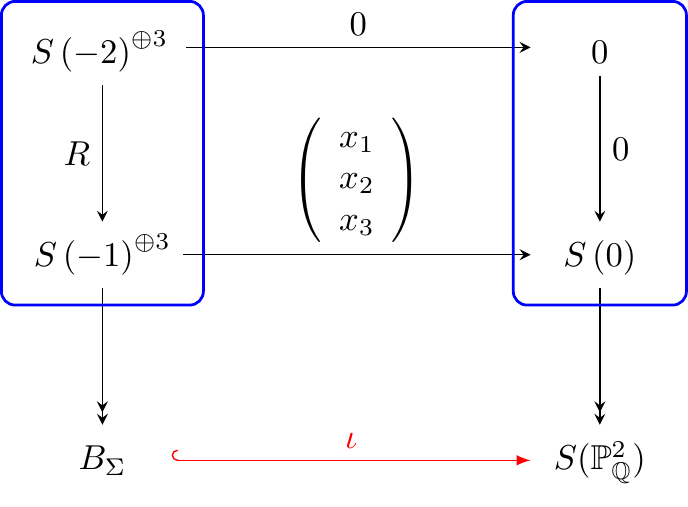} \]
It is of crucial importance to distinguish the black and red arrows in this diagram. The black ones are morphisms of projective graded $S$-modules, whilst the red ones  mediate between \fp graded $S$-modules. To avoid confusion we use snake lines for isomorphisms since the symbol $\widetilde{\phantom{m}}$ is reserved for the sheafification functor already.

Let us use this opportunity to point out that for a given \fp graded $S$-module, there exist numerous presentations. For example, the following is an isomorphism of two presentations that are both canonically isomorphic to the projective graded $S$-module $S ( 0 )$:
\[ \includegraphics[valign = c]{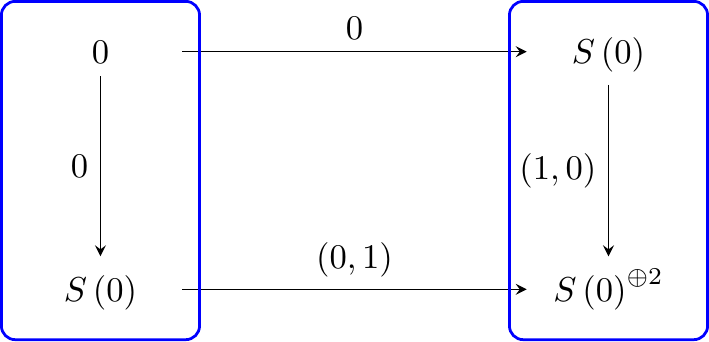} \]

Note also that the rank of generators and the rank of relations are utterly unrelated. We have already given examples of \fp graded $S$-modules for which the rank of the relations is either smaller than or identical to the rank of the generators. The ideal $\langle x_1^2, x_1 x_2, x_1 x_3, x_2^2, x_2 x_3 \rangle$ provides an example for which there are more relations than generators. Namely its standard embedding looks like
\[ \includegraphics[valign = c]{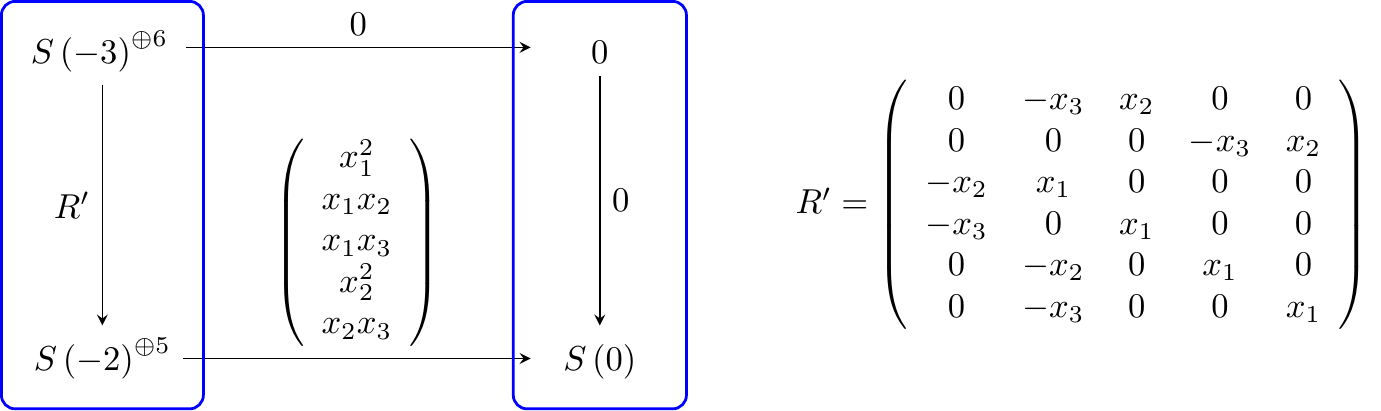}
\]
Hence, the five generators of this ideal satisfy six relations.

It can be proven that the category $S \mathrm{\textnormal{-}fpgrmod}$ is an Abelian monoidal category which is both strict and symmetric closed. See \cite{CAP, PosurDoktor, GutscheDoktor} for further details. This category being Abelian, kernel and cokernel exist for all morphisms in $S \mathrm{\textnormal{-}fpgrmod}$. Let us use this opportunity to display the kernel and cokernel of $\iota \colon B_\Sigma \hookrightarrow S$ in the following diagram:
\[ \label{equ:EmbeddingAndFactorObjectOfIrrelevantIdealOfP2}
\includegraphics[valign = c]{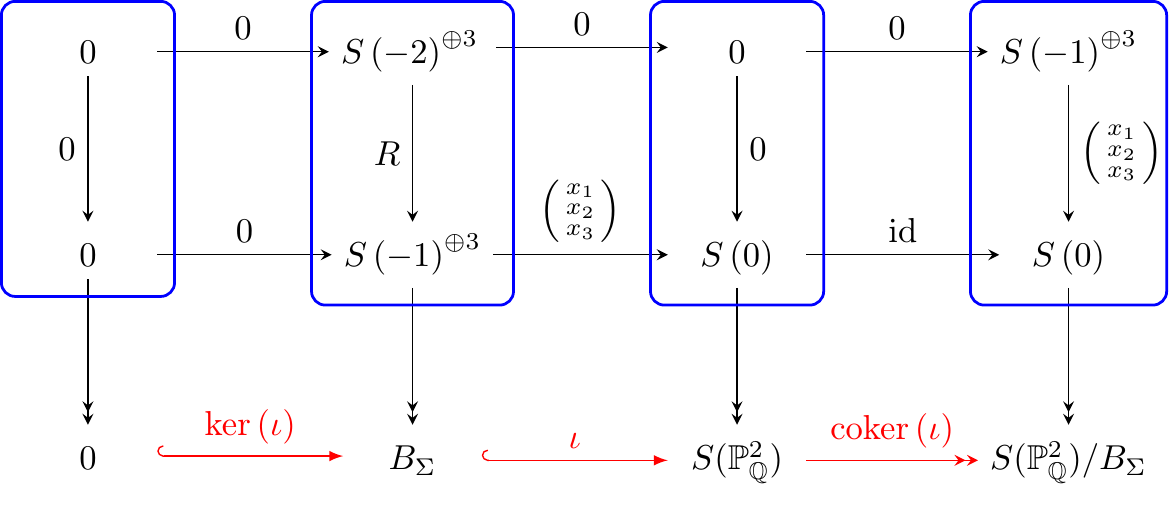}
\]
$S \mathrm{\textnormal{-}fpgrmod}$ being Abelian, a morphism is a monomorphism precisely if its kernel object is the zero object. The trivial box on the left hence reflects the fact that $\iota \colon B_\Sigma \hookrightarrow S ( 0 )$ is a monomorphism. The object boxed on the very right is a factor object. Such factor objects serve as models for structure sheaves of subloci of $X_\Sigma$ in \cref{subsec:LineBundlesFromModules}.

Quite generally, it is possible to associate to a given category $\mathcal{C}$ its \emph{Freyd category} \cite{Freyd1966}. An implementation of this mechanism along the lines of \cite{SeppFreydCategory} is provided in the \texttt{gap}-package \cite{CAPPresentationCategory}. Applying this technique to the category of projective graded $S$-modules, as introduced in \cref{sec:ProjSmodule}, provides the category $S \mathrm{\textnormal{-}fpgrmod}$. Along this philosophy, this very category is implemented in the language of \texttt{CAP} \cite{CAP, PosurDoktor, GutscheDoktor} in the software package \cite{PresentationsByProjectiveGradedModules}.

Finally a word on the terminology of the \emph{projective} graded $S$-modules (of finite rank). These modules are canonically embedded into the category $S \mathrm{\textnormal{-}fpgrmod}$. In the latter they constitute the projective objects. Hence their name.

\section{Sheafification of F.P.\ Graded \texorpdfstring{$\mathbf{S}$}{S}-Modules} \label{sec:Sheafification}

We assume that $X_\Sigma$ is a toric variety over $\mathbb{Q}$ without torus factor and start by briefly revising the important points of \cref{subsec:TowardsToricVarieties}. To this end, recall that $X_\Sigma$ is defined in terms of a fan $\Sigma$, whose ray generators we label as $\rho_1, \dots \rho_m$. To each of these ray generators, there is associated precisely one indeterminate of the Cox ring $S$. We assume $x_1 \leftrightarrow \rho_1$, $x_2 \leftrightarrow \rho_2$ and so on. Hence, $S = \mathbb{Q} [ x_1, \dots, x_m ]$.

To every cone $\sigma \in \Sigma$ we associate an affine variety: Form the monomial$x^{\hat{\sigma}} := \prod_{\rho \notin \sigma (1)}{x_\rho} \in S$, 
then the affine variety in question is $U_\sigma = \mathrm{Specm} ( ( S_{\hat{\sigma}} )_{0} )$. The $\{ U_\sigma \}_{\sigma \in \Sigma}$ glue together as a consequence of \cref{equ:Gluing}. Let us now pick an \fp graded $S$-module $M$ and turn it into a coherent sheaf $\tilde{M}$ on $X_\Sigma$. This works as follows:
\begin{itemize}
 \item The localisation $M_{x^{\hat{\sigma}}}$ turns out to be an \fp graded $S_{x^{\hat{\sigma}}}$-module. Therefore, $( M_{x^{\hat{\sigma}}} )_0$ is an \fp graded 
      $( S_{x^{\hat{\sigma}}} )_0$-module. Since $U_\sigma = \mathrm{Specm} ( ( S_{x^{\hat{\sigma}}} )_0 )$, we can follow the logic in \cref{subsec:CoherentSheavesOnVarieties}, to turn $( M_{x^{\hat{\sigma}}} )_0$ into a coherent sheaf $\tilde{ ( M_{x^{\hat{\sigma}}} )_0}$ on $U_\sigma$.
 \item \Cref{equ:Gluing} guarantees that the sheaves $\tilde{ ( M_{x^{\hat{\sigma}}} )_0}$ on $U_\sigma$ glue to form a sheaf on $X_\Sigma$.
\end{itemize}

\begin{table}[tbp]
\centering
\begin{tabular}{ccccc}
\toprule
cone & generator & indeterminate & affine variety & $x^{\hat{\sigma}}$ \\
\midrule
$\sigma_1$ & $1$ & $x_1$ & $U_1$ & $x_2$ \\
$\sigma_2$ & $-1$ & $x_2$ & $U_2$ & $x_1$ \\
\bottomrule
\end{tabular}
\caption[The fan $\Sigma$ of $\mathbb{P}^1_{\mathbb{Q}}$.]{The fan $\Sigma$ of $\mathbb{P}^1_{\mathbb{Q}}$ (Cox ring $\mathbb{Q} [ x_1, x_2 ]$) contains two maximal cones $\sigma_1$, $\sigma_2$.}
\label{table-N7}
\end{table}

\begin{figure}[tb]
\centering
\includegraphics{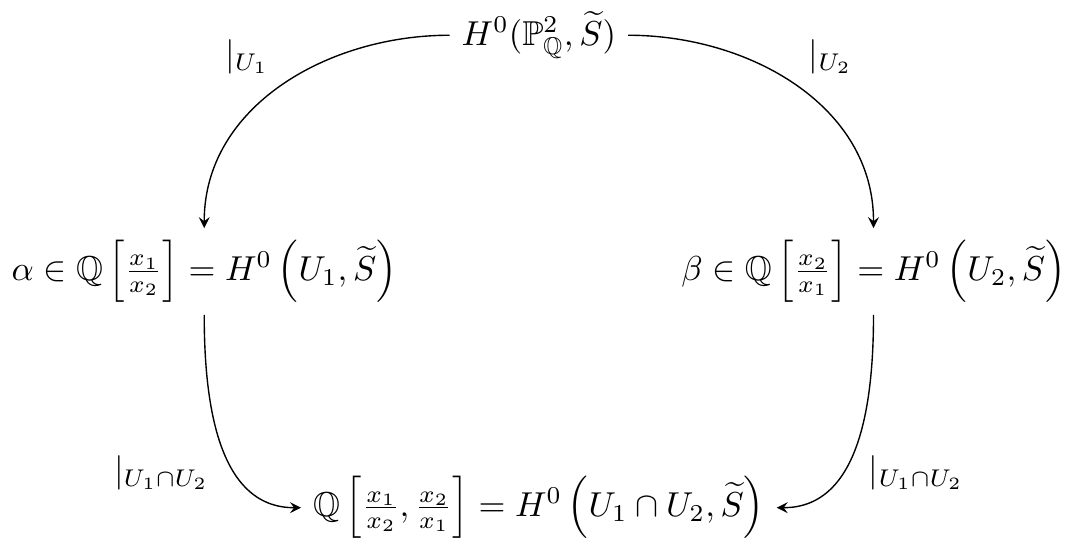}
\caption{The local sections of $\tilde{S}$ on $\mathbb{P}^1_{\mathbb{Q}}$.}
\label{figure:613098645} 
\end{figure}

\begin{figure}[tb]
\centering
\subfloat[The local sections of the sheaf $\tilde{S( 1 )}$ on $\mathbb{P}^1_{\mathbb{Q}}$.]{\label{figure-15923565} \includegraphics[width = 0.96 \textwidth]{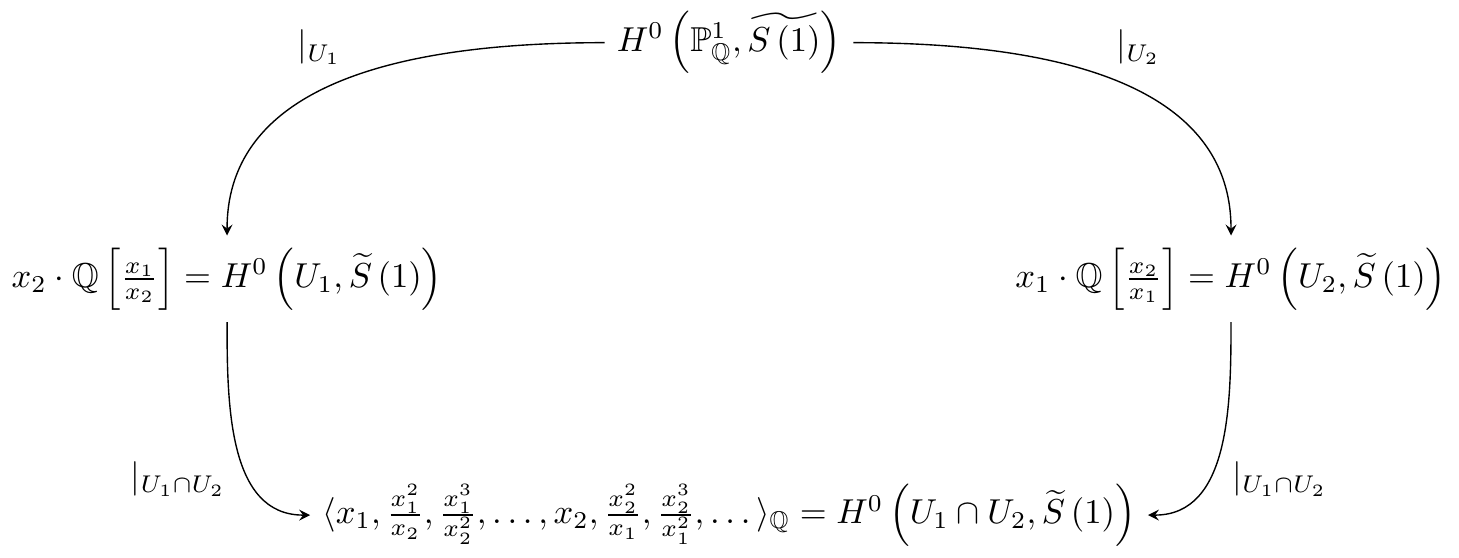} }

\subfloat[The local sections of the sheaf $\tilde{S( -1 )}$ on $\mathbb{P}^1_{\mathbb{Q}}$.]{\label{figure-1592352355} \includegraphics[width = 0.96 \textwidth]{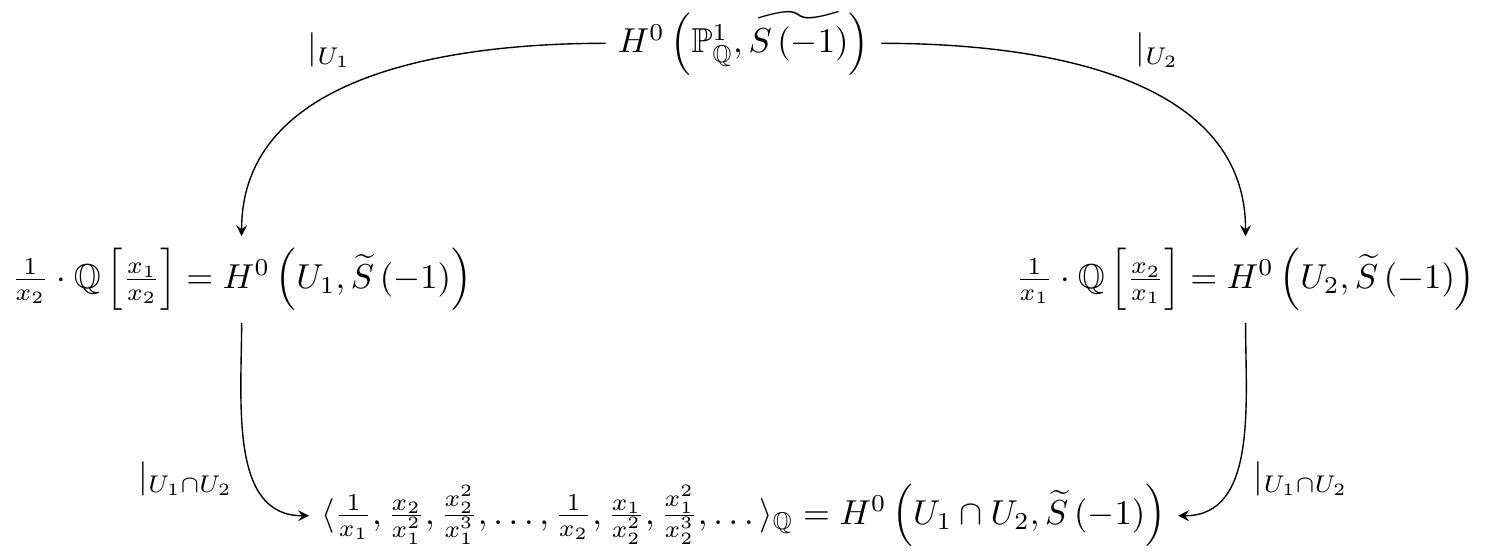} }
\caption[Identifying the global sections of the coherent sheaves $\tilde{S( -1 )}$, $\tilde{S( 1 )}$ on $\mathbb{P}^1_{\mathbf{Q}}$.]{The global sections of $\tilde{S( -1 )}$, $\tilde{S( 1 )}$ on $\mathbb{P}^1_{\mathbb{Q}}$ follow from the diagrams above.}
\label{figure-615235123515325}
\end{figure}

Let us illustrate these abstract words in an example. Consider the toric variety $\mathbb{P}^1_{\mathbb{Q}}$ with Cox ring $S = \mathbb{Q} [ x_1, x_2 ]$. Its fan $\Sigma \subseteq \mathbb{R}$ consists of the trivial cone and the maximal cones $\sigma_1$ and $\sigma_2$ listed in \cref{table-N7}. To compute the affine open cover $U_i \cong \mathrm{Specm} ( ( S_{\hat{\sigma}} )_0 )$ recall
\[ S ( n )_{(x_i)} \equiv \left( S \left( n \right)_{x_i} \right)_0 = \left\{ \left. \frac{f}{x_i^k} \; \right| \; f \in S \left( 0 \right) \mathrm{ homogeneous } \; , \; \mathrm{deg} \left( f \right) = n + k \right\} \, . \]
Consequently, $\mathbb{P}^1_{\mathbb{Q}}$ admits an affine open cover given by $\mathcal{U} = \{ U_1, U_2 \}$ with
\[ U_1 \cong \mathrm{Specm} \left( \mathbb{Q} \left[ \frac{x_1}{x_2} \right] \right), \qquad U_2 \cong \mathrm{Specm} \left( \mathbb{Q} \left[ \frac{x_2}{x_1} \right] \right) \, . \]
Upon homogenisation we can understand $U_1$ as the locus $\{ x_2 \neq 0 \}$, and $U_2$ as $\{ x_1 \neq 0 \}$. Let us now consider the Cox ring $S$ as \fp graded $S$-Module and investigate the sheaf $\tilde{S}$. On the affine patches $U_1$, $U_2$ we assign to this module $M$ the following coherent sheaves:
\[ \tilde{\left( S_{x^{\hat{\sigma_1}}} \right)_0} = \tilde{\mathbb{Q} \left[ \frac{x_1}{x_2}\right]}, \qquad \tilde{\left( S_{x^{\hat{\sigma_2}}} \right)_0} = \tilde{\mathbb{Q} \left[ \frac{x_2}{x_1}\right]} \, . \]
Note that we can also pick the trivial cone and localise at its monomial $x^{\hat{0}} = x_0 x_1$. This corresponds to the restriction to $U_1 \cap U_2$. On this intersection we therefore associate to the module $S$ the coherent sheaf $\tilde{( S_{x^{\hat{0}}} )_0} = \tilde{\mathbb{Q} [ \frac{x_1}{x_2}, \frac{x_2}{x_1} ]}$. The global sections of the coherent sheaves $\tilde{\mathbb{Q} [ \frac{x_1}{x_2}]}$, $\tilde{\mathbb{Q} [ \frac{x_2}{x_1}]}$ and $\tilde{\mathbb{Q} [ \frac{x_1}{x_2}, \frac{x_2}{x_1} ]}$ are given by $\mathbb{Q} [ \frac{x_1}{x_2}]$, $\mathbb{Q} [ \frac{x_2}{x_1}]$ and $\mathbb{Q} [ \frac{x_1}{x_2}, \frac{x_2}{x_1} ]$ (\cf \cref{subsec:CoherentSheavesOnVarieties}). We indicate this geometry in \cref{figure:613098645}. If we set $t \equiv \frac{x_1}{x_2}$, then a global section $s \in H^0 ( \mathbb{P}^1_{\mathbb{Q}}, \tilde{S} )$ can be regarded as a pair
\[ \left( \alpha, \beta \right) \in \tilde{S} \left( U_1 \right) \times \tilde{S} \left( U_2 \right) \cong \mathbb{Q} \left[ t \right] \times \mathbb{Q} \left[ t^{-1} \right] \]
with $\alpha |_{U_1 \cap U_2} = \beta |_{U_1 \cap U_2}$ (\cf sheaf property (S2) in \cref{sec:RevisionOnSheavesAndSheafCohomology}). It is readily verified that only the diagonal elements $( \alpha, \alpha ) \in \mathbb{Q} \times \mathbb{Q}$ satisfy this demand, and so $H^0 ( \mathbb{P}^1_{\mathbb{Q}}, \tilde{S} ) \cong \mathbb{Q}$. It is in fact well-known that $\tilde{S}$ is the structure sheaf $\mathcal{O}_{\mathbb{P}^1_{\mathbb{Q}}}$. Similarly, $S ( n )$ sheafifies to the twisted structure sheaf $\mathcal{O}_{\mathbb{P}^1_{\mathbb{Q}}} ( n )$. We list the corresponding diagrams of local sections for $n = \pm 1$ in \cref{figure-615235123515325} -- it follows $H^0 ( \mathbb{P}^1_{\mathbb{Q}}, \tilde{ S ( 1 )} ) \cong \langle x_1, x_2 \rangle_{\mathbb{Q}}$ and $H^0 ( \mathbb{P}^1_{\mathbb{Q}}, \tilde{S ( -1 )} ) \cong \{ 0 \}$.

The functor $\widetilde{\phantom{m}} \colon S \mathrm{\textnormal{-}fpgrmod} \to \mathfrak{Coh} X_\Sigma$ enables us to model coherent sheaves in a way suitable for computer manipulations -- namely as objects of the category $S \mathrm{\textnormal{-}fpgrmod}$. An important question is whether this description is unique. As it turns out, even for smooth toric varieties this is not the case. Rather for a smooth toric variety $X_\Sigma$ with irrelevant ideal $B_\Sigma \subseteq S$ it is known that an \fp graded $S$-module $M$ satisfies by proposition 5.3.10 in \cite{cox2011toric} \footnote{Proposition 5.3.10 of \cite{cox2011toric} fails for $\mathbb{P}_{\mathbb{C}} ( 2,3,1 )$, \ie is not applicable if $X_\Sigma$ is not smooth. See \cite{cox2011toric} for generalisations of this proposition.}
\[ \tilde{M} = 0 \quad \Leftrightarrow \quad B \left( \Sigma \right)^l M = 0 \text{ for } l \gg 0 \, . \label{equ:RedundancyInSmoothCase} \]
Let us exemplify this redundancy on $\mathbb{P}^2_{\mathbb{Q}}$. In \cref{subsec:FPGradedSModules} we described a presentation of the irrelevant ideal $B_\Sigma$ by means of \fp graded $S$-modules. Its embedding into $S$ and the corresponding factor object are given in \cref{equ:EmbeddingAndFactorObjectOfIrrelevantIdealOfP2}. We use the fact that the sheafification functor is exact, \ie turns the exact sequence $0 \to B_\Sigma \to S ( \mathbb{P}^2_{\mathbb{Q}} ) \to S (\mathbb{P}^2_{\mathbb{Q}}) / B_\Sigma \to 0$ of \fp graded $S$-modules into a short exact sequence of coherent sheaves on $\mathbb{P}^2_{\mathbb{Q}}$. We indicate this in the following commutative diagram:
\[ \includegraphics[valign = c]{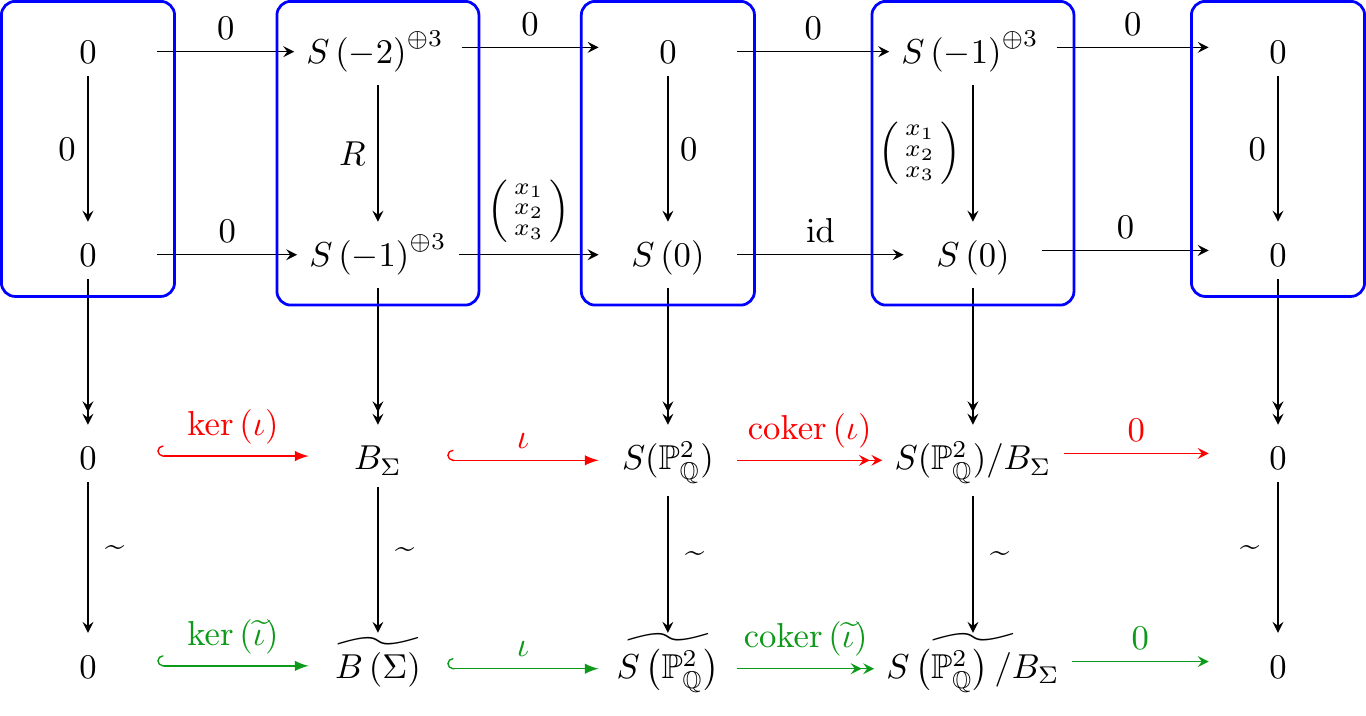} \]
By \cref{equ:RedundancyInSmoothCase} the sheaf $\tilde{S ( \mathbb{P}^2_{\mathbb{Q}} ) / B_\Sigma}$ is trivial. Exactness of the last row thus tells us $\tilde{B_\Sigma } \cong \tilde{S( \mathbb{P}^2_{\mathbb{Q}})}$. Hence, the computer can use both $B_\Sigma$ and $S( \mathbb{P}^2_{\mathbb{Q}} )$ as models of the structure sheaf. 
In \cref{subsec:SheafCohomologyFromFPGradedSModules} we will point out why $S(\mathbb{P}^2_{\mathbb{Q}} )$ is a better model for the structure sheaf of $\mathbb{P}^2_{\mathbb{Q}}$ than $B_\Sigma$. Nonetheless, ideals which sheafify to the structure sheaf are quite valuable to our sheaf-cohomology-algorithm as we explain in \cref{subsec:Ideals}.

In the smooth case we can say a little more about this redundancy. Namely let $S \mathrm{\textnormal{-}fpgrmod}^0$ be the thick subcategory of $S \mathrm{\textnormal{-}fpgrmod}$ consisting of those \fp graded $S$-modules which are supported on $V ( B_\Sigma )$. Then the equivalence of categories
\[ S \mathrm{\textnormal{-}fpgrmod} / S \mathrm{\textnormal{-}fpgrmod}^0 \xrightarrow{\sim} \mathfrak{Coh} X_\Sigma \, . \] 
gives a possible parametrisation of $\mathfrak{Coh} ( X_\Sigma )$ \cite{2012arXiv1210.1425B, 2011arXiv1110.0323P}.

\section{Massless Spectra in \emph{F-Theory} Toy Model} \label{sec:ComputingTheSpectra}

In this section we intend to apply our knowledge to compute the massless spectrum of a simple \emph{F-theory} compactification. First recall from \cref{sec:ComputingMasslessSpectraWithGAP} that our upshot is to compute the sheaf cohomologies of certain line bundles on the matter curves in question. If the base space $\mathcal{B}_6$ is embedded into a toric variety $X_\Sigma$, we can interpret these line bundles as coherent sheaves on $X_\Sigma$. The computation of the number of zero modes then reduces to the computation of sheaf cohomology dimensions on toric spaces, which is analysed in detail in \cref{chapter:MathDetailsSheafCohomologies}. For convenience to the reader we give a brief summary on our algorithmic approach in \cref{subsec:SheafCohomologyFromFPGradedSModules}. It is based on work of M.\ Barakat and collaborators \cite{2010arXiv1003.1943B, 2012arXiv1202.3337B, 2012arXiv1210.1425B, 2012arXiv1212.4068B, 2014arXiv1409.6100B, BL_GabrielMorphisms}. To adapt all this technology to the \emph{F-theory} setup, we are required to make a concrete choice of base space $\mathcal{B}_6$ and 7-brane divisor $W$ therein. Consequently, we fix the complex structure moduli which define the elliptic fibration at hand -- we apply our methods to an \emph{F-theory} geometry with $\mathcal{B}_6 = \mathbb P^3_{\mathbb{Q}}$. We describe this geometry in detail in \cref{subsec:Simplifying}. Finally, we demonstrate our computations in \cref{subsec:Spectrum1} for different choices of complex structure moduli of the matter curves. We observe jumps in the cohomology dimensions across the moduli space. Such phenomena are of relevance when it comes, for instance, to scanning for Standard-model-like spectra in \emph{F-theory} compactifications.

\subsection{Simplifying Assumptions and Geometric Consequences} \label{subsec:Simplifying}

\begin{table}[tbp]
\centering
\begin{tabular}{ccc@{\hskip 20pt}ccccc@{\hskip 20pt}cccc}
\toprule
$z_1$ & $z_2$ & $z_3$ & $z_4 \equiv e_0$ & $e_1$ & $e_2$ & $e_3$ & $e_4$ & x & y & z & s \\
\midrule
1 & 1 & 1 & 1 & 0 & 0 & 0 & 0 & 8 & 12 & 0 & 0 \\
\vspace{-0.5em} & \\
$\cdot$          & $\cdot$ &  $\cdot$    & -1        & 1 & $\cdot$ & $\cdot$ & $\cdot$ & -1 & -1 & $\cdot$ & $\cdot$ \\
$\cdot$          & $\cdot$ &  $\cdot$    & -1        & $\cdot$ & 1 & $\cdot$ & $\cdot$ & -2 & -2 & $\cdot$ & $\cdot$ \\
$\cdot$          & $\cdot$ &  $\cdot$    & -1        & $\cdot$ & $\cdot$ & 1 & $\cdot$ & -2 & -3 & $\cdot$ & $\cdot$ \\
$\cdot$          & $\cdot$ &  $\cdot$    & -1        & $\cdot$ & $\cdot$ & $\cdot$ & 1 & -1 & -2 & $\cdot$ & $\cdot$ \\
\vspace{-0.5em} \\
$\cdot$          & $\cdot$ &  $\cdot$    & $\cdot$   & $\cdot$ & $\cdot$ & $\cdot$ & $\cdot$ & 2 &  3 & 1 & $\cdot$ \\
$\cdot$          & $\cdot$ &  $\cdot$    & $\cdot$   & $\cdot$ & $\cdot$ & $\cdot$ & $\cdot$ & -1 & -1 & $\cdot$ & 1 \\
\bottomrule
\end{tabular}
\caption[The toric data of $\hat{Y}_\Sigma$ for a toric \emph{F-theory} toy-model.]{The elliptic fibration over $\mathcal{B}_6 = \mathbb P^3_{\mathbb{Q}}$ is described as complete intersection in a toric variety $\hat{Y}_\Sigma$, whose Cox ring is graded under $\mathbb{Z}^7$ according to this table.}
\label{table-N8}
\end{table}

To explicitly perform the computation of sheaf cohomologies we return to the fibration defined in \cref{subsec:SpecialFTheoryGUTModel} and specialise $\mathcal{B}_6 = \mathbb{P}_{\mathbb{Q}}^3$, with $[ z_1 \colon z_2 \colon z_3 \colon z_4 ]$ denoting its homogeneous coordinates. The Hodge diamond for this space is
\[ \begin{array}{ccccccc}
& & & 1 \\
& & 0 & & 0 \\
& 0 & & 1 & & 0 \\
0 & & 0 & & 0 & & 0 \\
& 0 & & 1 & & 0 \\
& & 0 & & 0 \\
& & & 1 
\end{array} \]
In addition, $\mathbb{P}^3_{\mathbb{Q}}$ is a connected 3-fold and its Picard group satisfies $\mathrm{Pic} ( \mathcal{B}_6 ) \cong \mathbb{Z}$, \ie is torsion-free. Hence, $\mathcal{B}_6 = \mathbb{P}^3_{\mathbb{Q}}$ is a 'valid` base space in the sense introduced in \cref{subsec:ConditionsOnBaseAndFibration}. To proceed, let us model the GUT-divisor as $W = V ( z_4 ) \cong \mathbb{P}_{\mathbb{Q}}^2$. The elliptic fibration $\hat{Y}_4$ is then a complete intersection in the toric ambient space $\hat{Y}_\Sigma$ whose Cox ring $S( \hat{Y}_\Sigma) = \mathbb{Q} [ z_1, z_2, z_3, e_0, e_1, e_2, e_3, e_4, x, y, z, s ]$ is graded by $\mathbb{Z}^7$ (\cf \cref{table-N8}) and satisfies
\begin{align}
\begin{split}
I_{\mathrm{SR}} ( \hat{Y}_\Sigma ) &= \left\langle xy, x e_4, z s, z e_1, z e_2, z e_3, z e_4, s e_0, s e_1, s e_2, s e_4, y e_1, y e_2,
e_0 e_3, \right. \\
& \hspace{5em} \left. e_1 e_3, e_0 e_2, e_0 z_1 z_2 z_3, e_1 z_1 z_2 z_3, e_2 z_1 z_2 z_3, e_3 z_1 z_2 z_3, e_4 z_1 z_2 z_3 \right\rangle \, .
\end{split}
\end{align}
This toric space $\hat{Y}_\Sigma$ is not smooth, but rather a toric orbifold. We have $\hat{Y}_4 = V ( P_T^\prime ) \subseteq \hat{Y}_\Sigma$ with
\[ P_T^\prime = y^2 s e_3 e_4 + a_1 x y z s + a_{3,2} y z^3 e_0^2 e_1 e_4 - x^3 s^2 e_1 e_2^2 e_3 - a_{2,1} x^2 z^2 s e_0 e_1 e_2 - a_{4,3} x z^4 e_0^3 e_1^2 e_2 e_4 \, . \]
Since $\overline{K}_{\mathcal{B}_6} \cong \mathcal{O}_{\mathbb{P}_{\mathbb{Q}}^3} ( 4 )$ and $W = V ( z_4 )$, the sections $a_{i,j}$ are taken according to
\begin{align*}
& a_{1,0} \in H^0 \left( \mathbb{P}_{\mathbb{Q}}^3, \mathcal{O}_{\mathbb{P}_{\mathbb{Q}}^3} \left( 4 \right) \right), & a_{2,1} \in H^0 \left( \mathbb{P}_{\mathbb{Q}}^3, \mathcal{O}_{\mathbb{P}_{\mathbb{Q}}^3} \left( 7 \right) \right), \\
& a_{3,2} \in H^0 \left( \mathbb{P}_{\mathbb{Q}}^3, \mathcal{O}_{\mathbb{P}_{\mathbb{Q}}^3} \left( 10 \right) \right), & a_{4,3} \in H^0 \left( \mathbb{P}_{\mathbb{Q}}^3, \mathcal{O}_{\mathbb{P}_{\mathbb{Q}}^3} \left( 13 \right) \right)
\end{align*}
For generic such sections $\hat{Y}_4$ is smooth and has Hodge diamond
\[ \begin{array}{ccccccccc}
& & & & 1 \\
& & & 0 & & 0 \\
& & 0 & & 7 & & 0 \\
& 0 & & 0 & & 0 & & 0 \\
1 & & 968 & & 3944 & & 968 & & 1 \\
& 0 & & 0 & & 0 & & 0 \\
& & 0 & & 7 & & 0 \\
& & & 0 & & 0 \\
& & & & 1
\end{array} \]
This again fits with the conditions phrased in \cref{subsec:ConditionsOnBaseAndFibration} and the `strict' Calabi--Yau condition $h^{1,0} ( \hat{Y}_4 ) = h^{2,0} ( \hat{Y}_4 ) = h^{3,0} ( \hat{Y}_4, 0 ) = 0$ as introduced in \cref{sec:StringTheory} is satisfied. The matter curves contained in $W$ are described in terms of these sections $a_{i,j}$ as
\[ C_{\mathbf{10}_{1}} = V \left( z_4, a_{1,0} \right), \qquad C_{\mathbf{5}_{3}} = V \left( z_4, a_{3,2} \right), \qquad C_{\mathbf{5}_{-2}} = V \left( z_4, a_{1,0} a_{4,3} - a_{2,1} a_{3,2} \right) \, . \]
From the computational point of view, this toy model has a very appealing feature -- the GUT-surface $W$ is itself a toric variety. In such a situation it is always favourable to apply the tools provided by \texttt{gap} \cite{GAP4} to this toric GUT-surface directly, rather than describing it as a subvariety of $\mathcal{B}_6$. In particular, we can model the matter curves in $W$ as curves in $\mathbb{P}_{\mathbb{Q}}^2$ with homogeneous coordinates $[ z_1 \colon z_2 \colon z_3 ]$ by use of the homogeneous polynomials
\begin{equation}
\begin{aligned}
& \tilde{a_{1,0}} \in H^0 \left( \mathbb{P}_{\mathbb{Q}}^2, \mathcal{O}_{\mathbb{P}_{\mathbb{Q}}^2} \left( 4 \right) \right) \, , \qquad & \tilde{a_{2,1}} \in H^0 \left( \mathbb{P}_{\mathbb{Q}}^2, \mathcal{O}_{\mathbb{P}_{\mathbb{Q}}^2} \left( 7 \right) \right) \, , \\
& \tilde{a_{3,2}} \in H^0 \left( \mathbb{P}_{\mathbb{Q}}^2, \mathcal{O}_{\mathbb{P}_{\mathbb{Q}}^2} \left( 10 \right) \right) \, , \qquad & \tilde{a_{4,3}} \in H^0 \left( \mathbb{P}_{\mathbb{Q}}^2, \mathcal{O}_{\mathbb{P}_{\mathbb{Q}}^2} \left( 13 \right) \right) \, .
\end{aligned}
\end{equation}
Then we have $C_{\mathbf{10}_{1}} \cong V ( \tilde{a_{1,0}} )$, $C_{\mathbf{5}_{3}} \cong V ( \tilde{a_{3,2}} )$ and $C_{\mathbf{5}_{-2}} \cong V ( \tilde{a_{1,0}} \tilde{a_{4,3}} - \tilde{a_{2,1}} \tilde{a_{3,2}} )$. To appreciate to what degree the defining polynomials for the matter curves simplify upon restriction to $W$ we compare the number of monomials defining $a_{i,j}$ and their restrictions $\tilde a_{i,j}$ displayed in \cref{table-N9}. In particular, whilst $a_{4,3} \in H^0( \mathbb{P}_{\mathbb{Q}}^3, \mathcal{O}_{\mathbb{P}_{\mathbb{Q}}^3}( 13 ))$ generically consists of 560 monomials, the corresponding $\tilde{a_{4,3}} \in H^0 ( \mathbb{P}_{\mathbb{Q}}^2, \mathcal{O}_{\mathbb{P}_{\mathbb{Q}}^2} ( 13 ) )$ merely consists of 105 monomials.

\begin{table}[tbp]
\centering
\begin{tabular}{c@{\hskip 20pt}cc}
\toprule
$d \in \mathbb{Z}$ & $h^0 \left( \mathbb{P}_{\mathbb{Q}}^3, \mathcal{O}_{\mathbb{P}_{\mathbb{Q}}^3} \left( d \right) \right)$ & $h^0 \left( \mathbb{P}_{\mathbb{Q}}^2, \mathcal{O}_{\mathbb{P}_{\mathbb{Q}}^2} \left( d \right) \right)$ \\
\midrule
4 & 35 & 15 \\
7 & 120 & 36 \\
10 & 286 & 66 \\
13 & 560 & 105 \\
\bottomrule
\end{tabular}
\caption{Comparison of sections over $\mathcal{B}_6 = \mathbb{P}^3_\mathbb{Q}$ and their restriction to $W = \mathbb{P}^2_\mathbb{Q}$.}
\label{table-N9}
\end{table}

With this preparation we now turn to the actual quantities to evaluate. For instance, the massless spectrum on $C_{\mathbf{5}_{-2}}$ induced by the gauge background $A ( \mathbf{10}_{1} ) ( - \lambda )$ is counted by the sheaf cohomologies of the line bundle 
\begin{align}
\begin{split}
L \left( S_{\mathbf{5}_{-2}} , A \left( \mathbf{10}_{1} \right) \left( \lambda \right)\right) &= \mathcal{O}_{C_{\mathbf{5}_{-2}}} \left( - \frac{3 \lambda Y_1 }{5} + \frac{2 \lambda}{5} Y_2 \right) \otimes \left. \mathcal{O}_{\mathbb{P}_{\mathbb{Q}}^2} \left( 7 \right) \right|_{C_{\mathbf{5}_{-2}}} \, \\
& \cong \mathcal{O}_{C_{\mathbf{5}_{-2}}} \left( - \lambda Y_1 \right) \otimes \left. \mathcal{O}_{\mathbb{P}_{\mathbb{Q}}^2} \left( 7 + \frac{8 \lambda}{5} \right) \right|_{C_{\mathbf{5}_{-2}}} \, .
\end{split}
\end{align}
As we outline in \cref{subsec:LineBundlesFromModules}, with the help of \texttt{gap} \cite{GAP4} we can in principle compute an \fp graded $S( \mathbb{P}_{\mathbb{Q}}^2)$-module which sheafifies to $L $ on $C_{\mathbf{5}_{-2}}$. The question whether this also works in practice strongly depends on the complexity of the involved polynomials $a_{i,j}$. Although the restriction from $\mathcal{B}_6 = \mathbb{P}_{\mathbb{Q}}^3$ to $W \cong \mathbb{P}_{\mathbb{Q}}^2$ removes many moduli from the polynomials $a_{i,j}$, we are still left with a huge polynomial $\tilde{a_{1,0}} \tilde{a_{4,3}} - \tilde{a_{3,2}} \tilde{a_{2,1}}$. For such a big polynomial, the currently available Gröbner basis algorithms come to their limits, which means that for such big polynomials defining the matter curve $C_{\mathbf{5}_{-2}}$ we are in practice unable to compute the \fp graded $S( \mathbb{P}_{\mathbb{Q}}^2)$-module which sheafifies to give the above line bundle.

To overcome this shortcoming, we will compute the massless spectrum for non-generic matter curves instead. In our first example, we pick
\[ a_{1,0} = c_1 \left( x_1 - x_2 \right)^4, \qquad a_{2,1} = c_2 x_1^7, \qquad a_{3,2} = c_3 x_2^{10}, \qquad a_{4,3} = c_4 x_3^{13} \,  \label{aijmodel1} \]
with $c_i \in \mathbb{N}_{>0}$ (pseudo-)random integers. Then the discriminant $\Delta$ of $P_T$ can be expanded in terms of the GUT-coordinate $w$ as
\begin{align}
\begin{split}
\Delta &= 16 a_{1,0}^4 a_{3,2} \left( -a_{2,1} a_{3,2} + a_{1,0} a_{4,3} \right) w^5 \\
& \qquad + 16 a_{1,0}^2 \left( -8 a_{2,1}^2 a_{3,2}^2 + 8 a_{1,0} a_{2,1} a_{3,2} a_{4,3} + a_{1,0} \left( a_{3,2}^3 + a_{1,0} a_{4,3}^2 \right) \right) w^6 + \mathcal{O} \left( w^7 \right) \,,
\end{split}
\end{align}
where for simplicity we have not written out the $a_{i,j}$ explicitly. No further factorisation occurs, and hence this choice of non-generic matter curves still leaves us with a $SU ( 5 ) \times U ( 1 )_X$-gauge theory. The curve $C_{\mathbf{5}_{-2}}$ is given by
\[ C_{\mathbf{5}_{-2}} = V \left( a_{1,0} a_{4,3} - a_{2,1} a_{3,2} \right) = V \left( \left( x_1 - x_2 \right)^4 x_3^{13} - \frac{c_2 c_3}{c_1 c_4} x_2^7 x_3^{10} \right) \, . \]
This curve $C_{\mathbf{5}_{-2}}$ is \emph{not} smooth. Let us therefore emphasise again that the techniques implemented in \texttt{gap} \cite{GAP4} are not limited to generic or smooth matter curves. In fact we are able to handle just about any subvariety of smooth and complete toric varieties, provided its defining polynomials are of reasonable size, so that the currently available Gröbner basis algorithms terminate in a timely fashion.

\subsection{Constructing the relevant Line Bundles from F.P.\ Graded \texorpdfstring{$\mathbf{S}$}{S}-Modules} \label{subsec:LineBundlesFromModules}

\begin{figure}[tbp]
\centering
\subfloat[Module $A_C$.]{\label{figure-92356} \includegraphics{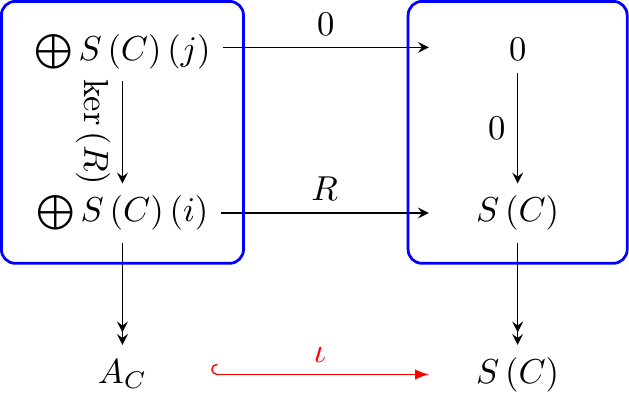} }
\hspace{3em}
\subfloat[Module $A$.]{\label{figure-92357} \includegraphics{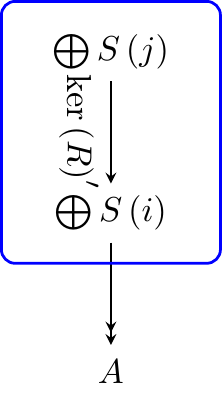} }
\hspace{3em}
\subfloat[Module $B$.]{\label{figure-92358} \includegraphics{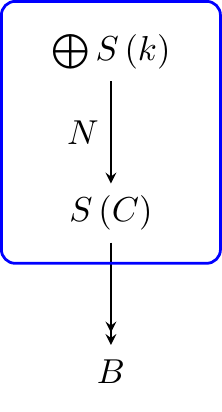} }
\caption{The modules $A_C$, $A$ and $B$ used during the computation of $M_{-}$.}
\label{figure-109385}
\end{figure}

To analyse this example geometry, one of the remaining tasks is as follows:
\ebox{Be $X_\Sigma$ a normal toric variety that is smooth and complete. Its Cox ring be $S$. Given a subvariety $C = V ( g_1, \dots, g_k ) \subseteq X_\Sigma$ and $D = V ( f_1, \dots, f_n ) \in \mathrm{Div} ( C )$ -- both not necessarily complete intersections -- we wish to construct \fp graded $S$-modules $M_{\pm}$ such that $\tilde{M_{\pm}} \in \mathfrak{Coh} \left( X_\Sigma \right)$ is supported only over $C$ and satisfies $\tilde{M_{\pm}} |_{C} \cong \mathcal{O}_{C} ( \pm D )$.}
To find $M_{-}$ we proceed as follows:
\begin{enumerate}
 \item The polynomials $f_i$ and $g_j$ are homogeneous. Consider the ring $S ( C ) := S / \langle g_1, \dots, g_k \rangle$. The canonical projection map $\pi \colon S 
      \twoheadrightarrow S ( C )$ allows us to consider the matrix
      \[ R = \left( \pi \left( f_1 \right), \dots, \pi \left( f_n \right) \right) \in M \left( 1 \times n, S \left( C \right) \right) \, . \]
      Let $\mathrm{ker} ( R )$ denote the kernel matrix of $R$. Then \cref{figure-92356} describes a monomorphism of \fp graded $S ( C )$-modules.\footnote{Note that $I, J \subseteq \mathrm{Cl} ( X_\Sigma )$ are \emph{finite} indexing sets.} By proposition 6.18 of \cite{hartshorne1977algebraic} it holds $\tilde{A_C} \cong \mathcal{O}_{C} \left( -D \right)$.
 \item Next we extend $\tilde{A_C}$ to a (proper) coherent sheaf on $X_\Sigma$. To this end, note that the entries of the matrix $\mathrm{ker} ( R )$ are 
      elements of $S ( C )$, \ie are equivalence classes of elements of the ring $S$. For each entry of $\mathrm{ker} \left( R \right)$ pick one representant in $S$. Thereby, we obtain a matrix $\mathrm{ker} ( R )^\prime$ with entries in $S$.\footnote{Of course this matrix $\mathrm{ker} ( R )^\prime$ is not unique. This ambiguity is taken care of by tensoring with the structure sheaf $\mathcal{O}_C$ of $C$, as explained in step 3.} Based on this we construct the module $A$ described in \cref{figure-92357}.
 \item The coherent sheaf $\tilde{A}$ could have support outside of $C$. To ensure for extension by zero outside of $C$, we tensor with the structure sheaf 
      $\mathcal{O}_C$ of $C$. Given the matrix $N = ( g_1, g_2, \dots, g_k )^T$, the \fp graded $S$-module $B$ given in \cref{figure-92358} sheafifies to $\mathcal{O}_C$.
 \item Finally, consider $M_{-} = A \otimes_S B \in S ( C ) \mathrm{\textnormal{-}fpgrmod}$. It defines $\tilde{M_{-}} \in \mathfrak{Coh} ( X_\Sigma )$ which is 
      supported only on $C$ and satisfies $\tilde{M_{-}} |_C \cong \mathcal{O}_C \left( -D \right)$.
\end{enumerate}

To obtain an \fp graded $S$-module $M_+$ such that $\tilde{M_+}$ is zero outside of $C$ and satisfies $\tilde{M_+} |_C \cong \mathcal{O}_C ( +D )$, we perform step 1 above. Now we dualise the so-obtained module $A_C$ in the category $S ( C ) \mathrm{\textnormal{-}fpgrmod}$. This means to compute (c.f. \cref{sec:ExtOfFPModules} for more details on the dualisation)
\[ A_C^\vee := \mathrm{Hom}_{S \left( C \right)} \left( S \left( C \right), A_C \right) \, . \]
With this \fp graded $S \left( C \right)$-module we now proceed just as before, \ie we turn $A_C^\vee$ into an $S$-module $A^\vee$ and finally consider $M_+ = A^\vee \otimes_S B$. This module $M_+$ has the desired properties.

\subsection{Computing Sheaf Cohomologies} \label{subsec:SheafCohomologyFromFPGradedSModules}

\paragraph{The Algorithm}
Let $X_\Sigma$ be a normal toric variety which is smooth and complete. Its Cox ring be $S$. By now we have described means to parametrise coherent sheaves $\mathcal{F} \in \mathfrak{Coh} ( X_\Sigma )$ by \fp graded $S$-modules. Given such an \fp graded $S$-module $M$, we can wonder how we extract the sheaf cohomologies of $\tilde{M}$ from the data defining $M$. Our algorithm is as follows:
\begin{enumerate}
 \item Given that $X_\Sigma$ is smooth, complete or simplicial, projective, the \emph{cohomCalg}-algorithm applies to it \cite{Blumenhagen:2010pv, Blumenhagen:2010ed, 
      Blumenhagen:2011xn, KoszulExtensionManual, cohomCalg:Implementation, 2011JMP....52c3506J, Rahn:2010fm}. This enables us to compute the vanishing sets
      \[ V^i \left( X_\Sigma \right) := \left\{ \left. D \in \mathrm{Cl} \left( X_\Sigma \right) \; \right| \; h^i \left( X_\Sigma, \mathcal{O}_{X_\Sigma} \left( D \right) \right) = 0 \right\} \]
      fairly rapidly. Note that the so-obtained vanishing sets serve as a properly refined version of the semigroup $\mathbb{\mathcal{K}}^{\mathrm{sat}}$ introduced in \cite{Maclagan03multigradedcastelnuovo-mumford} which was used in \cite{Oberwolfach} to propose a means to compute sheaf cohomology of coherent sheaves on smooth and projective toric varieties. We give details in \cref{sec:ImprovedVanishingSets} and use these results to derive vanishing theorems in \cref{sec:VanishingTheorems}.
 \item For every ample $u \in \text{Cl} ( X_\Sigma )$ and $e \in \mathbb{N}_{\geq 0}$ we consider the ideal $I ( u, e ) = \langle m_1^e, \dots, m_n^e   
      \rangle$, which is furnished from the e-th power of all monomials $m_i \in S$ of degree $u$. On smooth and complete toric varieties, our algorithm finds $u$ and $e$ such that $\tilde{I ( u, e )} \cong \mathcal{O}_{X_\Sigma}$ and all conditions of \cref{mytheorem} are fulfilled.\footnote{As pointed out in \cite[example 6.6]{2012arXiv1212.4068B}, there exist simplicial toric varieties $X_\Sigma$ and coherent sheaves $\mathcal{F}$ thereon, such that for $i > 1$ the sheaf cohomologies $h^i ( X_\Sigma, \mathcal{F} )$ cannot be obtained from (truncations of) extensions of \fp graded $S$-modules. In this work, the coherent sheaves of interest correspond to line bundles on compact, complex curves, for which reason their only non-zero sheaf cohomologies are $h^0 ( X_\Sigma, \mathcal{F} )$ and $h^1 ( X_\Sigma, \mathcal{F} )$. For such coherent sheaves, our methods extend to simplicial, projective toric varieties even.} Consequently the isomorphism
      \[ H^i ( X_\Sigma, \tilde{M} ) \cong \left[ \mathrm{Ext}^i_{S} \left( I \left( u, e \right), M \right) \right]_{0} \label{equ:CohoIso} \]
      allows us to identify the sheaf cohomologies from $\mathrm{Ext}^i_{S} ( I, M )_0$.
 \item The extensions $\mathrm{Ext}^i_{S} ( I (u ,e ), M )$ happen to be \fp graded $S$-modules also. In particular, we can truncate them to degree $0 \in \mathrm{Cl} ( 
      X_\Sigma )$. Since $\tilde{M}$ is a \emph{coherent} sheaf, this degree-$0$-layer happens to be a finite-dimensional $\mathbb{Q}$-vector space. Our algorithm returns its $\mathbb{Q}$-dimension. Details on the definition and computation of $\mathrm{Ext}^i_{S} ( I (u, e ), M )$ are provided in \cref{sec:ExtOfFPModules}.
\end{enumerate}

The packages \cite{CAPCategoryOfProjectiveGradedModules, CAPPresentationCategory, PresentationsByProjectiveGradedModules, TruncationsOfPresentationsByProjectiveGradedModules} provide the implementation of the category $S \mathrm{\textnormal{-}fpgrmod}$ in the language of \emph{categorical programming} of \texttt{CAP} \cite{CAP, PosurDoktor, GutscheDoktor}. In addition, basic functionality of toric varieties is provided by the \texttt{gap}-package \texttt{ToricVarieties} of \cite{homalg}. The package \cite{SheafCohomologyOnToricVarieties} extends this package and provides routines to find an ideal $I( u, e )$ as explained in the second step. In addition, this package provides implementations of algorithms which allow for a quick computation of $\mathrm{Ext}^i_{S} ( I( u, e ), M )$, as explained in \cref{sec:ExtOfFPModules}.

\paragraph{An Example}

Let us give an easy example on how these computations work in practise. Consider $\mathbb{P}^2_{\mathbb{Q}}$ and the \fp graded $S$-module $B_\Sigma$. We already found in \cref{sec:Sheafification} that $\tilde{B_\Sigma}$ is the structure sheaf of $\mathbb{P}^2_{\mathbb{Q}}$. So theory tells us that $h^i ( \mathbb{P}^2_{\mathbb{Q}}, \tilde{B_\Sigma} ) = ( 1, 0, 0 )$. We can use this as consistency check on the above implementations. Let us therefore compute these sheaf cohomologies with the \texttt{gap}-routines \cite{CAPCategoryOfProjectiveGradedModules, CAPPresentationCategory, PresentationsByProjectiveGradedModules, TruncationsOfPresentationsByProjectiveGradedModules, SheafCohomologyOnToricVarieties}. For $h^0 ( \mathbb{P}^2_{\mathbb{Q}}, \tilde{B_\Sigma} )$ our algorithm performs the following steps:
\begin{enumerate}
 \item Compute vanishing sets: $V^0 ( \mathbb{P}^2_{\mathbb{Q}} ) = \mathbb{Z}_{< 0}$, $V^1 ( \mathbb{P}^2_{\mathbb{Q}} ) = \mathbb{Z}$ and $V^2 ( 
      \mathbb{P}^2_{\mathbb{Q}} ) = \mathbb{Z}_{\geq -2}$.
 \item Find ideal: \texttt{gap} picked $I (u, e ) = \left\langle x_1, x_2, x_3 \right\rangle = B_\Sigma$.
 \item Compute truncation of global extension module: $H^0 ( \mathbb{P}^2_{\mathbb{Q}}, \tilde{B_\Sigma} ) \cong \mathrm{Hom}_S ( B_\Sigma, B_\Sigma )_0 \cong \mathbb{Q}$.
\end{enumerate}
Similarly, we found for the other sheaf cohomologies of $\tilde{B_\Sigma}$
\[ H^1 ( \mathbb{P}^2_{\mathbb{Q}}, \tilde{B_\Sigma} ) \cong \mathrm{Ext}^1_S \left( B_\Sigma, B_\Sigma \right)_0 \cong \mathbb{Q}^0 \, , \qquad H^2 ( \mathbb{P}^2_{\mathbb{Q}}, \tilde{B_\Sigma} ) \cong \mathrm{Ext}^2_S \left( S, B_\Sigma \right)_0 \cong \mathbb{Q}^0 \, . \]
So the sheaf cohomologies of $\tilde{B_\Sigma}$ indeed match those of the structure sheaf on $\mathbb{P}^2_{\mathbb{Q}}$.

\paragraph{Quality Assessment}

To tell how good or bad a model for $\mathcal{O}_{\mathbb{P}^2_{\mathbb{Q}}}$ this module $B_\Sigma$ really is, let us consider the sequence
\[ \mathfrak{h}^i \left( M, e \right) = \mathrm{dim}_{\mathbb{Q}} \left( \mathrm{Ext}^{(i)}_S \left( I \left( u, e \right) , M \right)_{0} \right) \, . \label{equ:QualitySeries} \]
With our choice of ideal $I( u, e )$ it is guaranteed that this sequence becomes constant for sufficiently large $e$. As the computations become harder and harder with increasing $e$, the upshot is to find the minimal integer $\check{e}$ such that the isomorphism \cref{equ:CohoIso} holds true. The quality of an estimate of this minimal integer $e$ and the \fp graded $S$-module $B_\Sigma$ to serve as model for $\mathcal{O}_{\mathbb{P}^2_{\mathbb{Q}}}$ can therefore be judged from looking at this sequence. In this particular example we have
\[ \mathfrak{h}^0 \left( e \right) = \left( 0, 1, 1, 1, \dots \right), \qquad \mathfrak{h}^1 \left( e \right) = \left( 0, 0, 0, 0, \dots \right), \qquad \mathfrak{h}^2 \left( e \right) = \left( 0, 0, 0, 0, \dots \right) \, . \]
So the perfect minimal integers are $\check{e}_0 = 1$, $\check{e}_1 = 0$ and $\check{e}_2 = 0$. Our estimates for $e_0$ and $e_2$ are therefore the ideal choices, but we overestimated $\check{e}_1$.

As for the quality of $B_\Sigma$ -- a perfect model of $\mathcal{O}_{\mathbb{P}^2_{\mathbb{Q}}}$ has $\check{e}_0 = \check{e}_1 = \check{e}_2 = 0$. For $B_\Sigma$ however $\check{e}_0 = \check{e}_1 = 1$. Hence, it is not a perfect model for $\mathcal{O}_{\mathbb{P}^2_{\mathbb{Q}}}$. In contrast the Cox ring $S$ sheafifies to the structure sheaf also and satisfies $\check{e}_0 = \check{e}_1 = \check{e}_2 = 0$. Consequently, this \fp graded $S$-module is, in absence of other constraints, \emph{the} best constructive model for $\mathcal{O}_{\mathbb{P}^2_{\mathbb{Q}}}$, which can be handled by \texttt{gap} \cite{GAP4}.

\subsection{Massless Spectrum on \emph{Non}-Generic Matter Curves} \label{subsec:Spectrum1}

As an example consider the gauge background
\[ \label{totalfluxchoice}
A \equiv A \left( \mathbf{10}_{1} \right) \left( -5 \right) + A_X \left( \frac{5}{2} \cdot H \right)
\]
with $H \in \mathrm{Cl} ( \mathbb{P}_{\mathbb{Q}}^3 )$ the hyperplane class on $\mathbb{P}_{\mathbb{Q}}^3$. This gauge background can be checked to satisfy the quantisation condition \cref{FW2}. This follows already from the analysis in \cite{oai:arXiv.org:1202.3138} around equ. (3.18) therein. The massless spectrum is counted by the sheaf cohomologies of the line bundles
\begin{align}
\begin{split}
L \left( A, C_{\mathbf{10}_{1}} \right) = \left. \mathcal{O}_{\mathbb{P}_{\mathbb{Q}}^2} \left( - 18 \right) \right|_{C_{\mathbf{10}_{1}}} \,,
\qquad L \left( A, C_{\mathbf{5}_{3}} \right) &= \left. \mathcal{O}_{\mathbb{P}_{\mathbb{Q}}^2} \left( 13 \right) \right|_{C_{\mathbf{5}_{3}}}  \,, \cr
L \left( A, C_{\mathbf{5}_{-2}} \right) \cong \mathcal{O}_{C_{\mathbf{5}_{-2}}} \left( - 5 Y_1 \right) \otimes \left. \mathcal{O}_{\mathbb{P}_{\mathbb{Q}}^2} \left( 14 \right) \right|_{C_{\mathbf{5}_{-2}}} \,,
\qquad L \left( A, C_{\mathbf{1}_{5}} \right) &= \left. \mathcal{O}_{\mathbb{P}_{\mathbb{Q}}^3} \left( 12 \right) \right|_{C_{\mathbf{1}_{5}}} \,.
\end{split}
\end{align}
The first two and the fourth line bundle manifestly arise by pullback of a line bundle on the toric base ${\mathbb{P}_{\mathbb{Q}}^3}$. Therefore, we can resolve these bundles by Koszul resolutions, formed from vector bundles on ${\mathbb{P}_{\mathbb{Q}}^3}$. For all of these vector bundles it is possible to compute the cohomology dimensions \eg via \emph{cohomCalg} \cite{Blumenhagen:2010pv, Blumenhagen:2010ed, Blumenhagen:2011xn, KoszulExtensionManual, cohomCalg:Implementation, 2011JMP....52c3506J, Rahn:2010fm}.

In general this information alone does not suffice to determine the cohomology dimensions of a pullback line bundle uniquely, rather the maps in the resolution need to be taken into account. However, in fortunate cases the exact sequences describing the resolution involve a sufficient number of zeros which allow one to predict the cohomology dimensions of the pullback line bundle without any knowledge about the involved mappings in the resolution. Indeed, the bundles on $C_{\mathbf{10}_{1}}$, $C_{\mathbf{5}_{3}}$ and $C_{\mathbf{1}_{5}}$ are such fortunate instances. Consequently, we are able to determine their cohomology dimension along the algorithms implemented in the \emph{Koszul extension of cohomCalg} \cite{Blumenhagen:2010pv, Blumenhagen:2010ed, Blumenhagen:2011xn, KoszulExtensionManual, cohomCalg:Implementation, 2011JMP....52c3506J, Rahn:2010fm}. The results are
\begin{align}
\begin{split}
& h^i \left( C_{\mathbf{10}_{1}}, L \left( A, C_{\mathbf{10}_{1}} \right) \right) = \left( 0, 74 \right) \, , \qquad
h^i \left( C_{\mathbf{5}_{3}}, L \left( A, C_{\mathbf{5}_{3}} \right) \right) = \left( 95, 0 \right) \, , \\
& h^i \left( C_{\mathbf{1}_{5}}, L \left( A, C_{\mathbf{1}_{5}} \right) \right) = \left( 445, 120 \right) \, .
\end{split}
\end{align}
Note that as a consequence of the zeros in the resolution, these values are independent of the complex structure moduli of the matter curves. In fact, if the matter curves in question were smooth, the above results for the cohomology groups on $C_{\mathbf{10}_{1}}$ and $C_{\mathbf{5}_{3}}$  would follow already from the Kodaira vanishing theorem and the  Riemann-Roch index theorem, which we discussed in \cref{subsec:LineBundlesOnRiemannSurfaces}.

By contrast, to determine the cohomology dimensions of the line bundle $L ( A, C_{\mathbf{5}_{-2}})$ we have to invoke the machinery described in  \cref{sec:ComputingMasslessSpectraWithGAP} as this line bundle does not descend from a line bundle on $\mathbb{P}^3_\mathbb{Q}$. As it turns out, the result is 
sensitive to the actual choice of Tate polynomials $\tilde{a_{ij}}$, \ie of complex structure moduli defining the elliptic fibration. 

\begin{table}[tb]
\begin{center}
\begin{tabular}{cccccc}
\toprule
Module & $\tilde{a_{1,0}}$ & $\tilde{a_{2,1}}$ & $\tilde{a_{3,2}}$ & $\tilde{a_{4,3}}$ & $h^i \left( C_{\mathbf{5}_{-2}}, L \left( A, C_{\mathbf{5}_{-2}} \right) \right)$ \\
\midrule
$M_1$ & $\left( x_1 - x_2 \right)^4$ & $x_1^7$ & $x_2^{10}$ & $x_3^{13}$ & $\left( 22, 43 \right)$ \\
$M_2$ & $\left( x_1 - x_2 \right) x_3^3$ & $x_1^7$ & $x_2^{10}$ & $x_3^{13}$ & $\left( 21, 42 \right)$ \\
$M_3$ & $x_3^4$ & $x_1^7$ & $x_2^7 \left( x_1 + x_2 \right)^3$ & $x_3^{12} \left( x_1 - x_2 \right)$ & $\left( 11, 32 \right)$ \\
$M_4$ & $\left( x_1 - x_2 \right)^3 x_3$ & $x_1^7$ & $x_2^{10}$ & $x_3^{13}$ & $\left( 9, 30 \right)$ \\
$M_5$ & $x_3^4$ & $x_1^7$ & $x_2^{8} \left( x_1 + x_2 \right)^2$ & $x_3^{11} \left( x_1 - x_2 \right)^2$ & $\left( 7, 28 \right)$ \\
$M_6$ & $x_3^4$ & $x_1^7$ & $x_2^{10}$ & $x_3^{8} \left( x_1 - x_2 \right)^5$ & $\left( 6, 27 \right)$ \\
$M_7$ & $x_3^4$ & $x_1^7$ & $x_2^9 \left( x_1 + x_2 \right)$ & $x_3^{10} \left( x_1 - x_2 \right)^3$ & $\left( 5, 26 \right)$ \\
\bottomrule
\end{tabular}
\end{center}
\caption[Moduli dependence of zero modes in \emph{F-theory} toy model.]{For a number of choices of the Tate polynomials $\tilde{a_{i,j}}$ we have computed the associated module $M_i$ and the sheaf cohomology dimensions of $\tilde{M_1}$.}
\label{table-resultsScanOverComplexStructureModuliSpace}
\end{table}

\begin{figure}[tb]
\centering
\subfloat[$\mathfrak{h}^0( M_1, e )$ for $0 \leq e \leq 50$.]{\label{figure 1001}
     \includegraphics{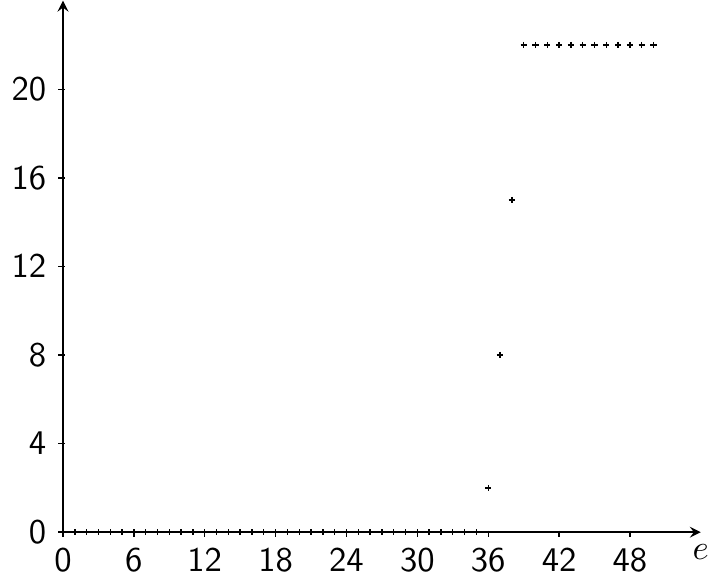}
}  
\subfloat[$\mathfrak{h}^1( M_1, e )$ for $0 \leq e \leq 50$.]{\label{figure 1002}
     \includegraphics{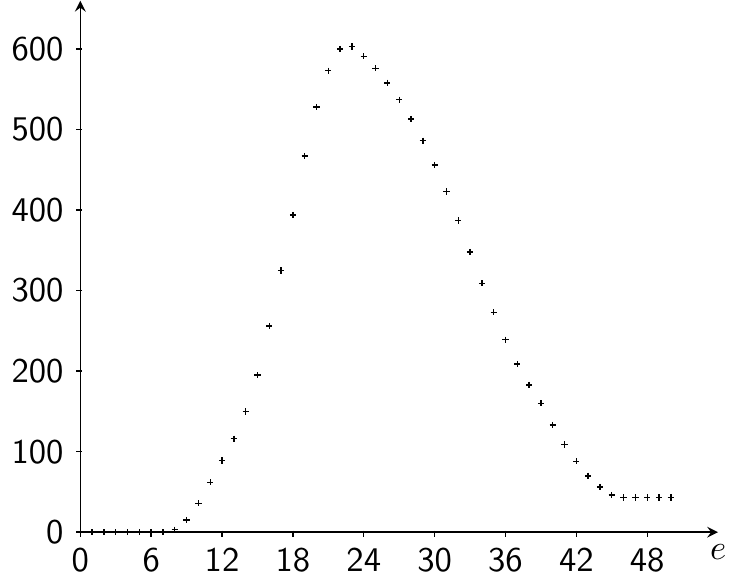}
}
\caption{Visualisation of the `quality series' $\mathfrak{h}^0( M_1, e )$ and $\mathfrak{h}^1( M_1, e )$.}
\label{fig1003}
\end{figure}

For a number of choices, we compute an \fp graded $S$-module $M$ and then deduce the cohomology dimension of $\tilde{M}$ by use of the technologies described in \cref{subsec:SheafCohomologyFromFPGradedSModules}. This leads to the results summarised in \cref{table-resultsScanOverComplexStructureModuliSpace}. In particular, we observe jumps in the cohomology dimensions of the line bundle on $C_{\mathbf{5}_{-2}}$ as we wander in the moduli space of the elliptic fibration $\hat{\pi} \colon \hat{Y}_4 \twoheadrightarrow \mathcal{B}_6$. E.g.\ moving from the first line to the second, we observe that a pair of a chiral and anti-chiral (super)-field becomes massive, and is therefore no longer accounted for by the massless spectrum. In moving to the last line, another 16 such pairs become massive.

Let us give a few more details on the involved computations. To this end, let us focus on module $M_1$. It has a minimal free resolution
\begin{align}
\begin{split}
0 \leftarrow M_1 & \leftarrow S \left( -6 \right) \oplus S \left( -21 \right) \\
                 &\leftarrow S \left( -34 \right) \oplus S \left( -35 \right) \oplus S \left( -45 \right) \oplus S \left( -23 \right) \oplus S \left( -38 \right) \\
                 & \leftarrow S \left( -36 \right) \oplus S \left( -47 \right) \oplus S \left( -48 \right) \leftarrow 0 \, .
\end{split}
\end{align}
From this resolution \texttt{plesken} first read off the Betti numbers of $M_1$, and then used them to compute $e_0 = e_1 = 46$ (\cf  \cref{subsec:SheafCohomologyFromFPGradedSModules}). From this we conclude $h^i ( C_{\mathbf{5}_{-2}}, \tilde{M_1} ) = ( 22, 43 )$. Note that we could have applied the so-called \emph{linear regularity} as well \cite{2014arXiv1409.6100B}. For this particular module $M_1$ this method predicts $e_0 = 45$, which is slightly better. Explicitly we computed with \texttt{plesken} the 'quality series`, whose values we list in \cref{section:QualitySeriesM1} and visualise in \cref{fig1003}. This shows that for this module the ideal values are $\check{e}_0 = 39$ and $\check{e}_1 = 46$, which shows that our estimates on $e_0$ and $e_1$ are not perfect, and present one possibility to improve this algorithm in the future. More example computations along these lines are given in \cite{Bies:2017fam}.

\section{Summary}

In this chapter we have explained \fp graded $S$-modules and how they encode coherent sheaves on a toric variety $X_\Sigma$ with coordinate ring $S$. In \cref{sec:ComputingTheSpectra} we have then added together the findings of \cref{chapter:MasslessSpectraAndSheafCohomology} and this knowledge about \fp graded $S$-modules. In working-out an \emph{F-theory} toy model, we have demonstrated how we can use the algorithm outlined in \cref{chapter:MathDetailsSheafCohomologies} to compute sheaf cohomologies and thereby zero modes in \emph{F-theory} vacua. In particular, we were able to demonstrate that the number of zero modes strongly depends on the complex structure of the matter curves. Along these lines we will discuss more refined \emph{F-theory} models in the next chapter. The studied geometries will indicate both the powers and limitations of our algorithm.

\chapter{Toric F-Theory GUT-Models} \label{chapter:GUTModels}
In this chapter we pick an explicit form of base space $\mathcal{B}_6$ and employ the $SU(5) \times U(1)_X$-top discussed in \cref{sec:ToricFTheoryGUTModels} and \cref{sec:SU5xU1Top} to construct a smooth resolution $\hat{\pi} \colon \hat{Y}_4 \twoheadrightarrow \mathcal{B}_6$. Together with a chosen gauge background, this geometry defines an \emph{F-theory} compactification to four dimensions with $SU(5) \times U(1)_X$ gauge symmetry.

As in all phenomenological applications of string compactifications, we try to make contact with the \emph{standard model}. This is achieved by interpreting the \emph{F-theory} vacuum in question as a Georgi-Glashow GUT, which we discussed in \cref{subsec:Georgi-Glashow}. In contrast to our original discussion, the breaking
\[ SU(5) \times U(1)_X \to SU(3) \times SU(2) \times U(1) \]
is achieved by giving a VEV to the \emph{hypercharge flux} $A_Y ( \mathcal{H} )$, which we discussed in \cref{chapter:MasslessSpectraAndSheafCohomology} already. The curve $\mathcal{H} \subseteq W$ is contained in the \emph{GUT-divisor} $W \subseteq \mathcal{B}_6$. To ensure a massless $U(1)_Y$ gauge boson, this curve is subject to a number of rather sophisticated conditions, which we discuss in \cref{sec:ChoiceOfHyperchargeFluxAndExotics} in much detail. In particular, it is found that this curve must not be pullback from $\mathcal{B}_6$. Therefore, the ability to handle a universal flux $A_Y ( \mathcal{H} )$ and subsequently compute its zero modes is possible only with the full technology presented in \cref{chapter:DetailsOnFPGradedSModules}. In particular, such models outline both the powers and limitations of our algorithm. Let us therefore try to compute the zero modes for the gauge background
\[ A = A_X \left( F \right) + A \left( \mathbf{10}_1 \right) \left( \lambda \right) + A_Y \left( \mathcal{H} \right) \in \mathrm{CH}^2 ( \hat{Y}_4 ) \]
which in addition to the hypercharge flux $A_Y ( \mathcal{H} )$ contains the $U(1)_X$-flux $A_X( F )$ and the universal flux $A( \mathbf{10}_1 )( \lambda )$.

We first study a geometry with $W \cong dP_3$ in \cref{sec:dP3-Example}. We analyse this geometry and perform a scan over the admissible fluxes. Thereby, we identify those fluxes which induce zero mode chiralities close to the ones required for the \emph{standard model}, lead to a supersymmetric \emph{F-theory} vacuum and at the same time satisfy the D3-tadpole cancellation. Whilst physically appealing, this demand is very restrictive. In the geometry in question there are no such fluxes, and we are looking at best at an \emph{F-theory} toy-model. Still this geometry forms a perfect setting to demonstrate the powers of the computer algorithm introduced in \cref{chapter:DetailsOnFPGradedSModules}. Consequently, we will simply compute zero modes of fluxes which satisfy the D3-tadpole constraint, whose chiralities exceed the required numbers in the \emph{standard model}, but will not lead to a supersymmetric \emph{F-theory} vacuum.

The fact that we can actually compute these zero modes is already a great achievement. It relies on the GUT-surface $W$ being a $dP_3$-surface, which consequently can be realised as toric variety directly. Experimental evidence indicates that our algorithmic powers are significantly stronger when performed on this toric model directly, than on the corresponding complete intersection in $\mathcal{B}_6$. Hence, the main simplification arises from constructing an isomorphism
\[ \varphi \colon \text{toric model of } dP_3 \xrightarrow{\sim} W \subseteq \mathcal{B}_6 \label{equ:Conjecture} \]
along which we can `pullback' many computations from $\mathcal{B}_6$. 

Unfortunately, we cannot construct such a morphism from first principles. Rather we look at how morphism of graded rings induce mappings of the associated projective schemes in algebraic geometry, and conjecture a natural generalisation thereof to smooth, projective toric schemes in \cref{conj:II}. We apply this conjecture to propose an embedding of the form \cref{equ:Conjecture}. In this particular example, we perform a number of rather non-trivial consistency checks, which involve comparison of intersection numbers, sheaf cohomologies and genera of matter curves. In all cases we found matching results. Therefore we are positive that, at least in this particular instance, our conjecture is correct.

Finally, in \cref{sec:dP7-Example}, we relax our assumptions on the GUT-surface and study a geometry with $W \cong dP_7$. Such a surface cannot be realised as a toric variety. This obstruction will turn out to limit our computational powers quite significantly.

\section{Gauge Group Breaking and the Choice Of Hypercharge Flux} \label{sec:ChoiceOfHyperchargeFluxAndExotics}

\subsection{Stückelberg Mass for the Hypercharge Flux} \label{subsec:StueckelbergMasses}

\paragraph{Breaking $\mathbf{SU(5)}$ with Hypercharge Flux}
We have studied the Georgi-Glashow model in \cref{subsec:Georgi-Glashow}. In particular, we explained how the Higgs mechanism can be employed to break the GUT-group. In short, we employed a field with potential $V$ such that its Lagrangian was invariant under the entire gauge group $SU(5)$, but used a ground state to break this symmetry in a controlled fashion.

So why not repeat the story here? The answer is that in general we are lacking knowledge of the Higgs potential -- in \emph{string theory} we do not simply add terms to the Lagrangian, as we did for the Higgs mechanism, but derive essentially all information (including potentials) from the geometry of the compactification. This is the reason why the parameter $\alpha^\prime$ is considered the only input parameter for \emph{string theory}. Unfortunately, it is in general quite involved to derive such potentials in compactifications, and so in general both knowledge and control over the Higgs potential are lacking. Luckily, despite the Higgs mechanism there are other means to break the gauge group of a GUT-model. 

An alternative way to break the GUT-group is the use of a so-called \emph{hypercharge flux}. Not only does it encompassing our ignorance of the Higgs potential, but it is also of major phenomenological relevance. Suffice it here to mention that it offers solutions to the proton decay problem (\cf \cref{subsec:Georgi-Glashow}) \cite{Beasley:2008dc, Beasley:2008kw, Donagi:2011jy, Donagi:2008ca, Donagi:2011dv}. In this sense hypercharge flux is really \emph{the} phenomenological backbone of \emph{F-theory} GUT-models. 

We can improve the situation even further by looking at so-called \emph{flipped GUT-models} whose gauge groups have additional $U(1)$-factors. For example, $SU(5) \times U(1)$ is flipped variant of the Georgi-Glashow model. Consequently, all particles, as being representations of the gauge group, have a $U(1)_X$-charge which puts constraints on the available Yukawa interactions. The resulting $U(1)$-selection rules can therefore help to prevent proton decay.

In this chapter we will embrace both options -- we look at a flipped $SU(5) \times U(1)_X$ GUT-model and employ the hypercharge flux to break this gauge group. To understand this breaking mechanism, let us think of a type IIB \emph{string theory} compactification. Recall that in such a compactification, spacetime is factored as $M_{10} = \mathcal{E}_4 \times \mathcal{B}_6$. Let us now consider a so-called \emph{spacetime-filling} D7-brane, \ie a D7-brane whose world volume covers the external space $\mathcal{E}_4$ completely and wraps a complex 2-cycle $\Sigma \subseteq \mathcal{B}_6$. The hypercharge flux on such a D7-brane is modelled by a line bundle $L_Y$ with structure group $U(1)$ on $\mathcal{E}_4 \times \Sigma$. This line bundle should encode the $U(1)$-subgroup of $SU(5)$ generated by (\cf \cref{subsec:Georgi-Glashow})
\[ T_1 = \frac{1}{2 \sqrt{15}} \cdot \mathrm{diag} \left( -2, -2, -2, 3, 3 \right) \, .  \]
We now pick a VEV for the $SU(5)$-field strength $F$ of the form $\langle F_{SU(5)} \rangle = c_1 ( L_Y ) \cdot T_1$. This choice then breaks the $SU(5)$-invariance to $SU(3) \times SU(2) \times U(1)_Y$ just as discussed in \cref{subsec:Georgi-Glashow}. But there is one important caveat here, namely the $U(1)_Y$ gauge boson can aquire a so-called \emph{Stückelberg} mass.

\paragraph{The Origin of the Stückelberg Mass}

In the following, our discussion follows in parts closely to \cite{Braun:2014pva} and \cite{Ibanez2014string}. The starting point of our analysis on Stückelberg masses is the type IIB 10-dimensional supergravity action. It induces the following three action terms
\[ \Mint_{\mathcal{M}_{1,3} \times \Sigma}{C_4 \wedge \mathrm{Tr} \left( F \wedge F \right)} \, , \qquad \Mint_{\mathcal{M}_{1,3} \times \Sigma}{\mathrm{Tr} \left( F \wedge \ast_8 F \right)} \, , \qquad  \Mint_{\mathcal{M}_{1,3} \times \mathcal{B}_6}{d C_4 \wedge \ast d C_4} \, \]
where $C_4$ is the RR 4-form field and $F$ the $SU(5)$ field strength. As a next step we perform a dimensional reduction of these terms to $\mathcal{M}_{1,3}$. To this end, we write $C_4 = \sum_{\alpha}{c_2^\alpha \wedge \omega_\alpha}$, where $\{ \omega_\alpha \}$ is a basis of $H^{1,1} ( \mathcal{B}_6 )$, and decompose the $SU(5)$ field strength $F = F_{\mathcal{E}} + F_{\Sigma}$ into an external and an internal piece. We use of the embedding $\iota \colon \Sigma \hookrightarrow \mathcal{B}_6$ to define
\[ K_\alpha^l = 2 \cdot \mathrm{tr} ( T^l T^l ) \cdot \Mint_{\Sigma}{\iota^\ast \omega_\alpha \wedge F_{\Sigma}^l} \, .  \]
Then the above three terms take the form
\[ \sum_{\alpha,l}{K_\alpha^l \cdot \Mint_{\mathcal{M}_{1,3}}{c_2^\alpha \wedge F_{\mathcal{E}}^l}} \, , \qquad \Mint_{\mathcal{M}_{1,3}}{\mathrm{Tr} \left( F_{\mathcal{E}} \wedge \ast_4 F_{\mathcal{E}} \right)} \, , \qquad \Mint_{\mathcal{M}_{1,3}}{d c_4 \wedge \ast d c_4} \, . \]
We can also rewrite these terms explicitly in the components of the involved harmonic forms. Let us focus on $F_{\mathcal{E}}^1$ only. By taking into account the corresponding prefactors also, the 4-dimensional Lagrangian takes the form
\begin{align}
\begin{split}
\mathcal{L}_{\text{Stückelberg}} &= \frac{1}{4} \epsilon^{\mu \nu \rho \sigma} \cdot \sum_{\alpha}{K^1_\alpha c^{(\alpha)}_{\mu \nu}} F^1_{\mathcal{E} \rho \sigma} - \frac{1}{4g^2} F_{\mathcal{E}}^{1 \mu \nu} F^1_{\mathcal{E} \mu \nu} - \frac{1}{12} \sum_{\alpha}{h^{\alpha, \mu \nu \rho} h^{\alpha}_{\mu \nu \rho}} \, , \\
h^{\alpha}_{\mu \nu \rho} &= \partial_{[ \mu} c^{\alpha}_{\nu \rho ]} = \partial_\mu c^{\alpha}_{\nu \rho} + \partial_{\nu} c^{\alpha}_{\rho \mu} + \partial_\rho c^{\alpha}_{\mu \nu} - \partial_\nu c^{\alpha}_{\mu \rho} - \partial_\mu c^{\alpha}_{\rho \nu} - \partial_\rho c^{\alpha}_{\nu \mu} \, .
\end{split}
\end{align}
This is the classical Stückelberg Lagrangian. To avoid too-heavy notation, let us drop the superscript $\alpha$. By construction we know $dh = d^2c = 0$. Nonetheless we can consider $h_{\mu \nu \rho}$ as an independent field and enforce $dh = 0$ by inserting a Lagrange multiplier $\eta$. The corresponding action takes the form
\[ \mathcal{L}_{\text{Stückelberg}} = - \frac{1}{12} h^{\mu \nu \rho} h_{\mu \nu \rho} - \frac{1}{4g^2} F_{\mathcal{E}}^{1 \mu \nu} F^1_{\mathcal{E} \mu \nu}
- \frac{K^1}{6} \epsilon^{\mu \nu \rho \sigma} h_{\mu \nu \rho} F^1_{\mathcal{E} \sigma} - \frac{1}{6} \eta \epsilon^{\mu \nu \rho \sigma} \partial_\mu h_{\nu \rho \sigma} \, . \label{equ:StückelbergII} \]
Let us integrate the last term by parts. This results into an action quadratic in $h_{\mu \nu \rho}$. A solution to the resulting equation of motion is given by
\[ h^{\mu \nu \rho} = - \epsilon^{\mu \nu \rho \sigma} \left( K^1 A^1_{\mathcal{E} \sigma} + \partial_\sigma \eta \right) \, . \]
By plugging this back into \cref{equ:StückelbergII} one finds
\[ \mathcal{L}_{\text{Stückelberg}} = - \frac{1}{4g^2} F_{\mathcal{E}}^{1 \mu \nu} F^1_{\mathcal{E} \mu \nu} - \frac{1}{2} \left( K^1 A^1_{\mathcal{E} \sigma} + \partial_\sigma \eta \right)^2 \, , \]
which indeed describes a massive gauge boson $A^1_{\mathcal{E}}$. This is the \emph{Stückelberg mechanism} \cite{2009esuf.book..273}.

\paragraph{Absence of the Stückelberg Mass}

A sufficient condition for $A^1_{\mathcal{E}}$ to remain massless is to demand
\[ K^1_\alpha = \Mint_{\Sigma}{\iota^\ast \omega_2 \wedge \mathcal{F}^1_{\Sigma}} = 0 \qquad \forall \omega_2 \in H^{1,1} \left( \mathcal{B}_6, \mathbb{Z} \right) \, ,  \label{equ:NoStückelbergMass} \]
where $\iota \colon \Sigma \hookrightarrow \mathcal{B}_6$ is the inclusion map. In particular, this forbids $L_Y$ to be pullback of a line bundle defined on $\mathcal{B}_6$.

This result easily generalises to \emph{F-theory}, owing to its close connection to type IIB \emph{string theory}. In this setup we switch on a $G_4$-flux, which is supported over a divisor $\mathcal{H} \subseteq W$ in the GUT-divisor $W$. By expressing the Lie-group generator $T_1$ in terms of the roots of $SU(5)$, one finds 
\[ A_Y \left( \mathcal{H} \right) =  \left. \left(2 E_1 + 4 E_2 + 6 E_3 + 3 E_4\right) \right|_{\mathcal{H}} \, , \label{equ:HyperchargeFluxII} \]
whose zero modes we discussed in \cref{subsec:ZeroModesOfHyperchargeFlux} already. We found that they are encoded in the sheaf cohomologies of (powers of) $\mathcal{O}_W ( \mathcal{H} ) |_{C_{\mathbf{R}}}$. Since the absence of a Stückelberg mass requires that $\mathcal{H}$ is not pullback from $\mathcal{B}_6$, computing these sheaf cohomologies requires the full machinery presented in \cref{chapter:DetailsOnFPGradedSModules}. 

If $\mathcal{H}$ is merely non-pullback, then \cref{equ:NoStückelbergMass} can be violated. As pointed out in \cite{Braun:2014pva}, it is sufficient to require that $\mathcal{H}$ is the \emph{difference} of (at least) two holomorphic curves in $W$, which all are non-pullback from $\mathcal{B}_6$. In the following examples we therefore ensure that $\mathcal{H} = C_1 - C_2$ where $\mathcal{H}$, $C_1$ and $C_2$ are all non-pullback from $\mathcal{B}_6$.

\subsection{Absence of Massless Exotics}

\paragraph{Origin and phenomenological significance}

We just found a sufficient condition on the hypercharge flux $L_Y$ to leave the $U(1)_Y$ gauge boson massless in the external space. Consequently such a hypercharge flux triggers the gauge group breaking
\[ SU \left( 5 \right) \times U \left( 1 \right)_X \to SU \left( 3 \right) \times SU \left( 2 \right) \times U \left( 1 \right)_X \times U \left( 1 \right)_Y \, . \]
Recall that this breaking induces the following decomposition of the adjoint representation:
\[ \mathbf{24}_{q_X} \to \left( \mathbf{8}, \mathbf{1} \right)_{q_X, 0_Y} \oplus \left( \mathbf{1}, \mathbf{3} \right)_{q_X, 0_Y} \oplus \left( \mathbf{1}, \mathbf{1} \right)_{q_X, 0_Y} \oplus \left( \mathbf{3}, \mathbf{2} \right)_{q_X, 5_Y} \oplus \left( \mathbf{\overline{3}}, \mathbf{2} \right)_{q_X,-5_Y} \, . \]
The gauge bosons in the representations $( \mathbf{3}, \mathbf{2} )_{q_X, 5_Y}$ and $( \overline{\mathbf{3}}, \mathbf{2} )_{q_X, -5_Y}$ are the $X$ and $Y$ gauge bosons of the Georgi-Glashow model. Recall that they trigger proton decay, which is one of the major phenomenological drawbacks of Georgi-Glashow GUT-model. The masses of these gauge bosons, which they acquire from the hypercharge flux breaking, suppress the proton decay. To match experimental findings, these masses must be huge. Consequently it is of phenomenological prime interest to rule out the existence of any chiral massless states in these representations. Can we therefore tweak the \emph{F-theory} geometry in such a way that all exotic states in these representations are massive?

\paragraph{A Criterion on del-Pezzo Surfaces}

Let us assume that the GUT-surface $W$ is a del-Pezzo surface. Then the absence of these exotic states has been studied in \cite{Beasley:2008kw}. It was found that this demand is equivalent to the requirement that for the 5-th and (-5)-th power of the hypercharge flux all sheaf cohomologies vanish, which is equivalent to
\[ \left( 5 \mathcal{H} \right) \cdot K_{dP_n} = 0 \quad \mathrm{ and } \quad \left( 5 \mathcal{H} \right)^2 = -2 \, . \label{equ:ProperHyperchargeflux} \]
In consequence, $\mathcal{H}$ must not be a $\mathbb{Z}$-Cartier, but a proper $\mathbb{Q}$-Cartier divisor. But we cannot easily associate a line bundle to a proper $\mathbb{Q}$-Cartier divisor. Is this the end of the story?

\begin{table}[tb]
\centering
\begin{tabular}{|c|c|c|c|c|}
\toprule
surface & $D \left( S^{a}_{\mathbf{10}_1}, A \right)$ & $D \left( S^{a}_{\mathbf{5}_3}, A \right)$ & $D \left( S^{a}_{\mathbf{5}_{-2}}, A \right)$ & $D \left( S^{a}_{\mathbf{1}_{5}}, A \right)$ \\
\hline \hline
$S_{*}^{(1)}$ & \multirow{3}{*}{\parbox{0.2\textwidth}{\centering $-4 \mathcal{H} + \frac{1}{5} F - \frac{3 \lambda}{5} Y_1 + \frac{4 \lambda}{5} Y_2 + \frac{1}{2} K_{C_{\mathbf{10}_1}}$}}
              & \multirow{3}{*}{\parbox{0.25\textwidth}{\centering $-2 \mathcal{H} + \frac{3}{5} F - \frac{2 \lambda}{5} Y_2 + \frac{1}{2} K_{C_{\mathbf{5}_3}}$}}
              & \multirow{3}{*}{\parbox{0.2\textwidth}{\centering $-2\mathcal{H} - \frac{2}{5} F + \frac{3 \lambda}{5} Y_1 - \frac{2}{5} Y_2 + \frac{1}{2} K_{C_{\mathbf{5}_{-2}}}$}}
              & $F + \frac{1}{2} K_{C_{\mathbf{1}_5}}$ \\
              \cline{5-5}
$S_{*}^{(2)}$ & & & \\
$S_{*}^{(3)}$ & & & \\ \cline{2-4}
$S_{*}^{(4)}$ & \multirow{6}{*}{\parbox{0.2\textwidth}{\centering $\mathcal{H} + \frac{1}{5} F - \frac{3 \lambda}{5} Y_1 + \frac{4 \lambda}{5} Y_2 + \frac{1}{2} K_{C_{\mathbf{10}_1}}$}}
              & \multirow{2}{*}{\parbox{0.25\textwidth}{\centering $3 \mathcal{H} + \frac{3}{5} F - \frac{2 \lambda}{5} Y_2 + \frac{1}{2} K_{C_{\mathbf{5}_3}}$}}
              & \multirow{2}{*}{\parbox{0.2\textwidth}{\centering $3 \mathcal{H} - \frac{2}{5} F + \frac{3 \lambda}{5} Y_1 - \frac{2}{5} Y_2 + \frac{1}{2} K_{C_{\mathbf{5}_{-2}}}$}} \\
$S_{*}^{(5)}$ & & & \\ \cline{3-4}
$S_{*}^{(6)}$ & \\
$S_{*}^{(7)}$ & \\
$S_{*}^{(8)}$ & \\
$S_{*}^{(9)}$ & \\\cline{2-2}
$S_{*}^{(10)}$ & \parbox{0.2\textwidth}{\centering $6 \mathcal{H} + \frac{1}{5} F - \frac{3 \lambda}{5} Y_1 + \frac{4 \lambda}{5} Y_2 + \frac{1}{2} K_{C_{\mathbf{10}_1}}$} \\
\cline{1-2}
\end{tabular}
\caption[Massless spectrum of $A$.]{The zero modes of the flux $A = A_X \left( F \right) + A \left( \mathbf{10}_1 \right) \left( \lambda \right) + A_Y \left( \mathcal{H} \right)$ are counted by the line bundles associated to the above divisors on $C_{\mathbf{10}_1}$, $C_{\mathbf{5}_3}$, $C_{\mathbf{5}_{-2}}$ and $C_{\mathbf{1}_5}$.}
\label{table-netspectrum}
\end{table}

Luckily, we are not interested in the zero modes of the hypercharge flux alone, but we are interested in the zero modes of the flux $A = A_X ( F ) + A ( \mathbf{10}_1 ) ( \lambda ) + A_Y ( \mathcal{H} )$, which is the sum of the $U(1)_X$-flux $A_X ( F )$, the matter surface flux $A ( \mathbf{10}_1 ) ( \lambda )$ and the hypercharge flux $A_Y ( \mathcal{H} )$. On every matter curve $C_{\mathbf{R}}$ we associated to these fluxes a $\mathbb{Q}$-Cartier divisor -- \cf \cref{table-N5} and \cref{table-N10}. By adding these results we obtain \cref{table-netspectrum}. Let us use this opportunity to point out that indeed, the states counted by the sheaf cohomologies of the divisors listed in \cref{table-netspectrum} can be interpreted in terms of (supersymmetry partners of) the \emph{standard model} particles according to \cref{table-brokenStates}.

As long as the sums of these $\mathbb{Q}$-divisors in \cref{table-netspectrum} add up to a $\mathbb{Z}$-Cartier divisor, nothing stops us from formally accepting a fractional divisor $\mathcal{H}$. The Freed-Witten quantisation condition $G_4 + \frac{1}{2} c_2(\hat{Y}_4) \in H^{2,2}_{{\mathbb Z}}(\hat{Y}_4)$ indeed ensures that these sums of $\mathbb{Q}$-Cartier divisor add up to a $\mathbb{Z}$-Cartier divisor. In practise however, it can be fairly involved to verify the linear equivalence of a sum of $\mathbb{Q}$-Cartier divisors to a $\mathbb{Z}$-Cartier divisor. If however one succeeds in overcoming this hurdle, then the technology from \cref{chapter:DetailsOnFPGradedSModules} can be applied to compute the massless spectrum.

\begin{table}[tb]
\centering
\begin{tabular}{|c|c|c|c|c|}
\toprule
surface & $C_{\mathbf{10}_1}$ & $C_{\mathbf{5}_3}$ & $C_{\mathbf{5}_{-2}}$ & $C_{\mathbf{1}_{5}}$ \\
\hline \hline
$S_{*}^{(1)}$ & \multirow{3}{*}{\parbox{0.2\textwidth}{\centering $U^c \colon \left( \overline{\mathbf{3}}, \mathbf{1} \right)_{1_X, -4_Y}$}}
              & \multirow{3}{*}{\parbox{0.25\textwidth}{\centering $D^c \colon \left( \mathbf{3}, \mathbf{1} \right)_{3_X, -2_Y}$}}
              & \multirow{3}{*}{\parbox{0.2\textwidth}{\centering $H_T \colon \left( \mathbf{3}, \mathbf{1} \right)_{-2_X, -2_Y}$}}
              & $N^c \colon \left( \mathbf{1}, \mathbf{1} \right)_{5_X, 0_Y}$ \\
              \cline{5-5}
$S_{*}^{(2)}$ & & & \\
$S_{*}^{(3)}$ & & & \\ \cline{2-4}
$S_{*}^{(4)}$ & \multirow{6}{*}{\parbox{0.2\textwidth}{\centering $Q \colon \left( \mathbf{3}, \mathbf{2} \right)_{1_X, 1_Y}$}}
              & \multirow{2}{*}{\parbox{0.25\textwidth}{\centering $L \colon \left( \mathbf{1}, \mathbf{2} \right)_{3_X, 3_Y}$}}
              & \multirow{2}{*}{\parbox{0.2\textwidth}{\centering $H_D \colon \left( \mathbf{1}, \mathbf{2} \right)_{-2_X, 3_Y}$}} \\
$S_{*}^{(5)}$ & & & \\ \cline{3-4}
$S_{*}^{(6)}$ & \\
$S_{*}^{(7)}$ & \\
$S_{*}^{(8)}$ & \\
$S_{*}^{(9)}$ & \\\cline{2-2}
$S_{*}^{(10)}$ & \parbox{0.2\textwidth}{\centering $E^c \colon \left( \mathbf{1}, \mathbf{1} \right)_{1_X, 6_Y}$} \\
\cline{1-2}
\end{tabular}
\caption[The representations of the states listed in \cref{table-netspectrum}.]{The hypercharge flux induces a breaking of the zero modes of $SU(5) \times U(1)_X$. The resulting states, listed in \cref{table-netspectrum}, allow for an interpretation in terms of the \emph{standard model} particles listed in \cref{my_old_table_2}. The only exceptions are $H_D$ and $H_T$, which are the supersymmetry partners of the Higgs doublet and triplet, respectively.}
\label{table-brokenStates}
\end{table}

\subsection{Choice Of Hypercharge Flux}

\paragraph{Tension between Computational Simplicity and Phenomenological Demands}

Let us recall that formally we allow each gauge background to give rise to $\mathbb{Q}$-Cartier divisors on the matter curves as long as their sums in \cref{table-netspectrum} add up to $\mathbb{Z}$-Cartier divisors. Let us now analyse under what conditions this is possible in the examples to follow.

By definition the anti-canonical bundle $\overline{K}_{\mathcal{B}_6}$ and $F$ are divisor classes on $\mathcal{B}_6$. By contrast, to avoid a Stückelberg mass for $A^1_{\mathcal{E}}$ the divisor $\mathcal{H}$ must not be a pullback from $\mathcal{B}_6$. We will model the GUT-surfaces by del-Pezzo surfaces, whose Picard group is free. Hence, $\mathcal{H}$ is linearly inequivalent to any linear combination of the pullbacks of $\overline{K}_{\mathcal{B}_6}$ and $F$. In consequence, any 'cancellation' among the divisor classes must involve the divisors induced by $A ( \mathbf{10}_1 ) ( \lambda )$. These divisor classes however can in general not even be expressed in terms of $\mathrm{Cl} ( W )$, but rather force us to engineer cancellations among divisors in $\mathrm{Cl} ( C_{\mathbf{R}} )$. This group is continuous, and therefore far more complicated than the Picard groups of smooth, complete toric varieties for example. This is the reason why the design of such quantisation conditions is at best very challenging.\footnote{For example, suppose that we guessed linear relations between two $\mathbb{Z}$-Cartier divisors on a matter curve, then the proof of this linear equivalence requires us to construct an isomorphism of the associated line bundles. This is anything but an algorithmic problem, and to date exceeds the structures implemented in \texttt{gap} \cite{GAP4} by far.} Currently, it exceeds our (algorithmic) abilities by far.

The only alternative is to switch on a `shift-flux' -- motivated from the corresponding situation in type IIB \cite{Blumenhagen:2008zz}. To this end, we are essentially left to construct a flux analogous to $A ( \mathbf{10}_1 )( \lambda )$, which in contrast is not supported by $C_{\mathbf{10}_1}$ but rather by $\mathcal{H}$. Unfortunately, such a shift-flux is unknown as of this writing, so that we cannot follow this approach either.

This observation leaves us with one option only -- we have to accept the presence of the exotics in the representations $( \mathbf{3}, \mathbf{2} )_{q_X, 5_Y}$ and $( \overline{\mathbf{3}}, \mathbf{2} )_{q_X, -5_Y}$. From a phenomenological point of view this renders our analysis into a toy model at best. In fact, in both examples to follow the zero modes will not even have the chiralities demanded for in the \emph{standard model}. Hence, our interest in studying these geometries is not to find a realistic \emph{F-theory} GUT-model, rather we intend to the test the limits of the tools presented in \cref{chapter:DetailsOnFPGradedSModules}. In addition, we will gather evidence for \cref{conj:II}.

\paragraph{Choice Of Hypercharge Flux}

This choice made, let us now pick an explicit parametrisation of $\mathcal{H}$. To this end, we first recall that for $1 \leq n \leq 8$, a $dP_n$-surface can be understood as blow-up of $\mathbb{P}_{\mathbb{Q}}^2$ in $n$ points $p_i$. The Class group of a $dP_n$-surface is isomorphic to $\mathbb{Z}^{n+1}$. It is freely generated by the hyperplane class $H$, inherited from the corresponding divisor class of $\mathbb{P}^2_{\mathbb{Q}}$, and the exceptional divisors $E_i$, each of which is associated to precisely one of the blow-up points.

Let us now pick the divisor $\mathcal{H}$ five times larger than phenomenologically appealing. This then amounts to the following conditions
\[ \mathcal{H} \cdot K_{dP_n} = 0 \qquad \mathrm{and} \qquad \mathcal{H}^2 = -2 \, . \label{equ:tophyperchargeflux} \]
As pointed out in \cite{Beasley:2008kw}, this implies that $\mathcal{H}$ is either of the form $E_i - E_j$ or $\pm ( H - E_i - E_j )$ for distinct $i,j$. Throughout this chapter we make the choice $\mathcal{H} = E_i - E_j$ for suitable exceptional divisors $E_i$ and $E_j$. To avoid a Stückelberg mass for $A^1_{\mathcal{E}}$, we then have to verify that $E_i$, $E_j$ and $E_i - E_j$ are not pullbacks from $\mathcal{B}_6$.

\paragraph{A Combinatorial Approach to Line Bundle Cohomology on del-Pezzo Surfaces}

Finally, we would like to identify the number of exotic bulk states, which result from our `non-proper' choice of hypercharge flux. To this end, we employ the results of \cref{sec:FromC3toL}. Thereby, we find that the chiral and anti-chiral exotic bulk states in representation $( \mathbf{3}, \mathbf{2} )_{q_X, 5_Y}$ are counted by
\begin{align}
\begin{split}
H^1 \left( dP_n, \mathcal{O}_W \left( 5 \mathcal{H} \right) \right) \oplus H^0 \left( dP_n, \mathcal{O}_W \left( 5 \mathcal{H} \right) \otimes K_{dP_n} \right) \, , \\
H^2 \left( dP_n, \mathcal{O}_W \left( 5 \mathcal{H} \right) \right) \oplus H^1 \left( dP_n, \mathcal{O}_W \left( 5 \mathcal{H} \right) \otimes K_{dP_n} \right) \, .
\end{split}
\end{align}
By Serre duality, it is equivalent to say that the chiral and anti-chiral exotic bulk states in representation $( \mathbf{3}, \mathbf{2} )_{q_X, 5_Y}$ are counted by
\begin{align}
\begin{split}
H^1 \left( dP_n, \mathcal{O}_W \left( 5 \mathcal{H} \right) \right) \oplus H^2 \left( dP_n, \mathcal{O}_W \left( -5 \mathcal{H} \right) \right) \, , \\
H^2 \left( dP_n, \mathcal{O}_W \left( 5 \mathcal{H} \right) \right) \oplus H^1 \left( dP_n, \mathcal{O}_W \left( -5 \mathcal{H} \right) \right) \, .
\label{equ:ChiralAndAntichiralExotics}
\end{split}
\end{align}
In addition, for supersymmetric fluxes the following sheaf cohomologies should vanish \cite{Blumenhagen:2008zz}
\[ H^0 \left( dP_n, \mathcal{O}_W \left( 5 \mathcal{H} \right) \right) \, , \qquad H^2 \left( dP_n, \mathcal{O}_W \left( 5 \mathcal{H} \right) \otimes K_{dP_n} \right) \cong H^2 \left( dP_n, \mathcal{O}_W \left( -5 \mathcal{H} \right) \right) \, , 
\label{equ:Consistency}
\]
where we used Serre duality once again. 

In \cite{Blumenhagen:2008zz} it was pointed out that line bundle cohomology on del-Pezzo surfaces $dP_n$ with $1 \leq n \leq 8$ can be computed, provided that the location of the blow-up points are known. Let us employ this formalism to compute the above sheaf cohomologies. To this end, let us recall that the Cox ring of $\mathbb{P}^2_{\mathbb{Q}}$, $S = \mathbb{Q} [ x_1, x_2, x_3 ]$, is $\mathbb{Z}$-graded by $\mathrm{deg} ( x_i ) = 1$. For $c_j, a \in \mathbb{N}_{\geq 0}$ let $A_{\sum{c_j p_j}} ( a )$ denote the $\mathbb{Q}$-dimension of the vector space of homogeneous polynomials of degree $a$ in the ring $S$ which vanish at the point $p_j$ to order $c_j$. Now we consider a line bundle $L \in \mathrm{Pic} ( dP_n )$ given by \footnote{For example $K_{dP_n} = \mathcal{O}_{dP_n} ( -3H + E_1 + E_2 + \dots + E_n )$.}
\[ L = \mathcal{O}_{dP_n} \left( a \cdot H + \sum_{i \in I}{b_i E_i} - \sum_{j \in J}{c_j E_j} \right) \label{equ:ParametrisationOfLineBundles} \]
where $I \cap J = \emptyset$, $I \cup J = \{ 1, 2, \dots, n \}$ and $b_i, c_j \in \mathbb{N}_{\geq 0}$ for all $i, j$. By use of Serre duality one can focus on the case $a \geq -2$. For such line bundles it was found in \cite{Blumenhagen:2008zz} that
\[ h^i \left( dP_n, L \right) = \left( A_{\sum{c_i p_i}} \left( a \right), - \chi \left( L \right) + A_{\sum{c_i p_i}} \left( a \right), 0 \right) \, , \label{equ:LineBundleCohomology} \]
where the Euler characteristic of $L$ is given by
\[ \chi \left( L \right) = \binom{a+2}{a} - \sum_{i \in I}{\frac{b_i \left( b_i - 1 \right)}{2}} - \sum_{j \in J}{\frac{c_j \left( c_j + 1 \right)}{2}}. \]

\paragraph{Counting of massless Exotics}

Let us employ this knowledge to compute the sheaf cohomologies in \cref{equ:Consistency} and \cref{equ:ChiralAndAntichiralExotics} for a hypercharge flux of the form $\mathcal{H} = E_i - E_j$. Consequently, it holds
\[ \mathcal{O}_{dP_n} \left( 5 \mathcal{H} \right) = \mathcal{O}_{dP_n} \left( 5 E_i - 5 E_j \right) \, . \]
Indeed, for this bundle it holds $a = 0 \geq -2$ (\cf \cref{equ:ParametrisationOfLineBundles}) and consequently it follows from \cref{equ:LineBundleCohomology} that $H^2 ( dP_n, \mathcal{O}_{dP_n} ( 5 \mathcal{H} ) ) = 0$. Even more, this implies that $h^0 ( dP_n, \mathcal{O}_{dP_n} ( 5 \mathcal{H} ) )$ is counted by the polynomials of degree $0$ in the Cox  ring of $\mathbb{P}^2_{\mathbb{Q}}$ which vanish at one blow-up point to order $5$. The only such polynomial is the one that vanishes identically. Since this polynomial furnishes a 0-dimensional $\mathbb{Q}$-vector space we find $h^i ( dP_n, \mathcal{O}_{dP_n} ( 5 E_i - 5 E_j ) ) = ( 0, 24, 0 )$. We are therefore lead to conclude from \cref{equ:ChiralAndAntichiralExotics} that there are 24 chiral and anti-chiral exotics in representation $( \mathbf{3}, \mathbf{2})_{q_X, 5_Y}$. In addition, the sheaf cohomologies in \cref{equ:Consistency} vanish, as they should do for supersymmetric fluxes \cite{Blumenhagen:2008zz}. Similarly, one finds that there are 24 chiral and anti-chiral exotics in representation $( \mathbf{3}, \mathbf{2})_{q_X, -5_Y}$ and that \cref{equ:Consistency} vanishes for these states also.

\section{An \emph{F-theory} GUT-Model on a \texorpdfstring{$\mathbf{dP_3}$}{dP3}-Surface} \label{sec:dP3-Example}

In this section we will describe the geometry of the elliptic fibration $\pi \colon \hat{Y}_4 \twoheadrightarrow \mathcal{B}_6$ which we will base our \emph{F-theory} GUT-model on. Together with a gauge background $A = A_X ( F ) + A ( \mathbf{10}_1 ) ( \lambda ) + A_Y( \mathcal{H} )$, which we subject to a number of more refined conditions in \cref{subsec:GaugeBackground}, this completes the data defining such an \emph{F-theory} vacuum. Finally, in \cref{subsec:MasslessSpectra} we set out to compute the massless spectra of such \emph{F-theory} vacua.

\subsection{The Geometry of \texorpdfstring{$\mathbf{\hat{Y}_4}$}{Y4}} \label{subsec:GeometryOfY4FordP3}

\paragraph{\emph{F-Theory} Base}

\begin{table}[tbp]
\begin{center}
\begin{tabular}{cccccc}
\toprule
$x_1$ & $x_2$ & $x_3$ & $x_4$ & $x_5$ & $x_6$ \\
\midrule
1 & 1 & 0 & 0 & 0 & 1 \\
0 & 1 & 1 & 1 & 1 & 1 \\
\bottomrule
\end{tabular}
\end{center}
\caption[Toric data of base ambient space of \emph{F-theory} GUT-model with $W \cong dP_3$.]{The Cox ring $\mathbb{Q} [ x_1, x_2, x_3, x_4, x_5, x_6 ]$ of the \emph{F-theory} base ambient space $X_\Sigma$ is graded by $\mathbb{Z}^2$. Its Stanley-Reisner ideal is $I_{\mathrm{SR}} ( X_\Sigma ) = \langle x_1 x_2 x_6, x_3 x_4 x_5 \rangle$. As explained in \cref{subsec:TowardsToricVarieties}, this defines a toric variety.}
\label{GradingOfCoxRingBasedP3Model}
\end{table}

We model the \emph{F-theory} base space $\mathcal{B}_6$ as a hypersurface in a smooth and projective (normal) toric variety $X_\Sigma$, whose Cox ring $S ( X_\Sigma ) = \mathbb{Q} [ x_1, \dots, x_6 ]$ is graded under $\mathrm{Cl} ( X_\Sigma ) = \mathbb{Z}^2$ according to \cref{GradingOfCoxRingBasedP3Model} and has Stanley-Reisner ideal $I_{\mathrm{SR}} ( X_\Sigma ) = \langle x_1 x_2 x_6, x_3 x_4 x_5 \rangle$.\footnote{Recall from \cref{subsec:TowardsToricVarieties} how this data defines the fan $\Sigma$ of $X_\Sigma$.} It holds
\[ \mathrm{Nef} \left( X_\Sigma \right) = \mathrm{Cone} \left( V \left( x_3 \right), V \left( x_6 \right) \right) \, . \]
The \emph{F-theory} base $\mathcal{B}_6$ is modelled as the hypersurface $V( P ) \subseteq X_\Sigma$ where
\begin{align}
\begin{split}
P = & x_1^2 \left( c_1 x_3^2 + c_2 x_4^2 + c_3 x_5^2 + c_4 x_3 x_4 + c_5 x_3 x_5 + c_6 x_4 x_5 \right) \\
    & + x_1 \left( c_7 x_2 x_3 + c_8 x_2 x_4 + c_9 x_2 x_5 + c_{10} x_6 x_3 + c_{11} x_6 x_4 + c_{12} x_6 x_5 \right) \\
    & + c_{13} x_2^2 + c_{14} x_2 x_6 + c_{15} x_6^2 \, .
\end{split}
\label{equ:most_general_P}
\end{align}
Since we model our geometries over the rational numbers $\mathbb{Q}$ to make them accessible to \emph{gap}, it holds $c_i \in \mathbb{Q}$. However, let us note that in working over $\mathbb{C}$ there exist analytic cycles which are not algebraic. To make such cycles appear in the `rigid' geometry over $\mathbb{Q}$, the coefficients $c_i$ are subject to tuning. For this very reason, and in order to simplify the following analysis, the coefficients $c_i \in \mathbb{Q}$ are subject to the condition that $P$ is irreducible and that
\begin{equation}
\begin{array}{rclcrclcrclcrclcrcl}
c_2 &=& c_8 \, , & & c_3 &=& 0 \, , & & c_6 &=& c_9 - 1 \, , & & c_8 &\neq& 0 \, , & & c_9 &\neq& 0 \, , \\
c_{11} &=& 1 \, , & & c_{12} &=& 0 \, , & & c_{13} &=& 0 \, , & & c_{14} &=& 1 \, , & & c_{15} &=& 0 \, .
\end{array}
\end{equation}
The remaining $c_1, c_4, c_5, c_7, c_8, c_9, c_{10}$ are taken pseudo-randomly. It can be verified \eg with the software \emph{Sage} \cite{sage} that $\mathcal{B}_6$ is smooth for this choice of parameters. In addition, we can use the results from the previous chapter to compute $H^0 ( \mathcal{B}_6, \mathcal{O}_{\mathcal{B}_6} )$. The dimension of this vector space counts the number of connected components of $\mathcal{B}_6$. The result $h^0 ( \mathcal{B}_6, \mathcal{O}_{\mathcal{B}_6} ) = 1$ therefore verifies that $\mathcal{B}_6$ is connected. The Hodge diamond of $\mathcal{B}_6$ is readily computed with \emph{cohomCalg} \cite{Blumenhagen:2010pv, Blumenhagen:2010ed, Blumenhagen:2011xn, KoszulExtensionManual, cohomCalg:Implementation, 2011JMP....52c3506J, Rahn:2010fm}. It reads
\[ \label{equ:HodgeBasedP3} \begin{array}{ccccccc}
& & & 1 \\
& & 0 & & 0 \\
& 0 & & 3 & & 0 \\
0 & & 0 & & 0 & & 0 \\
& 0 & & 3 & & 0 \\
& & 0 & & 0 \\
& & & 1 
\end{array} \]
Hence, $h^{1,0} ( \mathcal{B}_6 ) = h^{2,0} ( \mathcal{B}_6 ) = 0$. Now let us look at the fundamental group of $\mathcal{B}_6$. In general this is much harder to analyse. However, we can here apply \cite[theorem 1.6]{Batyrev:2005jc}, which implies
\[ \pi_1 \left( \mathcal{B}_6 \right) \cong N / \text{Span}_{\mathbb{Z}} \left\{ u_\rho \, \left| \, \rho \in \Sigma \left( 1 \right) \right. \right\} \cong 0 \, . \]
The last equality follows since $X_\Sigma$ is both smooth and complete. Consequently, also the Abelianization of $\pi_1 ( \mathcal{B}_6 ) = 0$, which is just $h_1 ( \mathcal{B}_6, \mathbb{Z} )$, vanishes also. In consequence, $\text{Pic} ( \mathcal{B}_6 )$ is torsion-free, all conditions stated in \cref{subsec:ConditionsOnBaseAndFibration} are satisfied and $\mathcal{B}_6$ is a bona fide \emph{F-theory} base space.

The Hodge diamond of $\mathcal{B}_6$ now shows $\mathrm{Pic} ( \mathcal{B}_6 ) \cong \mathbb{Z}^3$. We will eventually argue that the following following divisor classes of $\mathcal{B}_6$ are not linearly equivalent:
\[ G_1 = V \left( x_1, x_2 \right) \, , \qquad G_2 = V \left( x_1, x_6 \right) \, , \qquad G_3 = V \left( x_5, P \right) \, . \]
Hence, the line bundles associated to these divisors will provide free generators of $\mathrm{Pic} ( \mathcal{B}_6 )$. For the time-being suffice it to mention that $G_1 \cup G_2 = \left. V ( x_1 ) \right|_{\mathcal{B}_6}$, $G_3 = \left. V ( x_5 ) \right|_{\mathcal{B}_6}$ and that the triple-intersection numbers satisfy:
\begin{equation}
\label{equ:tripleIntersectionsBase}
\begin{array}{rclcrclcrclcrcl}
G_1^3      &=& +1, & & G_1^2 G_2  &=& 0,  & & G_1^2 G_3   &=& -1, \\
G_1 G_2 ^2 &=& 0,  & & G_1 G_2 G_3 &=& 0,  & & G_1 G_3 ^2 &=& +1, \\ 
G_2 ^3     &=& +1, & & G_2^2 G_3   &=& -1, & & G_2 G_3^2  &=& +1, & & G_3^3 &=& 0 \, .
\end{array}
\end{equation}

\paragraph{\emph{F-Theory} GUT-Surface}

The \emph{F-theory} GUT surface is given by $W = V ( P, x_3 ) \subseteq \mathcal{B}_6$. Its Hodge diamond satisfies
\[
\begin{array}{ccccc}
& & 1 \\
& 0 & & 0 & \\
0 & & 4 & & 0 \\
& 0 & & 0 & \\
& & 1
\end{array}
\]
We can employ the short exact sequence $0 \to T_W \hookrightarrow T_{X_\Sigma} |_{W} \twoheadrightarrow N_{W \subseteq X_\Sigma} \to 0$ to compute the Chern classes of $T_W$. This shows
\begin{align}
\begin{split}
\Mint_{W}{c_1 \left( T_W \right)^2} &= \left( G_1 + G_2 + 2 G_3 \right)^2 \cdot G_3 = 6 \, , \\
\Mint_{W}{c_2 \left( T_W \right)} &= \left( \left( G_1 + G_2 + G_3 \right)^2 + 2 \left( G_1 + G_2 \right) G_3 + 3 G_3^2 \right) \cdot G_3 = 6 \, ,
\end{split}
\end{align}
where we evaluated the triplet-intersections in $\mathcal{B}_6$ by use of \cref{equ:tripleIntersectionsBase}. This result identifies $W$ as a del-Pezzo surface $dP_3$ of degree $6$.\footnote{Conventionally this surface is denoted as $dP_3$, because it can be thought of as $\mathbb{P}^2_{\mathbb{Q}}$, blown up in three points.}

Let us try to make the geometry of this surface more explicit. To this end we first define $B ( x_4, x_5 ) = c_8 x_4 + c_9 x_5$. Then it holds $W = V ( \tilde{P}, x_3 )$ with
\begin{align}
\begin{split}
\tilde{P} &= \left[ x_1 x_4 + x_2 \right] \left[ B \left( x_4, x_5 \right) x_1 + x_6 \right] - x_1^2 x_4 x_5 \\
          &= \left( x_1, x_2, x_6 \right) \cdot \underbrace{\left( \begin{array}{ccc} - x_4 x_5 + B \left( x_4, x_5 \right) x_4 & \frac{B \left( x_4, x_5 \right)}{2} & \frac{x_4}{2} \\ \frac{B \left( x_4, x_5 \right)}{2} & 0 & \frac{1}{2} \\ \frac{x_4}{2} & \frac{1}{2} & 0 \end{array} \right)}_{:= M} \left( \begin{array}{c} x_1 \\ x_2 \\ x_6 \end{array} \right).
\end{split}
\end{align}
Note also that $\det M ( x_4, x_5 ) = \frac{1}{4} x_4 x_5$ and
\[ W \cap V \left( \det M \right) = V \left( x_3, x_4 x_5, \left[ B \left( x_4, x_5 \right) x_1 + x_6 \right] \left[ x_1 x_4 + x_2 \right] \right) \]
Based on this finding we define the following six divisors in W:
\[
\begin{aligned}
D_1 &= V \left( x_3, x_4, B \left( x_4, x_5 \right) x_1 + x_6 \right) \, , \\
D_3 &= V \left( x_3, x_5, B \left( x_4, x_5 \right) x_1 + x_6 \right) \, , \\
D_5 &= V \left( x_3, x_1, x_6 \right) \, , \\    
\end{aligned}
\hspace{5em}
\begin{aligned}
D_2 &= V \left( x_3, x_4, x_1 x_4 + x_2 \right) \, , \\
D_4 &= V \left( x_3, x_5, x_1 x_4 + x_2 \right) \, , \\
D_6 &= V \left( x_3, x_1, x_2 \right) \, . \\
\end{aligned}
\]
With \emph{Sage} \cite{sage} it can be verified that the GUT-surface $W$ and the above six divisors $D_i$ are smooth. With the \emph{gap}-package \cite{SheafCohomologyOnToricVarieties} (or \emph{cohomCalg} \cite{Blumenhagen:2010pv, Blumenhagen:2010ed, Blumenhagen:2011xn, KoszulExtensionManual, cohomCalg:Implementation, 2011JMP....52c3506J, Rahn:2010fm}) we can also identify the cohomologies of the structure sheaves $\mathcal{O}_{D_i}$. By use of the theorem of Riemann-Roch for curves (\cf \cref{subsec:LineBundlesOnRiemannSurfaces}) it is then found that all of these six divisors are $\mathbb{P}^1_{\mathbb{Q}}$s. The intersection numbers of the divisor classes $D_i$ follow from the formula \cite{cox2011toric, ATIT}
\[ D_i \cdot D_j = \mathrm{deg} \left( \left. \mathcal{O}_{W} \left( D_i \right) \right|_{D_j} \right). \]
which can be evaluated with the \emph{gap}-package \cite{SheafCohomologyOnToricVarieties}. This leads to the intersection numbers in \cref{table-N13}, which yield two relations among the divisor classes $D_i$, namely
\[ D_3 = D_1 + D_2 - D_4 \, , \qquad D_6 = D_1 + D_5 - D_4 \, . \]
Note that $\mathrm{Pic} ( dP_3 ) \cong \mathbb{Z}^4$. Also note that for the divisor classes
\[ H := D_1 + D_2 + D_5, \quad E_1 := D_2, \quad E_2 := D_4, \quad E_3 := D_5 \]
the intersection form takes the shape $H^2 = 1$, $H E_i = 0$ and $E_i E_j = - \delta_{ij}$ which happens to be the canonical intersection form of a $dP_3$-surface. Even more we can compute the sheaf cohomologies of $\mathcal{O}_{W} \left( \pm D_i \right)$. It turns out
\[ h^i \left( W, \mathcal{O}_{W} \left( D_i \right) \right) = \left( 1,0,0 \right) \, , \qquad h^i \left( W, \mathcal{O}_{W} \left( -D_i \right) \right) = \left( 0,0,0 \right) \, . \]
The corresponding computation on a toric model of a $dP_3$-surface uses the geometric data summarised in \cref{table-N14}. We focus on the six torus invariant prime divisors $A_i = V ( u_i ) \subseteq dP_3$ and identify their intersection numbers. These intersection numbers are listed in \cref{table-N15}. In addition, it is easily verified that
\[ h^i \left( dP_3, \mathcal{O}_{dP_3} \left( A_i \right) \right) = \left( 1,0,0 \right) \, , \qquad h^i \left( dP_3, \mathcal{O}_{dP_3} \left( -A_i \right) \right) = \left( 0,0,0 \right) \, . \]

\begin{table}[tb]
\centering
\begin{tabular}{c@{\hskip 15pt}cccccc}
\toprule
      & $D_1$ & $D_2$ & $D_3$ & $D_4$ & $D_5$ & $D_6$ \\
\midrule
$D_1$ & -1    & 1     & $\cdot$     & $\cdot$     & 1     & $\cdot$ \\ 
$D_2$ & 1     & -1    & $\cdot$     & $\cdot$     & $\cdot$     & 1 \\       
$D_3$ & $\cdot$     & $\cdot$     & -1    & 1     & 1     & $\cdot$ \\
$D_4$ & $\cdot$     & $\cdot$     & 1     & -1    & $\cdot$     & 1 \\
$D_5$ & 1     & $\cdot$     & 1     & $\cdot$     & -1    & $\cdot$ \\
$D_6$ & $\cdot$     & 1     & $\cdot$     & 1     & $\cdot$     & -1\\
\bottomrule
\end{tabular}
\caption{Intersection numbers of the divisor classes $D_i \in \mathrm{Cl} \left( W \right)$.}
\label{table-N13}
\end{table}

\begin{table}[tb]
\centering
\begin{tabular}{cccc}
      \toprule
      ray generators & name of variable & GLSM charge & divisor class \\
      \midrule
      $ \left( 1,0 \right)$ & $u_1$ & $\left( 1,0,0,1 \right)$ & $A_1$ \\
      $ \left( 1,1 \right)$ & $u_2$ & $\left( 0,1,0,0 \right)$ & $A_2$ \\
      $ \left( 0,-1 \right)$ & $u_3$ & $\left( 0,0,1,0 \right)$ & $A_3$ \\
      $ \left( 0,1 \right)$ & $u_4$ & $\left( 1,0,1,0 \right)$ & $A_4$ \\
      $ \left( -1,0 \right)$ & $u_5$ & $\left( 0,0,0,1 \right)$ & $A_5$ \\
      $ \left( -1,-1 \right)$ & $u_6$ & $\left( 1,1,0,0 \right)$ & $A_6$ \\
      \bottomrule
\end{tabular}
\caption[The defining data of a toric $dP_3$ surface.]{Summary of the defining data of a toric $dP_3$ surface. The Stanley-Reisner ideal is given by $I_{\mathrm{SR}} \left( \mathrm{dP}_3 \right) = \left\langle u_1 u_4, u_1 u_5, u_1 u_6, u_2 u_3, u_2 u_5, u_2 u_6, u_3 u_4, u_3 u_5, u_4 u_6 \right\rangle$.}
\label{table-N14}
\end{table}

\begin{table}[tb]
\centering
\begin{tabular}{c@{\hskip 15pt}cccccc}
\toprule
& $A_1$ & $A_2$ & $A_3$ & $A_4$ & $A_5$ & $A_6$ \\
\midrule
$A_1$ & -1 & 1 & 1 & $\cdot$ & $\cdot$ & $\cdot$ \\
$A_2$ & 1 & -1 & $\cdot$ & 1 & $\cdot$ & $\cdot$ \\
$A_3$ & 1 & $\cdot$ & -1 & $\cdot$ & $\cdot$ & 1 \\
$A_4$ & $\cdot$ & 1 & $\cdot$ & -1 & 1 & $\cdot$ \\
$A_5$ & $\cdot$ & $\cdot$ & $\cdot$ & 1 & -1 & 1 \\
$A_6$ & $\cdot$ & $\cdot$ & 1 & $\cdot$ & 1 & -1 \\
\bottomrule
\end{tabular}
\caption{Intersection numbers of the divisor classes $A_i \in \mathrm{Cl} \left( dP_3 \right)$ defined in \cref{table-N14}.}
\label{table-N15}
\end{table}

So the intersection numbers of the divisor classes $D_i$ closely resemble the intersection numbers of the divisor classes $A_i$. In addition, we have a perfect match between the vector space dimensions of the sheaf cohomologies of the associated line bundles. Based on this we are tempted to perform an identification $A_i \leftrightarrow D_j$. Let us pick $A_1 \leftrightarrow D_1$. Then the intersection numbers between the associated divisor classes force upon us, up to linear equivalence, the following 'dictionary':
\[ \label{equ:Dictionary}
\begin{aligned}
A_1 &\leftrightarrow D_1 \, , \\
A_4 &\leftrightarrow D_6 \, , \\
\end{aligned}
\qquad
\begin{aligned}
A_2 &\leftrightarrow D_2 \, , \\
A_5 &\leftrightarrow D_4 \, , \\
\end{aligned}
\qquad
\begin{aligned}
A_3 &\leftrightarrow D_5 \, , \\
A_6 &\leftrightarrow D_3 \, . \\
\end{aligned}
\]

Experience shows that we can compute sheaf cohomologies far more efficiently on the toric model of a $dP_3$-surface rather than on $X_\Sigma$. Thus, we may wonder if we can use the above knowledge to translate the necessary computations on $X_\Sigma$ into computations on the toric $dP_3$-surface. This ultimately leads to the question if we can express the embedding $\iota \colon dP_3 \hookrightarrow X_\Sigma$ explicitly. To this we turn next.

\paragraph{Conjecture on Morphisms Between Smooth, Projective Toric Varieties} 

To motivate our approach, let us start by looking at how morphisms of projective schemes arise in algebraic geometry. To this end, we consider the polynomial ring $S^{(n)} = R [ x_1, \dots, x_n ]$ over a (commutative and unitial) ring $R$. By setting $\mathrm{deg} ( x_i ) = 1$ this ring becomes $\mathbb{Z}$-graded. The irrelevant ideal of this graded ring is defined as $\mathcal{B}^{(n)} = \langle x_1, x_2, \dots, x_n \rangle$. In analogy to the discussion given in \cref{subsec:CoherentSheavesOnVarieties} we now consider the set of prime ideals
\[ \mathrm{Proj}_{\mathfrak{B}^{(n)}} ( S^{(n)} ) := \left\{ \left. \mathfrak{p} \in \mathrm{Spec} \left( S \right) \mathrm{ homogeneous} \; \right| \; \mathfrak{B}^{(n)} \not \subseteq \mathfrak{p} \right\} \, . \label{equ:NormalProj} \]
This set becomes a topological space once equipped with the Zariski topology. The Zariski closed subsets of $\mathrm{Proj}_{\mathfrak{B}^{(n)}} \left( S^{(n)} \right)$ are of the form
\[ V \left( \mathfrak{a} \right) = \left\{ \left. \mathfrak{p} \in \mathrm{Proj}_{\mathfrak{B}^{(n)}} \left( S^{(n)} \right) \right| \mathfrak{a} \subseteq \mathfrak{p} \right\} \, . \]
Given two $\mathbb{Z}$-graded rings $S^{(n)}$ and $S^{(m)}$, a ring homomorphism $\varphi \colon S^{(n)} \to S^{(m)}$ is said to satisfy this $\mathbb{Z}$-grading precisely if for all $d \in \mathbb{Z}$ it holds $\varphi ( S^{(n)}_d ) \subseteq S^{(m)}_d$. It is a well known result in algebraic geometry that such a ring homomorphism $\varphi \colon S^{(n)} \to S^{(m)}$ of $\mathbb{Z}$-graded rings induces a closed immersion of projective schemes
\[ f_\varphi \colon \mathrm{Proj}_{\mathcal{B}^{(m)}} ( S^{(m)} ) \to \mathrm{Proj}_{\mathcal{B}^{(n)}} ( S^{(n)} ) \, , \label{equ:MOrphismBetweenProjectiveSchemes} \]
provided that the truncated maps $\varphi_k \colon S^{(n)}_k \to S^{(m)}_k$ are surjective for sufficiently large $k$. More details can be found in \cite[p.80  ff]{hartshorne1977algebraic}.

We can hope that these results generalise to more general smooth, projective toric varieties. Unfortunately, to date, no result along these lines is known to the author. Therefore, we will now motivate a way, in which this result can be generalised. In this work, however, we will not try to prove such a generalised statement. We reserve it for future work to investigate these steps further. Therefore we phrase our `natural generalisation' as a conjecture.

We start by looking at a smooth, projective toric variety $X_\Sigma$. It comes equipped with its Cox ring $R$ and the irrelevant ideal $B_\Sigma \subseteq S$. In analogy to \cref{equ:NormalProj}, it seems natural to use this data to consider the set
\[ \mathrm{Proj}_{\mathfrak{B}} ( S ) := \left\{ \left. \mathfrak{p} \in \mathrm{Spec} \left( S \right) \mathrm{ homogeneous} \; \right| \; \mathfrak{B} \not \subseteq \mathfrak{p} \right\} \, , \]
and subsequently equip it with a Zariski topology. This leads to our first conjecture.

\begin{conj}
Be $( X_\Sigma, \mathcal{O}_{X_\Sigma} )$ a smooth, projective toric scheme (over $\mathbb{Q}$ or $\mathbb{C}$) with Cox ring $S$ and irrelevant ideal $\mathfrak{B}$. Then we claim that the following holds true:
\begin{itemize}
 \item$\mathrm{Proj}_{\mathfrak{B}} ( S )$ naturally admits a Zariski topology.
 \item $\mathrm{Proj}_{\mathfrak{B}} ( S )$ naturally admits a structure sheaf.
 \item There is a scheme isomorphism $( \mathrm{Proj}_{\mathfrak{B}} ( S ), \mathcal{O}_{\mathrm{Proj}_{\mathcal{B}} ( S )} ) \cong ( X_\Sigma, \mathcal{O}_{X_\Sigma} )$.
\end{itemize}
\end{conj}

That said, let us next look at a pair of smooth, projective toric varieties $X_\Sigma$, $X_{\Sigma^\prime}$ with Cox rings $S$, $S^\prime$ and irrelevant ideals $\mathfrak{B}$, $\mathfrak{B}^\prime$. For such varieties it holds $\mathrm{Cl} ( X_\Sigma ) \cong \mathbb{Z}^r$ and we first have to generalise the notion of a $\mathbb{Z}$-graded ring homomorphism to this situation. It seems particularly natural to state that a ring homomorphism $\varphi \colon S \to S^\prime$ respects the gradings of $S$ and $S^\prime$ if it induces a group homomorphism $\tilde{\varphi} \colon \mathrm{Cl} ( X_\Sigma ) \to \mathrm{Cl} ( X_{\Sigma^\prime} )$.

In the situation studied in \cite[p.80  ff]{hartshorne1977algebraic}, it holds $S_k \cong \mathfrak{B}_k$ for all $k \geq 1$. As the irrelevant ideals are particularly important to the geometry of a toric variety, let us focus on their truncations to generalise the surjectivity demand. In general, the image of an ideal under a ring homomorphism may not be an ideal. Still, inverse images of ideals are always ideals. 

With this motivation, let us return to our situation. Namely consider a pair of smooth, projective toric varieties $X_\Sigma$, $X_{\Sigma^\prime}$ with Cox rings $S$, $S^\prime$, irrelevant ideals $\mathfrak{B}$, $\mathfrak{B}^\prime$ and a ring homomorphism $\varphi \colon S^\prime \to S$, which respects the gradings of this rings in the sense described above. To formulate a supplement of the surjectivity demand in \cite[p.80  ff]{hartshorne1977algebraic}, we are lead to compare $B_{\Sigma^\prime}$ and $\varphi^{-1} ( B_\Sigma )$. More explicitly, we have to find an element $d \in \mathrm{Cl} ( X_{\Sigma^\prime} )$, such that the truncations $( B_{\Sigma^\prime} )_e$ and $( \varphi^{-1} ( B_\Sigma ) )_e$ match for all $e `\geq' d$. Hence the final piece to the story, is to find a suitable replacement for this inequality. As the ideals in question are formed from monomials, it seems sufficient to take into account their gradings in the Cox ring. This leads to a concept, which we call the \emph{cone of monomials}:

\begin{defi}
Be $S = k [ x_1, \dots, x_l ]$ a $\mathbb{Z}^n$-graded ring. Then we define its (integral) \emph{cone of monomials} as
\[ \mathcal{C} \left( S \right) := \mathrm{Span}_{\mathbb{Z}_{\geq 0}} \left\{ \mathrm{deg} \left( x_1 \right), \dots, \mathrm{deg} \left( x_n \right) \right\} \, . \]
\end{defi}

These pieces in place, we can finally formulate, what we believe to be a natural generalisation of the finding in \cite[p.80  ff]{hartshorne1977algebraic}. We are lead to the following:

\begin{conj} \label{conj:II}
Be $( X_\Sigma, \mathcal{O}_{X_\Sigma} )$, $( X_{\Sigma^\prime}, \mathcal{O}_{X_{\Sigma^\prime}} )$ two projective toric schemes with Cox rings $S$, $S^\prime$ and irrelevant ideals $\mathfrak{B}$, $\mathfrak{B}^\prime$. Be $\varphi \colon S^\prime \to S$ a ring homomorphism of the Cox rings such that
\begin{description}
 \item[A] it induces an \textbf{injective} group homomorphism of the degree groups of $S$ and $S^\prime$ and
 \item[B] there exists $D \in \mathrm{Deg} ( S^\prime )$ with $\mathfrak{B}^\prime_d = ( \varphi^{-1} ( \mathfrak{B} ) )_d$ for all $d \in D + \mathcal{C} ( S^\prime )$.
\end{description}
Under these assumptions we conjecture that the following holds true:
\begin{itemize}
 \item $\varphi$ induces a natural scheme morphism
      \[ 
      \resizebox{0.85\textwidth}{!}{$
      \left( X_\Sigma, \mathcal{O}_{X_\Sigma} \right) \cong \left( \mathrm{Proj}_{\mathfrak{B}} \left( S \right), \mathcal{O}_{\mathrm{Proj}_{\mathfrak{B}} \left( S \right)} \right) \stackrel{\varphi}{\rightarrow} \left( \mathrm{Proj}_{\mathfrak{B}^\prime} \left( S^\prime \right), \mathcal{O}_{\mathrm{Proj}_{\mathfrak{B}^\prime} \left( S^\prime \right)} \right) \cong \left( X_{\Sigma^\prime}, \mathcal{O}_{X_{\Sigma^\prime}} \right) \, .$}
      \]
 \item The map of the topological spaces is given by
      $ f_{\varphi} \colon \mathrm{Proj}_{\mathfrak{B}} \left( S \right) \to \mathrm{Proj}_{\mathfrak{B}^\prime} \left( S^\prime \right) \; , \; \mathfrak{p} \mapsto \varphi^{-1} \left( \mathfrak{p} \right)$.
\end{itemize}
\end{conj}

As we have highlighted, we believe that the ring homomorphism $\varphi \colon S^\prime \to S$ must not only induce a group homomorphism of $\mathrm{Cl} ( X_\Sigma )$ and 
$\mathrm{Cl} ( X_{\Sigma^\prime} )$, but must be a monomorphism even. Back in \cite[p.80  ff]{hartshorne1977algebraic}, the induced ring homomorphism $\mathbb{Z} \to \mathbb{Z}$ is an isomorphism even. But this seems to strong a demand, as the following examples show. Based on this evidence, we have supplemented the word `injective'.
Once again, let us emphasize that we have no proof for these conjectures as of this writing. Our motivation derives merely from the seemingly natural generalisation of the corresponding result in \cite[p.80  ff]{hartshorne1977algebraic}. We reserve a detailed analysis for future work.

Let us complete this discussion, but looking at the Segre embedding $\mathbb{P}^1_{\mathbb{Q}} \times \mathbb{P}^1_{\mathbb{Q}} \hookrightarrow \mathbb{P}_{\mathbb{Q}}^3$ as a non-trivial toric example. The relevant toric data is summarised in \cref{table-N11} and we focus on the ring homomorphism $\varphi \colon R = \mathbb{Q} [ z_0, \dots, z_3 ] \to S = \mathbb{Q} [ x_0, x_1, y_0, y_1 ]$ with
\[ z_0 \mapsto x_0 y_0, \quad z_1 \mapsto x_0 y_1, \quad z_2 \mapsto x_1 y_0, \quad z_3 \mapsto x_1 y_1. \]
This ring homomorphism $\varphi \colon R \to S$ has the following two properties:
\begin{description}
 \item[A] The induced homomorphism of the degree groups -- $\mathbb{Z} \mapsto \mathbb{Z}^2 \; , \; 1 \mapsto ( 1,1 )$ -- is injective.
 \item[B] It holds $\mathfrak{B}_R = \varphi^{-1} ( \mathfrak{B}_S )$, so in particular $( \mathfrak{B}_R)_d = (\varphi^{-1} ( \mathfrak{B}_S ))_d$ for all $d \in 
      \mathbb{Z}$.
\end{description}
So indeed all assumptions of \cref{conj:II} are satisfied.

\begin{table}[tbp]
\centering
\begin{tabular}{lcc}
\toprule
& $\mathbb{P}^1_{\mathbb{Q}} \times \mathbb{P}^1_{\mathbb{Q}}$ & $\mathbb{P}^3_{\mathbb{Q}}$ \\
\midrule
Cox ring & $S = \mathbb{Q} \left[ x_0, x_1, y_0, y_1 \right]$ & $R = \mathbb{Q} \left[ z_0, z_1, z_2, z_3 \right]$ \\
degree group & $\mathbb{Z}^2$ & $\mathbb{Z}$ \\
$\mathbb{Z}^k$-grading & $\mathrm{deg} \left( x_i \right) = \left( 1,0 \right)$, $\mathrm{deg} \left( y_i \right) = \left( 0,1 \right)$ & $\mathrm{deg} \left( z_i \right) = 1$ \\
irrelevant ideals & $\mathfrak{B}_S = \left\langle x_0 y_0, x_0 y_1, x_1 y_0, x_1 y_1 \right\rangle$ & $\mathfrak{B}_R = \left\langle z_0, z_1, z_2, z_3 \right\rangle$ \\
\bottomrule
\end{tabular}
\caption{Properties of the toric spaces $\mathbb{P}^1_{\mathbb{Q}} \times \mathbb{P}^1_{\mathbb{Q}}$ and $\mathbb{P}^3_{\mathbb{Q}}$.}
\label{table-N11}
\end{table}

\paragraph{Application to the GUT-Surface}

\begin{table}[tbp]
\centering
\resizebox{\textwidth}{!}{
\begin{tabular}{lcc}
\toprule
      & toric $dP_3$ & $X_{\Sigma}$ \\
\midrule
Cox ring & $S_{dP_3} = \mathbb{Q} \left[ u_1, u_2, u_3, u_4, u_5, u_6 \right]$ & $S_\Sigma = \mathbb{Q} \left[ x_1, x_2, x_3, x_4, x_5, x_6 \right]$ \\
degree group & $\mathbb{Z}^4$ & $\mathbb{Z}^2$ \\
$\mathbb{Z}^k$-grading & $\left( \begin{array}{cccccc} 1 & 0 & 0 & 1 & 0 & 1 \\ 
                                                 0 & 1 & 0 & 0 & 0 & 1 \\
                                                 0 & 0 & 1 & 1 & 0 & 0 \\
                                                 1 & 0 & 0 & 0 & 1 & 0 \end{array} \right)$ &
                   $\left( \begin{array}{cccccc} 1 & 1 & 0 & 0 & 0 & 1 \\ 
                                                 0 & 1 & 1 & 1 & 1 & 1 \end{array} \right)$ \\
irrelevant ideal & $\mathfrak{B}_{dP_3} = \left\langle u_1 u_2 u_3 u_4, u_1 u_2 u_4 u_5, u_1 u_2 u_3 u_6, \right.$ & $\mathfrak{B}_{\Sigma} = \left\langle x_1 x_3, x_1 x_4, x_1 
x_5, x_2 x_3, x_2 x_4, \right.$ \\
                  & \qquad \qquad \quad $\left. u_1 u_3 u_5 u_6, u_2 u_4 u_5 u_6, u_3 u_4 u_5 u_6 \right\rangle$ & \qquad \qquad \qquad $\left. x_2 x_5, x_3 x_6, x_4 x_6, x_5 x_6 \right\rangle$ \\
\bottomrule
\end{tabular}
}
\caption{Properties of the toric $dP_3$-surface and of the toric base ambient space $X_{\Sigma}$.}
\label{table-N12}
\end{table}

Let us now apply \cref{conj:II} to the analysis of the GUT-surface $W = V( P , x_3 ) \subseteq \mathcal{B}_6$. The relevant toric data is listed in \cref{table-N12}. Based on this let us consider the following family of ring homomorphisms
\[ \varphi_{c_8, c_9} \colon S_\Sigma = \mathbb{Q} \left[ x_1, \dots, x_6 \right] \to S_{dP_3} = \mathbb{Q} \left[ u_1, \dots, u_6 \right], \qquad  c_8, c_9 \in \mathbb{Q} \label{equ:ConjecturedEmbedding} \]
where
\[
\begin{aligned}
\varphi_{c_8, c_9} ( x_1 ) &= u_3 u_4 \, , \\
\varphi_{c_8, c_9} ( x_3 ) &= 0 \, , \\
\varphi_{c_8, c_9} ( x_5 ) &= u_5 u_6 \, ,
\end{aligned}
\hspace{3em}
\begin{aligned}
\varphi_{c_8, c_9} ( x_2 ) &= u_2 u_4 [ u_4 u_5 - u_1 u_3 ] \, , \\
\varphi_{c_8, c_9} ( x_4 ) &= u_1 u_2 \, , \\
\varphi_{c_8, c_9} ( x_6 ) &= - c_9 u_3 u_4 u_5 u_6 + u_1 u_3 [ u_3 u_6 - c_8 u_2 u_4 ] \, .
\end{aligned}
\]
This family of ring homomorphisms induces an injective group homomorphism 
\[ \tilde{\varphi} \colon D \left( S_\Sigma \right) = \mathbb{Z}^2 \to D \left( S_{dP_3} \right) = \mathbb{Z}^4 \; , \; \left( 1,0 \right) \mapsto \left( 1,0,2,0 \right), \quad \left( 0,1 \right) \mapsto \left( 1,1,0,1 \right) \, . \]
In addition, for all $c_8, c_9 \in \mathbb{Q}$ it holds $\varphi_{c_8, c_9}^{-1} ( \mathfrak{B}_{dP_3} ) = \langle x_1 x_4, x_2 x_4, x_1 x_5, x_2 x_5, x_4 x_6, x_5 x_6, x_3, x_2 x_6 \rangle$ and one can show
\[ \left( \varphi_{c_8, c_9}^{-1} \left( \mathfrak{B}_{dP_3} \right) \right)_d = \left( \mathfrak{B}_\Sigma \right)_d \; , \qquad \forall d \in \begin{psmallmatrix} 2 \\ 3 \end{psmallmatrix} + \mathcal{C} \left( S^\prime \right) = \begin{psmallmatrix} 2 \\ 3 \end{psmallmatrix} + \mathrm{Span}_{\mathbb{Z}} \left( \begin{psmallmatrix} 1 \\ 0 \end{psmallmatrix}, \begin{psmallmatrix} 0 \\ 1 \end{psmallmatrix} \right) \, . \]
Consequently, according to \cref{conj:II}, $\varphi_{c_8, c_9}$ induces a family of scheme morphisms $\iota_{c_8, c_9} \colon dP_3 \hookrightarrow X_{\Sigma}$ with
\[
\begin{aligned}
\iota_{c_8, c_9}^{-1} ( \langle u_1 \rangle ) &= \langle x_3, x_4, c_9 x_1 x_5 + x_6 \rangle \, , \\
\iota_{c_8, c_9}^{-1} ( \langle u_2 \rangle ) &= \langle x_3, x_4, x_1 x_4 + x_2 \rangle \, , \\
\iota_{c_8, c_9}^{-1} ( \langle u_3 \rangle ) &= \langle x_3, x_1, x_6 \rangle \, , \\
\end{aligned}
\hspace{3em}
\begin{aligned}
\iota_{c_8, c_9}^{-1} ( \langle u_4 \rangle ) &= \langle x_3, x_1, x_2 \rangle \, , \\
\iota_{c_8, c_9}^{-1} ( \langle u_5 \rangle ) &= \langle x_3, x_5, x_1 x_4 + x_2 \rangle \, , \\
\iota_{c_8, c_9}^{-1} ( \langle u_6 \rangle ) &= \langle x_3, x_5, c_8 x_1 x_4 + x_6 \rangle \, . \\
\end{aligned}
\]
In particular, the image of $\iota_{c_8, c_9}$ is given by
\[ \mathrm{Im} \left( \iota_{c_8, c_9} \right) = \varphi_{c_8, c_9}^{-1} ( \langle 0 \rangle ) = V \left( x_3, \left[ x_6 + c_8 x_1 x_4 + c_9 x_1 x_5 \right] \cdot \left[ x_1 x_4 + x_2 \right] - x_1^2 x_4 x_5 \right) \label{equ:ImageOfSchemeMorphismsFordP3Example}\]
which is exactly the GUT-surface in question! Hence, under the hypothesis that our conjecture is correct, we have just constructed an embedding $\iota \colon dP_3 \hookrightarrow X_\Sigma$ whose image is precisely $W$. Along this embedding $\iota$, the divisors $A_i$ and $D_i$ are mapped to each other according to \cref{equ:Dictionary}. We argued that the both the intersection numbers of these divisors and the sheaf cohomologies of their associated line bundles are respected by this dictionary. This is fairly non-trivial consistency check on the proposed embedding $\iota \colon dP_3 \hookrightarrow X_\Sigma$. Note also that the induced homomorphisms of the degree groups maps the canonical bundles of these spaces to each other. Ultimately, we envision to translate the matter curves from $W = V( P, x_3 )$ into curves in the toric model of a $dP_3$-surface along $\iota$. We will come to discuss this momentarily. Let us mention that the genera of the matter curves in $W$ matched with those of their cousins in the toric model of $dP_3$. Given this evidence, we are positive that at least in this particular instance \cref{conj:II} is correct.

Let us use this opportunity to mention that this `translation' makes use of the following identities:
\begin{itemize}
 \item $\left. \overline{K}_{\mathcal{B}_6} \right|_{W} = \left. \left( G_1 + G_2 + 3 G_3 \right) \right|_{W} \stackrel{\varphi}{\cong} \mathcal{O}_{dP_3} \left( 4,3,2,3 \right) = 4 H - E_1 - E_2 - 2 E_3$,
 \item $\left. W \right|_{W} = \left. ( D_3 + D_4 ) \right|_{W} \stackrel{\varphi}{\cong} \mathcal{O}_{dP_3} \left( 1,1,0,1 \right) = H - E_3$ and
 \item $\left. \mathcal{O}_{\mathcal{B}_6} \left( a G_1 + b G_2 + c G_3 \right) \right|_{W} \stackrel{\varphi}{\cong} \mathcal{O}_{dP_3} \left( a+c, c, a+b, c \right)$ 
      for all $a,b,c \in \mathbb{Z}$.
\end{itemize}

Finally let us come back to the divisors $G_1$, $G_2$, $G_3$ in $\mathcal{B}_6$. We claimed that they freely generate $\mathrm{Pic} ( \mathcal{B}_6 )$. Suppose this was wrong, \ie $\mathcal{O}_{\mathcal{B}_6} ( n G_i )$ and $\mathcal{O}_{\mathcal{B}_6} ( m G_j )$ were linearly equivalent on $\mathcal{B}_6$ for suitable $i \neq j$ and $n,m \in \mathbb{Z}^\ast$. Then also the pullbacks to $W$ were linearly equivalent. However it is readily confirmed along $dP_3 \stackrel{\varphi}{\cong} W$ that
\begin{itemize}
 \item $\left. \mathcal{O}_{\mathcal{B}_6} \left( G_1 \right) \right|_{W} \cong \mathcal{O}_{dP_3} \left( 1,0,1,0 \right)$,
 \item $\left. \mathcal{O}_{\mathcal{B}_6} \left( G_2 \right) \right|_{W} \cong \mathcal{O}_{dP_3} \left( 0,0,1,0 \right)$ and
 \item $\left. \mathcal{O}_{\mathcal{B}_6} \left( G_3 \right) \right|_{W} \cong \mathcal{O}_{dP_3} \left( 1,1,0,1 \right)$.
\end{itemize}
Now since $\mathrm{Pic} ( dP_3 )$ is freely generated by the canonical generators, the pullbacks are not linearly equivalent on $W$. Consequently, $G_1$, $G_2$, $G_3$ must freely generate $\mathrm{Pic} ( \mathcal{B}_6 )$.

\paragraph{The Elliptically Fibred 4-Fold \texorpdfstring{$\mathbf{\hat{Y}_4}$}{Y4} and the Matter Curves} 

We now employ the $SU(5) \times U(1)_X$-top geometry as revised in \cref{sec:ToricFTheoryGUTModels} to construct an elliptic fibration over the base space $\mathcal{B}_6$ described above. Let us use this opportunity to briefly recall this geometry.

First of all, since $\mathcal{B}_6$ is a toric variety, $\hat{Y}_4$ will be described as complete intersection of codimension $2$ in a toric ambient space $\hat{Y}_\Sigma$. Its Cox ring $S = \mathbb{Q} [ x_1, x_2, e_0, x_4, x_5, x_6, e_1, e_2, e_3, e_4, x, y, z, s ]$ is graded by $\mathrm{Cl} ( \hat{Y}_\Sigma ) = \mathbb{Z}^8$ according to \cref{GradingOfCoxRingOfFibreAmbientSpaedP3Example}. The Stanley-Reisner ideal $I_{\mathrm{SR}} ( \hat{Y}_\Sigma )$ satisfies
\begin{align}
\begin{split}
\label{SRIdealYSigmadP3}
I_{\mathrm{SR}} ( \hat{Y}_\Sigma ) =& \left\langle e_0 e_2, e_0 e_3, e_0 s, e_1 e_3, e_1 y, e_1 z, e_1 s, e_4 x, e_4 z, e_4 s, x y, e_2 y, e_2 z,e_3 z,z s, \right. \\
                               & \hspace{8em} \left. e_2 s, x_1 x_2 x_6, e_0 x_4 x_5, x_4 x_5 e_1, x_4 x_5 e_4, x_4 x_5 e_2, x_4 x_5 e_3 \right\rangle \, .
\end{split}
\end{align}
\begin{table}[tbp]
\centering
\begin{tabular}{cccccc@{\hskip 20pt}cccc@{\hskip 20pt}cccc}
\toprule
$x_1$ & $x_2$ & $e_0$ & $x_4$ & $x_5$ & $x_6$ & $e_1$ & $e_2$ & $e_3$ & $e_4$ & x & y & z & s \\
\midrule
1 & 1 & 0 & 0 & 0 & 1 & 0 & 0 & 0 & 0 & 2 & 3 & 0 & 0 \\
0 & 1 & 1 & 1 & 1 & 1 & 0 & 0 & 0 & 0 & 6 & 9 & 0 & 0 \\
\vspace{-0.5em} & \\
0 & 0 &-1 & 0 & 0 & 0 & 1 & 0 & 0 & 0 &-1 &-1 & 0 & 0 \\
0 & 0 &-1 & 0 & 0 & 0 & 0 & 1 & 0 & 0 &-2 &-2 & 0 & 0 \\
0 & 0 &-1 & 0 & 0 & 0 & 0 & 0 & 1 & 0 &-2 &-3 & 0 & 0 \\
0 & 0 &-1 & 0 & 0 & 0 & 0 & 0 & 0 & 1 &-1 &-2 & 0 & 0 \\
\vspace{-0.5em} & \\
0 & 0 & 0 & 0 & 0 & 0 & 0 & 0 & 0 & 0 &-1 &-1 & 0 & 1 \\
0 & 0 & 0 & 0 & 0 & 0 & 0 & 0 & 0 & 0 & 2 & 3 & 1 & 0 \\
\bottomrule
\end{tabular}
\caption[Toric data of $\hat{Y}_\Sigma$ in \emph{F-theory} GUT-model with $W \cong dP_3$.]{The Cox ring $\mathbb{Q} [ x_1, x_2, e_0, x_4, x_5, x_6, e_1, e_2, e_3, e_4, x, y, z, s]$ of the toric space $\hat{Y}_\Sigma$ has $\mathbb{Z}^8$-grading. Together with \cref{SRIdealYSigmadP3} this grading defines the geometry of $\hat{Y}_\Sigma$ (\cf \cref{subsec:TowardsToricVarieties}).}
\label{GradingOfCoxRingOfFibreAmbientSpaedP3Example}
\end{table}
In making contact with the notation used in \cite{Krause:2011xj} we have identified $x_3 \equiv e_0$. Note also that we have chosen this toric variety from a list of 243 triangulations because $I_{\mathrm{SR}} ( X_\Sigma ) \subseteq I_{\mathrm{SR}} ( \hat{Y}_\Sigma )$ and because $I_{\mathrm{SR}} ( \mathrm{top} ) \subseteq I_{\mathrm{SR}} ( \hat{Y}_\Sigma )$ where
\begin{align}
\begin{split}
I_{\mathrm{SR}} \left( \mathrm{top} \right) =& \left\langle xy, x e_0 e_3, x e_1 e_3, x e_4, y e_0 e_3, y e_1, y e_2, z s, z e_1 e_4, z e_2 e_4, \right. \\
                & \hspace{8em} \left. z e_3, s e_0, s e_1, s e_4, e_0 e_2, z e_4, z e_1, z e_2, s e_2, e_0 e_3, e_1 e_3 \right\rangle \, .
\end{split}
\end{align}
This means that we can recover $X_\Sigma$ and $\mathcal{B}_6$ as subvarieties of $\hat{Y}_\Sigma$ most easily.

$\hat{Y}_\Sigma$ is a complete and projective orbifold. In particular, it is not smooth. The resolved 4-fold $\hat{Y}_4$ is given as the complete intersection $\hat{Y}_4 = V ( P_T, P^\prime ) \subseteq Y_\Sigma$ where
\[ P_T = y^2 s e_3 e_4 + \tilde{a_1} x y z s + \tilde{a_{3,2}} y z^3 e_0^2 e_1 e_4 - x^3 s^2 e_1 e_2^2 e_3 - \tilde{a_{2,1}} x^2 z^2 s e_0 e_1 e_2 - \tilde{a_{4,3}} x z^4 e_0^3 e_1^2 e_2 e_4 \]
and $P^\prime$ is obtained from the base polynomial $P$ of \cref{equ:most_general_P} by replacing the original variable $x_3$ by $e_0 e_1 e_2 e_3 e_4$ \cite{Krause:2011xj}. The proper transform of the Tate polynomial $P_T$ depends on the coordinates $x_i$ of the \emph{F-theory} base space $\mathcal{B}_6$ via sections $a_i \in H^0 ( \mathcal{B}_6, \overline{K}_{\mathcal{B}_6}^{\otimes i} )$ which, to ensure for the $SU ( 5 ) \times U ( 1 )_X$ gauge symmetry, are factored according to
\[ a_{1,0} \equiv a_{1,0}, \qquad a_2 = a_{2,1} w, \qquad a_3 = a_{3,2} w^2, \qquad a_4 = a_{4,3} w^3 \, . \]
Note that in this expression $w \in H^0 ( \mathcal{B}_6, \mathcal{O}_{\mathcal{B}_6} )$ is the polynomial, which cuts out $W$ from the point of view of $\mathcal{B}_6$. So in the geometry at hand $w$ is the `pullback' of the polynomial $x_3$ onto $\mathcal{B}_6$. Consequently,
\begin{align*}
  a_{1,0} &\in H^0 \left( \mathcal{B}_6, \left. \mathcal{O}_{X_\Sigma} \left( 1, 3 \right) \right|_{\mathcal{B}_6} \right) \, , \quad & a_{2,1} &\in H^0 \left( \mathcal{B}_6, \left. \mathcal{O}_{X_\Sigma} \left( 2, 5 \right) \right|_{\mathcal{B}_6} \right) \, , \\
  a_{3,2} &\in H^0 \left( \mathcal{B}_6, \left. \mathcal{O}_{X_\Sigma} \left( 3, 7 \right) \right|_{\mathcal{B}_6} \right) \, , & a_{4,3} &\in H^0 \left( \mathcal{B}_6, \left. \mathcal{O}_{X_\Sigma} \left( 4, 9 \right) \right|_{\mathcal{B}_6} \right) \, .
\end{align*}
For generic such sections $\hat{Y}_4$ is smooth and its Hodge diamond reads
\[
\label{equ:HodgeOfY4dP3Example}
\begin{array}{ccccccccc}
& & & & 1 \\
& & & 0 & & 0 \\
& & 0 & & 9 & & 0 \\
& 0 & & 0 & & 0 & & 0 \\
1 & & 577 & & 2308 & & 557 & & 1 \\
& 0 & & 0 & & 0 & & 0 \\
& & 0 & & 9 & & 0 \\
& & & 0 & & 0 \\
& & & & 1
\end{array}
\]

Along the restriction maps $H^0 ( X_\Sigma, \mathcal{O}_{X_\Sigma} ( i, j ) ) \to H^0 ( \mathcal{B}_6, \left. \mathcal{O}_{X_\Sigma} ( i, j ) \right|_{\mathcal{B}_6} )$ 
we model these sections $a_{i,j}$ by polynomials $\tilde{a_{i,j}}$ on the base toric ambient space $X_\Sigma$ given as
\begin{align*}
  \tilde{a_{1,0}} &\in H^0 \left( X_\Sigma, \mathcal{O}_{X_\Sigma} \left( 1, 3 \right) \right), \quad & \tilde{a_{2,1}} &\in H^0 \left( X_\Sigma, \mathcal{O}_{X_\Sigma} \left( 2, 5 \right) \right), \\
  \tilde{a_{3,2}} &\in H^0 \left( X_\Sigma, \mathcal{O}_{X_\Sigma} \left( 3, 7 \right) \right),\quad & \tilde{a_{4,3}} &\in H^0 \left( X_\Sigma, \mathcal{O}_{X_\Sigma} \left( 4, 9 \right) \right).
\end{align*}
Explicitly we can perform this restriction by imposing the base polynomial $P$ alongside the $\tilde{a}_{i,j}$. For example, the matter curves satisfy
\begin{align*}
  C_{\mathbf{10}_1} &= V \left( P, x_3, \tilde{a_{1,0}} \right) \subseteq \mathcal{B}_6, \quad & C_{\mathbf{5}_3} &= V \left( P, x_3, \tilde{a_{3,2}} \right) \subseteq \mathcal{B}_6 \, , \\
  C_{\mathbf{5}_{-2}} &= V \left( P, x_3, \tilde{a_1} \tilde{a_{4,3}} - \tilde{a_{2,1}} \tilde{a_{3,2}} \right) \subseteq \mathcal{B}_6, & C_{\mathbf{1}_{5}} &= V \left( P, \tilde{a_{4,3}}, \tilde{a_{3,2}} \right) \subseteq \mathcal{B}_6 \, .
\end{align*}

This brings us to our final consistency check on the conjectured morphism $\iota \colon dP_3 \hookrightarrow X_\Sigma$. Namely, since $C_{\mathbf{10}_1}$, $C_{\mathbf{5}_3}$ and $C_{\mathbf{5}_{-2}}$ are subloci of $W$, we can use $W \stackrel{\varphi}{\cong} dP_3$ to express these matter curves in terms of the toric $dP_3$-surface. This implies
\begin{itemize}
 \item $C_{\mathbf{10}_1} \cong V \left( \varphi \left( \tilde{a_{1,0}} \right) \right)$ and $\mathrm{deg} \left( \varphi \left( \tilde{a_{1,0}} \right) \right) = 
      \left( 4,3,2,3 \right)$,
 \item $C_{\mathbf{5}_3} \cong V \left( \varphi \left( \tilde{a_{3,2}} \right) \right)$ and $\mathrm{deg} \left( \varphi \left( \tilde{a_{3,2}} \right) \right) = 
      \left( 10,7,6,7 \right)$ and
 \item $C_{\mathbf{5}_{-2}} \cong V ( \varphi \left( \tilde{a_{1,0}} \tilde{a_{4,3}} - \tilde{a_{2,1}} \tilde{a_{3,2}} \right) )$ and    
      $\mathrm{deg} \left( \tilde{a_{1,0}} \tilde{a_{4,3}} - \tilde{a_{2,1}} \tilde{a_{3,2}} \right) = \left( 17,12,10,12 \right)$.
\end{itemize}
As consistency check on our proposed embedding $\iota \colon dP_3 \hookrightarrow X_\Sigma$, we computed the genera of the matter curves $C_{\mathbf{10}_1}$, $C_{\mathbf{5}_3}$, $C_{\mathbf{5}_{-2}}$ both as subloci of $X_\Sigma$ and as subloci of the toric $dP_3$-surface. We found matching results, namely,
\[ g \left( C_{\mathbf{10}_1} \right) = 2, \qquad g \left( C_{\mathbf{5}_3} \right) = 24, \qquad g \left( C_{\mathbf{5}_{-2}} \right) = 79 \, . \]
Let us use this opportunity to recall that, as argued in \cref{subsec:LocalisedMatterAndTheSpinBundle}, the spin bundles are fixed by the embeddings of the matter curves in $\mathcal{B}_6$. Therefore, the spin bundles take the form
\begin{align}
\begin{split}
 \mathcal{O}_{\mathrm{spin},\mathbf{10}_1} &= \left. \mathcal{O}_{dP_3} \left( \frac{1}{2} \left( 1,1,0,1 \right) \right) \right|_{C_{\mathbf{10}_1}} \, , \qquad
 \mathcal{O}_{\mathrm{spin},\mathbf{5}_3} = \left. \mathcal{O}_{dP_3} \left( \frac{1}{2} \left( 7,5,4,5 \right) \right) \right|_{C_{\mathbf{5}_3}} \, , \\
 \mathcal{O}_{\mathrm{spin},\mathbf{5}_{-2}} &= \left. \mathcal{O}_{dP_3} \left( 7, 5, 4, 5 \right) \right|_{C_{\mathbf{5}_{-2}}} \, .
\end{split}
\end{align}
 
Finally note that the singlet curve $C_{\mathbf{1}_{5}}$ cannot be re-expression in terms of the toric $dP_3$ as $C_{\mathbf{1}_{5}} \not \subseteq W$. In terms of $X_\Sigma$ it holds $\mathcal{O}_{\mathrm{Spin}, C_{1_{-5_X}}} = \left. \mathcal{O}_{X_\Sigma} ( 3, \frac{13}{2} ) \right|_{C_{\mathbf{1}_{5}}}$ and we have $g ( C_{\mathbf{1}_{5}} ) = 680$.

\paragraph{The Nef-Cone Of The Base Space \texorpdfstring{$\mathbf{\mathcal{B}_6}$}{B6}}

In the next section, we will subject the available gauge backgrounds to various consistency checks. Among them is a condition for a supersymmetric \emph{F-theory} vacuum. To formulate this condition, we will need knowledge of $\mathrm{Nef} ( \mathcal{B}_6 )$. To identify this cone, let us start by constructing curves in $\mathcal{B}_6$:
\begin{itemize}
 \item $V \left( P, x_5 \right) = W$ and contains the six divisors $D_i$ described in the previous subsection. Those constitute 6 curves in $\mathcal{B}_6$.
 \item $V \left( P, x_1 \right) = G_1 \cup G_2$. Both $G_1$ and $G_2$ are found to be isomorphic to $\mathbb{P}^2_{\mathbb{Q}}$ and hence contain one class of curves 
      each. Representants for these classes are given by the curves $\mathcal{C}_{13} = G_1 \cap G_3$ and $\mathcal{C}_{23} = G_2 \cap G_3$.
 \item $V \left( P, x_2 \right) = G_1 \cup X$ where $X = V ( x_2, R + c_{10} x_3 x_6 )$ is given in terms of the polynomial $R = x_1 ( c_1 x_3^2 + c_2 x_4^2 + c_4 x_3 
      x_4 + c_5 x_3 x_5 + x_6 x_4 x_5 )$. It is now readily found that the following four curves lie in $X$:
      \[
      \begin{array}{lcl}
      \mathcal{C}_A = V \left( x_1, x_2, x_3 \right) \, , & & \mathcal{C}_B = V \left( x_2, x_3, x_4 \right) \, , \\
      \mathcal{C}_C = V \left( x_2, x_3, c_2 x_4 + c_6 x_5 \right) \, , & & \mathcal{C}_D = V \left( x_2, x_6, R \right) \, .
      \end{array}
      \]
\end{itemize}
The Hodge diamond of $\mathcal{B}_6$ in \cref{equ:HodgeBasedP3} shows $h_{1,1} ( \mathcal{B}_6 ) = 3$. Thus, the above curves more than suffice to describe all classes of curves in $\mathcal{B}_6$.

As a next step we compute the intersection numbers of the divisor classes $G_i \in \mathrm{Cl} ( \mathcal{B}_6 )$ with those 12 classes of curves. To do so recall that for divisor classes $D_i$ and $D_j$ in a surface $S$ it is known that \cite{cox2011toric, ATIT}
\[ D_i D_j = \mathrm{deg} \left( \left. \mathcal{O}_{S} \left( D_i \right) \right|_{D_j} \right). \]
In analogy to this formula we have \cite{ATIT} $D C = \mathrm{deg} ( \left. \mathcal{O}_{\mathcal{B}_6} ( D ) \right|_{C} )$, which we compute with the \texttt{gap}-package \cite{SheafCohomologyOnToricVarieties}. The so-computed intersection numbers are displayed in \cref{table-N16}. From knowledge of these intersection numbers, we can now easily extract those divisor classes which have non-negative intersection number with all of the 12 curves constructed above. One finds \footnote{We use the notation $( 1,1,1 )^T \widehat{=} G_1 + G_2 + G_3$ \dots to indicate the generators.}
\[ \label{equ:NefConeB6DP3-Example}
\mathrm{Nef} \left( \mathcal{B}_6 \right) = \mathrm{Span}_{\mathbb{Z}_{\geq 0}} \left\{ \left( \begin{array}{c} 1 \\ 1 \\ 1 \end{array} \right),
                                                                                           \left( \begin{array}{c} 0 \\ 0 \\ 1 \end{array} \right),
                                                                                           \left( \begin{array}{c} 1 \\ 0 \\ 1 \end{array} \right),
                                                                                           \left( \begin{array}{c} 0 \\ 1 \\ 1 \end{array} \right) \right\} \, . \]

\begin{table}[tbp]
\centering
\begin{tabular}{c@{\hskip 20pt}cccccc@{\hskip 20pt}cc@{\hskip 20pt}cccc}
      \toprule
      & $D_1$ & $D_2$ & $D_3$ & $D_4$ & $D_5$ & $D_6$ & $\mathcal{C}_{13}$ & $\mathcal{C}_{23}$ & $\mathcal{C}_A$ & $\mathcal{C}_B$ & $\mathcal{C}_C$ & $\mathcal{C}_d$ \\
      \midrule
      $G_1$ & 0 & 1 & 0 & 1 & 0 & -1 & -1 & 0 & -1 & 1 & 0 & 1 \\
      $G_2$ & 1 & 0 & 1 & 0 & -1 & 0 & 0 & -1 & 0 & 0 & 0 & 1 \\
      $G_3$ & 0 & 0 & 0 & 0 & 1 & 1 & 1 & 1 & 1 & 0 & 2 & 1 \\
      \bottomrule
\end{tabular}
\caption{Intersection numbers $G_i \cdot C$ for various curves $C \subseteq \mathcal{B}_6$.}
\label{table-N16}
\end{table}

\subsection{The Gauge Backgrounds} \label{subsec:GaugeBackground}

\paragraph{Flux Choice}

We now look at a gauge background of the form $A = A_X ( F ) + A( \mathbf{10}_1 )( \lambda ) + A_Y( \mathcal{H} )$. We already discussed our take on the hypercharge flux $A_Y ( \mathcal{H} )$ in great detail in \cref{sec:ChoiceOfHyperchargeFluxAndExotics}. In particular, we decided to choose $\mathcal{H} = E_i - E_j$ for distinct $i$ and $j$. These divisors $E_i, E_j, \mathcal{H} \in \mathrm{Pic} ( W )$ must all not be pullbacks from $\mathcal{B}_6$. Again, let us use our conjectured isomorphism $dP_3 \stackrel{\varphi}{\cong} W$ to identify which divisors satisfy this demand. We have
\[ \varphi^\ast \mathrm{Pic} \left( \mathcal{B}_6 \right) \cong \mathrm{Span}_{\mathbb{Z}} \left\{ \left( \begin{array}{c} 0 \\ 0 \\ 1 \\ 0 \end{array} \right), \left( \begin{array}{c} 1 \\ 0 \\ 0 \\ 0 \end{array} \right), \left( \begin{array}{c} 0 \\ 1 \\ 0 \\ 1 \end{array} \right) \right\} \subseteq \mathrm{Pic} \left( dP_3 \right). \]
Hence, in accordance to our discussion in \cref{sec:ChoiceOfHyperchargeFluxAndExotics} there are precisely two possible choice of $\mathcal{H}$, namely $\mathcal{H} = \pm E_1 - E_2$. For now we choose $\mathcal{O}_{dP_3} ( \mathcal{H} ) = \mathcal{O}_{dP_3} ( 0, 1, 0, -1 )$ to model the hypercharge flux, and come back to the other choice later.

Finally we need to make a choice on the divisor class $F \in \mathrm{Cl} ( \mathcal{B}_6 )$. As $\mathrm{Pic} ( \mathcal{B}_6 )$ is freely generated by $G_1$, $G_2$ and $G_3$ we simply set $F = a G_1 + b G_2 + c G_3$ with $a,b,c \in \mathbb{Q}$ and write $A ( a,b,c, \lambda )$ for such a flux background. 
Note that $a,b,c \in \mathbb{Q}$ defines a proper $\mathbb{Q}$-Cartier divisor and not a $\mathbb{Z}$-Cartier divisor. These rational parameter are subject to flux quantisation which we will come to discuss momentarily.

\paragraph{Chiralities Of The Broken Spectrum}

After our detailed discussion in \cref{chapter:MasslessSpectraAndSheafCohomology}, the chiralities of the massless spectrum immediately follow as
\[ \left( \begin{array}{c} \chi \left( \mathbf{10}_1 \right) \\ \chi \left( \mathbf{5}_3 \right) \\ \chi \left( \mathbf{5}_{-2} \right) \end{array} \right) = \frac{1}{5} \cdot \left( \begin{array}{cccc} 2 & 2 & 2 & 50 \\ 12 & 12 & 18 & -52 \\ -14 & -14 & -20 & 2  \end{array} \right) \cdot \left( \begin{array}{c} a \\ b \\ c \\ \lambda \end{array} \right) \, . \]
In terms of the broken spectrum listed in \cref{table-brokenStates} these chiralities encode
\begin{align}
\begin{split}
\chi \left( \mathbf{10}_1 \right) &= \chi \left( \left( \mathbf{3}, \mathbf{2} \right)_{1_X, -1_Y} \right) = \chi \left( \left( \overline{\mathbf{3}}, \mathbf{1} \right)_{1_X, 4_Y} \right) = \chi \left( \left( \mathbf{1}, \mathbf{1} \right)_{1_X, -6_Y} \right) \, . \\
\chi \left( \mathbf{5}_3 \right) &= \chi \left( \left( \mathbf{3}, \mathbf{1} \right)_{3_X, 2_Y} \right) = \chi \left( \left( \mathbf{1}, \mathbf{2} \right)_{3_X, -3_Y} \right) \, . \\
\chi \left( \mathbf{5}_{-2} \right) &= \chi \left( \left( \mathbf{3}, \mathbf{1} \right)_{-2_X, 2_Y} \right) = \chi \left( \left( \mathbf{1}, \mathbf{2} \right)_{-2_X, -3_Y} \right) \, . \\
\end{split}
\end{align}
Note that $\chi ( \mathbf{10}_1 ) + \chi ( \mathbf{5}_3 ) + \chi ( \mathbf{5}_{-2} ) = 0$, which is phenomenologically appealing, and that the chirality of the states on the singlet curve satisfies $\chi ( \mathbf{1}_{5} ) = \chi ( ( \mathbf{1}, \mathbf{1} )_{5_X, 0_Y} ) = 20a + 20b + 86c$.

\paragraph{D3-Tadpole Cancellation}

The $D3$-tadpole cancellation condition requires the existence of $N_{D_3} \in \mathbb{Z}_{\geq 0}$ such that
\[ N_{D_3} = \frac{\chi ( \hat{Y}_4 )}{24} - \frac{1}{2} \Mint_{\hat{Y}_4}{\left( A_X \left( F \right) + A \left( \mathbf{10}_1 \right) \left( \lambda \right) + A_Y \left( \mathcal{H} \right) \right)^2} \, . \]
The Euler characteristic $\chi ( \hat{Y}_4 )$ of the resolved 4-fold $\hat{Y}_4$ was analysed in \cite{oai:arXiv.org:1202.3138}. Additional details are summarised in \cref{subsec:TadpolesSU5xU1}. This then leads to \footnote{Recall that we have chosen $a,b,c, \lambda \in \mathbb{Q}$ thus far to parametrised $F \in \mathbb{Q} \mathrm{Pic} ( \mathcal{B}_6 )$. The demand $N_{D_3} \in \mathbb{Z}_{\geq 0}$ will serve as one integrability condition on these parameters. We will discuss this condition more in the context of flux quantisation.}
\[ N_{D_3} = \frac{287}{2} - \frac{7}{5} \left( a^2 + b^2 \right) + 2 c^2 + \frac{14}{5} c \left( a + b \right) - \frac{2 \lambda}{5} \left( a + b + c \right) - 5 \lambda^2 - 30 \stackrel{!}{\in} \mathbb{Z}_{> 0}. \label{equ:Tadpole} \]
Let us point out that the $-30$ in this expression stems from the self-intersection number of the hypercharge flux. Indeed, this term is expected to tighten the $D_3$-tadpole constraint since the hypercharge flux is supersymmetric.

\paragraph{D-Term}
The hypercharge flux $A_Y ( \mathcal{H} )$ is supersymmetric, and so its D-term vanishes. Therefore, the contributions to the D-term stem solemnly from $A_X ( F )$ and $A ( \mathbf{10}_1 ) ( \lambda )$. These contributions satisfy \cite{oai:arXiv.org:1202.3138} \footnote{In \cite{oai:arXiv.org:1202.3138} these results were derived up to an unspecified constant. In the following we will assume that the constants for $\xi_X ( A^X ( F ) )$ and $\xi_X ( A^\lambda ( \lambda ) )$ are non-zero and match.}
\begin{align}
\begin{split}
\xi_X \left( A_X \left( F \right) \right) &\simeq - \frac{2}{\mathcal{V}_B} \Mint_{\mathcal{B}_6}{J \wedge F \wedge \left( 3 W - 5 \overline{K}_{\mathcal{B}_6}\right)}, \\
\xi_X \left( A \left( \mathbf{10}_1 \right) \left( \lambda \right) \right) &\simeq - \frac{\lambda}{\mathcal{V}_B} \Mint_{\mathcal{B}_6}{J \wedge W \wedge \overline{K}} \, .
\end{split}
\end{align}
The supersymmetry constraint is the existence of a divisor class $J$ in the interior of $\mathrm{Nef} ( \mathcal{B}_6 )$ such that 
\[ 0 = \xi_X \left( A^X \left( F \right) \right) + \xi_X \left( A^\lambda \left( \lambda \right) \right) \, . \]
By looking at the generators $G_1$, $G_2$, $G_3$ we can write $J = \alpha G_1 + \beta G_2 + \gamma G_3$ -- $\alpha, \beta, \gamma \in \mathbb{Z}$ -- and work out the conditions on these parameters from the triple intersection numbers of the \emph{F-theory} base $\mathcal{B}_6$ (c.f. \cref{equ:tripleIntersectionsBase}). We find
\begin{align}
\begin{split}
 \mathcal{V}_B \xi &\equiv \mathcal{V}_B \left( \xi_X \left( A_X \left( F \right)\right) + \xi_X \left( A \left( \mathbf{10}_1 \right) \left( \lambda \right) \right) \right) \\
 &= \alpha \left( - 14a + 14c - 2 \lambda \right) + \beta \left( -14 b + 14c - 2 \lambda \right) + \gamma \left( 14 \left( a+b \right) + 20c - 2 \lambda \right) \, .
\end{split}
\end{align}
Finally recall that we computed the Nef cone $\mathrm{Nef} ( \mathcal{B}_6 )$. According to \cref{equ:NefConeB6DP3-Example} it is given by
\[ \mathrm{Nef} \left( \mathcal{B}_6 \right) = \mathrm{Span}_{\mathbb{Z}_{\geq 0}} \left\{ \left( \begin{array}{c} 1 \\ 1 \\ 1 \end{array} \right), \left( \begin{array}{c} 0 \\ 0 \\ 1 \end{array} \right), \left( \begin{array}{c} 1 \\ 0 \\ 1 \end{array} \right), \left( \begin{array}{c} 0 \\ 1 \\ 1 \end{array} \right) \right\} \, . \]
Thereby, we can formulate a necessary and sufficient supersymmetry condition. Namely:
\ebox{The \emph{F-theory} vacuum ($\hat{Y}_4$, $A ( a,b,c, \lambda )$ ) is supersymmetric iff there exist $n_i \in \mathbb{Z}_{> 0}$ such that
\begin{align*}
0 &= 6 n_1 \left( 8c - \lambda \right) + 2 n_2 \left( 7a + 7b + 10c - \lambda \right) \\
  & \hspace{13em} + 2 n_3 \left( 7b + 17c - 2 \lambda \right) + 2 n_4 \left( 7a + 17 c - 2 \lambda \right) \, .
\end{align*}
}
To check if ($\hat{Y}_4, A ( a,b,c, \lambda )$) is supersymmetric, it hence suffices to analyse the coefficients
\[ c_1 := 8c - \lambda, \; \; c_2 := 7a + 7b + 10c - \lambda, \; \; c_3 := 7b + 17c - 2 \lambda, \; \; c_4 := 7a + 17c - 2 \lambda \, . \]
The \emph{F-theory} vacuum is supersymmetric precisely if one of these parameters is positive and another negative.

\paragraph{Integrability Condition}

Let us take a first step towards a `proper' flux choice. To this end, we first need to work out the $\mathbb{Q}$-Cartier divisors whose associated `quasi'-line bundles encode the zero modes. These follow from \cref{table-netspectrum} and yield \cref{table-N20}. We are now looking for a quantisation condition, such that all of these divisors are actual $\mathbb{Z}$-Cartier divisors. As it turns out, this is the case precisely if
\[ \frac{a}{5} \in \mathbb{Z}, \quad \frac{b}{5} \in \mathbb{Z}, \quad \frac{c}{5} + \frac{1}{2} \in \mathbb{Z}, \quad \frac{\lambda}{5} \in \mathbb{Z} \, . \]
Therefore, we quantise $A ( a,b,c, \lambda)$ according to this rule and set ($\alpha, \beta, \gamma, \tilde{\lambda} \in \mathbb{Z}$)
\[ a = 5 \alpha, \quad b = 5 \beta, \quad c = 5 \left( \gamma + \frac{1}{2} \right), \quad \lambda = 5 \tilde{\lambda} \, . \]
It is worth mentioning that the quantisation conditions on $a,b,c$ is equivalent to the sufficient flux quantisation $\frac{1}{5} F + \frac{1}{2} W \in H^2 ( \mathcal{B}_6, \mathbb{Z} )$ \cite{oai:arXiv.org:1202.3138}.

\begin{table}[tbp]
\centering
\resizebox{\textwidth}{!}{
\begin{tabular}{lll}
\toprule
& Rep. & Line bundle on matter curve \\
\midrule
$C_{\mathbf{10}_1}$ & $\left( \mathbf{3}, \mathbf{2} \right)_{1_X, 1_Y}$ & 
$\left. \mathcal{O}_{dP_3} \left( \frac{a+c + 19 \lambda}{5} + \frac{1}{2}, \frac{c + 13 \lambda}{5} + \frac{3}{2}, \frac{a+b+12 \lambda}{5}, \frac{c + 13 \lambda}{5} - \frac{1}{2} \right) \right|_{C_{\mathbf{10}_1}}$ \\
& $\left( \overline{\mathbf{3}}, \mathbf{1} \right)_{1_X, -4_Y}$ & $\left. \mathcal{O}_{dP_3} \left( \frac{a+c + 19 \lambda}{5} + \frac{1}{2}, \frac{c + 13 \lambda}{5} - \frac{7}{2}, \frac{a+b+12 \lambda}{5}, \frac{c + 13 \lambda}{5} + \frac{9}{2} \right) \right|_{C_{\mathbf{10}_1}}$ \\
& $\left( \mathbf{1}, \mathbf{1} \right)_{1_X, 6_Y}$ & $ \left. \mathcal{O}_{dP_3} \left( \frac{a+c + 19 \lambda}{5} + \frac{1}{2}, \frac{c + 13 \lambda}{5} + \frac{13}{2}, \frac{a+b+12 \lambda}{5}, \frac{c + 13 \lambda}{5} - \frac{11}{2} \right) \right|_{C_{\mathbf{10}_1}}$ \\
\vspace{-0.3em} & \\
$C_{\mathbf{5}_3}$ & $\left( \mathbf{3}, \mathbf{1} \right)_{3_X, -2_Y}$ & $\left. \mathcal{O}_{dP_3} \left( \frac{3 \left( a+c \right) - 8 \lambda}{5} + \frac{7}{2}, \frac{3c - 6 \lambda}{5} + \frac{1}{2}, \frac{3 \left( a+b \right) - 4 \lambda + 10}{5}, \frac{3c - 6 \lambda}{5} + \frac{9}{2} \right) \right|_{C_{\mathbf{5}_3}}$ \\
& $\left( \mathbf{1}, \mathbf{2} \right)_{3_X, 3_Y}$ & $\left. \mathcal{O}_{dP_3} \left( \frac{3 \left( a+c \right) - 8 \lambda}{5} + \frac{7}{2}, \frac{3c - 6 \lambda}{5} + \frac{11}{2}, \frac{3 \left( a+b \right) - 4 \lambda + 10}{5}, \frac{3c - 6 \lambda}{5} - \frac{1}{2} \right) \right|_{C_{\mathbf{5}_3}}$ \\
\vspace{-0.3em} & \\
$C_{\mathbf{5}_{-2}}$ & $\left( \mathbf{3}, \mathbf{1} \right)_{-2_X, -2_Y}$ & $\left. \mathcal{O}_{dP_3} \left( 7 - \frac{2 \left( a+c \right)}{5}, 3 - \frac{2c}{5}, 4 - \frac{2 \left( a+b \right)}{5}, 7 - \frac{2c}{5} \right) \right|_{C_{\mathbf{5}_{-2}}} \otimes \mathcal{O}_{C_{\mathbf{5}_{-2}}} \left( \frac{\lambda ( 3 Y_1 - 2 Y_2 )}{5} \right)$ \\
& $\left( \mathbf{1}, \mathbf{2} \right)_{-2_X, 3_Y}$ & $\left. \mathcal{O}_{dP_3} \left( 7 - \frac{2 \left( a+c \right)}{5}, 8 - \frac{2c}{5}, 4 - \frac{2 \left( a+b \right)}{5}, 2 - \frac{2c}{5} \right) \right|_{C_{\mathbf{5}_{-2}}} \otimes \mathcal{O}_{C_{\mathbf{5}_{-2}}} \left( \frac{\lambda ( 3 Y_1 - 2 Y_2 )}{5} \right)$ \\
\vspace{-0.3em} & \\
$C_{\mathbf{1}_{5}}$ & $\left( \mathbf{1}, \mathbf{1} \right)_{5_X, 0_Y}$ & $\left. \mathcal{O}_{X_\Sigma} \left( 3, c + \frac{13}{2} \right) \right|_{C_{\mathbf{1}_{5}}} \otimes \left. \mathcal{O}_{\mathcal{B}_6} \left( a G_1 + b G_2 \right) \right|_{C_{\mathbf{1}_{5}}}$ \\
\bottomrule
\end{tabular}
}
\caption[Zero modes of broken GUT-model based on $W = dP_3$ via quasi line bundles.]{The sheaf cohomologies of the above line bundles count the zero modes of the flux $A (a,b,c, \lambda )$. No flux quantisation has been applied yet.}
\label{table-N20}
\end{table}

\paragraph{Suitable Chiralities}

As a next step we identify fluxes which produce chiralities of the form ($\Delta_1, \Delta_2 \in \mathbb{Z}$)
\[ \vec{\chi} \left( \Delta_1, \Delta_2 \right) = \left( \begin{array}{c} 3 + \Delta_1 \\ -3 - \Delta_2 \\ \Delta_2 - \Delta_1 \end{array} \right) \, . \]
It can be verified that we can find such fluxes whenever $\Delta_1$ and $\Delta_2$ are even. Hence, let us set 
\[ \Delta_1 = 2 \Delta_1^\prime, \qquad \Delta_2 = 2 \Delta_2^\prime \, . \]
Then all fluxes which give chiralities $\vec{\chi} \left( 2 \Delta_1^\prime, 2 \Delta_2^\prime \right)$ are of the form ($u,v \in \mathbb{Z}$)
\[ \left( \begin{array}{c} a \left( \Delta_1^\prime, \Delta_2^\prime, u, v \right) \\ b \left( \Delta_1^\prime, \Delta_2^\prime, u, v \right) \\ c \left( \Delta_1^\prime, \Delta_2^\prime, u, v \right) \\ \lambda \left( \Delta_1^\prime, \Delta_2^\prime, u, v \right) \end{array} \right) = \left( \begin{array}{c} 25 + 15 \Delta_1^\prime + 1675 \Delta_2^\prime \\ 0 \\ -\frac{35}{2} - 10 \Delta_1^\prime - 1175 \Delta_2^\prime \\ - 20 \Delta_2^\prime \end{array} \right) + u \cdot \left( \begin{array}{c} 1255 \\ 0 \\ -880 \\ - 15 \end{array} \right) + v \cdot \left( \begin{array}{c} -5 \\ 5 \\ 0 \\ 0 \end{array} \right) \, . \]

\paragraph{Scan for viable Fluxes}

For these fluxes the $D_3$-tadpole cancellation condition takes the form
\begin{align}
\begin{split}
N_{D_3} \left( \Delta_1, \Delta_2, u , v \right) =& -1374 - 1785 \frac{\Delta_1}{2} - 535 \frac{\Delta_1^2}{4} - 199265 \frac{\Delta_2}{2} \\
              & - 29890 \Delta_1 \Delta_2 - 6675375 \frac{\Delta_2^2}{4} - 149300 u - 44790 \Delta_1 u \\
              &- 5001550 \Delta_2 u - 3747430 u^2 + 350 v + 105 \Delta_1 v \\
              &+ 11725 \Delta_2 v + 17570 u v - 70 v^2 \, .
\end{split}
\end{align}
Instead of solving the constraint $N_{D_3} \in \mathbb{Z}_{\geq 0}$ analytically, we perform a scan with \emph{Mathematica} over the even pairs $( \Delta_1, \Delta_2 )$ in the window $-55 \leq \Delta_1, \Delta_2 \leq 55$. The results of this scan are listed in \cref{section:ListOfGoodFluxes}. Among these results, the fluxes whose chiralities come closest to the ones desired for a match with the \emph{standard model}, satisfy $\vec{\chi} = \pm ( 3, 21, -24 )$ and are parametrised by
$\left( a,b,c, \lambda \right) = \left( 5, 0, \frac{5}{2}, 0 \right)$ and $\left( a,b,c, \lambda \right) = \left( 0, 5, \frac{5}{2}, 0 \right)$. Apparently those chiralities are not quite what is needed for the \emph{standard model}. Consequently, this example can merely serve as toy model.\footnote{Recall also the existence of the exotic bulk states stemming from our choice of hypercharge flux. See \cref{sec:ChoiceOfHyperchargeFluxAndExotics} for more details.} In addition, these fluxes are not supersymmetric. In fact the only supersymmetric fluxes which we found satisfy $\chi = \pm ( 1, 3, -4 )$ -- even less phenomenologically appealing.

\subsection{Computing Zero Modes} \label{subsec:MasslessSpectra}

\paragraph{Massless Spectra on generic Matter Curves}

We look at the flux $A ( a,b,c, \lambda ) = A ( 5, 0, 2.5, 0 )$. Following \cref{table-N20}, its massless spectrum is counted by the sheaf cohomologies of the line bundles listed in \cref{table-N22}. In a number of cases the sheaf cohomologies could \emph{not} be obtained from application of \emph{cohomCalg} \cite{Blumenhagen:2010pv, Blumenhagen:2010ed, Blumenhagen:2011xn, KoszulExtensionManual, cohomCalg:Implementation, 2011JMP....52c3506J, Rahn:2010fm}, but required the full power of the technologies described in \cref{chapter:DetailsOnFPGradedSModules}. These line bundles are indicated by an asterix.

\begin{table}[tbp]
\centering
\begin{tabular}{llll}
\toprule 
& Rep. & relevant line bundle & Cohomologies \\
\midrule
$C_{\mathbf{10}_1}$ & $\left( \mathbf{3}, \mathbf{2} \right)_{1_X, 1_Y}$ &  $\left. \mathcal{O}_{dP_3} \left( 2, 2, 1, 0 \right) \right|_{C_{\mathbf{10}_1}}$ & $\left( 3,0 \right)$ \\
& $\left( \overline{\mathbf{3}}, \mathbf{1} \right)_{1_X, -4_Y}$ & $\left. \mathcal{O}_{dP_3} \left( 2, -3, 1, 5 \right) \right|_{C_{\mathbf{10}_1}}$ & $\left( 3,0 \right)^*$ \\
& $\left( \mathbf{1}, \mathbf{1} \right)_{1_X, 6_Y}$ & $\left. \mathcal{O}_{dP_3} \left( 2, 7, 1, -5 \right) \right|_{C_{\mathbf{10}_1}}$ & $\left( 3, 0 \right)^*$ \\
\vspace{-0.5em} \\
$C_{\mathbf{5}_3}$ & $\left( \mathbf{3}, \mathbf{1} \right)_{3_X, -2_Y}$ & $\left. \mathcal{O}_{dP_3} \left( 8, 2, 5, 6 \right) \right|_{C_{\mathbf{5}_3}}$ & $ \left( 21, 0 \right)$ \\
& $\left( \mathbf{1}, \mathbf{2} \right)_{3_X, 3_Y}$ & $\left. \mathcal{O}_{dP_3} \left( 8, 7, 5, 1 \right) \right|_{C_{\mathbf{5}_3}}$ & $\left( 21, 0 \right)^*$ \\
\vspace{-0.5em} \\
$C_{\mathbf{5}_{-2}}$ & $\left( \mathbf{3}, \mathbf{1} \right)_{-2_X, -2_Y}$ & $\left. \mathcal{O}_{dP_3} \left( 4, 2, 2, 6 \right) \right|_{C_{\mathbf{5}_{-2}}}$ & $\left( 9, 33 \right)$ \\
& $\left( \mathbf{1}, \mathbf{2} \right)_{-2_X, 3_Y}$ & $\left. \mathcal{O}_{dP_3} \left( 4, 7, 2, 1 \right) \right|_{C_{\mathbf{5}_{-2}}}$ & $\left( 6, 30 \right)$ \\
\vspace{-0.5em} \\
$C_{\mathbf{1}_{5}}$ & $\left( \mathbf{1}, \mathbf{1} \right)_{5_X, 0_Y}$ & $\left. \mathcal{O}_{\mathcal{B}_6} \left( 5 G_1 \right) \right|_{C_{\mathbf{1}_{5}}} \otimes \left. \mathcal{O}_{X_\Sigma} \left( 3, 9 \right) \right|_{C_{\mathbf{1}_{5}}}$ & $\left( 315 + A, A \right)$ \\
\bottomrule
\end{tabular}
\caption[Zero modes of broken GUT-model based on $W = dP_3$ for explicit flux choice.]{Zero modes of \emph{F-theory} GUT-model $( \hat{Y}_4, A (5, 0, 2.5, 0 )$ on $W \cong dP_3$. The breaking of the GUT-model is triggered by the hypercharge flux $\mathcal{H} = \mathcal{O}_{dP_3} ( 0, 1, 0, -1 )$.}
\label{table-N22}
\end{table}

As this is our first non-trivial example, let us pause a moment to elaborate on the computational resources required for this computation. First of all, in terms of the toric model of the $dP_3$-surface, the overall computation of the zero modes was completed within a few minutes. In contrast, if we perform the same computation on the base toric ambient space $X_\Sigma$, then our computations take significantly longer -- at best hours! The reason for this sharp increase in computational time is that the computations on $X_\Sigma$ take into account far more complex structure moduli, which in turn need to be processed by delicate Gröbner basis computations. We can phrase this finding in a short rule: smaller dimension is faster.\footnote{There is an obvious exception from this rule -- namely provided that the ambient space has additional structure, this additional datum can be used to simplify computations. For example, \cite{Anderson:2008ex} made use of flag varieties.} Consequently, this speed-up is a direct consequence of our conjectured isomorphism $\mathrm{dP}_3 \stackrel{\varphi}{\cong} W$.

Unfortunately, the singlet curve is not contained in $W$, so that we cannot use $\varphi$ to express this curve as a sublocus of a toric model of a $dP_3$-surface. This is the reason why we could not determine its massless spectrum. Thus, \cref{table-N22} is phrased in terms of the chiral index on the singlet curve and a parameter $A \in \mathbb{Z}$, which is determined by the complex structure moduli of the curve $C_{\mathbf{1}_{5}}$.\footnote{Let us mention that this constant is not fixed by the Kodaira vanishing theorem, as the line bundle in question has degree $364$ which is not greater or equal to $2 g ( C_{\mathbf{1}_{5}} ) - 1 = 2 \cdot 680 - 1 = 1359$.}
The spectra on $C_{\mathbf{10}_1}$ and $C_{\mathbf{5}_3}$ are minimal in that there exist no vector-like pairs. On $C_{\mathbf{10}_1}$ this is due to Kodaira's vanishing theorem (\cf \cref{subsec:LineBundlesOnRiemannSurfaces}) since
\[ \mathrm{deg} \left( \left. \mathcal{O}_{dP_3} \left( 2, 2, 1, 0 \right) \right|_{C_{\mathbf{10}_1}} \right) = 4 \geq 3 = 2 \cdot 2 - 1 = 2 g \left( C_{\mathbf{10}_1} \right) - 1 \, . \]
On $C_{\mathbf{5}_3}$ however, this is different since
\[ \mathrm{deg} \left( \left. \mathcal{O}_{dP_3} \left( 8, 2, 5, 6 \right) \right|_{C_{\mathbf{5}_3}} \right) = 44 \not \geq 47 = 2 \cdot 24 - 1 = 2 g \left( C_{\mathbf{5}_3} \right) - 1 \, . \]

We can repeat this very analysis for $A ( a,b,c, \lambda ) = A ( 0, 5, \frac{5}{2}, 0 )$. Although the divisors on the matter curves change, the massless spectrum is identical to the one of $A ( 5, 0, \frac{5}{2}, 0 )$. This holds true even if we repeat this computation with hypercharge flux $\mathcal{O}_{dP_3} ( 0, -1, 0, 1 )$.\footnote{Recall that we found in \cref{subsec:GaugeBackground} two possible choices, namely $\mathcal{H} = \pm ( E_1 - E_2 )$.} In this sense the massless spectrum is invariant under
\[ \mathcal{H} \leftrightarrow - \mathcal{H}, \qquad A (a,b,c, \lambda ) = A( 5,0,2.5,0) \leftrightarrow A(a,b,c, \lambda ) = A (0,5,2.5,0) \, . \]

\paragraph{Massless Spectra on non-generic Matter Curves}

So far the matter curves were taken maximally generic. As a consequence the matter curves were in particular smooth. However, let us now `tune' the matter curves and make quite special a choice as follows:
\[ \label{equ:non_generic_sections} \tilde{a_{1,0}} = 91 x_2 x_5^2, \qquad \tilde{a_{2,1}} = 11 x_1^2 x_4^5, \qquad \tilde{a_{3,2}} = 68 x_1^3 x_4^7, \qquad \tilde{a_{4,3}} = 67 x_2^4 x_5^5. \]
With these polynomials we repeat the computation of the zero modes for the flux $A ( a,b,c, \lambda ) = A ( 5, 0, 2.5, 0 )$ and $\mathcal{H} = \mathcal{O}_{dP_3} ( 0, 1, 0, -1 )$. This leads to the results summarised in \cref{table-N23}. Note that the sheaf cohomologies of the line bundles encoding the states $( \overline{\mathbf{3}}, \mathbf{1} )_{1_X, -4_Y}$ and $( \mathbf{1}, \mathbf{1} )_{1_X, 6_Y}$ changed quite drastically. This is not in contradiction with the Kodaira vanishing theorem or the Riemann-Roch theorem. Rather, the theorem of Riemann-Roch must be modified for singular curves, see \eg \cite{hartshorne1977algebraic}. For illustrative purposes let us repeat this analysis for yet another special choice of Tate polynomials, this time we take
\[
\label{equ:non_generic_sections2}
\begin{aligned}
\tilde{a_{1,0}} &= 88 x_1 x_2^2 + 91 x_1 x_5^2, \\
\tilde{a_{2,1}} &= 11 x_2^5 x_6^2 + 8 x_2^3 x_5^2 x_6^2, \\
\end{aligned}
\hspace{5em}
\begin{aligned}
\tilde{a_{3,2}} &= 68 x_2^7 x_6^3 + 14 x_2^5 x_5^2 x_6^3, \\
\tilde{a_{4,3}} &= 61 x_1^4 x_2^2 x_5^3 + 67 x_1^4 x_5^5. \\
\end{aligned}
\]
As a consequence the spectrum changes yet again. The result are given in \cref{table-N23}.

\begin{table}[tb]
\centering
\begin{tabular}{lllll}
\toprule 
& Rep. & relevant line bundle & \parbox{0.15\textwidth}{Zero modes for \cref{equ:non_generic_sections}} & \parbox{0.15\textwidth}{Zero modes for \cref{equ:non_generic_sections2}} \\
\midrule
$C_{\mathbf{10}_1}$ & $\left( \mathbf{3}, \mathbf{2} \right)_{1_X, 1_Y}$ &  $\left. \mathcal{O}_{dP_3} \left( 2, 2, 1, 0 \right) \right|_{C_{\mathbf{10}_1}}$ & $\left( 3,0 \right)$ & $\left( 3,0 \right)$ \\
& $\left( \overline{\mathbf{3}}, \mathbf{1} \right)_{1_X, -4_Y}$ & $\left. \mathcal{O}_{dP_3} \left( 2, -3, 1, 5 \right) \right|_{C_{\mathbf{10}_1}}$ & $\left( 9,6 \right)^*$ & $\left( 5,2 \right)$ \\
& $\left( \mathbf{1}, \mathbf{1} \right)_{1_X, 6_Y}$ & $\left. \mathcal{O}_{dP_3} \left( 2, 7, 1, -5 \right) \right|_{C_{\mathbf{10}_1}}$ & $\left( 16, 13 \right)^*$ & $\left( 7,4 \right)$ \\
\vspace{-0.5em} \\
$C_{\mathbf{5}_3}$ & $\left( \mathbf{3}, \mathbf{1} \right)_{3_X, -2_Y}$ & $\left. \mathcal{O}_{dP_3} \left( 8, 2, 5, 6 \right) \right|_{C_{\mathbf{5}_3}}$ & $ \left( 21, 0 \right)$ & $\left( 21,0 \right)$ \\
& $\left( \mathbf{1}, \mathbf{2} \right)_{3_X, 3_Y}$ & $\left. \mathcal{O}_{dP_3} \left( 8, 7, 5, 1 \right) \right|_{C_{\mathbf{5}_3}}$ & $\left( 21, 0 \right)^*$ & $\left( 21,0 \right)$ \\
\vspace{-0.5em} \\
$C_{\mathbf{5}_{-2}}$ & $\left( \mathbf{3}, \mathbf{1} \right)_{-2_X, -2_Y}$ & $\left. \mathcal{O}_{dP_3} \left( 4, 2, 2, 6 \right) \right|_{C_{\mathbf{5}_{-2}}}$ & $\left( 9, 33 \right)$ & $\left( 9,33 \right)$ \\
& $\left( \mathbf{1}, \mathbf{2} \right)_{-2_X, 3_Y}$ & $\left. \mathcal{O}_{dP_3} \left( 4, 7, 2, 1 \right) \right|_{C_{\mathbf{5}_{-2}}}$ & $\left( 6, 30 \right)$ & $\left( 6,30 \right)$ \\
\vspace{-0.5em} \\
$C_{\mathbf{1}_{5}}$ & $\left( \mathbf{1}, \mathbf{1} \right)_{5_X, 0_Y}$ & $\left. \mathcal{O}_{\mathcal{B}_6} \left( 5 G_1 \right) \right|_{C_{\mathbf{1}_{5}}} \otimes \left. \mathcal{O}_{X_\Sigma} \left( 3, 9 \right) \right|_{C_{\mathbf{1}_{5}}}$ & $\left( 315 + A, A \right)$ & $\left( 315 + A, A \right)$ \\
\bottomrule
\end{tabular}
\caption[Zero modes of broken GUT-model ($W = dP_3$) for non-generic complex structure.]{Zero modes of \emph{F-theory} GUT-model $( \hat{Y}_4, A (5, 0, 2.5, 0 )$ on $W \cong dP_3$ for matter curves modelled by non-generic Tate-sections \cref{equ:non_generic_sections}, \cref{equ:non_generic_sections2}.}
\label{table-N23}
\end{table}

\section{An \emph{F-Theory} GUT-Model on a \texorpdfstring{$\mathbf{dP_7}$}{dP7}-Surface} \label{sec:dP7-Example}

This geometry is taken from \cite{Braun:2014pva} and ensures $W \cong dP_7$. In contrast to a $dP_3$-surface, a $dP_7$-surface cannot be realised as a toric variety. This seemingly minor change will prove to limit our computational powers quite significantly.

The interested reader might wish to compare our findings to the original results of \cite{Braun:2014pva}. In particular, this applies to the defining data of the toric varieties involved. As explained in \cref{subsec:TowardsToricVarieties}, in the lingo of string phenomenology the latter is achieved by specifying both the Cox ring (including its grading) and the irrelevant ideal. To simplify this write-up, we the way \texttt{gap} uses it. Unfortunately, \texttt{gap} \cite{GAP4} permutes the input data of a toric variety. Hence, comparison with the original literature requires a permutation of the homogeneous variables.

\subsection{The Geometry of \texorpdfstring{$\mathbf{\hat{Y}_4}$}{Y4}} \label{subsec:GeometryOfY4FordP7}

\paragraph{\emph{F-Theory} Base} 

The base space $\mathcal{B}_6$ will be modelled as hypersurface in a smooth and projective (normal) toric variety $X_\Sigma$ whose Cox ring $S ( X_\Sigma ) = \mathbb{Q} [ x_1, \dots, x_6 ]$ is graded under $\mathrm{Cl} ( X_\Sigma ) = \mathbb{Z}^2$ according to \cref{GradingOfCoxringOfBasespacedP7} and has $I_{\mathrm{SR}} ( X_\Sigma ) = \langle x_1 x_2 x_6, x_3 x_4 x_5 \rangle \subseteq S$. It holds $\mathrm{Nef} ( X_\Sigma ) = \mathrm{Cone} ( V ( x_1 ), V ( x_3 ) )$. The \emph{F-theory} base $\mathcal{B}_6$ is given as $V( P ) \subseteq X_\Sigma$ where $P \in H^0 ( X_\Sigma, \mathcal{O}_{X_\Sigma} ( 5,2 ) )$. The most general such polynomial $P$ consists of 65 monomials. In analogy to \cite{Braun:2014pva} we make a special choice by setting ($c_i \in \mathbb{Q}$)
\begin{align}
\begin{split}
P =& x_1 x_2 x_5 \left( c_1 x_1 x_5 + c_2 x_2 x_5 + c_3 x_6 x_5 + c_4 x_3 \right) \\
   &+ x_6 x_3 \left( c_5 x_1 x_5 + c_6 x_2 x_5 + c_7 x_6 x_5 + c_8 x_3 \right) + x_4^2 R + x_4 x_5 T + x_4 x_3 U \\
  =& x_5^2 \left( c_1 x_1^2 x_2 + c_2 x_1 x_2^2 + c_3 x_1 x_2 x_6 \right) + x_5 x_3 \left( c_4 x_1 x_2 + c_5 x_1 x_6 + c_6 x_2 x_6 + c_7 x_6^2 \right) \\
   & + c_8 x_6 x_3^2 + x_4^2 R + x_4 x_5 T + x_4 x_3 U
\label{equ:most_general_P_second_model}
\end{split}
\end{align}
where $R \in H^0 ( X_\Sigma, \mathcal{O}_{X_\Sigma} ( 5, 0 ) )$, $T \in H^0 ( X_\Sigma, \mathcal{O}_{X_\Sigma} ( 4, 0 ) )$ and $U \in H^0 ( X_\Sigma, \mathcal{O}_{X_\Sigma} ( 3, 0 ) )$. The parameters $c_i \in \mathbb{Q} \backslash \{ 0 \}$ determine the complex structure of the base $\mathcal{B}_6$. Just as in \cref{subsec:GeometryOfY4FordP3}, we impose a number of conditions on these parameters to make some analytic cycles appear as algebraic cycles and to simplify the following analysis. In particular this choice of complex structure enables us to identify algebraic cycles whose associated line bundles furnish a generating set of $\mathrm{Pic} ( W )$. This is possible if we restrict the complex structure by the demand
\begin{align}
\begin{split}
c_2 &= \frac{\left( c_3 + c_7 \right) \left( c_4 + c_8 \right)}{2 \left( c_1 + c_5 \right)} - \frac{\left( c_3 - c_7 \right) \left( c_4 - c_8 \right)}{2 \left(  
c_1 - c_5 \right)} \, , \\
c_6 &= \frac{\left( c_3 + c_7 \right) \left( c_4 + c_8 \right)}{2 \left( c_1 + c_5 \right)} + \frac{\left( c_3 - c_7 \right) \left( c_4 - c_8 \right)}{2 \left(  c_1 - c_5 \right)} \, ,
\end{split}
\label{equ:crucial_assumption_in_dP7_example}
\end{align}
enforce $\sqrt{c_3^2 c_4^2 + \left( c_2 c_5 - c_1 c_6 \right)^2 - 2 c_3 c_4 \left( c_2 c_5 + c_1 c_6 \right) + 4 c_1 c_2 c_4 c_7} \in \mathbb{Q}$ and require
\[
\label{equ:ConditionsOnModuli}
\begin{aligned}
\alpha, \beta, \mu, \nu, \delta &\neq 0, \\
c_4 \pm c_8 &\neq 0, \\
c_2 \pm c_6 &\neq 0, \\
c_1 \pm c_5 &\neq 0, \\
c_4 c_6 - c_2 c_8 &\neq 0, \\
c_2 c_7 - c_3 c_6 &\neq 0, \\
c_1 + c_4 \pm \left( c_5 + c_8 \right) &\neq 0, \\
\nu \left( c_2 c_8 - c_4 c_6 \right) + \mu \left( c_4 c_7 - c_3 c_8 \right) &\neq 0, \\
\beta \left( c_4 c_5 - c_1 c_8 \right) + \left( c_4 c_7 - c_3 c_8 \right) &\neq 0, \\
\nu \left( c_3 c_7 - c_2 c_6 \right) + \delta \left( c_2 c_5 - c_1 c_7 \right) &\neq 0,
\end{aligned}
\qquad
\begin{aligned}
c_4 c_7 - c_3 c_8 &\neq 0, \\
c_4 c_5 - c_1 c_8 &\neq 0, \\
c_1 c_7 - c_3 c_5 &\neq 0, \\
c_1 c_5 - c_4 c_8 &\neq 0, \\
c_1 c_6 - c_2 c_5 &\neq 0, \\
c_2 + c_4 \pm \left( c_6 + c_8 \right) &\neq 0, \\
\alpha \left( c_4 c_6 - c_2 c_8 \right) + \left( c_4 c_7 - c_3 c_8 \right) &\neq 0, \\
\alpha \left( c_1 c_6 - c_2 c_5 \right) + \left( c_1 c_7 - c_3 c_5 \right) &\neq 0, \\
\beta \left( c_2 c_5 - c_1 c_6 \right) + \left( c_2 c_7 - c_3 c_6 \right) &\neq 0, \\
\mu \left( c_3 c_5 - c_1 c_7 \right) + \nu \left( c_1 c_6 - c_2 c_5 \right) &\neq 0.
\end{aligned}
\]
In these expression we have made use of the combinations
\[ \alpha = \frac{c_1 - c_5}{c_4 - c_8}, \quad \beta = \frac{c_2 - c_6}{c_4 - c_8}, \quad \mu = c_4 + c_8, \quad \nu = c_1 + c_5, \quad \delta = c_3 + c_7 \, . \]
\begin{table}[tbp]
\centering
\begin{tabular}{cccccc}
\toprule
$x_1$ & $x_2$ & $x_3$ & $x_4$ & $x_5$ & $x_6$ \\
\midrule
1 & 1 & 2 & 0 & 1 & 1 \\
0 & 0 & 1 & 1 & 1 & 0 \\
\bottomrule
\end{tabular}
\caption[Toric data of base ambient space of \emph{F-theory} GUT-model with $W \cong dP_7$.]{The base ambient space $X_\Sigma$ has Cox ring $\mathbb{Q}[ x_1, x_2, x_3, x_4, x_5, x_6]$ which is $\mathbb{Z}^2$-graded as described above. The Stanley-Reisner ideal is $I_{\mathrm{SR}} ( X_\Sigma ) = \langle x_1 x_2 x_6, x_3 x_4 x_5 \rangle$.}
\label{GradingOfCoxringOfBasespacedP7}
\end{table}
A possible such choice is $c_i = ( 26, \frac{1968}{7}, 90, 47, 16, \frac{1429}{7}, 68, 82 )$. In contrast to these rather strict conditions, we do not enforce any demands on the coefficients of the polynomials $R$, $T$ and $U$. This is because these polynomials vanish on the GUT-surface $W$ and are therefore of no importance to the structure of $W$. We merely use them to ensure that $\mathcal{B}_6$ is smooth. \emph{Sage} \cite{sage} indeed confirmed this smoothness for pseudo-randomly chosen rational coefficients of $R$, $T$ and $U$. In addition, the implementations \cite{CAPCategoryOfProjectiveGradedModules, CAPPresentationCategory, PresentationsByProjectiveGradedModules, TruncationsOfPresentationsByProjectiveGradedModules, SheafCohomologyOnToricVarieties} found $h^0 ( \mathcal{B}_6, \mathcal{O}_{\mathcal{B}_6} ) = 1$, which shows that $\mathcal{B}_6$ is connected. The Hodge diamond of $\mathcal{B}_6$ is computed from \emph{cohomCalg} \cite{Blumenhagen:2010pv, Blumenhagen:2010ed, Blumenhagen:2011xn, KoszulExtensionManual, cohomCalg:Implementation, 2011JMP....52c3506J, Rahn:2010fm} and reads
\[ \label{equ:HodgeDP7} \begin{array}{ccccccc}
& & & 1 \\
& & 0 & & 0 \\
& 0 & & 2 & & 0 \\
0 & & 27 & & 27 & & 0 \\
& 0 & & 2 & & 0 \\
& & 0 & & 0 \\
& & & 1 
\end{array} \]
As for the $dP_3$-example, we can apply \cite[theorem 1.6]{Batyrev:2005jc} to analyse the fundamental group $\pi_1 ( \mathcal{B}_6 )$. Again we find
$\pi_1 \left( \mathcal{B}_6 \right) \cong N / \text{Span}_{\mathbb{Z}} \left\{ u_\rho \, \left| \, \rho \in \Sigma \left( 1 \right) \right. \right\} \cong 0$,
so the fundamental group vanishes, its Abelianization $h_1 ( \mathcal{B}_6, \mathbb{Z} )$ also and the Picard group $\mathrm{Pic} ( \mathcal{B}_6 )$ has no torsion. This implies that all conditions phrased in \cref{subsec:ConditionsOnBaseAndFibration} are satisfied, and $\mathcal{B}_6$ is a bona fide \emph{F-theory} base space. Moreover it follows $\mathrm{Pic} ( \mathcal{B}_6 ) \cong \mathbb{Z}^2$ and we will eventually argue that the divisor classes $G_1 = V ( x_1, P )$ and $G_2 = V ( x_5, P )$ are not linearly equivalent. Therefore, we can employ them as free generators of $\mathrm{Pic} ( \mathcal{B}_6 )$. For the time being suffice it to note that the triple intersection numbers of these divisor classes satisfy
\[ G_1^3 = 0 \, , \qquad G_1^2 G_2 = 2 \, , \qquad G_1 G_2^2 = 3 \, , \qquad G_2^3 = 2 \, . \label{equ:TripleIntersectionsBasedP7} \]

\paragraph{\emph{F-Theory} GUT Surface} 
The \emph{F-theory} GUT surface is given by $W = V ( P, x_4 ) \subseteq \mathcal{B}_6$. As noted in the original literature \cite{Braun:2014pva} it happens to be a $dP_7$-surface. To find a generating free of $\mathrm{Pic} ( W ) \cong \mathbb{Z}^8$, let us first define the following expressions:
\[
\begin{tabular}{>{\centering}m{0.14\textwidth} >{\centering}m{0.13\textwidth} >{\centering}m{0.22\textwidth} >{\centering}m{0.13\textwidth} >{\centering}m{0.14\textwidth}}
\multicolumn{2}{c}{$\tilde{A} = c_1 x_1 + c_2 x_2 + c_6 x_6$,} & $C = c_4 x_1 x_2 x_5 + x_6 B$, & \multicolumn{2}{c}{$\tilde{B} = c_5 x_1 + c_6 x_2 + c_7 x_6$,} \\
\\[-0.5em]
\multicolumn{2}{c}{$A = x_5 \tilde{A} + c_4 x_3$,} & & \multicolumn{2}{c}{$B = x_5 \tilde{B} + c_8 x_3$,} \\
\\[-0.5em]
$\alpha = \frac{c_1 - c_5}{c_4 - c_8}$, & $\beta = \frac{c_2 - c_6}{c_4 - c_8}$, & $\delta = c_3 + c_7$, & $\nu = c_1 + c_5$, & $\mu = c_4 + c_8$.
\end{tabular}
\]
We use them to define the following divisors $D_i \in \mathrm{Cl} ( W )$ in $W$:
\[
\begin{array}{lcl}
D_1 = V \left( x_4, x_1, x_6 \right) \, , & \phantom{\hspace{3em}} & D_2 = V \left( x_4, x_1, x_3 \right) \, , \\
D_3 = V \left( x_4, x_1, B \right) \, ,   & & D_4 = V \left( x_4, x_2, x_6 \right) \, , \\
D_5 = V \left( x_4, x_2, x_3 \right) \, , & & D_6 = V \left( x_4, x_2, B \right) \, , \\
D_7 = V \left( x_4, x_5, x_6 \right) \, , & & D_8 = V \left( x_4, A, x_6 \right) \, , \\
D_9 = V ( x_4, \tilde{A}, x_3 ) \, , & & D_{10} = V \left( x_4, A, B \right) \, , \\
D_{11} = V ( x_4, \tilde{A}, C ) \, , & & D_{12} = V \left( x_4, A - B, x_2 - \alpha x_6 \right) \, , \\
D_{13} = V \left( x_4, A- B, x_1 - \beta x_6 \right) \, ,   & & D_{14} = V \left( x_4, A + B, \nu x_1 + \delta x_6 \right) \, , \\
D_{15} = V \left( x_4, A + B, \mu x_2 + \nu x_6 \right) \, . &
\end{array}
\label{equ:DivisorsIndP7}
\]
Note that the last four divisors can be defined ony upon use of \cref{equ:crucial_assumption_in_dP7_example}.

It can be verified that for pseudo-random coefficients $c_i$ subject to the conditions stated around \cref{equ:ConditionsOnModuli}, both $W$ and all of the above divisors $D_i$ are smooth. By use of the theorem of Riemann-Roch for (smooth, compact, connected) Riemann surfaces (\cf \cref{subsec:LineBundlesOnRiemannSurfaces}) and the \texttt{gap}-package \cite{SheafCohomologyOnToricVarieties} we then identify $g ( D_{i} ) = 0$, so the above divisors are all $\mathbb{P}_{\mathbb{Q}}^1$s. The intersection numbers of these divisors follow from $D_i D_j = \mathrm{deg} ( \left. \mathcal{O}_{W} ( D_i ) \right|_{D_j} )$ \cite{ATIT,cox2011toric}) which can be evaluated with the \texttt{gap}-package \cite{SheafCohomologyOnToricVarieties}. This leads to the intersection numbers summarised in \cref{table-N29}. From these intersection numbers it can be seen that these divisor satisfy precisely 7 relations, which leaves us with the following 8 generators of $\mathrm{Pic} ( W )$:
\[ \label{equ:HyperplaneClassAndExceptionalsdP7}
\begin{array}{lcl}
H = D_1 + D_7 + D_{13} + D_{14} \, , & \phantom{\hspace{2em}} & E_1 = D_{14} \, , \\
E_2 = D_{13} \, , & & E_3 = D_4 + D_5 - D_2 - 2 D_3 \\
                  & & \phantom{E_3 = D_4 + D_5 - D_2} + D_7 + D_{13} + D_{14}  \, , \\
E_4 = D_{13} - D_2 - D_3 + D_5 + D_{14} \, , & & E_5 = D_1 + D_3 - D_5 + D_7 \, , \\
E_6 = D_1 + D_2 + D_3 - D_4 - D_5 \, , & & E_7 = D_7 \, .
\end{array}
\]
The divisor classes $H$ and $E_i$ intersect according to $H^2 = 1$, $ HE_i = 0$ and $E_i E_j = - \delta_{ij}$ which is the canonical intersection form of a $dP_7$-surface. Even more we find
\[ \overline{K}_W = \left. G_1 \right|_{W} = D_1 + D_2 + D_3 = 3 H - \sum_{i = 1}^{7}{E_i} \]
and it is readily confirmed from \cref{table-N29} that $K_{W}^2 = 2$. All these results indicate $W \cong dP_7$. Finally, $\left. G_2 \right|_{W} = D_7 = E_7$ leads to two useful conclusions:
\begin{enumerate}
 \item $G_1$ and $G_2$ are \emph{free} generators of $\mathrm{Pic} \left( \mathcal{B}_6 \right)$: if they satisfied a relation on 
      $\mathcal{B}_6$, then so would $\left. G_1 \right|_{W}$ and $\left. G_2 \right|_{W}$ on $W$. However the above shows that the latter is impossible.
 \item It holds $\mathrm{Pic} ( W ) = \mathrm{Span}_{\mathbb{Z}} \left\{ H, E_1, E_2, E_3, E_5, E_6, K_{W}, E_7 \right\}$. Hence, we conclude that 
      $\mathrm{Pic} \left( W \right) - \iota^\ast \mathrm{Pic} \left( \mathcal{B}_6 \right) = \mathrm{Span}_{\mathbb{Z}} \left( H, E_1, E_2, E_3, E_5, E_6 \right)$, which will be quite useful to discuss the choice of hypercharge flux in \cref{sec:GaugeBackgroundsdP7}.
\end{enumerate}

\begin{table}[tbp]
\resizebox{\textwidth}{!}{
\begin{tabular}{c@{\hskip 20pt}ccccc@{\hskip 20pt}ccccc@{\hskip 20pt}ccccc}
\toprule
      & $D_1$ & $D_2$ & $D_3$ & $D_4$ & $D_5$ & $D_6$ & $D_7$ & $D_8$ & $D_9$ & $D_{10}$ & $D_{11}$ & $D_{12}$ & $D_{13}$ & $D_{14}$ & $D_{15}$ \\
\midrule
$D_1$    & -2 & 1  & 1  & $\cdot$ & $\cdot$ &  $\cdot$ & 1  & 1  & $\cdot$ & $\cdot$ & $\cdot$ & $\cdot$ & 1 & 1 & $\cdot$ \\ 
$D_2$    &    & -1 & 1  & $\cdot$ & 1  &  $\cdot$ & $\cdot$ & $\cdot$ & 1  & $\cdot$ & $\cdot$ & 1 & $\cdot$ & $\cdot$ & 1 \\
$D_3$    &    &    & -1 & $\cdot$ & $\cdot$ &  1 & $\cdot$ & $\cdot$ & $\cdot$ & 1 & 1 & $\cdot$ & $\cdot$ & $\cdot$ & $\cdot$ \\
$D_4$    &    &    &    & -2 & 1  &  1 & 1  & 1  & $\cdot$ & $\cdot$ & $\cdot$ & 1 & $\cdot$ & $\cdot$ & 1 \\
$D_5$    &    &    &    &    & -1 &  1 & $\cdot$ & $\cdot$ & 1 & $\cdot$ & $\cdot$ & $\cdot$ & 1 & 1 & $\cdot$ \\
\midrule
$D_6$    &    &    &    &    &    & -1 & $\cdot$ & $\cdot$ & $\cdot$ & 1 & 1 & $\cdot$ & $\cdot$ & $\cdot$ & $\cdot$ \\
$D_7$    &    &    &    &    &    &    & -1 & $\cdot$ & $\cdot$ & $\cdot$ & 1 & $\cdot$ & $\cdot$ & $\cdot$ & $\cdot$ \\
$D_8$    &    &    &    &    &    &    &    & -1 & 1 & 1 & $\cdot$ & $\cdot$ & $\cdot$ & $\cdot$ & $\cdot$ \\ 
$D_9$    &    &    &    &    &    &    &    &    & -1 & 1 & 2 & $\cdot$ & $\cdot$ & $\cdot$ & $\cdot$ \\
$D_{10}$ &    &    &    &    &    &    &    &    &    & -1 & $\cdot$ & 1 & 1 & 1 & 1 \\
\midrule
$D_{11}$ &    &    &    &    &    &    &    &    &    &    & -1 & 1 & 1 & 1 & 1 \\
$D_{12}$ &    &    &    &    &    &    &    &    &    &    &    & -1 & 1 & $\cdot$ & $\cdot$ \\
$D_{13}$ &    &    &    &    &    &    &    &    &    &    &    &    & -1 & $\cdot$ & $\cdot$ \\
$D_{14}$ &    &    &    &    &    &    &    &    &    &    &    &    &    & -1 & 1 \\
$D_{15}$ &    &    &    &    &    &    &    &    &    &    &    &    &    &    & -1 \\
\bottomrule
\end{tabular}}
\caption{Intersection numbers of the 15 divisors $D_i$ in W.}
\label{table-N29}
\end{table}

\paragraph{Location of Blow-Up Points of \texorpdfstring{$\mathbf{W \cong dP_7}$}{W=dP7} via Line Bundle Cohomology}

As mentioned already, a $dP_7$-surface cannot be realised as a toric variety. With an eye towards zero-mode-counting, this will significantly limit our computational powers. As discussed in \cref{sec:ChoiceOfHyperchargeFluxAndExotics} already, in \cite{Blumenhagen:2008zz} it was pointed out that there is a combinatorial approach to compute line bundle cohomology on a $dP_n$-surface, provided that the locations of the blow-up points are known. To improve the situation, let us therefore try to locate the blow up points by computing line bundle cohomologies on $W$ with \cite{SheafCohomologyOnToricVarieties}.

Let us briefly recall the approach from \cite{Blumenhagen:2008zz}: We look at the Cox ring $S = \mathbb{Q} [ x_1, x_2, x_3 ]$ of $\mathbb{P}^2_{\mathbb{Q}}$. For $c_j, a \in \mathbb{N}_{\geq 0}$, we denote by $A_{\sum{c_j p_j}} ( a )$ the dimension of the vector space of homogeneous polynomials of degree $a$ in $S$ which vanish to order $c_j$ at the blow-up point $p_j$. Now consider a line bundle of the form
\[ L = \mathcal{O}_{dP_n} \left( a \cdot H + \sum_{i \in I}{b_i E_i} - \sum_{j \in J}{c_j E_j} \right) \]
where $a \geq -2$, $I \cap J = \emptyset$, $I \cup J = \{ 1, 2, \dots, 7 \}$ and $b_i, c_j \in \mathbb{N}_{\geq 0}$ for all $i, j$. Then it holds
\[ h^i \left( dP_7, L \right) = \left( A_{\sum{c_i p_i}} \left( a \right), - \chi \left( L \right) + A_{\sum{c_i p_i}} \left( a \right), 0 \right) \]
where the Euler-characteristics $\chi ( L )$ is the Euler characteristic of this line bundle. Consequently, $h^0 ( W, L )$ is sensitive to the location of the blow-up points via the combinatorics of $A_{\sum{c_i p_i}} ( a )$.

By choosing coordinates for $\mathbb{P}^2_{\mathbb{Q}}$ appropriately, four of the blow-up points can always be located at $( 1 : 0 : 0 )$, $( 0 : 1 : 0 )$, $( 0 : 0 : 1 )$ and $( 1 : 1 : 1 )$. Therefore, we make the following ansatz for the location of the blow-up points:
\[ \begin{array}{llll}
p_1 = \left( 1 : 0 : 0 \right) \, , & p_2 = \left( 0 : 1 : 0 \right) \, , & p_3 = \left( 0 : 0 : 1 \right) \, , & p_4 = \left( 1 : 1 : 1 \right) \, , \\
p_5 = \left( A : B : C \right) \, , & p_6 = \left( D : E : F \right) \, , & p_7 = \left( G : H : I \right) \, . \label{equ:Ansatz}
\end{array} \]
To put constraints on the blow-up points, we first look at line bundles associated to the divisors of the form $H - E_i - E_j$ with $i \neq j$. By explicit computation with \cite{SheafCohomologyOnToricVarieties}, we found that all these line bundles satisfy $h^0 = 1$. By comparing this result to the combinatorics of $A_{\sum{c_i p_i}} ( a )$ it can be seen that for all $\lambda \in \mathbb{Q} - \left\{ 0 \right\}$ and any two distinct blow-up points $p_i, p_j$ it holds $p_i \neq \lambda p_j$.

As a next step we look at the line bundles associated to the divisors $H - E_i - E_j - E_k$ for distinct $i,j,k$. In two cases, namely $H - E_1 - E_3 - E_4$ and $H - E_2 - E_3 - E_4$, our algorithms were insufficient to determine the sheaf cohomologies. For almost all other line bundles we found that $h^0$ vanishes. The only exceptions are $H - E_1 - E_2 - E_7$ and $H - E_4 - E_6 - E_7$. In both cases it holds $h^0 = 1$! This puts non-trivial constraints on the location of the blow-up points. 

For example, the result for $D = H - E_1 - E_2 - E_7$ implies $I = 0$. Let us explain this finding in more detail: To identify $A_{\sum{c_i p_i} ( a )}$ we look at the vector space
\begin{align}
V &= \left\{ q = \alpha_1 x_1 + \alpha_2 x_2 + \alpha_3 x_3 \; \left| \; q \left( p_1 \right) = q \left( p_2 \right) = q \left( p_7 \right) = 0 \right. \right\} \\
&= \left\{ q = \alpha_3 x_3 \; \left| \; \alpha_3 \cdot I = 0 \right. \right\} \, . 
\end{align}
Our computation shows that this must be a vector space of dimension one. This in turn is the case precisely if $I = 0$. Similarly, the result on $D = H - E_4 - E_6 - E_7$ implies $D = \frac{E G  - F G}{H} + F$. Overall this constraints the original ansatz in \cref{equ:Ansatz} to take the shape ($H \neq 0$)
\[
\begin{array}{llll}
p_1 = \left( 1 : 0 : 0 \right) \, , & p_2 = \left( 0 : 1 : 0 \right) \, , & p_3 = \left( 0 : 0 : 1 \right) \, , & p_4 = \left( 1 : 1 : 1 \right) \, , \\
p_5 = \left( A : B : C \right) \, , & p_6 = ( F + \frac{E G -F G}{H} : E : F ) \, , & p_7 = \left( G : H : 0 \right) \, .
\end{array} \label{equ:RestrictedBlowupPoints}
\]

Unfortunately, with the currently available computational resources, we could not find any further constraints on the location of the blow-up points. Nonetheless, we see that it is indeed possible to narrow their location merely by looking at line bundle cohomology. In particular we remain optimistic that further improvements on the algorithms will improve the situation. For example, the results of \cite{StillmanStringPheno} which are  implemented in \cite{M2}, can be expected to improve this very analysis.

\paragraph{The Elliptically Fibred 4-Fold \texorpdfstring{$\mathbf{\hat{Y}_4}$}{Y4} and the Matter Curves}

\begin{table}[tb]
\centering
\begin{tabular}{cccccc@{\hskip 20pt}cccc@{\hskip 20pt}cccc}
\toprule
$x_1$ & $x_2$ & $x_3$ & $e_0$ & $x_5$ & $x_6$ & $e_1$ & $e_2$ & $e_3$ & $e_4$ & x & y & z & s \\
\midrule
1 & 1 & 2 & 0 & 1 & 1 & 0 & 0 & 0 & 0 & 2 & 3 & 0 & 0 \\
0 & 0 & 1 & 1 & 1 & 0 & 0 & 0 & 0 & 0 & 2 & 3 & 0 & 0 \\
\vspace{0.5em} & \\
0 & 0 &-1 & 0 & 0 & 0 & 1 & 0 & 0 & 0 &-1 &-1 & 0 & 0 \\
0 & 0 &-1 & 0 & 0 & 0 & 0 & 1 & 0 & 0 &-2 &-2 & 0 & 0 \\
0 & 0 &-1 & 0 & 0 & 0 & 0 & 0 & 1 & 0 &-2 &-3 & 0 & 0 \\
0 & 0 &-1 & 0 & 0 & 0 & 0 & 0 & 0 & 1 &-1 &-2 & 0 & 0 \\
\vspace{-0.5em} \\
0 & 0 & 0 & 0 & 0 & 0 & 0 & 0 & 0 & 0 &-1 &-1 & 0 & 1 \\
0 & 0 & 0 & 0 & 0 & 0 & 0 & 0 & 0 & 0 & 2 & 3 & 1 & 0 \\
\bottomrule
\end{tabular}
\caption[Toric data of $\hat{Y}_\Sigma$ in \emph{F-theory} GUT-model with $W \cong dP_7$.]{The Cox ring $\mathbb{Q} [ x_1, x_2, x_3, e_0, x_5, x_6, e_1, e_2, e_3, e_4, x, y, z, s]$ of the toric space $\hat{Y}_\Sigma$ has $\mathbb{Z}^8$-grading. Together with \cref{SRIdealYSigmadP7} this defines the geometry of $\hat{Y}_\Sigma$.}
\label{GradingOfCoxRingOfFibreAmbientSpacedP7Example}
\end{table}

We follow the same philosophy as in \cref{subsec:GeometryOfY4FordP3}. Here the Cox ring $S = \mathbb{Q} [ x_1, x_2, x_3, e_0, x_5, x_6, e_1, e_2, e_3, e_4, x, y, z, s]$ of $Y_\Sigma$ is graded by $\mathbb{Z}^8$ according to \cref{GradingOfCoxRingOfFibreAmbientSpacedP7Example} and its Stanley-Reisner ideal is given by
\begin{align}
\begin{split}
\label{SRIdealYSigmadP7}
I_{\mathrm{SR}} ( \hat{Y}_\Sigma ) =& \left\langle e_0 e_2, e_0 e_3, e_0 s, e_1 e_3, e_1 y, e_1 z, e_1 s, e_4 x, e_4 z, e_4 s, x y, e_2 y, e_2 z, e_3 z, \right. \\
& \hspace{8em} \left. z s, e_2 s, x_1 x_2 x_6, x_3 e_0 x_5, x_3 x_5 e_1, x_3 x_5 e_4, x_3 x_5 e_2, x_3 x_5 e_3 \right\rangle \, .
\end{split}
\end{align}
$\hat{Y}_4 = V( P_T^\prime )$ is smooth for generic Tate sections. With \emph{cohomCalg} \cite{Blumenhagen:2010pv, Blumenhagen:2010ed, Blumenhagen:2011xn, KoszulExtensionManual, cohomCalg:Implementation, 2011JMP....52c3506J, Rahn:2010fm} it can be verified that $h^{i,0} ( \hat{Y}_4 ) = 0$. Therefore, $\hat{Y}_4$ is a strict Calabi--Yau variety in the sense introduced in \cref{sec:StringTheory}. The Tate sections $a_i \in H^0 ( \mathcal{B}_6, \overline{K}_{\mathcal{B}_6}^{\otimes i} \otimes \mathcal{O}_{\mathcal{B}_6} ( - j W ) )$ are now given by
\begin{align*}
  a_{1,0} &\in H^0 \left( \mathcal{B}_6, \left. \mathcal{O}_{X_\Sigma} \left( 1, 1 \right) \right|_{\mathcal{B}_6} \right) \, , \quad & a_{2,1} &\in H^0 \left( \mathcal{B}_6, \left. \mathcal{O}_{X_\Sigma} \left( 2, 1 \right) \right|_{\mathcal{B}_6} \right) \, , \\
  a_{3,2} &\in H^0 \left( \mathcal{B}_6, \left. \mathcal{O}_{X_\Sigma} \left( 3, 1 \right) \right|_{\mathcal{B}_6} \right) \, , & a_{4,3} &\in H^0 \left( \mathcal{B}_6, \left. \mathcal{O}_{X_\Sigma} \left( 4, 1 \right) \right|_{\mathcal{B}_6} \right) \, .
\end{align*}
Upon use of $H^0 ( X_\Sigma, \mathcal{O}_{X_\Sigma} ( i, j ) ) \to H^0 ( \mathcal{B}_6, \left. \mathcal{O}_{X_\Sigma} ( i, j ) \right|_{\mathcal{B}_6} )$ we model these sections from polynomials $\tilde{a_{i,j}}$ on $X_\Sigma$. In consequence, the matter curves are given by
\begin{align*}
  C_{\mathbf{10}_1} &= V \left( P, x_4, \tilde{a_{1,0}} \right) \, , \quad & C_{\mathbf{5}_3} &= V \left( P, x_4, \tilde{a_{3,2}} \right) \, , \\
  C_{\mathbf{5}_{-2}} &= V \left( P, x_4, \tilde{a_1} \tilde{a_{4,3}} - \tilde{a_{2,1}} \tilde{a_{3,2}} \right) \, , & C_{\mathbf{1}_{5}} &= V \left( P, \tilde{a_{4,3}}, \tilde{a_{3,2}} \right) \, .
\end{align*}
They happen to be curves of genus $0$, $4$, $10$ and $48$, respectively.

\begin{table}[tbp]
\centering
\begin{tabular}{ccccc}
\toprule
$x_1$ & $x_2$ & $x_3$ & $x_5$ & $x_6$ \\
\midrule
1 & 1 & 2 & 1 & 1 \\
0 & 0 & 1 & 1 & 0 \\
\bottomrule
\end{tabular}
\label{table-tildeX}
\caption[The grading of the Cox ring of the toric space $\tilde{X}_{\Sigma}$.]{The Cox ring $\mathbb{Q}[ x_1, x_2, x_3, x_5, x_6]$ of the toric space $\tilde{X}_{\Sigma}$ is graded by $\mathbb{Z}^2$ as indicated above. Its Stanley-Reisner ideal is given by $I_{\mathrm{SR}} ( \tilde{X}_\Sigma ) = \langle x_1 x_2 x_6, x_3 x_5 \rangle \subseteq S$.}
\end{table}

The analysis of \cref{sec:dP3-Example} indicates that it is advantageous to perform the sheaf cohomology computations on a toric variety of small dimension. In the case at hand, we can not express $W$ as a toric variety directly, but as the hypersurface $V( P^\prime ) \subseteq \tilde{X}_\Sigma$, where $P^\prime = \left. P \right|_{x_4 = 0}$ and the toric data of $\tilde{X}_\Sigma$ is summarised in \cref{table-tildeX}. In particular the matter curves are now given as
\[ C_{\mathbf{10}_1} \cong V \left( P^\prime, \tilde{\tilde{a_{1,0}}} \right) \, , \quad C_{\mathbf{5}_3} \cong V \left( P^\prime, \tilde{\tilde{a_{3,2}}} \right) \, , \quad C_{\mathbf{5}_{-2}} \cong V \left( P^\prime, \tilde{\tilde{a_1}} \tilde{\tilde{a_{4,3}}} - \tilde{\tilde{a_{2,1}}} \tilde{\tilde{a_{3,2}}} \right) \]
where $\tilde{\tilde{a_{ij}}} = \left. \tilde{a_{ij}} \right|_{x_4 = 0}$. These `reduced' polynomials merely depend on $30$ complex structure moduli, which drastically simplifies the involved Gröbner basis computations. This should be contrasted to the analysis in \cref{sec:dP3-Example} where the matter curves depended on 788 complex structure moduli from the perspective of the base ambient space $X_\Sigma$. Even after restriction to the toric $dP_3$-surface, 148 moduli remained.

Let us use this opportunity to point out yet another advantage of this $dP_7$-model over the $dP_3$-example studied previously. Namely $\mathrm{Pic} ( \mathcal{B}_6 )$ can in the current geometry be obtained completely via pullback from $X_\Sigma$. This again simplifies the computation of the massless spectrum. Nonetheless, the major drawback is that the GUT-surface $W$ can only be realised as hypersurface. As we will discuss momentarily, this effect dominates in the current geometry. It leads to severe restrictions on our computational powers..

\paragraph{The Nef-Cone Of The Base Space \texorpdfstring{$\mathbf{\mathcal{B}_6}$}{B6}}

In the same spirit as in \cref{subsec:GeometryOfY4FordP3}, let us complete this discussion of the geometry of $\hat{Y}_4$ by working out the Nef-cone of $\mathcal{B}_6$. 
We see from $h_{1,1} ( \mathcal{B}_6 ) = 2$ (\cf \cref{equ:HodgeDP7}), that $\mathcal{B}_6$ admits two inequivalent classes of curves. These are for example furnished from the divisors $D_7$ and $D_{13}$, defined above. For completeness we list the intersection numbers of all 15 divisors $D_i$ with the generators $G_1$, $G_2$ of $\mathrm{Pic} ( \mathcal{B}_6 )$ in \cref{table-N30}. From this we can easily identify those divisor classes of $\mathcal{B}_6$ which have non-negative intersection number with all of of these curves. This shows $\mathrm{Nef} ( \mathcal{B}_6 ) = \mathrm{Cone} ( G_1, G_1 + G_2 )$.
\begin{table}[tbp]
\centering
\begin{tabular}{c@{\hskip 20pt}cccccccccccccccccc}
      \toprule
       & $D_1$ & $D_2$ & $D_3$ & $D_4$ & $D_5$ & $D_6$ & $D_7$ & $D_8$ & $D_9$ & $D_{10}$ & $D_{11}$ & $D_{12}$ & $D_{13}$ & $D_{14}$ & $D_{15}$ \\
      \midrule
      $G_1$ & 0 & 1 & 1 & 0 & 1 & 1 & 1  & 1 & 1 & 1 & 1 & 1 & 1 & 1 & 1 \\
      $G_2$ & 1 & 0 & 0 & 1 & 0 & 0 & -1 & 0 & 0 & 0 & 1 & 0 & 0 & 0 & 0 \\
      \bottomrule
\end{tabular}
\caption{Intersection numbers $G_i C$ for various curves $C \subseteq \mathcal{B}_6$.}
\label{table-N30}
\end{table}

\subsection{The Gauge Backgrounds} \label{sec:GaugeBackgroundsdP7}

\paragraph{The Choice of Flux}
We now make a choice of the hypercharge flux. To this end, recall from \cref{subsec:GeometryOfY4FordP7} that $\mathrm{Cl} \left( \mathcal{B}_6 \right)$ is freely generated by the divisors $G_1$, $G_2$ which satisfy
\[ \left. G_1 \right|_{W} = 3 H - \sum_{i = 1}^{7}{E_i} \, , \qquad \left. G_2 \right|_W = E_7 \, . \]
Consequently, the choice $\mathcal{H} = E_5 - E_6 = D_9 - D_6$ is in agreement with our discussion in \cref{sec:ChoiceOfHyperchargeFluxAndExotics}. Let us now express $F \in \mathbb{Q} \otimes_{\mathbb{Z}} \mathrm{Pic} ( \mathcal{B}_6 )$ as $F = a G_1 + b G_2$ with $a,b \in \mathbb{Q}$. Then the overall flux to be considered in the following analysis is given by
\[ A( a,b, \lambda ) = A_X( F ) + A ( \mathbf{10}_1 ) ( \lambda ) + A_Y( \mathcal{H} ) \, . \]

\paragraph{Chiralities Of The Broken Spectrum}
Let us set $\Sigma := E_1 + E_2 + \dots + E_6$. It then follows from \cref{table-netspectrum} that the sheaf cohomologies of the line bundles listed in \cref{table-N31} count the zero modes of the \emph{F-theory} vacuum $( \hat{Y}_4, A( a,b, \lambda )) $. Its is not too hard to identify the chiralities of these bundles. One finds
\[ 
\left( \begin{array}{c}
\chi \left( \mathbf{10}_1 \right) \\ 
\chi \left( \mathbf{5}_3 \right) \\ 
\chi \left( \mathbf{5}_{-2} \right) 
\end{array} \right)
= \frac{1}{5} \cdot \left( 
\begin{array}{ccc}
1 & -1 & -4 \\ 15 & 3 & 2 \\ -16 & -2 & 2 
\end{array} \right)
\cdot
\left( \begin{array}{c}
a \\ b \\ \lambda
\end{array} \right)
\, . \]
Note that $\chi ( \mathcal{C}_{\mathbf{10}_1} ) + \chi ( \mathcal{C}_{\mathbf{5}_3} ) + \chi ( \mathcal{C}_{\mathbf{5}_{-2}} ) = 0$ is required for anomaly cancellation -- we will have much to say about this in \cref{chapter:LocalAnomaliesInF-Theory}. For the singlet curve it holds $\chi ( \mathbf{1}_{5} ) = - 13 a - 29 b$.

\begin{table}[tbp]
\centering
\begin{tabular}{lll}
\toprule 
& Rep. & Divisor on $dP_7$ \\
\midrule
$C_{\mathbf{10}_1}$ & $\left( \mathbf{3}, \mathbf{2} \right)_{1_X, 1_Y}$ & $\left( \frac{3a}{5} - 3 \lambda - \frac{3}{2} \right) H + \left( \frac{b - a + 4 \lambda}{5} + 1 \right) E_7 + \left( \lambda + \frac{1}{2} - \frac{a}{5} \right) \Sigma + 1 \left( E_5 - E_6 \right)$ \\
& $\left( \overline{\mathbf{3}}, \mathbf{1} \right)_{1_X, -4_Y}$ & $\left( \frac{3a}{5} - 3 \lambda - \frac{3}{2} \right) H + \left( \frac{b - a + 4 \lambda}{5} + 1 \right) E_7 + \left( \lambda + \frac{1}{2} - \frac{a}{5} \right) \Sigma - 4 \left( E_5 - E_6 \right)$ \\
& $\left( \mathbf{1}, \mathbf{1} \right)_{1_X, 6_Y}$ & $\left( \frac{3a}{5} - 3 \lambda - \frac{3}{2} \right) H + \left( \frac{b - a + 4 \lambda}{5} + 1 \right) E_7 + \left( \lambda + \frac{1}{2} - \frac{a}{5} \right) \Sigma + 6 \left( E_5 - E_6 \right)$ \\
\vspace{-0.3em} & \\
$C_{\mathbf{5}_3}$ & $\left( \mathbf{3}, \mathbf{1} \right)_{3_X, -2_Y}$ & $\left( \frac{9a}{5} + \frac{3}{2} \right) H + \left( \frac{3}{5} \left( b - a \right) + \frac{2 \lambda}{5} \right) E_7 + \left( - \frac{3a}{5} - \frac{1}{2} \right) \Sigma - 2 \left( E_5 - E_6 \right)$ \\
& $\left( \mathbf{1}, \mathbf{2} \right)_{3_X, 3_Y}$ & $\left( \frac{9a}{5} + \frac{3}{2} \right) H + \left( \frac{3}{5} \left( b - a \right) + \frac{2 \lambda}{5} \right) E_7 + \left( - \frac{3a}{5} - \frac{1}{2} \right) \Sigma + 3 \left( E_5 - E_6 \right)$ \\
\vspace{-0.3em} & \\
$C_{\mathbf{5}_{-2}}$ & $\left( \mathbf{3}, \mathbf{1} \right)_{-2_X, -2_Y}$ & $\left( 3 -\frac{6a}{5} \right) H + \frac{2}{5} \left( a - b \right) E_7 + \left( \frac{2a}{5} - 1 \right) \Sigma - 2 \left( E_5 - E_6 \right) \textcolor{blue}{- \frac{3 \lambda}{5} Y_1 + \frac{2 \lambda}{5} Y_2}$ \\
& $\left( \mathbf{1}, \mathbf{2} \right)_{-2_X, 3_Y}$ & $\left( 3 -\frac{6a}{5} \right) H + \frac{2}{5} \left( a - b \right) E_7 + \left( \frac{2a}{5} - 1 \right) \Sigma + 3 \left( E_5 - E_6 \right) \textcolor{blue}{- \frac{3 \lambda}{5} Y_1 + \frac{2 \lambda}{5} Y_2}$ \\
\vspace{-0.3em} & \\
$C_{\mathbf{1}_{5}}$ & $\left( \mathbf{1}, \mathbf{1} \right)_{5_X, 0_Y}$ & $\left. \mathcal{O}_{\mathcal{B}_6} \left( a G_1 + b G_2 \right) \right|_{C_{\mathbf{1}_{5}}} \otimes \mathcal{O}_{C_{\mathbf{1}_{5}}} \left( \frac{1}{2} K_{C_{\mathbf{1}_{5}}} \right)$ \\
\bottomrule
\end{tabular}
\caption[Zero modes of broken GUT-model ($W \cong dP_7$) via quasi line bundles.]{The zero modes of the \emph{F-theory}-GUT model ($W \cong \mathrm{dP}_7$ and hypercharge flux $\mathcal{H} = E_5 - E_6$) are counted by the sheaf cohomologies of line bundles on the matter curves. These line bundles are associated to the above divisors, once they are restricted to matter curve in question. Note that the blue expressions indicate divisors on $C_{\mathbf{5}_{-2}}$ already.}
\label{table-N31}
\end{table}

\paragraph{D3-Tadpole Cancellation And The D-Term}

In this very geometry the D3-tadpole cancellation condition becomes (\cf \cref{subsec:TadpolesSU5xU1}) 
\[ N_{D_3} = - \frac{29}{4} + \frac{4 a^2}{5} - \frac{\lambda}{5} \left( a-b \right) + \frac{24 a b}{5} + \frac{13 b^2}{5} - \frac{2 \lambda^2}{5} \stackrel{!}{\in} \mathbb{Z}_{> 0} \, . \label{equ:TadpoleFordP7Model} \]
To identify a supersymmetry condition from the D-term, set $J = \alpha G_1 + \beta G_2$ with $\alpha, \beta \in \mathbb{Z}$ and $G_i$ the free generators of $\mathrm{Pic} ( \mathcal{B}_6 )$. Then it follows from the triple intersection numbers in $\mathcal{B}_6$ (c.f. \cref{equ:TripleIntersectionsBasedP7} and \cref{subsec:TadpolesSU5xU1}) that
\[ \mathcal{V}_B \xi \equiv \mathcal{V}_B \left( \xi_X \left( A_X \left( F \right)\right) + \xi_X \left( A \left( \mathbf{10}_1 \right) \left( \lambda \right) \right) \right) = \alpha \left( 8a + 24 b + \lambda \right) + \beta \left( 24a + 26b - \lambda \right) \, . \]
With the help of $\mathrm{Nef} ( \mathcal{B}_6 )$, which we computed before, we find the following necessary and sufficient supersymmetry condition.
\ebox{
$A_X ( F ) + A ( \mathbf{10}_1 ) ( \lambda ) + A_Y ( \mathcal{H} )$ is supersymmetric iff there exist $u_1, u_2 \in \mathbb{Z}_{> 0}$ such that
\[ \mathcal{V}_B \xi_X = u_1 \left( 8a + 24 b + \lambda \right) + u_2 \left( 32 a + 50 b \right) = 0. \]
}
This in turn is the case precisely if the coefficients $c_1 = 8a + 24 b + \lambda$ and $c_2 = 32 a + 50b$ either both vanish identically or are both non-zero but have different signs.

\paragraph{Integrability Condition}

In analogy to \cref{sec:dP3-Example} we enforce the (sufficient) quantisation condition
\[ \frac{1}{5} F + \frac{1}{2} W \in H^2 ( \mathcal{B}_6, \mathbb{Z} ). \]
We set $F = a G_1 + b G_2$ -- $G_1$, $G_2$ are the free generators of $\mathrm{Pic} ( \mathcal{B}_6 )$ -- and know $W = G_2 - G_1$. In consequence, this condition is equivalent to
\[ \frac{a}{5} - \frac{1}{2} \in \mathbb{Z}, \qquad \frac{b}{5} + \frac{1}{2} \in \mathbb{Z} \]
We quantise $A ( \mathbf{10}_1 ) ( \lambda )$-flux by the demand $\frac{\lambda}{5} \in \mathbb{Z}$, just as in \cref{sec:dP3-Example}. Overall this implies that we can express the flux quantisation as ($\alpha, \beta, \tilde{\lambda} \in \mathbb{Z}$)
\[ a = 5 \left( \alpha + \frac{1}{2} \right), \qquad b = 5 \left( \beta - \frac{1}{2} \right), \qquad \lambda = 5 \tilde{\lambda} \, . \label{equ:bundle_quantisation_dP7} \]
Let us mention that \cref{equ:bundle_quantisation_dP7} is the most general conditions under which the massless spectra on $C_{\mathbf{10}_1}$, $C_{\mathbf{5}_3}$ are counted by the sheaf cohomologies of pullbacks of \emph{proper} line bundles on $W$.

\paragraph{Suitable Chiralities}

Of the so-quantised fluxes, we wish to identify those which produce chiralities ($\Delta_1, \Delta_2 \in \mathbb{Z}$)
\[ \vec{\chi} \left( \Delta_1, \Delta_2 \right) = \left( 3 + \Delta_1, -3 - \Delta_2, \Delta_2 - \Delta_1 \right) \, . \]
It can be verified that such fluxes exist if either ($\Delta_1$ is even and $\Delta_2$ odd) or ($\Delta_1$ is odd and $\Delta_2$ even). In the first case, the following fluxes ($u \in \mathbb{Z}$) give the desired chiralities
\[ \vec{f}_{1} \left( \Delta_1, \Delta_2, u \right) = \left( \begin{array}{c} \alpha \\ \beta \\ \tilde{\lambda} \end{array} \right) = \left( \begin{array}{c} \frac{3}{2} + \Delta_1 + \frac{\Delta_2}{2}  \\ - \frac{25}{2} - 6 \Delta_1 - \frac{7}{2} \Delta_2 \\ 3 + \frac{3}{2} \Delta_1 + \Delta_2 \end{array} \right) + u \left( \begin{array}{c} 5 \\ -31 \\ 9 \end{array} \right) \label{equ:Flux1} \]
and in the second case the fluxes of interest are
\[ \vec{f}_{2} \left( \Delta_1, \Delta_2, u \right) = \left( \begin{array}{c} \alpha \\ \beta \\ \tilde{\lambda} \end{array} \right) = \left( \begin{array}{c} -1 + \Delta_1 + \frac{\Delta_2}{2} \\ 3 - 6 \Delta_1 - \frac{7}{2} \Delta_2 \\ - \frac{3}{2} + \frac{3}{2} \Delta_1  + \Delta_2 \end{array} \right) + u \left( \begin{array}{c} 5 \\ -31 \\ 9 \end{array} \right) \, . \]

\paragraph{Supersymmetry vs. \texorpdfstring{$\mathbf{D3}$}{D3}-Tadpole}

For the flux $\vec{f_1} ( \Delta_1, \Delta_2, u )$ the $D3$-tadpole condition takes the form
\begin{align}
\begin{split}
N_{D_3} &= \frac{30491}{4} + \frac{13825}{2} \Delta_1 + 1565 \Delta_1^2 + 4140 \Delta_2 + 1875 \Delta_1 \Delta_2 + \frac{2245 \Delta_2^2}{4} + 35795 u \\
        & \hspace{14em} + 16205 \Delta_1 u + 9705 \Delta_2 u + 41935 u^2 \stackrel{!}{\in} \mathbb{Z}_{\geq 0} \, .
\end{split}
\end{align}
Recall that for this flux $\Delta_1$ is even and $\Delta_2$ odd. Therefore, it holds $N_{D_3} \in \mathbb{Z}$ as required. For fixed $\Delta_1$ and $\Delta_2$, this integer takes the schematic form $N_{D_3} = c_1 + c_2 u + c_3 u^2$ with $c_3 > 0$. Hence, by taking $u$ sufficiently large, we can always find a gauge background which gives the chiralities
\[ \vec{\chi} \left( \Delta_1, \Delta_2 \right) = \left( 3 + \Delta_1, -3 - \Delta_2, \Delta_2 - \Delta_1 \right) \]
and has non-negative integer $D_3$-tadpole. However, in doing so we lose supersymmetry. This can be seen from \cref{equ:Flux1}. Namely, for $u \gg 1$ we have $\alpha \simeq 5 u$, $\beta \simeq - 31 u$ and $\tilde{\lambda} \simeq 9u$. Therefore the D-term becomes
\begin{align}
0 &\stackrel{!}{=} u_1 \left( -40 + 40 \alpha + 120 \beta + 5 \tilde{\lambda} \right) + u_2 \left( -45 + 160 \alpha + 250 \beta \right) \\
  &\simeq - 25 \cdot u \cdot \left( 139 u_1 + 278 u_2 \right) \, ,
\end{align}
which cannot be satisfied for $u_1, u_2 \in \mathbb{Z}_{> 0}$. More generally, for the flux $\vec{f}_1 ( \Delta_1, \Delta_2, u )$ the D-term demands
\begin{align}
\begin{split}
0 &\stackrel{!}{=} - u_1 \left( 293 + \frac{269}{2} \Delta_1 + 79 \Delta_2 + 695 u \right) - u_2 \left( 586 + 268 \Delta_1 + 159 \Delta_2 + 1390 u \right)
\end{split}
\end{align}
for $u_1, u_2 \in \mathbb{Z}_{\geq 0}$. If $\Delta_1 > \Delta_2$, it can be shown that this can only be satisfied by the integers $u$ which satisfy
\[ 0 < 2 \cdot p_1 \left( u \right) < \Delta_1 - \Delta_2 \, \qquad p_1 \left( u \right) \equiv 293 + \frac{269}{2} \Delta_1 + 79 \Delta_2 + 695 u \, . \]
Likewise, for $\Delta_2 > \Delta_1$ we are looking for the integer $u$ with $0 > 2 \cdot p_1 ( u ) > \Delta_1 - \Delta_2$.

A very similar analysis can be performed for the $\vec{f}_2$. We have implemented the resulting supersymmetry criteria in a \emph{Mathematica}-notebook, with which we performed a scan over the fluxes with $- 100 \leq \Delta_1, \Delta_2 \leq 100$. No supersymmetric solutions were found.

\subsection{Computing Zero Modes} \label{subsec:ComputeZeroModesdP7}

Given this lack of supersymmetric solutions, we will continue with a proof of principle and test the limits of computational powers of our algorithm. To this end, let us consider the flux $\vec{f}_2 ( 1,-5,0)$. This flux induces chiralities $\vec{\chi} = ( 7, 6, -13 )$ and has $N_{D_3} = 1014 > 0$. The massless spectrum for this choice of flux is encoded by the line bundles described in \cref{table-N33}.

\begin{table}[tbp]
\centering
\begin{tabular}{llll}
\toprule
& Rep. & Line bundle & Zero modes \\
\midrule
$C_{\mathbf{10}_1}$ & $\left( \mathbf{3}, \mathbf{2} \right)_{1_X, 1_Y}$ & $\left. \mathcal{O}_{\tilde{X}_{\Sigma}} \left( -4, -5 \right) \right|_{C_{\mathbf{10}_1}} \otimes \left. \mathcal{O}_{W} \left( D_9 - D_6 \right) \right|_{C_{\mathbf{10}_1}}$ & $ \left( 7,0 \right)$ \\
& $\left( \overline{\mathbf{3}}, \mathbf{1} \right)_{1_X, -4_Y}$ & $\left. \mathcal{O}_{\tilde{X}_{\Sigma}} \left( -4, -5 \right) \right|_{C_{\mathbf{10}_1}} \otimes \left. \mathcal{O}_{W} \left( - 4 D_9 + 4 D_6 \right) \right|_{C_{\mathbf{10}_1}}$ & $ \left( 7,0 \right)$ \\
& $\left( \mathbf{1}, \mathbf{1} \right)_{1_X, 6_Y}$ & $\left. \mathcal{O}_{\tilde{X}_{\Sigma}} \left( -4, -5 \right) \right|_{C_{\mathbf{10}_1}} \otimes \left. \mathcal{O}_{W} \left( 6 D_9 - 6 D_6 \right) \right|_{C_{\mathbf{10}_1}}$ & $ \left( 7,0 \right)$ \\
\vspace{-0.5em} & \\
$C_{\mathbf{5}_3}$ & $\left( \mathbf{3}, \mathbf{1} \right)_{3_X, -2_Y}$ & $\left. \mathcal{O}_{\tilde{X}_{\Sigma}} \left( -11, -16 \right) \right|_{C_{\mathbf{5}_3}} \otimes \left. \mathcal{O}_{W} \left( -2 D_9 + 2 D_6 \right) \right|_{C_{\mathbf{5}_3}}$ & $ \left( 6,0 \right)$ \\
& $\left( \mathbf{1}, \mathbf{2} \right)_{3_X, 3_Y}$ & $\left. \mathcal{O}_{\tilde{X}_{\Sigma}} \left( -11, -16 \right) \right|_{C_{\mathbf{5}_3}} \otimes \left. \mathcal{O}_{W} \left( 3 D_9 - 3 D_6 \right) \right|_{C_{\mathbf{5}_3}}$ & $ \left( 6,0 \right)$ \\
\vspace{-0.5em} & \\
$C_{\mathbf{5}_{-2}}$ & $\left( \mathbf{3}, \mathbf{1} \right)_{-2_X, -2_Y}$ & $\left. \mathcal{O}_{\tilde{X}_{\Sigma}} \left( 10, 12 \right) \right|_{C_{\mathbf{5}_{-2}}} \otimes \left. \mathcal{O}_{W} \left( -2 D_9 + 2 D_6 \right) \right|_{C_{\mathbf{5}_{-2}}}$ & $ \left( 0,13 \right)$ \\
& $\left( \mathbf{1}, \mathbf{2} \right)_{-2_X, 3_Y}$ & $\left. \mathcal{O}_{\tilde{X}_{\Sigma}} \left( 10, 12 \right) \right|_{C_{\mathbf{5}_{-2}}} \otimes \left. \mathcal{O}_{W} \left( 3 D_9 - 3 D_6 \right) \right|_{C_{\mathbf{5}_{-2}}}$ & $ \left( 0,13 \right)$ \\
\vspace{-0.5em} & \\
$C_{\mathbf{1}_{5}}$ & $\left( \mathbf{1}, \mathbf{1} \right)_{5_X, 0_Y}$ & $\left. \mathcal{O}_{\tilde{X}_{\Sigma}} \left( -17, -27 \right) \right|_{C_{\mathbf{1}_{5}}}$ & $ \left( 0,700 \right)$ \\
\bottomrule
\end{tabular}
\caption[Zero modes of broken GUT-model ($W = dP_7$) for explicit choice of flux.]{The zero modes of the \emph{F-theory} GUT model over $W \cong dP_7$ for hypercharge flux $\mathcal{H} = E_5 - E_6$ and the gauge background $\vec{f}_2 ( 1,-5,0)$.}
\label{table-N33}
\end{table}

We have chosen this particular flux because its entire massless spectrum can be deduced from Kodaira vanishing (\cf \cref{subsec:LineBundlesOnRiemannSurfaces}). The results are listed in \cref{table-N33}. Hence, this example can serve as non-trivial consistency check on the algorithms in \cite{SheafCohomologyOnToricVarieties}. At the same time, it serves as indication of the current computational limitations. Let us therefore try to verify the result for the state $( \mathbf{3}, \mathbf{2} )_{1_X, 1_Y}$ with the computer \texttt{plesken} at the \emph{University of Siegen}. 

The number of the (anti-)chiral states of the 4-dimensional low-energy effective theory is encoded in the sheaf cohomologies of the line bundle 
\[ L_ = \left. \mathcal{O}_{W} \left( D_9 - D_6 \right) \right|_{C_{\mathbf{10}_1}} \otimes \left. \mathcal{O}_{\tilde{X}_\Sigma} \left( -4, -5 \right) \right|_{C_{\mathbf{10}_1}} \in \mathrm{Pic} \left( C_{\mathbf{10}_1} \right) \, . \]
We thus compute an \fp graded $S$-module $M$ such that $\tilde{M}$ is supported over $C_{\mathbf{10}_1}$ only and satisfies
$\left. \tilde{M} \right|_{C_{\mathbf{10}_1}} \cong L$. For this module we computed the first entries of the quality series $\mathfrak{h}^0 ( M, e )$. For $e \leq 9$ we found
\[ \mathfrak{h}^0 ( M, e ) = \left( 0,0,0,0,0,21,34,9,7, \dots \right) \, . \]
In fact, \texttt{plesken} was able to deduce from a resolution of the module $M$ that for $e = 7$ the conditions of \cref{mytheorem} are satisfied. Consequently 
$h^0 ( C_{\mathbf{10}_1}, L ) = 7$, which matches the result obtained from the Kodaira vanishing theorem. Along the same lines we tried to compute the massless spectrum on the remaining matter curves. However, as of this writing, this exceeded the available computational resources of the computer \texttt{plesken} at \emph{University of Siegen}.

\section{Summary}

In this chapter we have studied two \emph{F-theory} GUT-models. The geometry of the elliptic fibration was derived from the $SU(5) \times U(1)_X$-top discussed in \cref{chapter:MasslessSpectraAndSheafCohomology} and \cref{sec:SU5xU1Top}. See also the original literature in \cite{Krause:2011xj}. We looked at the gauge background
\[ A = A_X \left( F \right) + A \left( \mathbf{10}_1 \right) \left( \lambda \right) + A_Y \left( \mathcal{H} \right) \, , \]
and set out to compute the zero modes of this flux. The results of \cref{chapter:MasslessSpectraAndSheafCohomology} allowed us to summarise the formal dependence of these zero modes in \cref{table-netspectrum}. The hypercharge flux $A_Y( \mathcal{H} )$ is used to induce the gauge group breaking
\[ SU( 5 ) \times U ( 1 )_X \to SU(3 ) \times SU(2 ) \times U(1)_X \times U(1)_Y \, . \label{equ:GaugeGroupBreaking} \]
This breaking can already be anticipated from \cref{table-netspectrum}. In particular, we can interpret the zero modes in terms of the \emph{standard model} particles as outlined in \cref{table-brokenStates}.

Ideally, the breaking \cref{equ:GaugeGroupBreaking} should leave the $U(1)_Y$ gauge boson massless in the external space $\mathcal{E}_4$. In addition, we argued that it is phenomenologically appealing to rule out the existence of massless exotic states in representation $( \mathbf{3}, \mathbf{2} )_{q_x, \pm 5_y}$ propagating along the GUT-divisor $W$. Combining these requirements poses a very strict condition on the divisor class $\mathcal{H} \subseteq W$, which supports the hypercharge flux. Our wish to compute the zero modes of such an \emph{F-theory} compactification comes, as a result of our lack of understanding the Picard group on the matter curves sufficiently well, with a number of sufficient (yet not necessary) structural demands. To date, these demands cannot be matched with the strict conditions on the curve $\mathcal{H}$, as we explained in \cref{sec:ChoiceOfHyperchargeFluxAndExotics}. To demonstrate the computational powers of our algorithm nonetheless, we have accepted the existence of exotic bulk states. With a view towards string phenomenology, the studied geometries can therefore merely serve as toy models.

That all said, we have set out to investigate an \emph{F-theory} GUT-model with $W \cong dP_3$ in \cref{sec:dP3-Example}. The involved geometries were discussed in \cref{subsec:GeometryOfY4FordP3}. To simplify the cohomology computations, we have noted that a $dP_3$-surface can be modelled as a toric variety directly. To take advantage thereof, we have recalled a basic result from algebraic geometry, according to which (special) graded ring homomorphisms encode morphisms between the associated projective schemes \cite[p.80  ff]{hartshorne1977algebraic}. We have conjectured, what we believe to be a natural generalisation thereof to smooth, projective toric varieties in \cref{conj:II}. Let us emphasize that we cannot prove the validity of this conjecture, as of this writing. We leave a detailed analysis thereof for future work. However, in the example geometry at hand, we were interested in just a single example of such an embedding, namely \cref{equ:ConjecturedEmbedding}. This morphism has therefore been put through a number of rather non-trivial consistency checks, ranging from intersections of divisors, over sheaf cohomology computations to genera of matter curves. As we found matching properties in all these cases, we remain positive that at least in this instance, our conjecture holds true.

Subsequently, we discussed the available gauge backgrounds in \cref{subsec:GaugeBackground} and identified those fluxes which come closest to being of phenomenological relevance. We list them in \cref{section:ListOfGoodFluxes}. Unfortunately, none of these fluxes satisfies the supersymmetry condition. This is yet another reason as to why this example geometry can merely serves as \emph{F-theory} toy-model.

Finally, in \cref{subsec:MasslessSpectra} we have put our algorithms to the test and computed zero modes for a few gauge backgrounds. Even for maximally generic matter curves we were able to compute all zero modes. This underlines the powers of our algorithms, as all complex structure moduli need to be taken into account explicitly during these computations. The results for such maximally generic matter curves are listed in \cref{table-N22}. Still, we can compute zero modes also for special choices of complex structure moduli. As the results in \cref{table-N22} show, the numbers of zero modes can change quite drastically for such special choices of complex structure.\footnote{A similar example has been studied in the previous chapter -- its results are summarised in \cref{table-resultsScanOverComplexStructureModuliSpace}.} 

To test the limits of our algorithm, we finally studied an example with $W \cong dP_7$ in \cref{sec:dP7-Example} along the same lines of \cref{sec:dP3-Example}. However, $W \cong dP_7$ prevents us from realising this surface as toric variety directly. In consequence, we had to compute the zero modes from sheaf cohomology of coherent sheaves on a toric variety $\tilde{X_\Sigma}$ which contains $W$ as subvariety. The relevant geometries were explained in \cref{subsec:GeometryOfY4FordP7}, and turned out to be far too demanding for our algorithms to complete in a timely fashion. For example, \cref{subsec:ComputeZeroModesdP7} we studied a gauge background whose zero modes can be predicted from the Kodaira vanishing theorem alone (\cf \cref{table-N33}). All, but the computation of the zero modes in representation 
$( \mathbf{3}, \mathbf{2} )_{1_X, 1_Y}$, exceeded the computational resources provided at \texttt{plesken}. However, our computation reproduced the result found from Kodaira vanishing. Thereby our implementation \cite{SheafCohomologyOnToricVarieties} passed a fairly non-trivial consistency check!

In \cite{Blumenhagen:2008zz} it was explained that one can compute line bundle cohomology on a del Pezzo surface from a combinatorial problem, once the locations of its blow-up points are known. Hence, in order to improve the situation in \cref{sec:dP7-Example} slightly, we tried to identify these blow-ups by computing line bundle cohomology on $W$ with \cite{SheafCohomologyOnToricVarieties}. We were able to derived a number of non-trivial conditions on the blow-up points. Thereby we parametrised their locations according to \cref{equ:RestrictedBlowupPoints}. Whilst, presumably due to computational limitations, we could not fix the blow-up points of the $ W \cong dP_7$ uniquely, we believe that this application nicely demonstrates the powers, but also limitations, of our algorithm.

Phenomenologically, most of the results presented in this chapter are at best insufficient. However, let us emphasize again, that we used these geometries primarily to demonstrate both the powers and limitations of our algorithm. We believe that we succeeded in doing so.

\chapter{Chow Groups and (Local) Anomalies in \emph{F-theory}} \label{chapter:LocalAnomaliesInF-Theory}

Up to this point, this thesis has focused on the study of zero modes in \emph{F-theory} vacua. In particular, in \cref{chapter:MasslessSpectraAndSheafCohomology} we have analysed matter surface fluxes in geometries derived form the $SU(5) \times U(1)_X$-top first introduced in \cite{Krause:2011xj} and discussed with more refined techniques, such as primary decompositions, in \cref{sec:SU5xU1Top}. We found that the divisors induced from the matter surfaces fluxes in this geometry satisfy a number of relations. Back in \cref{chapter:MasslessSpectraAndSheafCohomology} we therefore conjectured that these relations extend to give the following relations among Chow classes:
\begin{align}
\begin{split} \label{equ:MSFRelations}
A \left( \mathbf{10}_{1} \right) \left(  \lambda \right) &= A \left( \mathbf{5}_{3} \right) \left( -\lambda \right) + A \left( \mathbf{5}_{-2} \right) \left( - \lambda \right) \, , \\
A \left( \mathbf{5}_{-2} \right) \left( \lambda \right) &= A_X \left( \lambda W \right) \, , \\
A \left( \mathbf{1}_{5} \right) \left( \lambda \right) &= A_X \left( - \lambda \left[ 6 \overline{K}_{\mathcal{B}_6} - 5 W \right] \right) + A \left( \mathbf{10}_{1} \right) \left( \lambda \right) \, .
\end{split}
\end{align}
During this chapter, we will indeed prove this to be true. Even more, we will find that these relations are understood to originate from anomaly cancellation in \emph{F-theory}. Consequently, this chapter will investigate the role of Chow groups in the context of anomalies in \emph{F-theory}.

The constraints imposed by anomalies on globally consistent 6-dimensional \emph{F-theory} vacua have been studied in detail in  \cite{Kumar:2009us, Kumar:2009ac, Kumar:2010ru, Seiberg:2011dr, Park:2011wv} and references therein. Some extensions of this reasoning to 4-dimensional compactifications have appeared in \cite{Grimm:2012yq}. We start by revising anomalies in field theory in \cref{subsec:AnomalyCancellationFieldTheory}. In general, some of the 1-loop induced anomalies need not vanish by themselves, provided they can be consistently cancelled by a generalised Green--Schwarz mechanism. This is indeed the case in the effective field theory obtained by compactifications of \emph{string theory}. In the \emph{F-theory} context, the corresponding anomalies and counterterms have been worked out in \cite{Cvetic:2012xn}. We explain these results in \cref{sec:AnomalyCancellation4D}. Subsequently, we generalise these findings to relations in the Chow ring in \cref{sec:GeneralisationToChow}. These relations will be exemplify in the geometry of the $SU(5) \times U(1)_X$-top (\cf \cref{sec:SU5xU1Top}) in \cref{sec:ChowRelationsExamples}. For this class of toric F-theory vacua, we will indeed prove that these relations, including \cref{equ:MSFRelations}, hold true in $\mathrm{CH}^2 ( \hat{Y}_5 )$. We complete this chapter by generalising these findings to the anomaly cancellation in 6-dimensional \emph{F-theory} vacua in \cref{sec:Implications6d}.

\section{Generalities on Anomaly Cancellation}

\subsection{Anomalies in Field Theory} \label{subsec:AnomalyCancellationFieldTheory}

Let us start by looking at anomalies in quantum field theories. These anomalies are well understood. For example, \cite{Bilal:2008qx} provides a review on the topic. To revise the topic, let us for simplicity look at a quantum field theory whose gauge symmetry is mediated by a Lie group 
\[ G_{\text{tot}} = G \times U(1)_A \, , \]
where $G$ is a simple, connected Lie group with traceless generators of its Lie algebra. In particular, $G$ is different from $U(1)$. Moreover we assume that the field theory in question contains chiral fermions and allows for a perturbative description. 

\begin{figure}[tbp]
\begin{center}
\subfloat[$U(1)_A^3$-anomaly.]{\label{figure-Anomaly1}
\includegraphics{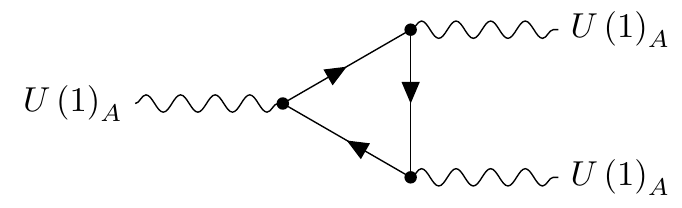}
}
\hspace{3em}
\subfloat[$U(1)_A$--$G^2$-anomaly.]{\label{figure-Anomaly2}
\includegraphics{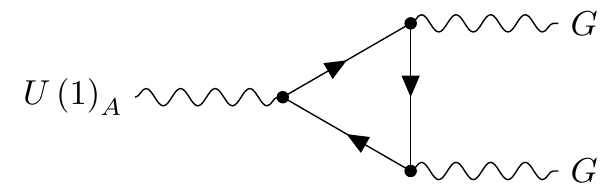}
}

\subfloat[$U(1)_A$--$\mathrm{graviton}^2$-anomaly.]{\label{figure-Anomaly3}
\includegraphics{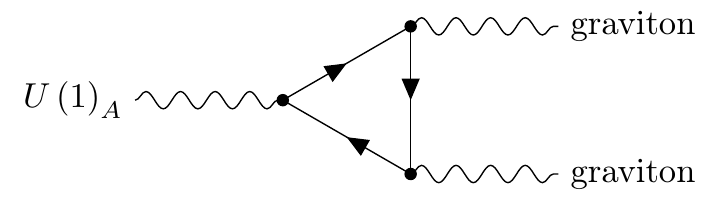}
}
\hspace{3em}
\subfloat[$G^3$-anomaly.]{\label{figure-Anomaly4}
\includegraphics{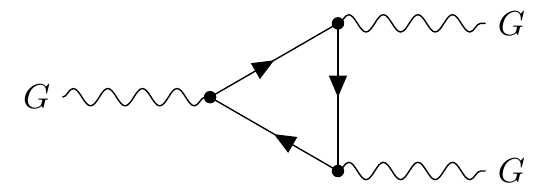}
}

\caption[Feynman diagrams which count the field theory anomalies.]{Feynman diagrams which count the anomalies of a field theory with gauge group $G \times U(1)_a$. These pictures are inspired by \cite{BiesBachelor}.}
\label{figure-SummaryOnAnomalies}
\end{center}
\end{figure}

The anomalies of such a quantum field theory are encoded in triangle Feynman diagrams, in which chiral fermions run along the triangular loop. The external legs are either formed from gauge bosons belonging to the group $G_{\text{tot}}$ or represent gravitons. To evaluate these Feynman diagrams, one applies the Feynman rules. These rules state to replace each of the external legs by an algebraic expression, which contains Lie algebra indices, and subsequently to trace over these indices. By assumption, the Lie algebra generators of the groups $G$ are traceless. Therefore, a triangle diagram in which only one gauge boson stems from $G$ will have vanishing contribution. By following this logic, the possible anomalies are found to be encoded by the diagrams listed in \cref{figure-SummaryOnAnomalies}.

The algebraic expressions of these Feynman diagrams are found to be sensitive only to the chiral index $\chi ( \mathbf{R} )$ of a given representation $\mathbf{R}$ of $G_{\mathrm{tot}}$, \ie the difference of the number of chiral fermions in representation $\mathbf{R}$ and the number of anti-chiral such fermions. Explicitly they evaluate to the following expressions:
\begin{align}
\begin{split}
 U(1)_A^3 \text{ anomaly: } & \quad \sum_{\mathbf{R}}{\mathrm{dim} \left( \mathbf{R} \right) \cdot q_A \left( \mathbf{R} \right)^3 \cdot \chi \left( \mathbf{R} \right) } \, , \\
 U(1)_A-G^2 \text{ anomaly: } & \quad \sum_{\mathbf{R}}{c_2 \left( \mathbf{R} \right) \cdot q_A \left( \mathbf{R} \right) \cdot \chi \left( \mathbf{R} \right) } \, , \\
 G^3 \text{ anomaly: } & \quad \sum_{\mathbf{R}}{c_3 \left( \mathbf{R} \right) \cdot \chi \left( \mathbf{R} \right) } \, , \\
 U(1)_A-\mathrm{graviton}^2 \text{ anomaly: } & \quad \sum_{\mathbf{R}}{\mathrm{dim} \left( \mathbf{R} \right) \cdot q_A \left( \mathbf{R} \right) \cdot \chi \left( \mathbf{R} \right) } \, .
 \label{equ:FieldTheoryAnomalies}
\end{split}
\end{align}
In these expressions $q_A ( \mathbf{R} )$ denotes the $U(1)_A$-charge of the representation $\mathbf{R}$. In addition, we are using the group theory constants $c_2 ( \mathbf{R} )$ and $c_3 ( \mathbf{R} )$. For any irreducible representation $\mathbf{R}$, they relate the trace over powers of the field strength to the corresponding trace in the fundamental representation, \ie
\[ \mathrm{tr}_{\mathbf{R}} \left( F^2 \right) = c_2 \left( \mathbf{R} \right) \cdot \mathrm{tr}_{\mathrm{fund}} \left( F^2 \right) \, , \qquad \mathrm{tr}_{\mathbf{R}} \left( F^3 \right) = c_3 \left( \mathbf{R} \right) \cdot \mathrm{tr}_{\mathrm{fund}} \left( F^3 \right) \, . \]

A priori, the expressions in \cref{equ:FieldTheoryAnomalies} need not vanish. However, for a gauge theory to be well-defined, these expressions must be cancelled by appropriate counterterms. This kind of cancellation strongly constrains the chiral spectrum of a quantum field theory.

\subsection{The Generalised Green--Schwarz mechanism}

\emph{String theory} is a consistent theory of quantum gravity, and indeed provides a mechanism to cancel anomalies. In 1984 this mechanism was discovered in type I \emph{string theory} by Michael Green and John H. Schwarz \cite{1984PhLB..149..117G}. This is the \emph{Green--Schwarz mechanism}, which can be generalised to all other string theories. Collectively these means of anomaly cancellation are termed the \emph{generalised Green--Schwarz mechanism} \cite{Sagnotti:1992qw} -- see also \cite{Blumenhagen:2006ci} for a review.

\begin{figure}[tbp]
\centering
\includegraphics{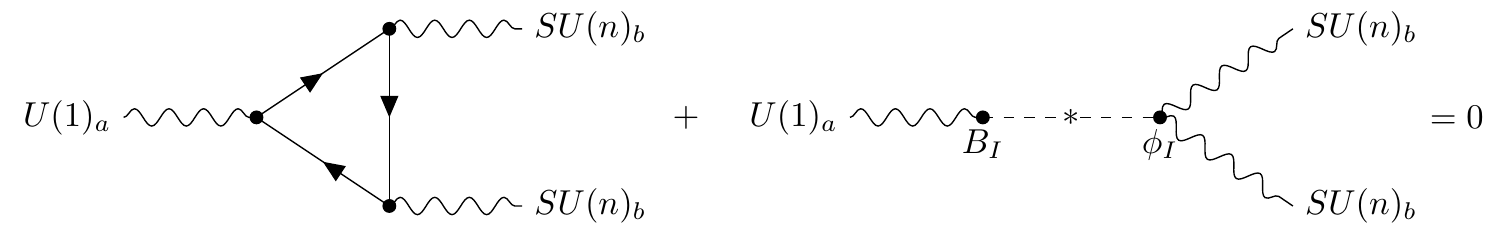}
\caption{4d cubic anomalies and their cancellation by Green--Schwarz counterterms.}
\label{figure:4DAnomaly}
\end{figure}

To gain some intuition for this mechanism in type IIB \emph{string theory}, let us look at an orientifold compactification on $\mathcal{E}_4 \times \mathcal{I}_6$ with two stacks of spacetime-filling D7-branes $\mathcal{D}_a$, $\mathcal{D}_b$ along $\mathcal{E}_4 \times \Sigma_a$ and $\mathcal{E}_4 \times \Sigma_b$, respectively. Let us assume that on $\mathcal{D}_a$ a $U(1)$-gauge theory with gauge field $C_a$ is supported. Similarly, let $\mathcal{D}_b$ support an $SU(n)$-gauge theory with field strength $F_b$. The counterterm to cancel the $U(1)_a$-$SU(n)_b^2$-anomaly then follows from the following steps:
\begin{itemize}
 \item First pick a basis $\{ \alpha^I \}$ of $H_4 ( \mathcal{I}_6, \mathbb{Z} )$ and expand the 4-cycles $\Sigma_a$, $\Sigma_b$ as
      \[ \Sigma_a = \sum_{I}{e_{I a} \alpha^I} \, , \qquad \Sigma_b = \sum_{I}{e_{I b} \alpha^I} \, . \]
 \item As discussed in \cref{subsec:TypeIIBStringTheory}, there exist both electric and magnetic couplings of D7-branes to the RR p-form fields $C_p$. In a democratic 
      formulation of type IIB \emph{string theory}, one therefore not only considers the form fields $C_0$, $C_2$ and $C_4$ but also $C_6$ and $C_8$. For anomaly cancellation the couplings of $C_4$ and $C_6$ are of ample importance.
 \item The RR form-field $C_6$ couples electrically to the $U(1)_a$ field strength via
      \[ S_1 = \Mint_{\mathcal{E}_4 \times \Sigma_a}{C_6 \wedge C_a} \, . \]
      Let us decompose $C_a$ into an external piece $C_a^{\text{ext}}$ and an internal piece. Then the dimensional reduction of $S_1$ leads to
      \[ S_1 \supseteq \sum_{I}{e_{I a} \Mint_{\mathcal{E}_4}{B_I \wedge C_a^{\text{ext}}}} \, , \qquad B_I \equiv \Mint_{\alpha^I}{C_6} \, . \]
 \item Similarly, the RR-form-field $C_4$ couples to the $SU(n)$-field strength $F_b$ via
      \[ S_2 = \Mint_{\mathcal{E}_4 \times \Sigma_a}{C_4 \wedge \mathrm{tr} \left( F_b \wedge F_b \right)} \supseteq \sum_{I}{e_{Ib} \Mint_{\mathcal{E}_4}{\phi_I \wedge \mathrm{Tr} \left( F_b^{\mathrm{ext}} \wedge F_b^{\mathrm{ext}} \right)} } \, , \qquad \phi_I \equiv \Mint_{\alpha^I}{C_4} \, . \]
 \item We thus realise that the dimensional reductions of $S_1$ and $S_2$ lead in the low-energy effective 4-dimensional theory to couplings. At tree-level, these 
      couplings allow for a diagram of the form
      \[ \includegraphics[valign = c]{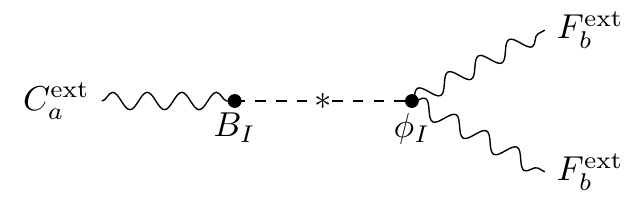} \]
      and this is precisely the Green--Schwarz counterterm needed to cancel the $U(1)_a$-$SU(n)_b^2$-anomaly. We summarise this finding in \cref{figure:4DAnomaly}.
\end{itemize}
Of course this counterterm must have the correct value to cancel the $U(1)_a$-$SU(n)_b^2$-anomaly. This is discussed in detail in \cite{Blumenhagen:2006ci}, to which we refer the interested reader. Let us finally point out that not only the $U(1)_a$-$SU(n)_b^2$-anomaly, but all anomalies which contain at least one $U(1)$ gauge bosons, can be cancelled in this way. For example, the cancellation of the $U(1)_a$-gravitational anomaly follows from the coupling
\[ \Mint_{\mathcal{E}_4 \times \Sigma_b}{C_4 \wedge \text{Tr} \left( R \wedge R \right)} \, . \]
Thus, all of the anomalies listed in \cref{equ:FieldTheoryAnomalies} can be cancelled, but the $SU(n)^3$-anomalies. This cubic non-Abelian anomaly must vanish by itself.

\section{Anomaly Cancellation in \emph{F-theory} as Intersections in Cohomology} \label{sec:AnomalyCancellation4D}

\subsection{\dots for gauge-invariant Fluxes}

Let us now look at the \emph{F-theory} analogue of these field theory anomalies and their Green--Schwarz counterterms. We assume that the gauge flux $G_4$ satisfies both \cref{transversality-gen1}, \cref{gaugeinvariantflux} and that the gauge group be $G_{\text{tot}} = G \times U(1)_A$ as before. Let us also recall that a matter surface $S^a_{\mathbf{R}}$, $a = 1, \ldots, {\mathrm{dim}}({\mathbf{R}})$, associated to one of the weights $\beta^a(\mathbf{R})$ of representation $\mathbf{R}$ satisfies $\chi \left( {\mathbf{R}} \right) = \left[S^a_{\mathbf{R}} \right] \cdot G_4$, as explained around \cref{chiRsecgen}. That said, an intuitive generalisation of \cref{equ:FieldTheoryAnomalies} comes from replacing $\chi \left( {\mathbf{R}} \right)$ by $\left[S^a_{\mathbf{R}} \right] \cdot G_4$. Indeed this is backed up by the results of \cite{Cvetic:2012xn}. In this work also the Green--Schwarz counterterms were identified. The results read as follows:
\begin{align}
 U(1)_A^3 \text{ anomaly: } & \quad \sum_{\mathrm{\mathbf{R}}}{ q^3_A \left( \mathbf{R} \right) \mathrm{dim} \left( \mathbf{R} \right) \left( \left[ S^a_{\mathbf{R}} 
                              \right] \cdot G_4 \right) } = G_4 \cdot [ U_A ] \cdot [ D^\mathbf{b}_{AAA} ] \, , \label{anom-111-a} \\
 U(1)_A-G^2 \text{ anomaly: } & \quad \sum_{\mathrm{\mathbf{R}}}{ q_A \left( {\mathrm{\mathbf{R}}} \right) \, c^{(2)}_{\mathrm{\mathbf{R}}} \, \left( \left[ 
                                S^a_{\mathbf{R}} \right] \cdot G_4 \right) } = G_4 \cdot [ U_A ] \cdot [ D^\mathbf{b}_{AGG} ] \, , \label{anom-G11-a} \\
 G^3 \text{ anomaly: } & \quad \sum_{\mathbf{R}}{ c^{(3)}_{\mathbf{R}} \, \left( \left[ S^a_{\mathbf{R}} \right] \cdot G_4 \right) } = 0 \,. \label{cubic-anom1} \\
 U(1)_A-\mathrm{graviton}^2 \text{ anomaly: } & \quad \sum_{\mathrm{\mathbf{R}}}{ q_A \left( {\mathbf{R}} \right) \mathrm{dim} \left( \mathrm{\mathbf{R}} \right) \left[ 
                                                S^a_{\mathbf{R}} \right] \cdot G_4 } = G_4 \cdot [ U_A ] \cdot [ D^\mathbf{b}_{ARR} ] \label{anom-1RR-a} \, .
\end{align}
The Green--Schwarz counterterms on the RHS are phrased in terms of the generator $U_A \in H^{1,1}(\hat{Y}_4)$ associated with gauge group $U(1)_A$ -- as defined in \cref{UAdefintion} -- and base divisor classes $D^\mathbf{b}_{AGG}, D^\mathbf{b}_{AAA}, D^\mathbf{b}_{ARR} \in \mathrm{CH}^1(\mathcal{B}_6)$, which we will discuss momentarily. Let us mention that for fluxes orthogonal to the subspace spanned by $\mathrm{span}_{\mathbb{C}} \{ U_A \wedge D^{\mathbf{b}}_\alpha \; | \; D^{\mathbf{b}}_\alpha \in H^{1,1} ( \mathcal{B}_6 ) \}$ all of these Green--Schwarz counterterms vanish. This is in agreement with the fact that such fluxes do not render $U(1)_A$ St\"uckelberg massive. Consequently all field theoretic anomalies in \cref{anom-111-a}, \cref{anom-G11-a}, \cref{cubic-anom1}, \cref{anom-1RR-a} vanish identically.

The results of \cite{Cvetic:2012xn} reach even further. Namely the above results have been generalised to \emph{F-theory} vacua with gauge group
\[ G_{\mathrm{tot}} = \prod_{I}{G_I} \times \prod_{A}{U(1)_A} \, , \qquad G_I \text{ a non-Abelian group} \]
and gauge-invariant $G_4$-flux. It was found that \cref{anom-111-a}, \cref{anom-G11-a}, \cref{cubic-anom1} -- generalised to this larger gauge group -- can elegantly be summarized in the single compact expression\footnote{Note that in our normalisation the symmetrisation of $\Gamma$, $\Lambda$, $\Sigma$ on the RHS comes with a factor of $1/3!$.}
\begin{align}
\label{anomalies-all} &\frac{1}{3} \sum_{\mathbf{R}} \sum_{a} n^a_{\Lambda \Sigma \Gamma} \left( \mathbf{R} \right) \, \left( G_4 \cdot \left[ S^a_{\mathbf{R}} \right] \right) = G_4 \cdot \left[ F_{ (\Gamma} \right] \cdot \left[ \hat{\pi}_* \left( F_{\Lambda} \cdot F_{\Sigma)} \right) \right] \,, \\
\label{naRdef} &n^a_{\Lambda \Sigma \Gamma} = \beta^a_{\Lambda} \left( \mathbf{R} \right) \, \beta^a_{\Sigma} \left( \mathbf{R} \right) \, \beta^a_{\Gamma} \left( \mathbf{R} \right)
\end{align}
$\beta^a_\Lambda(\mathbf{R})$ collectively denotes the Cartan charges, \ie the weights $\beta^a_{i_I} ( \mathbf{R} )$ introduced in \cref{weightdefinition}, and the Abelian $U(1)$-charges in \cref{qAdefinition}. The RHS of \cref{anomalies-all} represents the Green--Schwarz counterterms, with the indices $\Gamma$, $\Lambda$ and $\Sigma$ totally symmetrised. Recall that the operation $\hat{\pi}_*$ denotes projection of a complex 2-cycle class in $\hat{Y}_4$ onto the base $\mathcal{B}_6$, where it yields a divisor class. For instance if all $F_\Lambda$ are chosen to represent some of the non-Abelian resolution divisors, then the RHS of \cref{anomalies-all} vanishes identically since, by assumption, $G_4$ does not break the non-Abelian gauge group. Thereby, we recover \cref{cubic-anom1}.

Similarly, the mixed Abelian-gravitational anomaly in \cref{anom-1RR-a} generalises. In terms of the anti-canonical class $\overline{K}_{\mathcal{B}_6} = c_1 (\mathcal{B}_6)$ of the base $\mathcal{B}_6$ it takes the form \cite{Cvetic:2012xn}
\[ \frac{1}{3} \sum_{\mathbf{R}} \sum_{a} \beta^a_\Lambda \left( \mathbf{R} \right) \, \left( G_4 \cdot \left[ S^a_{\mathbf{R}} \right] \right) \, = - 2 \,G_4 \cdot [ \overline{K}_{\mathcal{B}_6} ] \cdot \left[F_{\Lambda} \right] \label{grav-anom-gen} \, . \]

\subsection{\dots For Non-Gauge Invariant Fluxes} \label{subsec:AnomaliesForNonGaugeInvariantFluxes}

It would be tempting to declare \cref{anomalies-all} and \cref{grav-anom-gen} to be valid as a relation among complex 2-cycles even without the need to project onto the flux $G_4$. This, however, cannot be correct because \cref{anomalies-all} and \cref{grav-anom-gen} do not hold if $G_4$ breaks the non-Abelian gauge group factors $G_I$ by violating \cref{gaugeinvariantflux}. The physical reason is that in such a situation it is not the anomalies of the original non-Abelian gauge group $G_I$ which need to be considered, but rather the anomalies of its subgroups to which it gets broken by the flux. Decomposing the massless spectrum into irreducible representations of these subgroups we find extra contributions to the anomalies which must be taken into account. These extra contributions are due to  matter states localised in the bulk of the 7-brane divisors $W_I$ associated with the non-Abelian gauge group $G_I$. While for gauge invariant flux, the bulk matter transforms in the adjoint representation of $G$ and hence does not contribute to the anomalies, more generally one finds extra chiral massless states which must not be neglected. 

To determine these extra contributions suppose that the flux $G_4$ breaks $G_I$ as
\begin{align}
\label{G-breaking}
G_I &\rightarrow H_I \times U \left( 1 \right)_{H_I} \\
\textbf{adj} \left( G_I \right) &\rightarrow \bigoplus_{m_I} {\mathbf r}_{m_I} \,,
\end{align}
where the irreducible representations ${\mathbf{r}}_{m_I}$ of the non-Abelian subgroup $H_I$ carry $U(1)_{H_I}$ charges $q_I({\mathbf r}_{m_I})$.
To each such ${\mathbf r}_{m_I}$ we associate the weight vector $\beta^a({\mathbf r}_{m_I})$.\footnote{For notational simplicity we do not introduce a new parameter to label the different weights of each subgroup $H_I$. It should always be clear from the context which values the index $a$ takes.} The state with weight $\beta^a({\mathbf r}_{m_I})$ arises from an M2-brane wrapping a linear combination of rational curves in the fibre over the 7-brane divisor $W_I$. Since the representation ${\mathbf r}_{m_I}$ descends from the adjoint representation of $G_I$, the rational curves in question are just suitable combinations of rational curves $\mathbb P^1_{i_I}$ associated with minus one times the simple roots $\alpha_{i_I}$. Let us denote the linear combination of $\mathbb P^1_{i_I}$ with charge $\beta^a({\mathbf r}_{m_I})$ as
\[ \label{rmfibre} C \left( \beta^a \left( {\mathbf{r}}_{m_I} \right) \right) = \sum_{i_I} a_{i_I} \left( {\mathbf{r}}_{m_I}, a \right) \mathbb{P}^1_{i_I} \, . \]

An important result for our analysis is that the chiral index of these states induced by the flux $G_4$ is
\[ \label{chi-bulkmatter} \chi \left( {\mathbf r}_{m} \right) = G_4 \cdot \left( - c_1 \left( W_I \right) \right) \cdot \sum_{i_I} \hat a_{i_I}({\mathbf r}_{m_I},a) \, \left[ E_{i_I} \right] \]
for any choice of weight $a$ associated with representation ${\mathbf r}_{m_I}$. Here $c_1(W_I)$ denotes the first Chern class of the divisor $W_I$. For simply laced Lie algebras, $\hat a_{i_I}({\mathbf r}_{m_I},a) = a_{i_I} \left( {\mathbf{r}}_{m_I}, a \right)$ because each $\mathbb P^1_{i_I}$ is the fibre of the resolution divisor $E_{i_I}$, but for non-simply laced Lie algebras, the fibre of $E_{i_I}$ splits locally into various rational curves, all homologous to $\mathbb P^1_{i_I}$, which are exchanged by a monodromy along $W_I$ \cite{Bershadsky:1996nh, Park:2011ji}. In this case $\hat a_{i_I}({\mathbf r}_{m_I},a)$ includes a fractional correction factor to account for this monodromy. 

To see this one first identifies the cohomology groups on $W_I$ counting massless bulk matter in representation  ${\mathbf r}_{m_I}$. If we denote by $L_{m_I}$ the line bundle induced by the flux background to which the bulk states in representation ${\mathbf r}_{m_I}$ couple, the results of \cite{Beasley:2008dc} imply for the associated chiral index 
\[ \label{indexrmlocal} \chi \left( \mathbf{r}_{m_I} \right) = - \Mint_{W_I}{c_1 \left( W_I \right)} \cdot c_1 \left( L_{m_I} \right) \,. \]

In order to connect this to the formula \cref{chi-bulkmatter} we need to extract the line bundle $L_{m_I}$ on $W_I$ from the flux data incorporated in the $G_4$ flux on $\hat{Y}_4$. This step has been spelled out in \cite{Bies:2017fam}, to which we refer for more details. The prescription is to intersect the algebraic complex 2-cycle class, \ie the element in $\mathrm{CH}^2(\hat{Y}_4)$, underlying the definition of $G_4$ with the linear combination $\sum_{i_I} \hat a_{i_I}({\mathbf r}_{m_I},a) \, E_{i_I}$ of resolution divisors. Projecting this onto $W_I$ defines an element in $\mathrm{CH}_1(W_I)$, \ie a line bundle on $W_I$, which we identify with $L_{m_I}$. Due to the projection onto $W_I$, the result is the same for each choice of weight vector $\beta^a({\mathbf r}_{m_I})$ in \cref{rmfibre}. The chiral index \cref{indexrmlocal} can then be rewritten as in \cref{chi-bulkmatter}.

For compactness of notation we define a complex 2-cycle associated with each representation ${\mathbf r}_{m_I}$ of the form
\[ \label{SarmI} S^{a}_{\mathbf{r}_{m_I}} = \sum_{i_I} \hat a_{i_I} \left( \mathbf{r}_{m_I},a \right) \, \left. E_{i_I} \right|_{K_{W_I}} \, . \]
Here $\left. E_{i_I} \right|_{K_{W_I}}$ denotes the restriction of the resolution divisor $E_{i_I}$, which is a fibration over $W_I$, to the canonical divisor $K_{W_I}$ on $W_I$. Let us point out that in general $S^{a}_{\mathbf{r}_{m_I}} $ defines a class in $\mathrm{CH}^2(\hat{Y}_4) \otimes \mathbb Q$, as do the matter surfaces $S^a(\mathbf{R})$ of the localised matter states. Note furthermore that $[K_{W_I}] = - c_1(W_I)$. Then the expression \cref{chi-bulkmatter} for the chiral index of the massless bulk matter can be rewritten as
\[ \chi \left( \mathbf{r}_{m}, \alpha_m \right) = G_4 \cdot [ S^{a}_{\mathbf{r}_{m_I}}  ] \,. \]

We are now in a position to generalise \cref{anomalies-all} such as to be valid also for fluxes $G_4$ not respecting the non-Abelian gauge algebra. If $G_4$ is responsible for a breaking of the form \cref{G-breaking}, the correct modification of \cref{anomalies-all} is to add on the LHS the contribution to the anomalies from the chiral bulk matter states in representation ${\mathbf r}_{m_I}$. Group theoretically, the set of all weights $\beta^a({\mathbf r}_{m_I})$ equals the set of all weights of the adjoint representation of $G_I$, \ie the complete set of roots of $G_I$. Labelling the roots of $G_I$ by $\rho_I = 1, \ldots, \mathrm{dim}(G_I)$, we define the complex 2-cycle
\[ \label{def-SrhoI} S^{\rho_I} = \sum_{i_I} \hat a_{i_I} \left( \rho_I \right) \, \left. E_{i_I} \right|_{K_{W_I}} \, , \]
which is to be viewed as the analogue of \cref{SarmI}, but directly for the adjoint representation of $G_I$. With this notation in place the generalisation of \cref{anomalies-all} is
\begin{align} \label{G4dotanomaly1}
& G_4 \cdot \left( \sum_{\mathbf{R} \neq \mathbf{adj}} \sum_{a} n^a_{\Lambda \Sigma \Gamma}(\mathbf{R}) \, \left[ S^a_{\mathbf{R}} \right] \, + \frac{1}{2} \sum_{\rho_I} n^{\rho_I}_{\Lambda \Sigma \Gamma} \, \left[ S^{\rho_I} \right] - 3 \, \cdot \left[ F_{(\Gamma} \right] \cdot \left[ \hat{\pi}_* \left( F_{\Lambda} \cdot F_{\Sigma)} \right) \right] \, \right) = 0  \,.
\end{align}
Here we are introducing, in analogy to \cref{naRdef}, the abbreviation
\[ n^{\rho_I}_{\Lambda \Sigma \Gamma} = \beta^{\rho_I}_{\Lambda} \, \beta^{\rho_I}_{\Sigma} \, \beta^{\rho_I}_{\Gamma}\,. \]
The extra factor of $1/2$ in the second line appears because the adjoint is a real representation and we are summing over all roots, positive and negative. 
Similarly, relation \cref{grav-anom-gen} generalises to 
\[ \label{G4dotanomaly2} G_4 \cdot \left( \sum_{\mathbf{R} \neq \mathbf{adj}} \sum_{a} \, \beta^a_\Lambda \left( \mathbf{R} \right) \, \left[ S^a_{\mathbf{R}} \right] + \frac{1}{2} \sum_{\rho_I} \, \beta^{\rho_I}_\Lambda \, \left[ S^{\rho_I} \right] + 6 \, [ \overline{K}_{\mathcal{B}_6} ] \cdot [ F_{\Lambda} ] \right) = 0 \,. \]

\section{Anomaly Cancellation in \emph{F-Theory} as Relation in the Chow Ring} \label{sec:GeneralisationToChow}

It was stressed already in \cite{Cvetic:2012xn} that \cref{anomalies-all} and \cref{grav-anom-gen} can be viewed as a set of relations satisfied by the matter surface classes $[S^a_{\mathbf{R}}]$ on any smooth elliptically fibred Calabi--Yau 4-fold $\hat{Y}_4$. More precisely it is a relation satisfied by the intersection products of matter surfaces with any element $G_4 \in H^{2,2}(\hat{Y}_4)$ subject to the conditions \cref{transversality-gen1} and \cref{gaugeinvariantflux}. In this spirit, the goal of this section is to understand \cref{G4dotanomaly1} and \cref{G4dotanomaly2} as relations in the Chow ring.

\subsection{Construction of Vertical Gauge Fluxes } \label{subsec:ConstructionOfVerticalGaugeFluxes}

Let us consider a smooth elliptically fibred Calabi--Yau 4-fold $\hat Y_4$ arising as the resolution of a singular Weierstrass model. We recall that there is the following decomposition
\[ H^{2,2} ( \hat{Y}_4 ) = H^{2,2}_{\mathrm{vert}} ( \hat{Y}_4 ) \oplus H^{2,2}_{\mathrm{hor}} ( \hat{Y}_4 ) \oplus H^{2,2}_{\mathrm{rem}} ( \hat{Y}_4 )  \, \label{H22decompo2} \]
where elements of the three subspaces are mutually orthogonal with respect to the homological intersection pairing. As discussed in \cref{sec:SystGaugeBack} we have
$H^{2,2}_{\mathrm{vert}}(\hat{Y}_4) = H^{1,1}(\hat{Y}_4) \wedge H^{1,1}(\hat{Y}_4)$. As a preparation for the step to follow, let us now prove the following assertion about $\hat{Y}_4$:

Given the decomposition of the primary vertical subspace $H^{2,2}_{\mathrm{vert}} ( \hat Y_4 ) = V_1 \cup V_2$ with 
\begin{align}
\begin{split}
V_1 &= \text{span} \left\{ [ E_{i_I} \cdot D^{\mathrm{b}}_\alpha ], \, [ D^{\mathrm{b}}_\alpha \cdot D^{\mathrm{b}}_\beta ], \, [ S_0 \cdot D^{\mathrm{b}}_\alpha ], \, [ S_A \cdot D^{\mathrm{b}}_\alpha ] \right\} \, , \\ \label{V3V4}
V_2 &= \text{span} \left\{ \left[E_{i_I} \cdot E_{j_J} \right], \, \left[S_A \cdot E_{i_I} \right], \, \left[S_A \cdot S_B \right], \, \left[S_0 \cdot S_A \right] \right\} \,,
\end{split}
\end{align}
one can associate to each single generator of $V_2$ as listed in \cref{V3V4} a 4-form flux $G_4 \in H^{2,2}_{\mathrm{ vert}}$ satisfying the conditions \cref{transversality-gen1} and \cref{gaugeinvariantflux} by adding elements of $V_1$.

The proof relies on general properties of the intersection numbers of the elements in $H^{2,2}_{\mathrm{vert}}(\hat Y_4)$. Namely, for each element $[X \cdot Y] \in V_2$, the following intersection numbers hold: 
\begin{align} \label{XYrelations}
[ X ] \cdot [ Y ] \cdot [S_0 ] \cdot [D_\alpha^\mathrm{b} ] &= [S_0 ] \cdot [C_{XY} ] \cdot [D_\alpha^\mathrm{b} ] \,, \\
[ X ] \cdot [ Y ] \cdot [D_\alpha^\mathrm{b} ] \cdot [D_\beta^\mathrm{b} ] &= [S_0 ] \cdot [D_{XY} ] \cdot [D_\alpha^\mathrm{b} ] \cdot [D_\beta^\mathrm{b} ] \,, \\
[ X ] \cdot [ Y ] \cdot [E_{i_I} ] \cdot [D_\alpha^\mathrm{b} ] &= [S_0 ] \cdot [F_{XY,i_I} ] \cdot [W_I ] \cdot [D_\alpha^\mathrm{b} ] \,,
\end{align}
where $C_{XY}$ describes a curve class on the base $B_3$ (which may well be zero as \eg for $X= E_{i_I}$, $Y = E_{j_J}$) and $D_{XY}$ and $F_{XY,i_I}$ are divisor classes on $B_3$. The rationale behind \cref{XYrelations} is simply that the LHS can be expressed as an intersection product entirely on the base $B_3$, which is the hypersurface in $\hat Y_4$ defined by the zero-section $S_0$. For every given fibration over a general base $B_3$ the curve and divisor classes on the base appearing on the RHS can be explicitly computed, but their concrete form will not be needed for our argument. 

In terms of the classes $C_{XY}$, $D_{XY}$ and $F_{XY,i_I}$ we can define the algebraic 4-cycle class 
\[ A \left(X \cdot Y \right) = X \cdot Y - C_{XY} - \left(S_0 + D^{\mathrm{b}}_0 \right) \cdot D_{XY} + {\mathfrak C}_{l_L m_M}^{-1} E_{l_L} \cdot F_{XY,m_M} \in \mathrm{CH}^2 ( \hat Y_4 ) \,, \label{AXYclass} \]
where the matrix ${\mathfrak{C}}_{l_L m_M}$ governs the intersections of the resolution divisors as in \cref{intersectionEiEj} and furthermore the divisor class $D_0^{\mathrm{b}}$ on $B_3$ is defined by the property that 
\[ [ S_0 ] \cdot [ S_0 + D^{\mathrm{b}}_0 ] \cdot [ D_\alpha^\mathrm{b} ] \cdot [ D_\beta^\mathrm{b} ] = 0 \,. \label{S0prop1} \]
The homology class associated with 4-cycle class $A(X \cdot Y)$ then gives rise to a transversal and gauge invariant flux
\[ G_4 \left( X \cdot Y \right) = \left[ A \left( X \cdot Y \right) \right] \in H^{2,2}_{\mathrm{ vert}} \,. \]
To show this we need to verify the transversality conditions \cref{transversality-gen1} and gauge invariance \cref{gaugeinvariantflux}. To this end, observe first that 
\[ G_4 \left( X \cdot Y \right) \cdot [ S_0 ] \cdot [ D_\alpha^\mathrm{b} ] = [ S_0 ] \cdot [ C_{XY} ] \cdot [ D_\alpha^\mathrm{b} ] - [ S_0 ] \cdot [ C_{XY} ] \cdot [ D_\alpha^\mathrm{b} ] = 0 \,, \label{transproof1} \]
where we used \cref{S0prop1} and $[S_0] \cdot [E_{l_L}] = 0$. Similarly,
\[ G_4 \left( X \cdot Y \right) \cdot [ D_\alpha^\mathrm{b} ] \cdot [ D_\beta^\mathrm{b} ] = [ D_{XY} ] \cdot [ D_\alpha^\mathrm{b} ] \cdot [ D_\beta^\mathrm{b} ] \cdot [ S_0 ] - [ D_{XY} ] \cdot [ D_\alpha^\mathrm{b} ] \cdot [ D_\beta^\mathrm{b} ] \cdot [ S_0 ] = 0 \label{transproof2} \]
with the help of
\[ [ D_\alpha^\mathrm{b} ] \cdot [ D_\beta^\mathrm{b} ] \cdot [ D_\gamma^\mathrm{b} ] \cdot [ D_\delta^\mathrm{b} ] = 0 \,, \qquad [ E_{i_I} ] \cdot [ D^{\mathrm{b}}_\alpha ] \cdot [ D^{\mathrm{b}}_\beta ] \cdot [ D^{\mathrm{b}}_\gamma ] = 0 \,. \]
\cref{transproof1} and \cref{transproof2} establish transversality of the flux.

Finally, gauge invariance of the flux, \cref{gaugeinvariantflux}, follows from \cref{intersectionEiEj}, which can also be written as 
\[
[ E_{i_I} ] \cdot [ E_{j_J} ] \cdot [ D_\alpha^{\mathrm{b}} ] \cdot [ D_\beta^{\mathrm{b}} ] = - \delta_{IJ} \, {\mathfrak C}_{i_I j_I} \, [ S_0 ] \cdot [ W_I ] \cdot [ D_\alpha^{\mathrm{b}} ] \cdot [ D_\beta^{\mathrm{b}} ],
\]
because
\begin{align}
\begin{split}
& G_4 \left( X \cdot Y \right) \cdot [ E_{i_I} ] \cdot [ D_\alpha^\mathrm{b} ] \\
& \quad \qquad = [ S_0 ] \cdot [ W_I ] \cdot [ F_{XY,i_I} ]\cdot [ D_\alpha^\mathrm{b} ] - {\mathfrak C}^{-1}_{l_L m_M} {\mathfrak C}_{l_L i_I} [ W_I ] \cdot [ F_{XY,m_M} ] \cdot [ D_\alpha^\mathrm{b} ] \cdot [ S_0 ] = 0 \,.
\end{split}
\end{align}
Note that the fluxes constructed in this way from elements of $V_2$ are in general not linearly independent as elements of $H^{2,2}_\mathrm{vert}(\hat Y_4)$. In particular, it is a priori not guaranteed that the class $[A(X\cdot Y)]$ is non-trivial in $H^{2,2}_\mathrm{vert}(\hat Y_4)$. What is important for us, however, is that the homology classes associated with the correction terms in \cref{AXYclass} needed to render  $[A(X\cdot Y)]$ transversal and gauge invariant indeed lie in the subspace $V_1$.

\subsection{Anomaly Cancellation as Cohomological Relations} \label{subsec:AnomalyAsCohomologicalRelations}

According to \cref{G4dotanomaly1} and \cref{G4dotanomaly2}, the sum of the expressions in the brackets is orthogonal to each element of $H^{2,2}(\hat{Y}_4,\mathbb Q)$ which satisfies the transversality condition \cref{transversality-gen1}. \footnote{In general $G_4$ is only half-integer quantised upon demanding that $G_4 + \frac{1}{2} c_2(\hat Y_4) \in H^{2,2}(\hat Y_4) \cap H^{4}(\hat Y_4,\mathbb Z)$. Unless stated otherwise, in the sequel we will always mean $H^{2,2}(\hat Y_4,\mathbb Q)$ when writing $H^{2,2}(\hat Y_4)$.} This is equivalent to a consistent cancellation of all anomalies in \emph{F-theory} including those of the $G_I$ subgroups in the presence of non-gauge-invariant flux. We will now promote \cref{G4dotanomaly1} and \cref{G4dotanomaly2} to relations among cohomology classes within the cohomology ring $H^{2,2}(\hat{Y}_4)$, and even to relations between equivalence classes of complex 2-cycles within $\mathrm{CH}^2(\hat{Y}_4)$. It had already been observed in \cite{Lin:2016vus} that the cancellation of gauge anomalies can be interpreted as a consequence of suitable homological relations between the matter surfaces up to terms orthogonal to all gauge invariant fluxes. The results of this section are a systematic generalisation of these observations to a considerably stronger set of relations which hold even up to rational equivalence.

To make the argument it suffices to focus on \cref{G4dotanomaly1}. As recalled above, the expression in brackets is orthogonal to all gauge fluxes $G_4$ satisfying \cref{transversality-gen1}. Combined with the manifest orthogonality properties of the various classes appearing in brackets, this implies orthogonality to the space
\[ \label{span2} V_1 = \mathrm{span}_{\mathbb{C}} \left\{ [ E_{i_I} \cdot D^{\mathrm{b}}_\alpha ], [ U_A \cdot D^{\mathrm{b}}_\alpha ],\, [ S_0 \cdot D^{\mathrm{b}}_\alpha ], \, [ D^{\mathrm{b}}_\alpha \cdot D^{\mathrm{b}}_\beta ] \right\} \subseteq H^{2,2}(\hat{Y}_4)\,. \]
Indeed, orthogonality to all elements of the first two types in \cref{span2} follows because these are examples of consistent $G_4$ fluxes. Furthermore, elements of the form  $[ D^{\mathrm{b}}_\alpha \cdot D^{\mathrm{b}}_\beta ] $ are manifestly orthogonal to any of the matter surfaces $S^a_\mathbf{R}$ and $S^{\rho_I}$ because the latter are fibrations over curves in the base and hence the topological intersection with two base divisors vanishes for dimensional reasons. As for terms of the form $[ S_0 \cdot D^{\mathrm{b}}_\alpha ]$, note that  the zero section $S_0$ does not intersect any of the fibres of the matter surfaces $S^a_\mathbf{R}$  or $S^{\rho_I}$, nor does it intersect the fibres of the non-Abelian resolution divisors. By construction of the Shioda map, we know in addition that $[ S_0 \cdot D^{\mathrm{b}}_\alpha ] \cdot [U_A \cdot D^{\mathrm{b}}_\beta] = 0$ with $U_A$ one of the generators of the non-Cartan $U(1)$ gauge group factors. Hence, $[ S_0 \cdot D^{\mathrm{b}}_\alpha ]$ is orthogonal to all types of terms appearing in the brackets of \cref{G4dotanomaly1}. This establishes the claim. 

$V_1$ is a subspace of the primary vertical subspace $H^{2,2}_{\mathrm{vert}}(\hat{Y}_4) = H^{1,1}(\hat{Y}_4) \wedge H^{1,1}(\hat{Y}_4)$ (\cf \cref{sec:SystGaugeBack}). By the Shioda-Tate-Wazir theorem \cite{Shioda, Tate1, Tate2, 2001math.....12259W}, $H^{1,1}(\hat{Y}_4)$ is given by
\[ \label{STW} H^{1,1} ( \hat{Y}_4 ) = \mathrm{span}_{\mathbb{C}} \left\{ [ S_0 ], \, [ F_\Sigma ], \, [D^{\mathrm{b}}_{\alpha} ] \right\} \,, \]
where, for future purposes, we have introduced the combined notation $F_\Sigma \in \{ E_{i_I}, U_A \}$.

Consequently we can express $H^{2,2}_{\mathrm{vert}}(\hat{Y}_4)$ as\footnote{Note that $S_0 \wedge E_{i_I}$ vanishes on $\hat{Y}_4$ since the zero-section does not intersect the resolution divisors.}
\[ \label{defV4} H^{2,2}_{\mathrm{vert}} ( \hat{Y}_4 ) = V_1 \cup V_2, \qquad V_2 = \mathrm{span}_{\mathbb{C}} \left\{ \left[ E_{i_I} \cdot E_{j_J} \right], \, \left[ S_A \cdot E_{i_I} \right], \, \left[ S_A \cdot S_B \right], \, \left[ S_0 \cdot S_A \right] \right\} \,. \]

This is exactly the situation analysed in the previous section: we argued that to any element in $V_2$ we can associate a gauge invariant 4-form flux element $G_4 \in H^{2,2}_{\mathrm{vert}}(\hat{Y}_4)$ by adding suitable correction terms, such that the flux satisfies \cref{transversality-gen1} as well as \cref{gaugeinvariantflux}. Even more, we found that the necessary correction terms lie in $V_1$. Together with the fact that the terms in brackets in \cref{G4dotanomaly1} are orthogonal to the set of all such $G_4$, this implies that they are orthogonal to all elements in $H^{2,2}_{\mathrm{vert}}(\hat{Y}_4)$. 

Let us now distinguish two qualitatively distinct situations. First, suppose that the cohomology groups associated with all matter surfaces lie in the primary vertical subspace, \ie suppose that
\[ \left[ S^a_{\mathbf{R}} \right] \in H^{2,2}_{\mathrm{vert}} ( \hat{Y}_4 ) \qquad \forall \, \, S^a_{\mathbf{R}} \,. \]
This is in fact the situation in most explicit examples of elliptically fibred Calabi--Yau 4-folds studied in the literature as of this writing, with the exception of the construction in \cite{Braun:2014pva}. Since the three subspaces in the decomposition \cref{H22decompo2} are mutually orthogonal, the expression in brackets in \cref{G4dotanomaly1} is orthogonal to $H^{2,2}_{\mathrm{hor}}(\hat{Y}_4)$ and $H^{2,2}_{\mathrm{rem}}(\hat{Y}_4)$. Furthermore, as just shown, it is orthogonal on $H^{2,2}_{\mathrm{vert}}(\hat{Y}_4)$. All of this together implies the following relations in cohomology
\begin{align}
\sum_{\mathbf{R} \neq \mathbf{adj}} \sum_{a} n^a_{\Lambda \Sigma \Gamma} \left( \mathbf{R} \right) \, \left[ S^a_{\mathbf{R}} \right] \, + \frac{1}{2} \sum_{\rho_I} n^{\rho_I}_{\Lambda \Sigma \Gamma} \, \left[ S^{\rho_I} \right] - 3 \, \cdot \left[ F_{(\Gamma} \right] \cdot \left[ \hat{\pi}_* \left( F_{\Lambda} \cdot F_{\Sigma)} \right) \right] &= 0 \label{Anomaly4dCohom1} \\
\sum_{\mathbf{R} \neq \mathbf{adj}} \sum_{a} \, \beta^a_\Lambda(\mathbf{R}) \, \left[ S^a_{\mathbf{R}} \right] + \frac{1}{2} \sum_{\rho_I} \, \beta^{\rho_I}_\Lambda \, \left[ S^{\rho_I} \right] + 6 \, \left[ \overline{K}_{\mathcal{B}_6} \right] \cdot \left[ F_{\Lambda} \right] &= 0 \label{Anomaly4dCohom2} \,, 
\end{align}
where \cref{Anomaly4dCohom2} follows by applying similar reasoning to \cref{G4dotanomaly2}.

The second situation corresponds to configurations where some of the matter surface classes have a part in the remainder $H^{2,2}_{\mathrm{rem}}(\hat{Y}_4)$. An example would be a situation in which the matter curve $C_{\mathbf{R}}$ on $W_I$ splits into several components which are individually not complete intersections of $W_I$ with a divisor from the base $B_3$ \cite{Braun:2014pva}. In this case, we can split the classes of the matter surfaces into orthogonal components
\[ \left[S^a(\mathbf{R})\right] = \left[S^a(\mathbf{R})\right]_{\mathrm{vert}} + \left[S^a(\mathbf{R})\right]_{\mathrm{rem}} \,. \]
For such a matter surface, \cref{G4dotanomaly1} gives
{\small
\begin{align}
\begin{split} \label{G4dotanomaly1vert}
0 &= G_4 \cdot \left( \sum_{\mathbf{R} \neq \mathbf{adj}} \sum_{a} n^a_{\Lambda \Sigma \Gamma} \left( \mathbf{R} \right) \, \left[ S^a_{\mathbf{R}} \right]_\mathrm{vert} \, + \frac{1}{2} \sum_{\rho_I} n^{\rho_I}_{\Lambda \Sigma \Gamma} \, \left[ S^{\rho_I} \right] - 3 \, \cdot \left[ F_{(\Gamma} \right] \cdot \left[ \hat{\pi}_* \left( F_{\Lambda} \cdot F_{\Sigma )} \right) \right] \right) \\
& \qquad + G_4 \cdot \sum_{\mathbf{R} \neq \mathbf{adj}} \sum_{a} n^a_{\Lambda \Sigma \Gamma} \left( \mathbf{R} \right) \left[ S^a_{\mathbf{R}} \right]_\mathrm{rem} \,.
\end{split}
\end{align}
}
The terms in brackets of the first line all lie in $H^{2,2}_{\mathrm{vert}}(\hat Y_4)$ and are thus orthogonal to $H^{2,2}_{\mathrm{rem}}(\hat Y_4)$ and $H^{2,2}_{\mathrm{hor}}(\hat Y_4)$, while the terms next to $G_4$ in the second line are orthogonal to $H^{2,2}_{\mathrm{vert}}(\hat Y_4)$ and $H^{2,2}_{\mathrm{hor}}(\hat Y_4)$. Repeating the arguments leading to \cref{Anomaly4dCohom1} in the following section, the expression in the first line is found to be orthogonal to the space $V_1$ and to all fluxes in $H^{2,2}_{\mathrm{vert}}(\hat Y_4)$, which is sufficient to conclude that the relations \cref{summary4dCoho1} hold in cohomology on $\hat Y_4$.
The second identity follows again from \cref{G4dotanomaly2}. In addition, anomaly cancellation implies
\[ G_4 \cdot \sum_{\mathbf{R} \neq \mathbf{adj}} \sum_{a} n^a_{\Lambda \Sigma \Gamma} \left( \mathbf{R} \right) \left[ S^a_{\mathbf{R}} \right]_\mathrm{rem} = 0 \,, \qquad
G_4 \cdot \sum_{\mathbf{R} \neq \mathbf{adj}} \sum_{a} \beta^a_\Lambda \left( \mathbf{R} \right) \left[ S^a_{\mathbf{R}} \right]_\mathrm{rem} = 0 \]
for every choice of transversal flux $G_4$. In order to determine if a stronger set of relations can be deduced directly at the level of cohomology, more information is needed about the nature of $[S^a(\mathbf{R})]_{\mathrm{rem}} $ in non-trivial situations. As noted already, there do currently not exist any examples of fibrations with non-trivial $[S^a_{\mathbf{R}}]_{\mathrm{rem}}$ except those where a single $[S^a_{\mathbf{R}}]$ splits into various components, each with some contribution from $H^{2,2}_{\mathrm{rem}}(\hat Y_4)$, but such that the net $[S^a_{\mathbf{R}}]$ remains purely vertical. In particular, if the only contribution to $[S^a_{\mathbf{R}}]_\mathrm{rem}$ comes from a splitting of $C_{\mathbf{R}}$ into several curves which are not all obtained by intersection with a divisor, then for fixed $\mathbf{R}$ the sum over all distinct components of $[S^a_{\mathbf{R}}]$ is again purely vertical because the different components belonging to the same representation $\mathbf{R}$ come with the same prefactor. Hence, in such examples the components along $H^{2,2}_{\mathrm{rem}}(\hat{Y}_4)$ trivially sum up to zero and there are no non-trivial relations among  $[S^a(\mathbf{R})]_{\mathrm{rem}}$ for different $\mathbf{R}$.

Overall we are lead to conclude the following: Given a matter surface $S^a ( \mathbf{R} )$, then anomaly cancellation implies that the vertical pice $[ S^a_{\mathbf{R}} ]_{\mathrm{vert}}$ of the cohomology class $[ S^a_{\mathbf{R}} ]$ satisfies the following two equations:
{ \small
\begin{subequations} \label{summary4dCoho1}
\begin{empheq}[box=\widefbox]{align}
\sum_{\mathbf{R} \neq \mathbf{adj}} \sum_{a} n^a_{\Lambda \Sigma \Gamma} \left( \mathbf{R} \right) \, \left[ S^a_{\mathbf{R}} \right]_\mathrm{vert} \, + \frac{1}{2} \sum_{\rho_I } n^{\rho_I}_{\Lambda \Sigma \Gamma} \, \left[ S^{\rho_I} \right] &= 3 \, \left[ F_{(\Gamma} \right] \cdot \left[ \hat{\pi}_* \left( F_{\Lambda} \cdot F_{\Sigma )} \right) \right] \,,  \label{summary4dCoho1-a} \\
\sum_{\mathbf{R} \neq \mathbf{adj}} \sum_{a} \, \beta^a_\Lambda \left( \mathbf{R} \right) \, \left[ S^a_{\mathbf{R}} \right]_\mathrm{vert} + \frac{1}{2} \sum_{\rho_I} \, \beta^{\rho_I}_\Lambda \, \left[ S^{\rho_I} \right] &= - 6 \, \left[ \overline{K}_{\mathcal{B}_6} \right] \cdot \left[ F_{\Lambda} \right] \label{summary4dCoho1-b} \,.
\end{empheq}
\end{subequations}
}

\subsection{Anomaly Cancellation as Cohomological Relations II}

In 4-dimensional \emph{F-theory} compactifications, the cohomological relation \cref{summary4dCoho1-b} fully describes the cancellation of mixed Abelian-gravitational anomalies whilst \cref{summary4dCoho1-a} captures all other anomalies. Interestingly, we can derive, in addition to \cref{summary4dCoho1}, another type of cohomological relations from \cref{G4dotanomaly1} and \cref{G4dotanomaly2} which are of some relevance by themselves. To arrive at these, note first that to each matter surface 2-cycle $S^a_{\mathbf{R}}$ one can associate a flux 2-cycle class
\[ \label{Aageneral1} A^a \left( \mathbf{R} \right) = S^a_{\mathbf{R}} + \Delta^a \left( \mathbf{R} \right) \in \mathrm{CH}^2 ( \hat{Y}_4 ) \,. \]
The correction factor $\Delta^a({\mathbf{R}})$ is chosen such that the cohomology class $[A^a({\mathbf{R}})] \in H^{2,2}(\hat{Y}_4)$ defines a bona-fide 4-form flux $G_4$ satisfying the two constraints \cref{transversality-gen1} and \cref{gaugeinvariantflux}. The transversality condition \cref{transversality-gen1} holds automatically by construction of the $S^a({\mathbf{R}})$.\footnote{In particular, $S^a_{\mathbf{R}}$ does not contain the components of the fibre intersected by the zero-section as the associated wrapped M2-brane states correspond to KK non-zero modes in the dual M-theory vacuum.} To implement \cref{gaugeinvariantflux}, it suffices to add the correction term \cite{Borchmann:2013hta,Bies:2017fam}
\[ \label{Aageneral2} \Delta^a \left( \mathbf{R} \right) = + \left( \beta^a \left( \mathbf{R} \right)^T_{i_I} {\mathfrak C}^{-1}_{i_I j_I} \right) \, \left. E_{j_I} \right|_{C_{\mathbf{R}}} \in \mathrm{CH}^2 ( \hat{Y}_4 ) \,. \]
Here we recall that ${\mathfrak C}_{i_I j_I}$ governs the intersection numbers of the resolution divisors as in \cref{intersectionEiEj}, with summation over repeated indices understood. The gauge anomaly relation \cref{G4dotanomaly1} can be rewritten in terms of \cref{Aageneral1} as\footnote{Note that for $\mathbf{R} = \mathbf{adj}$, $A^a(\mathbf{R}) = 0$ and therefore we can sum over all representations in the first line.}
\begin{align}
\begin{split}
\label{gaugeanomrel2}
0 &= G_4 \cdot \left( \sum_{\mathbf{R} } \sum_{a} n^a_{\Lambda \Sigma \Gamma} \left( \mathbf{R} \right) \, \left[ A^a \left( \mathbf{R} \right) \right] \, \right) \\
& \qquad + G_4 \cdot \left( \sum_{\mathbf{R} \neq \mathbf{adj}} \sum_a - n^a_{\Lambda \Sigma \Gamma} \left( \mathbf{R} \right) \, \left[ \Delta^a \left( \mathbf{R} \right) \right] \, + \frac{1}{2} \sum_{\rho_I} n^{\rho_I}_{\Lambda \Sigma \Gamma} \left[ S^{\rho_I} \right] \, \right) \\
& \qquad + G_4 \cdot \left( - 3 \, \left[ F_{(\Gamma} \right] \cdot \left[ \hat{\pi}_* \left( F_{\Lambda} \cdot F_{\Sigma)} \right) \right] \right) \,.
\end{split}
\end{align}
Similarly, the gravitational anomaly relation can be formulated as 
\begin{align}
\begin{split}
\label{gravanomrel2}
0 &= G_4 \cdot \left( \sum_{\mathbf{R} } \sum_{a} \beta^a_{\Lambda} \left( \mathbf{R} \right) \, \left[ A^a \left( \mathbf{R} \right) \right] \, \right) \\
& \qquad + G_4 \cdot \left( \sum_{\mathbf{R} \neq \mathbf{adj}} \sum_a - \beta^a_{\Lambda} \left( \mathbf{R} \right)  \, \left[ \Delta^a \left( \mathbf{R} \right) \right] \, + \frac{1}{2} \sum_{\rho_I} \beta^{\rho_I}_{\Lambda} \, \left[ S^{\rho_I} \right] \, \, \right) \\
& \qquad +G_4 \cdot \left( 6 \, \left[ F_{\Gamma} \right] \cdot \left[ \overline{K}_{B_3} \right] \right) \,.
\end{split}
\end{align}
The expression in the first line of \cref{gaugeanomrel2} is by construction orthogonal to the subspace $V_3$ of $H^{2,2}(\hat Y_4)$ given by the span
\begin{align} \label{span1}
V_3 = \text{span} \left\{ [ E_{i_I} \cdot D^{\mathrm{b}}_\alpha ], \, [ S_0 \cdot D^{\mathrm{b}}_\alpha ], \, [ D^{\mathrm{b}}_\alpha \cdot D^{\mathrm{b}}_\beta ] \right\} \subseteq H^{2,2}(\hat Y_4)
\end{align}
because the classes $[A^a(\mathbf{R})]$ represent gauge invariant fluxes which satisfy \cref{transversality-gen1} as well as \cref{gaugeinvariantflux}. The expression in  brackets in the second line, which involves the combinations \cref{Aageneral2}, is manifestly orthogonal to
\begin{align} \label{span4}
V_4 = \text{span} \left\{ [ U_A \cdot D^{\mathrm{b}}_\alpha ], \, [ S_0 \cdot D^{\mathrm{b}}_\alpha ], \, [ D^{\mathrm{b}}_\alpha \cdot D^{\mathrm{b}}_\beta ] \right\} \subseteq H^{2,2}(\hat Y_4) \,.
\end{align}
To see this note that  both $\Delta^a(\mathbf{R})$, defined in \cref{Aageneral2}, and each of the $S^\rho_I$, given in \cref{def-SrhoI}, can be expressed as a sum of terms each given by the restriction of some resolution divisor $E_{i_I}$ to a curve in the base. Such a 4-cycle class is 
 orthogonal to any class of the type listed in $V_4$.
The expression in  brackets in the second line is also manifestly orthogonal to the set of all gauge invariant fluxes $G_4$ satisfying \cref{transversality-gen1} as well as \cref{gaugeinvariantflux}.

The nature of the Green--Schwarz counterterms in the third line depends on the choice of indices $\Lambda, \Sigma, \Gamma$. Consider first the case where $\Lambda, \Sigma, \Gamma = i_I,j_J,k_K$ exclusively refer to the Cartan generators of the non-Abelian gauge group factors. As a result of \cref{intersectionEiEj} also the expression in brackets in the third line is orthogonal to $V_4$ and to the set of gauge invariant $G_4$. The sum of all three terms in brackets is orthogonal to the set of all gauge fluxes, which includes $ \text{span}_{\mathbb{C}} \{ [U_A \cdot D^{\mathrm{b}}_\alpha] \}$ and $ \text{span}_{\mathbb{C}} \{ [E_{i_I}] \cdot D^{\mathrm{b}}_\alpha] \}$. Combining these statements implies that  the sum of the terms in brackets in all three lines is orthogonal to $V_3 \cup V_4$. But since  
the sum of the terms in brackets in the second and third line is orthogonal to $V_4$ by itself, also the terms in brackets in the first line must be orthogonal to $V_4$ by themselves. Since we know from above that they are orthogonal to $V_3$ by themselves, they are hence orthogonal by themselves to $V_3 \cup V_4$. Note that this space coincides with the space $V_1$ we had defined in  \cref{span2}, $V_1 = V_3 \cup V_4$. By the same arguments as those given after \cref{span2}, orthogonality of the terms in brackets in the first line to $V_1$ implies their orthogonality to $H^{2,2}_\mathrm{vert}(\hat Y_4)$. With the same caveat as in the main text regarding the possibility that some of the matter surfaces might have a contribution in $H^{2,2}_\mathrm{rem}(\hat Y_4)$ we conclude
\[ \label{naijkA}
\sum_{\mathbf{R}} \sum_{a} n^a_{i_I j_K k_K} \left( \mathbf{R} \right) \, \left[ A^a \left( \mathbf{R} \right) \right]_{\mathrm{vert}} \, = 0 \quad \in H^{2,2} ( \hat Y_4 ) \,.
\]
If one or several of the generators $F_\Sigma$ are associated with a non-Cartan $U(1)_A$, the expression in brackets in the third line has the same orthogonality properties as those in the first line. The same arguments as before now imply that 
\[ \sum_{\mathbf{R} } \sum_{a} n^a_{A \Sigma \Gamma} \left( \mathbf{R} \right) \, \left[ A^a \left( \mathbf{R} \right) \right]_{\mathrm{vert}} \, - 3\, \left[ U_{(A} \right] \cdot \left[ \hat{\pi}_* \left( F_{\Sigma} \cdot F_{\Gamma)} \right) \right] = 0 \in H^{2,2} ( \hat Y_4 ) \label{mixedChow2} \]
and similarly
\[ \sum_{\mathbf{R} } \sum_{a} q_A \, \left[ A^a \left( \mathbf{R} \right) \right]_{\mathrm{vert}} \, + 6 \, \left[ U_A \right] \cdot \left[ \overline{K}_{B_3} \right] = 0 \in H^{2,2}( \hat Y_4 ) \,. \label{mixedChow3} \]

As promised, we have thus computed in addition to \cref{summary4dCoho1} another set of relations. Namely the following relations hold true in cohomology of $\hat{Y}_4$,\footnote{Since $A^a(\mathbf{adj}) = 0$ by construction we can sum over all representations.}
\begin{subequations} \label{summary4dCoho2}
\begin{empheq}[box=\widefbox]{align}
\sum_{\mathbf{R}} \sum_{a} n^a_{i_I j_K k_K} \left( \mathbf{R} \right) \, \left[ A^a(\mathbf{R}) \right]_\mathrm{vert} \, & = 0 \in H^{2,2} ( \hat{Y}_4 ) \label{naijkA-1} \\
\sum_{\mathbf{R}} \sum_{a} n^a_{A \Sigma \Gamma} \left( \mathbf{R} \right) \, \left[ A^a \left( \mathbf{R} \right) \right]_{\mathrm{vert}} \, - 3\, \left[ U_{(A} \right] \cdot \left[ \hat{\pi}_* \left( F_{\Sigma} \cdot F_{\Gamma)} \right) \right] &= 0 \in H^{2,2} ( \hat{Y}_4 ) \label{naijkA-2} \\
\sum_{\mathbf{R}} \sum_{a} q_A \, \left[ A^a \left( \mathbf{R} \right) \right]_{\mathrm{vert}} \, + 6 \,\left[ U_A \right] \cdot \left[ \overline{K}_{\mathcal{B}_6} \right] &= 0 \in H^{2,2} ( \hat{Y}_4 ) \,. \label{naijkA-3}
\end{empheq}
\end{subequations}
These cohomological relations \cref{summary4dCoho2} are in general independent of \cref{summary4dCoho1}. For instance pick $\Lambda, \Sigma, \Gamma = i_I, j_J, k_K$ and consider the difference of \cref{summary4dCoho1-a} and \cref{naijkA-1}. This gives
\[ \label{equ:DifferenceAnomalies}
\sum_{\mathbf{R} \neq \mathbf{adj}} \sum_a - n^a_{i_I j_J k_K} \left( \mathbf{R} \right) \, \left[ \Delta^a \left( \mathbf{R} \right) \right] \, + \frac{1}{2} \sum_{\rho_I} n^{\rho_I}_{i_I j_J k_K} \left[ S^{\rho_I} \right] \, - 3 \, \left[ F_{(i_I} \right] \cdot \left[ \hat{\pi}_* \left( F_{j_J} \cdot F_{k_K)} \right) \right] = 0  \,. \]
This relation is trivial when intersected with $[U_A] \wedge [D_\alpha^{\mathrm{b}}]$, but its intersection with $[E_{i_I}] \wedge [D_\alpha^{\mathrm{b}}]$ is equivalent to the intersection of $[E_{i_I}] \wedge [D_\alpha^{\mathrm{b}}]$ with \cref{summary4dCoho1-a}. In particular, it is \cref{summary4dCoho1-a} which encodes anomaly cancellation for the remnant gauge group after breaking $G_I \rightarrow H_I \times U(1)_{i_I}$ via $G_4 = [E_{i_I}] \wedge [D_\alpha^{\mathrm{b}}]$.

Note that the analogue of \cref{mixedChow3} with the index $A$ replaced by a Cartan index would be $\sum_{\mathbf{R} } \sum_{a} \beta^a_{i_I}(\mathbf{R}) \,  [A^a(\mathbf{R})]_{\mathrm{vert}} =0$ (without a Green--Schwarz term). This is a trivial relation since $A^a(\mathbf{R})$ is independent of $a$ and $\sum_{a} \beta^a_{i_I}(\mathbf{R})=0$ for a semi-simple Lie algebra.

Finally, let us point out that if we subtract \cref{summary4dCoho1-a} with $\Lambda = A$ referring to a non-Cartan $U(1)_A$ from \cref{mixedChow2} and use that $n^{\rho_I}_{A\Sigma \Gamma} = 0$ (since $\beta^{\rho_I}_A = 0$), we arrive at the relation
\[ \sum_{\mathbf{R} \neq \mathbf{adj}} \sum_a  n^a_{A \Sigma \Gamma} \left( \mathbf{R} \right) \, \left[ \Delta^a \left( \mathbf{R} \right) \right]  = 0 \,. \]
This is the analogue of the relation \cref{equ:DifferenceAnomalies} with  $\Lambda, \Sigma, \Gamma = i_I, j_J, k_K$. Similarly, it follows from the gravitational anomaly relations that
\[ \sum_{\mathbf{R} \neq \mathbf{adj}} \sum_a  \beta^a_{A} \left( \mathbf{R} \right) \, \left[ \Delta^a \left( \mathbf{R} \right) \right]  = 0 \,. \]

\subsection{Anomaly Cancellation as Relations in the Chow Ring}

To summarise anomaly cancellation in \emph{F-theory} compactified on a smooth, flat elliptically fibred Calabi--Yau 4-fold $\hat{Y}_4$ implies the relations \cref{summary4dCoho1} and \cref{summary4dCoho2} within the cohomology ring (over the rational numbers) of $\hat{Y}_4$. In fact, we conjecture that under suitable conditions these relations are valid not only in cohomology, but at the level of the Chow group, at least with rational coefficients. This means they hold as relations among rational equivalence classes of algebraic cycles of complex codimension two, which form the elements of  $\mathrm{CH}^2(\hat{Y}_4) \otimes \mathbb Q$. More precisely, our claim is that the relations
\begin{subequations} \label{summary4dChow1}
\begin{empheq}[box=\widefbox]{align}
\sum_{\mathbf{R} \neq \mathbf{adj}} \sum_{a} n^a_{\Lambda \Sigma \Gamma} \left( \mathbf{R} \right) \, \left. S^a_{\mathbf{R}} \right|_\mathrm{vert} \, + \frac{1}{2} \sum_{\rho_I} n^{\rho_I}_{\Lambda \Sigma \Gamma} \, S^{\rho_I} &= 3 F_{(\Gamma} \cdot \hat{\pi}_* \left( F_{\Lambda} \cdot F_{\Sigma)} \right) \label{summary4dChow1-a} \\
\sum_{\mathbf{R} \neq \mathbf{adj}} \sum_{a} \, \beta^a_\Lambda \left( \mathbf{R} \right) \, \left. S^a_{\mathbf{R}} \right|_\mathrm{vert} + \frac{1}{2} \sum_{\rho_I} \, \beta^{\rho_I}_\Lambda \, S^{\rho_I} &= - 6 \, \overline{K}_{\mathcal{B}_6} \cdot F_{\Lambda} \label{summary4dChow1-b} \,
\end{empheq}
\end{subequations}
and 
\begin{subequations} \label{summary4dChow2}
\begin{empheq}[box=\widefbox]{align}
\sum_{\mathbf{R}} \sum_{a} n^a_{i_I j_K k_K} \left( \mathbf{R} \right) \, \left. A^a \left( \mathbf{R} \right) \right|_{\mathrm{vert}} \, &= 0 \label{finalChow1} \\
\sum_{\mathbf{R}} \sum_{a} n^a_{A \Sigma \Gamma} \left( \mathbf{R} \right) \, \left. A^a \left( \mathbf{R} \right) \right|_{\mathrm{vert}} \, - 3\, U_{(A} \cdot \hat{\pi}_* \left( F_{\Sigma} \cdot F_{\Gamma)} \right) &= 0 \, \label{finalChow2} \\
\sum_{\mathbf{R}} \sum_{a} q_A \left( \mathbf{R} \right) \, \left. A^a \left( \mathbf{R} \right) \right|_{\mathrm{vert}} \, + 6 \,U_{A} \cdot \overline{K}_{\mathcal{B}_6} &= 0 \, \label{finalChow3}
\end{empheq}
\end{subequations}
hold as relations in $\mathrm{CH}^2(\hat{Y}_4) \otimes \mathbb Q$, at least in the following situation: The fibre of the smooth resolution $\hat{Y}_4$ of the elliptic fibration possesses a locally closed embedding into a toric fibre ambient space, possibly with orbifold singularities. This embedding must be such that the fibral component of every matter surface cohomology class $[S^a(\mathbf{R}) ]_{\mathrm{vert}}$ can be expressed as the pullback of a sum of cohomology classes on the fibre ambient space given by the product of ambient space divisor classes, and for these classes rational equivalence and cohomological equivalence on the fibre ambient space agree. These conditions are satisfied for instance when the underlying singular model $Y_4$ is given as a global Tate model which can be resolved crepantly, or generalisations over an arbitrary base $\mathcal{B}_6$ with non-Abelian divisors $W_I$ represented as arbitrary hypersurfaces on $\mathcal{B}_6$. We stress that a priori we only conjecture the above relations to be valid in the Chow group over the rational numbers, \ie ignoring potential torsional effects, but they may well be true more generally. The reason is that at least in our explicit proofs of the examples, we have to allow for rational coefficients and hence cannot detect potential torsion effects, but this may well be overcome more generally. From now on, whenever we write $\mathrm{CH}^\bullet(\hat{Y}_4)$ we understand that this space is taken over the rational numbers.

The importance of \cref{summary4dChow2} lies in the fact that it provides us with relations between different gauge backgrounds represented by the cycle classes $A^a(\mathbf{R})$. For instance, in \cref{chapter:MasslessSpectraAndSheafCohomology} these relations have been used in order to facilitate the computation of massless matter states in the presence of such gauge backgrounds. While we do not have a general proof that \cref{summary4dChow1} and \cref{summary4dChow2} hold at the level of the Chow ring, and not merely in cohomology, we will exemplify the validity of our conjecture in the following section in geometries derived from the $SU(5) \times U(1)_X$-top described in \cref{chapter:MasslessSpectraAndSheafCohomology} and \cref{sec:SU5xU1Top}. What simplifies this analysis is that all $A^a(\mathbf{R})$ give rise to the same 2-cycle class, \ie $A^a(\mathbf{R}) \equiv A(\mathbf{R})$ for all $a$. In addition, the gauge group is of the form $G \times U(1)_A$. This allows us to simplify the expressions further. Consider for instance \cref{finalChow2}: Let us take $\Sigma, \Gamma$ to be non-Abelian indices, \ie $\Sigma = i$ and $\Gamma = j$. Then \cref{finalChow2} takes the form
\[ \sum_{\mathbf{R}}\sum_{a}{n^a_{Aij} \left( \mathbf{R} \right) }   { \left. A \left( \mathbf{R} \right) \right|_{\mathrm{vert}}}   - U_A \cdot \hat{\pi}_\ast \left( E_i \cdot E_j \right) = 0 \in {\mathrm{CH}}^2 ( \hat{Y}_4 ) \,. \label{equ:3.16bForSU(5)xU(1)}\]
From the definition \cref{naRdef}, the trace relation \cref{trace-coroot} and the intersection numbers \cref{intersectionEiEj} it follows that
\[ \sum_a n^a_{A ij} \left( \mathbf{R} \right) = q_{\mathbf{R}} \, \mathrm{tr}_{\mathbf{R}} \mathcal{T}_i \mathcal{T}_j = q_{\mathbf{R}} c_{\mathbf{R}}^{(2)} \, \lambda \, \mathfrak{C}_{ij} \, , \qquad \hat{\pi}_\ast \left( E_i \cdot E_j \right) = - \mathfrak{C}_{ij} \cdot W \, . \]
In consequence, \cref{equ:3.16bForSU(5)xU(1)} is equivalent to
\[ \sum_{\mathbf{R}} q_{\mathbf{R}} \, c^{(2)}_{\mathbf{R}} \, \left. A \left( \mathbf{R} \right) \right|_{\mathrm{vert}} - \frac{U_A}{\lambda} \cdot (-W) = 0 \in {\mathrm{CH}}^2 ( \hat{Y}_4 ) \, . \]
By evaluating \cref{finalChow2} for $\Sigma = A, \Gamma = A$ and for $\Sigma = i$, $\Gamma = A$ and applying similar reasoning to the other two equations in \cref{summary4dChow2}, it is found that these three equations give rise to the following four anomaly conditions:
\begin{subequations} \label{anomGU1all}
\begin{align}
\sum_{\mathbf{R}} c^{(3)}_{\mathbf{R}} \, \left. A \left( \mathbf{R} \right) \right|_{\mathrm{vert}} &= 0 \in {\mathrm{CH}}^2 ( \hat{Y}_4 ) \, , \label{anom-cub-c} \\
\sum_{\mathrm{\mathbf{R}}} q_{\mathrm{\mathbf{R}}} \, c^{(2)}_{\mathbf{R}} \, \left. A \left( \mathrm{\mathbf{R}} \right) \right|_{\mathrm{vert}} + \frac{1}{\lambda} U_A \cdot W &= 0 \in {\mathrm{CH}}^2 ( \hat{Y}_4 ) \, , \label{anom-mixed2a} \\
\sum_{\mathrm{\mathbf{R}}} q^3_{\mathrm{\mathbf{R}}} \, {\mathrm{dim}} \left( \mathrm{\mathbf{R}} \right) \left. A \left( \mathrm{\mathbf{R}} \right) \right|_{\mathrm{vert}} - U_A \cdot 3 \hat{\pi}_* \left( U_A \cdot U_A \right) &= 0 \in {\mathrm{CH}}^2 ( \hat{Y}_4 ) \, , \label{anom-mixed2b} \\
\sum_{\mathrm{\mathbf{R}}} q_{\mathrm{\mathbf{R}}} \, {\mathrm{dim}} \left( \mathrm{\mathbf{R}} \right) \left. A \left( \mathrm{\mathbf{R}} \right) \right|_{\mathrm{vert}} + U_A \cdot (6 \overline{K}_{\mathcal{B}_6} ) &= 0 \in {\mathrm{CH}}^2 ( \hat{Y}_4 ) \label{anom-mixed2c} \,.
\end{align}
\end{subequations}
We will verify these relations in \cref{sec:ChowRelationsExamples}.

Finally, let us point out that the relation \cref{anom-cub-c} derived from cancellation of the cubic non-Abelian anomalies can be only as powerful as the underlying constraints from absence of anomalies themselves. A representation $\mathbf{R}$ can only contribute to the cubic non-Abelian gauge anomalies if it is complex and if the anomaly coefficient $c^{(3)}_{\mathbf{R}}$ is non-vanishing. As is well-known, the only non-Abelian gauge groups with representations satisfying both of these conditions are $SU(N)$ with $N \geq 3$ and $SO(6)$. This does not mean that there may not exist similar interesting relations between complex 2-cycles in fibrations featuring different gauge groups, but their relation to constraints in the 4-dimensional effective field theory would necessarily have to be a different one.

\section{Chow Relations Exemplified} \label{sec:ChowRelationsExamples}

In this section we prove the Chow relations \cref{summary4dChow1} and \cref{summary4dChow2} in the prototypical class of elliptically fibred Calabi--Yau 4-folds $\hat{Y}_4$ based on the top geometry discussed in \cref{sec:ToricFTheoryGUTModels} and \cref{sec:SU5xU1Top}. Recall that in this setup there are four matter curves
\begin{align}
  C_{\mathbf{10}_1} &= V \left( e_0, a_{1,0} \right) \, , \quad & C_{\mathbf{5}_3} &= V \left( e_0, a_{3,2} \right) \, , \cr
  C_{\mathbf{5}_{-2}} &= V \left( e_0, a_1 a_{4,3} - a_{2,1} a_{3,2} \right) \, , & C_{\mathbf{1}_{5}} &= V \left( a_{4,3}, a_{3,2} \right) \, .
\end{align}
which support the matter surface fluxes discussed in introduced in \cref{sec:ToricFTheoryGUTModels}. In particular, we explained that we can express the corresponding cycles as elements $\mathcal{A} \in \mathrm{CH}^2 ( \hat{Y}_5)$ such that $\left. \mathcal{A} \right|_{\hat{Y}_4} = A$ describes the original flux. In this sense the admissible fluxes are given by
\begin{align}
\begin{split}
\label{equ:Fluxes-Y5}
\mathcal{A} \left( \mathbf{10}_1 \right) &= - \frac{\lambda}{5} \, \left( 2 \cE_1 - \cE_2 + \cE_3 - 2 \cE_4 \right) \cdot\overline{\mathcal{K}}_{\mathcal{B}_6} - \cE_2 \cdot \cE_4 \, , \\
\mathcal{A} \left( \mathbf{5}_3 \right) &= - \frac{\lambda}{5} \, \left( \cE_1 + 2 \cE_2 - 2 \cE_3 - \cE_4 \right) \cdot\left( 3 \overline{\mathcal{K}}_{\mathcal{B}_6} - 2 \mathcal{W} \right) - \cE_3 \cdot\mathcal{X} \, , \\
\mathcal{A} \left( \mathbf{5}_{-2} \right) &= - \frac{\lambda}{5} \, \left( \cE_1 + 2 \cE_2 + 3 \cE_3 - \cE_4 \right) \cdot\left( 5 \overline{\mathcal{K}}_{\mathcal{B}_6} - 3 \mathcal{W} \right) \\
& \hspace{14em} + \left( \cE_3 \, \overline{\mathcal{K}}_{\mathcal{B}_6} + \cE_3 \cdot\mathcal{Y} - \cE_3 \cdot\cE_4 \right) \, , \\
\mathcal{A} \left( \mathbf{1}_{5} \right)  &=\lambda \cdot \mathcal{S} \cdot \left( 3 \overline{\mathcal{K}}_{\mathcal{B}_6} - 2 \mathcal{W} \right) - \lambda \cdot \mathcal{S} \cdot \mathcal{X} \, , \\
\mathcal{A}_X \left( \mathcal{F} \right) &= - \frac{1}{5} \, \mathcal{F} \cdot \left( 5 \mathcal{S} - 5 \mathcal{Z} - 5 \overline{\mathcal{K}}_{\mathcal{B}_6} + 2 \cE_1 + 4 \cE_2 + 6 \cE_3 + 3 \cE_4 \right) \, ,
\end{split}
\end{align}
where $\mathcal{F}$ is such that $\left. \mathcal{F} \right|_{\hat{Y}_4} = F \in \text{Cl} ( \mathcal{B}_6 )$.

We will work over a generic base $\mathcal{B}_6$ in this section. Thereby, we are able to exemplify the full structure of the relations \cref{summary4dChow1} and \cref{summary4dChow2}. Apart from supporting our conjecture concerning the general validity of these equations within the Chow ring, the following analysis will illustrate the usefulness of \cref{summary4dChow1} and \cref{summary4dChow2} for studying gauge backgrounds in \emph{F-theory}. In \cref{subsec:SU4example} we will exemplify how the relations \cref{summary4dChow2} explain the aforementioned absence of vertical fluxes for $I_n$ Tate models with $n<4$, focussing for concreteness on the most interesting case $n=4$. For this analysis we use the $SU(4)$-top, whose geometry is summarised in \cref{sec:DetailsOfSU4Top}.

\subsection{\texorpdfstring{\Cref{summary4dChow2}}{(6.60)} for \texorpdfstring{$\mathbf{SU(5) \times U(1)_X}$}{SU(5)xU(1)}}

Let us start by evaluating \cref{summary4dChow2}. We have worked these relation out in more concrete terms for a gauge group $G \times U(1)_A$ in \cref{anomGU1all}. It thus remains to evaluate these expressions. For $G= SU(N)$, the relevant values for group theoretic constants $c_{\mathbf{R}}^{(n)}$ and $\lambda$, defined in \cref{cRn-def} and \cref{lambdadef}, are (see e.g. \cite{Erler:1993zy}) \footnote{The group theoretic factor $\lambda$ must not be confused with the parameter $\lambda \in \mathbb{Q}$ used in the definition of the matter surface fluxes.}
\[ c^{(3)}_{\mathbf{\Lambda^2 N}} = N-4, \qquad c^{(2)}_{\mathbf{\Lambda^2 N}} = N-2, \qquad \lambda  = 1 \,. \]
For the $U(1)_X$ generator in the model under consideration
$\hat{\pi}_* \left( U_X \cdot U_X \right) = 30 \, W - 50 \, \overline{K}_{\mathcal{B}_6} \label{projectionUX}$.\footnote{In particular, for every base divisor $D^{\mathbf{b}}_\alpha, D^{\mathbf{b}}_\beta$, $[U_X] \cdot_{\hat{Y}_4} [U_X] \cdot_{\hat{Y}_4} [D^{\mathbf{b}}_\alpha] \cdot_{\hat{Y}_4} [D^{\mathbf{b}}_\beta] = [ \hat{\pi}_*(U_X \cdot U_X )] \cdot_{\mathcal{B}_6} [D^{\mathbf{b}}_\alpha] \cdot_{\mathcal{B}_6} [D^{\mathbf{b}}_\beta]$. This integral has been evaluated for the model at hand in \cite{Krause:2011xj}, leading to $\hat{\pi}_* \left( U_X \cdot U_X \right) = 30 \, W - 50 \, \overline{K}_{\mathcal{B}_6}$.}
With this information \cref{anomGU1all} becomes 
\begin{align}
&A \left( \mathbf{10}_1 \right) \left( \lambda \right) + A \left( \mathbf{5}_3 \right) \left( \lambda \right) + A \left( \mathbf{5}_{-2} \right) \left( \lambda \right) = 0 \, , \label{equ:Chow1ForSU5xU1}  \\
& 3  A \left( \mathbf{10}_1 \right) \left( \lambda \right) + 3 A \left( \mathbf{5}_3 \right) \left( \lambda \right) - 2  A \left( 5_{-2} \right) \left( \lambda \right) + U_A \cdot W = 0 \, , \label{equ:Chow2ForSU5xU1} \\
& 2 A \left( \mathbf{10}_1 \right) \left( \lambda \right) + 27  A \left( \mathbf{5}_3 \right) \left( \lambda \right) - 8 A \left( \mathbf{5}_{-2} \right) \left( \lambda \right) + 25  A \left( \mathbf{1}_5 \right) \left( \lambda \right) + U_A \cdot (30  \overline{K}_{\mathcal{B}_6} - 18 W) = 0 \, , \label{equ:Chow3ForSU5xU1} \\
&10 A \left( \mathbf{10}_1 \right) \left( \lambda \right) + 15  A \left( \mathbf{5}_3 \right) \left( \lambda \right) - 10  A \left( \mathbf{5}_{-2} \right) \left( \lambda \right) + 5  A \left( \mathbf{1}_5 \right) \left( \lambda \right) + 6 \cdot U_A \cdot \overline{K}_{\mathcal{B}_6} = 0\, . \label{equ:Chow4ForSU5xU1} 
\end{align}
It is readily seen that \cref{equ:Chow1ForSU5xU1} - \cref{equ:Chow4ForSU5xU1} are not independent. Rather they are equivalent to the following three linearly independent relations within $\mathrm{CH}^2(\hat{Y}_4)$:
\begin{align}
A \left( \mathbf{5}_3 \right) \left( \lambda \right) &= A \left( \mathbf{5}_{-2} \right) \left( - \lambda \right) + A \left( \mathbf{10}_1 \right) \left( - \lambda \right) \, , \label{relation1} \\
A \left( \mathbf{5}_{-2} \right) \left( \lambda \right) &= {A}_X \left( \lambda W \right) \, , \label{relation2} \\
A \left( \mathbf{1}_{5} \right) \left( \lambda \right) &= {A}_X \left( - \lambda \left( 6 \overline{K}_{\mathcal{B}_6} - 5 {W} \right) \right) + {A} \left( \mathbf{10}_1 \right) \left( \lambda \right) \label{relation3} \,.
\end{align}
These relations have been used in \cref{subsec:MasslessSpectraMSFSU5xU1} to simplify the computation of zero modes for the matter surface fluxes introduced in \cref{chapter:MasslessSpectraAndSheafCohomology}. In \cref{subsec:ProofOfChowRelationsSU5xU1} we prove that \cref{relation1}, \cref{relation2} and \cref{relation3} indeed hold true as relations between equivalence classes of algebraic 2-cycles modulo rational equivalence, \ie between elements of $\mathrm{CH}^2(\hat{Y}_4)$. The proof rests on two important properties of the Chow groups.

First, on any algebraic variety $X$, two cycles $C_1, C_2$ are rationally equivalent if and only if one can find a rationally parametrized family of cycles which interpolates between $C_1$ and $C_2$. This means that we can find a cycle $\Gamma(t)$ on $\mathbb{P}^1 \times X$ such that 
\[ \label{Gammdefintion-general} \Gamma \left( t=t_1 \right) = C_1, \qquad \quad \Gamma \left( t=t_2 \right) = C_2 \]
with $t \in \mathbb{P}^1$ parametrising the interpolation between $C_1$ and $C_2$. In other words, $C_1$ and $C_2$ are related by a `rational homotopy'. We have pictured this idea in \cref{figure-takenFrom2014Paper}.

The second property we are using is specific to the fact that the elliptic fibre of $\hat{Y}_4$ is embedded into a toric fibre ambient space. Given a regular embedding $\iota \colon X \hookrightarrow Y$ between two algebraic varieties, the pullback map is a well-defined linear map \cite{FultonInt}
\[ \label{pullbackformula} \iota^*: \mathrm{CH}^p \left( Y \right) \rightarrow \mathrm{CH}^p \left( X \right) \,. \]
This means that if two algebraic cycles $C_1, C_2 \subseteq Y$ are rationally equivalent on $Y$, then their pullbacks to $X$ are rationally equivalent on $X$.
The importance of this property of Chow groups is that for cycles arising as pullbacks from the ambient space we can use rational equivalence in the ambient space $Y$ to check for rational equivalence on $X$. This leads to drastic simplifications if the space $Y$ is a complete toric variety which in addition is simplicial or even smooth \cite{cox2011toric}: For a smooth and complete toric variety $Y$, rational equivalence coincides with homological equivalence, \ie $\mathrm{CH}^\bullet(Y)_{\mathbb{Z}}  \cong H^\bullet ( Y, \mathbb{Z} )$, and to  check rational equivalence for pullback cycles on $X$ we are hence allowed to perform operations on the pre-image of the cycle on $Y$ as long as these preserve its homology class on $Y$. Even if $Y$ is merely a complete and simplicial toric variety, we still have $\mathrm{CH}^\bullet(Y)_{\mathbb{Q}} \cong H^{\bullet}(Y, \mathbb{Q})$. These properties, which indeed hold in the example studied in this section, motivate the conditions stated after \cref{summary4dChow2} for our more general conjecture. 

Note that even though anomaly cancellation alone suffices to show that the relations \cref{summary4dChow2} and hence \cref{anomGU1all} hold true as relations in $H^\bullet(\hat{Y}_4)$, it is in general not the case that they hold already for the homology classes of the cycles $\mathcal{A}$ on the ambient space $\hat{Y}_5$ from which the cycles on $\hat{Y}_4$ descend via pullback. If this were the case, then by the above reasoning it would be immediate that the relations hold in $\mathrm{CH}^\bullet(\hat{Y}_4)$. Instead, we have to have work harder to show this latter, stronger statement.

\subsection{\texorpdfstring{\Cref{summary4dChow1}}{(6.59)} for \texorpdfstring{$\mathbf{SU(5) \times U(1)_X}$}{SU(5)xU(1)}} \label{subsec:AnomalyDifferencesForSU(5)xU(1)}

We now turn to the proof of \cref{summary4dChow1}. Since \cref{summary4dChow2} has already been established, it suffices to analyse the difference of both types of equations with compatible index structures. If all gauge indices are purely non-Abelian, the relevant expression is \cref{equ:DifferenceAnomalies}. In the present example, the LHS of \cref{equ:DifferenceAnomalies} simplifies to 
\begin{align} 
\Xi_{ijk} = &\sum_{\mathbf{R} \neq \mathbf{adj}}{ \sum_{a}{- n^a_{ijk} \left( \mathbf{R} \right) \Delta^a \left( \mathbf{R} \right) } } + \frac{1}{2} \sum_{\rho}{\beta^\rho_i \beta^\rho_j \beta^\rho_k S^\rho}  - 3 E_{(i} \cdot \hat{\pi}_\ast \left( E_j \cdot E_{k)} \right). \label{equ:RelationsDifference}
\end{align}
We will exemplify this computation for $i = j = k = 1$ and start by evaluating the first sum. With the help of the explicit form of the matter surfaces $S^a_{\mathbf{R}}$ and their associated weight vectors tabulated in \cref{subsec:FibreStructureSU5xU1} we arrive at 
\[ \sum_{\mathbf{R} \neq \mathbf{adj}}{ \sum_{a}{- \beta^a_1 \left( \mathbf{R} \right) \beta^a_1 \left( \mathbf{R} \right) \beta^a_1 \left( \mathbf{R} \right) \Delta^a \left( \mathbf{R} \right) } } = -3 \mathbb{P}^1_{14} \left( \mathbf{10}_1 \right) - \mathbb{P}^1_1 \left( \mathbf{5}_3 \right) - \mathbb{P}^1_1 \left( \mathbf{5}_{-2} \right) \, . \]
The second sum in \cref{equ:RelationsDifference} refers to the adjoint representation of $SU(5) \times U(1)_X$. The matter surfaces associated with the negative roots of $SU(5)$ are given by
\begin{align}
\begin{split}
S^\rho_{-} \left( SU \left( 5 \right) \times U( 1 )_X \right) =& \left\{ E_1, E_2, E_3, E_4, \right. \\
                                                               & \; \left. E_1 + E_2, E_2 + E_3, E_3 + E_4, \right. \\
                                                               & \; \left. E_1 + E_2 + E_3, E_2 + E_3 + E_4, \right. \\
                                                               & \;  \left. \left. E_1 + E_2 + E_3 + E_4 \right\} \right|_{K_W} \, ,
\end{split}
\end{align}
and the 10 positive roots satisfy $S_+^\rho = - S_-^\rho$. This leads to
\[ \frac{1}{2} \sum_{\rho}{\beta^\rho_1 \beta^\rho_1 \beta^\rho_1 \, S^\rho} = -11 \, E_1 \cdot \left( W - \overline{K}_{\mathcal{B}_6} \right) \, . \]
As for the third term in \cref{equ:RelationsDifference}, the analogue of \cref{intersectionEiEj} in the Chow ring implies 
\[ - 3 E_{(1} \cdot \hat{\pi}_\ast \left( E_1 \cdot E_{1)} \right) = - 3 E_1 \cdot \hat{\pi}_\ast \left( E_1 \cdot E_1 \right) = - 3 E_1 \cdot \left( -2 \right) W = 6 E_1 \cdot W \]
and altogether we find
\[ \Xi_{111} = 11 \, E_1 \cdot  \overline{K}_{\mathcal{B}_6} - 5 \, E_1 \cdot W -3 \mathbb{P}^1_{14} \left( \mathbf{10}_1 \right) - \mathbb{P}^1_1 \left( \mathbf{5}_3 \right) - \mathbb{P}^1_1 \left( \mathbf{5}_{-2} \right) \, . \]
Finally recall
\[ \mathbb{P}^1_{14} \left( \mathbf{10}_1 \right) = E_1 \cdot \overline{K}_{\mathcal{B}_6}, \quad \mathbb{P}^1_{1} \left( \mathbf{5}_3 \right) = E_1 \cdot  \left( 3 \overline{K}_{\mathcal{B}_6} - 2 W \right), \quad \mathbb{P}^1_{1} \left( \mathbf{5}_{-2} \right) = E_1 \cdot \left( 5 \overline{K}_{\mathcal{B}_6} - 3 W \right) \, . \]
This leads to $\Xi_{111} = 0 \in \mathrm{CH}^\bullet(\hat{Y}_4)$. It is simple to repeat this analysis for all $( i,j,k ) \in \{ 1, 2, 3, 4 \}^3$ and to convince oneself that $\Xi_{ijk}$ is the trivial cycle in $\mathrm{CH}^{\bullet} ( \hat{Y}_4 )$ for any such indices.

The analogue of \cref{equ:DifferenceAnomalies} for the mixed Abelian-non-Abelian, cubic Abelian and mixed gravitational anomalies are the relations
\begin{align}
\begin{split}
\displaystyle \sum_{\mathbf{R} \neq \mathrm{adj}}{ \sum_{a}{ n_{ijA}^a \left( \mathbf{R} \right) \Delta^a \left( \mathbf{R} \right) } } &= 0 \, , \qquad
\displaystyle \sum_{\mathbf{R} \neq \mathrm{adj}}{ \sum_{a}{ n_{AAA}^a \left( \mathbf{R} \right) \Delta^a \left( \mathbf{R} \right) } } = 0 \, , \\
\displaystyle \sum_{\mathbf{R} \neq \mathrm{adj}}{ \sum_{a}{ q_{\mathbf{R}} \, \Delta^a \left( \mathbf{R} \right) } } &= 0 \, .
\end{split}
\end{align}
In these equations the label `A' refers to the Abelian $U(1)_X$-group. Along the same strategy as outlined above, these expressions can be shown to vanish in $\mathrm{CH}^\bullet ( \hat{Y}_4 )$. This completes the proof of \cref{summary4dChow1} in the model at hand -- for any base $\mathcal{B}_6$.

\subsection{Fluxless \texorpdfstring{$\mathbf{SU(4)}$}{SU(4)}} \label{subsec:SU4example}

As a further amusing application we now exemplify that the absence of certain matter surface fluxes for \emph{F-theory} models in $I_n$ Tate models with $n < 5$, observed already in \cite{oai:arXiv.org:1202.3138}, can be traced back to the relation \cref{anom-cub-c}.

For brevity we only consider the model with $n=4$ in detail. A brief summary of the geometry of the $SU(4)$ Tate model is provided in  \cref{subsec:FibreStructureSU4}.
The  $SU(4)$  divisor $W$ contains two matter curves associated with massless matter in representations $\mathbf{4}$ and $\mathbf{6}$ of $SU(4)$. From the form of the $SU(N)$ index in the anti-symmetric representation, $c^{(3)}_{\mathbf{\Lambda^2 N}} = N-4$,  with $N=4$ it is clear that the matter surface flux $A(\mathbf{6})$ constructed from the matter surface $S^a_\mathbf{6}$ does not appear in \cref{anom-cub-c}. Hence, the relation in homology derived from absence of cubic anomalies is simply\footnote{It should be noted that the matter surface flux $A \left( \mathbf{4} \right)$ is a vertical flux by construction, see \cref{subsec:systematicsvertical} for more details. Hence, it is not necessary to indicate the restriction to the vertical subspace as in \cref{summary4dCoho2}.}
\[ \label{A4=0} \left[ A \left( \mathbf{4} \right) \right] = 0 \in H^{2,2} ( \hat{Y}_4 ) \,. \]
In fact, we show in \cref{subsec:ProofSU(4)} that this relation does hold also at the level of Chow groups, as stated by our more general conjecture. 

The result fits with the aforementioned observation of \cite{oai:arXiv.org:1202.3138} that this model does not allow for any matter surface fluxes. Indeed we will show in \cref{subsec:ProofSU(4)} that, in addition to \cref{A4=0} the only other candidate for a  matter surface flux $A(\mathbf{6}) = 0 \in \mathrm{CH}^2(\hat{Y}_4)$. This result, or rather its a priori weaker version in cohomology, is also related to absence of cubic anomalies, but not quite in the way written in our relation \cref{anom-cub-c}. Namely, since $\mathbf{6} = \mathbf{\overline{6}}$ the chiral index
\[ \chi(\mathbf{6}) = \left[ A \left( \mathbf{6} \right) \right] \cdot \left[ S \left( \mathbf{6} \right) \right] = 0 \, . \]
Then by anomaly cancellation clearly also $\chi(\mathbf{4}) =  [ A ( \mathbf{6} ) ] \cdot [ S ( \mathbf{4} ) ] = 0$. We can now show that this implies
\[ \left[ A \left( \mathbf{6} \right) \right] = 0 \in H^{2,2} ( \hat{Y}_4 ) \label{A6is0} \, . \]
By construction $[A(\mathbf{6})] $ is orthogonal on the subspace $V_1$ defined in \cref{span2} (recall that there are no extra sections). Furthermore, each of the generators of $V_2$ appearing in \cref{defV4} can be interpreted as the class of a matter surface plus a sum of terms in $V_1$. This is because each such element can be written as a vertical flux plus terms in $V_1$. In absence of a $U(1)$ the only vertical fluxes are the matter surface fluxes, and their explicit form $A({\mathbf{R}}) = S^a(\mathbf{R}) + \Delta^a(\mathbf{R})$ with $\Delta^a(\mathbf{R}) \in V_1$ then implies the statement. Hence, \cref{A6is0} even follows without explicit computation.

\section{Implications for Calabi--Yau 3-Folds} \label{sec:Implications6d}

The derivation of \cref{summary4dCoho1} and \cref{summary4dCoho2} for smooth, flat elliptically fibred Calabi--Yau 4-folds $\hat{Y}_4$ as presented in \cref{sec:GeneralisationToChow} rests on the absence of anomalies in the 4-dimensional low-energy effective action obtained from compactification of \emph{F-theory} on $\hat{Y}_4$. While we do not have a purely mathematical proof of these relations in general, we have verified the stronger set of relations \cref{summary4dChow1}, \cref{summary4dChow2} for explicit non-trivial examples in \cref{sec:ChowRelationsExamples}. This analysis does not rely on specific properties of the base $\mathcal{B}_6$, but only on the structure of the fibration over it. In particular, the complex dimension of the base of the fibration does not enter explicitly. Therefore, \cref{summary4dChow1} and \cref{summary4dChow2} continue to hold as geometric relations in $\mathrm{CH}^2(\hat Y_{n+1})$ over a base $B_{n}$. We conjecture that this is the case not only for the explicit examples of \cref{sec:ChowRelationsExamples}, but more generally as long as the fibration satisfies the conditions stated after \cref{summary4dChow2}.

\begin{figure}[tbp]
\centering
\includegraphics{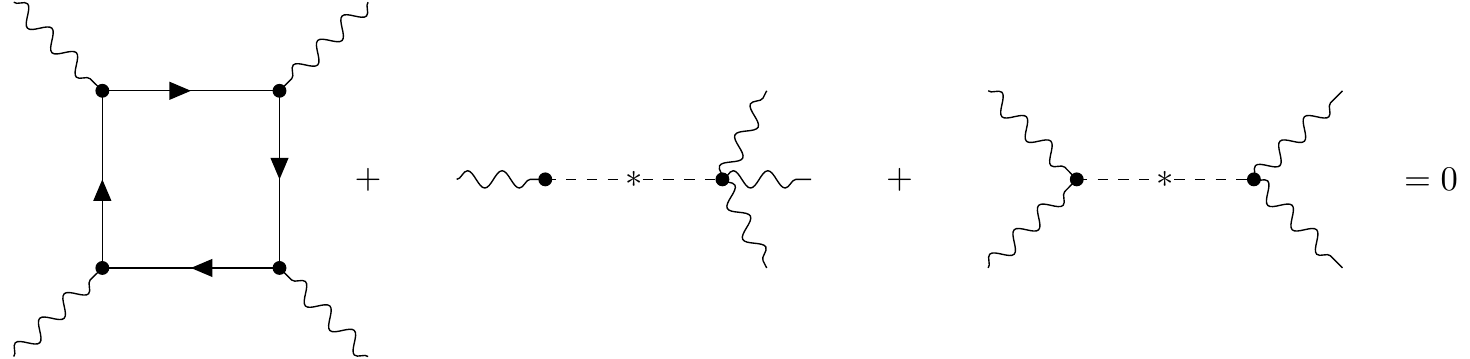}
\caption[6d quartic anomalies and their cancellation by Green--Schwarz counterterms.]{6-dimensional quartic anomalies and their cancellation via Green--Schwarz counterterms.}
\label{figure:6DAnomaly}
\end{figure}

This raises the interesting question how to interpret these relations for general complex dimension $n$ of the base $B_n$. As we will now show, for $n=2$ the weaker version \cref{summary4dCoho1} is equivalent to the consistent cancellation of all gauge and mixed gauge-gravitational anomalies in the 6-dimensional $\mathcal{N}=(1,0)$ theory obtained by compactification of \emph{F-theory} on $\hat{\pi} \colon \hat Y_3 \twoheadrightarrow B_2$. This is an intriguing result because a priori the structure of loop-induced anomalies and their Green--Schwarz counterterms in six and four dimensions is very different (compare \cref{figure:4DAnomaly} and \cref{figure:6DAnomaly}). 
The geometric manifestation of anomaly cancellation in 6-dimensional \emph{F-theory} models has been studied in great detail in the literature, most notably in \cite{Grassi:2000we,Grassi:2011hq} and \cite{Park:2011ji}, as reviewed already in the Introduction. The fact that the anomaly equations  in four and six dimensions are governed by a universal set of cohomological relations among algebraic codimension-2 cycles, however, is a new and stronger result.

For simplicity, assume a gauge group of the special form $G \times U(1)_A$. Generalisations to several gauge group factors will be immediate. The second Chow class $\mathrm{CH}^2(\hat Y_3)$ in which \cref{summary4dChow1} is valued now describes the rational equivalence class of curves on $\hat Y_3$. Specifically, the curve classes appearing in \cref{summary4dChow1} are located in the fibre over isolated points on $B_2$. For $\mathbf{R} \neq \mathbf{adj}(G)$, $S^a_{\mathbf{R}}$ describes such fibral curves over isolated points, whose Chow class we collectively denote by $p_{\mathbf{R}} \in \mathrm{CH}_0(B_2)$. M2-branes wrapping the fibre of $S^a_{\mathbf{R}}$ (in both orientations) give rise to one hypermultiplet over each of these points with weight vector $\beta^a(\mathbf{R})$ and $U(1)_A$ charge $q_A$. It will turn out useful to take the intersection product of \cref{summary4dChow1} with the divisors of $\hat Y_3$, beginning with the resolution divisors $E_i$ associated with the Cartan generators of $G$. The intersection product of $S^a_{\mathbf{R}}$ with $E_i$ within the Chow ring  on $\hat Y_3$ gives an element in $\mathrm{CH}_0(\hat Y_3)$, the Chow group of points on $\hat Y_3$. Its projection to the base describes the point class $p_{\mathbf{R}} \in \mathrm{CH}_0(B_2)$ with a multiplicity given by the intersection in the fibre. Since this fibral intersection reproduces the weight vector, we find 
\[ \label{Chowint1-6d} \hat{\pi}_\ast \left( E_i \cdot S^a_{\mathbf{R}} \right) = \beta^a \left( \mathbf{R} \right) \, p_{\mathbf{R}} \,. \]
Similarly, the intersection with the $U(1)_A$ divisor $U_A$ gives
\[ \label{Chowint2-6d} \hat{\pi}_\ast \left( U_A \cdot S^a_{\mathbf{R}} \right) = q_{\mathbf{R}} \, p_{\mathbf{R}} \,. \]
The cohomology class $x_{\mathbf{R}}  = [p_{\mathbf{R}}] \in H_0(B_2,\mathbb Z)$ equals the number of points in the Chow class $p_{\mathbf{R}}$. It coincides with the number of hypermultiplets in representation $\mathbf{R}$. Hence, at the level of cohomology,
\[ \left[ E_{i} \right] \cdot \left[ S^a_{\mathbf{R}} \right] = \beta^a_{i} \left( \mathbf{R} \right) \, x_{\mathbf{R}}, \qquad \left[ U_A \right] \cdot \left[ S^a_{\mathbf{R}} \right] = q_{\mathbf{R}} \, x_{\mathbf{R}} \,. \]
By construction, all remaining divisor classes have trivial intersection with $S^a_{\mathbf{R}}$,
\[ S_0 \cdot S^a_{\mathbf{R}} = 0 \,, \qquad D_\alpha^{\mathrm{b}} \cdot S^a_{\mathbf{R}} = 0 \qquad \forall D_\alpha^{\mathbf{b}} \in \mathrm{CH}^{1} \left( B_2 \right) \,. \]

The cycle classes $S^\mathrm{\rho}$ associated with the adjoint representation are defined as in \cref{def-SrhoI}, with $W$ now representing a curve on the base $B_2$. Hence, $S^\mathrm{\rho}$ defines a class of fibral curves located over the canonical divisor $K_W$ of $W$. For a genus $g$ curve $W$, $K_W$ is the divisor class of degree
\[ \mathrm{deg} \left( K_W \right) = - \Mint_{W}{c_1 \left( W \right) } = 2 g - 2 \,. \]
The intersections within the Chow ring are\footnote{$\iota \colon W \hookrightarrow B_2$ is the embedding of $W$ into the base $B_2$ of the elliptic fibration $\hat{Y}_3$.}
\[ \label{Chowint3-6d} \hat{\pi}_\ast \left( E_i \cdot S^\rho \right) = \rho_i \, \iota_{\ast} \left( K_W \right) \,, \qquad \hat{\pi}_\ast \left( U_A \cdot S^\rho \right) = 0 \]
and at the level of cohomology
\[ \left[ E_{i} \right] \cdot \left[ S^\rho \right] = \rho_{i} \, (2 g-2), \qquad \left[ U_A \right] \cdot \left[ S^\rho \right] = 0 \,. \]
Here $\rho_i$ is the component of the root $\rho$ with respect to the coroot $\mathcal{T}_i$ defined in \cref{subsec:GaugeTheoriesInFTheory}.

After this preparation, consider the intersection product of $E_l$ with \cref{summary4dChow1-a}, for indices $\Lambda, \Sigma, \Gamma = i, j, k$. At the level of cohomology, this yields
\[ \sum_{\mathbf{R} \neq \mathbf{adj}} \sum_a n^a_{lijk} \left( \mathbf{R} \right) \, x_{\mathbf{R}} + (g-1) \sum_{\rho} \, n^{\rho}_{lijk} - 3 \, \left[ \hat{\pi}_\ast \left( E_l \cdot E_{(i} \right) \right] \cdot_{B_2} \left[ \hat{\pi}_\ast \left( E_j \cdot E_{k)} \right) \right] = 0 \,, \]
where we have defined
\[ n^a_{lijk}(\mathbf{R}) = \beta^a_l \left( \mathbf{R} \right) \, \beta^a_i \left( \mathbf{R} \right) \, \beta^a_j \left( \mathbf{R} \right) \, \beta^a_k \left( \mathbf{R} \right), \qquad n^{\rho}_{lijk} = \rho_l \, \rho_i \rho_j \rho_k \,. \]
Summation over all weights and roots gives\footnote{The last identity is often written rather as $\mathrm{tr}_\mathbf{R} F^4 = c_\mathbf{R}^{(4)} \mathrm{tr}_\mathrm{fund} F^4 + d_\mathbf{R}^{(2)} \left(\mathrm{tr}_\mathrm{fund} F^2 \right)^2 \,$.}
\begin{align}
\begin{split}
\sum_{a=1}^{\mathrm{dim} \left( \mathbf{R} \right) } n^a_{lijk} \left( \mathbf{R} \right) &= \mathrm{tr}_{\mathbf{R}} {\mathcal T}_l \, {\mathcal T}_i \, {\mathcal T}_j \, {\mathcal T}_k = c_{\mathbf{R}}^{(4)} \mathrm{tr}_\mathrm{fund} {\mathcal T}_l \, {\mathcal T}_i \, {\mathcal T}_j \, {\mathcal T}_k + d_{\mathbf{R}}^{(2)} \mathrm{tr}_\mathrm{fund} {\mathcal T}_l \, {\mathcal T}_{(i} \, \mathrm{tr}_\mathrm{fund} {\mathcal T}_j \, {\mathcal T}_{k)} \, , \\
\sum_{\rho} n^{\rho}_{lijk} &= \mathrm{tr}_{\mathbf{adj}} {\mathcal T}_l \, {\mathcal T}_i \, {\mathcal T}_j \, {\mathcal T}_k = c_{\mathbf{adj}}^{(4)} \mathrm{tr}_\mathrm{fund} {\mathcal T}_l \, {\mathcal T}_i \, {\mathcal T}_j \, {\mathcal T}_k + d_{\mathbf{adj}}^{(2)} \mathrm{tr}_\mathrm{fund} {\mathcal T}_l \, {\mathcal T}_{(i} \, \mathrm{tr}_\mathrm{fund} {\mathcal T}_j \, {\mathcal T}_{k)} \,.
\end{split}
\end{align}
For generic indices $l,i,j,k$, the quartic and the quadratic traces are independent. Separating them yields the following two equations
\begin{align}
\begin{split} \label{xRcRequ}
\sum_{\mathbf{R} \neq \mathbf{adj}} x_{\mathbf{R}} \, c_{\mathbf{R}}^{(4)} + (g-1) \, c_\mathbf{adj}^{(4)} &= 0 \, , \\
\sum_{\mathbf{R} \neq \mathbf{adj}} x_{\mathbf{R}} \, d_{\mathbf{R}}^{(2)} + (g-1)  \,  d_\mathbf{adj}^{(2)} &= 3 \, \frac{\left[ W \right]}{\lambda} \cdot_{B_2} \frac{\left[ W \right]}{\lambda} \,,
\end{split}
\end{align}
where we used \cref{intersectionEiEj} and \cref{trace-coroot} to bring the second equation into this form. \Cref{xRcRequ} coincides with the conditions for cancellation of the non-factorisable and, respectively, factorisable quartic non-Abelian anomalies, as listed \eg in section 2 of \cite{Park:2011ji}. The RHS of the last equation is the Green--Schwarz counterterm for the factorisable part. If we intersect $[U_A]$ with \cref{summary4dChow1-a}, for $\Lambda, \Sigma, \Gamma = i, j, k$, the same logic gives, at the level of cohomology,
\[ \sum_{\mathbf{R}} q_{\mathbf{R}} \, c^{(3)}_{\mathbf{R}} \, x_{\mathbf{R}} = 0 \,. \]
This is nothing but the condition for cancellation of the mixed $U(1)_A - G^3 $ gauge anomaly in the 6-dimensional effective action.

This logic can be repeated for all conditions \cref{summary4dCoho1} with the final result 
\begin{subequations} \label{Y3gaugeweak}
\begin{align}
\sum_{\mathbf{R} \neq \mathbf{adj}} x_{\mathbf{R}} \, c_{\mathbf{R}}^{(4)} + (g-1) \, c_{\mathbf{adj}}^{(4)} &= 0 \,, \\
\sum_{\mathbf{R} \neq \mathbf{adj}} x_{\mathbf{R}} \, d_{\mathbf{R}}^{(2)} + (g-1) \, c_{\mathbf{adj}}^{(2)} - 3 \, \frac{\left[ W \right]}{\lambda} \cdot_{B_2} \frac{ \left[ W \right] }{\lambda} &= 0 \,, \\
\sum_{\mathbf{R}} q_{\mathbf{R}} \, c^{(3)}_{\mathbf{R}} \, x_{\mathbf{R}} &= 0 \,, \\
\sum_{\mathbf{R}} q_{\mathbf{R}}^2 c_{\mathbf{R}}^{(2)}  \, x_{\mathbf{R}} + \frac{1}{\lambda} \hat{\pi}_\ast \left( \left[ U_A \right] \cdot \left[ U_A \right] \right) \cdot_{B_2} \left[ W \right] &= 0 \,, \\
\sum_{\mathbf{R}} q_{\mathbf{R}}^4 \mathrm{dim} \left( \mathbf{R} \right) \, x_{\mathbf{R}} + 3 \, \hat{\pi}_\ast \left( \left[ U_A \right] \cdot \left[ U_A \right] \right) \cdot_{B_2} \hat{\pi}_\ast \left( \left[ U_A \right] \cdot \left[ U_A \right] \right) &= 0
\end{align}
\end{subequations}
and
\begin{subequations} \label{Y3gaugebravweak}   
\begin{align}
\sum_{\mathbf{R} \neq \mathbf{adj}} c_{\mathbf{R}}^{(2)} \, x_{\mathbf{R}} + (g-1) \, c_\mathbf{adj}^{(2)} + 6 \frac{\left[ W \right]}{\lambda} \cdot_{B_2} \left[ \overline{K}_{\mathcal{B}_6} \right] &= 0 \,, \\
\sum_{\mathbf{R}} q_{\mathbf{R}}^2 \mathrm{dim} \left( \mathbf{R} \right) \, x_{\mathbf{R}} + 6 \hat{\pi}_\ast \left( \left[ U_A \right] \cdot \left[ U_A \right] \right) \cdot_{B_2} \left[ \overline{K}_{\mathcal{B}_6} \right] &= 0 \,.
\end{align}
\end{subequations}
These equations describe precisely the conditions for cancellation of the gauge and the mixed gauge-gravitational anomalies including the correct Green--Schwarz counterterms \cite{Park:2011ji}. Generalisations to several Abelian and non-Abelian gauge group factors are straightforward.

Let us summarize the logic so far: 4-dimensional anomaly cancellation implies \cref{summary4dCoho1} on any smooth elliptically fibred Calabi--Yau 4-fold $\hat{Y}_4$. Assuming that \cref{summary4dCoho1} holds more generally on any smooth elliptically fibred Calabi--Yau 4-fold $\hat Y_{n+1}$, as suggested by the considerations of \cref{sec:ChowRelationsExamples}, we have derived the 6-dimensional anomaly cancellation conditions from \cref{summary4dCoho1} interpreted as relations in $H^{2,2}(\hat Y_3)$. We can now turn tables round and take anomaly cancellation in six dimensions as the \emph{starting point} to \emph{derive} \cref{summary4dCoho1} on $\hat Y_3$: Namely, the cohomology class of a curve in $H^{2,2}(\hat Y_3)$ is trivial if and only if its cohomological intersection with every element in $H^{1,1}(\hat Y_3)$ vanishes. By construction, \cref{summary4dCoho1} on $\hat Y_3$ is orthogonal to any base divisor class $[D_\alpha^{\mathrm b}]$ as well as to the zero-section $S_0$. Intersection with $[U_A]$ and $[E_{i_I}]$ gives rise to a set of equations which, as just shown, are nothing but the 6-dimensional anomaly equations. Hence, anomaly cancellation implies \cref{summary4dCoho1} at the level of $H^{2,2}(\hat Y_3)$. This statement is of course in the spirit of the intersection theoretic identities derived from 6-dimensional anomaly cancellation in \cite{Park:2011ji}.

\begin{figure}[tbp]
\centering
\includegraphics{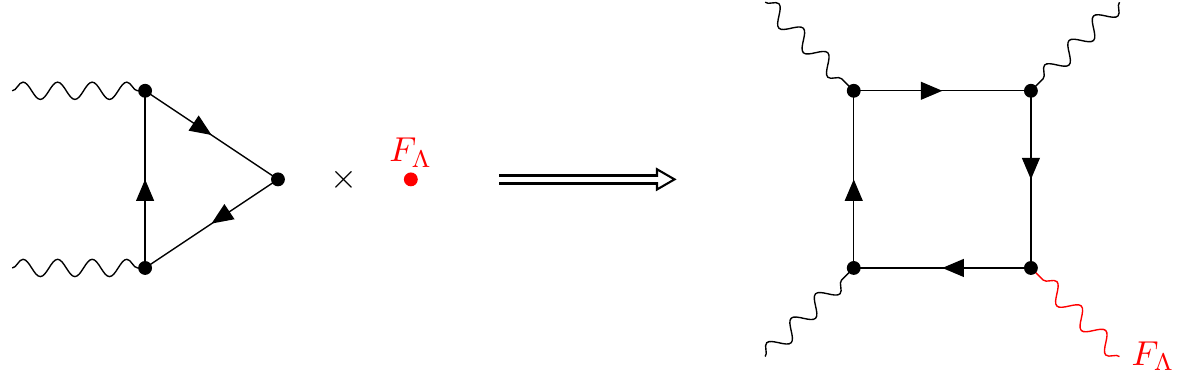}
\caption[Transition from 4-dimensional anomalies to 6-dimensional anomalies.]{The transition from the 4-dimensional to the 6-dimensional anomalies: Intersecting Chow classes encoding the structure of 4-dimensional cubic anomalies with the $U(1)_\Lambda$ divisor $F_\Lambda$  corresponds to adding an external leg to the loop diagram. A similar relation exists for the Green--Schwarz counterterms.}
\label{figure:4Dto6DAnomaly}
\end{figure}

The fact that the same cohomological relations govern anomaly cancellation in four and six dimensions is intriguing, but in retrospect maybe not completely surprising: The gauge and mixed gauge gravitational anomalies in four dimensions are generated by cubic Feynman diagrams. Intuitively speaking, intersecting the cohomological relations on $\hat{Y}_4$ (which incorporate these diagrams geometrically) with the divisors $E_{i_I}$ and $U_A$ adds an extra external leg for the associated gauge field (see \cref{figure:4Dto6DAnomaly}). The resulting relation hence encodes  the information of the quartic box diagrams underlying the structure of anomalies in six dimensions. 

Particularly interesting is furthermore the role of the cycles $S^{\rho}$ appearing in the relations \cref{summary4dCoho1}: In four dimensions these terms are required for cancellation of all anomalies associated with the possible subgroups of a geometrically realized gauge group $G_I$ once it is broken by gauge flux. In six dimensions, no such breaking of the gauge group by fluxes can occur, but  at the same the states in the adjoint representation, which are accounted for by $S^\rho$, contribute to the $G_I$ anomalies due to their quartic nature. In this sense the more complicated structure of the anomalies in six dimensions is shadowed by the possibility of gauge group breaking flux in four dimensions. 

We can, however, go even further: According to our conjecture, on any $\hat Y_3$ \cref{summary4dChow1} holds at the level of Chow classes (at least with rational coefficients and subject to the conditions stated after \cref{summary4dChow2}. We can now consider the intersection of \cref{summary4dChow1} with $E_i$ and $U_A$ within the Chow ring and project the result to the base $B_2$, using \cref{Chowint1-6d}, \cref{Chowint2-6d} and \cref{Chowint3-6d}. This gives rise to a set of relations analogous to \cref{Y3gaugeweak} and \cref{Y3gaugebravweak} (and their obvious generalisations) valued in $\mathrm{CH}_0(B_2)$, the Chow group of points on the base. These are obtained from  \cref{Y3gaugeweak} and \cref{Y3gaugebravweak} by replacing $x_\mathbf{R}$ by the point class $p_\mathbf{R}$ and $g-1$ by $\frac{1}{2} K_W$, interpreted via pushforward as a Chow class on $B_2$. Finally, the cohomological intersection product $\cdot_{B_2}$ in the Green--Schwarz terms is to be replaced by the intersection in the Chow ring of $B_2$. Though formulated slightly differently, a set of relations among Chow classes of points on $B_2$ has been discussed in \cite{Grassi:2011hq} and shown to imply (non-Abelian) anomaly cancellation.

Note that once \cref{summary4dCoho1} is established to hold on elliptic Calabi--Yau 3-folds, their cousin relations \cref{summary4dCoho2} follow in the same way as on elliptic 4-folds. On Calabi--Yau 3-folds, however, \cref{summary4dCoho2} does not have an interpretation as relations among valid gauge backgrounds as \emph{F-theory} compactifications to six dimensions do not allow for such backgrounds. It would be interesting to see if one can nonetheless attribute a physical meaning to these equations (and their conjectured stronger version \cref{summary4dChow2} also on Calabi--Yau 3-folds.

\section{Summary}

In this chapter we have studied anomaly cancellation in \emph{F-theory} compactifications. We have taken the results of \cite{Cvetic:2012xn} and generalised them in several directions. First we have generalised these results of non-gauge-invariant fluxes in \cref{sec:AnomalyCancellation4D}. We found that the gauge anomalies are captured by \cref{G4dotanomaly1} and the gravitational-anomalies are described by \cref{G4dotanomaly2}. These results state that the homological intersection pairing $C \cdot G_4$ of a certain (co-)homology class $C$ with any $G_4$-fluxes satisfying \cref{transversality-gen1} will vanish.

In \cref{sec:GeneralisationToChow} we then argued that actually this cohomology class $C$ is trivial by itself. In this sense \cref{summary4dCoho1-a} describes the gauge anomalies and \cref{summary4dCoho1-b} the gravitational anomalies. Although these equation are the natural analogues of \cref{G4dotanomaly1}, \cref{G4dotanomaly2} we found that anomaly cancellation implies yet another set of relations, which are in general independent of \cref{summary4dCoho1-a}, \cref{summary4dCoho1-b}. These `new' relations are summarised in \cref{summary4dCoho2}. To understand the meaning of these additional relations, we first conjectured a yet more strict condition on anomaly cancellation. Rather than stating that cohomologically \cref{summary4dCoho1-a}, \cref{summary4dCoho1-b} and the `new' relations \cref{summary4dCoho2} hold true, we actually conjecture that the defining cycles are zero in the Chow ring. We then conjecture that \cref{summary4dChow1} and \cref{summary4dChow2} are to describe anomaly cancellation. In \cref{sec:Implications6d} we even generalised these ideas to 6-dimensional \emph{F-theory} compactifications.

To date, we cannot give a mathematical proof of our conjecture. However, we can gather some evidence by studying examples. Therefore we have investigate a class of F-theory compactifications in \cref{sec:ChowRelationsExamples}. The geometries of these compactifications were derived from the $SU(5) \times U(1)_X$-top discussed in \cref{chapter:MasslessSpectraAndSheafCohomology} and \cref{sec:SU5xU1Top} (\cf \cite{Krause:2011xj}). For this class of compactifications we proved our conjecture and argued that the `new' relations \cref{summary4dCoho2} imply \cref{relation1}, \cref{relation2}, and \cref{relation3} which describe relations among the admissible gauge backgrounds.

\chapter{Conclusion and Outlook}

\paragraph{Chow Groups as Gauge Backgrounds}
In this thesis we have investigated \emph{F-theory} compactifications to four dimensions. A compactification like this is encoded in the geometry of a singular torus fibration $\pi \colon Y_4 \twoheadrightarrow \mathcal{B}_6$. Throughout this work we have assumed that this is an elliptic fibration which allows for a flat, smooth and crepant resolution $\hat{\pi} \colon \hat{Y}_4 \twoheadrightarrow \mathcal{B}_6$ which is Calabi--Yau.

An \emph{F-theory} vacuum is not specified by the geometry of $Y_4$ alone, but additionally a gauge background has to be specified. In M-theory this is taken care of by the 3-form field $C_3$. As \emph{F-theory} is defined as a special limit of M-theory, which we explained in \cref{subsec:DefinitionOfFTheory}, we can invoke from $C_3$ the existence of a 4-form flux $G_4 \in H^{2,2} ( \hat{Y}_4 )$ in \emph{F-theory} \cite{Becker:1996gj, Sethi:1996es, Dasgupta:1999ss}. 

Unfortunately, this flux $G_4$ merely describes the `field strength' of the gauge background present in an \emph{F-theory} compactification. To obtain the full gauge data, there exist essentially two approaches. On the one hand Cheeger-Simons differential characters were proposed to fulfil this role \cite{Cheeger-Simons}, and on the other hand Deligne cohomology $H^4_D ( \hat{Y}_4, \mathbb{Z} ( 2 ) )$ was applied to described the full gauge data -- in M-theory in \cite{Diaconescu:2003bm, Freed:2004yc, oai:arXiv.org:hep-th/0409158, oai:arXiv.org:1203.6662} and in \emph{F-theory} compactifications in \cite{Donagi:2011jy, Clingher:2012rg, Anderson:2013rka}.

We follow the latter approaches and make use of Deligne cohomology. As explained in \cite{Bies:2014sra} a subset of the Deligne cohomology can be accessed from the Chow group $\mathrm{CH}^2 ( \hat{Y}_4 )$. We revised this subject in \cref{sec:TheChowRing}. The overall process is nicely summarised in the following commutative diagram:
\[ \includegraphics[valign = c]{Pic70.pdf} \]

\paragraph{Divisors on Matter Curves induced from Chow Classes}
In \cite{Bies:2014sra} it was argued that a gauge background modelled by $A \in CH^2( \hat{Y}_4 )$ leads to a divisor $D( S^a_{\mathbf{R}}, A )$ on the matter curve $C_{\mathbf{R}}$. In this expression $S^a_{\mathbf{R}}$ is the matter surface over $C_{\mathbf{R}}$ corresponding to the a-th state in representation $\mathbf{R}$ (\cf \cref{subsec:GaugeTheoriesInFTheory}). The identification of this divisor makes use of the intersection product in the Chow ring, which we explained in \cref{sec:TheChowRing}. Even more it was argued in \cite{Bies:2014sra} that the sheaf cohomologies of the line bundle
\[ L \left( S^a_{\mathbf{R}}, A \right) = \mathcal{O}_{C_{\mathbf{R}}} \left( D \left( S^a_{\mathbf{R}}, A \right) + \frac{1}{2} K_{C_{\mathbf{R}}} \right) \label{equ:BundleCountingZeroModes} \]
counts the chiral- and anti-chiral zero modes in the a-th state of representation $\mathbf{R}$, which are localised on $C_{\mathbf{R}}$ in the presence of the gauge background $A$. In this expression $\frac{1}{2} K_{C_{\mathbf{R}}}$ denotes -- in an abuse of terminology -- a spin divisor on $C_{\mathbf{R}}$. On a Riemann surface $M_g$ of genus $g$ there exist $2^{2g}$ different spin structures \cite{atiyah1971riemann, mumford1971theta}. Hence, this spin bundle is a priori not unique! This ambiguity was discussed in \cite{Bies:2014sra}. It was argued that the global consistency of the compactification requires the spin bundle to be pullback from $\mathcal{B}_6$. Therefore, the holomorphic embeddings $C_{\mathbf{R}} \hookrightarrow \mathcal{B}_6$ fix this spin bundle uniquely. For convenience of the reader we have explained this conclusion in \cref{subsec:LocalisedMatterAndTheSpinBundle}.

\paragraph{The Need for Coherent Sheaves}
In this work we have systematically worked out the divisors $D( S^a_{\mathbf{R}}, A )$ for an extended class of vertical fluxes. In \cref{subsec:systematicsvertical} we have introduced the notion of a matter surface flux, for which an intuitive approach to the divisors $D( S^a_{\mathbf{R}}, A )$ exists. This is explained in  \cref{subsec:NewStrategyForMasslessSpectra}. In this section we have also explained how we can compute the divisors $D( S^a_{\mathbf{R}}, A )$ for other types of vertical fluxes. For example, this includes gauge backgrounds in the presence of $U(1)$ gauge group factors -- such fluxes were already considered in \cite{Bies:2014sra}. In addition, we can look at slightly more exotic fluxes than vertical fluxes. This includes the hypercharge flux $A_Y ( \mathcal{H} )$ which gives rise to an element $G_4 \in H^{2,2}_{\mathrm{rem}} ( \hat{Y}_4, \mathbb{R} )$\footnote{Recall that the cohomology group $H^{2,2}(\hat{Y}_4)$ enjoys a decomposition into vertical, horizontal and remaining fluxes via $H^{2,2}( \hat{Y}_4,\mathbb{R} ) =  H^{2,2}_\mathrm{vert}(\hat{Y}_4,\mathbb{R} ) \oplus H^{2,2}_\mathrm{hor}(\hat{Y}_4,\mathbb{R} )  \oplus H^{2,2}_\mathrm{rem}(\hat{Y}_4, \mathbb{R})$.}.

In order to study explicit examples of \emph{F-theory} compactifications, we have worked out the divisors $D( S^a_{\mathbf{R}}, A )$ for all fluxes present in geometries induced by two so-called toric tops. This includes the $SU(5) \times U(1)_X$-top -- originally presented in \cite{Krause:2011xj} and discussed with more refined techniques, such as primary decompositions, in \cref{sec:SU5xU1Top}. This geometry allows for a $U(1)_X$-flux and four matter surface fluxes. The associated divisors are summarised in \cref{table-N5}. Also the hypercharge flux $A_Y ( \mathcal{H} )$ can be considered in this geometry. In the form presented in \cite{Donagi:2008kj,Mayrhofer:2013ara,Braun:2014pva} this flux leads to the divisors listed in \cref{table-N10}.

The results in \cref{table-N5} showed that the line bundle $L ( S^a_{\mathbf{R}}, A )$ can in general not be described as pullback line bundle \eg from $\mathcal{B}_6$. Even more, \emph{F-theory} GUT-models require to break the gauge group down to $SU(3) \times SU(2) \times U(1)$. This is achieved by giving a VEV to the hypercharge flux. As argued in \cite{Braun:2014pva}, this flux must be modelled by a special non-pullback line bundle on the GUT-surface $W$.

The methods of \emph{cohomCalg} \cite{Blumenhagen:2010pv, Blumenhagen:2010ed, Blumenhagen:2011xn, KoszulExtensionManual, cohomCalg:Implementation, 2011JMP....52c3506J, Rahn:2010fm} or the techniques developed for heterotic compactifications in \cite{Anderson:2008ex} unfortunately both assume that we are interested in pullback line bundles only. Consequently, these techniques do not apply, and we need to investigate more general techniques.

To resolve this problem, the prime assumption in this thesis is that $\mathcal{B}_6$ is embedded into a toric space $X_\Sigma$. Any line bundle on a subvariety $C \subseteq \mathcal{B}_6$ -- be it pullback from $\mathcal{B}_6$ or not -- can be extended by zero outside of $C$. The so-constructed object is known in the mathematics literature as a coherent sheaf $\mathcal{F}$ on $X_\Sigma$. Coherent sheaves include line bundles and vector bundles, but also encompass sky-scraper sheaves, ideal sheaves and even T-branes in the language of \cite{Collinucci:2014qfa}. This observation hence leads to the following question: How can we compute the sheaf cohomologies of coherent sheaves $\mathcal{F}$ on a toric space $X_\Sigma$?

\paragraph{Algorithmic Approach to Sheaf Cohomologies of Coherent Sheaves}

Our approach is based on the works \cite{2010arXiv1003.1943B, 2012arXiv1202.3337B, 2012arXiv1210.1425B, 2012arXiv1212.4068B, 2014arXiv1409.6100B, BL_GabrielMorphisms} of Mohamed Barakat and his collaborators at the mathematics department of the \emph{University of Siegen}. In collaboration with this research group, we have added software packages to the \emph{homalg\_project}: In the language of categorical programming \cite{PosurDoktor, GutscheDoktor, CAP} we have implemented the 
category of finitely presented (f.p.) graded $\mathbf{S}$-modules -- see \cite{CAPCategoryOfProjectiveGradedModules, CAPPresentationCategory, PresentationsByProjectiveGradedModules, TruncationsOfPresentationsByProjectiveGradedModules} -- which enables us to model coherent sheaves in the computer via the sheafification functor \cite{cox2011toric}
\begin{align*} \widetilde{\phantom{m}} \colon S \mathrm{\textnormal{-}fpgrmod} \to \mathfrak{Coh} X_\Sigma \, , \label{redundant_description} \end{align*}
where $S$ is the coordinate ring of $X_\Sigma$, commonly referred to as the \emph{Cox ring}. We have explained the category $S \mathrm{\textnormal{-}fpgrmod}$ and the sheafification functor in much detail in \cref{chapter:DetailsOnFPGradedSModules}. The final task is now twofold: 
\begin{itemize}
 \item Given a divisor $D \in \text{Cl} ( C_{\mathbf{R}} )$, how can we construct \fp graded $S$-modules $M_{pm}$ such that $\tilde{M_{\pm}}$ is supported on 
      $C_{\mathbf{R}}$ only and satisfies $\left. \tilde{M_{\pm}} \right|_{C_{\mathbf{R}}} \cong \mathcal{O}_{C_{\mathbf{R}}} ( \pm D )$? The answer to this question is given in \cref{subsec:LineBundlesFromModules}.
 \item Given a \fp graded $S$-module $M$, how can we compute the sheaf cohomologies of $\tilde{M}$ from the data defining $M$? To answer this question, we have 
      formulated an algorithm to compute the sheaf cohomologies of coherent sheaves. Our general philosophy follows \cite{1998math......7170S, Maclagan03multigradedcastelnuovo-mumford, Oberwolfach}. In contrast however, we rely on \emph{cohomCalg} \cite{Blumenhagen:2010pv, Blumenhagen:2010ed, Blumenhagen:2011xn, KoszulExtensionManual, cohomCalg:Implementation, 2011JMP....52c3506J, Rahn:2010fm} to compute vanishing sets, both on smooth, complete and simplicial, projective toric varieties. The so-obtained vanishing sets form properly refined versions of the semigroup $\mathbb{\mathcal{K}}^{\mathrm{sat}}$ which was introduced in \cite{Maclagan03multigradedcastelnuovo-mumford} and put to use in \cite{Oberwolfach} to propose a means to compute sheaf cohomologies on smooth, projective toric varieties. Eventually, this approach lead to \cref{mytheorem} which applies on smooth and complete toric varieties.

      As outlined in \cref{sec:ComputeCohomologies}, we have specialised this theorem to compute sheaf cohomologies of coherent sheaves on smooth and complete normal toric varieties $X_\Sigma$. The corresponding algorithm is implemented in \cite{SheafCohomologyOnToricVarieties}. More details on this algorithm can be found in \cref{chapter:MathDetailsSheafCohomologies}.
\end{itemize}
In \cref{sec:ComputingTheSpectra} we have used this implementation to compute the zero modes of an \emph{F-theory} toy model. In this example geometry, we compute the number of zero modes localised on \emph{singular} matter curves -- our approach is not bound to smooth matter curves. Even more, this algorithm explicitly takes all complex structure moduli of the matter curves into account. By repeating this very computation for different choices of complex structure moduli we observed that the number of zero modes strongly depends on these moduli, as summarised in \cref{table-resultsScanOverComplexStructureModuliSpace}. These jumps in the moduli space are expected -- general field theory reasoning identifies them with the lifting of vector-like pairs as we vary the vacuum expectation value of some of the chiral fields of the model. In heterotic compactifications, such effects have been studied in \cite{Braun:2005xp, Bouchard:2005ag, Anderson:2013qca}. It is particularly interesting to determine the minimal number of vector-like pairs as we vary the complex structure moduli and to interpret this result in terms of an effective field theory.

The computation of $M_{\pm}$ as explained in \cref{subsec:LineBundlesFromModules} involves Gröbner basis computations. Similarly, computing a generating set of the cohomologies of the associated coherent sheaf $\tilde{M_{\pm}}$ involves such computations. Both complete in a timely fashion only if the polynomials defining the matter curves $C_{\mathbf{R}}$ and the divisors $D( S^a_{\mathbb{R}}, A )$ are simple. For example, the computations summarised in \cref{table-resultsScanOverComplexStructureModuliSpace} have been performed at non-generic points in moduli space, at which the corresponding polynomials became simple enough for our computations to complete in a timely fashion. Hence, this explicit moduli dependence is both a curse and a blessing:
\begin{itemize}
 \item It is a blessing as it allows us to handle \emph{singular} matter curves with ease. In particular, we can investigate how the number of zero modes depends on the 
      complex structure.
 \item It is a curse, since the defining polynomials often consist of many (\eg hundreds of) monomials. In consequence, the Gröbner basis computations require a lot 
      of computational time and resources.
\end{itemize}

If we are interested in computing a generating set of the sheaf cohomologies, then many Gröbner basis computations are required. If in contrast, only the number of generators is required, then a number of the involved Gröbner basis computations can be replaced by Gau{\ss} eliminations. Experimental evidence indicates that the latter approach outperforms the former, owing in essence to state-of-the-art implementations of Gau{\ss} eliminations in \texttt{MAGMA} \cite{MR1484478}. Details can be found in \cref{sec:ExtOfFPModules}.

\paragraph{Application to \emph{F-Theory} GUT-Models}

In \cref{chapter:GUTModels} we have applied the implementation \cite{SheafCohomologyOnToricVarieties} to study \emph{F-theory} GUT-models. The involved geometries were derived from the $SU(5) \times U(1)_X$-top presented in \cite{Krause:2011xj} (see \cref{sec:SU5xU1Top} for more details on the involved geometry). The most general flux in this geometry, which we know how to access as of this writing, is of the form
\[ A = A_X \left( F \right) + A \left( \mathbf{10}_1 \right) \left( \lambda \right) + A_Y \left( \mathcal{H} \right) \, \]
where $A_X( F )$ is the flux associated to the $U(1)_X$-gauge symmetry, $A ( \mathbf{10}_1 ) ( \lambda )$ the matter surface flux of $C_{\mathbf{10}_1}$ \footnote{All other matter surface fluxes are equivalent to this flux in a sense explained in \cref{chapter:LocalAnomaliesInF-Theory}. We will discuss this equivalence below.} and $A_Y( \mathcal{H} )$ the hypercharge flux used to induce the gauge group breaking
\[ SU( 5 ) \times U ( 1 )_X \to SU(3 ) \times SU(2 ) \times U(1)_X \times U(1)_Y \, . \label{equ:GaugeGroupBreakingII} \]
This breaking can already be anticipated from \cref{table-netspectrum}. In particular, we can interpret the zero modes in terms of the \emph{standard model} particles as outlined in \cref{table-brokenStates}.

Ideally, the gauge group reduction in \cref{equ:GaugeGroupBreakingII} should leave the $U(1)_Y$ gauge boson massless in the external space $\mathcal{E}_4$. In addition, it is phenomenologically appealing to rule out the existence of exotic states propagating along the GUT-divisor $W$. Combining these requirements poses a very strict condition on the divisor class $\mathcal{H} \subseteq W$, which supports the hypercharge flux. Our wish to compute the zero modes of such an \emph{F-theory} compactification comes, as a result of our lack of understanding of the Picard group on the matter curves sufficiently well, with a number of sufficient (yet not necessary) structural demands. These structural demands cannot be matched with the strict conditions on the curve $\mathcal{H}$, as we explained in \cref{sec:ChoiceOfHyperchargeFluxAndExotics}. To demonstrate the computational powers of our algorithm nonetheless, we have accepted exotic bulk states. With a view towards string phenomenology, the studied geometries can therefore merely serve as toy models.

We studied two \emph{F-theory} GUT-models -- in \cref{sec:dP3-Example} one in which the GUT-surface $W$ is a $dP_3$-surface, and in \cref{sec:dP7-Example} one with $W \cong dP_7$. In both examples we have then taken $\mathcal{H}$ as difference $E_i - E_j$. This choice leaves the $U(1)_Y$-gauge boson massless provided that $E_i$, $E_j$ and their difference $E_i - E_j$ are not pullbacks from $\mathcal{B}_6$ \cite{Braun:2014pva}. This choice of hypercharge flux leads to 24 chiral- and anti-chiral exotic bulk states in representation $( \mathbf{3}, \mathbf{2} )_{q_X, 5_Y}$ and $( \overline{\mathbf{3}}, \mathbf{2} )_{q_X, -5_Y}$. More details are given in \cref{sec:ChoiceOfHyperchargeFluxAndExotics}.

A $dP_3$-surface can be realised as toric variety directly. We have conjectured an embedding
\[ \iota \colon \text{ toric model of } dP_3 \text{ surface } \hookrightarrow \mathcal{B}_6 \]
in \cref{sec:dP3-Example}, along which we translate many computations from the more complicated toric ambient space $X_\Sigma$ of $\mathcal{B}_6$ onto this toric model of the $dP_3$-surface. As a consequence, we were able to compute zero modes for many fluxes and different choices of complex structures, including maximally generic ones. 

Our conjecture is based on a basic finding in algebraic geometry -- a (special) ring homomorphism of $\mathbb{Z}$-graded rings induces a mapping of the associated projective schemes, see for example \cite[p.80  ff]{hartshorne1977algebraic}. In this sense, we conjecture that (special) ring homomorphisms of the Cox rings of smooth, projective toric varieties induce mappings between these varieties, which need not be toric in general. We formulated this conjecture in \cref{conj:II}. Whilst we cannot provide a proof for this conjecture, as of this writing, we were merely interested in a single such embedding in \cref{sec:dP3-Example}. This is the mapping proposed around \cref{equ:ConjecturedEmbedding}. This morphism has undergone a number of rather non-trivial consistency checks, ranging from intersections of divisors, over sheaf cohomology computations to genera of matter curves. As we found matching properties in all these cases, we remain positive that at least in this instance our conjecture holds true. We reserve a more detailed analysis for future research.

The limits our algorithm were tested in \cref{sec:dP7-Example} -- here $W \cong dP_7$ and so the GUT-surface cannot be realised as a toric variety directly. Consequently, we had to perform the necessary computations from a toric ambient space $\tilde{X}_\Sigma$ of strictly greater dimension than $W$. As a consequence, our predictive powers were severely limited. In \cref{subsec:ComputeZeroModesdP7} we looked at a flux, whose zero modes can be determined from the Kodaira vanishing theorem alone -- the computations with \cite{SheafCohomologyOnToricVarieties} required so very many computational resources that we were only able to compute the exact charged massless spectrum of states in representation $( \mathbf{3}, \mathbf{2} )_{1_X, 1_Y}$. The computed result matched the finding from the Kodaira vanishing theorem.

Altogether, these findings indicate the types of geometries in which we can hope to compute the exact charged massless spectrum with the currently available technologies. These setups are as follows:
\begin{enumerate}
 \item Pick a smooth and projective toric variety $X_\Sigma$ of dimension $4$.
 \item Take $\mathcal{B}_6 = V \left( P \right) \subseteq X_\Sigma$ such that all line bundles on $\mathcal{B}_6$ are pullbacks of $\text{Pic} ( X_\Sigma )$.
 \item Pick $W = V( P, Q ) \subseteq \mathcal{B}_6 \subseteq X_\Sigma$ such that $W \cong dP_3$ and such that the Tate sections are of minimal degree.
 \item Work out the embedding $\varphi$ of the toric model of $dP_3$-surface into $W$ explicitly.
 \item Now employ $\varphi$ to count the zero modes localised on curves $C_{\mathbf{R}} \subseteq W$ from sheaf cohomologies on the toric model of the $dP_3$-surface. 
      The latter computations can be performed with \cite{SheafCohomologyOnToricVarieties}.
\end{enumerate}

Let us mention again that the geometry discussed in \cref{sec:dP7-Example} is taken from \cite{Braun:2014pva}. In this original work, the complex structure moduli of the matter curve $C_{\mathbf{5}_{-2}}$ were tuned in such a way that it split into two curves $C_{D}$ and $C_{T}$. As a consequence, the supersymmetry partners of the Higgs doublet and triplet, which are located on $C_{\mathbf{5}_{-2}}$, are now separately supported on $C_{D}$ and $C_T$, respectively. Therefore, this approach simplifies  the search for gauge backgrounds, in which no (supersymmetry partners of the) Higgs triplet are present. It would be interesting to perform such an analysis in future work with the methods presented in \cref{chapter:DetailsOnFPGradedSModules}. For example this could be done in the example geometry discussed in \cref{sec:dP3-Example}

In fact, it is also interesting to go in the other direction, \ie have some matter curves coalesce. For example, assume that we start with a geometry derived from the $SU(5) \times U(1)_X$-top (\cf \cite{Krause:2011xj} and \cref{sec:SU5xU1Top}). In this geometry there are three matter curves $C_{\mathbf{10}_{1_X}}$, $C_{\mathbf{5}_{3_X}}$ and $C_{\mathbf{5}_{-2_X}}$. We can employ the Higgs mechanism to remove the $U(1)_X$-symmetry. In the resulting geometry -- derived from a corresponding $SU(5)$-top -- only two matter curves are present, because $C_{\mathbf{5}_{3_X}}$, $C_{\mathbf{5}_{-2_X}}$ join to form the matter curve $C_{\mathbf{5}}$. During such a \emph{brane recombination process}, vector-like pairs can acquire masses. Whilst it is conceptionally interesting to study which pairs exactly become massive, such a process could in principle also improve the spectrum of the GUT-model in question. Therefore, also for phenomenology, these processes are of relevance. We leave such applications for future work.

\paragraph{Software Improvements and extended Search for the \emph{Standard Model}}
To summarise the program up to now -- the techniques in \cite{CAPCategoryOfProjectiveGradedModules, CAPPresentationCategory, PresentationsByProjectiveGradedModules, TruncationsOfPresentationsByProjectiveGradedModules, SheafCohomologyOnToricVarieties} are tailor-made for \emph{F-theory} GUT models. Therefore, phenomenological applications along the examples given in \cref{chapter:GUTModels} are immediate. In particular, we can investigate a number of rather interesting quantities:
\begin{itemize}
 \item Our methods take the complex structure directly into account. In particular, we are not limited to smooth matter curves -- they may well be singular.
 \item We can compute the zero modes for different choices of complex structure. In this sense we can probe how the number of zero-modes depends on the 
      complex-structure moduli. An example along these lines is given in \cite{Bies:2017fam}.
 \item We can investigate \emph{F-theory} GUT-models: The hypercharge flux used to break the GUT gauge group must be modelled by non-pullback line bundles
      \cite{Braun:2014pva}. Their handling requires the full technology of \cite{CAPCategoryOfProjectiveGradedModules, CAPPresentationCategory, PresentationsByProjectiveGradedModules, TruncationsOfPresentationsByProjectiveGradedModules, SheafCohomologyOnToricVarieties}.
\end{itemize}
Unfortunately, the \emph{F-theory} vacua studied in \cref{chapter:GUTModels} cannot satisfy the physical demands posed on realistic \emph{string theory} models. Whilst this is in part due to our lack of theoretical understanding (\eg our ignorance to understand and access the structure of the Picard group of the matter curves $C_{\mathbf{R}}$) also algorithmic limitations show prominently. Consequently, future work should try to improve the involved algorithms. Such improvements include the following:
\begin{itemize}
 \item \textbf{Functoriality of \emph{cohomCalg} \cite{Blumenhagen:2010pv, Blumenhagen:2010ed, Blumenhagen:2011xn, KoszulExtensionManual, cohomCalg:Implementation, 
      2011JMP....52c3506J, Rahn:2010fm}:} Given an \fp graded $S$-module $M$, the package \cite{SheafCohomologyOnToricVarieties} identifies an 
      ideal $I \subseteq S$ with $H^0 ( X_\Sigma, \tilde{M} ) \cong \mathrm{Hom}^0_S ( I, M )_ 0$, and then computes the RHS with \cite{CAPCategoryOfProjectiveGradedModules, CAPPresentationCategory, PresentationsByProjectiveGradedModules, TruncationsOfPresentationsByProjectiveGradedModules}. An ideal $I$ is suitable to establish this isomorphism if it degenerates a number of spectral sequences. Currently, we use a sufficient, but in general far too strong condition for this degeneracy -- we demand that a number of sheaf cohomologies vanish, which we check by use of \emph{cohomCalg} \cite{Blumenhagen:2010pv, Blumenhagen:2010ed, Blumenhagen:2011xn, KoszulExtensionManual, cohomCalg:Implementation, 2011JMP....52c3506J, Rahn:2010fm}. This vanishing is achieved if the conditions in \cref{mytheorem} are satisfied.
      
      A weaker yet sufficient condition is to merely demand that the cokernel of some maps of sheaf cohomologies vanishes. Following such a criterion, it is plausible that we can find simpler ideals $I$ to establish the above isomorphism, and thereby compute the sheaf cohomologies quicker. For example, for the module $M_1$ discussed in \cref{subsec:Spectrum1}, our algorithm picked the $46$-th Frobenius power of the irrelevant ideal $B_\Sigma$ for the computation of $h^0 ( X_\Sigma, \tilde{M}_1 )$. After these improvements, we could hope that it takes the $39$-th Frobenius power of $B_\Sigma$ instead, and consequently performs the resulting computation quicker.
      
      Consequently, we should investigate the functoriality of the \emph{cohomCalg}-algorithm \cite{Blumenhagen:2010pv, Blumenhagen:2010ed, Blumenhagen:2011xn, KoszulExtensionManual, cohomCalg:Implementation, 2011JMP....52c3506J, Rahn:2010fm}: Given a morphism $\varphi \colon L_1 \to L_2$ of two line bundles, can the existing \emph{cohomCalg}-algorithm be extended to \emph{easily} predict the induced maps $H^i ( X_\Sigma, L_1 ) \to H^i( X_\Sigma, L_2 )$?      
 \item \textbf{Serre quotients for `nice' computer models:} Even for smooth $X_\Sigma$ the functor $\widetilde{\phantom{m}} \colon S \mathrm{\textnormal{-}fpgrmod} \to 
      \mathfrak{Coh} ( X_\Sigma )$ is no equivalence of categories. This means that for a given coherent sheaf $\mathcal{F}$ on $X_\Sigma$ two modules $M_1$, $M_2$ with $M_1 \not \cong M_2$ may well satisfy $\tilde{M}_1 \cong \tilde{M}_2$. The algorithms in \cite{CAPCategoryOfProjectiveGradedModules, CAPPresentationCategory, PresentationsByProjectiveGradedModules, TruncationsOfPresentationsByProjectiveGradedModules} strongly depend on the chosen computer model $M_1$, $M_2$ -- for $M_1$ the computations may complete within seconds, whilst $M_2$ might lead to weeks of computations. It is therefore crucial to pick a computer model of the coherent sheaf $\mathcal{F}$ in question, which is as simple as possible.
      
      In the smooth case the redundancy of the sheafification functor was identified. By dividing out this redundancy, the category of coherent sheaves was found to be equivalent to a Serre quotient \cite{2012arXiv1210.1425B}. By now Serre quotients are supported in \texttt{CAP} \cite{PosurDoktor, GutscheDoktor, CAP}, but are not yet being used in \cite{SheafCohomologyOnToricVarieties}. Is it possible to take advantage of these Serre quotients in order to find computer models for coherent sheaves, from which the corresponding sheaf cohomologies can be deduced with minimal computational effort?
 
      As of this writing \texttt{gap} \cite{GAP4} identifies in a special inverse system of ideal $\mathcal{F}$, of which each sheafifies to the structure sheaf $\mathcal{O}_{X_\Sigma}$, an ideal $I$ such that the conditions of \cref{mytheorem} and satisfied and consequently it holds 
      \[ H^0 ( X_\Sigma, \tilde{M} ) \cong \mathrm{Hom}^0_S ( I, M )_0 \, . \]
      Given the knowledge of Serre quotients, we can perform the same search in a bigger inverse system of ideals $\hat{\mathcal{F}}$. Presumably, this bigger search  yields an ideal $\hat{I} \in \hat{\mathcal{F}}$, which allows to compute the sheaf cohomologies of interest with less computational resources.
 \item \textbf{Include other approaches also:} Recall that our algorithm brings together the ideas of Gregory G. Smith \cite{1998math......7170S, 
      Maclagan03multigradedcastelnuovo-mumford, Oberwolfach} and of \emph{cohomCalg} \cite{Blumenhagen:2010pv, Blumenhagen:2010ed, Blumenhagen:2011xn, KoszulExtensionManual, cohomCalg:Implementation, 2011JMP....52c3506J, Rahn:2010fm}. In the same spirit, it is beneficial to include the ideas of \cite{StillmanStringPheno} also. Similarly, the techniques of \cite{Anderson:2008ex} add valuable improvements and options. To analyse how the number of zero modes depends on the complex structure moduli of the matter curves $C_{\mathbf{R}}$, it is of particular interest to include techniques from \emph{machine learning}. Such an approach to the \emph{cohomCalg}-algorithm \cite{Blumenhagen:2010pv, Blumenhagen:2010ed, Blumenhagen:2011xn, KoszulExtensionManual, cohomCalg:Implementation, 2011JMP....52c3506J, Rahn:2010fm} has been presented in \cite{RuehleStringPheno}. See also \cite{NelsonStringPheno} for yet another application of machine learning in \emph{string theory}.
\end{itemize}

Of course the presented technology is not limited to \emph{F-theory} GUT-models. Whilst GUT-models are definitely very appealing from a phenomenological point of view, motivation for them comes from the unification of the \emph{standard model} couplings at high energies. For this unification to happen to high accuracy, one must focus on a \emph{minimally supersymmetric extension of the standard model} (MSSM). And even in such an MSSM model, gauge coupling unification happens only if supersymmetry is broken at energies in the TeV range. Unfortunately, the LHC has not yet found direct evidence for supersymmetry. Consequently, the idea of a grand unified theory is under severe tension. This was the motivation for the works presented in \cite{Lin:2014qga, Lin:2016vus, Lin:2016zha}, which study different means to realise the \emph{standard model} in \emph{F-theory}. Also these models can in principle be handled with the techniques presented in this thesis. To date however, practical limitations arise from too large defining polynomials, which then exceed the abilities of the currently available Gröbner basis algorithms. It would be very interesting to explore this direction further in future work.

While our methods are inspired by \emph{F-theory}, they are by no means restricted to it. For example, the technology presented in \cref{chapter:DetailsOnFPGradedSModules} allows to investigate more general classes of GUT models in heterotic compactifications. This includes the models studied in \cite{Anderson:2011ns}, which employed a vector bundle of the form
\[ V = L_1 \oplus L_2 \oplus L_3 \oplus L_4 \oplus L_5 \, . \]
This vector bundle has structure group $S( U(1)^5 ) \subseteq SU(5)$, and this embedding can be used to break the $SU(5)$ gauge group to \emph{standard model}-like groups. As a consequence of this simple form of $V$, the zero modes are counted by line bundle cohomologies only. However, with the technology implemented in \cite{SheafCohomologyOnToricVarieties} we can handle more general choices of $\mathbf{V}$. This application therefore allows one to extend the works of \cite{Anderson:2011ns, Anderson:2012yf, Buchbinder:2014qda, Anderson:2013qca}.

Slightly more exotic are applications in topological \emph{string theory}: First recall that a coherent sheaf $\mathcal{F}$ on a Calabi--Yau 3-fold admits a locally free resolution
\[ \mathcal{F}_\bullet \colon \dots 0 \to 0 \dots \mathcal{V}_{0} \to \mathcal{V}_{1} \to \dots \to \mathcal{V}_{n} \to \mathcal{F} \to 0 \to 0 \to \dots \]
where each $\mathcal{V}_i$ is a vector bundle on $\mathcal{B}_6$. This is a special object in $D^b \mathfrak{Coh} ( \mathcal{B}_3 )$ -- the (bounded) derived category of coherent sheaves on $\mathcal{B}_6$ -- in which (bounded) complexes of coherent sheaves serve as objects. As reviewed in \eg \cite{Aspinwall:2004jr}, in topological \emph{string theory} the category $D^b \mathfrak{Coh} ( \mathcal{B}_6 )$ models the category of B-branes on $\mathcal{B}_6$. In this sense a coherent sheaf $\mathcal{F}$ encodes a B-brane $\mathbf{\mathcal{F}_{\bullet}}$.

Open strings between two B-branes $\mathcal{F}_{\bullet}$, $\mathcal{G}_\bullet$ are paralleled by morphisms in $D^b \mathfrak{Coh} ( \mathcal{B}_6 )$, which in turn can be understood as (global) Extension groups in $\mathfrak{Coh} ( \mathcal{B}_3 )$ \cite{Aspinwall:2004jr}. For example, open strings from $\mathcal{F}_{\bullet}$ to $\mathcal{G}_{\bullet}$ (of ghost number $q$) are counted by $\mathbf{\mathrm{\textbf{Ext}}^q} \mathbf{( \mathcal{F}, \mathcal{G} )}$. Although not yet implemented, the theoretical foundations of \cite{SheafCohomologyOnToricVarieties} allow one to compute these groups (\cf \cref{sec:ComputingGlobalBivariateExt}). Implementing the required technology into \texttt{gap} \cite{GAP4} and using it to investigate spectra of B-brane configurations is interesting. A constructive approach to B-brane stability is equally exciting.

\paragraph{Extensions to more general \emph{F-theory} Vacua}

It is desirable to deepen our understanding of the second Chow group $\mathrm{CH}^2 ( \hat{Y}_4 )$ of the flat, smooth resolution $\hat{\pi} \colon \hat{Y}_4 \twoheadrightarrow \mathcal{B}_6$ of the singular elliptic fibration $\pi \colon Y_4 \twoheadrightarrow \mathcal{B}_6$. Regarding the situations studied in this thesis, we can achieve this in two ways:
\begin{itemize}
 \item The examples studied in \cref{sec:ComputingTheSpectra} and \cref{sec:dP3-Example} satisfy $h^{2,1} ( \hat{Y}_4 ) = 0$.\footnote{In the example studied in 
      \cref{sec:dP3-Example} \emph{cohomCalg} could not determine $h^{2,1} ( \hat{Y}_4 )$, but matched it with a non-negative integer $A$. This includes the possibility of $A = 0$. An explicit computation of $A$ can in principle be done with \cite{CAPCategoryOfProjectiveGradedModules, CAPPresentationCategory, PresentationsByProjectiveGradedModules, TruncationsOfPresentationsByProjectiveGradedModules, SheafCohomologyOnToricVarieties}, but turned out to exceed the currently available computational resources.} As a consequence the intermediate Jacobian
      $J^2 ( \hat{Y}_4 )$ in the short exact sequence
      \[ 0 \to J^2( \hat{Y}_4 ) \hookrightarrow H_D^4( \hat{Y}_4, \mathbb{Z} ( 2 ) ) \xtwoheadrightarrow{\hat{c}_2} H^{2,2}_{\mathbb{Z}} ( \hat{Y}_4 ) \to 0 \]
      is trivial in these examples. Consequently, Deligne cohomology and ordinary cohomology match identically. It seems natural to wonder what happens in the presence of a non-trivial $J^2( \hat{Y}_4 )$. Various aspects of this intermediate Jacobian have been studied in \cite{Greiner:2015mdm,Greiner:2017ery} for 4-dimensional \emph{F-theory} compactifications. In extending this work, it is quite interesting to study gauge backgrounds $A \in H^4_D( \hat{Y}_4, \mathbb{Z} ( 2 ) )$ which are non-trivial and flat, \ie satisfy $A \neq 0$ but $\hat{c}_2 ( A ) = 0$. By a generalised Abel-Jacobi map, such gauge data is related to the kernel of the cycle map $\mathrm{CH}^2(\hat{Y}_4) \rightarrow H^{2,2}_{\mathbb{Z}}(\hat{Y}_4)$. In particular, effects from torsional $H^{2,1}(\hat{Y}_4)$ can influence such gauge backgrounds.
 \item Recall also that we demanded the base $\mathcal{B}_6$ to be torsion-free. Thus, no effects from torsional 4-cycles were encountered. It would be quite an 
      interesting task to adapt our computations to situations including such torsional cycles.
\end{itemize}

Recall that we assumed throughout this thesis the existence of a flat, smooth resolution $\hat{\pi} \colon \hat{Y}_4 \twoheadrightarrow \mathcal{B}_6$ of the singular elliptic fibration $\pi \colon Y_4 \twoheadrightarrow \mathcal{B}_6$. This resolution process is paralleled in 3-dimensional M-theory vacua by moving along the Coulomb branch. Therefore, we are limited to studying Abelian gauge backgrounds. To access more general gauge backgrounds, it would be interesting to investigate \emph{F-theory} and Deligne cohomology on singular spaces \cite{Anderson:2013rka, Collinucci:2014taa}. Similar situations arise from studying T-brane configurations, which obstruct a resolution \cite{Collinucci:2014qfa, Collinucci:2016hpz, Anderson:2017rpr}.

Following the proposal in \cite{Collinucci:2014taa, Collinucci:2014qfa}, we can try to understand \emph{F-theory} on singular spaces and T-branes therein from Eisenbud's matrix factorisations. Although the results were formulated on affine spaces only, the description of T-branes in \cite{Collinucci:2014qfa} is strikingly similar to that of  \fp graded $S$-modules. This observation leads to all sorts of questions. Provided that the \emph{F-theory} geometry is cast in the language of toric varieties, can a T-brane then be understood as a complex $\tilde{M}_{\bullet}$ of the coherent sheaves associated to \fp graded $S$-modules? If yes, what physical properties of this T-brane are encoded in the modules $M$? Can we use the works of \cite{Bridgeland, 2003math......7164B, 2006math......2129B} -- which investigate stability conditions in triangulated categories and thus lend themselves nicely to the concept of \emph{categorical programming} \texttt{CAP} \cite{PosurDoktor, GutscheDoktor, CAP} -- to formulate and algorithmically study decay channels of T-branes? Can generalised \emph{Harder-Narasimhan filtrations} of \fp graded $S$-modules be used to predict the decay channel? And if so, can physical insights help to compute such filtrations algorithmically? Answering these questions is a thrilling, but also very demanding challenge.

In any case, we are optimistic that the formalism regarding Chow groups can be generalised to \emph{F-theory} on singular spaces -- intersection theory within the Chow ring can be formulated on non-smooth varieties. A natural question is to ask how far one can actually push this formalism. This is a thrilling research project for the future.

Finally, note that we have mostly focused on 4-dimension \emph{F-theory} compactifications in this thesis. Still, one can consider \emph{F-theory} compactifications to other dimensions as well. For example, \cite{Schafer-Nameki:2016cfr,Apruzzi:2016iac} studied 2-dimensional \emph{F-theory} compactifications. It was found that in such string vacua, the zero modes are described by structures very similar to the ones encountered for 4-dimensional compactifications. It is certainly an interesting research topic to extend the presented formalism to \emph{F-theory} compactifications described by Calabi--Yau 5-folds.

\paragraph{Chow Groups and Local Anomalies}

In \cref{chapter:LocalAnomaliesInF-Theory} we investigated anomaly cancellation in \emph{F-theory} compactifications. At first sight, this may seem utterly unrelated to the previous analysis, which focused on the meaning of Chow groups for gauge backgrounds. Nonetheless, it turned out that Chow groups can help to shape our understanding of anomalies in \emph{F-theory} also. Let us summarise how we reached this conclusion.

We started our analysis by extending the works of \cite{Cvetic:2012xn} to non-gauge-invariant gauge fluxes in \emph{F-theory}. For such a $G_4$-flux we argued that both the gauge and gauge-gravitational anomalies are summarised by \cref{G4dotanomaly1} and \cref{G4dotanomaly2}. We argued that these results hold true for any $G_4 \in H^{2,2}_{\text{vert}} ( \hat{Y}_4 )$, which led us to generalise these results to the cohomological relations in \cref{summary4dCoho1}. We even extended these results to 6-dimensional compactifications and thus identified relations in $H^{2,2}(\hat{Y}_{n+1},\mathbb{Q})$, where $n=2$ and $n=3$ refer to to six- and four-dimensional compactifications, respectively. This is quite remarkable -- the cancellation of local anomalies in both 4- and 6-dimensional \emph{F-theory} compactifications is governed by the same type of cohomological relations!

Our results reproduce the identities of \cite{Park:2011ji} and generalise the observations made in \cite{Lin:2016vus} for explicit \emph{F-theory} examples. Whilst our derivation is motivated from \emph{string theory}, for any smooth, flat and crepant resolution of an elliptically fibred Calabi--Yau 3- and 4-fold, the relations in \cref{summary4dCoho1} must hold as mathematical identities. It is desirable to find a formal mathematical proof, to supplement our stringy arguments.

As pointed out, anomalies in 4- and 6-dimensional \emph{F-theory} compactifications are governed by the same type of cohomological relations. How about 2-dimensional $\mathcal{N} = (0,2)$ supersymmetric compactifications? Such \emph{F-theory} vacua are described by Calabi--Yau 5-folds and have been investigated in \cite{Schafer-Nameki:2016cfr, Apruzzi:2016iac, Apruzzi:2016nfr, Lawrie:2016rqe}. It would be interesting to relate the cohomological identities of this work to the anomaly structure of such 2-dimensional vacua. In such 2-dimensional compactifications additional contributions to the gauge and gravitational anomalies stem from \emph{chiral} fermions located at the intersection of D3- and D7-branes, which wrap curves in the base $\mathcal{B}_4$ of the elliptically fibred 5-fold $\hat{Y}_5$ \cite{Schafer-Nameki:2016cfr,Apruzzi:2016iac,Lawrie:2016rqe}. Such contributions are unparalleled in 4- and 6-dimensional compactifications. To date, they remain somewhat mysterious (\cf \cite{Lawrie:2016axq} and references therein).

The results of \cref{chapter:MasslessSpectraAndSheafCohomology} pointed towards cohomological relations enjoyed by matter surface fluxes in geometries derived from the $SU(5) \times U(1)_X$-top (see \cref{sec:SU5xU1Top} or \cite{Krause:2011xj, oai:arXiv.org:1202.3138} for details on these tops). Our results tie these interrelations to anomaly cancellation in \emph{F-theory}. Namely, in addition to \cref{summary4dCoho1} we were able to derive an additional set of cohomological relations, which we summarised in \cref{summary4dCoho2}. These new relations are in general independent of \cref{summary4dCoho1}. Most importantly, for the $SU(5) \times U(1)_X$-top these equations govern the cohomological relations among the matter surface fluxes.

Even more, our analysis in \cref{chapter:MasslessSpectraAndSheafCohomology} indicates that these relations should not only hold in cohomology, but even in the Chow ring. In fact, such rational equivalences of matter surface fluxes have been used in \cref{chapter:MasslessSpectraAndSheafCohomology} to express self-intersections as transverse intersections. Thereby, many computations simplified significantly. Based on this observation, we conjecture that \cref{summary4dCoho1} and \cref{summary4dCoho2} hold true even at the level of the Chow ring. The assumptions under which we expect the resulting Chow relations \cref{summary4dChow1},  \cref{summary4dChow2} to hold true are summarised below \cref{summary4dChow2}.

A very interesting question concerns the difference between the cohomological relations in \cref{summary4dCoho1}, \cref{summary4dCoho2} and their cousins \cref{summary4dChow1}, \cref{summary4dChow2} in the Chow ring. As pointed out, the stronger set of relations \cref{summary4dChow2} encodes rational equivalences of matter surface fluxes for \emph{F-theory} compactifications based on the $SU(5) \times U(1)_X$-top \cite{Krause:2011xj}. Also, they explain the absence of vertical gauge backgrounds in a number of classes of elliptic fibrations \cite{oai:arXiv.org:1202.3138}. Then again, the relations \cref{summary4dCoho1} within $H^{2,2}( \hat{Y}_{n+1},\mathbb Q)$ are both necessary and sufficient to ensure anomaly cancellation. Of course, the Chow ring contains much more geometric information than $H^{2,2}(\hat Y_{n+1},\mathbb{Q})$. However, to date we cannot fully rule out the possibility that they coincide under assumptions phrased after \cref{summary4dChow2}. We leave it for future work to settle this question. In particular, if they do not coincide, the meaning of the additional information in the Chow ring is to be investigated.

To avoid matter states in addition to the ones studied in \cref{chapter:MasslessSpectraAndSheafCohomology}, our analysis focused on Weierstrass models which admit a flat, smooth and crepant resolution with section. An interesting task concerns the generalisation of our results to more general geometries. A first pointer into this direction follows from \cite{Lin:2015qsa}, where analogous constructions for fluxes were described, but in the absence of a section. In consequence, we expect that it is indeed possible to generalise our results to \emph{F-theory} without section, as studied in \cite{Witten:1996bn, Braun:2014oya, Anderson:2014yva, Klevers:2014bqa, Garcia-Etxebarria:2014qua, Mayrhofer:2014haa, Mayrhofer:2014laa, Morrison:2014era, Cvetic:2015moa, Lin:2015qsa, Kimura:2016crs}.

\paragraph{Chow Groups and Global Anomalies?}
In perturbative D-brane models, the cancellation of all K-theory charges carried by D-branes has been found to encode the absence of \emph{global} $SU(2)$ Witten anomalies \cite{Uranga:2000xp}. Likewise, the absence of global Witten anomalies in \emph{F-theory} is related to a suitable quantisation of the $G_4$-flux \cite{GarciaEtxebarria:2005qc}. More explicitly, the absence of such anomalies is described by divisibility properties of the second Chern class $c_2 ( \hat{Y}_4 )$, which in turn enter $G_4$-flux quantisation via the the Freed-Witten condition $G_4 + \frac{1}{2} c_2 ( \hat{Y}_4 ) \in H^4( \hat{Y}_4, \mathbb{Z} )$. This program has been studied more extensively in \cite{Lin:2016vus} -- explicitly such divisibility properties have been related to the absence of global Witten anomalies in \emph{F-theory} with gauge group $SU(2)$. It would be interesting to study the absence of global anomalies in \emph{F-theory} further and to investigate to which extent they can be described by cohomology ring and/or the Chow ring of the elliptically fibred Calabi--Yau n-fold $\hat{Y}_n$.

\paragraph{Significance of Chow Groups}
To summarise this program, gauge backgrounds in \emph{F-theory} can be understood as elements of Deligne cohomology, and special such gauge backgrounds can in turn be described by algebraic cycles modulo rational equivalence, \ie Chow groups \cite{Curio:1998bva, Donagi:1998vw, EsnaultDeligne, oai:arXiv.org:1203.6662, Clingher:2012rg, Anderson:2013rka}. This we exploited in \cite{Bies:2014sra, Bies:2017fam}. While putting the finishing touches to \cite{Bies:2017fam}, these results were found to point towards relations in the Chow ring, enjoyed by the matter surface fluxes discussed in \cref{chapter:MasslessSpectraAndSheafCohomology}. In \cite{Bies:2017abs} these relations were finally found to be intricately related to (local) anomaly cancellation in \emph{F-theory}. As Chow groups encode rich geometric data and \emph{F-theory} is governed by an intricate interplay between physics and geometry, the future will tell what else Chow groups can tell about \emph{F-theory} vacua.

\newpage

\section*{Acknowledgements}

First, I am grateful to my family and friends for their support and love during my PhD studies. In particular, let me thank my friends Fabian Klein and Randolf Beerwerth for proofreading this thesis.

I am also very grateful to my scientific family at the \emph{Institut für theoretische Physik (ITP)} at \emph{Heidelberg university}, of which I was a member not only during my  PhD studies, but also for my master's thesis. During this time I enjoyed a relaxed, yet exciting research atmosphere. As a list of names is almost certain to be incomplete, I would like to express my words of gratitude collectively to all members of the ITP. Special thanks however, go to Philipp Henkenjohann, Craig Lawrie and Christian Reichelt for proofreading and bringing forward lots of useful comments.

I am also very thankful to my long-time collaborator Christoph Mayrhofer. It is truly a pleasure to witness the ease with which he visualises abstract toric geometries. Therefore I am grateful for our many toric discussions!

This research project would not have been possible without the mathematician Mohamed Barakat. Despite my lack of degrees in the mathematics, he accepted me for supervision during this thesis. His support, eye for detail and patient explanations showed me a new level of scientific precision and taught me the beauty of category theory. I very much enjoyed the resulting collaboration with his research group, for which I would like to thank its members. Special thanks go to Sebastian Posur for our many helpful discussions on algebraic geometry and category theory, and to Sebastian Gutsche for his superhuman abilities with the programming language \texttt{gap}. Also the computer \texttt{plesken} at \emph{University of Siegen} deserves recognition -- its computational powers derived many of the results in this thesis.

Finally let me thank my supervisor Timo Weigand. Not only for my PhD, but also for my bachelor's and master's thesis, I was happy enough to benefit from his supervision. In my opinion it mixes precisely the right amount of scientific guidance with sufficient freedom to explore own research interests. It is this mixture which allowed me to turn to fairly mathematical topics during my master's thesis already, and thereby set the foundations for the collaboration with Mohamed Barakat during this PhD thesis. Thank you very much indeed!

\appendix

\chapter{Two Toric Tops and their Properties} \label{chapter:ToricTops}
\section{The Geometry of an \texorpdfstring{$\mathbf{SU(5) \times U(1)_X}$}{SU(5)xU(1)}-Top} \label{sec:SU5xU1Top}

In this section we analyse the fibre structure of the resolved 4-fold $\hat{\pi} \colon \hat{Y}_4 \twoheadrightarrow \mathcal{B}_6$ of the \emph{F-theory} GUT models described in \cref{subsec:SpecialFTheoryGUTModel}. This toric top was originally introduced in \cite{Krause:2011xj}.

The general philosophy to analyse this geometry is to identify $\mathbb{P}^1$-fibrations -- of which some will be present over 'special' subloci of $W$ only, namely the matter curves and the Yukawa points -- and then to work out their intersection numbers in the fibre. The discussion of the massless spectrum of matter surface fluxes in \cref{sec:ToricFTheoryGUTModels} heavily relies on this information. This is the main reason why we present this material here. We refine the original analysis of \cite{Krause:2011xj} in two important respects:
\begin{itemize}
 \item A $\mathbb{P}^1$-fibration present over generic points of $W$ -- that is points which are not contained in any matter curve -- in general splits into 
      a formal linear combination of $\mathbb{P}^1$-fibrations over the matter curves. Similarly, a $\mathbb{P}^1$-fibration present over generic points of the matter curves -- \ie non-Yukawa points of the matter curves -- in general splits into a formal linear combination of $\mathbb{P}^1$-fibrations. In order to deduce these splittings we use primary decompositions of the relevant ideals. \\
      This differs from the original approach in \cite{Krause:2011xj}. In consequence, we find different defining equations for the fibrations $\mathbb{P}^1_{3G} ( \mathbf{5}_{-2} )$, $\mathbb{P}^1_{3H} ( \mathbf{5}_{-2} )$ and different splittings for these fibrations once restricted to the Yukawa locus $Y_1$, namely
      \[ \mathbb{P}^1_{3G} \left( \mathbf{5}_{-2} \right) \to \mathbb{P}^1_{34} \left( Y_1 \right), \qquad \mathbb{P}^1_{3H} \left( \mathbf{5}_{-2} \right) \to \mathbb{P}^1_{3J} \left( Y_1 \right) \, . \]
 \item In the original work \cite{Krause:2011xj}, the intersections of the $\mathbb{P}^1$-fibrations were said `to have the \emph{structure} of certain Dynkin diagrams'. 
      However, the actual intersection numbers -- in particular the self-intersection numbers -- were not stated. Here we work out these numbers. \\
      To describe our findings let us introduce the notation $T^2 ( C_{\mathbf{R}} )$ to indicate the total elliptic fibre over a matter curve $C_{\mathbf{R}}$. Similarly, $T^2 ( Y_i )$ is to indicate the total elliptic fibre over the Yukawa locus $Y_i$. Furthermore let $\mathbb{P}_i ( C_{\mathbf{R}} )$ and $\mathbb{P}_i ( Y_j )$ denote a $\mathbb{P}^1$-fibration over a matter curve $C_{\mathbf{R}}$ and a Yukawa locus $Y_j$ respectively. Then for all matter curves $C_{\mathbf{R}}$ and all Yukawa loci $Y_j$ in the current geometry, the following holds true:
      \[ T^2 \left( Y_j \right) \cdot \mathbb{P}^1_i \left( Y_j \right) = T^2 \left( C_{\mathbf{R}} \right) \cdot \mathbb{P}^1_i \left( C_{\mathbf{R}} \right) = 0 \, . \]
      Next suppose $Y_j \subseteq C_{\mathbf{R}}$ and assume that the splitting of $\mathbb{P}^1_i ( C_{\mathbf{R}} )$ onto $Y_j$ takes the form
      \[ \mathbb{P}^1_i \left( C_{\mathbf{R}} \right) \to \sum_{k = 1}^{N \left( i \right)}{\alpha_{k}^{(i)} \mathbb{P}^1_{k} \left( Y_j \right)} \, . \]
      Naively we might expect
      \[ \mathbb{P}^1_i \left( C_{\mathbf{R}} \right) \cdot \mathbb{P}^1_j \left( C_{\mathbf{R}} \right) = \left( \sum_{k = 1}^{N \left( i \right)}{\alpha_{k}^{(i)} \mathbb{P}^1_{k} \left( Y_j \right)} \right) \cdot \left( \sum_{l = 1}^{N \left( j \right) }{\alpha_{l}^{(j)} \mathbb{P}^1_{l} \left( Y_j \right)} \right) \, . \label{equ:IntersectionsStableWRTIntersections} \]
      As it turns out, this is not the case for the geometry described by this $SU ( 5 ) \times U ( 1 )_X$-top. \Cref{equ:IntersectionsStableWRTIntersections} fails 
      precisely for the restrictions of $C_{\mathbf{10}_1}$ to $Y_1$ which involve either $\mathbb{P}^1_{24} ( Y_1 )$ or $\mathbb{P}^1_{2B} ( Y_1 )$. The reason for this failure is that these two $\mathbb{P}^1$-fibrations encounter a $\mathbb{Z}_2$-singularity over their intersection point. This parallels the situation studied for the enhancement $A_5 \hookrightarrow E_6$ in \cite{Morrison:2011mb}. As a consequence we find half-integer intersection numbers
      \[ \mathbb{P}^1_{24} \left( Y_1 \right)^2 = \mathbb{P}^1_{2B} \left( Y_1 \right)^2 = - \frac{3}{2}, \qquad \mathbb{P}^1_{24} \left( Y_1 \right) \cdot \mathbb{P}^1_{2B} \left( Y_1 \right) = \frac{1}{2} \, . \]
      Therefore, a particular example in which \cref{equ:IntersectionsStableWRTIntersections} fails is the following:
      \[ \mathbb{P}_{24} \left( \mathbf{10}_{1} \right)^2 = -2 \neq - \frac{3}{2} = \mathbb{P}^1_{24} \left( Y_1 \right)^2 \, . \]
\end{itemize}
The Cox ring $\mathbb{Q}[ e_0, e_1, e_2, e_3, e_4, x, y, z, s]$ of the toric fibre ambient space is graded according to \cref{table:N34}. Throughout this work we apply triangulation $T_{11}$ of \cite{Krause:2011xj}, \ie we take the Stanley-Reisner ideal as
\begin{align}
\begin{split}
I_{\mathrm{SR}} \left( \mathrm{top} \right) = & \left\{ xy, x e_0 e_3, x e_1 e_3, x e_4, y e_0 e_3, y e_1, y e_2, z s, z e_1 e_4, z e_2 e_4, \right. \\
  & \hspace{10em} \left. z e_3, s e_0, s e_1, s e_4, e_0 e_2, z e_4, z e_1, z e_2, s e_2, e_0 e_3, e_1 e_3 \right\} \, . \label{equ:SRSU5xU1InAppendix}
\end{split}
\end{align}

\begin{table}[tbp]
\centering
\begin{tabular}{c@{\hskip 20pt}ccccc@{\hskip 20pt}cccc}
\toprule
& $e_0$ & $e_1$ & $e_2$ & $e_3$ & $e_4$ & x & y & z & s \\
\midrule
$\overline{K}_{\mathcal{B}_6}$ & $\cdot$ & $\cdot$ & $\cdot$ & $\cdot$ & $\cdot$ & 2 & 3 & $\cdot$ & $\cdot$ \\
W                    & 1 & $\cdot$ & $\cdot$ & $\cdot$ & $\cdot$ & $\cdot$ & $\cdot$ & $\cdot$ & $\cdot$ \\
\vspace{-0.5em} & \\
$E_1$ & -1        & 1 & $\cdot$ & $\cdot$ & $\cdot$ & -1 & -1 & $\cdot$ & $\cdot$ \\
$E_2$ & -1        & $\cdot$ & 1 & $\cdot$ & $\cdot$ & -2 & -2 & $\cdot$ & $\cdot$ \\
$E_3$ & -1        & $\cdot$ & $\cdot$ & 1 & $\cdot$ & -2 & -3 & $\cdot$ & $\cdot$ \\
$E_4$ & -1        & $\cdot$ & $\cdot$ & $\cdot$ & 1 & -1 & -2 & $\cdot$ & $\cdot$ \\
\vspace{-0.5em} & \\
Z & $\cdot$   & $\cdot$ & $\cdot$ & $\cdot$ & $\cdot$ & 2 &  3 & 1 & $\cdot$ \\
S & $\cdot$   & $\cdot$ & $\cdot$ & $\cdot$ & $\cdot$ & -1 & -1 & $\cdot$ & 1 \\
\bottomrule
\end{tabular}
\caption[Toric data of $SU(5) \times U(1)_X$-top.]{The Cox ring $\mathbb{Q}[ e_0, e_1, e_2, e_3, e_4, x, y, z, s]$ of the $SU(5) \times U(1)_X$-top is graded according to this table. Together with the Stanley-Reisner ideal \cref{equ:SRSU5xU1InAppendix} this defines the geometry of this toric space.}
\label{table:N34}
\end{table}

We can use this top to construct a space $\hat{Y}_5$ over a base space $\mathcal{B}_6$. Suppose that it is our goal to design an $SU(5)$-singularity over a divisor $W \subseteq \mathcal{B}_6$, then we pick sections 
\[ a_{i,j} \in H^0 \left( \mathcal{B}_6, \overline{K}_{\mathcal{B}_6}^{\otimes i} \otimes \mathcal{O}_{\mathcal{B}_6} \left( - j W \right) \right) \, . \]
The fibre of $\hat{Y}_4 \xtwoheadrightarrow{\pi} \mathcal{B}_6$ over $p \in \mathcal{B}_6$ is then given as the vanishing locus of the following proper transform of the Tate polynomial in the above top:
\begin{align}
\begin{split}
P_T^\prime &= y^2 s e_3 e_4 + a_1 \left( p \right) x y z s + a_{3,2} \left( p \right) y z^3 e_0^2 e_1 e_4 - x^3 s^2 e_1 e_2^2 e_3 \\
           & \hspace{13em} - a_{2,1} \left( p \right) x^2 z^2 s e_0 e_1 e_2 - a_{4,3} \left( p \right) x z^4 e_0^3 e_1^2 e_2 e_4 \, .
\end{split}
\end{align}

In this section we proceed as follows: First we analyse the intersection structure of this toric space in \cref{subsec:FibreStructureSU5xU1}. Subsequently, 
\cref{subsec:MasslessSpectraMSFSU5xU1} discusses the zero modes of the matter surface fluxes
\begin{align}
\begin{split}
\mathcal{A} \left( \mathbf{10}_{1} \right) \left( \lambda \right) &= - \frac{\lambda}{5} \cdot \left( 2 \mathcal{E}_1 - \mathcal{E}_2 + \mathcal{E}_3 - 2 \mathcal{E}_4 \right) \cdot \overline{\mathcal{K}}_{\mathcal{B}_6} - \lambda \cdot \mathcal{E}_2 \cdot \mathcal{E}_4 \, , \\
\mathcal{A} \left( \mathbf{5}_{3} \right) \left( \lambda \right) &= - \frac{\lambda}{5} \cdot \left( \mathcal{E}_1 + 2 \mathcal{E}_2 - 2 \mathcal{E}_3 - \mathcal{E}_4 \right) \cdot \left( 3 \overline{\mathcal{K}}_{\mathcal{B}_6} - 2 \mathcal{W} \right) - \lambda \cdot \mathcal{E}_3 \cdot \mathcal{X} \, , \\
\mathcal{A} \left( \mathbf{5}_{-2} \right) \left( \lambda \right) &= - \frac{\lambda}{5} \cdot \left( \mathcal{E}_1 + 2 \mathcal{E}_2 + 3 \mathcal{E}_3 - \mathcal{E}_4 \right) \cdot \left( 5 \overline{\mathcal{K}}_{\mathcal{B}_6} - 3 \mathcal{W} \right) \\
& \hspace{14em} + \lambda \cdot \left( \mathcal{E}_3 \cdot \overline{\mathcal{K}}_{\mathcal{B}_6} + \mathcal{E}_3 \cdot \mathcal{Y} - \mathcal{E}_3 \cdot \mathcal{E}_4 \right) \, , \\
\mathcal{A} \left( \mathbf{1}_{5} \right) \left( \lambda \right) &= \lambda \cdot \mathcal{S} \cdot \left( 3 \overline{\mathcal{K}}_{\mathcal{B}_6} - 2 \mathcal{W} \right) - \lambda \cdot \mathcal{S} \cdot \mathcal{X} \, .
\end{split}
\end{align}

These fluxes were introduced in \cref{subsec:VerticalAndGaugeInvariantMatterSurfaceFluxes}, where we gave an intuitive derivation of their massless spectra. In contrast to this intuitive approach, we will in section make use of the intersection theory on $\hat{Y}_5$ to derive these results anew. Of ample importance for this are the generators of the ideal $I_{\mathrm{LR}}$ of this top geometry (\cf \cref{subsec:TowardsToricVarieties}). In the case at hand these generators are given by
\begin{align}
\begin{split}
\mathcal{X} - \mathcal{Y} + \mathcal{Z} + \mathcal{E}_0 + \mathcal{E}_1 + \mathcal{E}_2 - \mathcal{W} + \overline{\mathcal{K}}_{\mathcal{B}_6} &= 0 \in \mathrm{CH}^1 ( \hat{Y}_5 ) \, , \\
-3 \mathcal{X} + 2 \mathcal{Y} - \mathcal{S} - \mathcal{E}_1 -2 \mathcal{E}_2 + \mathcal{E}_4 &= 0 \in \mathrm{CH}^1 ( \hat{Y}_5 ) \, , \\
2 \mathcal{X} - \mathcal{Y} - \mathcal{Z} + \mathcal{S} + \mathcal{E}_1 + 2 \mathcal{E}_2 + \mathcal{E}_3 - \overline{\mathcal{K}}_{\mathcal{B}_6} &= 0 \in \mathrm{CH}^1 ( \hat{Y}_5 ) \, .
\end{split}
\end{align}
The results obtained along this alternative approach match the earlier findings, which were summarised in \cref{table-N5}. These results imply that the massless spectra of the above matter surface fluxes satisfy the following relations:
\begin{align}
\begin{split}
A \left( \mathbf{5}_3 \right) \left( \lambda \right) &= A \left( \mathbf{5}_{-2} \right) \left( - \lambda \right) + A \left( \mathbf{10}_1 \right) \left( - \lambda \right) \, , \\
A \left( \mathbf{5}_{-2} \right) \left( \lambda \right) &= {A}_X \left( \lambda W \right) \, , \\
A \left( \mathbf{1}_{5} \right) \left( \lambda \right) &= {A}_X \left( - \lambda \left( 6 \overline{K}_{\mathcal{B}_6} - 5 {W} \right) \right) + {A} \left( \mathbf{10}_1 \right) \left( \lambda \right) \,.
\end{split}
\end{align}
In \cref{subsec:ProofOfChowRelationsSU5xU1} we will show that these relations actually hold true in the Chow ring of $\hat{Y}_4$. This has implications for the understanding of anomaly cancellation in \emph{F-theory}. The latter is discussed in detail in \cref{chapter:LocalAnomaliesInF-Theory}.

\subsection{Fibre Structure} \label{subsec:FibreStructureSU5xU1}

\subsubsection{Intersection Structure away from Matter Curves}

We start by looking at the five divisors $E_i := V ( P_T^\prime, e_i ) \subseteq \hat{Y}_4$ for $0 \leq i \leq 4$. Note that $e_i = 0$ automatically implies that the 'new' GUT-coordinate $e_0 e_1 e_2 e_3 e_4$ vanishes. Hence, $E_i$ indeed is a subset of $\hat{Y}_4$. These subsets can be understood as fibration of the $i$-th exceptional divisor over the GUT-surface $W$.

Now let $p \in W$ a point which is not contained in any matter curve. By means of the projection map $\hat{\pi} \colon \hat{Y}_4 \twoheadrightarrow \mathcal{B}_6$ we can describe the fibre over the point $p$ as $\hat{\pi}^{-1} ( p )$. We now wish to work out the intersection structure of the divisor classes $E_i$ in $\hat{\pi}^{-1} ( p )$. For simplicity we merely focus on the set-theoretic intersection $E_0 \cap E_1$. By use of the Stanley-Rei{\ss}ner ideal of the top -- see \cref{subsec:SpecialFTheoryGUTModel} -- it is readily confirmed that
\begin{align}
\begin{split}
E_0 \cap E_1 \cap \hat{\pi}^{-1} \left( p \right) &= V \left( e_0, e_1, e_3 e_4 s y^2 + a_1 \left( z_i \right) s x y z \right) \cap \pi^{-1} \left( p \right) \\
           &= \left\{ p \right\} \times \underbrace{\left\{ \left[ 0 : 1 : 1 : - a_1 \left( p \right) : 1 : 1 : 1 : 1 \right] \right\}}_{\mathrm{ fibre coordinates }} \, ,
\end{split}
\end{align}
where $p = [ p_1 : p_2 : \dots : p_{n-1} : e_0 = 0 ]$ are inhomogeneous coordinates of the point $p$. Hence, we have found a single intersection point in $\hat{\pi}^{-1} ( p )$. This finding can be made precise to state that in $\hat{\pi}^{-1} ( p )$ the divisor classes $E_0$ and $E_1$ intersect with intersection number 1 \cite{Krause:2011xj}. Moreover this analysis is easily repeated for all intersections of the divisor classes $E_i$, $0 \leq i \leq 4$. 

To compute $U ( 1 )_X$-charges, intersection numbers with the $U ( 1 )_X$-generator $U_X$ are required. These intersections involve \cite{Krause:2011xj}
\[ E_5 = V \left( P_T^\prime, s \right) - V \left( P_T^\prime, z \right) - V \left( P_T^\prime, k_{\mathcal{B}_6} \right) \] 
where $k_{\mathcal{B}_6}$ is a polynomial in the coordinate ring of $\hat{Y}_5$ such that its degree matches $\overline{\mathcal{K}}_{\mathcal{B}_6}$. To simplify notation we set $\alpha = V ( P_T^\prime, s )$, $\beta = V ( P_T^\prime, z )$ and $\gamma = V ( P_T^\prime, k_{\mathcal{B}_6} )$. The intersection numbers are then as follows:
\begin{align}
\begin{tabular}{c@{\hskip 20pt}ccccc@{\hskip 20pt}cccc}
\toprule
& $E_0$ & $E_1$ & $E_2$ & $E_3$ & $E_4$ & $E_5$ & $\alpha$ & $\beta$ & $\gamma$ \\
\midrule
$E_0$ & -2 & 1 & $\cdot$ & $\cdot$ & 1  & -1 & $\cdot$ & 1 & $\cdot$ \\
$E_1$ & 1 & -2 & 1 & $\cdot$ & $\cdot$ & $\cdot$ & $\cdot$ & $\cdot$ & $\cdot$ \\
$E_2$ & $\cdot$ & 1 & -2 & 1 & $\cdot$ & $\cdot$ & $\cdot$ & $\cdot$ & $\cdot$ \\
$E_3$ & $\cdot$ & $\cdot$ & 1 & -2 & 1 & 1 & 1 & $\cdot$ & $\cdot$ \\
$E_4$ & 1 & $\cdot$ & $\cdot$ & 1 & -2 & $\cdot$ & $\cdot$ & $\cdot$ & $\cdot$ \\
\bottomrule
\end{tabular}
\end{align}

\subsubsection{Intersection Structure over Matter Curves away from Yukawa Loci}

\paragraph{Over $\mathbf{C_{\mathbf{10}_{1}}}$}
Over the matter curves singularity enhancements occur. This expresses itself geometrically in the presence of new $\mathbb{P}^1$-fibrations, of which linear combinations eventually serve as matter surfaces. Over $C_{\mathbf{10}_{1}} = V( w, a_{1,0} ) \subseteq \mathcal{B}_6$ the following six $\mathbb{P}^1$-fibrations are present:
\begin{subequations}
\begin{align}
\mathbb{P}_{0,A}^1 \left( \mathbf{10}_{1} \right) &= V \left( a_{1,0}, e_0, y^2 e_4 - x^3 s e_1 e_2^2 \right) \, , \\
\mathbb{P}_{14}^1 \left( \mathbf{10}_{1} \right) &= V \left( a_{1,0}, e_1, e_4 \right) \, , \\
\mathbb{P}_{24}^1 \left( \mathbf{10}_{1} \right) &= V \left( a_{1,0}, e_2, e_4 \right) \, , \\
\mathbb{P}_{2B}^1 \left( \mathbf{10}_{1} \right) &= V \left( a_{1,0}, e_2, y s e_3 + a_{3,2} z^3 e_0^2 e_1 \right) \, , \\
\mathbb{P}_{3C}^1 \left( \mathbf{10}_{1} \right) &= V \left( a_{1,0}, e_3, a_{3,2} y z e_0 e_4 - a_{2,1} x^2 s e_2 - a_{4,3} x z^2 e_0^2 e_1 e_2 e_4 \right) \, , \\
\mathbb{P}^1_{4D} \left( \mathbf{10}_{1} \right) &= V \left( a_{1,0}, e_4, x s e_2 e_3 + a_{2,1} z^2 e_0 \right) \, .
\end{align}
\end{subequations}
The total elliptic fibre over $C_{\mathbf{10}_{1}}$ is given by
\[ 
\resizebox{0.9\textwidth}{!}{$
T^2 \left( \mathbf{10}_{1} \right) = \mathbb{P}_{0A}^1 \left( \mathbf{10}_{1} \right) + 2 \mathbb{P}_{14}^1 \left( \mathbf{10}_{1} \right) +  2 \mathbb{P}_{24}^1 \left( \mathbf{10}_{1} \right) +  \mathbb{P}_{2B}^1 \left( \mathbf{10}_{1} \right) + \mathbb{P}_{3C}^1 \left( \mathbf{10}_{1} \right) +  \mathbb{P}_{4D}^1 \left( \mathbf{10}_{1} \right) \, .$}
\]
The above $\mathbb{P}^1$-fibrations originate from the divisors $E_i$ according to the following splitting:
\begin{align}
\label{splitting101}
\begin{tabular}{cc}
\toprule
Original & Split components \\
\midrule
$E_0$ & $\mathbb{P}_{0A}^1 \left( \mathbf{10}_{1} \right)$ \\
$E_1$ & $\mathbb{P}_{14}^1 \left( \mathbf{10}_{1} \right)$ \\
$E_2$ & $\mathbb{P}_{24}^1\left( \mathbf{10}_{1} \right)$ + $\mathbb{P}_{2B}^1\left( \mathbf{10}_1 \right)$ \\
$E_3$ & $\mathbb{P}_{3C}^1\left( \mathbf{10}_1 \right)$ \\
$E_4$ & $\mathbb{P}_{14}^1\left( \mathbf{10}_1 \right)$ + $\mathbb{P}_{24}^1\left( \mathbf{10}_1 \right)$ + $\mathbb{P}_{4D}^1\left( \mathbf{10}_1 \right)$ \\
\bottomrule
\end{tabular}
\end{align}
Over $p \in C_{\mathbf{10}_1}$ which is not a Yukawa point, these $\mathbb{P}^1$-fibrations intersect in $\hat{\pi}^{-1} \left( p \right)$ as follows:
\begin{align}
\resizebox{0.85\textwidth}{!}{
\begin{tabular}{c@{\hskip 20pt}cccccc}
\toprule
& $\mathbb{P}_{0A}^1 \left( \mathbf{10}_1 \right)$ & $\mathbb{P}_{14}^1 \left( \mathbf{10}_1 \right)$ & $\mathbb{P}_{24}^1 \left( \mathbf{10}_1 \right)$
& $\mathbb{P}_{2B}^1 \left( \mathbf{10}_1 \right)$ & $\mathbb{P}_{3C}^1 \left( \mathbf{10}_1 \right)$ & $\mathbb{P}^1_{4D} \left( \mathbf{10}_1 \right)$ \\
\midrule
$\mathbb{P}_{0A}^1 \left( \mathbf{10}_1 \right)$ & -2 & 1  & $\cdot$ & $\cdot$ & $\cdot$ & $\cdot$ \\
$\mathbb{P}_{14}^1 \left( \mathbf{10}_1 \right)$ & 1 & -2 & 1 & $\cdot$ & $\cdot$ & 1 \\
$\mathbb{P}_{24}^1 \left( \mathbf{10}_1 \right)$ & $\cdot$ & 1 & -2 & 1 & 1 & $\cdot$ \\
$\mathbb{P}_{2B}^1 \left( \mathbf{10}_1 \right)$ & $\cdot$ & $\cdot$ & 1 & -2 & $\cdot$ & $\cdot$ \\
$\mathbb{P}_{3C}^1 \left( \mathbf{10}_1 \right)$ & $\cdot$ & $\cdot$ & 1 & $\cdot$ & -2 & $\cdot$ \\
$\mathbb{P}_{4D}^1 \left( \mathbf{10}_1 \right)$ & $\cdot$ & 1 & $\cdot$ & $\cdot$ & $\cdot$ & -2 \\
\bottomrule
\end{tabular}
}
\end{align}
The intersection numbers between the $\mathbb{P}^1$-fibrations over $C_{\mathbf{10}_1}$ and the pullbacks of the divisors $E_i$ -- including $E_5$ -- onto the fibration over $C_{\mathbf{10}_1}$ are readily computed. It holds:
\begin{align}
\begin{tabular}{c@{\hskip 20pt}ccccc@{\hskip 20pt}cccc}
\toprule
 & $E_0$ & $E_1$ & $E_2$ & $E_3$ & $E_4$ & $E_5$ & $\alpha$ & $\beta$ & $\gamma$ \\
\midrule
$\mathbb{P}_{0A}^1 \left( \mathbf{10}_1 \right)$ & -2 & 1 & $\cdot$ & $\cdot$ & 1 & -1 & $\cdot$ & 1 & $\cdot$ \\
$\mathbb{P}_{14}^1 \left( \mathbf{10}_1 \right)$ & 1 & -2 & 1 & $\cdot$ & $\cdot$ & $\cdot$ & $\cdot$ & $\cdot$ & $\cdot$ \\
$\mathbb{P}_{24}^1 \left( \mathbf{10}_1 \right)$ & $\cdot$ & 1 & -1 & 1  & -1 & $\cdot$ & $\cdot$ & $\cdot$ & $\cdot$ \\
$\mathbb{P}_{2B}^1 \left( \mathbf{10}_1 \right)$ & $\cdot$ & $\cdot$ & -1 & $\cdot$ & 1 & $\cdot$ & $\cdot$ & $\cdot$ & $\cdot$ \\
$\mathbb{P}_{3C}^1 \left( \mathbf{10}_1 \right)$ & $\cdot$ & $\cdot$ & 1 & -2 & 1 & 1  & 1 & $\cdot$ & $\cdot$ \\
$\mathbb{P}_{4D}^1 \left( \mathbf{10}_1 \right)$ & $\cdot$ & 1 & $\cdot$ & $\cdot$ & -1 & $\cdot$ & $\cdot$ & $\cdot$ & $\cdot$ \\
\bottomrule
\end{tabular}
\end{align}
The matter surfaces $S_{\mathbf{10}_1}^{(a)}$ over $C_{\mathbf{10}_1}$ are linear combinations of the above $\mathbb{P}^1$-fibrations. We use $\vec{P}$ to denote such a linear combination compactly. To this end, $\vec{P}$ is a list of the multiplicities with which these $\mathbb{P}^1$-fibrations appear in the above order. Hence,
$\vec{P} = ( 0, 1, 0, 4, 0, 0 )$ corresponds to $1 \cdot \mathbb{P}^1_{14} ( \mathbf{10}_1 ) + 4 \cdot \mathbb{P}^1_{2B} ( \mathbf{10}_1 )$. $\vec{\beta}$ indicates the Cartan charges of such a linear combination, \ie lists the intersection numbers with the resolution divisors $E_i$, $1 \leq i \leq 4$. We will adopt these notations also for the other matter curves. All that said, the matter surfaces over $C_{\mathbf{10}_1}$ take the following form:
\begin{align}
\resizebox{0.87\textwidth}{!}{
\begin{tabular}{ccc@{\hskip 20pt}ccc}
\toprule
Label & $\vec{P}$ & $\vect{\beta}$ & Label & $\vec{P}$ & $\vect{\beta}$ \\
\midrule
$S_{\mathbf{10}_1}^{(1)}$ & $\left( 0, -1, -2, -1, -1, 0 \right)$ & $\left( 0, 1, 0, 0 \right)$ & $S_{\mathbf{10}_1}^{(6)}$ & $\left( 0, 0, 0, 0, 0, 1 \right)$ & $\left( 1, 0, 0, -1 \right)$ \\
$S_{\mathbf{10}_1}^{(2)}$ & $\left( 0, -1, -1, 0, -1, 0 \right)$ & $\left( 1, -1, 1, 0 \right)$ & $S_{\mathbf{10}_1}^{(7)}$ & $\left( 0, 0, 0, 1, 0, 0 \right)$ & $\left( 0, -1, 0, 1 \right)$ \\
$S_{\mathbf{10}_1}^{(3)}$ & $\left( 0, 0, -1, 0, -1, 0 \right)$ & $\left( -1, 0, 1, 0 \right)$ & $S_{\mathbf{10}_1}^{(8)}$ & $\left( 0, 1, 0, 0, 0, 1 \right)$ & $\left( -1, 1, 0, -1 \right)$ \\
$S_{\mathbf{10}_1}^{(4)}$ & $\left( 0, -1, -1, 0, 0, 0 \right)$ & $\left( 1, 0, -1, 1 \right)$ & $S_{\mathbf{10}_1}^{(9)}$ & $\left( 0, 1, 1, 1, 0, 1 \right)$ & $\left( 0, -1, 1, -1 \right)$ \\
$S_{\mathbf{10}_1}^{(5)}$ & $\left( 0, 0, -1, 0, 0, 0 \right)$ & $\left( -1, 1, -1, 1 \right)$ & $S_{\mathbf{10}_1}^{(10)}$ & $\left( 0, 1, 1, 1, 1, 1 \right)$ & $\left( 0, 0, -1, 0 \right)$ \\
\bottomrule
\end{tabular}} \label{app_ S10}
\end{align}

\paragraph{Over $\mathbf{C_{\mathbf{5}_{3}}}$}
Over $C_{\mathbf{5}_{3}} = V( w, a_{3,2} ) \subseteq \mathcal{B}_6$ the following six $\mathbb{P}^1$-fibrations are present:
\begin{subequations} \label{53curvesplitting}
\begin{align}
\mathbb{P}_{0}^1 \left( \mathbf{5}_3 \right) &= V \left( a_{3,2}, e_0, a_{1,0} x y z - e_1 e_2^2 e_3 s x^3 + e_3 e_4 y^2 \right) \, , \\
\mathbb{P}_{1}^1 \left( \mathbf{5}_3 \right) &= V \left( a_{3,2}, e_1, e_3 e_4 y + a_{1,0} x z \right) \, , \\
\mathbb{P}_{2E}^1 \left( \mathbf{5}_3 \right) &= V \left( a_{3,2}, e_2, e_3 e_4 y + a_{1,0} x z \right) \, , \\
\mathbb{P}_{3x}^1 \left( \mathbf{5}_3 \right) &= V \left( a_{3,2}, e_3, x \right) \, , \\
\mathbb{P}_{3F}^1 \left( \mathbf{5}_3 \right) &= V \left( a_{3,2}, e_3, a_{1,0} s y - a_{2,1} e_0 e_1 e_2 s x z - a_{4,3} e_0^3 e_1^2 e_2 e_4 z^3 \right) \, , \\ 
\mathbb{P}_{4}^1 \left( \mathbf{5}_3 \right) &= V \left( a_{3,2}, e_4, a_{1,0} y z - e_1 e_2^2 e_3 s x^2 - a_{2,1} e_0 e_1 e_2 x z^2 \right) \, .
\end{align}
\end{subequations}
The total elliptic fibre over $C_{\mathbf{5}_3}$ is given by
\[ T^2 \left( \mathbf{5}_3 \right) = \mathbb{P}_{0}^1 \left( \mathbf{5}_3 \right) + \mathbb{P}_{1}^1 \left( \mathbf{5}_3 \right) +  \mathbb{P}_{2E}^1 \left( \mathbf{5}_3 \right) +  \mathbb{P}_{3x}^1 \left( \mathbf{5}_3 \right) +  \mathbb{P}_{3F}^1 \left( \mathbf{5}_3 \right) +  \mathbb{P}_{4}^1 \left( \mathbf{5}_3 \right) \, . \]
The above $\mathbb{P}^1$-fibrations emerge from the $E_i$ according to the following table.
\begin{align} \label{splitting53}
\begin{tabular}{cc}
\toprule
Original & Split components \\
\midrule
$E_0$ & $\mathbb{P}_{0}^1 \left( \mathbf{5}_3 \right)$ \\
$E_1$ & $\mathbb{P}_{1}^1 \left( \mathbf{5}_3 \right)$ \\
$E_2$ & $\mathbb{P}_{2E}^1 \left( \mathbf{5}_3 \right)$ \\
$E_3$ & $\mathbb{P}_{3x}^1 \left( \mathbf{5}_3 \right)$ + $\mathbb{P}_{3F}^1 \left( \mathbf{5}_3 \right)$ \\
$E_4$ & $\mathbb{P}_{4}^1 \left( \mathbf{5}_3 \right)$ \\
\bottomrule
\end{tabular}
\end{align}
Over $p \in C_{\mathbf{5}_3}$ which is not a Yukawa point, these $\mathbb{P}^1$-fibrations intersect in $\hat{\pi}^{-1} \left( p \right)$ as follows:
\begin{align}
\begin{tabular}{c@{\hskip 20pt}cccccc}
\toprule
 & $\mathbb{P}_{0}^1 \left( \mathbf{5}_3 \right)$ & $\mathbb{P}_{1}^1 \left( \mathbf{5}_3 \right)$ & $\mathbb{P}_{2E}^1 \left( \mathbf{5}_3 \right)$ & $\mathbb{P}_{3x}^1 \left( \mathbf{5}_3 \right)$ & $\mathbb{P}_{3F}^1 \left( \mathbf{5}_3 \right)$ & $\mathbb{P}^1_{4} \left( \mathbf{5}_3 \right)$ \\
\midrule
$\mathbb{P}_{0}^1 \left( \mathbf{5}_3 \right)$  & -2 &  1 & $\cdot$ & $\cdot$ & $\cdot$ & 1 \\
$\mathbb{P}_{1}^1 \left( \mathbf{5}_3 \right)$  & 1 & -2 & 1  & $\cdot$ & $\cdot$ & $\cdot$ \\
$\mathbb{P}_{2E}^1 \left( \mathbf{5}_3 \right)$ & $\cdot$ & 1 & -2 & 1 & $\cdot$ & $\cdot$ \\
$\mathbb{P}_{3x}^1 \left( \mathbf{5}_3 \right)$ & $\cdot$ & $\cdot$ & 1 & -2 & 1 & $\cdot$ \\
$\mathbb{P}_{3F}^1 \left( \mathbf{5}_3 \right)$ & $\cdot$ & $\cdot$ & $\cdot$ & 1 & -2 & 1 \\
$\mathbb{P}^1_{4} \left( \mathbf{5}_3 \right)$  & 1 & $\cdot$ & $\cdot$ & $\cdot$ & 1 & -2 \\
\bottomrule
\end{tabular}
\end{align}
The intersections with the pullbacks of the divisors $E_i$ onto the fibre over $C_{\mathbf{5}_3}$ are as follows.
\begin{align}
\begin{tabular}{c@{\hskip 20pt}ccccc@{\hskip 20pt}cccc}
\toprule
 & $E_0$ & $E_1$ & $E_2$ & $E_3$ & $E_4$ & $E_5 = \alpha - \beta - \gamma$ & $\alpha$ & $\beta$ & $\gamma$ \\
\midrule
$\mathbb{P}_{0}^1 \left( \mathbf{5}_3 \right)$  & -2 & 1  & $\cdot$ & $\cdot$ & 1 & -1 & $\cdot$ & 1 & $\cdot$ \\
$\mathbb{P}_{1}^1 \left( \mathbf{5}_3 \right)$  &  1 & -2 & 1  & $\cdot$ & $\cdot$ & $\cdot$ & $\cdot$ & $\cdot$ & $\cdot$ \\
$\mathbb{P}_{2E}^1 \left( \mathbf{5}_3 \right)$ &  $\cdot$ & 1 & -2 & 1 & $\cdot$ & $\cdot$ & $\cdot$ & $\cdot$ & $\cdot$ \\
$\mathbb{P}_{3x}^1 \left( \mathbf{5}_3 \right)$ &  $\cdot$ & $\cdot$ & 1 & -1 & $\cdot$ & 1 & 1 & $\cdot$ & $\cdot$ \\
$\mathbb{P}_{3F}^1 \left( \mathbf{5}_3 \right)$ &  $\cdot$ & $\cdot$ & $\cdot$ & -1 & 1 & $\cdot$ & $\cdot$ & $\cdot$ & $\cdot$ \\
$\mathbb{P}_{4}^1 \left( \mathbf{5}_3 \right)$  &  1 & $\cdot$ & $\cdot$ & 1 & -2 & $\cdot$ & $\cdot$ & $\cdot$ & $\cdot$ \\
\bottomrule
\end{tabular}
\end{align}
The matter surfaces over $C_{\mathbf{5}_3}$ are:
\begin{align} \label{app_S53}
\begin{tabular}{ccc@{\hskip 20pt}ccc}
\toprule
Label & $\vec{P}$ & $\vect{\beta}$ & Label & $\vec{P}^1$ & $\vect{\beta}$ \\
\midrule
$S_{\mathbf{5}_3}^{(1)}$ & $\left( 0, -1, -1, -1, 0, 0 \right)$ & $\left( 1, 0, 0, 0 \right)$ & $S_{\mathbf{5}_3}^{(4)}$ & $\left( 0, 0, 0, 0, 1, 0 \right)$ & $\left( 0, 0, -1, 1 \right)$ \\
$S_{\mathbf{5}_3}^{(2)}$ & $\left( 0, 0, -1, -1, 0, 0 \right)$ & $\left( -1, 1, 0, 0 \right)$ & $S_{\mathbf{5}_3}^{(5)}$ & $\left( 0, 0, 0, 0, 1, 1 \right)$ & $\left( 0, 0, 0, -1 \right)$ \\
$S_{\mathbf{5}_3}^{(3)}$ & $\left( 0, 0, 0, -1, 0, 0 \right)$ & $\left( 0, -1, 1, 0 \right)$ & & & \\
\bottomrule
\end{tabular}
\end{align}

\paragraph{Over $\mathbf{C_{\mathbf{5}_{-2}}}$}
By primary decompositions it is readily found that over $C_{\mathbf{5}_{-2}} = V( w, a_{1,0} a_{4,3} - a_{2,1} a_{3,2} ) \subseteq \mathcal{B}_6$ the following $\mathbb{P}^1$-fibrations are present:
\begin{subequations}
\label{52curvesplitting}
\begin{align}
\mathbb{P}_{0}^1 \left( \mathbf{5}_{-2} \right) &= V \left( a_{3,2} a_{2,1} - a_{4,3} a_{1,0}, e_0, e_3 e_4 y^2 + a_{1,0} x y z - e_1 e_2^2 e_3 s x^3 \right) \, , \\
\mathbb{P}_{1}^1 \left( \mathbf{5}_{-2} \right) &= V \left( a_{3,2} a_{2,1} - a_{4,3} a_{1,0}, e_1, e_3 e_4 y + a_{1,0} x z \right) \, , \\
\mathbb{P}_{2}^1 \left( \mathbf{5}_{-2} \right) &= V \left( a_{3,2} a_{2,1} - a_{4,3} a_{1,0}, e_2, e_0^2 z^3 e_1 e_4 a_{3,2} + y s e_3 e_4 + a_{1,0} x z s, \right. \\
      & \hspace{12em} \left. a_{1,0} a_{4,3} e_0^2 z^3 e_1 e_4 + a_{2,1} y s e_3 e_4 + a_{1,0} a_{2,1} x z s \right) \, , \nonumber \\
\mathbb{P}_{3G}^1 \left( \mathbf{5}_{-2} \right) &= V \left( a_{3,2} a_{2,1} - a_{4,3} a_{1,0}, e_3, a_{4,3} e_0^2 z^2 e_1 e_4 + a_{2,1} x s, a_{3,2} e_0^2 z^2 e_1 e_4 + a_{1,0} x s \right) \, , \\
\mathbb{P}_{3H}^1 \left( \mathbf{5}_{-2} \right) &= V \left( a_{3,2} a_{2,1} - a_{4,3} a_{1,0}, e_3, a_{4,3} e_0 x z e_1 e_2 - a_{3,2} y, a_{2,1} e_0 x z e_1 e_2 - a_{1,0} y  \right) \, , \\
\mathbb{P}_{4}^1 \left( \mathbf{5}_{-2} \right) &= V \left( a_{3,2} a_{2,1} - a_{4,3} a_{1,0}, e_4, a_{1,0} y z - a_{2,1} e_0 e_1 e_2 x z^2 - e_1 e_2^2 e_3 s x^2 \right) \, .
\end{align}
\end{subequations}
Note that these results differ from \cite{Krause:2011xj}, where primary decomposition was not applied. The total elliptic fibre over $C_{\mathbf{5}_{-2}}$ is given by
\[ T^2 \left( \mathbf{5}_{-2} \right) = \mathbb{P}_{0}^1 \left( \mathbf{5}_{-2} \right) + \mathbb{P}_{1}^1 \left( \mathbf{5}_{-2} \right) +  \mathbb{P}_{2}^1 \left( \mathbf{5}_{-2} \right) +  \mathbb{P}_{3G}^1 \left( \mathbf{5}_{-2} \right) +  \mathbb{P}_{3H}^1 \left( \mathbf{5}_{-2} \right) +  \mathbb{P}_{4}^1 \left( \mathbf{5}_{-2} \right) \, . \]
The above $\mathbb{P}^1$-fibrations emerge from the $E_i$ according to the following table.
\begin{align} \label{splitting5-2}
\begin{tabular}{cc}
\toprule
Original & Split components \\
\midrule
$E_0$ & $\mathbb{P}_{0}^1 \left( \mathbf{5}_{-2} \right)$ \\
$E_1$ & $\mathbb{P}_{1}^1 \left( \mathbf{5}_{-2} \right)$ \\
$E_2$ & $\mathbb{P}_{2}^1 \left( \mathbf{5}_{-2} \right)$ \\
$E_3$ & $\mathbb{P}_{3G}^1\left( \mathbf{5}_{-2} \right)$ + $\mathbb{P}_{3H}^1 \left( \mathbf{5}_{-2} \right)$ \\
$E_4$ & $\mathbb{P}_{4}^1 \left( \mathbf{5}_{-2} \right)$ \\
\bottomrule
\end{tabular}
\end{align}
Over $p \in C_{\mathbf{5}_3}$ which is not a Yukawa point, these $\mathbb{P}^1$-fibrations intersect in $\hat{\pi}^{-1} \left( p \right)$ as follows:
\begin{align}
\begin{tabular}{c@{\hskip 20pt}cccccc}
\toprule
 & $\mathbb{P}_{0}^1 \left( \mathbf{5}_{-2} \right)$ & $\mathbb{P}_{1}^1 \left( \mathbf{5}_{-2} \right)$ & $\mathbb{P}_{2}^1 \left( \mathbf{5}_{-2} \right)$ 
 & $\mathbb{P}_{3G}^1 \left( \mathbf{5}_{-2} \right)$ & $\mathbb{P}_{3H}^1 \left( \mathbf{5}_{-2} \right)$ & $\mathbb{P}^1_{4} \left( \mathbf{5}_{-2} \right)$ \\
\midrule
$\mathbb{P}_{0}^1 \left( \mathbf{5}_{-2} \right)$  & -2 &  1 & $\cdot$ & $\cdot$ & $\cdot$ & 1\\
$\mathbb{P}_{1}^1 \left( \mathbf{5}_{-2} \right)$  & 1 & -2 & 1 & $\cdot$ & $\cdot$ & $\cdot$ \\
$\mathbb{P}_{2}^1 \left( \mathbf{5}_{-2} \right)$  & $\cdot$ & 1 & -2 & 1 & $\cdot$ & $\cdot$ \\
$\mathbb{P}_{3G}^1 \left( \mathbf{5}_{-2} \right)$ & $\cdot$ & $\cdot$ & 1 & -2 & 1 & $\cdot$ \\
$\mathbb{P}_{3H}^1 \left( \mathbf{5}_{-2} \right)$ & $\cdot$ & $\cdot$ & $\cdot$ & 1 & -2 & 1 \\
$\mathbb{P}^1_{4} \left( \mathbf{5}_{-2} \right)$  & 1 & $\cdot$ & $\cdot$ & $\cdot$ & 1 & -2 \\
\bottomrule
\end{tabular}
\end{align}
The intersections with the pullbacks of the divisors $E_i$ onto the fibre over $C_{\mathbf{5}_{-2}}$ are as follows.
\begin{align}
\begin{tabular}{c@{\hskip 20pt}ccccc@{\hskip 20pt}cccc} 
\toprule
 & $E_0$ & $E_1$ & $E_2$ & $E_3$ & $E_4$ & $E_5$ & $\alpha$ & $\beta$ & $\gamma$ \\
\midrule
$\mathbb{P}_{0}^1 \left( \mathbf{5}_{-2} \right)$  & -2 & 1  & $\cdot$ & $\cdot$ & 1 & -1 & $\cdot$ & 1 & $\cdot$ \\
$\mathbb{P}_{1}^1 \left( \mathbf{5}_{-2} \right)$  & 1 & -2 & 1 & $\cdot$ & $\cdot$ & $\cdot$ & $\cdot$ & $\cdot$ & $\cdot$ \\
$\mathbb{P}_{2}^1 \left( \mathbf{5}_{-2} \right)$ & $\cdot$ & 1 & -2 & 1 & $\cdot$ & $\cdot$ & $\cdot$ & $\cdot$ & $\cdot$ \\
$\mathbb{P}_{3G}^1 \left( \mathbf{5}_{-2} \right)$ &  $\cdot$ & $\cdot$ & 1 & -1 & $\cdot$ & $\cdot$ & $\cdot$ & $\cdot$ & $\cdot$ \\
$\mathbb{P}_{3H}^1 \left( \mathbf{5}_{-2} \right)$ &  $\cdot$ & $\cdot$ & $\cdot$ & -1 & 1 & 1 & 1 & $\cdot$ & $\cdot$ \\
$\mathbb{P}_{4}^1 \left( \mathbf{5}_{-2} \right)$  &  1 & $\cdot$ & $\cdot$ & 1 & -2 & $\cdot$ & $\cdot$ & $\cdot$ & $\cdot$ \\
\bottomrule
\end{tabular}
\end{align}
The matter surfaces over $C_{\mathbf{5}_{-2}}$ are:
\begin{align}
\begin{tabular}{ccc@{\hskip 20pt}ccc}
\toprule
Label & $\vec{P}$ & $\vect{\beta}$ & Label & $\vec{P}^1$ & $\vect{\beta}$ \\
\midrule
$S_{\mathbf{5}_{-2}}^{(1)}$ & $\left( 0, -1, -1, -1, 0, 0 \right)$ & $\left( 1, 0, 0, 0 \right)$ & $S_{\mathbf{5}_{-2}}^{(4)}$ & $\left( 0, 0, 0, 0, 1, 0 \right)$ & $\left( 0, 0, -1, 1 \right)$ \\
$S_{\mathbf{5}_{-2}}^{(2)}$ & $\left( 0, 0, -1, -1, 0, 0 \right)$ & $\left( -1, 1, 0, 0 \right)$ & $S_{\mathbf{5}_{-2}}^{(5)}$ & $\left( 0, 0, 0, 0, 1, 1 \right)$ & $\left( 0, 0, 0, -1 \right)$ \\
$S_{\mathbf{5}_{-2}}^{(3)}$ & $\left( 0, 0, 0, -1, 0, 0 \right)$ & $\left( 0, -1, 1, 0 \right)$ & & & \\
\bottomrule
\end{tabular} \label{app:MatterSurfaces52}
\end{align}

\paragraph{Over $\mathbf{C_{\mathbf{1}_{5}}}$}
Over the singlet curve $C_{\mathbf{1}_{5}} = V \left( a_{3,2}, a_{4,3} \right)$ the following two $\mathbb{P}^1$-fibrations are present:
\begin{subequations}
\begin{align}
\mathbb{P}_{A}^1 \left( \mathbf{1}_{5} \right) &= V \left( a_{3,2}, a_{4,3}, s \right) \, , \\
\mathbb{P}_{B}^1 \left( \mathbf{1}_{5} \right) &= V \left( a_{3,2}, a_{4,3}, y^2 e_3 e_4 + a_{1,0} x y z - x^3 s e_1 e_2^2 e_3 - a_{2,1} x^2 z^2 e_0 e_1 e_2 \right) \, .
\end{align}
\end{subequations}
The intersection numbers of these fibrations in $\hat{\pi}^{-1} \left( p \right)$ are as follows:
\[ \mathbb{P}^1_A \left( \mathbf{1}_5 \right)^2 = \mathbb{P}^1_B \left( \mathbf{1}_5 \right)^2 = -2 \, , \qquad \mathbb{P}^1_A \left( \mathbf{1}_5 \right) \mathbb{P}^1_B \left( \mathbf{1}_5 \right) = 2 \, . \]
The intersection numbers with the divisors $E_i$ are given by
\begin{align}
\begin{tabular}{c@{\hskip 20pt}ccccc@{\hskip 20pt}cccc@{\hskip 20pt}c}
\toprule
 & $E_0$ & $E_1$ & $E_2$ & $E_3$ & $E_4$ & $E_5$ & $\alpha$ & $\beta$ & $\gamma$ & $q_X$ \\
\midrule
$\mathbb{P}_{A}^1 \left( \mathbf{1}_{5} \right) = S_{\mathbf{1}_{5}}^{(1)}$ & $\cdot$ & $\cdot$ & $\cdot$ & $\cdot$ & $\cdot$ & -1 & -1 & $\cdot$ & $\cdot$ & $5$ \\
\vspace{-0.8em} & \\
$\mathbb{P}_{B}^1 \left( \mathbf{1}_{5} \right) = S_{\mathbf{1}_{5}}^{(2)}$ & $\cdot$ & $\cdot$ & $\cdot$ & $\cdot$ & $\cdot$ & 1 & 2 & 1 & $\cdot$ & $-5$ \\
\bottomrule
\end{tabular}
\end{align}
Only $\mathbb{P}^1_A \left( \mathbf{1}_5 \right)$ has vanishing intersection with the zero-section $\beta = V \left( z \right)$
and thus defines the only viable matter surface. As this fibration has $q_X = 5$, we denote the singlet curve as $C_{\mathbf{1}_5}$.

\subsubsection{Intersection Structure over Yukawa Loci}

\paragraph{Over $\mathbf{Y_1}$}
Over $Y_1 = V ( w, a_{1,0}, a_{2,1} )$ the following six $\mathbb{P}^1$-fibrations are present:
\begin{subequations}
\begin{align}
 \mathbb{P}_{0A}^1 \left( Y_1 \right) &= V \left( a_{1,0}, a_{2,1}, e_0, y^2 e_4 - x^3 s e_1 e_2^2 \right) \, , \\
 \mathbb{P}_{14}^1 \left( Y_1 \right) &= V \left( a_{1,0}, a_{2,1}, e_1, e_4 \right) \, , \\
 \mathbb{P}_{24}^1 \left( Y_1 \right) &= V \left( a_{1,0}, a_{2,1}, e_2, e_4 \right) \, , \\
 \mathbb{P}_{2B}^1 \left( Y_1 \right) &= V \left( a_{1,0}, a_{2,1}, e_2, y s e_3 + a_{3,2} z^3 e_0^2 e_1 \right) \, , \\
 \mathbb{P}_{34}^1 \left( Y_1 \right) &= V \left( a_{1,0}, a_{2,1}, e_3, e_4 \right) \, , \\
 \mathbb{P}_{3J}^1 \left( Y_1 \right) &= V \left( a_{1,0}, a_{2,1}, e_3, a_{3,2} y - a_{4,3} x z e_0 e_1 e_2 \right) \, .
\end{align}
\end{subequations}
The total elliptic fibre over $Y_1$ is given by
\[ T^2 \left( Y_1 \right) = \mathbb{P}_{0A}^1 \left( Y_1 \right) + 2 \mathbb{P}_{14}^1 \left( Y_1 \right) + 3 \mathbb{P}_{24}^1 \left( Y_1 \right)
+ \mathbb{P}_{2B}^1 \left( Y_1 \right) + 2 \mathbb{P}_{34}^1 \left( Y_1 \right) + \mathbb{P}_{3J}^1 \left( Y_1 \right) \, . \]
Starting from $C_{\mathbf{10}_1}$, the above fibrations emerge from the following splitting process:
\begin{align}
\begin{tabular}{cc}
\toprule
Split components over $C_{10_{1}}$ & Split components over $Y_1$ \\
\midrule
$\mathbb{P}_{0A}^1 \left( \mathbf{10}_1 \right)$ & $\mathbb{P}_{0A}^1 \left( Y_1 \right)$ \\
$\mathbb{P}_{14}^1 \left( \mathbf{10}_1 \right)$ & $\mathbb{P}_{14}^1 \left( Y_1 \right)$ \\
$\mathbb{P}_{24}^1\left( \mathbf{10}_1 \right)$ & $\mathbb{P}_{24}^1\left( Y_1 \right)$ \\
$\mathbb{P}_{2B}^1\left( \mathbf{10}_1 \right)$ & $\mathbb{P}_{2B}^1\left( Y_1 \right)$ \\
$\mathbb{P}_{3C}^1\left( \mathbf{10}_1 \right)$ & $\mathbb{P}_{34}^1\left( Y_1 \right)$ + $\mathbb{P}_{3J}^1\left( Y_1 \right)$ \\
$\mathbb{P}_{4D}^1\left( \mathbf{10}_1 \right)$ & $\mathbb{P}_{24}^1\left( Y_1 \right)$ + $\mathbb{P}_{34}^1\left( Y_1 \right)$ \\
\bottomrule
\end{tabular}
\label{tab:Splitting10ToY1SU(5)xU(1)}
\end{align}
The splitting behaviour, when approached from $C_{\mathbf{5}_{-2}}$, is different:
\begin{align}
\begin{tabular}{cc}
\bottomrule
Split components over $C_{5_{-2}}$ & Split components over $Y_1$ \\
\midrule
$\mathbb{P}_{0}^1 \left( \mathbf{5}_{-2} \right)$ & $\mathbb{P}_{0A}^1 \left( Y_1 \right)$ \\
$\mathbb{P}_{1}^1 \left( \mathbf{5}_{-2} \right)$ & $\mathbb{P}_{14}^1 \left( Y_1 \right)$ \\
$\mathbb{P}_{2}^1 \left( \mathbf{5}_{-2} \right)$ & $\mathbb{P}^{1}_{24} \left( Y_1 \right)$ + $\mathbb{P}^{1}_{2B} \left( Y_1 \right)$ \\
$\mathbb{P}_{3G}^1\left( \mathbf{5}_{-2} \right)$ & $\mathbb{P}^{1}_{34} \left( Y_1 \right)$ \\
$\mathbb{P}_{3H}^1\left( \mathbf{5}_{-2} \right)$ & $\mathbb{P}^{1}_{3J} \left( Y_1 \right)$ \\
$\mathbb{P}_{4}^1 \left( \mathbf{5}_{-2} \right)$ & $\mathbb{P}_{14}^1\left( Y_1 \right)$ + $2 \mathbb{P}_{24}^1\left( Y_1 \right)$ + $\mathbb{P}_{34}^1\left( Y_1 \right)$ \\
\bottomrule
\end{tabular}
\label{tab:Splitting5M2ToY1SU(5)xU(1)}
\end{align}
The intersection numbers in the fibre over $Y_1$ are as follows:
\begin{align}
\begin{tabular}{c@{\hskip 20pt}cccccc}
\toprule
 & $\mathbb{P}_{0A}^1 \left( Y_1 \right)$ & $\mathbb{P}_{14}^1 \left( Y_1 \right)$ & $\mathbb{P}_{24}^1 \left( Y_1 \right)$ & $\mathbb{P}_{2B}^1 \left( Y_1 \right)$ & $\mathbb{P}_{34}^1 \left( Y_1 \right)$ & $\mathbb{P}^1_{3J} \left( Y_1 \right)$ \\
\midrule
$\mathbb{P}_{0A}^1 \left( Y_1 \right)$  & -2 & 1 & $\cdot$ & $\cdot$ & $\cdot$ & $\cdot$ \\
$\mathbb{P}_{14}^1 \left( Y_1 \right)$  & 1 & -2 & 1 & $\cdot$ & $\cdot$ & $\cdot$ \\
$\mathbb{P}_{24}^1 \left( Y_1 \right)$  & $\cdot$ & 1 & $-\frac{3}{2}$ & $\frac{1}{2}$ & 1 & $\cdot$ \\
$\mathbb{P}_{2B}^1 \left( Y_1 \right)$  & $\cdot$ & $\cdot$ & $\frac{1}{2}$ & $-\frac{3}{2}$ & $\cdot$ & $\cdot$ \\
$\mathbb{P}_{34}^1 \left( Y_1 \right)$  & $\cdot$ & $\cdot$ & 1 & $\cdot$ & -2 & 1 \\
$\mathbb{P}_{3J}^1 \left( Y_1 \right)$  & $\cdot$ & $\cdot$ & $\cdot$ & $\cdot$ & 1 & -2 \\
\bottomrule
\end{tabular}
\label{tab:IntersectionsOfFibralCurvesOverY1SU(5)xU(1)}
\end{align}

\paragraph{Over $\mathbf{Y_2}$}
Over $Y_2 = V ( w, a_{1,0}, a_{3,2} )$ the following seven $\mathbb{P}^1$-fibrations are present:
\begin{subequations}
\begin{align}
 \mathbb{P}_{0A}^1 \left( Y_2 \right) &= V \left( a_{1,0}, a_{3,2}, e_0, y^2 e_4 - x^3 s e_1 e_2^2 \right) \, , \\
\mathbb{P}_{14}^1 \left( Y_2 \right) &= V \left( a_{1,0}, a_{3,2}, e_1, e_4 \right) \, , \\
 \mathbb{P}_{24}^1 \left( Y_2 \right) &= V \left( a_{1,0}, a_{3,2}, e_2, e_4 \right) \, , \\
 \mathbb{P}_{23}^1 \left( Y_2 \right) &= V \left( a_{1,0}, a_{3,2}, e_2, e_3 \right) \, , \\
 \mathbb{P}_{3x}^1 \left( Y_2 \right) &= V \left( a_{1,0}, a_{3,2}, e_3, x \right) \, , \\
 \mathbb{P}_{3K}^1 \left( Y_2 \right) &= V \left( a_{1,0}, a_{3,2}, e_3, a_{2,1} x s + a_{4,3} z^2 e_0^2 e_1 e_4 \right) \, , \\
 \mathbb{P}_{4D}^1 \left( Y_2 \right) &= V \left( a_{1,0}, a_{3,2}, e_4, x s e_2 e_3 + a_{2,1} z^2 e_0 \right) \, .
\end{align}
\end{subequations}
The total elliptic fibre over $Y_2$ is given by
\[ T^2 \left( Y_2 \right) = \mathbb{P}_{0A}^1 \left( Y_2 \right) + 2 \mathbb{P}_{14}^1 \left( Y_2 \right) + 2 \mathbb{P}_{24}^1 \left( Y_2 \right)
+ 2 \mathbb{P}_{23}^1 \left( Y_2 \right) + \mathbb{P}_{3x}^1 \left( Y_2 \right) + \mathbb{P}_{3K}^1 \left( Y_2 \right) + \mathbb{P}_{4D}^1 \left( Y_2 \right) \, . \]
The above $\mathbb{P}^1$-fibrations result from splittings of the corresponding fibrations over $C_{\mathbf{10}_1}$:
\begin{align}
\begin{tabular}{cc}
\toprule
Split components over $C_{10_{1}}$ & Split components over $Y_2$ \\
\midrule
$\mathbb{P}_{0A}^1 \left( \mathbf{10}_1 \right)$ & $\mathbb{P}_{0A}^1 \left( Y_2 \right)$ \\
$\mathbb{P}_{14}^1 \left( \mathbf{10}_1 \right)$ & $\mathbb{P}_{14}^1 \left( Y_2 \right)$ \\
$\mathbb{P}_{24}^1\left( \mathbf{10}_1 \right)$ & $\mathbb{P}_{24}^1\left( Y_2 \right)$ \\
$\mathbb{P}_{2B}^1\left( \mathbf{10}_1 \right)$ & $\mathbb{P}_{23}^1\left( Y_2 \right)$ \\
$\mathbb{P}_{3C}^1\left( \mathbf{10}_1 \right)$ & $\mathbb{P}_{3x}^1\left( Y_2 \right)$ + $\mathbb{P}_{23}^1 \left( Y_2 \right)$ + $\mathbb{P}_{3K}^1 \left( Y_2 
\right)$ \\
$\mathbb{P}_{4D}^1\left( \mathbf{10}_1 \right)$ & $\mathbb{P}_{4D}^1\left( Y_2 \right)$ \\
\bottomrule
\end{tabular}
\label{tab:Splitting10ToY2SU(5)xU(1)}
\end{align}
When approached from $C_{\mathbf{5}_3}$ the splittings are different, namely
\begin{align}
\begin{tabular}{cc}
\toprule
Split components over $C_{5_{3}}$ & Split components over $Y_2$ \\
\midrule
$\mathbb{P}_{0}^1 \left( \mathbf{5}_3 \right)$ & $\mathbb{P}_{0A}^1 \left( Y_2 \right)$ \\
$\mathbb{P}_{1}^1 \left( \mathbf{5}_3 \right)$ & $\mathbb{P}_{14}^1 \left( Y_2 \right)$ \\
$\mathbb{P}_{2E}^1 \left( \mathbf{5}_3 \right)$ & $\mathbb{P}^{1}_{23} \left( Y_2 \right)$ + $\mathbb{P}^{1}_{24} \left( Y_2 \right)$ \\
$\mathbb{P}_{3x}^1\left( \mathbf{5}_3 \right)$ & $\mathbb{P}^{1}_{3x} \left( Y_2 \right)$ \\
$\mathbb{P}_{3F}^1\left( \mathbf{5}_3 \right)$ & $\mathbb{P}^{1}_{23} \left( Y_2 \right)$ + $\mathbb{P}^{1}_{3K} \left( Y_2 \right)$ \\
$\mathbb{P}_{4}^1 \left( \mathbf{5}_3 \right)$ & $\mathbb{P}_{14}^1\left( Y_2 \right)$ + $\mathbb{P}_{24}^1\left( Y_2 \right)$ + $\mathbb{P}_{4D}^1\left( Y_2 \right)$ \\
\bottomrule
\end{tabular}
\end{align}
Finally, the splitting as seen from $C_{\mathbf{5}_{-2}}$, is yet again different:
\begin{align}
\begin{tabular}{cc}
\toprule
Split components over $C_{5_{-2}}$ & Split components over $Y_2$ \\
\midrule
$\mathbb{P}_{0}^1 \left( \mathbf{5}_{-2} \right)$ & $\mathbb{P}_{0A}^1 \left( Y_2 \right)$ \\
$\mathbb{P}_{1}^1 \left( \mathbf{5}_{-2} \right)$ & $\mathbb{P}_{14}^1 \left( Y_2 \right)$ \\
$\mathbb{P}_{2}^1 \left( \mathbf{5}_{-2} \right)$ & $\mathbb{P}^{1}_{23} \left( Y_2 \right)$ + $\mathbb{P}^{1}_{24} \left( Y_2 \right)$ \\
$\mathbb{P}_{3G}^1\left( \mathbf{5}_{-2} \right)$ & $\mathbb{P}_{3K}^1\left( Y_2 \right)$ \\
$\mathbb{P}_{3H}^1\left( \mathbf{5}_{-2} \right)$ & $\mathbb{P}_{23}^1\left( Y_2 \right)$ + $\mathbb{P}_{3x}^1\left( Y_2 \right)$ \\
$\mathbb{P}_{4}^1 \left( \mathbf{5}_{-2} \right)$ & $\mathbb{P}_{14}^1\left( Y_2 \right)$ + $\mathbb{P}_{24}^1\left( Y_2 \right)$ + $\mathbb{P}_{4D}^1\left( Y_2 \right)$ \\
\bottomrule
\end{tabular}
\label{tab:Splitting5M2ToY2SU(5)xU(1)}
\end{align}
The intersection numbers in the fibre over $Y_2$ are as follows:
\begin{align}
\resizebox{0.9\textwidth}{!}{
\begin{tabular}{c@{\hskip 20pt}ccccccc}
\toprule
 & $\mathbb{P}_{0A}^1 \left( Y_2 \right)$ & $\mathbb{P}_{14}^1 \left( Y_2 \right)$ & $\mathbb{P}_{24}^1 \left( Y_2 \right)$ & $\mathbb{P}_{23}^1 \left( Y_2 \right)$ & $\mathbb{P}_{3x}^1 \left( Y_2 \right)$ & $\mathbb{P}^1_{3K} \left( Y_2 \right)$ & $\mathbb{P}^1_{4D} \left( Y_2 \right)$ \\
\midrule
$\mathbb{P}_{0A}^1 \left( Y_2 \right)$  & -2 & 1 & $\cdot$ & $\cdot$ & $\cdot$ & $\cdot$ & $\cdot$ \\
$\mathbb{P}_{14}^1 \left( Y_2 \right)$  & 1 & -2 & 1 & $\cdot$ & $\cdot$ & $\cdot$ & 1 \\
$\mathbb{P}_{24}^1 \left( Y_2 \right)$  & $\cdot$ & 1 & -2 & 1 & $\cdot$ & $\cdot$ & $\cdot$ \\
$\mathbb{P}_{23}^1 \left( Y_2 \right)$  & $\cdot$ & $\cdot$ & 1 & -2 & 1 & 1 & $\cdot$ \\
$\mathbb{P}_{3x}^1 \left( Y_2 \right)$  & $\cdot$ & $\cdot$ & $\cdot$ & 1 & -2 & $\cdot$ & $\cdot$ \\
$\mathbb{P}_{3K}^1 \left( Y_2 \right)$  & $\cdot$ & $\cdot$ & $\cdot$ & 1 & $\cdot$ & -2 & $\cdot$ \\
$\mathbb{P}_{4D}^1 \left( Y_2 \right)$  & $\cdot$ & 1 & $\cdot$ & $\cdot$ & $\cdot$ & $\cdot$ & -2 \\
\bottomrule
\end{tabular}}
\end{align}

\paragraph{Over $\mathbf{Y_3}$}
Over the Yukawa locus $Y_3 = V ( w, a_{32}, a_{43} )$, the following seven $\mathbb{P}^1$-fibrations are present:
\begin{subequations}
\begin{align}
 \mathbb{P}_{0A}^1 \left( Y_3 \right) &= V \left( a_{3,2}, a_{4,3}, e_0, a_{1,0} x y z - e_1 e_2^2 e_3 s x^3 + e_3 e_4 y^2 \right) \, , \\
 \mathbb{P}_{1}^1 \left( Y_3 \right) &= V \left( a_{3,2}, a_{4,3}, e_1, e_3 e_4 y + a_{1,0} x z \right) \, , \\
 \mathbb{P}_{2E}^1 \left( Y_3 \right) &= V \left( a_{3,2}, a_{4,3}, e_2, e_3 e_4 y + a_{1,0} x z \right) \, , \\
 \mathbb{P}_{3x}^1 \left( Y_3 \right) &= V \left( a_{3,2}, a_{4,3}, e_3, x \right) \, , \\
 \mathbb{P}_{3s}^1 \left( Y_3 \right) &= V \left( a_{3,2}, a_{4,3}, e_3, s \right) \, , \\
 \mathbb{P}_{3L}^1 \left( Y_3 \right) &= V \left( a_{3,2}, a_{4,3}, e_3, a_{1,0} y - a_{2,1} e_0 e_1 e_2 x z \right) \, , \\
 \mathbb{P}_{4}^1 \left( Y_3 \right) &= V \left( a_{3,2}, a_{4,3}, e_4, a_{1,0} y z - e_1 e_2^2 e_3 s x^2 - a_{2,1} e_0 e_1 e_2 x z^2 \right) \, .
\end{align}
\end{subequations}
The total elliptic fibre over $Y_3$ is given by
\[ T^2 \left( Y_3 \right) = \mathbb{P}_{0A}^1 \left( Y_3 \right) + \mathbb{P}_{1}^1 \left( Y_3 \right) + \mathbb{P}_{2E}^1 \left( Y_3 \right) + \mathbb{P}_{3x}^1 \left( Y_3 \right) + \mathbb{P}_{3s}^1 \left( Y_3 \right) + \mathbb{P}_{3L}^1 \left( Y_3 \right)  + \mathbb{P}_{4}^1 \left( Y_3 \right) \, . \]
The individual $\mathbb{P}^1$-fibrations appear from the split components over $C_{\mathbf{5}_3}$ as follows:
\begin{align}
\begin{tabular}{cc}
\toprule
Split components over $C_{5_{3}}$ & Split components over $Y_3$ \\
\midrule
$\mathbb{P}_{0}^1 \left( \mathbf{5}_3 \right)$ & $\mathbb{P}_{0A}^1 \left( Y_3 \right)$ \\
$\mathbb{P}_{1}^1 \left( \mathbf{5}_3 \right)$ & $\mathbb{P}_{1}^1 \left( Y_3 \right)$ \\
$\mathbb{P}_{2E}^1 \left( \mathbf{5}_3 \right)$ & $\mathbb{P}^{1}_{2E} \left( Y_3 \right)$ \\
$\mathbb{P}_{3x}^1\left( \mathbf{5}_3 \right)$ & $\mathbb{P}^{1}_{3x} \left( Y_3 \right)$ \\
$\mathbb{P}_{3F}^1\left( \mathbf{5}_3 \right)$ & $\mathbb{P}^{1}_{3s} \left( Y_3 \right)$ + $\mathbb{P}^{1}_{3L} \left( Y_3 \right)$ \\
$\mathbb{P}_{4}^1 \left( \mathbf{5}_3 \right)$ & $\mathbb{P}_{4}^1\left( Y_3 \right)$ \\
\bottomrule
\end{tabular}
\end{align}
However, if we approach $Y_3$ from $C_{\mathbf{5}_{-2}}$ we have the following behaviour.
\begin{align}
\begin{tabular}{cc}
\toprule
Split components over $C_{5_{-2}}$ & Split components over $Y_3$ \\
\midrule
$\mathbb{P}_{0}^1 \left( \mathbf{5}_{-2} \right)$ & $\mathbb{P}_{0A}^1 \left( Y_3 \right)$ \\
$\mathbb{P}_{1}^1 \left( \mathbf{5}_{-2} \right)$ & $\mathbb{P}_1^1 \left( Y_3 \right)$ \\
$\mathbb{P}_{2}^1 \left( \mathbf{5}_{-2} \right)$ & $\mathbb{P}^{1}_{2E} \left( Y_3 \right)$ \\
$\mathbb{P}_{3G}^1\left( \mathbf{5}_{-2} \right)$ & $\mathbb{P}^{1}_{3x} \left( Y_3 \right)$ + $\mathbb{P}^{1}_{3s} \left( Y_3 \right)$ \\
$\mathbb{P}_{3H}^1\left( \mathbf{5}_{-2} \right)$ & $\mathbb{P}^{1}_{3L} \left( Y_3 \right)$ \\
$\mathbb{P}_{4}^1 \left( \mathbf{5}_{-2} \right)$ & $\mathbb{P}_{4}^1\left( Y_3 \right)$ \\
\bottomrule
\end{tabular}
\end{align}
Finally, when approached from the singlet curve $C_{\mathbf{1}_{5}}$ we have the following splitting:
\begin{align}
\begin{tabular}{cc} 
\toprule
Split components over $C_{1_{5}}$ & Split components over $Y_3$ \\
\midrule
$\mathbb{P}_{A}^1 \left( \mathbf{1}_{5} \right)$ & $\mathbb{P}_{3s}^1 \left( Y_3 \right)$ \\
$\mathbb{P}_{B}^1 \left( \mathbf{1}_{5} \right)$ & \pbox{20cm}{$\mathbb{P}_{0A}^1 \left( Y_3 \right) + \mathbb{P}_{1}^1 \left( Y_3 \right) + \mathbb{P}_{2E}^1 \left( Y_3 \right)$ \\ $+\mathbb{P}_{3x}^1 \left( Y_3 \right) + \mathbb{P}_{3L}^1 \left( Y_3 \right) + \mathbb{P}_{4}^1 \left( Y_3 \right)$} \\
\bottomrule
\end{tabular}
\end{align}
The intersection numbers in the fibre over $Y_3$ are:
\begin{align}
\begin{tabular}{c@{\hskip 20pt}ccccccc}
\toprule
 & $\mathbb{P}_{0A}^1 \left( Y_3 \right)$ & $\mathbb{P}_{1}^1 \left( Y_3 \right)$ & $\mathbb{P}_{2E}^1 \left( Y_3 \right)$ & $\mathbb{P}_{3x}^1 \left( Y_3 \right)$ & $\mathbb{P}_{3s}^1 \left( Y_3 \right)$ & $\mathbb{P}^1_{3L} \left( Y_3 \right)$ & $\mathbb{P}^1_{4} \left( Y_3 \right)$ \\
\midrule
$\mathbb{P}_{0A}^1 \left( Y_3 \right)$  & -2 & 1 & $\cdot$ & $\cdot$ & $\cdot$ & $\cdot$ & 1 \\
$\mathbb{P}_{1}^1 \left( Y_3 \right)$   & 1 & -2 & 1 & $\cdot$ & $\cdot$ & $\cdot$ & $\cdot$ \\
$\mathbb{P}_{2E}^1 \left( Y_3 \right)$  & $\cdot$ & 1 & -2 & 1 & $\cdot$ & $\cdot$ & $\cdot$ \\
$\mathbb{P}_{3x}^1 \left( Y_3 \right)$  & $\cdot$ & $\cdot$ & 1 & -2 & 1 & $\cdot$ & $\cdot$ \\
$\mathbb{P}_{3s}^1 \left( Y_3 \right)$  & $\cdot$ & $\cdot$ & $\cdot$ & 1 & -2 & 1 & $\cdot$ \\
$\mathbb{P}_{3L}^1 \left( Y_3 \right)$  & $\cdot$ & $\cdot$ & $\cdot$ & $\cdot$ & 1 & -2 & 1 \\
$\mathbb{P}_{4}^1 \left( Y_3 \right)$   & 1 & $\cdot$ & $\cdot$ & $\cdot$ & $\cdot$ & 1 & -2 \\
\bottomrule
\end{tabular}
\end{align}

\subsection{Zero Modes from Ambient Space Intersections} \label{subsec:MasslessSpectraMSFSU5xU1}

In \cref{subsec:VerticalAndGaugeInvariantMatterSurfaceFluxes} we analysed the matter surface fluxes in the geometry of this $SU(5) \times U( 1 )_X$-top. We found the following fluxes:
\begin{align}
\begin{split}
\mathcal{A} \left( \mathbf{10}_{1} \right) \left( \lambda \right) &= - \frac{\lambda}{5} \cdot \left( 2 \mathcal{E}_1 - \mathcal{E}_2 + \mathcal{E}_3 - 2 \mathcal{E}_4 \right) \cdot \overline{\mathcal{K}}_{\mathcal{B}_6} - \lambda \cdot \mathcal{E}_2 \cdot \mathcal{E}_4 \, , \\
\mathcal{A} \left( \mathbf{5}_{3} \right) \left( \lambda \right) &= - \frac{\lambda}{5} \cdot \left( \mathcal{E}_1 + 2 \mathcal{E}_2 - 2 \mathcal{E}_3 - \mathcal{E}_4 \right) \cdot \left( 3 \overline{\mathcal{K}}_{\mathcal{B}_6} - 2 \mathcal{W} \right) - \lambda \cdot \mathcal{E}_3 \cdot \mathcal{X} \, , \\
\mathcal{A} \left( \mathbf{5}_{-2} \right) \left( \lambda \right) &= - \frac{\lambda}{5} \cdot \left( \mathcal{E}_1 + 2 \mathcal{E}_2 + 3 \mathcal{E}_3 - \mathcal{E}_4 \right) \cdot \left( 5 \overline{\mathcal{K}}_{\mathcal{B}_6} - 3 \mathcal{W} \right) \\
& \hspace{14em} + \lambda \cdot \left( \mathcal{E}_3 \cdot \overline{\mathcal{K}}_{\mathcal{B}_6} + \mathcal{E}_3 \cdot \mathcal{Y} - \mathcal{E}_3 \cdot \mathcal{E}_4 \right) \, , \\
\mathcal{A} \left( \mathbf{1}_{5} \right) \left( \lambda \right) &= \lambda \cdot \mathcal{S} \cdot \left( 3 \overline{\mathcal{K}}_{\mathcal{B}_6} - 2 \mathcal{W} \right) - \lambda \cdot \mathcal{S} \cdot \mathcal{X} \, .
\end{split}
\end{align}
In this section we will compute the line bundles induced by these matter surface fluxes from ambient space intersections. By this we mean that given an algebraic cycle $\mathcal{A}$ of the above form, we perform the following steps for all matter curves $C_{\mathbf{R}}$ presented in \cref{subsec:FibreStructureSU5xU1}:
\begin{enumerate}
 \item Pick a matter surface $S^a_{C_\mathbf{R}} \in \mathrm{CH}^3 ( \hat{Y}_5 )$ over $C_{\mathbf{R}}$ from the ones listed in 
      \cref{subsec:FibreStructureSU5xU1}. \footnote{The result of the following steps does not depend on which matter surface over $C_{\mathbf{R}}$ since the matter surface fluxes listed in \cref{equ:Fluxes-Y5} are gauge invariant. Thus, we drop the superscript `a' from now on.}
 \item The $SU ( 5 ) \times U ( 1 )_X$-top induces the following linear relations on the ambient space $\hat{Y}_5$:
      \begin{align}
      \begin{split} \label{linearrelationsambientII}
      \mathcal{X} - \mathcal{Y} + \mathcal{Z} + \mathcal{E}_0 + \mathcal{E}_1 + \mathcal{E}_2 - \mathcal{W} + \overline{\mathcal{K}}_{\mathcal{B}_6} &= 0 \in \mathrm{CH}^1 ( \hat{Y}_5 ) \, , \\
      -3 \mathcal{X} + 2 \mathcal{Y} - \mathcal{S} - \mathcal{E}_1 -2 \mathcal{E}_2 + \mathcal{E}_4 &= 0 \in \mathrm{CH}^1 ( \hat{Y}_5 ) \, , \\ 
      2 \mathcal{X} - \mathcal{Y} - \mathcal{Z} + \mathcal{S} + \mathcal{E}_1 + 2 \mathcal{E}_2 + \mathcal{E}_3 - \overline{\mathcal{K}}_{\mathcal{B}_6} &= 0 \in \mathrm{CH}^1 ( \hat{Y}_5 ) \, .
      \end{split}
      \end{align}
      In order to obtain an algebraic cycle $\mathcal{A}^\prime \in \mathrm{CH}^2 ( \hat{Y}_5 )$ which is both equivalent to $\mathcal{A}$ and has transverse intersection $\mathcal{A}^\prime \cdot S_{C_\mathbf{R}}$, we apply these linear relations to $\mathcal{A}$. 
 \item We can now work out the transverse intersection $\mathcal{A}^\prime \cdot S_{C_\mathbf{R}}$ from looking at the set-theoretic intersection 
      $\mathcal{A}^\prime \cap S_{C_\mathbf{R}}$. The latter is achieved by the use of primary decompositions.
 \item Project the locus $\mathcal{A}^\prime \cap S_{C_\mathbf{R}}$ onto $C_{\mathbf{R}}$. This gives a divisor $D ( S_{C_\mathbf{R}}, \mathcal{A} ) \in 
      \mathrm{CH}^1 ( C_{\mathbf{R}} )$.
 \item Tensor the line bundle $\mathcal{O}_{C_{\mathbf{R}}} ( D ( S_{C_\mathbf{R}} , \mathcal{A} ) )$ with the spin bundle on $C_{\mathbf{R}}$. The latter 
      is identified as $\sqrt{K_{C_{\mathbf R}}}$ from the embedding of the matter curve $C_{\mathbf{R}}$ into $\hat{Y}_4$. 
 \item The so-obtained line bundle $L( S_{C_{\mathbf{R}}}, \mathcal{A})$ counts the zero modes localised on $C_{\mathbf{R}}$ in the presence of the gauge flux 
      $\mathcal{A}$ via its sheaf cohomologies $H^i(C_{\mathbf{R}}, L( S_{C_{\mathbf{R}}},\mathcal{A}))$.
\end{enumerate}
Let us mention that the following computations make us of rational equivalence on $\mathcal{B}_6$ to (re)express cycles in the classes $\overline{K}_{\mathcal{B}_6}$ and $W$ whenever this simplifies the computations. Furthermore, we stress again that we are assuming that $\mathcal{B}_6$ is torsion-free (\cf \cref{sec:SummaryChapter2}).

\subsubsection{Line Bundles induced by \texorpdfstring{$\mathbf{\Delta \mathcal{A} ( \lambda )}$}{Delta A(lambda)}}

To compute the massless spectra of the matter surface fluxes in \cref{equ:Fluxes-Y5}, it will be useful to show that the following gauge background induced the trivial line bundle on all matter curves:
\[ \resizebox{0.9\textwidth}{!}{$
\Delta \mathcal{A} \left( \lambda \right) = \lambda \left[ \mathcal{E}_2 \mathcal{E}_4 + \overline{\mathcal{K}}_{\mathcal{B}_6} \left( \mathcal{E}_1 + \mathcal{E}_2 - \mathcal{E}_3 - \mathcal{E}_4 \right) + 2 \mathcal{E}_3 \mathcal{W} + \mathcal{E}_4 \mathcal{W} - \overline{\mathcal{K}}_{\mathcal{B}_6} \mathcal{W} + \mathcal{S} \mathcal{W} + \mathcal{E}_3 \mathcal{X} - \mathcal{W} \mathcal{Z} \right]$}
\]
\begin{itemize} 
 \item On $C_{\mathbf{10}_1}$: \\
	By use of the linear relations \cref{linearrelationsambientII} we have
	\begin{align}
	\begin{split}
	\Delta \mathcal{A} \left( \lambda \right) &= \lambda \left[ \mathcal{E}_0 \mathcal{E}_3 + \mathcal{E}_1 \mathcal{E}_3 + \mathcal{E}_0 \mathcal{S} + \mathcal{E}_1 \mathcal{S} + \mathcal{E}_2 \mathcal{S} + 2 \mathcal{E}_0 \mathcal{X} + \mathcal{E}_1 \mathcal{X} + \mathcal{E}_2 \mathcal{X} \right. \\
	& \hspace{4em} \left. + \mathcal{E}_3 \mathcal{X} - \mathcal{W} \mathcal{X} - \mathcal{E}_0 \mathcal{Y} - \mathcal{E}_0 \mathcal{Z} - 2 \mathcal{E}_1 \mathcal{Z} - 2 \mathcal{E}_1 \mathcal{Z} - 2 \mathcal{E}_2 \mathcal{Z} + \mathcal{W} \mathcal{Z} \right] \, .
	\end{split}
	\end{align}
	The Stanley-Reisner ideal \cref{equ:SRSU5xU1InAppendix} then shows $\mathbb{P}^1_{4D} ( \mathbf{10}_1 ) \cdot \Delta \mathcal{A} ( \lambda ) = \emptyset.$
 \item On $C_{\mathbf{5}_3}$: \\
        The linear relations \cref{linearrelationsambientII} allow us to write
	\begin{align}
	\begin{split}
	\Delta \mathcal{A} \left( \lambda \right) &= \lambda \left[ - \frac{1}{6} \mathcal{E}_0 \mathcal{E}_1 - \frac{4}{3} \mathcal{E}_0 \mathcal{E}_2 - \frac{1}{2} \mathcal{E}_1 \mathcal{E}_2 + \frac{2}{3} \mathcal{E}_0 \mathcal{E}_4 + \frac{1}{2} \mathcal{E}_1 \mathcal{E}_4 + \mathcal{E}_2 \mathcal{E}_4 + \frac{1}{3} \mathcal{E}_0 \mathcal{S} \right. \\
	& \hspace{3em} \left. + \frac{1}{2} \mathcal{E}_1 \mathcal{S} + \mathcal{E}_0 \overline{\mathcal{K}}_{\mathcal{B}_6} + \frac{3}{2} \mathcal{E}_1 \overline{\mathcal{K}}_{\mathcal{B}_6} + \mathcal{E}_4 \overline{\mathcal{K}}_{\mathcal{B}_6} + \mathcal{E}_2 \mathcal{S} - \frac{1}{2} \mathcal{E}_1 \mathcal{W} - \mathcal{E}_4 \mathcal{W} \right. \\
	& \hspace{6em} \left. - \frac{2}{3} \mathcal{E}_0 \mathcal{Y} - \frac{1}{2} \mathcal{E}_1 \mathcal{Y} - \mathcal{E}_4 \mathcal{Y} - \frac{1}{2} \mathcal{E}_1 \mathcal{Z} - 2 \mathcal{E}_2 \mathcal{Z} + \mathcal{E}_4 \mathcal{Z} + \mathcal{W} \mathcal{Z} \right]
	\end{split}
	\end{align}
	and the Stanley-Reisner ideal \cref{equ:SRSU5xU1InAppendix} then shows $-1 \cdot \mathbb{P}^1_{3x} ( \mathbf{5}_3 ) \cdot \Delta \mathcal{A} ( \lambda )  = \emptyset$.
 \item On $C_{\mathbf{5}_{-2}}$: \\
	By use of the linear relations \cref{linearrelationsambientII} we find
	\begin{align}
	\begin{split}
	\Delta \mathcal{A} \left( \lambda \right) &= \resizebox{0.73\textwidth}{!}{$\lambda \left[ \mathcal{E}_2 \mathcal{E}_4 + \mathcal{E}_0 \overline{\mathcal{K}}_{\mathcal{B}_6} + 2 \mathcal{E}_1 \overline{\mathcal{K}}_{\mathcal{B}_6} + 2 \mathcal{E}_2 \overline{\mathcal{K}}_{\mathcal{B}_6} - 2 \mathcal{E}_1 \mathcal{W} - \mathcal{S} \mathcal{W} + 2 \mathcal{E}_0 \mathcal{X} - 6 \mathcal{W} \mathcal{X} \right.$} \\
	& \hspace{2.5em} \resizebox{0.70\textwidth}{!}{$\left. + \mathcal{E}_1 \mathcal{X} + 3 \overline{\mathcal{K}}_{\mathcal{B}_6} \mathcal{X} - \mathcal{S} \mathcal{X} + 2 \mathcal{W} \mathcal{Y} - \mathcal{X} \mathcal{Y} + \mathcal{W} \mathcal{Z} + 3 \mathcal{X} \mathcal{Z} - 4 \mathcal{E}_2 \mathcal{W} + \mathcal{E}_4 \mathcal{W} \right] \, .$}
	\end{split}
	\end{align}
	Now the Stanley-Reisner ideal \cref{equ:SRSU5xU1InAppendix} implies
	\[\mathbb{P}^1_{3H} \left( \mathbf{5}_{-2} \right) \cdot  \Delta \mathcal{A} \left( \lambda \right) = V \left( a_{1,0}, a_{3,2}, e_2, e_3, e_4 \right) - V \left(  a_{1,0}, a_{3,2}, s, e_3, s \right) \, . \]
	Consequently, $\pi_{C_{\mathbf{5}_{-2}} \ast} (  \mathbb{P}^1_{3H} ( \mathbf{5}_{-2} ) \cdot \Delta \mathcal{A} ( \lambda ) ) = Y_2 - Y_2 = 0 \in \mathrm{CH}^1 ( C_{\mathbf{5}_{-2}} )$.
 \item On $C_{\mathbf{1}_{5}}$: \\
	By use of the linear relations \cref{linearrelationsambientII} we can write
	\begin{align}
	\begin{split}
	\Delta \mathcal{A} \left( \lambda \right) &= \lambda \left[ \mathcal{E}_2 \mathcal{E}_4 + \mathcal{E}_1 \overline{\mathcal{K}}_{\mathcal{B}_6} + \mathcal{E}_2 \overline{\mathcal{K}}_{\mathcal{B}_6} - \mathcal{E}_3 \overline{\mathcal{K}}_{\mathcal{B}_6} - \mathcal{E}_4 \overline{\mathcal{K}}_{\mathcal{B}_6} - \mathcal{E}_1 \mathcal{W} - 2 \mathcal{E}_2 \mathcal{W} \right. \\
	& \hspace{5em} \left. + 2 \mathcal{E}_3 \mathcal{W} + 2 \mathcal{E}_4 \mathcal{W} - \mathcal{W} \overline{\mathcal{K}}_{\mathcal{B}_6} + \mathcal{E}_3  \mathcal{X} - 3 \mathcal{W} \mathcal{X} + 2 \mathcal{W} \mathcal{Y} - \mathcal{W} \mathcal{Z} \right] \, .
	\end{split}
	\end{align}
	Let $k_{\mathcal{B}_6}$ and $w$ be polynomials in the coordinate ring of $\hat{Y}_5$ with degree of $\overline{\mathcal{K}}_{\mathcal{B}_6}$ and $\mathcal{W}$ respectively. Then it follows by use of the Stanley-Reisner ideal \cref{equ:SRSU5xU1InAppendix} that
	\begin{align}
	\begin{split}
	\mathbb{P}^1_{A} \left( \mathbf{1}_{5} \right) \cdot \Delta \mathcal{A} \left( \lambda \right) &= - V \left( k_{\mathcal{B}_6}, w, a_{3,2}, a_{4,3}, s \right) + V \left( e_3, x, s, a_{3,2}, a_{4,3} \right) \\
	& \hspace{5em} - 3 V \left( w, x, s, a_{3,2}, a_{4,3} \right) + 2 V \left( w, y, s, a_{3,2}, a_{4,3} \right) \, .
	\end{split}
	\end{align}
	Consequently, we find $\pi_{C_{\mathbf{1}_{5}} \ast} ( \mathbb{P}^1_{A} ( \mathbf{1}_{5} ) \Delta \mathcal{A} ( \lambda ) ) = Y_3 - 3 Y_3 + 2 Y_3 = 0 \in \mathrm{CH}_0 ( C_{\mathbf{1}_{5}} )$.
\end{itemize}

\subsubsection{Line Bundles Induced by \texorpdfstring{$\mathbf{\mathcal{A} ( \mathbf{10}_1 ) ( \lambda )}$}{A(10)(lambda)} and \texorpdfstring{$\mathbf{\mathcal{A} ( \mathbf{5}_3 ) ( \lambda )}$}{A(53)(lambda)}}

Let us compute the massless spectra of $\mathcal{A} \left( \mathbf{10}_1 \right) \left( \lambda \right)$ and $\mathcal{A} \left( \mathbf{5}_3 \right) \left( \lambda \right)$. We begin with $\mathcal{A} \left( \mathbf{10}_1 \right) \left( \lambda \right)$:
\begin{itemize}
 \item On $C_{\mathbf{5}_3}$: \\
	By use of the linear relations in \cref{linearrelationsambientII} we can write
	\[ \mathcal{A} \left( \mathbf{10}_1 \right) \left( \lambda \right) = - \frac{\lambda}{5} \left[ 5 \mathcal{E}_2 \mathcal{E}_4 + \left( - \mathcal{E}_0 + \mathcal{E}_1 - 2 \mathcal{E}_2 - 3 \mathcal{E}_4 + \mathcal{W} \right) \overline{\mathcal{K}}_{\mathcal{B}_6} \right] \, . \]
	From \cref{equ:SRSU5xU1InAppendix} it now follows $\mathbb{P}^1_{3x} ( \mathbf{5}_3 ) \cdot \mathcal{A} ( \mathbf{10}_1 ) ( \lambda ) = - \frac{2 \lambda}{5} \cdot V ( k_{\mathcal{B}_6}, a_{3,2}, e_2, e_3, x )$. As above, let $k_{\mathcal{B}_6}$ be a polynomial in the coordinate ring of $\hat{Y}_5$ whose degree matches $\overline{\mathcal{K}}_{\mathcal{B}_6}$. Consequently, $\pi_{C_{\mathbf{5}_3} \ast} ( - \mathbb{P}^1_{3x} ( \mathbf{5}_3 ) \cdot \mathcal{A} ( \mathbf{10}_1 ) ( \lambda ) ) = - \frac{2 \lambda}{5} Y_2 \in \mathrm{CH}^1 ( C_{\mathbf{5}_3} )$, which implies
	\[ L \left(  S_{\mathbf{5}_3},\mathcal{A} \left( \mathbf{10}_1 \right) \right) = \mathcal{O}_{C_{\mathbf{5}_3}} \left( - \frac{2 \lambda}{5} Y_2 \right) \otimes \sqrt{K_{C_{\mathbf{5}_3}}} \, . \]
 \item On $C_{\mathbf{5}_{-2}}$: \\
	The linear relations \cref{linearrelationsambientII} enable us to write
	\[ \mathcal{A} \left( \mathbf{10}_1 \right) \left( \lambda \right) = - \frac{\lambda}{5} \left[ 5 \mathcal{E}_2 \mathcal{E}_4 + \left( - \mathcal{E}_0 + \mathcal{E}_1 - 2 \mathcal{E}_2 - 3 \mathcal{E}_4 + \mathcal{W} \right) \overline{\mathcal{K}}_{\mathcal{B}_6} \right] \, . \]
	From this and \cref{equ:SRSU5xU1InAppendix} it follows
	\begin{align}
	\begin{split}
	\mathcal{A} \left( \mathbf{10}_1 \right) \left( \lambda \right) & \cdot \mathbb{P}^1_{3H} \left( \mathbf{5}_{-2} \right) = - \lambda \cdot V \left( a_{1,0}, a_{3,2}, e_2, e_3, e_4 \right) \\
	& + \frac{3 \lambda}{5} \cdot V \left( k_{\mathcal{B}_6}, e_3, e_4, a_{2,1} e_0 x z e_1 e_2 - a_{1,0} y, a_{4,3} e_0 x z e_1 e_2 - a_{3,2} y  \right) \\
	& - \frac{\lambda}{5} \cdot V \left( w, k_{\mathcal{B}_6}, e_3, a_{21} e_0 x z e_1 e_2 - a_{1,0} y, a_{43} e_0 x z e_1 e_2 - a_{3,2} y \right) \, . 
	\end{split}
	\end{align}
	Again, $k_{\mathcal{B}_6}$ and $w$ are polynomials in the coordinate ring of $\hat{Y}_5$ whose degree match $\overline{\mathcal{K}}_{\mathcal{B}_6}$ and $\mathcal{W}$ respectively. Consequently,
	\[ \pi_{C_{\mathbf{5}_{-2}} \ast} \left( \mathbb{P}^1_{3H}  \left(\mathbf{5}_{-2} \right) \cdot \mathcal{A} \left( \mathbf{10}_1 \right) \left( \lambda \right) \right) = - \lambda Y_2 + \frac{3 \lambda}{5} \left( Y_1 + Y_2 \right) = \frac{3 \lambda}{5} Y_1 - \frac{2 \lambda}{5} Y_2 \, . \]
	This leads us to conclude that $L ( S_{\mathbf{5}_{-2}}, \mathcal{A} ( \mathbf{10}_1 ) ) = \mathcal{O}_{C_{\mathbf{5}_{-2}}} ( \frac{3 \lambda}{5} Y_1 - \frac{2 \lambda}{5} Y_2 ) \otimes \sqrt{K_{C_{\mathbf{5}_{-2}}}}$
 \item On $C_{\mathbf{1}_{5}}$: \\
	With \cref{linearrelationsambientII} it follows
	\[ \mathcal{A} \left( \mathbf{10}_1 \right) \left( \lambda \right) = - \frac{\lambda}{5} \left[ 5 \mathcal{E}_2 \cdot \mathcal{E}_4 + \left( - \mathcal{E}_0 + \mathcal{E}_1 - 2 \mathcal{E}_2 - 3 \mathcal{E}_4 + \mathcal{W} \right) \cdot \overline{\mathcal{K}}_{\mathcal{B}_6} \right] \, . \]
	The Stanley-Reisner ideal \cref{equ:SRSU5xU1InAppendix} now implies ($k_{\mathcal{B}_6}$, $w$ as before)
	\[ \mathbb{P}^1_{A} \left( \mathbf{1}_{5} \cdot \mathcal{A} \left( \mathbf{10}_1 \right) \left( \lambda \right)  \right) = - \frac{\lambda}{5} \cdot V \left( k_{\mathcal{B}_6}, w, s, a_{3,2}, a_{4,3} \right) = \emptyset \, . \]
	Hence, $\pi_{C_{\mathbf{1}_{5}} \ast} ( \mathbb{P}^1_{A} ( \mathbf{1}_{5} ) \cdot \mathcal{A} ( \mathbf{10}_1 ) ( \lambda ) ) = 0 \in \mathrm{CH}^1 ( C_{\mathbf{1}_{5}} )$, \ie $L (  S_{\mathbf{1}_{5}}, \mathcal{A} ( \mathbf{10}_1 ) ) = \sqrt{K_{C_{\mathbf{1}_{5}}}}$.
 \item On $C_{\mathbf{10}_1}$: \\
	To compute the massless spectrum on $C_{\mathbf{10}_1}$ we note that 
	\[ \mathcal{A} \left( \mathbf{10}_1 \right) \left( \lambda \right) = \mathcal{A} \left( \mathbf{5}_3 \right) \left( - \lambda \right) + \mathcal{A}_X \left( - \lambda W \right) - \Delta \mathcal{A} \left( \lambda \right) \, . \label{equ:RelationOfFluxes} \]
	We have shown already that $\Delta \mathcal{A} ( \lambda )$ implies a trivial divisor on $C_{\mathbf{10}_1}$. The massless spectrum of $\mathcal{A}_X ( - \lambda W )$ follows from \cref{subsec:ExampleOnMasslessSpectrumComputation}. In particular, $D ( S_{\mathbf{10}_1}, \mathcal{A}_X ( - \lambda W ) ) = - \frac{\lambda}{5} W |_{C_{\mathbf{10}_1}}$. Note that $W = 3 \cdot ( 2 \overline{K}_{\mathcal{B}_6} - W ) - 2 \cdot ( 3 \overline{K}_{\mathcal{B}_6} - 2 W )$. Therefore,
	\[ D \left(  S_{\mathbf{10}_1},\mathcal{A}_X \left( - \lambda W \right) \right)  = - \frac{3 \lambda}{5} Y_1 + \frac{2 \lambda}{5} Y_2 \in \mathrm{Div} \left( C_{\mathbf{10}_1} \right) \, . \]
	To identify the massless spectrum of $\mathcal{A} ( \mathbf{10}_1 ) ( \lambda )$ on $C_{\mathbf{10}_1}$ we are therefore just short of $D ( S_{\mathbf{10}_1},\mathcal{A} ( \mathbf{5}_3 ) ( \lambda ) )$. So let us compute this divisor now. Upon use of \cref{linearrelationsambientII} we write
	\[ \mathcal{A} \left( \mathbf{5}_3 \right) \left( \lambda \right) = - \frac{\lambda}{5} \cdot \left( \mathcal{E}_0 + 2 \mathcal{E}_1 + 3 \mathcal{E}_2 - \mathcal{E}_3 - \mathcal{W} \right) \cdot \left( 3 \overline{\mathcal{K}}_{\mathcal{B}_6} - 2 \mathcal{W} \right) - \lambda \mathcal{E}_3 \cdot \mathcal{X} \, . \]
	From this it follows $\mathcal{A} ( \mathbf{5}_3 ) ( \lambda ) \cdot \mathbb{P}^1_{4D} ( \mathbf{10}_1 ) = - \frac{2 \lambda}{5} \cdot V (  a_{1,0}, a_{3,2}, e_0, e_4, x s e_2 e_3 + a_{2,1} z^2 e_0 )$, which implies $L ( S_{\mathbf{10}_1},\mathcal{A} ( \mathbf{5}_3 ) ( \lambda ) ) = \mathcal{O}_{C_{\mathbf{10}_1}} ( - \frac{2 \lambda}{5} Y_2 ) \otimes \sqrt{K_{C_{\mathbf{10}_1}}}$. With this we are lead to conclude by use of \cref{equ:RelationOfFluxes} that $D ( S_{\mathbf{10}_1},\mathcal{A} ( \mathbf{10}_1 ) ( \lambda ) ) = - \frac{3 \lambda}{5} Y_1 + \frac{4 \lambda}{5} Y_2$.
\end{itemize}
Finally we use \cref{equ:RelationOfFluxes} to compute the massless spectrum of $\mathcal{A} ( \mathbf{5}_3 ) ( \lambda )$ from the zero modes of the other fluxes. The results are summarised in \cref{table-N5}.

\subsubsection{Line Bundles Induced by \texorpdfstring{$\mathbf{\mathcal{A} ( \mathbf{5}_{-2} ) ( \lambda )}$}{A(5-2)(lambda)}}

We now turn to $\mathcal{A} ( \mathbf{5}_{-2} ) ( \lambda )$. Its massless spectrum looks as follows:
\begin{itemize}
 \item On $C_{\mathbf{10}_1}$: \\
	Upon use of the linear relations \cref{linearrelationsambientII} it follows
	\[ \mathcal{A} \left( \mathbf{5}_{-2} \right) \left( \lambda \right) = - \frac{\lambda}{5} \left( 5 \overline{\mathcal{K}}_{\mathcal{B}_6} - 3 \mathcal{W} \right) \left( 3 \mathcal{E}_3 + 2 \mathcal{Y} \right) - \lambda \left( 3 \mathcal{E}_2 \mathcal{E}_3 - \mathcal{W} \mathcal{E}_3 - 4 \mathcal{E}_3 \mathcal{Y} \right) \]
	which is readily seen to imply $L ( S_{\mathbf{10}_1}, \mathcal{A} ( \mathbf{5}_{-2} ) ( \lambda ) ) = \mathcal{O}_{C_{\mathbf{5}_{-2}}} ( \frac{3 \lambda}{5} Y_1 - \frac{2 \lambda}{5} Y_2 ) \otimes \sqrt{K_{C_{\mathbf{10}_1}}}$.
 \item On $C_{\mathbf{5}_3}$: \\
	The linear relations \cref{linearrelationsambientII} imply
	\begin{align}
	\begin{split}
	\mathcal{A} \left( \mathbf{5}_{-2} \right) \left( \lambda \right) &= - \frac{\lambda}{5} \left( 5 \overline{\mathcal{K}}_{\mathcal{B}_6} - 3 \mathcal{W} \right) \left( - 3 \mathcal{E}_0 - 2 \mathcal{E}_1 - \mathcal{E}_2 - 4 \mathcal{E}_4 + 3 \mathcal{W} \right) \\
	& \hspace{2em} - \lambda \left( 2 \mathcal{E}_0 \mathcal{E}_4 + \mathcal{E}_1 \mathcal{E}_4 + \mathcal{E}_0 \overline{\mathcal{K}}_{\mathcal{B}_6} + \mathcal{E}_1 \overline{\mathcal{K}}_{\mathcal{B}_6} + \mathcal{E}_2 \overline{\mathcal{K}}_{\mathcal{B}_6} + 4 \mathcal{E}_4 \overline{\mathcal{K}}_{\mathcal{B}_6} - \mathcal{E}_4 \mathcal{S} \right. \\
	& \hspace{4.9em} \left. - 2 \mathcal{E}_4 \mathcal{W} - \overline{\mathcal{K}}_{\mathcal{B}_6} \mathcal{W}  + \mathcal{E}_0 \mathcal{Y} + \mathcal{E}_1 \mathcal{Y} + \mathcal{E}_2 \mathcal{Y} - \mathcal{W} \mathcal{Y} + 3 \mathcal{E}_4 \mathcal{Z} \right) \, .
	\end{split}
	\end{align}
	From \cref{equ:SRSU5xU1InAppendix} it now follows $L (S_{\mathbf{5}_3}, \mathcal{A} ( \mathbf{5}_{-2} ) ( \lambda ) ) = \mathcal{O}_{C_{\mathbf{5}_3}} ( - \frac{\lambda}{5} Y_3 + \frac{4 \lambda}{5} Y_2 ) \otimes \sqrt{K_{C_{\mathbf{10}_1}}}$.
 \item On $C_{\mathbf{1}_5}$: \\
	Here the intersections with the matter surface $\mathbb{P}^1_{A} ( \mathbf{1}_{5} )$ are automatically transverse and one finds $L ( S_{\mathbf{1}_{5}}, \mathcal{A} ( \mathbf{5}_{-2} ) ( \lambda ) ) = \mathcal{O}_{C_{\mathbf{1}_{5}}} ( \lambda Y_3 ) \otimes \sqrt{K_{C_{\mathbf{1}_{5}}}}$.
 \item On $C_{\mathbf{5}_{-2}}$: \\
	To compute this spectrum we look at the difference $\Delta \mathcal{A}^{(2)} ( \lambda ) := \mathcal{A} ( \mathbf{5}_{-2} ) ( \lambda ) - \mathcal{A}^X ( \lambda W )$. Upon use of \cref{linearrelationsambientII} we write
	\begin{align}
	\begin{split}
	\Delta \mathcal{A}^{(2)} \left( \lambda \right) &= - \lambda \left[ - \mathcal{E}_1 \mathcal{E}_4 - 2 \mathcal{E}_2 \mathcal{E}_4 - 2 \mathcal{E}_0 \overline{\mathcal{K}}_{\mathcal{B}_6} - 3 \mathcal{E}_1 \overline{\mathcal{K}}_{\mathcal{B}_6} - 4 \mathcal{E}_2 \overline{\mathcal{K}}_{\mathcal{B}_6} - \mathcal{E}_4 \mathcal{S} + \mathcal{X} \mathcal{Y} \right. \\
	& \hspace{1em} \left. -2 \overline{\mathcal{K}}_{\mathcal{B}_6} \mathcal{S} + 2 \mathcal{E}_1 \mathcal{W} + 4 \mathcal{E}_2 \mathcal{W} + 2 \mathcal{S} \mathcal{W} -2 \mathcal{E}_4 \mathcal{X} - 6 \overline{\mathcal{K}}_{\mathcal{B}_6} \mathcal{X} + \mathcal{E}_4 \mathcal{Z} - 2 \mathcal{W} \mathcal{Z}  \right. \\
	& \hspace{2.5em} \left. + 6 \mathcal{W} \mathcal{X} - \mathcal{E}_0 \mathcal{Y} + \mathcal{E}_2 \mathcal{Y} + \mathcal{E}_4 \mathcal{Y} + 2 \overline{\mathcal{K}}_{\mathcal{B}_6} \mathcal{Y} + \mathcal{S} \mathcal{Y} - 2 \mathcal{W} \mathcal{Y} - 2 \mathcal{Y} \mathcal{Z} \right] \, .
	\end{split}
	\end{align}
	Now \cref{equ:SRSU5xU1InAppendix} implies
	\begin{align*}
	\Delta \mathcal{A}^{(2)} \left( \lambda \right) \cdot \mathbb{P}^1_{3H} \left( \mathbf{5}_{-2} \right) &= - \lambda \left[ - 2 \mathcal{E}_2 \mathcal{E}_4 + 2 \left( \mathcal{W} - \overline{\mathcal{K}}_{\mathcal{B}_6} \right) \mathcal{S} + \mathcal{E}_4 \mathcal{Y} + \mathcal{S} \mathcal{Y} \right] \cdot \mathbb{P}^1_{3H} \left( \mathbf{5}_{-2} \right).
	\end{align*}
	Upon projection onto $C_{\mathbf{5}_{-2}}$ we then find $\pi_{C_{\mathbf{5}_{-2}} \ast} ( \mathbb{P}^1_{3H} ( \mathbf{5}_{-2} ) \cdot \Delta \mathcal{A}^{(2)} ( \lambda ) ) = 0$. Consequently,
	\[ L \left(S_{\mathbf{5}_{-2}}, \mathcal{A} \left( \mathbf{5}_{-2} \right) \left( \lambda \right) \right) = \mathcal{O}_{C_{\mathbf{5}_{-2}}} \left( - \frac{3 \lambda}{5} Y_1 - \frac{2 \lambda}{5} Y_2 + \frac{\lambda}{5} Y_3 \right) \otimes \sqrt{K_{C_{\mathbf{5}_{-2}}}} \, . \]
\end{itemize}

\subsubsection{Line Bundles Induced by \texorpdfstring{$\mathbf{\mathcal{A} ( \mathbf{1}_{5} ) ( \lambda )}$}{A(15)(lambda)}}

The massless spectrum of the flux $\mathcal{A} ( \mathbf{1}_{5} ) ( \lambda )$ is as follows:
\begin{itemize}
 \item On $C_{\mathbf{10}_1}$: \\
      The intersections with $\mathbb{P}^1_{4D} ( \mathbf{10}_1 )$ are transverse already. From \cref{equ:SRSU5xU1InAppendix} it then follows $\mathcal{A} ( \mathbf{1}_{5} ) ( \lambda ) \cdot \mathbb{P}^1_{4D} ( \mathbf{10}_1 ) = \emptyset$, so $L ( S_{\mathbf{10}_1}, \mathcal{A} ( \mathbf{1}_{5} ) ( \lambda ) ) = \sqrt{K_{C_{\mathbf{10}_1}}}$.
 \item On $C_{\mathbf{5}_3}$: \\
        By use of \cref{linearrelationsambientII} we can write $\mathcal{A} ( \mathbf{1}_{5} ) ( \lambda ) = \lambda [ 4 \mathcal{S} \overline{\mathcal{K}}_{\mathcal{B}_6} - 3 \mathcal{S} \mathcal{W} + \mathcal{S} \mathcal{E}_0 + \mathcal{S} \mathcal{E}_1 + \mathcal{S} \mathcal{E}_2 - \mathcal{S} \mathcal{Y} + \mathcal{S} \mathcal{Z}]$. From \cref{equ:SRSU5xU1InAppendix} it now follows
	\[ S \cdot \mathcal{A} \left( \mathbf{1}_{5} \right) \left( \lambda \right)  = - \lambda \left[ 4 V \left( k_{\mathcal{B}_6}, a_{3,2}, s, e_3, x \right) - 3 V \left( w, a_{3,2}, s, e_3, x \right) \right] \, . \]
	To work out the projection onto $C_{\mathbf{5}_3}$ first write $W = \frac{4}{3} \overline{K}_{\mathcal{B}_6} - \frac{1}{3} ( 4 \overline{K}_{\mathcal{B}_6} - 3 W )$, which implies $\left. W \right|_{C_{\mathbf{5}_3}} = \frac{4}{3} Y_2 - \frac{1}{3} Y_3$. Therefore, $\pi_{C_{\mathbf{5}_3} \ast} (  S \cdot \mathcal{A} ( \mathbf{1}_{5} ) ( \lambda ) ) = - \lambda Y_3$ and we conclude
	\[ L \left(S_{\mathbf{5}_3} , \mathcal{A} \left( \mathbf{1}_{5} \right) \left( \lambda \right) \right) = \mathcal{O}_{C_{\mathbf{5}_3}} \left( - \lambda Y_3 \right) \otimes \sqrt{K_{C_{\mathbf{5}_3}}} \, . \]
 \item On $C_{\mathbf{5}_{-2}}$: \\
	The intersections with $\mathbb{P}^1_{3H} ( \mathbf{5}_{-2} ) = V ( a_{21} e_0 x z e_1 e_2 - a_{10} y, a_{32} y - a_{43} e_0 e_1 e_2 x z )$ are transverse automatically and \cref{equ:SRSU5xU1InAppendix} implies
	\begin{align}
	\begin{split}
	&\mathbb{P}^1_{3H} \left( \mathbf{5}_{-2} \right) \cdot \mathcal{A} \left( \mathbf{1}_{5} \right) \left( \lambda \right) \\
	&\hspace{2em}=\lambda V \left( a_{3,2}, s, e_3, a_{2,1} x - a_{1,0} y, a_{4,3} x - a_{3,2} y \right) - \lambda V \left( a_{1,0}, a_{3,2}, e_3, s, x \right) \, .
	\end{split}
	\end{align}
	So $\pi_{C_{\mathbf{5}_{-2}} \ast} (S \cdot  \mathcal{A} ( \mathbf{1}_{5} ) ( \lambda ) ) = \lambda Y_3$ and $L ( S_{\mathbf{5}_{-2}} \cdot \mathcal{A} ( \mathbf{1}_{5} ) ( \lambda ) ) = \mathcal{O}_{C_{\mathbf{5}_{-2}}} ( \lambda Y_3 ) \otimes \sqrt{K_{C_{\mathbf{5}_{-2}}}}$.
 \item On $C_{\mathbf{1}_{5}}$: \\
	To compute this spectrum we introduce
	\[ \Delta \mathcal{A}^{(3)} \left( \lambda \right) := \mathcal{A} \left( \mathbf{1}_{5} \right) \left( \lambda \right) - \mathcal{A}^X \left( - \lambda \left[ 6 \overline{K}_{\mathcal{B}_6} - 5 W \right] \right) - \mathcal{A} \left( \mathbf{10}_1 \right) \left( \lambda \right) \, . \label{equ:SpectrumOfMSFOnSingletCurveOntoSingletCurve} \]
	By use of the linear relations \cref{linearrelationsambientII} we can write
	\begin{align}
	\begin{split}
	\Delta \mathcal{A}^{(3)} \left( \lambda \right) &= \frac{\lambda}{5} \left[ -31 \mathcal{E}_0 \overline{\mathcal{K}}_{\mathcal{B}_6} -26 \mathcal{E}_1 \overline{\mathcal{K}}_{\mathcal{B}_6} -36 \mathcal{E}_3 \overline{\mathcal{K}}_{\mathcal{B}_6} -36 \mathcal{E}_4 \overline{\mathcal{K}}_{\mathcal{B}_6} -5 \mathcal{E}_1 \mathcal{W} + 5 \mathcal{X} \mathcal{Y} \right. \\
	& \hspace{0em} \left. + 30 \mathcal{E}_4 \mathcal{W} + 6 \overline{\mathcal{K}}_{\mathcal{B}_6} \mathcal{W} + 5 \mathcal{E}_1 \mathcal{X} + 15 \mathcal{E}_3 \mathcal{X} + 10 \mathcal{E}_4 \mathcal{X} + 30 \mathcal{E}_3 \mathcal{W} + 30 \mathcal{W} \mathcal{Y} \right. \\
	& \hspace{0em} \left. + \mathcal{E}_2 \left( 5 \mathcal{E}_4 - 26 \overline{\mathcal{K}}_{\mathcal{B}_6} - 10 \mathcal{W} + 10 \mathcal{X} \right) - 5 \mathcal{Z} \left( 5 \mathcal{W} + 3 \mathcal{X} \right) - 45 \mathcal{W} \mathcal{X} \right] \, .
	\end{split}
	\end{align}
	Now the Stanley-Reisner ideal \cref{equ:SRSU5xU1InAppendix} implies
	\begin{align}
	\begin{split}
	\mathbb{P}^1_A \left( \mathbf{1}_{5} \right) \cdot \Delta \mathcal{A}^{(3)} \left( \lambda \right)  &= \frac{\lambda}{5} \left[ 15 V \left( s, x, e_3, a_{3,2}, a_{4,3} \right) - 45 V \left( s, x, w, a_{3,2}, a_{4,3} \right) \right. \\
	& \hspace{14em} \left. + 30 V \left( s, y, w, a_{3,2}, a_{4,3} \right) \right] \, .
	\end{split}
	\end{align}
	By projecting this quantity onto $C_{\mathbf{1}_{5}}$ we obtain $\pi_{C_{\mathbf{1}_{5}} \ast} ( \Delta \mathcal{A}^{(3)} ( \lambda ) \cdot \mathbb{P}^1_A ( \mathbf{1}_{5} ) ) = 0 \in \mathrm{Pic} ( C_{\mathbf{1}_{5}} )$. In consequence, \cref{equ:SpectrumOfMSFOnSingletCurveOntoSingletCurve} yields
	$L ( \mathcal{A} ( \mathbf{1}_{5} ) ( \lambda ), C_{\mathbf{1}_{5}} ) = \mathcal{O}_{\mathcal{B}_6} ( - \lambda ( 6 \overline{K}_{\mathcal{B}_6} - 5 W ) ) 
	|_{C_{\mathbf{1}_{5}}} \otimes \sqrt{K_{C_{\mathbf{1}_{5}}}}$.
\end{itemize}

\subsection{D3-Tadpole and D-term} \label{subsec:TadpolesSU5xU1}

In \cref{chapter:GUTModels} we consider fluxes of the form $A = A_X ( F ) + A ( \mathbf{10}_1 ) ( \lambda ) + A_Y ( \mathcal{H} )$. For such a flux the D3-tadpole cancellation is the demand that there exists $N_{D_3} \geq 0$ with
\[ \frac{1}{2} \Mint_{\hat{Y}_4}{ \left( A_X \left( F \right) + A \left( \mathbf{10}_1 \right) \left( \lambda \right) + A_Y \left( \mathcal{H} \right) \right)^2} = \frac{\chi ( T_{\hat{Y}_4} )}{24} \, . \]
$\chi ( T_{\hat{Y}_4} )$ was worked out in \cite{oai:arXiv.org:1202.3138} and it holds \footnote{In comparison to \cite{oai:arXiv.org:1202.3138}, we have included the hypercharge flux $A_Y (\mathcal{H} )$ and chosen an additional $-1$ for the universal flux $A ( \mathbf{10}_1 ) ( \lambda )$. Dropping $A_Y ( \mathcal{H} )$ and replacing $\lambda$ by $- \lambda$ should thus recover the old result. Due to a typo in \cite{oai:arXiv.org:1202.3138} this is not the case.}
\begin{align} 
\begin{split}
&\frac{1}{2} \cdot \Mint_{\hat{Y}_4}{ \left( A_X \left( F \right) + A \left( \mathbf{10}_1 \right) \left( \lambda \right) + A_Y \left( \mathcal{H} \right) \right)^2} = - \frac{30}{2} \Mint_{W}{\mathcal{H}^2} \\
&\hspace{6em} - \Mint_{B_3}{\left[ F^2 \left( \overline{K}_{B_3} - \frac{3}{5} W \right) - \frac{\lambda}{5} W F \overline{K}_{B_3}
                                 - \frac{\lambda^2}{2} W \overline{K}_{B_3} \left( \frac{6}{5} \overline{K}_{B_3} - W \right) \right]} \, .
\end{split}
\end{align}
Since $A_Y ( \mathcal{H} )$ is supersymmetric, the D-term for $A = A_X ( F ) + A ( \mathbf{10}_1 ) ( \lambda ) + A_{Y} ( \mathcal{H} )$ matches the D-term for $A_X ( F ) + A ( \mathbf{10}_1 ) ( \lambda )$, which was worked out \cite{oai:arXiv.org:1202.3138}. It holds
\[ \resizebox{0.9\textwidth}{!}{$\displaystyle
\xi_X \left( A_X \left( F \right) \right) \simeq - \frac{2}{\mathcal{V}_B} \Mint_{\mathcal{B}_6}{J \wedge F \wedge \left( 3 W - 5 \overline{K} \right)}, \qquad \xi_X \left( A \left( \mathbf{10}_1 \right) \left( \lambda \right) \right) \simeq \frac{\lambda}{\mathcal{V}_B} \Mint_{\mathcal{B}_6}{J \wedge W \wedge \overline{K}} \, . 
$} \]

\subsection{Proving the Chow Relations for \texorpdfstring{$\mathbf{SU(5) \times U(1)_X}$}{SU(5)xU(1)}} \label{subsec:ProofOfChowRelationsSU5xU1}

In this appendix we prove the relations \cref{summary4dChow2} for the $SU(5) \times U(1)_X$ model considered in the main text. In the concrete geometry at hand, \cref{summary4dChow2} reduces to the system of equations \cref{relation1}, \cref{relation2}, \cref{relation3}. To prepare us for these proofs, let us recall the linear relations of the $SU(5) \times U(1)_X$-top. These are
\begin{align}
\begin{split}
\mathcal{X} - \mathcal{Y} + \mathcal{Z} + \mathcal{E}_0 + \mathcal{E}_1 + \mathcal{E}_2 - \mathcal{W} + \overline{\mathcal{K}}_{\mathcal{B}_6} &= 0 \in \mathrm{CH}^1 ( \hat{Y}_5 ) \, , \\
-3 \mathcal{X} + 2 \mathcal{Y} - \mathcal{S} - \mathcal{E}_1 -2 \mathcal{E}_2 + \mathcal{E}_4 &= 0 \in \mathrm{CH}^1 ( \hat{Y}_5 ) \, , \\
2 \mathcal{X} - \mathcal{Y} - \mathcal{Z} + \mathcal{S} + \mathcal{E}_1 + 2 \mathcal{E}_2 + \mathcal{E}_3 - \overline{\mathcal{K}}_{\mathcal{B}_6} &= 0 \in \mathrm{CH}^1 ( \hat{Y}_5 ) \, .
\end{split}
\end{align}
Below we will often make use of these relations to show equivalence of algebraic cycles in the ambient space $\hat{Y}_5$. Let us explain in an example what we mean by this. We look at
\[ C = \mathcal{E}_2 \cdot \mathcal{W} \in \mathrm{CH}^2 ( \hat{Y}_5 ) \, . \]
In working over the rational numbers, the linear relations yield
\[ \mathcal{W} = \frac{1}{2} \left( 2 \mathcal{E}_0 + \mathcal{E}_1 - \mathcal{E}_3 + 3 \overline{\mathcal{K}}_{\mathcal{B}_6} - \mathcal{S} - \mathcal{Y} + 3 \mathcal{Z} \right) \, . \]
Consequently, we find that $C$ is rationally equivalent to
\[ C^\prime = \mathcal{E}_2 \cdot \frac{1}{2} \left( 2 \mathcal{E}_0 + \mathcal{E}_1 - \mathcal{E}_3 + 3 \overline{\mathcal{K}}_{\mathcal{B}_6} - \mathcal{S} - \mathcal{Y} + 3 \mathcal{Z} \right) \, . \]
In going from $C$ to $C^\prime$ we have eliminated the class $\mathcal{W}$ and have avoided self-intersections. This is the rational along which we employ the linear relations. In particular, most of these transformations require us to work over $\mathbb{Q}$. This is what made us excluded torsional effects in the first place.

\subsubsection{Trivial Restrictions}

Consider the cycle $V ( e_1, x ) \in Z^2 ( \hat{Y}_5 )$. This cycle is non-trivial in $\hat{Y}_5$. Nonetheless we have $V ( P_T^\prime, e_1, x ) = V ( e_1, x, e_3 e_4 s y^2 ) = \emptyset$, so this cycle pulls back to give the trivial cycle in $\hat{Y}_4$. Similarly, $V ( P_T^\prime, x, z ) = V ( P_T^\prime, y, z ) = V ( P_T^\prime, e_0, x ) = V ( P_T^\prime, e_0, y ) = \emptyset$. Hence, the restriction of all these non-trivial cycles in $\hat{Y}_5$ gives trivial cycles on $\hat{Y}_4$.

\subsubsection{A Useful Identity}

In this subsection we justify the following identity, which we will make use of in the proof of \cref{relation3} below:
\[ \left. \left( \mathcal{W} - 2 \overline{\mathcal{K}}_{\mathcal{B}_6} \right) \left( \mathcal{E}_1 + 2 \mathcal{E}_2 - \mathcal{E}_4 \right) \right|_{\hat{Y}_4} = \left. \mathcal{E}_3 \mathcal{E}_4 \right|_{\hat{Y}_4} - \left. \mathcal{E}_2 \mathcal{X} \right|_{\hat{Y}_4} - 2 \left. \overline{\mathcal{K}}_{\mathcal{B}_6} \mathcal{E}_2 \right|_{\hat{Y}_4} + 2 \left. \mathcal{E}_2 \mathcal{E}_4 \right|_{\hat{Y}_4} \, . \label{usefulrelation} \]
We proceed in a number of steps to justify this result.
\begin{enumerate}
 \item First of all we exploit the linear relations induced from the $\mathrm{SU} ( 5 ) \times \mathrm{U} ( 1 )_X$-top and its Stanley-Reisner ideal. Thereby, one can 
      show that \cref{usefulrelation} is equivalent to
      \[ \left. \mathcal{E}_1 \mathcal{W} \right|_{\hat{Y}_4} + 4 \left. \mathcal{E}_2 \mathcal{W} \right|_{\hat{Y}_4} - 2 \left. \mathcal{E}_1 \overline{\mathcal{K}}_{\mathcal{B}_6} \right|_{\hat{Y}_4} - 6 \left. \mathcal{E}_2 \overline{\mathcal{K}}_{\mathcal{B}_6} \right|_{\hat{Y}_4} = \left. \mathcal{E}_1 \mathcal{E}_2 \right|_{\hat{Y}_4} - 2 \left. \mathcal{E}_2 \mathcal{E}_3 \right|_{\hat{Y}_4} + \left. \mathcal{E}_0 \mathcal{E}_4 \right|_{\hat{Y}_4} \, . \label{equ:ProofUsefulIdentityEqu1} \]
 \item By use of $\mathcal{W} = 2 \overline{\mathcal{K}}_{\mathcal{B}_6} + ( \mathcal{W} - 2 \overline{\mathcal{K}}_{\mathcal{B}_6} )$ and $\mathcal{W} = \frac{3}{2} 
      \overline{\mathcal{K}}_{\mathcal{B}_6} + \frac{1}{2} ( 2 \mathcal{W} - 3 \overline{\mathcal{K}}_{\mathcal{B}_6} )$ we find that \cref{equ:ProofUsefulIdentityEqu1} is equivalent to
      \[
      \resizebox{0.82\textwidth}{!}{$
      V \left( P_T^\prime, e_1, a_{2,1} \right) + 2 V \left( P_T^\prime, e_2, a_{3,2} \right) = 2 V \left( P_T^\prime, e_2, e_3 \right) - V \left( P_T^\prime, e_1, e_2 \right) - V \left( P_T^\prime, e_0, e_4 \right) \, .$}
      \label{equ:ProofUsefulIdentityEqu2}
      \]
 \item The $\mathrm{SU} ( 5 ) \times U ( 1 )_X$-\emph{top} yields a complete and simplicial toric variety. Thus, there exists an isomorphism $\mathrm{CH}^\bullet ( 
      \mathrm{top} )_{\mathbb{Q}} \cong H^\bullet ( \mathrm{top}, \mathbb{Q} )$  \cite{cox2011toric} which enables us to find the following identities in $\mathrm{CH}^\bullet ( \hat{Y}_5 )_{\mathbb{Q}}$: 
      \begin{align}
      \begin{split}
      V \left( e_1, e_2 e_3 s x \right) &= V \left( e_1, e_0 a_{2,1} z^2 \right) \\
      \Leftrightarrow V \left( e_1, e_2 \right) + V \left( e_1, x \right) &= V \left( e_1, e_0 \right) + V \left( e_1, a_{2,1} \right) \\
      \Leftrightarrow V \left( e_1, e_2 \right) &= V \left( e_1, e_0 \right) + V \left( e_1, a_{2,1} \right) - V \left( e_1, x \right) \, .
      \end{split}
      \end{align}
      Along the same lines it can be shown that $V ( e_2, e_0^2 e_1 a_{3,2} z^3 ) = V ( e_2, e_3 s y )$ is equivalent to $V ( e_2, e_3 ) = V ( e_2, e_1 ) + V ( e_2, a_{3,2} )$. By use of these two identities it then follows that \cref{equ:ProofUsefulIdentityEqu2} is equivalent to
      \[ 0 = V \left( P_T^\prime, e_0, e_4 \right) + V \left( P_T^\prime, e_1, x \right) - V \left( P_T^\prime, e_1, e_0 \right) \, . \label{equ:ProofUsefulIdentityEqu3} \]
 \item Next recall that $V ( P_T^\prime, e_1, x ) = \emptyset$ is a `trivial restriction' as discussed previously. Hence, all we are left to show is that
      $V ( P_T^\prime, e_0, e_1 ) = V ( P_T^\prime, e_0, e_4 )$ holds true in $\mathrm{CH}^2( \hat{Y}_4 )$. To see this we start with a primary decomposition, which shows
      \[ V \left( P_T^\prime, e_0, e_1 \right) = V \left( e_0, e_1, y e_3 e_4 + x z a_{1,0} \right) = V \left( P_T^\prime, e_0, y s e_3 e_4 + x z a_{1,0} \right) \, . \]
      By use of a suitable interpolating cycle in $\mathbb{P}^1 \times \hat{Y}_4$ we thus arrive at $V ( P_T^\prime, e_0, e_1 ) = V ( P_T^\prime, e_0, x z a_{1,0} )$. A primary decomposition of $V ( P_T^\prime, e_0, x z a_{1,0} )$ now shows
      \[ V \left( P_T^\prime, e_0, e_1 \right) = V \left( a_{1,0}, e_0, x^3 s e_1 e_2^2 - y^2 e_4 \right) + V \left( z, e_0, x^3 s e_1 e_2^2 - y^2 e_4 \right) \, . \]
      For $V ( P_T^\prime, e_0, e_4 )$ we proceed along the very same lines. Namely first we obtain from primary decomposition
      \[ \resizebox{0.83\textwidth}{!}{$ \displaystyle
      V \left( P_T^\prime, e_0, e_4 \right) = V \left( P_T^\prime, e_0, e_4, x^2 s e_1 e_2^2 e_3 - y z a_{1,0} \right) = V \left( P_T^\prime, e_0, x^2 s e_1 e_2^2 e_3 - y z a_{1,0} \right) \, . $} \]
      By use of an interpolating cycle in $\mathbb{P}^1 \times \hat{Y}_4$ we see $V ( P_T^\prime, e_0, e_4 ) = V ( P_T^\prime, e_0, y z a_{1,0} )$. Finally, a primary decomposition of $V ( P_T^\prime, e_0, y z a_{1,0} )$ yields
      \[ V \left( P_T^\prime, e_0, e_4 \right) = V \left( a_{1,0}, e_0, x^3 s e_1 e_2^2 - y^2 e_4 \right) + V \left( z, e_0, x^3 s e_1 e_2^2 - y^2 e_4 \right) \, \label{equ:AlmostEndOfProofOfAUsefulIdentity}. \]
      Indeed, these results imply that $V ( P_T^\prime, e_0, e_1 ) = V( P_T^\prime, e_0, e_4)$ holds true in $\mathrm{CH}^2 ( \hat{Y}_4 )$, which completes our proof.
\end{enumerate}

\subsubsection{Proof of \texorpdfstring{\Cref{relation1}}{7.72}}

Let us recall that \cref{relation1} states $A ( \mathbf{5}_3 ) ( \lambda ) = - A ( \mathbf{5}_{-2} ) ( \lambda ) - A ( \mathbf{10}_1 ) ( \lambda )$. To modify this identity, we use the explicit representation of the involved fluxes. As stated in \cref{equ:Fluxes-Y5} we can express them in terms of the ambient space $\hat{Y}_5$ as
\begin{align}
\begin{split} \label{equ:FluxesInAmbientSpace}
\mathcal{A} \left( \mathbf{10}_1 \right) \left( \lambda \right) &= - \frac{\lambda}{5} \, \left( 2 \cE_1 - \cE_2 + \cE_3 - 2 \cE_4 \right) \cdot\overline{\mathcal{K}}_{B_3} - \lambda \cE_2 \cdot \cE_4 \, , \\
\mathcal{A} \left( \mathbf{5}_3 \right) \left( \lambda \right) &= - \frac{\lambda}{5} \, \left( \cE_1 + 2 \cE_2 - 2 \cE_3 - \cE_4 \right) \cdot\left( 3 \overline{\mathcal{K}}_{B_3} - 2 \mathcal{W} \right) - \lambda \cE_3 \cdot\mathcal{X} \, , \\
\mathcal{A} \left( \mathbf{5}_{-2} \right) \left( \lambda \right) &= - \frac{\lambda}{5} \, \left( \cE_1 + 2 \cE_2 + 3 \cE_3 - \cE_4 \right) \cdot\left( 5 \overline{\mathcal{K}}_{B_3} - 3 \mathcal{W} \right) + \lambda \mathcal{E}_3 \cdot \left( \overline{\mathcal{K}}_{B_3} + \mathcal{Y} - \cE_4 \right) \, .
\end{split}
\end{align}
We transform these expression by use of \cref{splitting101}, \cref{splitting53}, \cref{splitting5-2}. For example, we have
\[ \left. - \mathcal{E}_2 \cdot \mathcal{E}_4 \right|_{\hat{Y}_4} = - V \left( P_T^\prime, e_2, e_4 \right) = \left. - \mathcal{E}_2 \cdot \mathcal{K}_{B_3} \right|_{\hat{Y}_4} + \mathbb{P}^1_{2B} \left( \mathbf{10}_1 \right) \, . \]
By use of \cref{splitting53} and \cref{splitting5-2} one also find
\begin{align}
\begin{split}
\left. \mathcal{E}_3 \cdot \mathcal{X} \right|_{\hat{Y}_4} &= \left. \mathcal{E}_3 \cdot \left( 3 \mathcal{K}_{B_3} - 2 \mathcal{W} \right) \right|_{\hat{Y}_4} - \mathbb{P}^1_{3F} \left( \mathbf{5}_{3} \right) \, , \\
\left. \mathcal{E}_3 \cdot \mathcal{K}_{B_3} + \mathcal{E}_3 \mathcal{Y} - \mathcal{E}_3 \mathcal{E}_4 \right|_{\hat{Y}_4} &= \left. \mathcal{E}_3 \cdot \left( 5 \mathcal{K}_{B_3} - 3 \mathcal{W} \right) \right|_{\hat{Y}_4} - \mathbb{P}^1_{3G} \left( \mathbf{5}_{-2} \right) \, .
\end{split}
\end{align}
With these results it is readily found that \cref{equ:FluxesInAmbientSpace} is equivalent to
\begin{align}
\begin{split}
A \left( \mathbf{10}_1 \right) &= - \left. \frac{\lambda}{5} \left[ 2 \mathcal{E}_1 + 4 \mathcal{E}_2 + \mathcal{E}_3 - 2 \mathcal{E}_4 \right] \cdot \overline{K}_{B_3} \right|_{\hat{Y}_4} + \lambda \mathbb{P}^1_{2B} \left( \mathbf{10}_1 \right) \, , \\
A \left( \mathbf{5}_3 \right) &= - \left. \frac{\lambda}{5} \left[ \mathcal{E}_1 + 2 \mathcal{E}_2 + 3 \mathcal{E}_3 - \mathcal{E}_4 \right] \cdot \left( 3 \overline{K}_{B_3} - 2 \mathcal{W} \right) \right|_{\hat{Y}_4} + \lambda \mathbb{P}^1_{3F} \left( \mathbf{5}_3 \right) \, , \\
A \left( \mathbf{5}_{-2} \right) &= - \left. \frac{\lambda}{5} \left[ \mathcal{E}_1 + 2 \mathcal{E}_2 - 2 \mathcal{E}_3 - \mathcal{E}_4 \right] \cdot \left( 5 \overline{K}_{B_3} - 3 \mathcal{W} \right) \right|_{\hat{Y}_4} - \lambda \mathbb{P}^1_{3G} \left( \mathbf{5}_{-2} \right) \, .
\end{split}
\end{align}
Consequently, \cref{relation1} is equivalent to
\[ \lambda \cdot \left( \mathbb{P}^1_{3G} \left( \mathbf{5}_{-2} \right) - \mathbb{P}^1_{3F} \left( \mathbf{5}_3 \right) - \mathbb{P}^1_{2B} \left( \mathbf{10}_1 \right)  \right) = \lambda \cdot \left. \left( \mathcal{W} - 2 \overline{\mathcal{K}}_{B_3} \right) \cdot \left( \mathcal{E}_1 + 2 \mathcal{E}_2 - \mathcal{E}_4 \right) \right|_{\hat Y_4} \, . \label{relation1b} \]
We are thus left to connect the matter surfaces $\mathbb{P}^1_{3G} ( \mathbf{5}_{-2} )$, $\mathbb{P}^1_{3F} ( \mathbf{5}_3 )$ and $\mathbb{P}^1_{2B} ( \mathbf{10}_1 )$, whose explicit form in terms of vanishing loci of certain ideals is given in \cref{subsec:FibreStructureSU5xU1}, to the pullback on the RHS. To prove that \cref{relation1b} holds true in $\mathrm{CH}^\bullet ( \hat{Y}_4)$ for all $\lambda \in \mathbb{C}$, it suffices to verify the statement for $\lambda = 1$. We proceed in various steps.
\begin{enumerate}
\item As we justify below, up to rational equivalence in $\hat{Y}_4$ the involved surfaces satisfy
     \[ \mathbb{P}^1_{3G} \left( \mathbf{5}_{-2} \right) - \mathbb{P}^1_{3F} \left( \mathbf{5}_3 \right) + \mathbb{P}^1_{2B} \left( \mathbf{10}_1 \right) =  V \left(P_T^\prime, e_3, e_4 \right)  -  V \left(P_T^\prime, e_2, x \right) \, . \label{starting-rel1} \]
 \item Let us now subtract $2 \mathbb{P}^1_{2B} ( \mathbf{10}_1 )$ from both sides. By use of the explicit representation of $\mathbb{P}^1_{2B} ( \mathbf{10}_1 )$ given   
      in \cref{subsec:FibreStructureSU5xU1}, we  arrive at
      \begin{align}
      \begin{split}
      \label{relation1c}
      &\mathbb{P}^1_{3G} \left( \mathbf{5}_{-2} \right) - \mathbb{P}^1_{3F} \left( \mathbf{5}_3 \right) - \mathbb{P}^1_{2B} \left( \mathbf{10}_1 \right) 
      =  V \left( P_T^\prime, e_3, e_4 \right) \\ 
      & \hspace{8em}- V \left(P_T^\prime, e_2, x \right) -2  V \left( P_T^\prime, \overline{K}_{\mathcal{B}_6}, e_2 \right) +2  V \left( P_T^\prime, e_2, e_4 \right) \,.
      \end{split}
      \end{align}
 \item The final step consists in showing that the RHS can  be simplified further. By use of the `useful identity' proven above, we have the following 
      identity in $\mathrm{CH}^2 ( \hat{Y}_5)$
      \[ \left( \mathcal{W} - 2 \overline{\mathcal{K}}_{\mathcal{B}_6} \right) \cdot \left( \mathcal{E}_1 + 2 \mathcal{E}_2 - \mathcal{E}_4 \right) = \mathcal{E}_3 \cdot \mathcal{E}_4 - \mathcal{E}_2 \cdot \mathcal{X} - 2 \overline{\mathcal{K}}_{\mathcal{B}_6} \cdot \mathcal{E}_2 + 2 \mathcal{E}_2 \cdot \mathcal{E}_4 \, . \label{usefulidentity} \]
      According to the discussion around \cref{pullbackformula}, pulling back all terms to $\hat{Y}_4$ gives a corresponding relation in $\mathrm{CH}^2(\hat{Y}_4)$. In particular, we showed that after pullback to $\hat{Y}_4$ the RHS of \cref{usefulidentity} and of \cref{relation1c} coincide. Thereby, we arrive at \cref{relation1b}.
\end{enumerate}

To complete the proof, it remains to justify \cref{starting-rel1}. By performing a primary ideal decomposition we obtain the following two identities in rational equivalence of $\hat{Y}_4$,
\begin{align}
\begin{split}
V \left(P_T^\prime, e_3, a_{3,2} e_0^2 z^2 e_1 e_4 + a_{1,0} x s \right) &= \mathbb{P}^1_{3x} \left( \mathbf{5}_3 \right) + \mathbb{P}^1_{3G} \left( \mathbf{5}_{-2} \right) \\
& \hspace{6em} + V \left( e_3, e_2, e_0^2 z^2 e_1 e_4 a_{3,2} + x s a_{1,0} \right)\,, \\
V \left(P_T^\prime, e_2, e_0^2 z^2 e_1 e_4 a_{3,2} + x s a_{1,0} \right) &= \mathbb{P}^1_{24} \left( \mathbf{10}_1 \right) + V \left(e_3, e_2, e_0^2 z^2 e_1 e_4 a_{3,2} + x s a_{1,0} \right).
\end{split}
\end{align}
As a consequence, we learn
\begin{align}
\begin{split}
\mathbb{P}^1_{3G} \left( \mathbf{5}_{-2} \right) + \mathbb{P}^1_{3x} \left( \mathbf{5}_3 \right) - \mathbb{P}^1_{24} \left( \mathbf{10}_1 \right) &= V \left(P_T^\prime, e_3, a_{3,2} e_0^2 z^2 e_1 e_4 + a_{1,0} x s \right) \\
& \hspace{6em} - V \left( P_T^\prime, e_2, a_{3,2} e_0^2 z^2 e_1 e_4 + a_{1,0} x s \right) \, . \label{equ:RelationAmongMatterSurfacesForInteractionY2}
\end{split}
\end{align}
At this stage we make use of the very definition of rational equivalence as recalled around equation \cref{Gammdefintion-general}. To this end, we define a cycle
$\Gamma_1(t)$ on $\mathbb P^1 \times \hat{Y}_4$ parametrized by $t \in \mathbb P^1$ such that $\Gamma_1(t) = V (P_T^\prime, e_3, a_{3,2} e_0^2 z^2 e_1 e_4 + t a_{1,0} x s )$. By definition, $\Gamma_1(t=0) = V ( P_T^\prime, e_3, a_{3,2} e_0^2 z^2 e_1 e_4 )$ and $\Gamma_1(t=1)$ are rationally equivalent cycles. Note also that
\[ V \left( P_T^\prime, e_3, a_{3,2} e_0^2 z^2 e_1 e_4\right) = \left. V \left( e_3, a_{3,2} e_0^2 z^2 e_1 e_4 \right) \right|_{\hat{Y}_4} = \left. V \left( e_3, a_{3,2} e_4 \right) \right|_{\hat{Y}_4} \]
where the second equality makes use of the Stanley-Reisner ideal \cref{SRideal} of the \emph{top}. Consequently, we learn that in rational equivalence of $\hat{Y}_4$
\[ V \left( P_T^\prime, e_3, a_{3,2} e_0^2 z^2 e_1 e_4 + a_{1,0} x s \right) = V \left( P_T^\prime, e_3, a_{3,2} \right) + V \left( P_T^\prime, e_3, e_4 \right) \, . \label{zwischen-rel1a} \]
Similarly, we can consider the cycle $\Gamma_2(t) = V ( P_T^\prime, e_2, t \, a_{3,2} e_0^2 z^2 e_1 e_4 + a_{1,0} x s )$ and exploit the rational equivalence of $\Gamma_2(t=1)$ and $\Gamma_2(t=0)$ to conclude
\[ V \left(P_T^\prime, e_2, a_{3,2} e_0^2 z^2 e_1 e_4 + a_{1,0} x s \right) = V \left( P_T^\prime, e_2, a_{1,0} \right) + V \left( P_T^\prime, e_2, x \right) \, . \label{zwischen-rel1b}  \]
The first terms on the RHS of \cref{zwischen-rel1a} and \cref{zwischen-rel1b} represent two roots restricted to the matter curves $C_{\mathbf{5}_3}$ and $C_{\mathbf{10}_1}$, respectively. As detailed in \cref{subsec:FibreStructureSU5xU1}, these are related to the relevant surfaces in \cref{starting-rel1} as follows 
\[ - \mathbb{P}_{24}^1\left( \mathbf{10}_1 \right) = \mathbb{P}_{2B}^1\left( \mathbf{10}_1 \right) -  V(P_T^\prime, e_2, a_{1,0}) \, , \qquad
\mathbb{P}_{3x}^1 \left( \mathbf{5}_3 \right) = - \mathbb{P}_{3F}^1 \left( \mathbf{5}_3 \right) + V(P_T^\prime, e_3, a_{3,2}) \, . \]
Plugging everything into \cref{equ:RelationAmongMatterSurfacesForInteractionY2}, we recover \cref{starting-rel1} as claimed.

\subsubsection{Proof of \texorpdfstring{\Cref{relation2}}{7.73}}

The second relation, \cref{relation2}, is equivalent to the following statement in $\mathrm{CH}^2( \hat{Y}_4)$:
\begin{align}
\begin{split}
\lambda \cdot \mathbb{P}^1_{3H} \left( \mathbf{5}_{-2} \right) &= \lambda \cdot \left. \overline{\mathcal{K}}_{\mathcal{B}_6} \cdot \left( \mathcal{E}_1 + 2 \mathcal{E}_2 + 3 \mathcal{E}_3 - \mathcal{E}_4 \right) \right|_{\hat{Y}_4} - \lambda \cdot \left. \mathcal{W} \cdot \left( \mathcal{E}_1 + 2 \mathcal{E}_2 + 3 \mathcal{E}_3 \right) \right|_{\hat{Y}_4} \\
& \hspace{12em} + \lambda \cdot \left. \overline{\mathcal{K}}_{\mathcal{B}_6} \mathcal{W} \right|_{\hat{Y}_4} - \lambda \cdot \left. \mathcal{S} \mathcal{W} \right|_{\hat{Y}_4} + \lambda \cdot \left. \mathcal{Z} \mathcal{W} \right|_{\hat{Y}_4} \, . \label{relation2-form2}
\end{split}
\end{align}
It suffices to treat the case $\lambda = 1$. We proceed in several steps for the proof.
\begin{enumerate}
 \item First, a primary decomposition of $V \left( P_T^\prime, e_3, a_{2,1} e_0 x z e_1 e_2 - a_{1,0} y \right)$ yields
      \[ \mathbb{P}^1_{3H} \left( \mathbf{5}_{-2} \right) = V \left( P_T^\prime, e_3, a_{2,1} e_0 x z e_1 e_2 - a_{1,0} y \right) - V \left( P_T^\prime, e_3, e_4 \right) \, . \]
      By the method of rational homotopy on $\mathbb P^1 \times \hat{Y}_4$ we find the relation
      \[ V \left( P_T^\prime, e_3, a_{2,1} e_0 x z e_1 e_2 - a_{1,0} y \right) = V(P_T^\prime,e_3,a_{1,0}) + V(P_T^\prime,e_3,y) \in \mathrm{CH}^2(\hat{Y}_4) \,. \]
      We combine these finding to conclude that in rational equivalence on $\hat{Y}_4$
      \[ \mathbb{P}^1_{3H} \left( \mathbf{5}_{-2} \right) = \left. \mathcal{E}_3 \cdot \overline{\mathcal{K}}_{\mathcal{B}_6} \right|_{\hat{Y}_4} + \left. \mathcal{E}_3 \cdot \mathcal{Y} \right|_{\hat{Y}_4} - \left. \mathcal{E}_3 \cdot \mathcal{E}_4 \right|_{\hat{Y}_4} \,. \label{P13hrel2} \]
 \item To compare this with \cref{relation2-form2}, first note that 
      \begin{align}
      \begin{split}
      \left. \mathcal{E}_3 \cdot \left( \mathcal{Y} - \mathcal{E}_4 \right) \right|_{\hat{Y}_4} &= \left. \overline{\mathcal{K}}_{\mathcal{B}_6} \cdot \left( \mathcal{E}_1 + 2 \mathcal{E}_2 + 2 \mathcal{E}_3 - \mathcal{E}_4 \right) \right|_{\hat{Y}_4} - \left. \mathcal{W} \cdot \left( \mathcal{E}_1 + 2 \mathcal{E}_2 + 3 \mathcal{E}_3 \right) \right|_{\hat{Y}_4} \\
      & \hspace{10.5em} + \left. \overline{\mathcal{K}}_{\mathcal{B}_6} \cdot \mathcal{W} \right|_{\hat{Y}_4} - \left. \mathcal{S} \cdot \mathcal{W} \right|_{\hat{Y}_4} + \left. \mathcal{Z} \cdot \mathcal{W} \right|_{\hat{Y}_4} \label{Ve3y}
      \end{split}
      \end{align}
      holds in rational equivalence on $\hat{Y}_4$ if and only if 
      \[ 0 = - \left. \mathcal{E}_0 \mathcal{E}_1 \right|_{\hat{Y}_4} + \left. \mathcal{E}_0 \overline{\mathcal{K}}_{\mathcal{B}_6} \right|_{\hat{Y}_4} + 
      \left. \mathcal{E}_0 \mathcal{Z} \right|_{\hat{Y}_4} \,. \label{zwischen-rel2a} \]
      The equivalence of \cref{Ve3y} and \cref{zwischen-rel2a} follows readily from the linear relations induced by the $\mathrm{SU} ( 5 ) \times U ( 1 )_X$-top and the trivial restrictions introduced previously. Since \cref{Ve3y} and \cref{P13hrel2} imply \cref{relation2-form2}, it remains to prove \cref{zwischen-rel2a}.
 \item Indeed, $V ( P_T^\prime, e_0, x^3 s e_1 e_2^2 - y^2 e_4 ) = V ( P_T^\prime, e_0, a_{1,0} ) + V ( P_T^\prime, e_0, z )$ follows from primary decomposition. In 
      addition, a suitable rational homotopy shows the following identity in $\mathrm{CH}^2 ( \hat{Y}_4 )$
      \[ V \left( P_T^\prime, e_0, x^3 s e_1 e_2^2 - y^2 e_4 \right) = V \left( P_T^\prime, e_0, x^3 s e_1 e_2^2 \right) = V \left( P_T^\prime, e_0, e_1 \right) \, , \label{equ:proof2last2} \]
      where the last step uses the Stanley-Reisner ideal on $\hat{Y}_4$. Combining these results implies \cref{zwischen-rel2a}, as desired.
\end{enumerate}

\subsubsection{Proof of \texorpdfstring{\Cref{relation3}}{7.74}}

\begin{enumerate}
 \item First we make use of \cref{equ:Fluxes-Y5} to write the final relation \cref{relation3} as
      \begin{align}
      \begin{split}
             & - \left. \lambda \cdot \mathcal{S} \cdot \left( 3 \overline{\mathcal{K}}_{\mathcal{B}_6} - 2 \mathcal{W} - \mathcal{X} \right) \right|_{\hat{Y}_4} - \left. \frac{\lambda}{5} \left( 2 \mathcal{E}_1 - \mathcal{E}_2 + \mathcal{E}_3 - 2 \mathcal{E}_4 \right) \overline{\mathcal{K}}_{\mathcal{B}_6} \right|_{\hat{Y}_4} - \left. \lambda \mathcal{E}_2 \mathcal{E}_4 \right|_{\hat{Y}_4} \\
             & \hspace{3em} = - \left. \frac{\lambda}{5} \left( 6 \overline{\mathcal{K}}_{\mathcal{B}_6} - 5 \mathcal{W} \right) \cdot \left( 2 \mathcal{E}_1 + 4 \mathcal{E}_2 + 6 \mathcal{E}_3 + 3 \mathcal{E}_4 + 5 \mathcal{S} - 5 \mathcal{Z} - 5 \overline{\mathcal{K}}_{\mathcal{B}_6} \right) \right|_{\hat{Y}_4} \, .
      \end{split}
      \end{align}
      We focus on $\lambda = 1$. Then this equation is equivalent to
      \begin{align}
      \begin{split}
      0 &= - \left. \mathcal{E}_2 \mathcal{E}_4 \right|_{\hat{Y}_4} + 2 \left. \mathcal{E}_1 \overline{\mathcal{K}}_{\mathcal{B}_6} \right|_{\hat{Y}_4} + 5 \left. \mathcal{E}_2 \overline{\mathcal{K}}_{\mathcal{B}_6} \right|_{\hat{Y}_4} + 7 \left. \mathcal{E}_3 \overline{\mathcal{K}}_{\mathcal{B}_6} \right|_{\hat{Y}_4} + 4 \left. \mathcal{E}_4 \overline{\mathcal{K}}_{\mathcal{B}_6} \right|_{\hat{Y}_4} \\
      & \hspace{2em} - 6 \left. \overline{\mathcal{K}}_{\mathcal{B}_6}^2 \right|_{\hat{Y}_4} + 3 \left. \overline{\mathcal{K}}_{\mathcal{B}_6} \mathcal{S} \right|_{\hat{Y}_4} - \left. 2 \mathcal{E}_1 \mathcal{W} \right|_{\hat{Y}_4} - \left. 4 \mathcal{E}_2 \mathcal{W} \right|_{\hat{Y}_4} - \left. 6 \mathcal{E}_3 \mathcal{W} \right|_{\hat{Y}_4} \\
      & \hspace{2em} - \left. 3 \mathcal{E}_4 \mathcal{W} \right|_{\hat{Y}_4} + \left. 5 \overline{\mathcal{K}}_{\mathcal{B}_6} \mathcal{W} \right|_{\hat{Y}_4} - 3 \left. \mathcal{S} \mathcal{W} \right|_{\hat{Y}_4} + \left. \mathcal{S} \mathcal{X} \right|_{\hat{Y}_4} - \left. 6 \overline{\mathcal{K}}_{\mathcal{B}_6} \mathcal{Z} \right|_{\hat{Y}_4} + \left. 5 \mathcal{W} \mathcal{Z} \right|_{\hat{Y}_4} \, . \label{equ:Proof3Equ1}
      \end{split}
      \end{align}
 \item By use of the linear relations induced from the $\mathrm{SU} ( 5 ) \times \mathrm{U} ( 1 )_X$-top and the trivial restrictions discussed previously, it can be 
      seen that \cref{equ:Proof3Equ1} is equivalent to
      \begin{align}
      \begin{split}
      0 &= 3 V \left( P_T^\prime, e_0, e_1 \right) -2 V \left( P_T^\prime, e_2, e_3 \right) - 3 V \left( P_T^\prime, e_0, e_4 \right) + V \left( P_T^\prime, e_3, e_4 \right) \\
      & \hspace{12em} - V \left( P_T^\prime, e_3, s \right) - 3 V \left( P_T^\prime, e_3, x \right) + 2 V \left( P_T^\prime, e_3, y \right) \, .
      \label{equ:Proof3Equ2}
      \end{split}
      \end{align}
      As argued around \cref{equ:AlmostEndOfProofOfAUsefulIdentity}, $V ( P_T^\prime, e_0, e_1 ) = V ( P_T^\prime, e_0, e_4 )$ in rational equivalence of $\hat{Y}_4$. Therefore, \cref{equ:Proof3Equ2} is equivalent to
      \[ \resizebox{0.82\textwidth}{!}{$ \displaystyle
      0 = 2 V \left( P_T^\prime, e_2, e_3 \right) - V \left( P_T^\prime, e_3, e_4 \right) + V \left( P_T^\prime, e_3, s \right) + 3 V \left( P_T^\prime, e_3, x \right) - 2 V \left( P_T^\prime, e_3, y \right) \, . \label{equ:Proof3Equ3} $} \]
 \item We can rewrite \cref{equ:Proof3Equ3} as $0 = - \left. \mathcal{E}_3 \left( 2 \mathcal{E}_2 - \mathcal{E}_4 + \mathcal{S} + 3 \mathcal{X} - 2 \mathcal{Y} \right) 
      \right|_{\hat{Y}_4}$. Upon use of the linear relations on $\hat{X}_\Sigma$ induced from the $\mathrm{SU} ( 5 ) \times U ( 1 )_X$-top we see that this in turn is equivalent to $0 = \left. \mathcal{E}_1 \mathcal{E}_3 \right|_{\hat{Y}_4}$. And indeed $V ( P_T, e_1, e_3 ) = \emptyset$ which completes the proof.
\end{enumerate}
\section{The Geometry of an \texorpdfstring{$\mathbf{SU(4)}$}{SU(4)}-Top} \label{sec:DetailsOfSU4Top}

\subsection{Fibre Structure} \label{subsec:FibreStructureSU4}

\begin{table}[tbp]
\centering
\begin{tabular}{c@{\hskip 20pt}ccc@{\hskip 20pt}cccc}
\toprule
& $x$ & $y$ & $z$ & $w \equiv e_0$ & $e_1$ & $e_2$ & $e_3$ \\
\midrule
$\overline{K}_{\mathcal{B}_6}$ & 2 & 3 & $\cdot$ & $\cdot$ & $\cdot$ & $\cdot$ & $\cdot$ \\
$W$ & $\cdot$ & $\cdot$ & $\cdot$ & 1 & $\cdot$ & $\cdot$ & $\cdot$ \\
\vspace{-0.5em} & \\
& -1 & -1 & $\cdot$ & -1 & 1 & $\cdot$ & $\cdot$ \\
& -1 & -2 & $\cdot$ & -1 & $\cdot$ & $\cdot$ & 1 \\
& -2 & -2 & $\cdot$ & -1 & $\cdot$ & 1 & $\cdot$ \\
\vspace{-0.5em} & \\
Z &  2 &  3 & 1 & $\cdot$ & $\cdot$ & $\cdot$ & $\cdot$ \\
\bottomrule
\end{tabular}
\caption[Toric data of $SU(4)$-top.]{The Cox ring $\mathbb{Q}[ e_0, e_1, e_2, e_3, x, y, z]$ of the $SU(4)$-top is graded according to this table. Together with \cref{equ:SRSU4} this defines the geometry of this toric space.}
\label{table-N35}
\end{table}

We consider the resolved $SU(4)$ Tate model defined by $\hat{Y}_4 = V( P_T^\prime ) \subseteq \hat{Y}_5$ with 
\begin{align}
\begin{split}
P_T^\prime &= y^2 e_3 + a_{1,0} x y z + a_{3,2} y z^3 e_0^2 e_1 e_3 - x^3 e_1 e_2^2 - a_{2,1} x^2 z^2 e_0 e_1 e_2 \\
           & \hspace{17em} - a_{4,2} x z^4 e_0^2 e_1 - a_{6,4} z^6 e_0^4 e_1^2 e_3 \, .
\end{split}
\end{align}
The divisors $E_i= V (e_i)$, $i=1,2,3$, denote the $SU(4)$ Cartan divisors, and $E_0$ represents the fibration of the affine node over the $SU(4)$ divisor $W = V(w) \subseteq \mathcal{B}_6$. The Stanley-Reisner ideal 
\[ I_{\mathrm{SR}} \left( \mathrm{top} \right) = \left\langle x y z, x y e_0, z e_1, z e_2, z e_3, y e_1, x e_3, e_0 e_2 \right\rangle \label{equ:SRSU4} \]
and the gradings in \cref{table-N35} define the fibre toric ambient space, as explained in \cref{subsec:TowardsToricVarieties}. In particular, we find the linear relations \footnote{These are simply the generators of the ideal $I_{\mathrm{LR}}$ as introduced in \cref{subsec:TowardsToricVarieties}.}
\begin{align}\label{eq:linear_relations-su4}
\begin{split}
\mathcal{X} - 2 \mathcal{Z} - \mathcal{E}_0 + \mathcal{E}_2 - 2 \overline{\mathcal{K}}_{\mathcal{B}_6} + \mathcal{W} &= 0 \in \mathrm{CH}^{\bullet} ( \hat{Y}_4 ) \, , \\
- \mathcal{X} + \mathcal{Y} - \mathcal{Z} + \mathcal{E}_3 - \overline{\mathcal{K}}_{\mathcal{B}_6} &= 0 \in \mathrm{CH}^{\bullet} ( \hat{Y}_4 ) \, , \\
- \mathcal{Y} + 3 \mathcal{Z} + 2 \mathcal{E}_0 + \mathcal{E}_1 + 3 \overline{\mathcal{K}}_{\mathcal{B}_6} -2 \mathcal{W} &= 0 \in \mathrm{CH}^{\bullet} (\hat{Y}_4) \, .
\end{split}
\end{align}
From the discriminant of $P_T^\prime$ one can read off the enhancement loci. They are given by the following complete intersections in $\mathcal{B}_6$:
\[ C_{\mathbf{6}} = V \left(w, a_{1,0} \right), \quad  C_{\mathbf{4}} = V \left( w, a_{4,2} \left(a_{4,2} + a_{1,0} a_{3,2} \right) - a_{1,0}^2 a_{6,4} \right) \, . \label{C46SU4} \]

\paragraph{Fibre Structure over $\mathbf{C_{\mathbf{6}}}$}
Over generic points of the matter curve $C_{\mathbf{6}} = V(w, a_{1,0})$ the resolution divisors behave as
\begin{align}
\begin{split}
E_0 |_{C_\mathbf{6}} &= \mathbb{P}_{0}^1 \left( \mathbf{6} \right) \, , \qquad \qquad 
E_1 |_{C_\mathbf{6}} = \mathbb{P}_{13}^1 \left( \mathbf{6} \right) \, , \\
E_2 |_{C_\mathbf{6}} &= \mathbb{P}_{2}^1 \left( \mathbf{6} \right) \, , \qquad \qquad 
E_3 |_{C_\mathbf{6}} = \mathbb{P}_{13}^1 \left( \mathbf{6} \right) + \mathbb{P}_{3A}^1 \left( \mathbf{6} \right) \, , \\
\end{split}
\end{align}
where the $\mathbb{P}^1$-fibrations are given by
\begin{subequations}
\begin{align}
\mathbb{P}^1_0 \left( \mathbf{6} \right) &= V \left( a_{1,0}, e_0, x^3 e_1 e_2^2 - y^2 e_3 \right) \, , \\
\mathbb{P}^1_{13} \left( \mathbf{6} \right) &= V \left(a_{1,0}, e_1, e_3 \right) \, , \\
\mathbb{P}^1_2 \left( \mathbf{6} \right) &= V \left( a_{1,0}, e_2, e_0^4 z^6 e_1^2 e_3 a_{6,4} - e_0^2 y z^3 e_1 e_3 a_{3,2} + e_0^2 x z^4 e_1 a_{4,2} - y^2 e_3 \right) \, , \\
\mathbb{P}^1_{3A} \left( \mathbf{6} \right) &= V \left( a_{1,0}, e_3, e_0^2 z^4 a_{4,2} + e_0 x z^2 e_2 a_{2,1} + x^2 e_2^2 \right) \, .
\end{align}
\end{subequations}
The resulting intersection numbers in the fibre with the Cartan divisors are as follows:
\begin{align}
\begin{tabular}{c@{\hskip 20pt}cccc}
\toprule
& $\mathbb{P}^1_0 \left( \mathbf{6} \right)$ & $\mathbb{P}^1_{13} \left( \mathbf{6} \right)$ & $\mathbb{P}^1_2 \left( \mathbf{6} \right)$ & $\mathbb{P}^1_{3A} \left( \mathbf{6} \right)$ \\
\midrule
$E_0$ & -2 & 1  & $\cdot$  & $\cdot$ \\
$E_1$ & 1  & -2 & 1  & 2 \\
$E_2$ & $\cdot$ & 1  & -2 & $\cdot$ \\
$E_3$ & 1  & $\cdot$ & 1  & -2 \\
\bottomrule
\end{tabular}
\end{align}
Consequently, the weight vector $\vect{\beta}$ associated with $\mathbb{P}^1_{3A} ( \mathbf{6} )$ satisfies $\vect{\beta} ( \mathbb{P}^1_{3A} ( \mathbf{6} ) ) = ( 2, 0, -2 )$.\footnote{Note that for this matter surface there are actually two $\mathbb P^1$'s over every point of the $\mathbf 6$-curve of the base. Since these $\mathbf P^1$'s are exchanged around the points $w=a_{1,0}=a_{2,1}^2-4\,a_{4,2}=0$ they are indistinguishable and  become identified. This is also the reason why we put a one-half in front of $\mathbb{P}^1_{3A} ( \mathbf{6} )$ to define the `right' matter surface. }
The matter surface associated with the $\mathbf{6}$ weight $( 1, 0, -1 )$ is therefore $\frac{1}{2} \mathbb{P}^1_{3A} ( \mathbf{6} )$, defined as an element in $\mathrm{CH}^2(\hat{Y}_4) \otimes \mathbb Q$.

\paragraph{Fibre Structure over $\mathbf{C_{\mathbf{4}}}$}
Over $C_{\mathbf{4}} = V(Q, w)$ with $Q = a_{4,2} [ a_{4,2} + a_{1,0,} a_{3,2} ] - a_{1,0}^2 a_{6,4}$, the fibral structure is 
\begin{align}
\begin{split}
E_0 |_{C_\mathbf{4}} &= \mathbb{P}_{0}^1 \left( \mathbf{4} \right) \, , \qquad \qquad \qquad \qquad \quad 
E_1 |_{C_\mathbf{4}} = \mathbb{P}_{1}^1 \left( \mathbf{4} \right) \, , \\
E_2 |_{C_\mathbf{4}} &= \mathbb{P}_{2B}^1 \left( \mathbf{4} \right) + \mathbb{P}_{2C}^1 \left( \mathbf{4} \right) \, , \qquad \qquad 
E_3 |_{C_\mathbf{4}} = \mathbb{P}_{3}^1 \left( \mathbf{4} \right) \, , \\
\end{split}
\end{align}
where the $\mathbb{P}^1$-fibrations are given by
\begin{subequations}
\begin{align}
\mathbb{P}^1_0 \left( \mathbf{4} \right) &= V \left( e_0, Q, x^3 e_1 e_2^2 - x y z a_{1,0} - y^2 e_3 \right) \, , \\
\mathbb{P}^1_1 \left( \mathbf{4} \right) &= V \left(e_1, Q, x z a_{1,0} + y e_3, y e_3 a_{3,2} a_{4,2} - x z a_{4,2}^2 - y e_3 a_{1,0} a_{6,4} 
      \right) \, , \\
\mathbb{P}^1_{2B} \left( \mathbf{4} \right) &= V \left(e_2, Q, e_0^2 z^3 e_1 a_{4,2} - y a_{1,0}, e_0^2 z^3 e_1 a_{1,0} a_{6,4} - y a_{1,0} a_{3,2} - y a_{4,2}, \right. \\
  & \hspace{16em} \left. e_0^4 z^6 e_1^2 a_{6,4} - e_0^2 y z^3 e_1 a_{3,2} - y^2 \right) \, , \nonumber \\
\mathbb{P}^1_{2C} \left( \mathbf{4} \right) &= V \left(e_2, Q, e_0^2 z^3 e_1 e_3 a_{4,2} a_{6,4} - y e_3 a_{3,2} a_{4,2} + x z a_{4,2}^2 + y e_3 a_{1,0} a_{6,4}, \right. \nonumber \\
& \qquad \quad e_0^2 z^3 e_1 e_3 a_{1,0} a_{6,4} + x z a_{1,0} a_{4,2} + y e_3 a_{4,2}, \\
& \qquad \quad e_0^2 z^3 e_1 e_3 a_{1,0} a_{3,2} + e_0^2 z^3 e_1 e_3 a_{4,2} + x z a_{1,0}^2 + y e_3 a_{1,0} \, , \nonumber \\
& \hspace{6em} \left. e_0^4 z^6 e_1^2 e_3 a_{6,4} - e_0^2 y z^3 e_1 e_3 a_{3,2} + e_0^2 x z^4 e_1 a_{4,2} - x y z a_{1,0} - y^2 e_3 \right) \, \nonumber , \\
\mathbb{P}^1_3 \left( \mathbf{4} \right) &= V \left( e_3, Q, e_0^2 z^4 e_1 a_{4,2} + e_0 x z^2 e_1 e_2 a_{2,1} + x^2 e_1 e_2^2 - y z a_{1,0}, \right. \nonumber \\
& \left. e_0^2 z^4 e_1 a_{1,0}^2 a64 + e_0 x z^2 e_1 e_2 a_{1,0} a_{2,1} a_{3,2} +  e_0 x z^2 e_1 e_2 a_{2,1} a_{4,2} + x^2 e_1 e_2^2 a_{1,0} a_{3,2} \right. \\
& \hspace{15.5em} \left. + x^2 e_1 e_2^2 a_{4,2} - y z a_{1,0}^2 a_{3,2} - y z a_{1,0} a_{4,2} \right) \, . \nonumber
\end{align}
\end{subequations}
The fibral intersection numbers with the divisors $E_i$ are
\begin{align}
\begin{tabular}{c@{\hskip 20pt}ccccc}
\toprule
& $\mathbb{P}^1_0 \left( \mathbf{4} \right)$ & $\mathbb{P}^1_{1} \left( \mathbf{4} \right)$ & $\mathbb{P}^1_{2B} \left( \mathbf{4} \right)$ & $\mathbb{P}^1_{2C} \left( \mathbf{4} \right)$ & $\mathbb{P}^1_{3} \left( \mathbf{4} \right)$ \\
\midrule
$E_0$ & -2 & 1 & $\cdot$ & $\cdot$ & 1 \\
$E_1$ & 1 & -2 & $\cdot$ & 1 & $\cdot$ \\
$E_2$ & $\cdot$ & 1 & -1 & -1 & 1 \\
$E_3$ & 1 & $\cdot$ & 1 & $\cdot$ & -2 \\
\bottomrule
\end{tabular} 
\end{align}
The resulting weight vectors associated with the split surfaces,
\[ \vect{\beta} \left( \mathbb{P}^1_{2B} \left( \mathbf{4} \right) \right) = \left( 0, -1, 1 \right), \qquad \vect{\beta} \left( \mathbb{P}^1_{2C} \left( \mathbf{4} \right) \right) = \left( 1, -1, 0 \right) \,, \]
identify the latter as matter surfaces for the $\mathbf{4}$ and $\mathbf{\overline{4}}$ representations, respectively.

\subsection{Proof of Fluxlessness} \label{subsec:ProofSU(4)}

To verify explicitly that the matter surface fluxes $A(\mathbf{4})$ and $A(\mathbf{6})$ are trivial in the Chow ring, we use the fibral structure discussed in \cref{subsec:FibreStructureSU4}. Over $C_{\mathbf{6}}$, the fibre of $E_3$ splits into $\mathbb{P}^1_{13} ( \mathbf{6} ) + \mathbb{P}^1_{3A} ( \mathbf{6} )$, and the weight vector associated with $S^1(\mathbf{6}) = \frac{1}{2} \mathbb{P}^1_{3A} ( \mathbf{6} )$ is $ \boldsymbol{\beta} \left( S^1_{\mathbf{6}} \right) = \left( 1,0,-1 \right)$. The associated matter surface flux is therefore indeed trivial in $\mathbb{Q} \otimes \mathrm{CH}^2(\hat{Y}_4)$ because
\begin{align}
\begin{split}
{A}\left( \mathbf{6} \right) &= \frac{1}{2} \mathbb{P}^1_{3A} \left( \mathbf{6} \right) + \mathfrak{\beta}^T({\mathbf{6}}) \, C^{-1} \cdot \left( \begin{array}{c} \mathbb{P}^1_{13} \left( \mathbf{6} \right) \\ \mathbb{P}^1_2 \left( \mathbf{6} \right) \\ \mathbb{P}^1_{13} \left( \mathbf{6} \right) + \mathbb{P}^1_{3A} \left( \mathbf{6} \right) \end{array} \right) \\
&= \frac{1}{2} \left( \mathbb{P}^1_{3A} \left( \mathbf{6} \right) - \left( - \mathbb{P}^1_{13} \left( \mathbf{6} \right) + \mathbb{P}^1_{13} \left( \mathbf{6} \right) + \mathbb{P}^1_{3A} \left( \mathbf{6} \right) \right)    \right)= 0 \, .
\end{split}
\end{align}
Here $\beta^T$ denotes the weights of the $\mathbf{6}$ representation. Over $C_\mathbf{4}$ it is $E_2$ which splits into two surfaces with weight-vectors
\[ \vect{\beta} \left( \mathbb{P}^1_{2B} \left( \mathbf{4} \right) \right) = \left( 0, -1, 1 \right), \qquad \vect{\beta} \left( \mathbb{P}^1_{2C} \left( \mathbf{4} \right) \right) = \left( 1, -1, 0 \right) \, . \]
We can for instance start with $S^1_{\mathbf{4}} = \mathbb{P}^1_{2B} ( \mathbf{4} )$ and deduced the gauge invariant flux
\[ {A} \left( \mathbf{4} \right)  = \frac{1}{4} \left( \mathbb{P}^1_{3} \left( \mathbf{4} \right) - \mathbb{P}^1_{1} \left( \mathbf{4} \right) \right) + \frac{1}{2} \left( \mathbb{P}^1_{2B} \left( \mathbf{4} \right) - \mathbb{P}^1_{2C} \left( \mathbf{4} \right) \right) \, . \]
Recall that $Q = a_{4,2} [ a_{4,2} + a_{1,0,} a_{3,2} ] - a_{1,0}^2 a_{6,4}$ denotes the polynomial which cuts out $C_{\mathbf{4}}$ in \cref{C46SU4}. Thereby, we can write down the following identities:
\[ \mathbb{P}^1_1 \left( \mathbf{4} \right) = V \left( P_T, Q, e_1 \right) \,, \quad
\mathbb{P}^1_3 \left( \mathbf{4} \right) = V \left( P_T, Q, e_3 \right) \,, \quad 
\mathbb{P}^1_{2C} \left( \mathbf{4} \right) = V \left( P_T, Q, e_2, \right) - \mathbb{P}^1_{2B} \left( \mathbf{4} \right) \,. \]
In addition, we have $\mathbb{P}^1_{2B} ( \mathbf{4} ) = V ( P_T, e_2, e_0^2 z^3 e_1 a_{42} - y a_{1,0} ) - V ( P_T, e_2, e_3 )$. By use of these identifies we can express the matter surface $A( \mathbf{4} )$ as restriction from $\hat{Y}_5$ to $\hat{Y}_4$,
\begin{align}\label{eq:su4-flux-generator-A4}
\begin{split}
A \left( \mathbf{4} \right) &= \left( - \cE_2 \cdot \cE_3  - 2 \cE_1 \cdot \overline{\mathcal{K}}_{\mathcal{B}_6} - 3  \cE_2 \cdot \overline{\mathcal{K}}_{\mathcal{B}_6} + 2 \cE_3 \cdot  \overline{\mathcal{K}}_{\mathcal{B}_6} \right. \\
& \hspace{8em} \left. + \cE_1 \cdot \mathcal{W} + 2 \cE_2 \cdot \mathcal{W} - \cE_3 \cdot \mathcal{W} + \cE_2 \cdot \mathcal{Y} \right) |_{\hat{Y}_4} \,.
\end{split}
\end{align}
We are now free to use the Chow relations on $\hat{Y}_5$ to manipulate this expression without changing the Chow class of $A(\mathbf{4})$ on $\hat{Y}_4$. By use of the  linear relations \eqref{eq:linear_relations-su4} we thus rewrite $A\left( \mathbf{4} \right)$ as
\begin{equation}
A \left( \mathbf{4} \right) =  \left(  \cE_0 \cdot (\cE_1 - \cE_3) - \mathcal{X} \cdot \cE_1  \right) |_{\hat{Y}_4} \,.
\end{equation}
By applying the linear relations \eqref{eq:linear_relations-su4} of the ambient space once more to rewrite $\cE_1 - \cE_3 = 2\,\mathcal{Y}-3\,\mathcal{X}-2\,\cE_2$, we find that $A ( \mathbf{4}) = 0 \in \mathrm{CH}^2(\hat{Y}_4)$. This is a consequence of the Stanley-Reisner ideal together with the fact that the cycles $V ( e_0,x )$, $V ( e_0,y )$, $V ( x,z )$, $V ( e_1,x )$ and $V ( y,z )$ are empty once restricted to the hypersurface $\hat{Y}_4$.

\chapter{Computing Sheaf Cohomologies on Toric Varieties with \texttt{GAP}} \label{chapter:MathDetailsSheafCohomologies}
In this chapter we describe an algorithm which computes sheaf cohomologies of coherent sheaves on normal toric varieties $X_\Sigma$ over $\mathbb{Q}$ which are smooth and complete. Our algorithm rests on the \emph{cohomCalg}-algorithm, which was proposed in \cite{Blumenhagen:2010pv}, subsequently proven in \cite{Rahn:2010fm, 2011JMP....52c3506J}, implemented in \cite{cohomCalg:Implementation, KoszulExtensionManual} and finally applied to \emph{string theory} in \cite{Blumenhagen:2011xn, Blumenhagen:2010ed}. We will briefly revise this algorithm in \cref{subsec:CohomCalgAlgorithm}. In \cref{subsec:VanishingSets} we apply this algorithm to find vanishing sets for line bundle cohomology. The so-obtained vanishing sets improve the semigroup $\mathbb{\mathcal{K}}^{\mathrm{sat}}$ introduced in \cite{Maclagan03multigradedcastelnuovo-mumford}.

In \cref{sec:VanishingTheorems} and \cref{sec:ComputingGlobalBivariateExt} we follow \cite{1998math......7170S, Oberwolfach} \footnote{See  \cite{2000math......1159E} for more background information.} except that we rest our analysis on the vanishing sets obtained from \emph{cohomCalg}, which we utilise to establish degeneracy of spectral sequences. Eventually this leads to \cref{mytheorem}, which is \emph{the} central building block in our computations of global sheaf extensions. In \cref{sec:ComputeCohomologies} we apply this theorem to compute sheaf cohomologies of coherent sheaves on $X_\Sigma$. The resulting algorithm requires to identify ample divisors in $X_\Sigma$. This we explain in \cref{sec:IdentifyAmpleDivisors}. All algorithms presented in this article have been implemented in the \texttt{gap}-package \cite{SheafCohomologyOnToricVarieties}, which in turn rests on the packages \cite{CAPCategoryOfProjectiveGradedModules, CAPPresentationCategory, PresentationsByProjectiveGradedModules, TruncationsOfPresentationsByProjectiveGradedModules}. All of these have been written in the language of \emph{categorical programming (CAP)} as introduced in \cite{PosurDoktor, GutscheDoktor, CAP}. Together they enlarge the functionality of the \emph{homalg\_project} \cite{homalg}.

\section{Vanishing Sets from \emph{cohomCalg}} \label{sec:ImprovedVanishingSets}

\subsection{Revision: Line Bundle Cohomology from \emph{cohomCalg}} \label{subsec:CohomCalgAlgorithm}

Before we start, let us recall that the Stanley-Reisner ideal $I_{\mathrm{SR}}$ and the irrelevant ideal $B_\Sigma$ of a toric variety $X_\Sigma$ must not be confused. We explained their definition in \cref{subsec:TowardsToricVarieties}. For the discussion on the \emph{cohomCalg} algorithm, the Stanley-Reisner ideal will be of ample importance.

\subsubsection{The \emph{cohomCalg}-Algorithm}
\begin{enumerate}
 \item Combinatorics of the Stanley-Reisner ideal $I_{\mathrm{SR}}$: \\
      Consider the Stanley-Reisner ideal $I_{\mathrm{SR}} ( X_\Sigma ) = \langle S_1, \dots, S_N \rangle$, $k \in \{ 1, 2, \dots, N \}$ and let $A = \{ \alpha_1, \dots, \alpha_k \} \subseteq \{ 1, \dots, N \}$. By $Q_A^k$ we denote the set of all coordinates that appear in $\{ S_{\alpha_1}, \dots , S_{\alpha_k} \}$. These sets are of prime interest. We analyse their structure by introducing the so-called \emph{c-degree} $c ( Q_A^k ) := | Q_A^k | - k$. 
      
      Note that $A = \{ \alpha_1, \dots, \alpha_k \}$, $B = \{ \beta_1, \dots, \beta_k \}$ with $A,B \subseteq \{ 1, \dots, N \}$ and $A \neq B$ can lead to $Q_A^k = Q_B^k$ and hence share the same $c$-degree. That said, let $c^i ( Q_A^k )$ be the number of all subsets $B \subseteq \{ 1, \dots, N \}$ which satisfy
      \[ \left| B \right| = k, \qquad Q_A^k = Q_{B}^k, \qquad i = c ( Q_B^k ) \, . \]
      These integers are the dimensions of $\mathbb{Q}$-vector spaces $\mathcal{C}^i \left( Q_A^k \right)$, which form the complex
      \footnote{We do not give further details of this complex here. See \cite{Rahn:2010fm, 2011JMP....52c3506J} for details.}
      \[ \dots \to 0 \to \mathcal{C}^0 ( Q_A^k ) \to \mathcal{C}^1 ( Q_A^k ) \to \dots \to \mathcal{C}^d ( Q_A^k ) \to \dots \, . \]
      The dimension of the homology of this complex at position $i$ is denoted by $h^i ( Q_A^k )$. \emph{cohomCalg} computes $h^i ( Q_A^k )$ for all $Q_A^k$.
\item Line bundle cohomology dimension $h^i ( X_\Sigma, \mathcal{O}_{X_\Sigma} ( \alpha ) )$: \\
     Pick $\alpha \in \mathrm{Cl} ( X_\Sigma )$, $0 \leq i \leq \mathrm{dim} ( X_\Sigma )$. All $Q_A^k = \{ y_1, y_2, \dots, y_k \}$ with $h^i ( Q_A^k ) \neq 0$ contribute to $h^i ( X_\Sigma, \mathcal{O}_{X_\Sigma} ( \alpha ) )$ \cite{Blumenhagen:2010pv,Rahn:2010fm, 2011JMP....52c3506J}. The sum of all contributions gives the final result. A set $Q_A^k$ with $h^i ( Q_A^k ) \neq 0$ contributes to $h^i ( X_\Sigma, \mathcal{O}_{X_\Sigma} ( \alpha ) )$ as follows: \\
     Let $Q_A^k = \{ y_1, \dots, y_k \}$ and denote the homogeneous coordinates not contained in $Q_A^k$ by $\{ x_1, \dots, x_m \}$. The contribution of $Q_A^k$ is given by multiplying $h^i ( Q_A^k )$ with the number of rationoms that are of degree $\alpha$ and can be written as 
     \[ \frac{T \left( x_1 \dots, x_n \right)}{y_1 \cdot y_2 \cdot \ldots \cdot y_k \cdot W \left( y_1, \dots, y_k \right)} \]
     for suitable monomials $T$ and $W$.
\end{enumerate}

\subsubsection{Combinatorics Data of $\mathbf{dP_1}$}

We exemplify the \emph{cohomCalg} algorithm by computing the cohomologies of $\mathcal{O}_{dP_1} ( 5, -2 )$ on a $dP_1$-surface. To this end, note that a $dP_1$-surface can be described as toric variety with Cox ring $S ( dP_1 ) = \mathbb{Q} [ x_1, x_2, x_3, x_4 ]$ which is graded with respect to $\mathrm{Cl} ( dP_1 ) = \mathbb{Z}^2$ by
\[ \mathrm{deg} ( x_1 ) = \mathrm{deg} ( x_2 ) = ( 1,0 ) \, , \qquad \mathrm{deg} ( x_3 ) = ( 1,1 ) \, , \qquad \mathrm{deg} ( x_4 ) = ( 0, 1 ) \, , \]
and has Stanley-Reisner ideal $I_{\mathrm{SR}} ( dP_1 ) = \langle x_1 x_2, x_3 x_4 \rangle$. We label the generators of $I_{\Sigma}$ by $S_1 = x_1 x_2$ and $S_2 = x_3 x_4$. The set $\{ S_1 \}$ consisting of the first generator only -- hence denoted by $\{ 1 \}$ below -- contains the homogeneous variables $x_1$ and $x_2$. So we write $Q^1_{\{ 1 \}} = \{ x_1, x_2 \}$. Repeating this process for all other subsets of $\{ S_1, S_2 \}$ leads to
\[ Q_{\emptyset}^0 = \emptyset, \quad Q_{\left\{ 1 \right\}}^1 = \left\{ x_1, x_2 \right\}, \quad Q_{\left\{ 2 \right\}}^1 = \left\{ x_3, x_4 \right\}, \quad
Q_{\left\{ 1,2 \right\}}^2 = \left\{ x_1, x_2, x_3, x_4 \right\} \, . \]
\emph{cohomCalg} analyses the combinatorics of these sets. This leads to the $c$-degrees, $c^i ( Q^k_A )$ and $h^i ( Q^k_A )$, which we introduced above. We state their values in \cref{table-N36}.
\begin{table}[tb]
\centering
\begin{tabular}{c@{\hskip 20pt}ccc}
\toprule
sets & c-degree & $c^i \left( Q_A^k \right)$ & $h^i \left( Q_A^k \right)$ \\
\midrule
$Q_{\emptyset}^0 = \emptyset$ & 0 & $\left( 1,0,0 \right)$ & $\left( 1,0,0 \right)$ \\
$Q_{\left\{ 1 \right\}}^1 = \left\{ x_1, x_2 \right\}$ & 1 & $\left( 0,1,0 \right)$ & $\left( 0,1,0 \right)$ \\
$Q_{\left\{ 2 \right\}}^1 = \left\{ x_3, x_4 \right\}$ & 1 & $\left( 0,1,0 \right)$ & $\left( 0,1,0 \right)$ \\
$Q_{\left\{ 1,2 \right\}}^2 = \left\{ x_1, x_2, x_3, x_4 \right\}$ & 2 & $\left( 0,0,1 \right)$ & $\left( 0,0,1 \right)$ \\
\bottomrule
\end{tabular}
\caption{\emph{cohomCalg}'s combinatorics data of a $dP_1$-surface.}
\label{table-N36}
\end{table}

\begin{itemize}
 \item $h^0$: \\
      Only $Q_\emptyset^0$ has $h^0 ( Q_\emptyset^0 ) \neq 0$. Consequently, it holds
      \[
         h^0 \left( dP_1, \mathcal{O}_{dP_1} \left( 5,-2 \right) \right) = h^0 \left( Q_{\emptyset}^0 \right) \cdot \left| \left\{ W \in S ( dP_1 ) \; | \; \mathrm{deg} \left( W \right) = \left( 5, -2 \right) \right\} \right| \, .
      \]
      The grading of $S ( dP_1 )$ now leads us to look for $a_i \in \mathbb{Z}_{\geq 0}$ such that
      \[ \left( 5,-2 \right) = a_1 \left( 1,0 \right) + a_2 \left( 1,0 \right) + a_3 \left( 1,1 \right) + a_4 \left( 0,1 \right) \, . \]
      As there are no such solutions, it follows $h^0 ( dP_1, \mathcal{O}_{dP_1} ( 5,-2 ) ) = 0$.
 \item $h^1$: \\
      $Q_{{1}}^1$ and $Q_{{2}}^1$ both satisfy $h^1 ( Q_A^k ) \neq 0$ and so they both contribute to $h^1 ( dP_1, \mathcal{O}_{dP_1} ( 5,-2 ) )$. We discuss them in turns.
      \begin{itemize}
       \item $Q_{(2)}^{1}$: \\
            In this case the contribution is given by
            \[ \underbrace{h^1 \left( Q_{{1}}^1 \right)}_{=1} \cdot \underbrace{\left| \left. \left\{ R \left( x_i \right) = \frac{W \left( x_1, x_2 \right)}{x_3 x_4 T \left( x_3, x_4 \right)} \; \right| \; \mathrm{deg} \left( R \left( x_i \right) \right) = \left( 5, -2 \right) \right\} \right|}_{=R}\]
            where $T, W$ must be suitable monomials. Hence, the rationoms in question are of the form $R \left( x_i \right) = x_1^{a_1} x_2^{a_2} x_3^{-1-a_3} x_4^{-1-a_4}$ (with $a_i \in \mathbb{Z}_{\geq 0}$) and it holds
            \begin{align}
            \begin{split}
            \mathrm{deg} \left( R \left( x_i \right) \right) &= a_1 \mathrm{deg} \left( x_1 \right) + a_2 \mathrm{deg}\left( x_2 \right) + a_3 \left( - \mathrm{deg} \left( x_3 \right) \right) + a_4 \left( - \mathrm{deg} \left( x_4 \right) \right) \\
            & \qquad - \left( \mathrm{deg} \left( x_3 \right) + \mathrm{deg} \left( x_4 \right) \right) \, .
            \end{split}
            \end{align}
            Therefore, our task is to find the $a_i \in \mathbb{Z}_{\geq 0}$ with
            \[ \left( 6,0 \right) = a_1 \left( 1, 0 \right) + a_2 \left( 1,0 \right) + a_3 \left( -1, -1 \right) + a_4 \left( 0, -1 \right) \, . \]
            There are 7 such solutions and they correspond to the rationoms
            \[ \left\{ \frac{x_1^6}{x_3 x_4}, \frac{x_2 x_1^5}{x_3 x_4}, \frac{x_2^2 x_1^4}{x_3 x_4}, \frac{x_2^3 x_1^3}{x_3 x_4}, \frac{x_2^4 x_1^2}{x_3 x_4}, \frac{x_2^5 x_1}{x_3 x_4}, \frac{x_2^6}{x_3 x_4} \right\} \, . \]
            Consequently, $Q_{{(2)}}^1$ contributes a $7$ to $h^1 ( dP_1, \mathcal{O}_{dP_1} ( 5, -2 ) )$.
       \item $Q_{(1)}^{1}$: \\
            The logic is identical and eventually requires to find $a_i \in \mathbb{Z}_{\geq 0}$ such that
            \[ \left( 5,-2 \right) = \left( -2,0 \right) + a_1 \left( -1, 0 \right) + a_2 \left( -1,0 \right) + a_3 \left( 1,1 \right) + a_4 \left( 0,1 \right) \, . \]
            There are no such solutions, so $Q_{{(1)}}^1$ contributes a $0$ to $h^1 ( dP_1, \mathcal{O}_{dP_1} ( 5,-2 ) )$.
      \end{itemize}
      Consequently, $h^1 ( dP_1, \mathcal{O}_{dP_1} ( 5,-2 ) ) = h^1 ( Q_{{1}}^1 ) \cdot 0 + h^1 ( Q_{{2}}^1 ) \cdot 7 = 7$.
 \item $h^2$: \\
      Only $h^2 ( Q_{{1,2}}^2 ) \neq 0$. Hence, we focus on $Q_{{1,2}}^2$ and are lead to count the number of solutions $a_i \in \mathbb{Z}_{\geq 0}$ with
      $( 5,-2 ) \notin (-3, -2 ) + a_1 ( -1,0 ) + a_2 ( -1,0 ) + a_3 ( -1,-1 ) + a_4 ( -1,0 )$. As there are no such solutions, we conclude $h^2 ( dP_1, \mathcal{O}_{dP_1} ( 5,-2 ) ) = 0$.
\end{itemize}

\subsection{Vanishing Sets for Line Bundle Cohomology from \emph{cohomCalg}} \label{subsec:VanishingSets}

\subsubsection{Algorithm}

\begin{itemize}
 \item Input:
      \begin{itemize}
       \item normal toric variety $X_\Sigma$ which is smooth and complete,
       \item integer $i \in \{ 0, 1, \dots, \mathrm{dim} ( X_\Sigma ) \}$.
      \end{itemize}
 \item Output: \\
      Vanishing set $V^i ( X_\Sigma ) = \{ D \in \mathrm{Cl} ( X_\Sigma ) \; | \; h^i ( X_\Sigma, \mathcal{O}_{X_\Sigma} ( D ) ) = 0 \}$.
 \item Processing:
      \begin{enumerate}
       \item Compute the Stanley-Reisner ideal $I_{\mathrm{SR}} ( X_\Sigma )$ and the Cox ring $S ( X_\Sigma )$.
       \item Send $S ( X_\Sigma )$, $I_\Sigma ( X_\Sigma )$ to \emph{cohomCalg}. This software iterates over all sets $Q_A^k$, computes $h^i ( Q_A^k )$ and returns a  
            list of these results.\footnote{In fact only those sets $Q_A^k$ are important for which at least one cohomology $h^k ( Q_A^k )$ is non-zero. Thus, \emph{cohomCalg} only returns those important sets $Q_A^k$ and their cohomologies $h^i ( Q_A^k )$.}
       \item Identify the list $\{ \mathrm{set}_1, \mathrm{set}_2, \dots \mathrm{set}_l \}$ consisting of all sets $Q_A^k$ for which $h^i ( Q_A^k ) \neq 0$.
       \item For all $m \in \left\{ 1, 2, \dots, l \right\}$ compute $( \mathcal{B}_m, \mathbf{o}_m) $ given as follows: \\
            Let $y_1, \dots, y_k$ be the variables in $\mathrm{set}_m$ and $x_{k+1}, \dots, x_n$ the remaining variables of $S ( X_\Sigma )$. There is no
            contribution to $h^i ( X_\Sigma, \mathcal{O}_{X_\Sigma} ( D ) )$ precisely if
            \[ D \notin \mathbf{o}_m + \mathrm{Span}_{\mathbb{Z}_{\geq 0}} \left( \mathcal{B}_m \right), \qquad \mathbf{o}_m := \sum_{i = 1}^{k}{\left( -\mathrm{deg} \left( y_i \right) \right)} \]
            and $\mathcal{B}_m := \{ - \mathrm{deg} ( y_1 ), \dots, - \mathrm{deg} ( y_k ), + \mathrm{deg} ( x_{k+1} ), \dots, \mathrm{deg} ( x_n ) \}$.
       \item Return $\{ ( \mathcal{B}_1, \mathbf{o}_1 ), ( \mathcal{B}_2, \mathbf{o}_2 ), \ldots, ( \mathcal{B}_l, \mathbf{o}_l ) \}$.
      \end{enumerate}
\end{itemize}

\subsubsection{Example: Vanishing Sets on $\mathbf{dP_1}$}

\paragraph{Vanishing Set for $\mathbf{h^0}$}

In \cref{subsec:CohomCalgAlgorithm} we gathered combinatorics data of $dP_1$ and studied an example computation of sheaf cohomology. From this discussion we learn
\[ h^0 \left( dP_1, \mathcal{O}_{dP_1} \left( D \right) \right) = h^0 \left( Q_\emptyset^0 \right) \cdot \underbrace{\left| \left\{ a_i \in \mathbb{Z}_{\geq 0} \; \left| \; D = a_1 \begin{psmallmatrix} 1 \\ 0 \end{psmallmatrix} + a_2 \begin{psmallmatrix} 1 \\ 0 \end{psmallmatrix} + a_3 \begin{psmallmatrix} 1 \\ 1 \end{psmallmatrix} + a_4 \begin{psmallmatrix} 0 \\ 1 \end{psmallmatrix} \right. \right\} \right|}_{S^0_\emptyset \left( D \right)} \, . \]
Thus, $h^0 ( dP_1, \mathcal{O}_{dP_1} ( D ) ) = 0$ if and only if $S^0_\emptyset ( D ) = 0$. This identifies the vanishing set for $h^0$.

\paragraph{Vanishing Set for $\mathbf{h^1}$}

To $h^1 ( dP_1, \mathcal{O}_{dP_1} ( D ) )$ the rationoms
\[ R_{(1)} = \frac{x_3^{a_3} x_4^{a_4}}{x_1^{a_1 + 1} x_2^{a_2 + 1}}, \qquad R_{(2)} = \frac{x_1^{a_1} x_2^{a_2}}{x_3^{a_3 + 1} x_4^{a_4 + 1}} \, . \]
contribute and it holds $h^1 ( dP_1, \mathcal{O}_{dP_1} ( D ) ) = h^1 ( Q_{\left\{1\right\}}^1 ) \cdot S^1_{\left\{1\right\}} ( D ) + h^1 ( Q_{\left\{2\right\}}^1 ) \cdot S^1_{\left\{2\right\}} ( D )$, where
\begin{align}
\begin{split}
S^1_{\left\{1\right\}} ( D ) &= \left| \left\{ a_i \in \mathbb{Z}_{\geq 0} \; \left| \; 
D = \begin{psmallmatrix} -2 \\ 0 \end{psmallmatrix}
+ a_1 \cdot \begin{psmallmatrix} -1 \\ 0 \end{psmallmatrix}
+ a_2 \cdot \begin{psmallmatrix} -1 \\ 0 \end{psmallmatrix}
+ a_3 \cdot \begin{psmallmatrix} 1 \\ 1 \end{psmallmatrix}
+ a_4 \cdot \begin{psmallmatrix} 0 \\ 1 \end{psmallmatrix}
\right. \right\} \right| \, , \\
S^1_{\left\{2\right\}} ( D ) &= \left| \left\{ a_i \in \mathbb{Z}_{\geq 0} \; \left| \; 
D = \begin{psmallmatrix} -1 \\ -2 \end{psmallmatrix}
+ a_1 \cdot \begin{psmallmatrix} 1 \\ 0 \end{psmallmatrix}
+ a_2 \cdot \begin{psmallmatrix} 1 \\ 0 \end{psmallmatrix}
+ a_3 \cdot \begin{psmallmatrix} -1 \\ -1 \end{psmallmatrix}
+ a_4 \cdot \begin{psmallmatrix} 0 \\ -1 \end{psmallmatrix}
\right. \right\} \right| \, .
\end{split}
\end{align}
Consequently, $h^1 \left( X_\Sigma, \mathcal{O}_{X_\Sigma} \left( \alpha_1, \alpha_2 \right) \right) = 0$ if and only if $0 = S^1_{\left\{1\right\}} \left( D \right) = S^1_{\left\{2\right\}} \left( D \right)$.

\paragraph{Vanishing Set for $\mathbf{h^2}$}

It holds $h^2 ( dP_1, \mathcal{O}_{dP_1} ( D ) ) = h^2 ( Q_{\{ 1,2 \}}^2 ) \cdot S^2_{\{ 1,2 \}} ( D )$ where
\[ S^2_{\left\{ 1,2 \right\}} \left( D \right) = \left| \left\{ a_i \in \mathbb{Z}_{\geq 0} \; \left| \; D = \begin{psmallmatrix} -3 \\ -2 \end{psmallmatrix} - a_1 \cdot  \begin{psmallmatrix} 1 \\ 0 \end{psmallmatrix} - a_2 \cdot  \begin{psmallmatrix} 1 \\ 0 \end{psmallmatrix} - a_3 \cdot \begin{psmallmatrix} 1 \\ 1 \end{psmallmatrix} - a_4 \cdot \begin{psmallmatrix} 0 \\ 1 \end{psmallmatrix} \right. \right\} \right| \, . \]
Consequently, $h^2 ( dP_1, \mathcal{O}_{dP_1} ( D ) ) = 0$ if and only if $S^2_{\{ 1,2 \}} ( D ) = 0$.

\begin{figure}[tbp]
\centering
\includegraphics{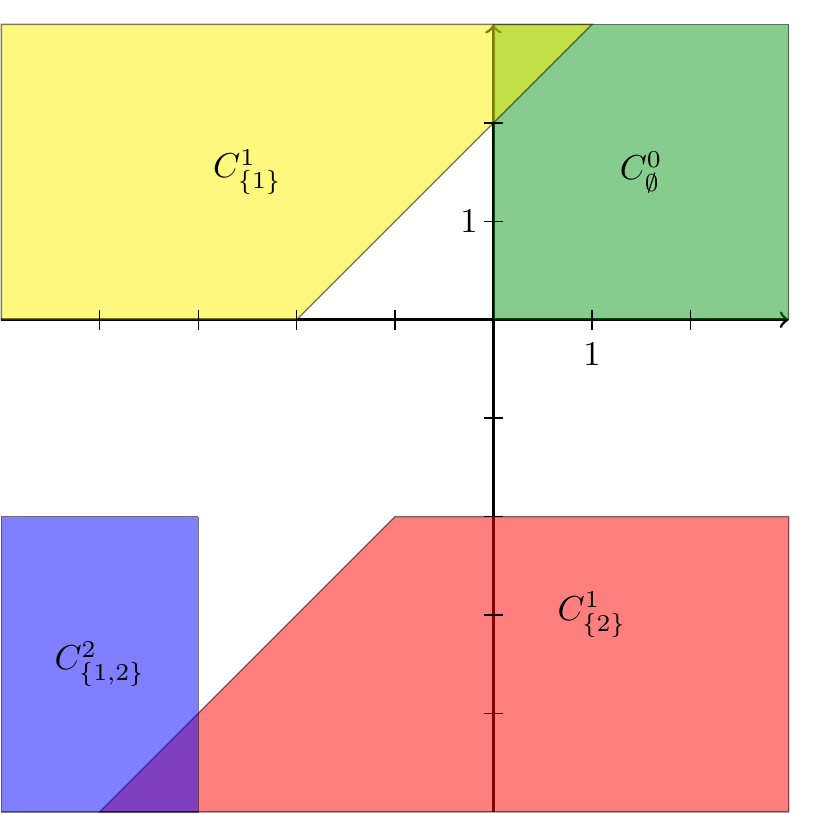}
\caption{The complements of $C^0_\emptyset$, $C_{\left\{1\right\}}^{1}$, $C_{\left\{2\right\}}^{1}$ and $C^2_{\left\{1,2\right\}}$ define the vanishing sets of $dP_1$.}
\label{figure-1x}
\end{figure}

\paragraph{Visualisation}
In this example, the affine cones defining $S^0_\emptyset$, $S^1_{\{ 1 \}}$, $S^1_{\{ 2 \}}$ and $S^2_{\{ 1,2 \}}$ are generated by Hilbert bases. Therefore, we can write:
\begin{itemize}
 \item $h^0 ( dP_1, \mathcal{O}_{dP_1} ( D ) ) = 0$ iff $D \notin C_\emptyset^0$ where $C^{0}_\emptyset = \mathrm{Cone} \left( \left( 1,0 \right), \left( 0,1 
      \right) \right)$.
 \item $h^1 ( dP_1, \mathcal{O}_{dP_1} ( D ) ) = 0$ iff $D \notin \{ C_{\{ 1 \}}^1 \cup C_{\{ 2 \}}^1 \}$ where
      \begin{align*}
        C^{1}_{\left\{1\right\}} = \left( -2, 0 \right) + \mathrm{Cone} \left( \left( 1,1 \right), \left( -1,0 \right) \right) \, , \quad
        C^{1}_{\left\{2\right\}} = \left( -1, -2 \right) + \mathrm{Cone} \left( \left( -1,-1 \right), \left( 1,0 \right) \right).
      \end{align*}
 \item $h^2 ( dP_1, \mathcal{O}_{dP_1} ( D ) ) = 0$ iff $D \notin C_{\{ 1,2 \}}^2 = ( -3, -2 ) + \mathrm{Cone}_{\mathbb{R}_{\geq 0}} ( ( -1, 0 ), ( 0,-1 ) )$.
\end{itemize}
We plot these affine cones in \cref{figure-1x}. The integral points in the complements of $C^{0}_\emptyset$, $C^{1}_{\{1\}} \cup C^{1}_{\{2\}}$ and $C^{2}_{\{ 1,2 \}}$ correspond to the line bundles on $dP_1$ for which $h^0$, $h^1$ and $h^2$ vanish, respectively. In general one cannot expect to find the vanishing sets defined as complements of cones generated by their Hilbert basis. In consequence, we perform the membership test in \texttt{gap} \cite{GAP4} by \emph{4ti2} \cite{4ti2}.

\subsubsection{Another Example}

\begin{figure}[tbp]
\centering
\includegraphics{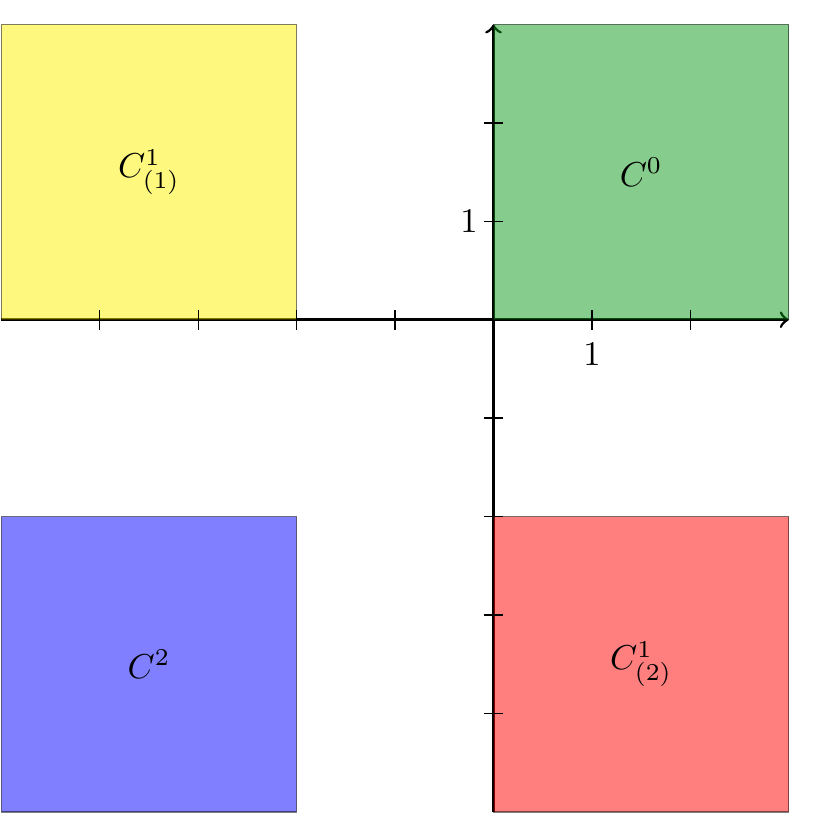}
\caption{The complements of $C^0$, $C_{(1)}^{1}$, $C_{(2)}^{1}$ and $C^2$ define the vanishing sets of $\mathbb{P}^1_{\mathbb{Q}} \times \mathbb{P}^1_{\mathbb{Q}}$.}
\label{figure-2}
\end{figure}

The following \texttt{gap}-code computes the vanishing sets of $\mathbb{P}^1_{\mathbb{Q}} \times \mathbb{P}^1_{\mathbb{Q}}$:
\begin{gapConsole}
!gapprompt@gap>& !gapinput@LoadPackage( "SheafCohomologyOnToricVarieties" );&
true
!gapprompt@gap>& !gapinput@P1 := ProjectiveSpace( 1 );&
<A projective toric variety of dimension 1>
!gapprompt@gap>& !gapinput@P1xP1 := P1*P1;&
<A projective toric variety of dimension 2 which is a product of 2 toric 
varieties>
!gapprompt@gap>& !gapinput@v := VanishingSets( P1xP1 );&
rec( 0 := <A non-full vanishing set in Z^2 for cohomological index 0>, 
     1 := <A non-full vanishing set in Z^2 for cohomological index 1>, 
     2 := <A non-full vanishing set in Z^2 for cohomological index 2> )
!gapprompt@gap>& !gapinput@Display( v.0 );&
A non-full vanishing set in Z^2 for cohomological index 0 formed from the 
points NOT contained in the following affine semigroup: 
 
A non-trivial affine cone-semigroup in Z^2
Offset: [ 0, 0 ]
Hilbert basis: [ [ 1, 0 ], [ 0, 1 ] ]
!gapprompt@gap>& !gapinput@Display( v.1 );&
A non-full vanishing set in Z^2 for cohomological index 1 formed from the 
points NOT contained in the following 2 affine semigroups: 
 
Affine semigroup 1: 
A non-trivial affine cone-semigroup in Z^2
Offset: [ 0, -2 ]
Hilbert basis: [ [ 1, 0 ], [ 0, -1 ] ]

Affine semigroup 2: 
A non-trivial affine cone-semigroup in Z^2
Offset: [ -2, 0 ]
Hilbert basis: [ [ -1, 0 ], [ 0, 1 ] ]
!gapprompt@gap>& !gapinput@Display( v.2 );&
A non-full vanishing set in Z^2 for cohomological index 2 formed from the 
points NOT contained in the following affine semigroup: 
 
A non-trivial affine cone-semigroup in Z^2
Offset: [ -2, -2 ]
Hilbert basis: [ [ -1, 0 ], [ 0, -1 ] ]
\end{gapConsole}
We picture the so-computed vanishing sets in \cref{figure-2}. On $\mathbb{P}^1_{\mathbb{Q}}$ we know $h^1 ( \mathbb{P}^1_{\mathbb{Q}}, \mathcal{O}_{\mathbb{P}^1_{\mathbb{Q}}} ( n ) ) = 0$ for $n \geq -1$. In \cref{figure-2} we see how this statement generalises to $\mathbb{P}^1_{\mathbb{Q}} \times \mathbb{P}^1_{\mathbb{Q}}$. Serre duality relates the vanishing sets $V^0 ( \mathbb{P}^1_{\mathbb{Q}} \times \mathbb{P}^1_{\mathbb{Q}} )$ and $V^2 ( \mathbb{P}^1_{\mathbb{Q}} \times \mathbb{P}^1_{\mathbb{Q}} )$. For smooth toric varieties, \texttt{gap} exploits this duality to provide faster computation.

\section{Vanishing Theorems from \emph{cohomCalg}} \label{sec:VanishingTheorems}

Finitely presented (f.p.) graded $S$-modules model coherent sheaves on $X_\Sigma$ via the sheafification functor \cite{cox2011toric}
\[ \widetilde{\phantom{m}} \colon S \mathrm{\textnormal{-}fpgrmod} \to \mathfrak{Coh} X_\Sigma \, . \label{redundant_description} \]
The software packages \cite{CAPCategoryOfProjectiveGradedModules, CAPPresentationCategory, PresentationsByProjectiveGradedModules, TruncationsOfPresentationsByProjectiveGradedModules, SheafCohomologyOnToricVarieties} are designed to use \fp graded $S$-modules as computer models for coherent sheaves on $X_\Sigma$. Thus, we phrase our findings in terms of \fp graded $S$-modules. 

Even for smooth $X_\Sigma$ with irrelevant ideal $B_\Sigma \subseteq S$, \cref{redundant_description} is in general not an equivalence of categories. Let $S \mathrm{\textnormal{-}fpgrmod}^0$ denote the thick subcategory of $S \mathrm{\textnormal{-}fpgrmod}$ consisting of \fp graded $S$-modules which are supported on $V ( B_\Sigma )$. Then, for smooth $X_\Sigma$, there is an equivalence of categories which establishes a possible parametrisation of $\mathfrak{Coh} ( X_\Sigma )$ by the Serre quotient \cite{2012arXiv1210.1425B, 2011arXiv1110.0323P} 
\[ S \mathrm{\textnormal{-}fpgrmod} / S \mathrm{\textnormal{-}fpgrmod}^0 \stackrel{\sim}{\rightarrow} \mathfrak{Coh} X_\Sigma \, . \] 
The \emph{cohomCalg}-algorithm is proven to work for smooth, complete toric varieties in \cite{Rahn:2010fm} and for simplicial, projective toric varieties in \cite{2011JMP....52c3506J}. Therefore, we can compute the vanishing sets $V^i ( X_\Sigma )$ for smooth, complete and simplicial, projective toric varieties. Along these lines we could now hope to obtain methods for sheaf cohomologies of coherent sheaves on all smooth, complete and simplicial projective toric varieties. Consequently, the remainder of this chapter will try to relate the defining data of a \fp graded $S$-module $M$ to the sheaf cohomologies of the corresponding coherent sheaf $\tilde{M}$. 
More explicitly, we will find that (truncation of) extensions of such modules encode these sheaf cohomologies.

As pointed out in \cite[example 6.6]{2012arXiv1212.4068B}, there exist simplicial toric varieties and coherent sheaves thereon, such that for these sheaves the sheaf cohomologies cannot be obtained from (truncations of) extensions of \fp graded $S$-modules. Hence, one has to be careful. In the main part of this thesis we are interested in coherent sheaves which correspond to line bundles on smooth, complete curves and are zero otherwise. For such sheaves $\mathcal{F}$ only $h^0 ( X_\Sigma, \mathcal{F} )$ and $h^1 ( X_\Sigma, \mathcal{F} )$ are non-zero, and those can also be obtained from (truncations of) module extension even for simplicial, projective toric varieties. In this sense the following results apply to smooth, complete and simplicial, projective toric varieties. However, as we wish to phrase our findings for all sheaf cohomologies, we restrict our attention in the remainder of this chapter to smooth, complete toric varieties $X_\Sigma$.

\subsection{\dots for Sheaf Cohomology} \label{subsec:VanishingTheoremForSheafCohomology}

\paragraph{Two Spectral Sequences For Hypercohomology}
The $p$-th right hyperderived functor of the global section functor $\Gamma ( X_\Sigma, - )$ is called the $p$-th hypercohomology functor (\cf \cite{weibel1995introduction}). In agreement with \cite{1998math......7170S} we denote it by $\mathbf{H}^p ( X_\Sigma, - )$. Let $\mathcal{E}^\bullet$ be a cochain complex of $\mathcal{O}_{X_\Sigma}$-modules and $\mathcal{H}^q ( \mathcal{E}^\bullet )$ its $q$-th cohomology, \ie $\mathcal{H}^q ( \mathcal{E}^\bullet )$ is a coherent sheaf. For this sheaf let $H^p ( X_\Sigma, \mathcal{H}^q ( \mathcal{E}^\bullet ) )$ denote its $p$-th sheaf cohomology. Then the following spectral sequence relates these sheaf cohomologies to the hypercohomology \cite{weibel1995introduction, griffiths2011principles}:
\[ ^{\prime \prime} E_2^{p,q} = H^p \left( X_\Sigma, \mathcal{H}^q \left( \mathcal{E}^\bullet \right) \right) \Rightarrow \mathbf{H}^{p+q} \left( X_\Sigma, \mathcal{E}^\bullet \right) \, . \]
Likewise we can form the complex of sheaf cohomologies $H^q ( X_\Sigma, \mathcal{E}^\bullet )$ and denote the cohomology of this complex at position $p$ by
$\mathfrak{H}^p ( H^q ( X_\Sigma, \mathcal{E}^\bullet ) )$. Then also the following spectral sequence computes the hypercohomology \cite{weibel1995introduction, griffiths2011principles}
\[ ^\prime E_2^{p,q} = \mathfrak{H}^p \left( H^q \left( X_\Sigma, \mathcal{E}^\bullet \right) \right) \Rightarrow \mathbf{H}^{p+q} \left( X_\Sigma, \mathcal{E}^\bullet \right) \, . \label{equ:E2Prime} \]

\paragraph{An Example}
Let us consider a \fp graded $S$ module $N$ with minimal free resolution
\[ 0 \leftarrow F_0 \left( N \right) \leftarrow F_{1} \left( N \right) \leftarrow \dots \leftarrow F_{L \left( N \right)} \left( N \right) \leftarrow 0 \, , \qquad F_i \left( N \right) = \bigoplus_{j = 1}^{b_i \left( N \right)}{S \left( - a_{i,j} \left( N \right) \right)} \, . \]
Since the sheafification functor is exact, we can form this a complex $\mathcal{E}^\bullet$ of coherent sheaves, which is given by
\[ \mathcal{E}^\bullet \colon 0 \leftarrow \tilde{F_{1} \left( N \right)} \leftarrow \tilde{F_{1} \left( N \right)} \leftarrow \dots \leftarrow \tilde{F_{L \left( N \right)} \left( N \right)} \to 0 \, . \]
Of this complex we wish to compute the hypercohomology $\mathbf{H}^{p+q} ( X_\Sigma, \mathcal{E}^\bullet )$. To this end, let us employ the spectral sequence $^{\prime \prime} E^{p,q}$. This means that we have to compute the cohomologies $\mathcal{H}^q ( \mathcal{E}^\bullet )$. But since $\mathcal{E}^\bullet$ is exact everywhere but at $q = 0$, the only non-trivial such cohomology is $\mathcal{H}^0 ( \mathcal{E}^\bullet ) \cong \tilde{N}$. Consequently, the sheet $^{\prime \prime} E_2^{p,q}$ takes the following form:
\[ \includegraphics[valign = c]{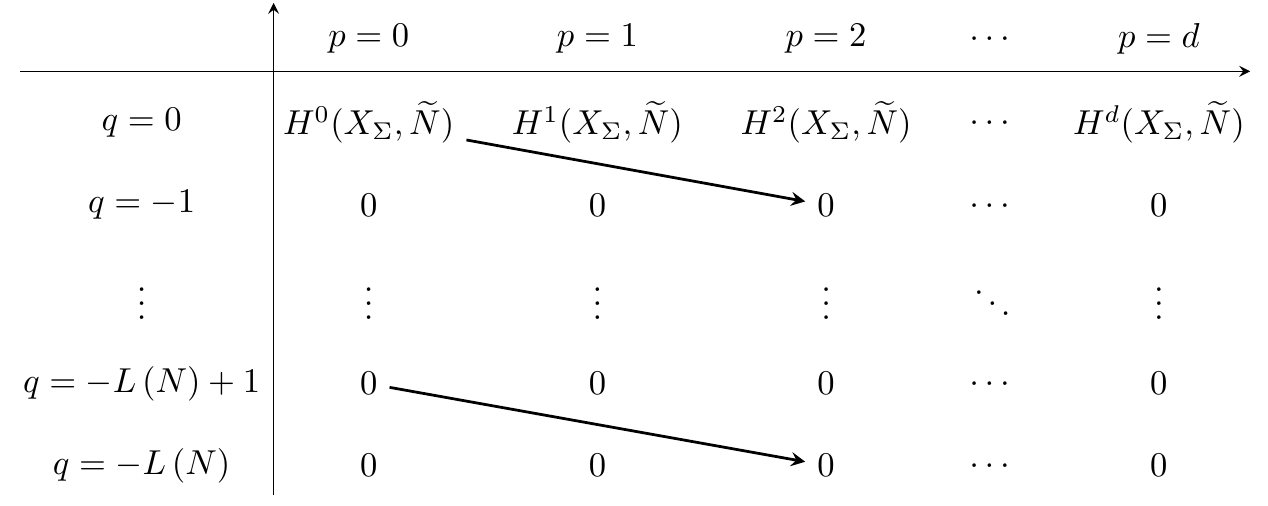} \]
From this it follows $\mathbf{H}^{p+q} ( X_\Sigma, \mathcal{E}^\bullet ) = H^{p+q} ( X_\Sigma, \tilde{N} )$, indeed connecting hypercohomology and sheaf cohomology. Since $^\prime E^{p,q}$ and $^{\prime \prime} E^{p,q}$ compute the same hypercohomology, also $^\prime E^{p,q}$ computes the sheaf cohomologies of $\tilde{N}$. Let $d = \mathrm{dim} ( X_\Sigma )$ and consider for $0 \leq q \leq d$ the cochain complex
\begin{align*}
0 \to H^q \left( X_\Sigma, \tilde{F_{L \left( N \right)}} \left( N \right) \right) \to H^q \left( X_\Sigma, \tilde{F_{L \left( N \right) + 1}} \left( N \right) \right) 
\to \dots \to H^q \left( X_\Sigma, \tilde{F_0} \left( N \right) \right) \to 0 \, .
\end{align*}
In agreement with \cref{equ:E2Prime}, let us abbreviate its cohomology at position $-L ( N ) \leq p \leq 0$ by $\mathfrak{H}_2^{p,q}$. Thus, $^\prime E_2^{p,q}$ is given by
\[ \includegraphics[valign = c]{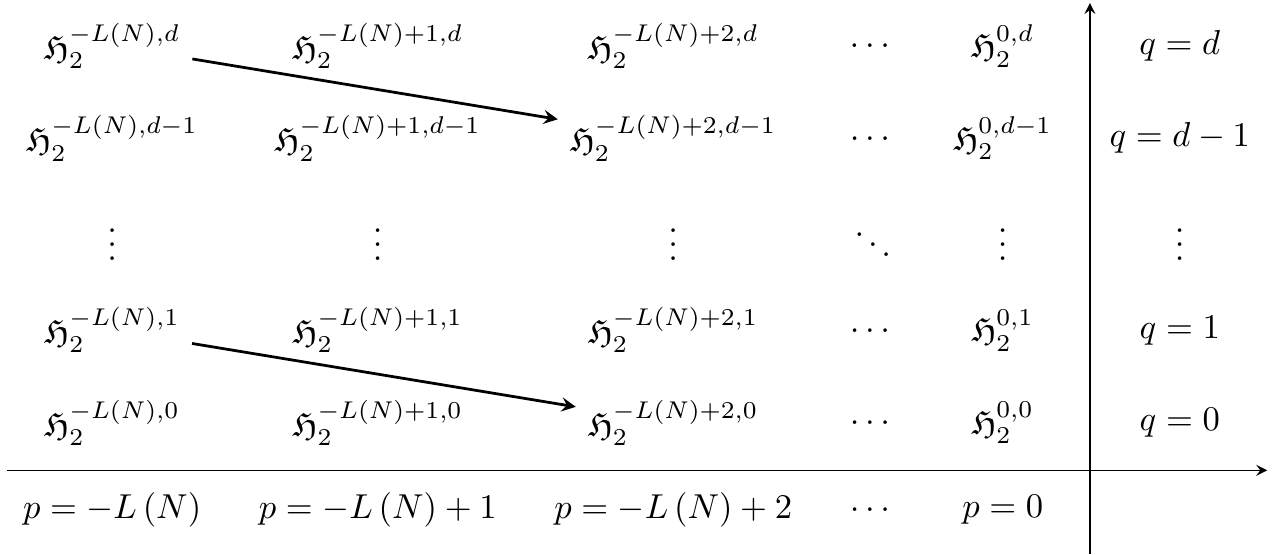} \]
From this we learn that for $0 \leq m \leq d$ it holds $\displaystyle H^m ( X_\Sigma, \tilde{N} ) \cong \bigoplus_{u = 0}^{\mathrm{min} \left\{L \left( N \right), d-m \right\}}{\mathfrak{h}_\infty^{-u,u+m}}$.

\begin{cor} \label{cor:VanishingTheoremForSheafCohomology}
Let $X_\Sigma$ be a smooth and complete normal toric variety of dimension $d$ with Cox ring $S$ and vanishing sets $V^i ( X_\Sigma ) = \{ D \in \mathrm{Cl} ( X_\Sigma ) \; , \; H^i ( X_\Sigma, \mathcal{O}_{X_\Sigma} ( D ) ) = 0 \}$. Consider a \fp graded $S$-module $N$ with minimal free resolution
\[ 0 \leftarrow F_0 \left( N \right) \leftarrow F_{1} \left( N \right) \leftarrow \dots \leftarrow F_{L \left( N \right)} \left( N \right) \leftarrow 0 \, , \qquad F_i \left( N \right) = \bigoplus_{j = 1}^{b_i \left( N \right)}{S \left( - a_{i,j} \left( N \right) \right)} \]
and let $v \in \mathrm{Cl} ( X_\Sigma )$. Pick $m \in \{ 0, 1, \dots, d \}$ and assume that for $0 \leq u \leq \mathrm{min} \{ d-m, L ( N ) \}$
\[ \left\{ -a_{u,j} \left( N \right) + v \; \left| \; 1 \leq j \leq b_{u} \left( N \right) \right. \right\} \subseteq V^{u+m} \left( X_\Sigma \right) \, . \]
Then it holds $h^m ( X_\Sigma, \tilde{N} ( v ) ) = 0$.
\end{cor}

\begin{myproof}
By assumption it holds
\[ 0 = h^{u+m} \left( X_\Sigma, \tilde{F_{u}} \left( N \right) \left( v \right) \right) \qquad 0 \leq u \leq \mathrm{min} \left\{ L \left( N \right), d-m \right\} \, . \]
Hence, $\mathfrak{H}_2^{-u,u+m} = 0$ for $0 \leq u \leq \mathrm{min} \{ L ( N ), d-m \}$, which implies $h^m ( X_\Sigma, \tilde{N} ( v ) ) = 0$.
\end{myproof}

\subsection{\dots for Local Cohomology}

For a \fp graded $S$-module $N$ we can consider the local cohomology groups $H^i_{B_\Sigma} ( N )$. For our purposes we can define these group via their relation to the 
(Zariski) sheaf cohomology:
\begin{enumerate}
 \item There is an exact sequence of \fp graded $S$-modules
      \[ 0 \to H^0_{B_\Sigma} \left( N \right) \to N \to \bigoplus_{p \in \mathrm{Cl} \left( X_\Sigma \right)}{H^0 ( X_\Sigma, \tilde{N} ( p ) )} \to H^1_{B_\Sigma} \left( N \right) \to 0 \, . \label{equ:connection1} \]
 \item For $1 \leq i \leq d$ there is an isomorphism $\displaystyle H^{i+1}_{B_\Sigma} ( N ) \cong \bigoplus_{p \in \mathrm{Cl} ( X_\Sigma )}{H^i ( X_\Sigma, \tilde{N} ( p ) )}$.
\end{enumerate}
For completeness let us mention that ordinarily these groups are introduced as \cite{brodmann2007local}
\[ H^i_{B_\Sigma} \left( N \right) := \varinjlim_{l} \mathrm{Ext}_\bullet^i \left( S / B_\Sigma^l, N \right) \, . \]
and the above connection to sheaf cohomology then follows from \cite[proposition 2.3]{2000math......1159E}.

\begin{remark}
In the following we use $N_p$ to denote the truncation of a \fp graded $S$-module $N$ to $p \in \mathrm{Cl} ( X_\Sigma )$. $N_p$ must not be confused with any sort of localisation.
\end{remark}

\begin{cor}
Let $X_\Sigma$ be a smooth and complete normal toric variety with Cox ring $S$. Based on \emph{cohomCalg} we then see the following:
\begin{itemize}
 \item $S_p \cong H^0 ( X_\Sigma, \mathcal{O}_{X_\Sigma} ( p ) )$ for all $p \in \mathrm{Cl} ( X_\Sigma )$. Thus, $H^0_{B_\Sigma} ( S ) =  H^1_{B_\Sigma} ( S ) = 0$.
 \item For $2 \leq i \leq d+1$ have $H^i_{B_\Sigma} ( S )_p = 0$ iff $p \in V^{i-1} ( X_\Sigma )$. This extends corollary 3.6 of 
      \cite{Maclagan03multigradedcastelnuovo-mumford}.
\end{itemize}
\end{cor}

\begin{cor}
Let $X_\Sigma$ be a smooth and complete normal toric variety of dimension $d$ with Cox ring $S$ and vanishing sets $V^i ( X_\Sigma ) = \{ D \in \mathrm{Cl} ( X_\Sigma ) \; , \; H^i ( X_\Sigma, \mathcal{O}_{X_\Sigma} ( D ) ) = 0 \}$. Consider a \fp graded $S$-module $N$ with minimal free resolution
\[ 0 \leftarrow F_0 \left( N \right) \leftarrow F_{1} \left( N \right) \leftarrow \dots \leftarrow F_{L \left( N \right)} \left( N \right) \leftarrow 0 \, , \qquad F_i \left( N \right) = \bigoplus_{j = 1}^{b_i \left( N \right)}{S \left( - a_{i,j} \left( N \right) \right)} \, , \]
let $v \in \mathrm{Pic} ( X_\Sigma )$, pick $m \in \{ 1, 2, \dots, d \}$ and assume that for $0 \leq u \leq \mathrm{min} \{ d-m, L ( N ) \}$
\[ \left\{ -a_{u,j} \left( N \right) + v \; \left| \; 1 \leq j \leq b_{u} \left( N \right) \right. \right\} \subseteq V^{u+m} \left( X_\Sigma \right) \, . \]
Then $H^{m+1}_{B_\Sigma} \left( N \left( v \right) \right)_{0} = 0$.
\end{cor}

\begin{myproof}
Use $H^{m+1}_{B_\Sigma} \left( N \left( v \right) \right)_{0} \cong H^m ( X_\Sigma, \tilde{N} \left( v \right) )$ and \cref{cor:VanishingTheoremForSheafCohomology}.
\end{myproof}

To establish vanishing theorems for $H^0_{B_\Sigma} ( N )$ and $H^1_{B_\Sigma} ( N )$ consider the minimal free resolution
\[ \mathbf{F} \left( N \right) \colon 0 \leftarrow F_0 \left( N \right) \leftarrow F_{1} \left( N \right) \leftarrow \dots \leftarrow F_{L \left( N \right)} \left( N \right) \leftarrow 0 \, , \quad F_i \left( N \right) = \bigoplus_{j = 1}^{b_i \left( N \right)}{S \left( - a_{i,j} \left( N \right) \right)} \]
and use it to compute the local cohomology groups of $N$ from the spectral sequence \cite{Maclagan03multigradedcastelnuovo-mumford}
\[ E_2^{i,j} := H^i \left( H^j_{B_\Sigma} \left( \mathbf{F} \left( N \right) \right) \right) \Rightarrow H_{B_\Sigma}^{i+j} \left( N \right). \label{spectralSequenceLocalCohomologies} \]
For every $0 \leq j \leq d+1$ we thus consider the cochain complex
\[ 0 \leftarrow H^j_{B_\Sigma} \left( F_{0} \left( N \right) \right) \leftarrow H^j_{B_\Sigma} \left( F_{1} \left( N \right) \right) \leftarrow \dots \leftarrow H^j_{B_\Sigma} \left(  F_{L \left( N \right)} \left( N \right) \right) \leftarrow 0 \]
Its cohomology at $-L ( N ) \leq i \leq 0$ be $h_2^{i,j}$. Then the $E_2$-sheet of \cref{spectralSequenceLocalCohomologies} looks as follows:
\[ \includegraphics[valign = c]{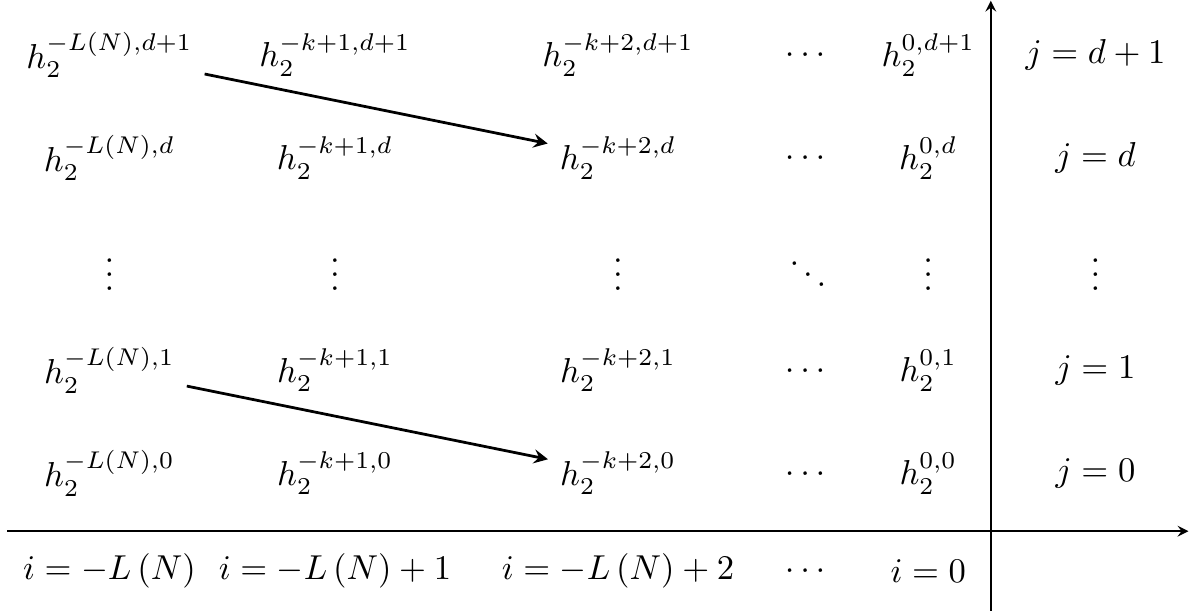} \]
This shows $\displaystyle H_{B_\Sigma}^m \left( N \right) \cong \bigoplus_{u=0}^{\mathrm{min} \left\{d+1-m, L \left( N \right) \right\}}{h_\infty^{-u,m+u}}$.

\begin{cor} \label{cor:localCohomologyHi}
Let $X_\Sigma$ be a smooth and complete normal toric variety of dimension $d$ with Cox ring $S$ and vanishing sets $V^i ( X_\Sigma ) = \{ D \in \mathrm{Cl} ( X_\Sigma ) \; , \; H^i ( X_\Sigma, \mathcal{O}_{X_\Sigma} ( D ) ) = 0 \}$. Consider a \fp graded $S$-module $N$ with minimal free resolution
\[ 0 \leftarrow F_0 \left( N \right) \leftarrow F_{1} \left( N \right) \leftarrow \dots \leftarrow F_{L \left( N \right)} \left( N \right) \leftarrow 0 \, , \qquad F_i \left( N \right) = \bigoplus_{j = 1}^{b_i \left( N \right)}{S \left( - a_{i,j} \left( N \right) \right)} \, . \]
Let $v \in \mathrm{Pic} ( X_\Sigma )$. Then the following holds true:
\begin{enumerate}
 \item Assume $\{ -a_{u,j} ( N ) + v \; | \; 1 \leq j \leq b_{u} ( N ) \} \subseteq V^{u-1} ( X_\Sigma )$ for $2 \leq u \leq \mathrm{min} \{ d+1, L ( N ) \}$. Then 
      $H^0_{B_\Sigma} ( X_\Sigma, \tilde{N} \left( v \right) )_{0} = 0$.
 \item Assume $\{ -a_{u,j} ( N ) + v \; | \; 1 \leq j \leq b_{u} ( N ) \} \subseteq V^{u} ( X_\Sigma )$ for $1 \leq u \leq \mathrm{min} \{ d, L ( N ) \}$. Then 
      $H^1_{B_\Sigma} ( X_\Sigma, \tilde{N} \left( v \right) )_{0} = 0$.
\end{enumerate}
\end{cor}

\begin{myproof}
Let us prove the first statement. By assumption it holds
\[ H^{u-1} \left( X_\Sigma, \tilde{F_{u}} \left( N \right) \left( v \right) \right) = 0 \qquad 2 \leq u \leq \mathrm{min} \left\{ d+1, L \left( N \right) \right\} \, . \]
Since $H_{B_\Sigma}^0 ( F_{u} ( N ) ( v ) ) = H_{B_\Sigma}^1 ( F_{u} ( N ) ( v ) ) = 0$ this is equivalent to
\[ H_{B_\Sigma}^u \left( F_{u} \left( N \right) \left( v \right) \right)_{0} = 0 \qquad 0 \leq u \leq \mathrm{min} \left\{ d+1, L \left( N \right) \right\} \, . \]
Consequently, $( h_2^{-u,u} )_{0} = 0$ for $0 \leq u \leq \mathrm{min} \{ d+1, L ( N ) \}$ which implies $H^0_{B_\Sigma} ( X_\Sigma, \tilde{N} ( v ) )_{0} = 0$. The second statement follows along the same lines.
\end{myproof}

\section{Computing (global) Bivariate Extensions of Coherent Sheaves} \label{sec:ComputingGlobalBivariateExt}

In this section we will formulate a theorem for the computation of the (global) bivariate $\mathrm{Ext}^q ( \mathcal{M}, \mathcal{N} )$ on smooth and complete normal toric varieties $X_\Sigma$. Our approach follows \cite{1998math......7170S, Oberwolfach}, but uses the more refined vanishing sets and vanishing theorems presented in the previous two sections.

\begin{cor} \label{cor:step1}
Let $X_\Sigma$ be a smooth and complete normal toric variety of dimension $d$ with Cox ring $S$. Consider two \fp graded $S$-modules $M,N$. A minimal free resolution of $M$ is
\[
\resizebox{0.9\textwidth}{!}{$ \displaystyle
\mathbf{F} \left( M \right) \colon 0 \leftarrow F_0 \left( M \right) \leftarrow F_1 \left( M \right) \leftarrow \dots \leftarrow F_{L \left( M \right)} \left( M \right) \leftarrow 0, \quad F_i ( M ) = \bigoplus_{j = 1}^{b_i ( M )}{S ( - a_{i,j} ( M ) )} \, .$}
\]
For $v \in \mathrm{Cl} ( X_\Sigma )$ we form the following cochain complex 
\begin{align}
\begin{split}
\mathbf{C}^0 \colon 0 & \to H^0 \left( X_\Sigma, \tilde{F}^\vee_{0} \left( M \right) \otimes_{\mathcal{O}_X} \tilde{N} \left( v \right) \right) \to H^0 \left( X_\Sigma, \tilde{F}^\vee_{1} \left( M \right) \otimes_{\mathcal{O}_X} \tilde{N} \left( v \right) \right) \to \dots \\
& \hspace{13em} \dots \to H^0 \left( X_\Sigma, \tilde{F}^\vee_{L \left( M \right)} \left( M \right) \otimes_{\mathcal{O}_X} \tilde{N} \left( v \right) \right) \to 0 \, .
\end{split}
\end{align}
Now pick $m \in \{ 0, 1, \dots, d \}$ and require that
\begin{itemize}
 \item $\displaystyle H^{m-u} ( X_\Sigma, \mathcal{F}_{u}^\vee ( M ) \otimes_{\mathcal{O}_{X_\Sigma}} \mathcal{N} ( v ) ) = 0$ for all 
      $0 \leq u \leq \mathrm{min} \{ L ( M ), m-1 \}$ and
 \item $\displaystyle H^{m-1-u} ( X_\Sigma, \mathcal{F}_{u}^\vee ( M ) \otimes_{\mathcal{O}_{X_\Sigma}} \mathcal{N} ( v ) ) = 0$ for all 
      $0 \leq u \leq \mathrm{min} \{ L ( M ), m-2 \}$.
\end{itemize}
Then it holds $\mathrm{Ext}^{m}_{\mathcal{O}_{X_\Sigma}} ( \mathcal{M}, \mathcal{N} ( v ) ) \cong H^m ( \mathbf{C}^0 )$.
\end{cor}

\begin{myproof}
By dualising $\tilde{\mathbf{F}} ( M )$ and tensoring with $\tilde{N} ( v )$ we obtain the cochain complex
\[ 
 \resizebox{0.9\textwidth}{!}{$
 \mathcal{E}^\bullet \colon 0 \to \tilde{F_0}^\vee \left( M \right) \otimes_{\mathcal{O}_{X_\Sigma}} \tilde{N} \left( v \right) \to \tilde{F_1}^\vee \left( M \right) \otimes_{\mathcal{O}_X} \tilde{N} \left( v \right) \to \dots \to \tilde{F_{L \left( M \right)}}^\vee \left( M \right) \otimes_{\mathcal{O}_X} \tilde{N} \left( v \right) \to 0 \,.$}
\]
General arguments show that the hypercohomology of this complex matches $\mathrm{Ext}_{\mathcal{O}_{X_\Sigma}}^{p+q} ( \mathcal{M}, \mathcal{N} )$ \cite{grothendieck1957}. To compute this hypercohomology we apply the spectral sequence $^\prime E_2^{p,q}$ from \cite{weibel1995introduction}. Therefore, for $0 \leq q \leq d$, we consider the cochain complex
\begin{align} 
\begin{split}
 & 0 \to H^q \left( X_\Sigma, \tilde{F_0}^\vee \left( M \right) \otimes_{\mathcal{O}_{X_\Sigma}} \tilde{N} \left( v \right) \right) \to H^q \left( X_\Sigma, \tilde{F_1}^\vee \left( M \right) \otimes_{\mathcal{O}_X} \tilde{N} \left( v \right) \right) \to \dots \\
 & \hspace{17em} \to H^q \left( X_\Sigma, \tilde{F_{L \left( M \right)}}^\vee \left( M \right) \otimes_{\mathcal{O}_X} \tilde{N} \left( v \right) \right) \to 0
\end{split}
\end{align}
and denote its cohomology at $0 \leq p \leq L(M)$ by $\mathfrak{H}_{2}^{p,q}$. Consequently, the sheet $^\prime E_2^{p,q} ( \mathcal{E}^\bullet )$ of the spectral sequence $^\prime E^{p,q} ( \mathcal{E}^\bullet )$ introduced in \cref{equ:E2Prime} takes the form:
\[ \includegraphics[valign = c]{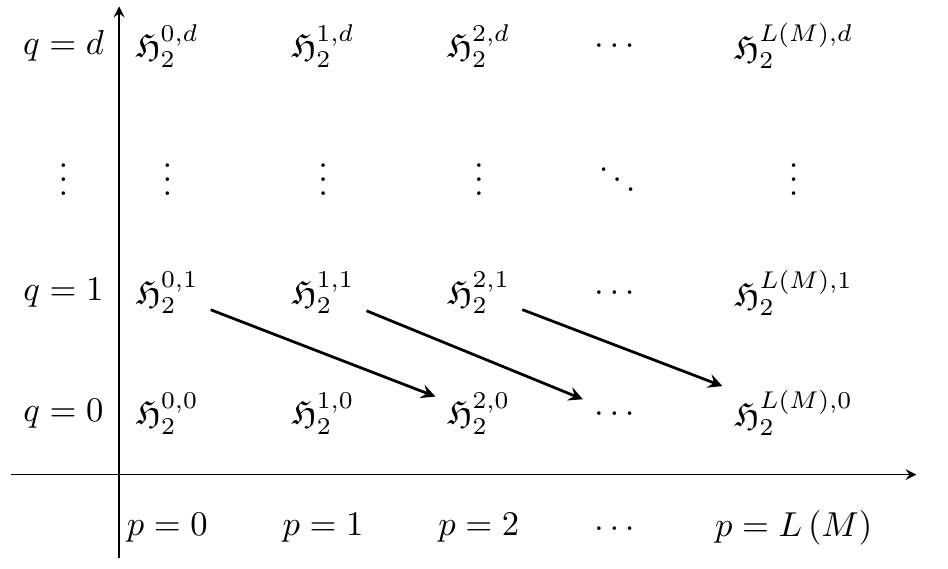} \]
By assumption the following objects vanish:
\begin{itemize}
 \item $\mathfrak{H}_2^{u,m-u} = 0$ for $0 \leq u \leq \mathrm{min} \{ L ( M ), m-1 \}$ and
 \item $\mathfrak{H}_2^{u,m-1-u} = 0$ for $0 \leq u \leq \mathrm{min} \{ L ( M ), m-2 \}$.
\end{itemize}
Consequently, $\mathrm{Ext}^{m}_{\mathcal{O}_{X_\Sigma}} ( \mathcal{M}, \mathcal{N} ( v ) ) = \mathfrak{H}_{2}^{m,0}$ and by definition $\mathfrak{H}_2^{m,0} = \mathfrak{H}^m ( \mathbf{C}^0 )$.
\end{myproof}

\begin{lemma} \label{lemma1}
Let $\mathcal{C}$ be an Abelian category. In the commutative diagram
\[ \includegraphics[valign = c]{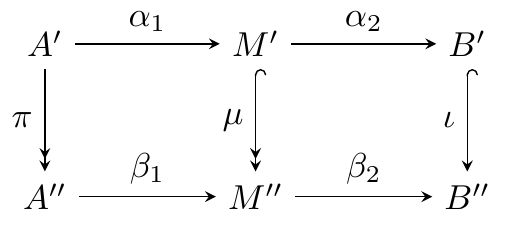} \]
let $\pi$ be an epimorphism, $\mu$ an isomorphism and $\iota$ a monomorphism. Then $\mathrm{im} ( \alpha_1 ) \cong \mathrm{im} ( \beta_1 )$ and $\mathrm{ker} ( \alpha_2 ) \cong \mathrm{ker} ( \beta_2 )$.
\end{lemma}

\begin{myproof}
\begin{itemize}
 \item $\alpha_1 = \mu^{-1} \circ \beta_1 \circ \pi$. Since $\mu$ is an isomorphism and $\pi$ an epimorphism, it follows $\mathrm{im} (\alpha_1) \cong \mathrm{im} (\beta_1)$.
 \item $\beta_2 = \iota \circ \alpha_2 \circ \mu^{-1}$. Since $\mu$ is an isomorphism and $\iota$ a monomorphism, it follows $\mathrm{ker} (\alpha_2) \cong \mathrm{ker} (\beta_2)$.
\end{itemize}
This completes the proof.
\end{myproof}

\begin{cor} \label{cor:step2}
Let $X_\Sigma$ be a smooth and complete normal toric variety of dimension $d$ with Cox ring $S$. Consider two \fp graded $S$-modules $M,N$. A minimal free resolution of $M$ is
\[ 
\resizebox{0.9\textwidth}{!}{$\displaystyle
\mathbf{F} \left( M \right) \colon 0 \leftarrow F_0 \left( M \right) \leftarrow F_1 \left( M \right) \leftarrow \dots \leftarrow F_{L \left( M \right)} \left( M \right) \leftarrow 0, \quad F_i ( M ) = \bigoplus_{j = 1}^{b_i ( M )}{S ( - a_{i,j} ( M ) )} \, .$}
\]
For $v \in \mathrm{Cl} ( X_\Sigma )$ form the cochain complex 
\begin{align}
\begin{split}
\mathbf{C}^0 \colon 0 & \to H^0 \left( X_\Sigma, \tilde{F}^\vee_{0} \left( M \right) \otimes_{\mathcal{O}_X} \tilde{N} \left( v \right) \right) \to H^0 \left( X_\Sigma, \tilde{F}^\vee_{1} \left( M \right) \otimes_{\mathcal{O}_X} \tilde{N} \left( v \right) \right) \to \dots \\
& \hspace{15em} \dots \to H^0 \left( X_\Sigma, \tilde{F}^\vee_{L \left( M \right)} \left( M \right) \otimes_{\mathcal{O}_X} \tilde{N} \left( v \right) \right) \to 0 \, .
\end{split}
\end{align}
and assume that $H^1_{B_\Sigma} \left( F_{m-1}^\vee \left( M \right) \otimes_S N \right)_v$, $H^0_{B_\Sigma} \left( F_m^\vee \left( M \right) \otimes_S N \right)_v$, $H^1_{B_\Sigma} \left( F_m^\vee \left( M \right) \otimes_S N \right)_v$ and $H^1_{B_\Sigma} \left( F_{m+1}^\vee \left( M \right) \otimes_S N \right)_v$ vanish. Then $\mathfrak{H}^m \left( \mathbf{C}^0 \right) \cong \mathrm{Ext}^m_S \left( M, N \right)_v$.
\end{cor}

\begin{figure}[tb]
\centering
\includegraphics{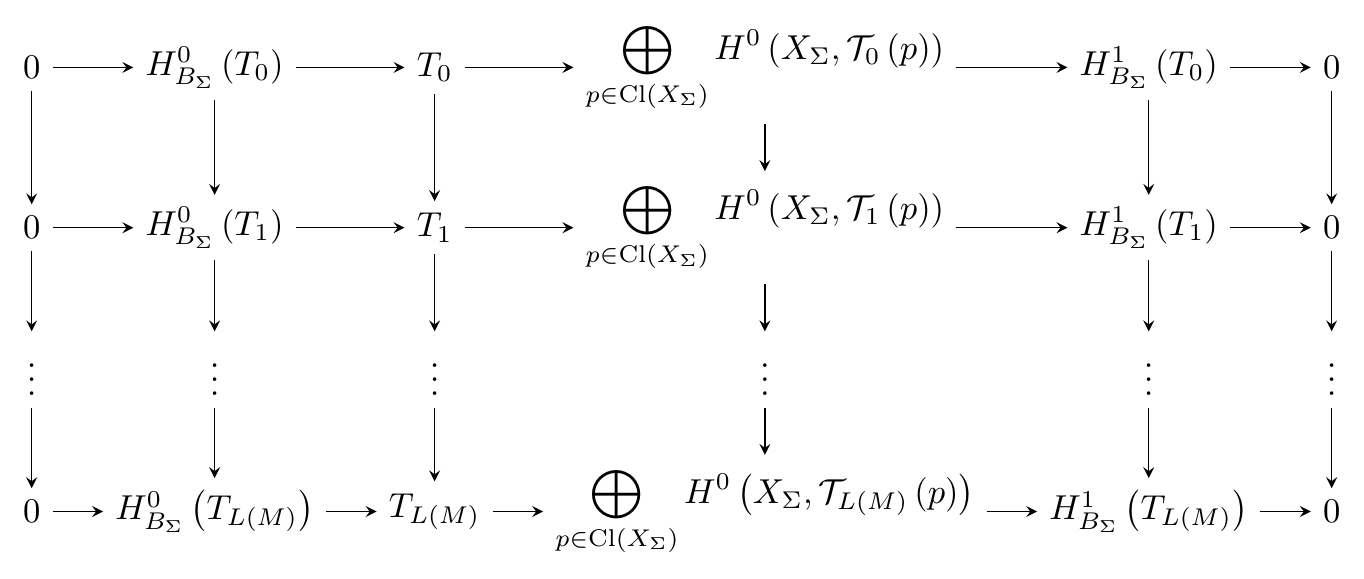}
\caption{Commutative diagram used in the proof of \cref{cor:step2}.}
\label{figure-1092835209563}
\end{figure}

\begin{figure}[tb]
\centering
\includegraphics{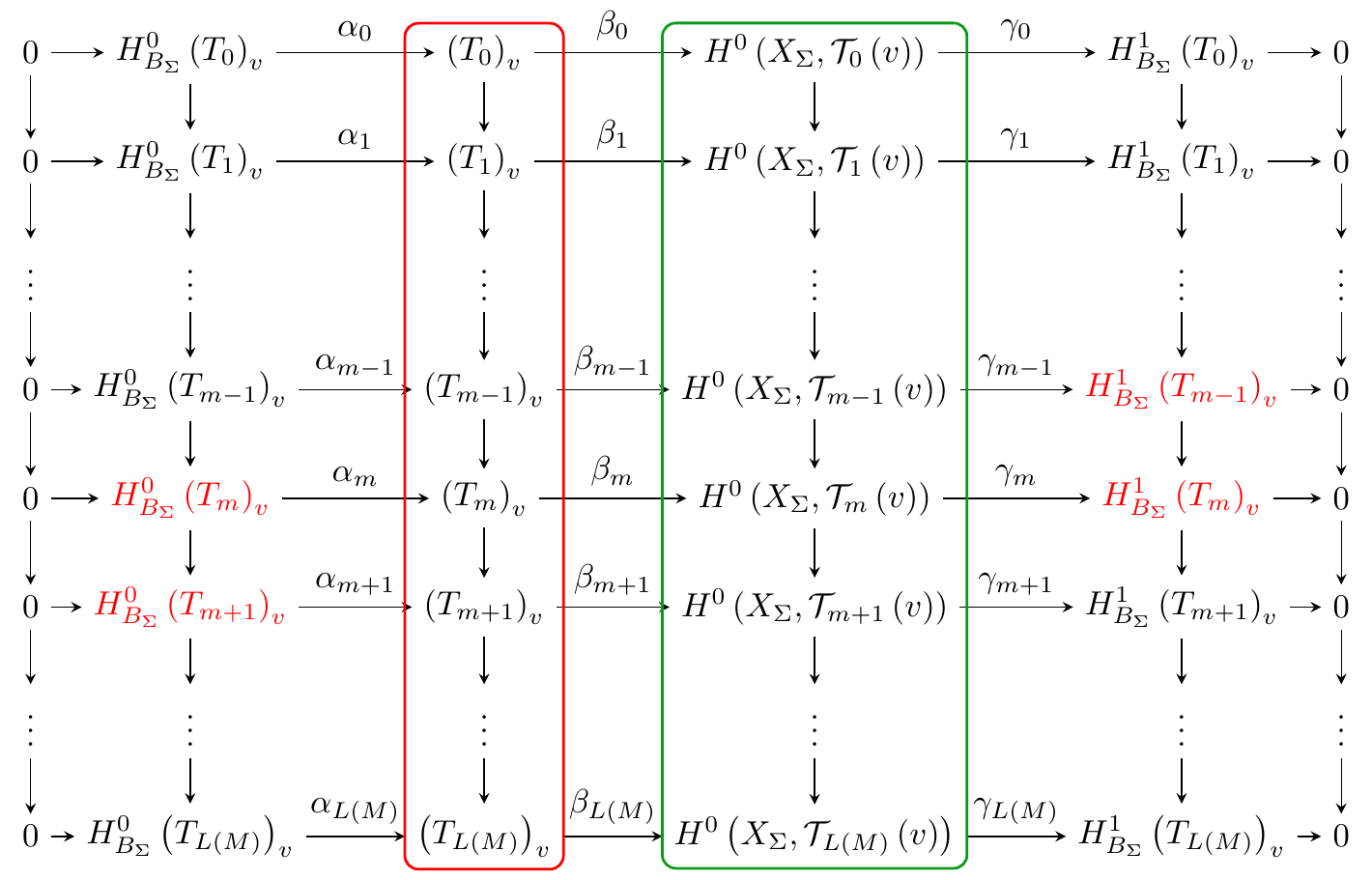}
\caption{Another commutative diagram used in the proof of \cref{cor:step2}.}
\label{figure-1098235098302830}
\end{figure}

\begin{myproof}
For $0 \leq i \leq L ( M )$ and $p \in \mathrm{Cl} ( X_\Sigma )$ define
\[ 
\resizebox{0.9\textwidth}{!}{$
T_i = F_i^\vee \left( M \right) \otimes_{S} N = \displaystyle \bigoplus_{j = 1}^{b_i \left( M \right)}{N \left( a_{ij} \left( M \right) \right)}, \quad \mathcal{T}_i \left( p \right) = \tilde{F}_i^\vee \left( M \right) \otimes \tilde{N} \left( p \right) = \displaystyle \bigoplus_{j = 1}^{b_i \left( M \right)}{\tilde{N} \left( p + a_{ij} \left( M \right) \right)}$}
\]
and look at the commutative diagram in \cref{figure-1092835209563}. Truncate this diagram to $v \in \mathrm{Cl} ( X_\Sigma )$. This gives the commutative diagram of finite dimensional $\mathbb{Q}$-vector spaces in \cref{figure-1098235098302830}. The m-th homology of the red column matches $\mathrm{Ext}_S^m \left( M, N \right)_{v}$, the $m$-th homology of the green column $\mathfrak{H}^m \left( \mathbf{C}^0 \right)$. Since the four red objects vanish by assumption, it follows from the previous lemma that indeed $\mathrm{Ext}_S^m \left( M, N \right)_{v} \cong \mathfrak{h}^m \left( \mathbf{C}^0 \right)$.
\end{myproof}

\begin{theorem} \label{mytheorem}
Let $X_\Sigma$ be a smooth, complete normal toric variety of dimension $d$, Cox ring $S$ and let $M, N$ be two \fp graded $S$-modules. Let $0 \leftarrow M \leftarrow F_0 ( M ) \leftarrow F_1 ( M ) \leftarrow \dots \leftarrow F_{L ( M )} ( M ) \leftarrow 0$ be a minimal free resolution of $M$, where $F_i ( M ) = \bigoplus_{j = 1}^{b_i ( M )}{S ( - a_{i,j} ( M ) )}$. Pick $m \in \{ 0, 1, \dots, d \}$, $v \in \mathrm{Cl} ( X_\Sigma )$ and require
\begin{enumerate}
\itemsep-0.5em
 \item $v + a_{u,j} \left( M \right) - a_{k,l} \left( N  \right) \in V^{k+m-u} \left( X_\Sigma \right)$ for all
      \[ \begin{array}{lcl} 
      0 \leq u \leq \min \left\{ L \left( M \right), m-1 \right\} \, , & & 1 \leq j \leq b_u \left( M \right) \, ,\\
      0 \leq k \leq \min \left\{ d - m + u, L \left( N \right) \right\} \, , & & 1 \leq l \leq b_k \left( N \right) \, .
      \end{array} \]
 \item $v + a_{u,j} \left( M \right) - a_{k,l} \left( N  \right) \in V^{k+m-u-1} \left( X_\Sigma \right)$ for all
      \[ \begin{array}{lcl}
      0 \leq u \leq \min \left\{ L \left( M \right), m-2 \right\} \, , & & 1 \leq j \leq b_u \left( M \right) \, , \\
      0 \leq k \leq \min \left\{ d - m + 1 + u, L \left( N \right) \right\} \, , & & 1 \leq l \leq b_k \left( N \right) \, .
      \end{array} \]
 \item $v + a_{m-1,j} \left( M \right) - a_{u,l} \left( N \right) \in V^u \left( X_\Sigma \right)$ for all 
      \[ \begin{array}{ccc}
      1 \leq j \leq b_{m-1} \left( M \right) \, , & 1 \leq u \leq \min \left\{ d, L \left( N \right) \right\} \, , & 1 \leq l \leq b_u \left( N \right) \, .
      \end{array} \]
 \item $v + a_{m,j} \left( M \right) - a_{u,l} \left( N \right) \in V^u \left( X_\Sigma \right)$ for all
      \[ \begin{array}{ccc} 
      1 \leq j \leq b_{m} \left( M \right) \, , & 1 \leq u \leq \min \left\{ d, L \left( N \right) \right\} \, , & 1 \leq l \leq b_u \left( N \right) \, .
      \end{array} \]
 \item $v + a_{m,j} \left( M \right) - a_{u,l} \left( N \right) \in V^{u-1} \left( X_\Sigma \right)$ for all
      \[ \begin{array}{ccc}
      1 \leq j \leq b_{m} \left( M \right) \, , & 2 \leq u \leq \min \left\{ d+1, L \left( N \right) \right\} \, , & 1 \leq l \leq b_u \left( N \right) \, .
      \end{array} \]
 \item $v + a_{m+1,j} \left( M \right) - a_{u,l} \left( N \right) \in V^{u-1} \left( X_\Sigma \right)$ for all
      \[ \begin{array}{ccc}
      1 \leq j \leq b_{m+1} \left( M \right) \, , & 2 \leq u \leq \min \left\{ d+1, L \left( N \right) \right\} \, , & 1 \leq l \leq b_u \left( N \right) \, .
      \end{array} \]
\end{enumerate}
Then it holds $\mathrm{Ext}^{m}_{\mathcal{O}_{X_\Sigma}} ( \tilde{M}, \tilde{N} ( v ) ) \cong \mathrm{Ext}^m_{S} ( M, N )_{v}$.
\end{theorem}

\begin{myproof}
We combine \cref{cor:step1} and \cref{cor:step2}. Then we express the vanishing conditions in terms of the Betti numbers of $M$ and $N$ along the lines presented in \cref{sec:VanishingTheorems}.
\end{myproof}

\begin{note}[Comparison to work in \cite{Oberwolfach}]
Let us compare this result with the findings in \cite{Oberwolfach}. In the latter work, the focus was on smooth, projective toric varieties $X_\Sigma$ of dimension $d$ and $\mathrm{Cl} ( X_\Sigma ) \cong \mathbb{Z}^r$. For such varieties, in \cite{Maclagan03multigradedcastelnuovo-mumford} a cone $\mathbf{C} \subseteq \mathrm{Cl} ( X_\Sigma )$ was identified such that for every $\mathbf{v} \in C$ it holds 
\[ h^i \left( X_\Sigma, \mathcal{O}_{X_\Sigma} \left( \mathbf{v} \right) \right) = 0 \, \qquad \forall \, 1 \leq i \leq d \, . \]
Back in the original work this cone is referred to as the semigroup $\mathcal{K}^{\mathrm{sat}}$, which we already commented on before. We can recover this cone via
\[ \mathbf{C} = V^1 \left( X_\Sigma \right) \cap V^2 \left( X_\Sigma \right) \cap \dots \cap V^d \left( X_\Sigma \right) \, . \]
This shows that the vanishing sets employed in this work serve as refined versions of this cone $\mathbf{C}$. In particular, our theorem above can be understood as a refined version of the following result from \cite{Oberwolfach}:
\begin{quote}
`Let $M$ and $N$ be finitely presented $\mathbb{Z}^r$-graded $S$-modules, let $m$ be an integer, and let $\mathbf{v}, \mathbf{u} \in \mathbb{Z}^r$ be two vectors such that $\mathcal{O}_X ( \mathbf{u} )$ is ample. If $e \in \mathbb{N}$ satisfies $\mathbf{v} + \mathbf{a}_{m-k,l} ( S_{e \mathbf{u}} M ) - a_{i,j} ( N ) \in \mathbf{C}$ for all $0 \leq k \leq m$, $1 \leq l \leq b_{m-k} ( M )$, $0 \leq i \leq d- m$ and $1 \leq j \leq b_i ( N )$, then there is a canonical isomorphism $\mathrm{Ext}^m_X ( \mathcal{M}, \mathcal{N} ( \mathbf{v} )) = \mathrm{Ext}_S^m ( S_{e \mathbf{u}} M, N )_{\mathbf{V}}$.'
\end{quote}
Note that $S_{e \mathbf{u}}$ denotes the ideal formed from the $e$-th power of all monomials of degree $\mathbf{u}$ in the Cox ring $S$ of $X_\Sigma$. The significance of these ideals is that they furnish special models for the structure sheaf $\mathcal{O}_{X_\Sigma}$, such that the demand $\mathbf{v} + \mathbf{a}_{m-k,l} ( S_{e \mathbf{u}} M ) - a_{i,j} ( N ) \in \mathbf{C}$ is always satisfied for sufficiently large integer $e$. The latter is a consequence of $\mathbf{u} \in \mathrm{Cl} ( X_\Sigma )$ being ample. We will discuss these ideals in detail in \cref{subsec:Ideals}.
\end{note}

\section{Identifying Ample Divisors} \label{sec:IdentifyAmpleDivisors}

\subsection{Intersections of Divisors and Curves} \label{subsec:IntersectionComputationToric}

\begin{theorem}[Theorem 4.2.8 of \cite{cox2011toric}]
Let $X_\Sigma$ be the toric variety of a fan $\Sigma$ and $D = \sum_{\rho}{a_\rho D_\rho}$ a Weil divisor on $X_\Sigma$. Then the following are equivalent:
\begin{itemize}
 \item $D$ is Cartier.
 \item For each $\sigma \in \Sigma_{\mathrm{max}}$, there is $m_\sigma \in M$ such that $\langle m_\sigma, u_\rho \rangle = - a_\rho$ for all $\rho \in \sigma ( 1 )$.
\end{itemize}
\end{theorem}

\begin{defi}[Cartier Data]
For a Cartier divisor $D$ on a toric variety $X_\Sigma$, we term $\{ m_\sigma \}_{\sigma \in \Sigma_{\mathrm{max}}}$ its \textbf{Cartier data}.
\end{defi}

\begin{note}
Intersection products in toric spaces can be computed along \cite{cox2011toric} page 289. Namely:
\begin{quote}
``In the toric case, $D \cdot C$ is easy to compute when $D$ and $C$ are torus-invariant in $X_\Sigma$. In order for $C$ to be torus-invariant and complete, we must have $C = V ( \tau ) = \overline{ O ( \tau )}$ where $\tau = \sigma \cap \sigma^\prime \in \Sigma ( n-1 )$ is the wall separating cones $\sigma, \sigma^\prime \in \Sigma ( n )$, $n = \mathrm{dim} ( X_\Sigma )$. We do not assume $\Sigma$ is complete. \\
In this situation, we have the sublattice $N_\tau = \mathrm{Span} ( \tau ) \cap N \subseteq N$ and the quotient $N ( \tau ) = N / N_\tau$. Let $\overline{\sigma}$ and $\overline{\sigma}^\prime$ be the images of $\sigma$ and $\sigma^\prime$ in $N ( \tau )_{\mathbb{R}}$. Since $\tau$ is a wall, $N ( \tau ) \simeq \mathbb{Z}$ and $\overline{\sigma}$, $\overline{\sigma}^\prime$ are rays that correspond to rays in the usual fan of $\mathbb{P}^1$. In particular, $V ( \tau ) \simeq \mathbb{P}^1$ is smooth, so no normalization is needed when computing the intersection product.''
\end{quote}
This leads to the following proposition.
\end{note}

\begin{proposition}[Prop. 6.3.8 in \cite{cox2011toric}]
Let $X_\Sigma$ be the toric variety associated to the fan $\Sigma$ and $C = V ( \tau )$ the complete torus-invariant curve in $X_\Sigma$ coming from the wall $\tau = \sigma \cap \sigma^\prime$. Let $D$ be a Cartier divisor with Cartier data $m_\sigma, m_{\sigma^\prime} \in M$ corresponding to $\sigma, \sigma^\prime \in \Sigma ( n )$. Also pick $u \in \sigma^\prime \cap N$ that 
maps to the minimal generator of $\overline{\sigma}^\prime \subseteq N ( \tau )_{\mathbb{R}}$. Then $D \cdot C = \langle m_\sigma - m_{\sigma^\prime}, u \rangle \in \mathbb{Z}$.
\end{proposition}

\begin{construc}
The following algorithm is implemented in \cite{SheafCohomologyOnToricVarieties}.
\begin{itemize}
 \item Input:
      \begin{itemize}
       \item Normal toric variety $X_\Sigma$ which is smooth and complete.
       \item Cone $\tau = \sigma^\prime \cap \sigma^{\prime \prime} \in \Sigma$, where $\sigma^\prime, \sigma^{\prime \prime} \in \Sigma ( n )$ for $n = \mathrm{dim} ( \Sigma )$.
       \item Cartier divisor $D$ on $X_\Sigma$ given by its Cartier data $\{ m_\sigma \}_{\sigma \in \Sigma_{\mathrm{max}}}$.
      \end{itemize}
 \item Output: \\
      The intersection number $V ( \tau ) \cdot D$ of $V ( \tau )$ and $D$.
 \item Processing: \\
      \begin{enumerate}
      \item Let $R = \{ u_\rho \}_{\rho \in \Sigma ( 1 )}$ be the ray generators of $\Sigma$. Then there exist $R^\prime, R^{\prime \prime} \subseteq R$ with
           \[ \sigma^\prime = \mathrm{Cone} \left( R^\prime \right), \qquad \sigma^{\prime \prime} = \mathrm{Cone} \left( R^{\prime \prime} \right) \, . \]
           In particular, $\tau = \mathrm{Cone} ( G )$ for $G = R^\prime \cap R^{\prime \prime}$ -- write $G = \{ v_1, \dots, v_{|G|} \}$, $v_i \in \mathbb{Z}^{\mathrm{rk} ( N )}$.
      \item We wish to describe the projection map $\pi \colon N \to N ( \tau ) \cong \mathbb{Z}$ with
           \[ N_{\tau} = \mathrm{Span} \left( \tau \right) \cap N \subseteq N, \qquad N \left( \tau \right) = N / N_{\tau} \cong \mathbb{Z} \, . \]
           Since $v_1, \dots, v_{|G|}$ need not be linearly independent over $\mathbb{Z}$ we look at
           \[ \includegraphics[valign = c]{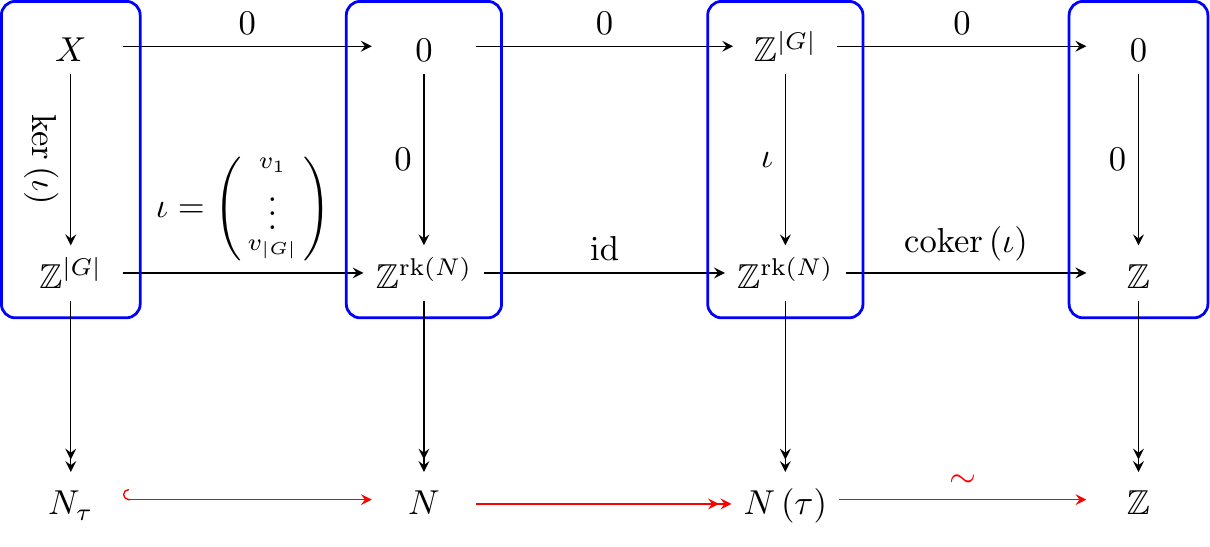} \]
           Consequently, we utilise $\mathrm{coker} ( \iota )$ to describe the projection $\pi \colon N \to N ( \tau ) \cong \mathbb{Z}$.
      \item Compute $U := \pi^{-1} ( 1 ) \cap \sigma^{\prime \prime}$ and differ the following cases:
           \begin{itemize}
            \item $U \neq \emptyset$: pick a $u \in \pi^{-1} ( 1 ) \cap \sigma^{\prime \prime}$.
            \item $U = \emptyset$: pick a $u \in \pi^{-1} ( -1 ) \cap \sigma^{\prime \prime}$.
           \end{itemize}
      \item Finally, use the Cartier data $\{ m_\sigma \}_{\sigma \in \Sigma_{\mathrm{max}}}$ of $D$ to compute $\langle m_{\sigma^\prime} - m_{\sigma^{\prime \prime}}, u 
           \rangle$ which matches $V ( \tau ) \cdot D$.
      \end{enumerate}
\end{itemize}
\end{construc}

\subsection{Nef-Cone} \label{subsec:NefCone}

\begin{theorem}[6.3.12 in \cite{cox2011toric}]
Let $D$ be a Cartier divisor on a toric variety $X_\Sigma$ whose fan $\Sigma$ has convex support of full dimension. Then the following are equivalent:
\begin{enumerate}
 \item $D$ is nef.
 \item $D \cdot C \geq 0$ for all torus-invariant irreducible complete curves $C \subseteq X_\Sigma$.
\end{enumerate}
\end{theorem}

\begin{conseq}
For a toric variety $X_\Sigma$ whose fan $\Sigma$ has convex support of full dimension, $\mathrm{Nef} ( X )$ is the cone in $\mathrm{Cl} ( X_\Sigma ) \otimes_{\mathbb{Z}} \mathbb{R} = \mathrm{Pic} ( X_\Sigma ) \otimes_{\mathbb{Z}} \mathbb{R}$ generated by all nef divisor classes $D \in \mathrm{Pic} ( X_\Sigma )$.
\end{conseq}

\begin{remark}
Before we formulate our algorithm, we need some preparation. Set $n := \mathrm{Rank} ( N )$, $k = | \Sigma_{\mathrm{max}} |$ and label the maximal cones as $\sigma_1, \dots, \sigma_k$. Then any collection
\[ \mathbf{m} = \left\{ m_\sigma \in N \right\}_{\sigma \in \Sigma_{\mathrm{max}}} = \left( m_1, \dots, m_{k} \right) \in \mathbb{Z}^{n \cdot k} \]
will be referred to as \emph{rare} Cartier data. This terminology is to indicate that the Cartier data of every Cartier divisor is of the form $\mathbf{m}$, but the converse is not quite true. \\
The third step of our algorithm identifies those $\mathbf{m}$ which are associated to Cartier divisors in $X_\Sigma$. This is achieved by the following observation:
\ebox{Let $\rho \in \Sigma ( 1 )$. Then for every $\sigma \in \Sigma_{\mathrm{max}}$ with $u_{\rho} \in \sigma ( 1 )$ the inner product $\langle m_\sigma, u_\rho \rangle$ must yield the same value $-a_\rho$.}
Unfortunately, the general procedure is very hard to describe. To gain some understanding of this step nonetheless, let us discuss the example of $X_\Sigma = \mathbb{P}_{\mathbb{Q}}^2$. The ray generators are
\[ u_1 = \left( 1,0 \right), \qquad u_2 = \left( 0,1 \right), \qquad u_3 = \left( -1,-1 \right) \]
and maximal cones are given as
\[ \sigma_1 = \mathrm{Cone} \left( u_1, u_2 \right), \quad \sigma_2 = \mathrm{Cone} \left( u_1, u_3 \right), \quad \sigma_3 = \mathrm{Cone} \left( u_2, u_3 \right) \, . \]
Thus, $\mathbf{m} = ( m_{\sigma_1}, m_{\sigma_2}, m_{\sigma_3} ) \in \mathbb{Z}^6$ constitute the \emph{rare} Cartier data. The following linear map computes all inner products which we use to deduce the values $-a_{\rho}$:
\begin{align} 
\begin{split}
\varphi \colon \mathbb{Z}^6 \to \mathbb{Z}^6 \; , \; \mathbf{m} \mapsto \left( \begin{array}{cccccc} 1 & 0 \\ & & 1 & 0 \\ 0 & 1 & \\ & & & & 0 & 1 \\ & & -1 & -1 \\ & & & & -1 & -1 \end{array} \right) \cdot \mathbf{m} = \left( \begin{array}{c} \left\langle u_1, m_{\sigma_1} \right\rangle \\ \left\langle u_1, m_{\sigma_2} \right\rangle \\ \left\langle u_2, m_{\sigma_1} \right\rangle \\ \left\langle u_2, m_{\sigma_3} \right\rangle \\ \left\langle u_3, m_{\sigma_2} \right\rangle \\ \left\langle u_3, m_{\sigma_3} \right\rangle \end{array} \right).
\end{split}
\end{align}
The differences between the respective inner products are computed upon multiplication with
\[ d = \left( \begin{array}{cccccc} 1 & -1 & \cdot & \cdot & \cdot & \cdot \\ \cdot & \cdot & 1 & -1 & \cdot & \cdot \\ \cdot & \cdot & \cdot & \cdot & 1 & -1 \end{array} \right) \, . \]
So overall we consider the map
\begin{align}
\begin{split}
\psi \colon \mathbb{Z}^6 \to \mathbb{Z}^3 \; , \; \mathbf{m} \mapsto \left( \begin{array}{cccccc} 1 & \cdot & -1 & \cdot & \cdot & \cdot \\ \cdot & 1 & \cdot & \cdot & \cdot & -1 \\ \cdot & \cdot & -1 & -1 & 1 & 1 \end{array} \right) \cdot \mathbf{m}.
\end{split}
\end{align}
The kernel of $\psi$ defines the \emph{proper} Cartier data as subset of $\mathbb{Z}^6$. Explicitly we have
\[ \mathrm{ker} \left( \psi \right) = \mathrm{Cone} \left( \pm \left( \begin{array}{c} -1 \\ \cdot \\ -1 \\ 1 \\ \cdot \\ \cdot \end{array} \right), \pm \left( \begin{array}{c} 1 \\ \cdot \\ 1 \\ \cdot \\ 1 \\ \cdot \end{array} \right), \pm \left( \begin{array}{c} 1 \\ 1 \\ 1 \\ \cdot \\ \cdot \\ 1 \end{array} \right) \right) \cap \mathbb{Z}^6 \equiv C_2 \cap \mathbb{Z}^6 \, . \]
\end{remark}

\begin{construc}
The following algorithm is implemented in \cite{SheafCohomologyOnToricVarieties}.
\begin{itemize}
 \item Input: \\
      A smooth and complete normal toric variety $X_\Sigma$.
 \item Output: \\
      The Nef-cone $\mathrm{Nef} ( X_\Sigma ) \subseteq \mathrm{Pic} ( X_\Sigma ) \otimes_{\mathbb{Z}} \mathbb{R}$.
 \item Processing:
      \begin{enumerate}
       \item Look at a torus-invariant, irreducible and complete curve $C \subseteq X_\Sigma$. Thus, $C = V \left( \tau \right)$ where $\tau = \sigma^{\prime} \cap 
            \sigma^{\prime \prime}$ with $\sigma^\prime, \sigma^{\prime \prime} \in \Sigma_{\mathrm{max}}$. The algorithm described in \cref{subsec:IntersectionComputationToric} then computes $u_{\sigma^{\prime}\sigma^{\prime \prime}} \in \sigma^{\prime \prime}$. This $u_{\sigma^\prime \sigma^{\prime \prime}}$ is such that for a given Cartier divisor $D$ on $X_\Sigma$ with Cartier data $\left\{ m_\sigma \right\}_{\sigma \in \Sigma \left( n \right)}$, the intersection number $D \cdot C$ matches $\langle m_{\sigma^\prime} - m_{\sigma^{\prime \prime}}, u_{\sigma^{\prime}\sigma^{\prime \prime}} \rangle$. By repeating this step we obtain the collection
            \[ \mathcal{U} := \left\{ \left. u_{\sigma^{\prime} \sigma^{\prime \prime}} \in \sigma^{\prime \prime} \; \right| \; \sigma^{\prime}, \sigma^{\prime \prime} \in \Sigma_{\mathrm{max}} \text{ s.t. } \mathrm{dim} \left( \sigma^\prime \cap \sigma^{\prime \prime} \right) = \mathrm{dim} \left( \sigma^\prime \right) - 1 \right\} \, . \]
       \item Consider \emph{rare} Cartier data $\mathbf{m} = ( m_1, \dots, m_{k} ) \in \mathbb{Z}^{n \cdot k}$ and apply the linear map $\Delta \colon \mathbb{Z}^{n 
            \cdot k} \to \mathbb{Z}^{n \cdot \binom{k}{2}} \; , \; \mathbf{m} \to \Delta \mathbf{m}$ given by
            \[ 
            \resizebox{0.79\textwidth}{!}{$
            \Delta \mathbf{m} = \left( m_{\sigma_1} - m_{\sigma_2}, m_{\sigma_1} - m_{\sigma_3}, \dots, m_{\sigma_1} - m_{\sigma_N}, m_{\sigma_2} - m_{\sigma_3}, \dots, m_{\sigma_{N-1}} - m_{\sigma_N} \right) \, .$}
            \]
            Note that $\Delta \mathbf{ m } \cdot \left( u_{\sigma_1 \sigma_2}, 0, \dots, 0 \right) = \left\langle m_{\sigma_1} - m_{\sigma_2}, u_{\sigma_1 \sigma_2} \right\rangle$. Prepare $\mathcal{U}$ accordingly as
            \[ \tilde{\mathcal{U}} := \left\{ \left. \tilde{u}_{\sigma^{\prime} \sigma^{\prime \prime}} \in \mathbb{Z}^{n \cdot \binom{k}{2}} \; \right| \; \sigma^{\prime}, \sigma^{\prime \prime} \in \Sigma_{\mathrm{max}} \mathrm{ s.t. } \mathrm{dim} \left( \sigma^\prime \cap \sigma^{\prime \prime} \right) = \mathrm{dim} \left( \sigma^\prime \right) - 1 \right\} \]
            and phrase the nef condition as $\left\langle \Delta \mathbf{m}, \tilde{u} \right\rangle \geq 0$ for all $\tilde{u} \in \tilde{\mathcal{U}}$. This leads to consider
            \[ C := \left\{ \left. p \in \mathbb{R}^{n \cdot \binom{k}{2}} \; \right| \; \left\langle p, \tilde{u} \right\rangle \geq 0 \; \; \forall \tilde{u} \in \tilde{\mathcal{U}} \right\} \, . \]
            Its preimage defines a cone $C_1 = \Delta^{-1} \left( C \right) \subseteq \mathbb{R}^{n \cdot k}$ given by
            \[ C_1 = \Delta^{-1} \left( C \right) = \left\{ \left. p \in \mathbb{Z}^{n \cdot k} \; \right| \; \langle p, \tilde{\tilde{u}} \rangle \geq 0 \; \; \forall \tilde{\tilde{u}} \in \tilde{\tilde{\mathcal{U}}} \right\}, \quad \tilde{\tilde{\mathcal{U}}} = \left\{ \left. M_\Delta^T \tilde{u} \; \right| \; \tilde{u} \in \tilde{\mathcal{U}} \right\} \]
            where $M_{\Delta}$ is the mapping matrix of $\Delta$.
       \item Identify proper Cartier data. As exemplified above this leads to a cone $C_2 \subseteq \mathbb{R}^{n \cdot k}$. 
       \item Compute the cone $\hat{C} = C_1 \cap C_2$ and denote its ray generators by $\left\{ w_1, \dots, w_l \right\} \subseteq \mathbb{Z}^{n \cdot k}$.
       \item Now compute the image of $\hat{C}$ in $\mathrm{Cl} ( X_\Sigma )$. Proper Cartier data $\left\{ m_\sigma \right\}_{\sigma \in 
            \Sigma_{\mathrm{max}}}$ corresponds to the divisor $D = \sum{a_\rho D_\rho}$ with $a_\rho = - \left\langle m_\sigma, u_\rho \right\rangle$. In consequence, the matrix
            \[ M = \left( \begin{array}{cccc} - u_1 & & & \\ & - u_2 & & \\ & & \ddots & \\ & & & - u_N \end{array} \right) \]
            provides a linear map $\mathbb{Z}^{n \cdot k} \to \mathrm{Div}_T \left( X_\Sigma \right)$. We compose it with $\mathrm{Div}_T \left( X_\Sigma \right) \to \mathrm{Cl} \left( X_\Sigma \right)$ to obtain $\varphi \colon \mathbb{Z}^{n \cdot k} \to \mathrm{Cl} \left( X_\Sigma \right)$. It holds $\mathrm{Nef} ( X_\Sigma ) = \varphi ( \hat{C} ) = \mathrm{Cone} ( \varphi ( w_1 ), \dots, \varphi ( w_l ) )$.
      \end{enumerate}      
\end{itemize}
\end{construc}

\begin{examp}
The following lines compute $\mathrm{Nef} ( \mathbb{P}^1 ) $ and $\mathrm{Nef} ( \mathbb{P}^1 \times \mathbb{P}^1 )$.
\begin{gapConsole}
!gapprompt@gap>& !gapinput@LoadPackage( "ToricVarieties" );&
true
!gapprompt@gap>& !gapinput@P1 := ProjectiveSpace( 1 );&
<A projective toric variety of dimension 1>
!gapprompt@gap>& !gapinput@NefCone( P1 );&
[ [ 1 ] ]
!gapprompt@gap>& !gapinput@P1xP1 := P1*P1;&
<A projective smooth toric variety of dimension 2 which is a 
                                                     product of 2 toric varieties>
!gapprompt@gap>& !gapinput@NefCone( P1xP1 );&
[ [ 0, 1 ], [ 1, 0 ] ]
\end{gapConsole}
\end{examp}

\subsection{Ample Divisors} \label{subsec:AmpleDivsiors}

\begin{theorem}[Theorem 6.3.13 of \cite{cox2011toric} -- Toric Kleiman Criterion]
Let $D$ be a Cartier divisor on a complete toric variety $X_\Sigma$. Then $D$ is ample if an only if $D \cdot C > 0$ for all torus-invariant, irreducible curves $C \subseteq X_\Sigma$.
\end{theorem}

\begin{conseq}
Given a normal toric variety $X_\Sigma$ which is smooth and complete, we can thus identify a small ample divisor $u \in \mathrm{Cl} ( X_\Sigma )$ as follows:
\begin{enumerate}
 \item Compute $\mathrm{Nef} ( X_\Sigma )$ as explained in \cref{subsec:NefCone}.
 \item Use the toric Kleiman theorem as stated above, to relate the integral internal points of $\mathrm{Nef} ( X_\Sigma )$ to the ample divisors of $X_\Sigma$. 
 \item Choose as small as possible an integral and internal point of $\mathrm{Nef} ( X_\Sigma )$ to identify a small ample divisor $u \in \mathrm{Cl} ( X_\Sigma )$. Note 
      that in general this choice is far from unique. To date, the implementations in \cite{SheafCohomologyOnToricVarieties} simply pick one such divisor.
\end{enumerate}
\end{conseq}

\section{Algorithmic Approach to Sheaf Cohomology} \label{sec:ComputeCohomologies}

\subsection{Ideals as special Models for the Structure Sheaf} \label{subsec:Ideals}

Given a \fp graded $S$-module $N$, the sheaf cohomologies of $\tilde{N}$ satisfy
\[ H^i ( X_\Sigma, \tilde{N} ) \cong \mathrm{Ext}_{\mathcal{O}_{X_\Sigma}}^i ( \mathcal{O}_{X_\Sigma}, \tilde{N} ) \, . \]
Hence, if we find a \fp graded $S$-module $M$ with $\tilde{M} \cong \mathcal{O}_{X_\Sigma}$ and such that all conditions in \cref{mytheorem} are satisfied, then we can compute these sheaf cohomologies via
\[ H^i ( X_\Sigma, \tilde{N} ) \cong \mathrm{Ext}_{\mathcal{O}_{X_\Sigma}}^i ( \mathcal{O}_{X_\Sigma}, \tilde{N} ) \cong \mathrm{Ext}_{S}^i ( M, N ) \, . \]
$\tilde{B_\Sigma} \cong \mathcal{O}_{X_\Sigma}$ on a smooth toric variety \cite[proposition 5.3.10]{cox2011toric}, but there are more such ideals.

\begin{defi}
Let $X_\Sigma$ be a normal toric variety which is smooth and complete. Consider $u \in \mathrm{Cl} ( X_\Sigma )$ and let $\{ m_1, \dots, m_k \}$ denote all monomials of degree $u$ in the Cox ring $S$ of $X_\Sigma$. For $e \in \mathbb{N}_{\geq 0}$ we then define the ideal $I \left( u, e \right) = \left\langle m_1^e, \dots, m_k^e \right\rangle \subseteq S$.
\end{defi}

\begin{cor}
Let $X_\Sigma$ be a smooth, complete normal toric variety, $u \in \text{Nef} ( X_\Sigma )$ and $e \geq 0$. Then $\tilde{I ( u , e )} \cong \mathcal{O}_{X_\Sigma}$.
\end{cor}

\begin{myproof}
Note that on a complete toric variety, Nef divisors and basepoint-free divisors coincide \cite[theorem 6.3.12]{cox2011toric}. Hence, for $D \in \text{Nef} ( X_\Sigma )$, the global sections of $\mathcal{O}_{X_\Sigma} ( D )$ are furnished from the $\mathbb{Q}$-span of all monomials in the Cox ring of degree $D$. Let us denote these by $\{ m_1, \dots, m_k \}$.  Then, since $\mathcal{O}_{X_\Sigma} ( D )$ is basepoint-free, $\emptyset = V( m_1, \dots, m_k ) \subseteq X_\Sigma$. As the ideal $I ( u, e )$ encodes the ideal sheaf associated to the empty locus $V( m_1^e, \dots, m_k^e )$, it follows $\tilde{I (u, e)} \cong \mathcal{O}_{X_\Sigma}$.
\end{myproof}

To apply \cref{mytheorem}, also the conditions of this theorem need to be satisfied. This can be achieved by focusing on special divisors $D \in \text{Nef} ( X_\Sigma )$, namely the ample divisors.

\begin{cor} \label{cor:MinimalE}
Let $X_\Sigma$ be a normal toric variety which is smooth, complete, consider a \fp graded $S$-module $N$ and an ample $u \in \mathrm{Cl} ( X_\Sigma )$. Then there exists $\check{e} \in \mathbb{N}_{\geq 0}$ such that the pair $( I( u, e ), N )$ satisfies the conditions of \cref{mytheorem} for all $e \geq \check{e}$.
\end{cor}

\begin{figure}
\centering
\includegraphics[valign = c]{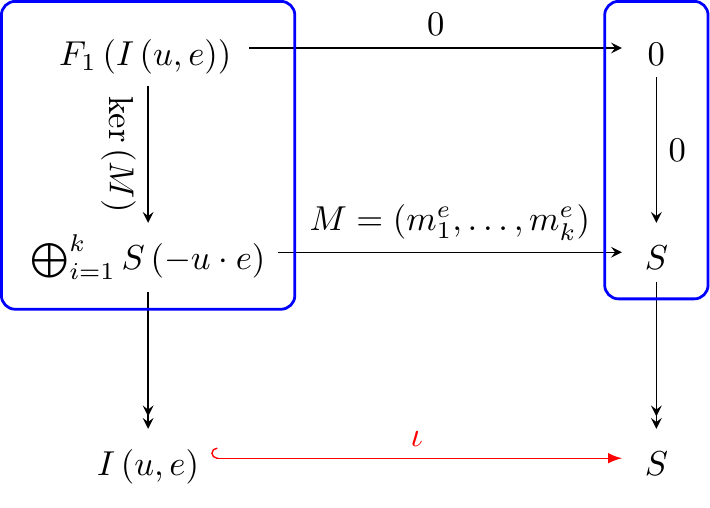}
\caption{The defining data of the ideal $I( u, e )$.}
\label{fig:DefinitionOfIue}
\end{figure}

\begin{myproof}
The ideal $I ( e, u )$ is described by the commutative diagram in \cref{fig:DefinitionOfIue}.\footnote{See \cref{chapter:DetailsOnFPGradedSModules} for more background on this notation.} This shows that a minimal free resolution of $I \left( u, e \right)$ looks like
\[
\resizebox{0.9\textwidth}{!}{$\displaystyle
0 \leftarrow \bigoplus_{i = 1}^{k}{S \left( - u \cdot e \right)} \xleftarrow{\mathrm{ker} \left( M \right)} F_1 \left( I \left( u, e \right) \right) \leftarrow \dots, \qquad \tilde{F_i \left( I \left( u, e \right) \right)} = \bigoplus_{k \in I_i}{\mathcal{O}_{X_\Sigma} \left( - u \cdot e - \Delta_k \right) }$}
\]
for suitable indexing sets $I_i$ and $\Delta_k \in \mathrm{Cl} ( X_\Sigma )$. For simplicity, let us assume $L ( I ( u, e ) ) \geq m - 1$. Now note
\[ \tilde{F_i^\vee} \left( I \left( u, e \right) \right) \otimes \tilde{N} = \bigoplus_{k \in I_i}{\mathcal{O}_{X_\Sigma} \left( u \cdot e + \Delta_k \right)} \otimes \tilde{N} \cong \mathcal{O}_{X_\Sigma} \left( u \right)^{\otimes e} \otimes \bigoplus_{k \in I_i}{\mathcal{O}_{X_\Sigma} \left( \Delta_k \right)} \otimes \tilde{N} \, . \]
Since $\mathcal{O}_{X_\Sigma} ( u )$ is an ample line bundle, there exists $\check{e} \in \mathbb{N}_{\geq 0}$ such that for all $e \geq \check{e}$ it holds
\[ h^k \left( X_\Sigma, \tilde{F_i^\vee} \left( I \left( u, e \right) \right) \otimes \tilde{N} \right) = 0 \qquad \mathrm{for } k \in \{ 1, 2, \dots, \mathrm{dim} ( X_\Sigma ) \} \, . \label{equ:VanishingCohomologyClasses} \]
\Cref{cor:localCohomologyHi} now implies vanishing local cohomologies $H^0_{B_\Sigma}$ and $H^1_{B_\Sigma}$ for sufficiently large $e$. In addition, the conditions of \cref{cor:step1} are
\begin{align}
\begin{split}
h^m \left( X_\Sigma, \tilde{F_0^\vee} \left( I \left( u, e \right) \right) \otimes \tilde{N} \right) &= h^{m-1} \left( X_\Sigma, \tilde{F_0^\vee} \left( I \left( u, e \right) \right) \otimes \tilde{N} \right) = 0, \\
h^{m-1} \left( X_\Sigma, \tilde{F_1^\vee} \left( I \left( u, e \right) \right) \otimes \tilde{N} \right) &= h^{m-2} \left( X_\Sigma, \tilde{F_1^\vee} \left( I \left( u, e \right) \right) \otimes \tilde{N} \right) = 0, \\
                                                                    & \vdots \\
h^2 \left( X_\Sigma, \tilde{F_{m-2}^\vee} \left( I \left( u, e \right) \right) \otimes \tilde{N} \right) &= h^{1} \left( X_\Sigma, \tilde{F_{m-2}^\vee} \left( I \left( u, e \right) \right) \otimes \tilde{N} \right) = 0, \\
h^1 \left( X_\Sigma, \tilde{F_{m-1}^\vee} \left( I \left( u, e \right) \right) \otimes \tilde{N} \right) &= 0.
\end{split}
\end{align}
These are satisfied for sufficiently large $e$ as a consequence of \cref{equ:VanishingCohomologyClasses}. So there exists $\check{e}$ such that for $e \geq \check{e}$ the conditions of \cref{mytheorem} are met for $( I( u, e ), N )$. This completes the proof.
\end{myproof}

\subsection{Algorithm for Computation of Sheaf Cohomologies}

 \begin{itemize}
 \item Input:
      \begin{itemize}
       \item A normal toric variety $X_\Sigma$ which is smooth and complete.
       \item An integer $i \in \{ 0, 1, \dots, \mathrm{dim} ( X_\Sigma ) \}$.
       \item A \fp graded $S$-module $N$.
      \end{itemize}
 \item Output: \\
      A $\mathbb{Q}$-vector space presentation of $H^i ( X_\Sigma, \tilde{N} )$.
 \item Processing:
      \begin{enumerate}
        \item Apply \emph{cohomCalg} to compute the vanishing sets $V^i ( X_\Sigma )$.
        \item Pick small ample $u \in \mathrm{Cl} ( X_\Sigma )$ as explained in \cref{subsec:AmpleDivsiors}.
        \item Find smallest $\check{e}$ such that the pair $( I( u, \check{e} ), N )$ satisfies the conditions of \cref{mytheorem}.
        \item Compute $\mathrm{Ext}_S^i \left( I ( \check{e} ), N \right)_0$.
      \end{enumerate}
\end{itemize}

\section{Computing Extension Modules in \texorpdfstring{$\mathbf{S\text{\textnormal{-}fpgrmod}}$}{S-fpgrmod}} \label{sec:ExtOfFPModules}

We have just convinced ourselves that the computation of sheaf cohomologies on smooth, complete toric varieties $X_\Sigma$ boils down to the computation of truncations of extension modules $\mathrm{Ext}^i_S ( M, N )_0$ for \fp graded $S$-modules $M, N$. Let us therefore explain in this section how we compute these modules.\footnote{Recall that for our setups it is sufficient to assume that the Cox ring $S = \mathbb{Q} \left[ x_0, \dots, x_n \right]$ is a $\mathbb{Z}^n$-graded ring.} For sake of simplicity we start by investigating $\mathrm{Hom}_S ( M, N)_0$.

First recall that $\mathrm{Hom}_S \left( M, N \right)$ is the module formed from all morphisms $\varphi \colon M \to N$ of \fp graded $S$-modules. To gain some intuition for this module and its truncations, we first consider the category of \emph{projective graded $S$-modules}. So pick two projective graded $S$-modules $M$, $N$ given by
\[ M \cong \bigoplus_{i \in I}{S( i )}, \qquad N \cong \bigoplus_{j \in J}{S( j )} \, . \]
Then the module of all homomorphism $\varphi \colon M \to N$ of projective graded $S$-modules is given by
\[ \mathrm{Hom}_S \left( M, N \right) \cong \bigoplus_{i \in I, j \in J}{S \left( -i + j \right)} \, . \]
So indeed $\mathrm{Hom}_S ( M, N )$ is a projective $\mathbb{Z}^n$-graded $S$-module. Therefore, $\mathrm{Hom}_S ( M, N )$ can be truncated to any $d \in \mathbb{Z}^n$. The dual module of the projective graded $S$-module $M = \bigoplus_{i \in I}{S \left( i \right)}$ is defined as
\[ M^\vee := \mathrm{Hom}_S \left( M, S \right) \cong \bigoplus_{i \in I}{S \left( - i \right)} \, . \]
Intuitively, we write $\mathrm{Hom}_S \left( M, N \right) \cong M^\vee \otimes_S N$.

Let us now turn back to $S\mathrm{\textnormal{-}fpgrmod}$ in order to generalise the above procedure to compute $\mathrm{Hom}_S ( M, N )_0$ for (proper) \fp graded $S$-modules $M$ and $N$. We assume that $M$, $N$ are presented as $\rho_M \colon R_M \to G_M$ and $\rho_N \colon R_N \to G_N$. Then $\mathrm{Hom}_S ( M, N )$ is the kernel object of the following morphism of \fp graded $S$-modules:
\[ \label{equ:HomEmbedding} \includegraphics[valign = c]{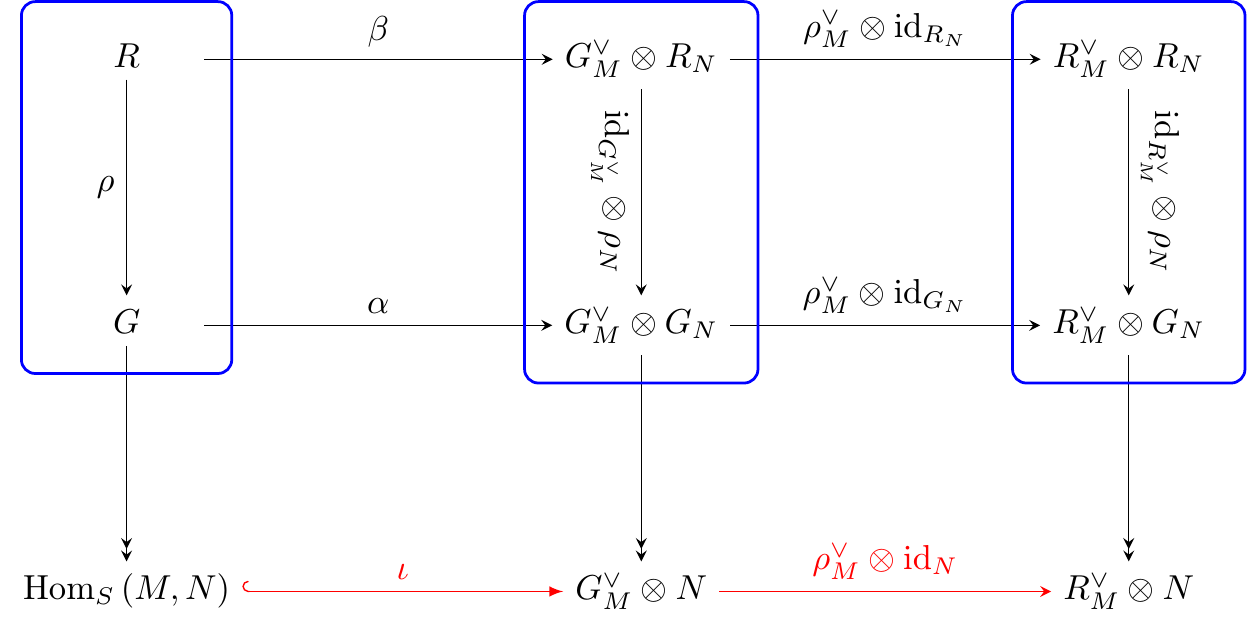} \]
The monomorphism $\iota$ is termed the \emph{Hom-embedding}.

One way to compute $\mathrm{Hom}_S ( M, N )_0$ is therefore to first identify $\mathrm{Hom}_S ( M, N )$ as the domain of the Hom-embedding $\iota$ and then truncate it to degree $0$. This process proceeds in the following 3 steps:
\begin{enumerate}
 \item First compute the following pullback diagram in the category of projective graded $S$-modules:
      \[ \label{equ:pullback1} \includegraphics[valign = c]{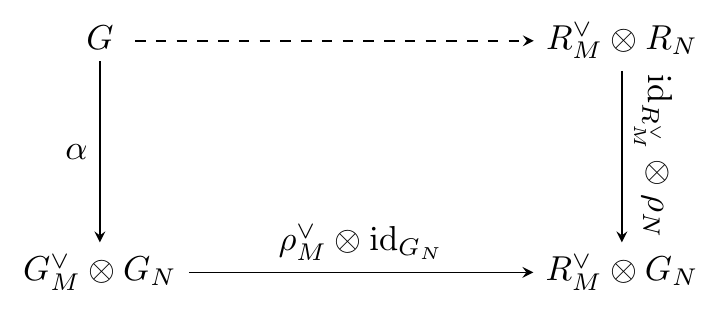} \]
      Hence, $G$ is the pullback object of the morphisms $( \rho_M^\vee \otimes \mathrm{id}_{G_N}, \mathrm{id}_{R_M^\vee} \otimes \rho_N )$ and $\alpha$ is its canonical projection onto $G_M^\vee \otimes G_N$.\footnote{$G$ and $\alpha$ are unique up to isomorphism by general category theory. The identification of $\alpha$ and $G$ in the computer involves Gr\"obner basis algorithms to be applied to the matrices representing the morphisms $\rho_M^\vee \otimes \mathrm{id}_{G_N}$ and $\mathrm{id}_{R_M^\vee} \otimes \rho_N$.}
 \item Next compute the following pullback diagram in the category of projective graded $S$-modules:
      \[ \label{equ:pullback2} \includegraphics[valign = c]{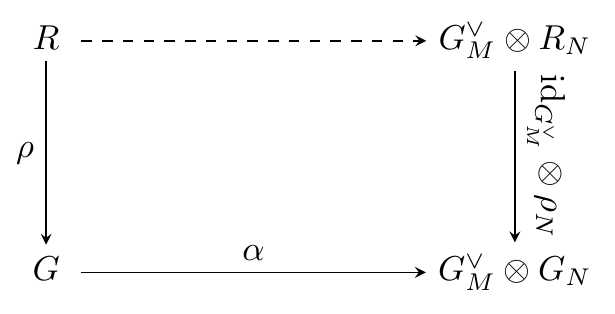} \]
      So $R$ is the pullback object of $( \alpha, \mathrm{id}_{G_M^\vee} \otimes \rho_N )$ and $\rho$ is its projection onto $G$.\footnote{The computation of $R$ and $\rho$ in the computer involves Gr\"obner basis algorithms.} \\
      Let us mention that the computation of kernel embeddings in the category $S \mathrm{\textnormal{-}fpgrmod}$ always proceeds along these first two steps. The interested reader might find it illustrative to use this knowledge to check that all kernel embeddings presented in \cref{subsec:FPGradedSModules} are indeed obtained along this strategy.
 \item Finally, truncate the morphism $R \xrightarrow{\rho} G$ of projective graded $S$-modules to degree $0$. Thereby, we obtain a morphism of finite-dimensional 
      vector spaces over $\mathbb{Q}$. The so-obtained object
      \[ \label{equ:presentationOfHomOfIAndMTruncatedToDegreeZero} \includegraphics[valign = c]{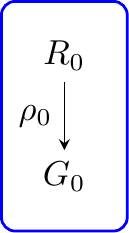} \]
      is an \fp $\mathbb{Q}$-vector space with the property $\mathrm{Hom}_S ( M, N )_0 \cong \mathrm{coker} ( \rho_0 )$. Thus, it is now a simple matter to determine the $\mathbb{Q}$-dimension of $\mathrm{Hom}_S ( M, N )_0$ from $\mathrm{rk} ( \rho_0 )$.
\end{enumerate}

This algorithm requires a lot of computational resources and time, due to the use of Gr\"obner basis algorithms for the computation of the pullback diagrams \cref{equ:pullback1} and \cref{equ:pullback2} in the first two steps. To overcome this shortcoming we have developed an alternative algorithm. This algorithm applies the truncation functor $\mathrm{tr} \colon S \mathrm{\textnormal{-}fpgrmod} \to \mathbb{Q} \mathrm{\textnormal{-}}fpvec$ to \cref{equ:HomEmbedding} directly. Thereby, we obtain the following diagram:
\[ \label{equ:HomEmbeddingTruncated} \includegraphics[valign = c]{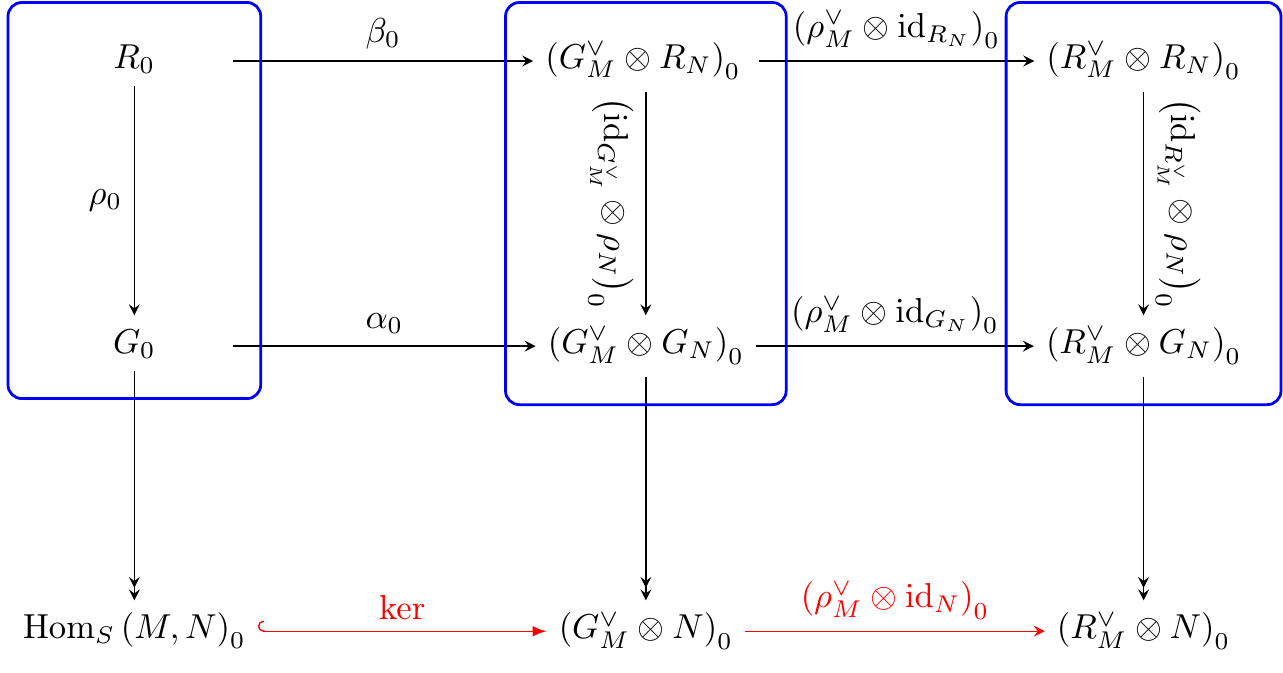} \]
This is a diagram in the category $\mathbb{Q} \mathrm{\textnormal{-}}fpvec$ of \fp $\mathbb{Q}$-vector spaces -- the black arrows are morphisms of $\mathbb{Q}$-vector spaces and the red ones mediate between \fp $\mathbb{Q}$-vector spaces. The crucial observation is that $\mathrm{tr} \colon S \mathrm{\textnormal{-}fpgrmod} \to \mathbb{Q} \mathrm{\textnormal{-}}fpvec$ is an exact functor. Therefore, $\mathrm{Hom}_S ( M, N )_0$ is the kernel object of the map $( \rho_M^\vee \otimes \mathrm{id}_N )_0$ of \fp $\mathbb{Q}$-vector spaces. This now comes with two advantages:
\begin{itemize}
 \item For once, the application of the truncation to \cref{equ:HomEmbedding} is easily prepared for parallel computing by applying the functor $\mathrm{tr} \colon S 
      \mathrm{\textnormal{-}fpgrmod} \to \mathbb{Q} \mathrm{\textnormal{-}}fpvec$ to the morphisms $\mathrm{id}_{G_M^\vee} \otimes \rho_N$, $\mathrm{id}_{R_M^\vee} \otimes \rho_N$ and $\rho_M^\vee \otimes \mathrm{id}_{G_N}$ in parallel.
 \item Subsequently, we compute the following two pullback diagrams:
      \begin{center}
      \includegraphics{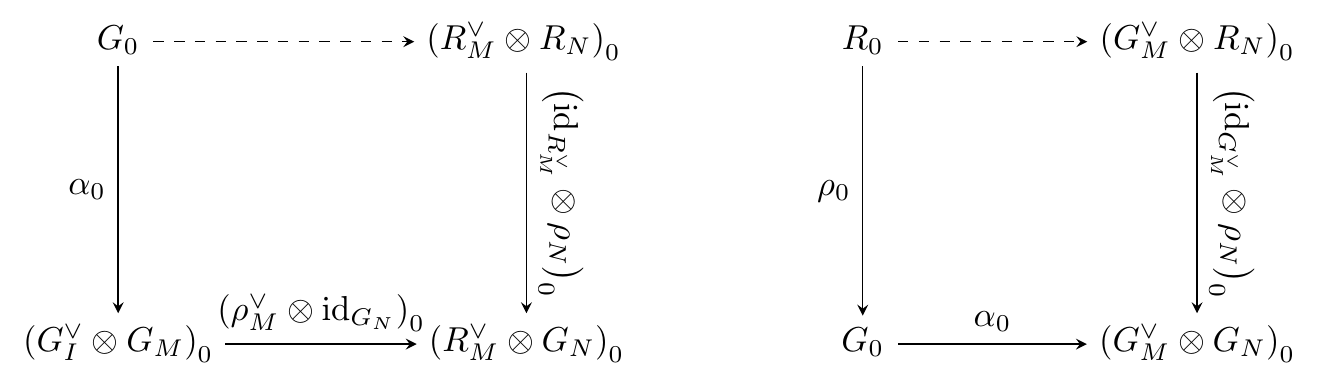}
      \end{center}
      Crucially though, this time the involved matrices are $\mathbb{Q}$-valued. Consequently, the pullbacks can be computed from Gauss eliminations. For the latter there exist state-of-the-art computer implementations \eg in \texttt{MAGMA} \cite{MR1484478}, which are prepared for parallel computing. Our experimental evidence indicates that the computations with Gau{\ss} eliminations in general outperform the corresponding approach with Gröbner basis computations.
\end{itemize}
For these two reasons, this alternative algorithm allows us to identify the object \cref{equ:presentationOfHomOfIAndMTruncatedToDegreeZero} faster. In consequence, we can also identify $\mathrm{dim}_{\mathbb{Q}} ( \mathrm{coker} ( \rho_0 ) )$ more efficiently along these lines.

The examples in the main text are indeed computed by applying $\mathrm{tr} \colon S \mathrm{\textnormal{-}fpgrmod} \to \mathbb{Q} \mathrm{\textnormal{-}}fpvec$ to \cref{equ:HomEmbedding} first and use \texttt{MAGMA} \cite{MR1484478} to perform the subsequent Gau{\ss} eliminations for huge matrices over $\mathbb{Q}$ -- they easily reach sizes of more than 10.000 x 10.000. The overall algorithm is provided by the \texttt{gap}-package \cite{SheafCohomologyOnToricVarieties}.

Let us use this opportunity to point out a crucial fact. As we explained in \cref{chapter:DetailsOnFPGradedSModules}, we can understand the category of \fp graded $S$-modules as a derivation of the category of projective graded $S$-modules. One can generalise this procedure in order to associate to a given Abelian category $\mathcal{C}$ its so-called \emph{Freyd category} \cite{Freyd1966}, and an implementation of this mechanism along the lines of \cite{SeppFreydCategory} is provided in the software-package \cite{CAPPresentationCategory}. As we learned in \cref{chapter:DetailsOnFPGradedSModules}, the category of projective graded $S$-modules and its Freyd category are \emph{not} equivalent categories. However, the truncation diagram \cref{equ:HomEmbeddingTruncated} shows that we are using a presentation $\rho_0 \colon R_0 \to G_0$ to describe the $\mathbb{Q}$-vector space $\text{Hom}_S ( M, N )_0$. This however is only meaningful if the Frey category $\mathbb{Q} \mathrm{\textnormal{-}}fpvec$ is equivalent to the category of finite-dimensional $\mathbb{Q}$-vector spaces. Luckily this indeed happens to be the case, as pointed out in \cite{SeppFreydCategory}.

That all said, let us finally turn to the computation of the extension modules $\mathrm{Ext}^i_S ( M, N )$ and their truncations. To this end, we first recall the theoretical foundations of this bivariate functor $\mathrm{Ext}_S^n ( -, - )$. To this end, let us pick a \fp graded $S$-module $B$ and use it to define an endofunctor $G_B$ of $S\mathrm{\textnormal{-}fpgrmod}$ as follows:
\begin{itemize}
 \item $A \mapsto \mathrm{Hom}_S \left( A,B \right)$
 \item $\left[ \varphi \colon A_1 \to A_2 \right] \mapsto \tilde{\varphi} \equiv \left[ \mathrm{Hom}_S \left( A_2 , B \right) \to \mathrm{Hom}_S \left( A_1, B \right) \; , \; \alpha \mapsto \alpha 
      \circ \varphi \right]$
\end{itemize}
It can then be verified that $G_B$ is a contravariant left-exact endofunctor of $S\mathrm{\textnormal{-}fpgrmod}$. Now the abstract mathematical definition of 
$\mathrm{Ext}_S^n ( A,B )$ is to say
\[ \mathrm{Ext}_S^n \left( A,B \right) := \left( R^n G_B \right) \left( A \right) \, , \]
which means that $\mathrm{Ext}_S^n ( - , - )$ is the n-th right-derived functor of the functor $G_B$. This abstract statement can be made far more explicit. Namely, to compute $\mathrm{Ext}_S^n ( A , B )$ ($n \geq 0$) for two \fp graded $S$-modules $A,B$ we perform the following steps:
\begin{enumerate}
 \item Consider a minimal free resolution $F^\bullet ( A )$ of $A$ by projective objects in $S\mathrm{\textnormal{-}fpgrmod}$. Use this opportunity to recall 
      that the projective objects in $S\mathrm{\textnormal{-}fpgrmod}$ are the projective graded $S$-modules of finite rank. In addition, we state it as a fact that such a resolution exists for every $A \in S\mathrm{\textnormal{-}fpgrmod}$. The resolution $F^\bullet ( A )$ will be denoted as follows:
      \[ \dots \xrightarrow{\alpha_4} P^3 \xrightarrow{\alpha_3} P^2 \xrightarrow{\alpha_2} P^1 \xrightarrow{\alpha_1} P^0 \to 0 \, . \label{equ:MinimalFreeResolution}\]
 \item Now apply the functor $G_B$ to \cref{equ:MinimalFreeResolution}. Recall that this functor is a contravariant left-exact endofunctor of 
      $S\mathrm{\textnormal{-}fpgrmod}$. Hence, we obtain the following complex in $S\mathrm{\textnormal{-}fpgrmod}$:
      \[ \dots \xleftarrow{\tilde{\alpha_3}} \mathrm{Hom}_S \left( P^2, B \right) \xleftarrow{\tilde{\alpha_2}} \mathrm{Hom}_S \left( P^1, B \right) \xleftarrow{\tilde{\alpha_1}} \mathrm{Hom}_S \left( P^0, B \right) \leftarrow 0 \, . \label{equ:complex} \]
 \item Finally, recall that $S\mathrm{\textnormal{-}fpgrmod}$ is an Abelian category. Thus, we can compute the homology of this complex at position $n$ and the 
      result will be a \fp graded $S$-module. For $n > 0$ we say that $( R^n G_B ) ( A )$ is the homology of this complex \cref{equ:complex} at position $n$. For $n = 0$ we set
     \[ \left( R^0 G_B \right) \left( A \right) = G_B \left( A \right) = \mathrm{Hom}_S \left( A,B \right) \, . \]
\end{enumerate}

\begin{figure}[tbp]
\centering
\includegraphics[width=0.9\textwidth]{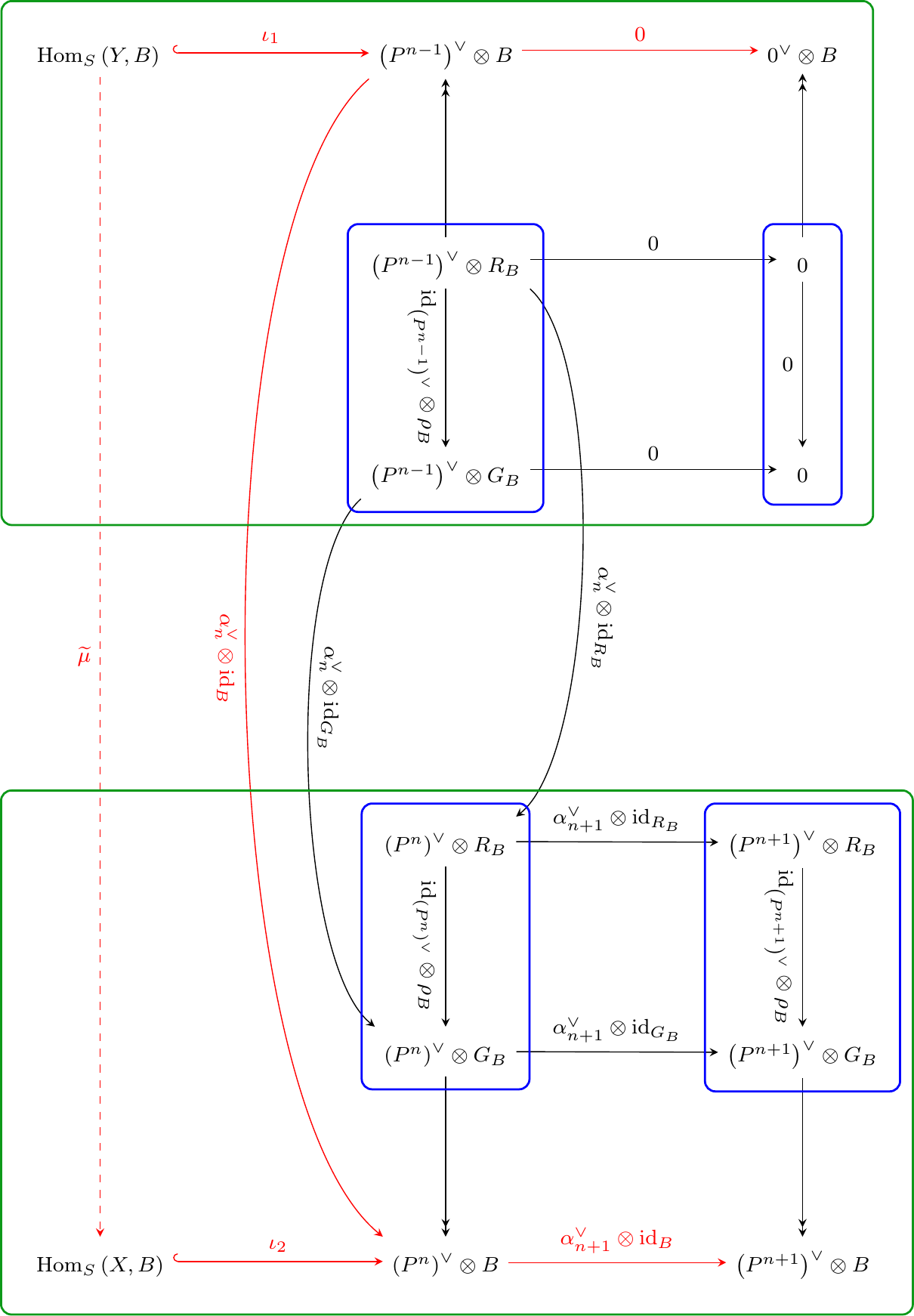}
\caption[A morphism $\mu \colon X \to Y$ induces a map $\tilde{\mu} \colon \mathrm{Hom}_S \left( Y, B \right) \to \mathrm{Hom}_S \left( X, B \right)$.]{For given $\mu \colon X \to Y$ we can compute $\tilde{\mu} \colon \mathrm{Hom}_S \left( Y, B \right) \to \mathrm{Hom}_S \left( X, B \right)$.}
\label{DiagramForComputationOfGradedHomOnMorphisms}
\end{figure}

That all said, we are finally in the position to explain how we compute $\mathrm{Ext}_S^n ( A,B )$ in the package \cite{SheafCohomologyOnToricVarieties}. The computation of $\mathrm{Ext}_S^0 ( A,B )$ of course makes use of the fact
\[ \mathrm{Ext}_S^0 \left( A,B \right) = \mathrm{Hom}_S \left( A, B \right) \]
so that we simply have to compute $\mathrm{Hom}_S ( A, B )$ as outlined above. The interesting case is therefore $\mathrm{Ext}_S^n ( A,B )$ with $n > 0$. In this case we compute a minimal free resolution of $A$. It takes the following form:
\[ \includegraphics[valign = c]{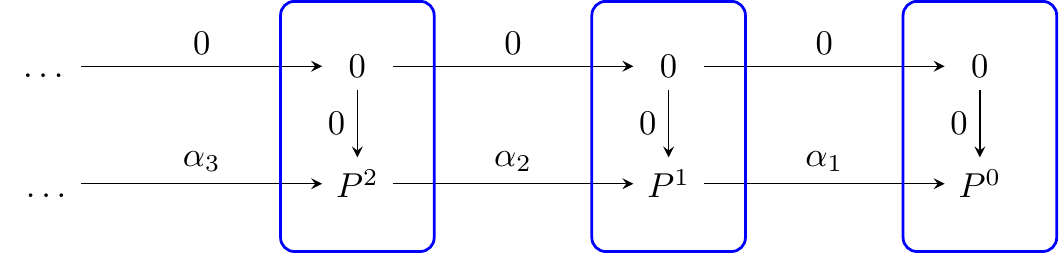} \]
Recall that we aim to compute the homology of the complex \cref{equ:complex} at position $n$. Therefore, we need to take into account both $\alpha_n$ and $\alpha_{n+1}$. To this end, we compute the kernel embedding of the cokernel projection of the $n$-th morphism in the above resolution, \ie of the morphism:
\[ \includegraphics[valign = c]{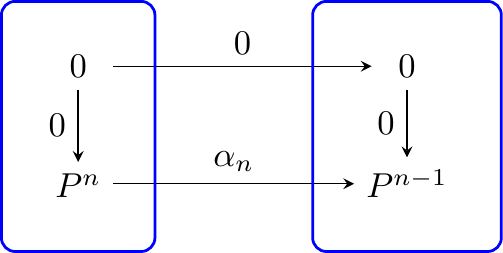} \]
It is readily verified that the so-obtained morphism $\mu \colon X \to Y$ is given by
\[ \includegraphics[valign = c]{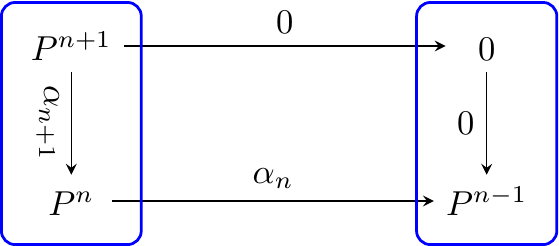} \]
The functor $G_B$ now induces the morphism $\tilde{\mu} \colon \mathrm{Hom}_S \left( Y, B \right) \to \mathrm{Hom}_S \left( X, B \right)$. It turns out that the cokernel object of this morphism is isomorphic to the n-th homology of the complex \cref{equ:complex} at position $n$. So $\mathrm{coker} ( \tilde{\mu} ) \cong \mathrm{Ext}^n_S ( A, B )$. We compute $\tilde{\mu}$ based on the commutative diagram in \cref{DiagramForComputationOfGradedHomOnMorphisms}. The green boxes denote the Hom-embeddings of $\mathrm{Hom}_S ( Y, B )$ and $\mathrm{Hom}_S ( X, B )$ (c.f. \cref{equ:HomEmbedding}). As outlined above we can therefore compute the monomorphisms $\iota_1$ and $\iota_2$. The ranges of these maps are connected by $\alpha_n^\vee \otimes \mathrm{id}_B$. The map $\tilde{\mu}$ is now the lift of the pair of morphisms $\left( \alpha_n^\vee \otimes \mathrm{id}_B \circ \iota_1, \iota_2 \right)$.

The computation of cohomology dimension will involve $\mathrm{Ext}^n_S \left( I, M \right)_0$ for ideals $I \subseteq S$ and `special' modules $M$. Instead of computing $\tilde{\mu}$ along \cref{DiagramForComputationOfGradedHomOnMorphisms}, determining its cokernel module and then truncate the latter to degree $0$, it turns out to be much more efficient computational-wise to truncate the diagram in \cref{DiagramForComputationOfGradedHomOnMorphisms} to degree $0$, compute $\tilde{\mu}_0$ and then determine its cokernel vector space (presentation). This parallels the philosophy employed for $\mathrm{Hom}_S ( A,B )_0$, which allows us to replace Gr\"obner basis computations by Gau{\ss} eliminations. The corresponding algorithms are implemented in the \texttt{gap}-package \cite{SheafCohomologyOnToricVarieties}.

\chapter{Technical Data}

\section{`Quality Series' for Module \texorpdfstring{$\mathbf{M_1}$}{M1}} \label{section:QualitySeriesM1}

\begin{center}
\begin{tabular}{ccc@{\hskip 31pt}ccc@{\hskip 31pt}ccc}
\toprule
$\mathfrak{h}^0( M_1, e )$ & $e$ & $\mathfrak{h}^1( M_1, e )$ & $\mathfrak{h}^0( M_1, e )$ & $e$ & $\mathfrak{h}^1 ( M_1, e )$ & $\mathfrak{h}^0( M_1, e )$ & $e$ & $\mathfrak{h}^1 ( M_1, e )$ \\
\midrule
0 & 0 & 0  & 0 & 17 & 325 & 0 & 34 & 309 \\
0 & 1 & 0  & 0 & 18 & 394 & 0 & 35 & 273 \\
0 & 2 & 0  & 0 & 19 & 467 & 2 & 36 & 239 \\
0 & 3 & 0  & 0 & 20 & 528 & 8 & 37 & 209 \\
0 & 4 & 0  & 0 & 21 & 573 & 15 & 38 & 183 \\
0 & 5 & 0  & 0 & 22 & 600 & 22 & 39 & 160 \\
0 & 6 & 0  & 0 & 23 & 603 & 22 & 40 & 133 \\
0 & 7 & 0  & 0 & 24 & 591 & 22 & 41 & 109 \\
0 & 8 & 3  & 0 & 25 & 576 & 22 & 42 & 88 \\
0 & 9 & 15 & 0 & 26 & 558 & 22 & 43 & 70 \\
0 & 10 & 36 & 0 & 27 & 537 & 22 & 44 & 56 \\
0 & 11 & 62 & 0 & 28 & 513 & 22 & 45 & 46 \\
0 & 12 & 89 & 0 & 29 & 486 & 22 & 46 & 43 \\
0 & 13 & 116 & 0 & 30 & 456 & 22 & 47 & 43 \\
0 & 14 & 150 & 0 & 31 & 423 & 22 & 48 & 43 \\
0 & 15 & 195 & 0 & 32 & 387 & 22 & 49 & 43 \\
0 & 16 & 256 & 0 & 33 & 348 & 22 & 50 & 43 \\
\bottomrule
\end{tabular}
\end{center}

\section{Fluxes In \texorpdfstring{$\mathbf{\mathrm{dP}_3}$}{dP3}-Example} \label{section:ListOfGoodFluxes}

\begin{longtable}{cccc@{\hskip 20pt}cc@{\hskip 20pt}l@{\hskip 20pt}l} 
\toprule
$\Delta_1$ & $\Delta_2$ & $u$ & $v$ & $N_{\mathrm{D3}}$ & $\vect{\chi}$ & $\left( a,b,c,\lambda \right)$ & SUSY? \\
\midrule
\endhead
\endfoot
-10 & 42 & -28 & -2 & 56 &$\left(-7,-45,52 \right)$& $\left(-5,-10,-\frac{5}{2},0\right)$ & {False} \\
-10 & 42 & -28 & -1 & 56 &$\left(-7,-45,52 \right)$& $\left(-10,-5,-\frac{5}{2},0\right)$ & {False} \\[0.65em]

 -10 & 48 & -32 & -3 & 86 &$\left(-7,-51,58 \right)$& $\left(5,-15,-\frac{15}{2},0\right)$ & {False} \\
 -10 & 48 & -32 & 1 & 86 &$\left(-7,-51,58 \right)$& $\left(-15,5,-\frac{15}{2},0\right)$ & {False} \\
 -10 & 48 & -32 & -2 & 296 &$\left(-7,-51,58 \right)$& $\left(0,-10,-\frac{15}{2},0\right)$ & {False} \\
 -10 & 48 & -32 & 0 & 296 &$\left(-7,-51,58 \right)$& $\left(-10,0,-\frac{15}{2},0\right)$ & {False} \\
 -10 & 48 & -32 & -1 & 366 &$\left(-7,-51,58 \right)$& $\left(-5,-5,-\frac{15}{2},0\right)$ & {False} \\[0.65em]

 -10 & 54 & -36 & -3 & 146 &$\left(-7,-57, 64 \right)$& $\left(10,-15,-\frac{25}{2},0\right)$ & {False} \\
 -10 & 54 & -36 & 2 & 146 &$\left(-7,-57, 64 \right)$& $\left(-15,10,-\frac{25}{2},0\right)$ & {False} \\
 -10 & 54 & -36 & -2 & 426 &$\left(-7,-57,64 \right)$& $\left(5,-10,-\frac{25}{2},0\right)$ & {False} \\
 -10 & 54 & -36 & 1 & 426 &$\left(-7,-57,64 \right)$& $\left(-10,5,-\frac{25}{2},0\right)$ & {False} \\
 -10 & 54 & -36 & -1 & 566 &$\left(-7,-57,64 \right)$& $\left(0,-5,-\frac{25}{2},0\right)$ & {False} \\
 -10 & 54 & -36 & 0 & 566 &$\left(-7,-57,64 \right)$& $\left(-5,0,-\frac{25}{2},0\right)$ & {False} \\[0.65em]

 -8 & 30 & -20 & -2 & 56 &$\left(-5,-33,38 \right)$& $\left(0,-10,-\frac{5}{2},0\right)$ & {False} \\
 -8 & 30 & -20 & 0 & 56 &$\left(-5,-33,38 \right)$& $\left(-10,0,-\frac{5}{2},0\right)$ & {False} \\
 -8 & 30 & -20 & -1 & 126 &$\left(-5,-33,38 \right)$& $\left(-5,-5,-\frac{5}{2},0\right)$ & {False} \\[0.65em]

 -8 & 36 & -24 & -2 & 156 &$\left(-5,-39,44\right)$& $\left(5,-10,-\frac{15}{2},0\right)$ & {False} \\
 -8 & 36 & -24 & 1 & 156 &$\left(-5,-39,44\right)$& $\left(-10,5,-\frac{15}{2},0\right)$ & {False} \\
 -8 & 36 & -24 & -1 & 296 &$\left(-5,-39,44\right)$& $\left(0,-5,-\frac{15}{2},0\right)$ & {False} \\
 -8 & 36 & -24 & 0 & 296 &$\left(-5,-39,44\right)$& $\left(-5,0,-\frac{15}{2},0\right)$ & {False} \\[0.65em]

 -8 & 42 & -28 & -2 & 146 &$\left(-5,-45,50\right)$& $\left(10,-10,-\frac{25}{2},0\right)$ & {False} \\
 -8 & 42 & -28 & 2 & 146 &$\left(-5,-45,50\right)$& $\left(-10,10,-\frac{25}{2},0\right)$ & {False} \\
 -8 & 42 & -28 & -1 & 356 &$\left(-5,-45,50\right)$& $\left(5,-5,-\frac{25}{2},0\right)$ & {False} \\
 -8 & 42 & -28 & 1 & 356 &$\left(-5,-45,50\right)$& $\left(-5,5,-\frac{25}{2},0\right)$ & {False} \\
 -8 & 42 & -28 & 0 & 426 &$\left(-5,-45,50\right)$& $\left(0,0,-\frac{25}{2},0\right)$ & {False} \\[0.65em]

 -8 & 48 & -32 & -2 & 26 &$\left(-5,-51,56\right)$& $\left(15,-10,-\frac{35}{2},0\right)$ & {False} \\
 -8 & 48 & -32 & 3 & 26 &$\left(-5,-51,56\right)$& $\left(-10,15,-\frac{35}{2},0\right)$ & {False} \\
 -8 & 48 & -32 & -1 & 306 &$\left(-5,-51,56\right)$& $\left(10,-5,-\frac{35}{2},0\right)$ & {False} \\
 -8 & 48 & -32 & 2 & 306 &$\left(-5,-51,56\right)$& $\left(-5,10,-\frac{35}{2},0\right)$ & {False} \\
 -8 & 48 & -32 & 0 & 446 &$\left(-5,-51,56\right)$& $\left(5,0,-\frac{35}{2},0\right)$ & {False} \\
 -8 & 48 & -32 & 1 & 446 &$\left(-5,-51,56\right)$& $\left(0,5,-\frac{35}{2},0\right)$ & {False} \\[0.65em]

 -8 & 54 & -36 & -1 & 146 &$\left(-5,-57,62\right)$& $\left(15,-5,-\frac{45}{2},0\right)$ & {False} \\
 -8 & 54 & -36 & 3 & 146 &$\left(-5,-57,62\right)$& $\left(-5,15,-\frac{45}{2},0\right)$ & {False} \\
 -8 & 54 & -36 & 0 & 356 &$\left(-5,-57,62\right)$& $\left(10,0,-\frac{45}{2},0\right)$ & {False} \\
 -8 & 54 & -36 & 2 & 356 &$\left(-5,-57,62\right)$& $\left(0,10,-\frac{45}{2},0\right)$ & {False} \\
 -8 & 54 & -36 & 1 & 426 &$\left(-5,-57,62\right)$& $\left(5,5,-\frac{45}{2},0\right)$ & {False} \\[0.65em]

 -6 & 18 & -12 & -1 & 126 &$\left(-3,-21,24\right)$& $\left(0,-5,-\frac{5}{2},0\right)$ & {False} \\
 -6 & 18 & -12 & 0 & 126 &$\left(-3,-21,24\right)$& $\left(-5,0,-\frac{5}{2},0\right)$ & {False} \\[0.65em]

 -6 & 24 & -16 & -1 & 156 &$\left(-3,-27,30\right)$& $\left(5,-5,-\frac{15}{2},0\right)$ & {False} \\
 -6 & 24 & -16 & 1 & 156 &$\left(-3,-27,30\right)$& $\left(-5,5,-\frac{15}{2},0\right)$ & {False} \\
 -6 & 24 & -16 & 0 & 226 &$\left(-3,-27,30\right)$& $\left(0,0,-\frac{15}{2},0\right)$ & {False} \\[0.65em]

 -6 & 30 & -20 & -1 & 76 &$\left(-3,-33,36\right)$& $\left(10,-5,-\frac{25}{2},0\right)$ & {False} \\
 -6 & 30 & -20 & 2 & 76 &$\left(-3,-33,36\right)$& $\left(-5,10,-\frac{25}{2},0\right)$ & {False} \\
 -6 & 30 & -20 & 0 & 216 &$\left(-3,-33,36\right)$& $\left(5,0,-\frac{25}{2},0\right)$ & {False} \\
 -6 & 30 & -20 & 1 & 216 &$\left(-3,-33,36\right)$& $\left(0,5,-\frac{25}{2},0\right)$ & {False} \\[0.65em]

 -6 & 36 & -24 & 0 & 96 &$\left(-3,-39,42\right)$& $\left(10,0,-\frac{35}{2},0\right)$ & {False} \\
 -6 & 36 & -24 & 2 & 96 &$\left(-3,-39,42\right)$& $\left(0,10,-\frac{35}{2},0\right)$ & {False} \\
 -6 & 36 & -24 & 1 & 166 &$\left(-3,-39,42\right)$& $\left(5,5,-\frac{35}{2},0\right)$ & {False} \\[0.65em]

 -6 & 42 & -28 & 1 & 6 &$\left(-3,-45,48\right)$& $\left(10,5,-\frac{45}{2},0\right)$ & {False} \\
 -6 & 42 & -28 & 2 & 6 &$\left(-3,-45,48\right)$& $\left(5,10,-\frac{45}{2},0\right)$ & {False} \\[0.65em]
 
 \textcolor{red}{\textbf{-4}} & \textcolor{red}{\textbf{0}} & \textcolor{red}{\textbf{0}} & \textcolor{red}{\textbf{-1}} & \textcolor{red}{\textbf{56}} & \textcolor{red}{$\mathbf{\left(-1,-3,4\right)}$} & \textcolor{red}{$\mathbf{\left(0,-5,\frac{5}{2},0\right)}$} & \textcolor{red}{\textbf{True}} \\
 \textcolor{red}{\textbf{-4}} & \textcolor{red}{\textbf{0}} & \textcolor{red}{\textbf{0}} & \textcolor{red}{\textbf{0}} & \textcolor{red}{\textbf{56}} & \textcolor{red}{$\mathbf{\left(-1,-3,4\right)}$} & \textcolor{red}{$\mathbf{\left(-5,0,\frac{5}{2},0\right)}$} & \textcolor{red}{\textbf{True}} \\[0.65em]
 
 -4 & 6 & -4 & -1 & 56 &$\left(-1,-9,10\right)$& $\left(5,-5,-\frac{5}{2},0\right)$ & {False} \\
 -4 & 6 & -4 & 1 & 56 &$\left(-1,-9,10\right)$& $\left(-5,5,-\frac{5}{2},0\right)$ & {False} \\
 -4 & 6 & -4 & 0 & 126 &$\left(-1,-9,10\right)$& $\left(0,0,-\frac{5}{2},0\right)$ & {False} \\[0.65em]

 -4 & 12 & -8 & 0 & 86 &$\left(-1,-15,16\right)$& $\left(5,0,-\frac{15}{2},0\right)$ & {False} \\
 -4 & 12 & -8 & 1 & 86 &$\left(-1,-15,16\right)$& $\left(0,5,-\frac{15}{2},0\right)$ & {False} \\[0.65em]

 -4 & 18 & -12 & 1 & 6 &$\left(-1,-21,22\right)$& $\left(5,5,-\frac{25}{2},0\right)$ & {False} \\[0.65em]

 -2 & -24 & 16 & -1 & 6 &$\left(1,21,-22\right)$& $\left(-5,-5,\frac{25}{2},0\right)$ & {False} \\[0.65em]

 -2 & -18 & 12 & -1 & 86 &$\left(1,15,-16\right)$& $\left(0,-5,\frac{15}{2},0\right)$ & {False} \\
 -2 & -18 & 12 & 0 & 86 &$\left(1,15,-16\right)$& $\left(-5,0,\frac{15}{2},0\right)$ & {False} \\[0.65em]

 -2 & -12 & 8 & -1 & 56 &$\left(1,9,-10\right)$& $\left(5,-5,\frac{5}{2},0\right)$ & {False} \\
 -2 & -12 & 8 & 1 & 56 &$\left(1,9,-10\right)$& $\left(-5,5,\frac{5}{2},0\right)$ & {False} \\
 -2 & -12 & 8 & 0 & 126 &$\left(1,9,-10\right)$& $\left(0,0,\frac{5}{2},0\right)$ & {False} \\[0.65em]

 \textcolor{red}{\textbf{-2}} & \textcolor{red}{\textbf{-6}} & \textcolor{red}{\textbf{4}} & \textcolor{red}{\textbf{0}} & \textcolor{red}{\textbf{56}} & \textcolor{red}{$\mathbf{\left(1,3,-4\right)}$} & \textcolor{red}{$\mathbf{\left(5,0,-\frac{5}{2},0\right)}$} & \textcolor{red}{\textbf{True}} \\ 
 \textcolor{red}{\textbf{-2}} & \textcolor{red}{\textbf{-6}} & \textcolor{red}{\textbf{4}} & \textcolor{red}{\textbf{1}} & \textcolor{red}{\textbf{56}} & \textcolor{red}{$\mathbf{\left(1,3,-4\right)}$} & \textcolor{red}{$\mathbf{\left(0,5,-\frac{5}{2},0\right)}$} & \textcolor{red}{\textbf{True}} \\[0.65em]

 0 & -48 & 32 & -2 & 6 &$\left(3,45,-48\right)$& $\left(-5,-10,\frac{45}{2},0\right)$ & {False} \\
 0 & -48 & 32 & -1 & 6 &$\left(3,45,-48\right)$& $\left(-10,-5,\frac{45}{2},0\right)$ & {False} \\[0.65em]

 0 & -42 & 28 & -2 & 96 &$\left(3,39,-42\right)$& $\left(0,-10,\frac{35}{2},0\right)$ & {False} \\
 0 & -42 & 28 & 0 & 96 &$\left(3,39,-42\right)$& $\left(-10,0,\frac{35}{2},0\right)$ & {False} \\
 0 & -42 & 28 & -1 & 166 &$\left(3,39,-42\right)$& $\left(-5,-5,\frac{35}{2},0\right)$ & {False} \\[0.65em]

 0 & -36 & 24 & -2 & 76 &$\left(3,33,-36\right)$& $\left(5,-10,\frac{25}{2},0\right)$ & {False} \\
 0 & -36 & 24 & 1 & 76 &$\left(3,33,-36\right)$& $\left(-10,5,\frac{25}{2},0\right)$ & {False} \\
 0 & -36 & 24 & -1 & 216 &$\left(3,33,-36\right)$& $\left(0,-5,\frac{25}{2},0\right)$ & {False} \\
 0 & -36 & 24 & 0 & 216 &$\left(3,33,-36\right)$& $\left(-5,0,\frac{25}{2},0\right)$ & {False} \\[0.65em]

 0 & -30 & 20 & -1 & 156 &$\left(3,27,-30\right)$& $\left(5,-5,\frac{15}{2},0\right)$ & {False} \\
 0 & -30 & 20 & 1 & 156 &$\left(3,27,-30\right)$& $\left(-5,5,\frac{15}{2},0\right)$ & {False} \\
 0 & -30 & 20 & 0 & 226 &$\left(3,27,-30\right)$& $\left(0,0,\frac{15}{2},0\right)$ & {False} \\[0.65em]

 0 & -24 & 16 & 0 & 126 &$\left(3,21,-24\right)$& $\left(5,0,\frac{5}{2},0\right)$ & {False} \\
 0 & -24 & 16 & 1 & 126 &$\left(3,21,-24\right)$& $\left(0,5,\frac{5}{2},0\right)$ & {False} \\[0.65em]

 2 & -54 & 36 & -3 & 26 &$\left(5,51,-56\right)$& $\left(10,-15,\frac{35}{2},0\right)$ & {False} \\
 2 & -54 & 36 & 2 & 26 &$\left(5,51,-56\right)$& $\left(-15,10,\frac{35}{2},0\right)$ & {False} \\
 2 & -54 & 36 & -2 & 306 &$\left(5,51,-56\right)$& $\left(5,-10,\frac{35}{2},0\right)$ & {False} \\
 2 & -54 & 36 & 1 & 306 &$\left(5,51,-56\right)$& $\left(-10,5,\frac{35}{2},0\right)$ & {False} \\
 2 & -54 & 36 & -1 & 446 &$\left(5,51,-56\right)$& $\left(0,-5,\frac{35}{2},0\right)$ & {False} \\
 2 & -54 & 36 & 0 & 446 &$\left(5,51,-56\right)$& $\left(-5,0,\frac{35}{2},0\right)$ & {False} \\[0.65em]

 2 & -48 & 32 & -2 & 146 &$\left(5,45,-50\right)$& $\left(10,-10,\frac{25}{2},0\right)$ & {False} \\
 2 & -48 & 32 & 2 & 146 &$\left(5,45,-50\right)$& $\left(-10,10,\frac{25}{2},0\right)$ & {False} \\
 2 & -48 & 32 & -1 & 356 &$\left(5,45,-50\right)$& $\left(5,-5,\frac{25}{2},0\right)$ & {False} \\
 2 & -48 & 32 & 1 & 356 &$\left(5,45,-50\right)$& $\left(-5,5,\frac{25}{2},0\right)$ & {False} \\
 2 & -48 & 32 & 0 & 426 &$\left(5,45,-50\right)$& $\left(0,0,\frac{25}{2},0\right)$ & {False} \\[0.65em]

 2 & -42 & 28 & -1 & 156 &$\left(5,39,-44\right)$& $\left(10,-5,\frac{15}{2},0\right)$ & {False} \\
 2 & -42 & 28 & 2 & 156 &$\left(5,39,-44\right)$& $\left(-5,10,\frac{15}{2},0\right)$ & {False} \\
 2 & -42 & 28 & 0 & 296 &$\left(5,39,-44\right)$& $\left(5,0,\frac{15}{2},0\right)$ & {False} \\
 2 & -42 & 28 & 1 & 296 &$\left(5,39,-44\right)$& $\left(0,5,\frac{15}{2},0\right)$ & {False} \\[0.65em]

 2 & -36 & 24 & 0 & 56 &$\left(5,33,-38\right)$& $\left(10,0,\frac{5}{2},0\right)$ & {False} \\
 2 & -36 & 24 & 2 & 56 &$\left(5,33,-38\right)$& $\left(0,10,\frac{5}{2},0\right)$ & {False} \\
 2 & -36 & 24 & 1 & 126 &$\left(5,33,-38\right)$& $\left(5,5,\frac{5}{2},0\right)$ & {False} \\[0.65em]

 4 & -54 & 36 & -1 & 86 &$\left(7,51,-58\right)$& $\left(15,-5,\frac{15}{2},0\right)$ & {False} \\
 4 & -54 & 36 & 3 & 86 &$\left(7,51,-58\right)$& $\left(-5,15,\frac{15}{2},0\right)$ & {False} \\
 4 & -54 & 36 & 0 & 296 &$\left(7,51,-58\right)$& $\left(10,0,\frac{15}{2},0\right)$ & {False} \\
 4 & -54 & 36 & 2 & 296 &$\left(7,51,-58\right)$& $\left(0,10,\frac{15}{2},0\right)$ & {False} \\
 4 & -54 & 36 & 1 & 366 &$\left(7,51,-58\right)$& $\left(5,5,\frac{15}{2},0\right)$ & {False} \\[0.65em]

 4 & -48 & 32 & 1 & 56 &$\left(7,45,-52\right)$& $\left(10,5,\frac{5}{2},0\right)$ & {False} \\
 4 & -48 & 32 & 2 & 56 &$\left(7,45,-52\right)$& $\left(5,10,\frac{5}{2},0\right)$ & {False} \\[0.65em]

 50 & 10 & -7 & -1 & 26 &$\left(53,-13,-40\right)$& $\left(-5,-5,\frac{35}{2},5\right)$ & {False} \\[0.65em]

 50 & 16 & -11 & -1 & 76 &$\left(53,-19,-34\right)$& $\left(0,-5,\frac{25}{2},5\right)$ & {False} \\
 50 & 16 & -11 & 0 & 76 &$\left(53,-19,-34\right)$& $\left(-5,0,\frac{25}{2},5\right)$ & {False} \\[0.65em]
 
 50 & 22 & -15 & -1 & 16 &$\left(53,-25,-28\right)$& $\left(5,-5,\frac{15}{2},5\right)$ & {False} \\
 50 & 22 & -15 & 1 & 16 &$\left(53,-25,-28\right)$& $\left(-5,5,\frac{15}{2},5\right)$ & {False} \\
 50 & 22 & -15 & 0 & 86 &$\left(53,-25,-28\right)$& $\left(0,0,\frac{15}{2},5\right)$ & {False} \\[0.65em]

 52 & -14 & 9 & -3 & 6 &$\left(55,11,-66\right)$& $\left(0,-15,\frac{55}{2},5\right)$ & {False} \\
 52 & -14 & 9 & 0 & 6 &$\left(55,11,-66\right)$& $\left(-15,0,\frac{55}{2},5\right)$ & {False} \\
 52 & -14 & 9 & -2 & 146 &$\left(55,11,-66\right)$& $\left(-5,-10,\frac{55}{2},5\right)$ & {False} \\
 52 & -14 & 9 & -1 & 146 &$\left(55,11,-66\right)$& $\left(-10,-5,\frac{55}{2},5\right)$ & {False} \\[0.65em]

 52 & -8 & 5 & -2 & 206 &$\left(55,5,-60\right)$& $\left(0,-10,\frac{45}{2},5\right)$ & {False} \\
 52 & -8 & 5 & 0 & 206 &$\left(55,5,-60\right)$& $\left(-10,0,\frac{45}{2},5\right)$ & {False} \\
 52 & -8 & 5 & -1 & 276 &$\left(55,5,-60\right)$& $\left(-5,-5,\frac{45}{2},5\right)$ & {False} \\[0.65em]

 52 & -2 & 1 & -2 & 156 &$\left(55,-1,-54\right)$& $\left(5,-10,\frac{35}{2},5\right)$ & {False} \\
 52 & -2 & 1 & 1 & 156 &$\left(55,-1,-54\right)$& $\left(-10,5,\frac{35}{2},5\right)$ & {False} \\
 52 & -2 & 1 & -1 & 296 &$\left(55,-1,-54\right)$& $\left(0,-5,\frac{35}{2},5\right)$ & {False} \\
 52 & -2 & 1 & 0 & 296 &$\left(55,-1,-54\right)$& $\left(-5,0,\frac{35}{2},5\right)$ & {False} \\[0.65em]

 52 & 4 & -3 & -1 & 206 &$\left(55,-7,-48\right)$& $\left(5,-5,\frac{25}{2},5\right)$ & {False} \\
 52 & 4 & -3 & 1 & 206 &$\left(55,-7,-48\right)$& $\left(-5,5,\frac{25}{2},5\right)$ & {False} \\
 52 & 4 & -3 & 0 & 276 &$\left(55,-7,-48\right)$& $\left(0,0,\frac{25}{2},5\right)$ & {False} \\[0.65em]

 52 & 10 & -7 & -1 & 6 &$\left(55,-13,-42\right)$& $\left(10,-5,\frac{15}{2},5\right)$ & {False} \\
 52 & 10 & -7 & 2 & 6 &$\left(55,-13,-42\right)$& $\left(-5,10,\frac{15}{2},5\right)$ & {False} \\
 52 & 10 & -7 & 0 & 146 &$\left(55,-13,-42\right)$& $\left(5,0,\frac{15}{2},5\right)$ & {False} \\
 52 & 10 & -7 & 1 & 146 &$\left(55,-13,-42\right)$& $\left(0,5,\frac{15}{2},5\right)$ & {False} \\[0.65em]

 52 & 16 & -11 & 0 & 206 &$\left(55,-19,-36\right)$& $\left(10,0,\frac{5}{2},5\right)$ & {False} \\
 52 & 16 & -11 & 2 & 206 &$\left(55,-19,-36\right)$& $\left(0,10,\frac{5}{2},5\right)$ & {False} \\
 52 & 16 & -11 & 1 & 276 &$\left(55,-19,-36\right)$& $\left(5,5,\frac{5}{2},5\right)$ & {False} \\[0.65em]

 54 & -44 & 29 & -3 & 136 &$\left(57,41,-98\right)$& $\left(-10,-15,\frac{85}{2},5\right)$ & {False} \\
 54 & -44 & 29 & -2 & 136 &$\left(57,41,-98\right)$& $\left(-15,-10,\frac{85}{2},5\right)$ & {False} \\[0.65em]

 54 & -38 & 25 & -4 & 106 &$\left(57,35,-92\right)$& $\left(0,-20,\frac{75}{2},5\right)$ & {False} \\
 54 & -38 & 25 & 0 & 106 &$\left(57,35,-92\right)$& $\left(-20,0,\frac{75}{2},5\right)$ & {False} \\
 54 & -38 & 25 & -3 & 316 &$\left(57,35,-92\right)$& $\left(-5,-15,\frac{75}{2},5\right)$ & {False} \\
 54 & -38 & 25 & -1 & 316 &$\left(57,35,-92\right)$& $\left(-15,-5,\frac{75}{2},5\right)$ & {False} \\
 54 & -38 & 25 & -2 & 386 &$\left(57,35,-92\right)$& $\left(-10,-10,\frac{75}{2},5\right)$ & {False} \\[0.65em]

 54 & -32 & 21 & -4 & 106 &$\left(57,29,-86\right)$& $\left(5,-20,\frac{65}{2},5\right)$ & {False} \\
 54 & -32 & 21 & 1 & 106 &$\left(57,29,-86\right)$& $\left(-20,5,\frac{65}{2},5\right)$ & {False} \\
 54 & -32 & 21 & -3 & 386 &$\left(57,29,-86\right)$& $\left(0,-15,\frac{65}{2},5\right)$ & {False} \\
 54 & -32 & 21 & 0 & 386 &$\left(57,29,-86\right)$& $\left(-15,0,\frac{65}{2},5\right)$ & {False} \\
 54 & -32 & 21 & -2 & 526 &$\left(57,29,-86\right)$& $\left(-5,-10,\frac{65}{2},5\right)$ & {False} \\
 54 & -32 & 21 & -1 & 526 &$\left(57,29,-86\right)$& $\left(-10,-5,\frac{65}{2},5\right)$ & {False} \\[0.65em]

 54 & -26 & 17 & -3 & 346 &$\left(57,23,-80\right)$& $\left(5,-15,\frac{55}{2},5\right)$ & {False} \\
 54 & -26 & 17 & 1 & 346 &$\left(57,23,-80\right)$& $\left(-15,5,\frac{55}{2},5\right)$ & {False} \\
 54 & -26 & 17 & -2 & 556 &$\left(57,23,-80\right)$& $\left(0,-10,\frac{55}{2},5\right)$ & {False} \\
 54 & -26 & 17 & 0 & 556 &$\left(57,23,-80\right)$& $\left(-10,0,\frac{55}{2},5\right)$ & {False} \\
 54 & -26 & 17 & -1 & 626 &$\left(57,23,-80\right)$& $\left(-5,-5,\frac{55}{2},5\right)$ & {False} \\[0.65em]

 54 & -20 & 13 & -3 & 196 &$\left(57,17,-74\right)$& $\left(10,-15,\frac{45}{2},5\right)$ & {False} \\
 54 & -20 & 13 & 2 & 196 &$\left(57,17,-74\right)$& $\left(-15,10,\frac{45}{2},5\right)$ & {False} \\
 54 & -20 & 13 & -2 & 476 &$\left(57,17,-74\right)$& $\left(5,-10,\frac{45}{2},5\right)$ & {False} \\
 54 & -20 & 13 & 1 & 476 &$\left(57,17,-74\right)$& $\left(-10,5,\frac{45}{2},5\right)$ & {False} \\
 54 & -20 & 13 & -1 & 616 &$\left(57,17,-74\right)$& $\left(0,-5,\frac{45}{2},5\right)$ & {False} \\
 54 & -20 & 13 & 0 & 616 &$\left(57,17,-74\right)$& $\left(-5,0,\frac{45}{2},5\right)$ & {False} \\[0.65em]

 54 & -14 & 9 & -2 & 286 &$\left(57,11,-68\right)$& $\left(10,-10,\frac{35}{2},5\right)$ & {False} \\
 54 & -14 & 9 & 2 & 286 &$\left(57,11,-68\right)$& $\left(-10,10,\frac{35}{2},5\right)$ & {False} \\
 54 & -14 & 9 & -1 & 496 &$\left(57,11,-68\right)$& $\left(5,-5,\frac{35}{2},5\right)$ & {False} \\
 54 & -14 & 9 & 1 & 496 &$\left(57,11,-68\right)$& $\left(-5,5,\frac{35}{2},5\right)$ & {False} \\
 54 & -14 & 9 & 0 & 566 &$\left(57,11,-68\right)$& $\left(0,0,\frac{35}{2},5\right)$ & {False} \\[0.65em]

 54 & -8 & 5 & -1 & 266 &$\left(57,5,-62\right)$& $\left(10,-5,\frac{25}{2},5\right)$ & {False} \\
 54 & -8 & 5 & 2 & 266 &$\left(57,5,-62\right)$& $\left(-5,10,\frac{25}{2},5\right)$ & {False} \\
 54 & -8 & 5 & 0 & 406 &$\left(57,5,-62\right)$& $\left(5,0,\frac{25}{2},5\right)$ & {False} \\
 54 & -8 & 5 & 1 & 406 &$\left(57,5,-62\right)$& $\left(0,5,\frac{25}{2},5\right)$ & {False} \\[0.65em]

 54 & -2 & 1 & 0 & 136 &$\left(57,-1,-56\right)$& $\left(10,0,\frac{15}{2},5\right)$ & {False} \\
 54 & -2 & 1 & 2 & 136 &$\left(57,-1,-56\right)$& $\left(0,10,\frac{15}{2},5\right)$ & {False} \\
 54 & -2 & 1 & 1 & 206 &$\left(57,-1,-56\right)$& $\left(5,5,\frac{15}{2},5\right)$ & {False} \\
\bottomrule
\end{longtable}

\newpage
\bibliographystyle{custom1}
\bibliography{papers}

\end{document}